\documentclass[final,epj,nopacs]{svjour}
\usepackage{graphicx}
\usepackage[english]{babel}
\usepackage{amsmath,amssymb}
\usepackage{wasysym}
\usepackage{appendix}
\usepackage{rotating}
\usepackage{longtable}
\usepackage[verbose]{placeins}
\usepackage[b]{esvect}
\usepackage{bm}
\usepackage{sidecap}
\usepackage{floatrow}
\usepackage{morefloats}

\let\vec\bm 

\title{Determining the dominant partial wave contributions from angular distributions of single- and double-po\-la\-ri\-za\-tion observables in pseudoscalar meson photoproduction}
\author{Y. Wunderlich, F. Afzal, A. Thiel and R. Beck}
\institute{Helmholtz-Institut f\"ur Strahlen- und Kernphysik,\\ Universit\"at Bonn,\\ 53115 Bonn, Germany\\ \email{wunderlich@hiskp.uni-bonn.de, afzal@hiskp.uni-bonn.de, thiel@hiskp.uni-bonn.de, \newline \hspace*{25.5pt} beck@hiskp.uni-bonn.de}}

\date{February 6, 2017}

\abstract{
This work presents a simple method to determine the significant partial wave contributions to experimentally determined observables in pseudoscalar meson photoproduction. First, fits to angular distributions are presented and the maximum orbital angular momentum $\text{L}_{\mathrm{max}}$ needed to achieve a good fit is determined. Then, recent po\-la\-ri\-za\-tion measurements for $\gamma p \rightarrow \pi^{0} p$ from ELSA, GRAAL, JLab and MAMI are investigated according to the proposed method. This method allows us to project high-spin partial wave contributions to any observable as long as the measurement has the necessary statistical accuracy. \newline We show, that high precision and large angular coverage in the polarization data are needed in order to be sensitive to high-spin resonance-states and thereby also for the finding of small resonance contributions. This task can be achieved via interference of these resonances with the well-known states. For the channel $\gamma p \rightarrow \pi^{0} p$, those are the $N(1680)\frac{5}{2}^{+}$ and $\Delta(1950)\frac{7}{2}^{+}$, contributing to the $F$-waves.
}

\begin{document}

\authorrunning{Wunderlich, Afzal, Thiel, and Beck}

\titlerunning{Dominant partial wave contributions from fits to angular distributions}

\maketitle



\section{Introduction} \label{sec:Introduction}

\begin{figure}[htb]
\RawFloats
\centering
\begin{minipage}{0.48\linewidth}
\centering
\hspace*{-9pt}
\includegraphics[width=0.99999\textwidth]{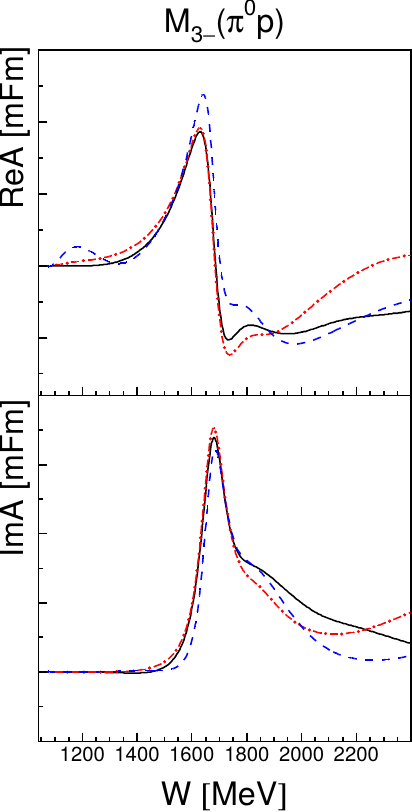}
\end{minipage}%
\begin{minipage}{0.48\linewidth} \centering \hspace*{0pt} \includegraphics[width=0.99999\textwidth]{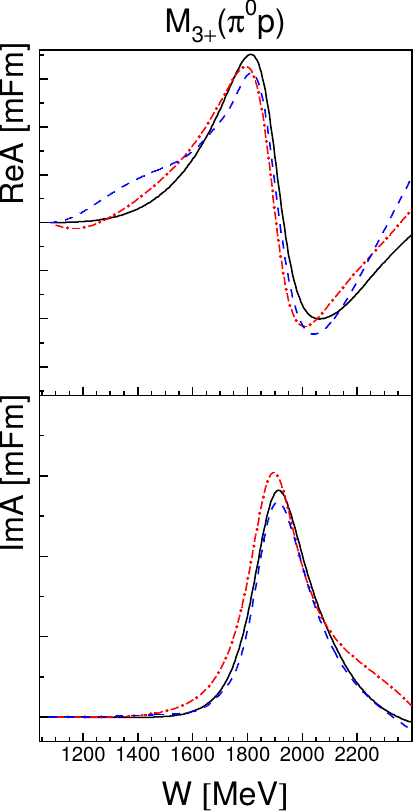} \end{minipage}
\caption{Left: The real- and imaginary part of the $M_{3-}$-multipole for the $\pi^{0} p$ channel, coming from recent energy-dependent fits, are shown \cite{CommonPaper}. Compared are results from BnGa (black solid line), SAID (red dash-dotted line) and J\"uBo (blue dashed line). Right: The same for the $M_{3+}$-multipole. \newline Both multipoles contain contributions from two dominant $F$-wave resonances. The $N(1680)\frac{5}{2}^{+}$ $\left(F_{15} (1680)\right)$ couples to $M_{3-}$, while the $\Delta(1950)\frac{7}{2}^{+}$ $\left(F_{37} (1950)\right)$ couples to $M_{3+}$.}
\label{fig:MultipoleComparisons}
\end{figure}

The study of the nucleon excitation spectrum provides an ideal testing ground for the investigation of the theory describing the strong interactions, QCD. In the region of strong coupling, i.e. for low momentum transfers, QCD can\-not be solved by elementary methods of quantum field theory \cite{Olive:2016xmw}. The binding of quarks into protons and neutrons is a non-perturbative phenomenon. In order to derive predictions for the ground and excited states of nucleons (the $N^{\ast}$ spectrum) that are closely linked to QCD several approaches have been developed.\newline
One of the first methods consists of phenomenological quark models (cf. \cite{CapstickRoberts} and \cite{BONNModel}), solving a bound state equation for a system consisting of three constituent quarks and modeling the strong interactions by a suitable approximation of full QCD. Strong QCD can also be attacked in the numerical ab initio approach of Lattice QCD and there have been first attempts in the direction of predicting the $N^{\ast}$ spectrum \cite{EdwardsEtAl}. \newline
Once these predictions for the $N^{\ast}$ spectrum are compared to resonances extracted from experimental scattering data, the phenomenon known as the so-called missing resonances becomes present \cite{Klempt}. In the high mass region above $1800$ $\mathrm{MeV}$, many more states have been predicted than measured until now. \newline
A lot of the available information on resonant nucleon states has been extracted from data of the  pion nucleon ($\pi N$) elastic scattering process. A persisting hope is that the investigation of alternative processes may yield signals of $N^{\ast}$ states that couple only weakly to the $\pi N$ process. A particular example for such a process is the photoproduction of a single pseudoscalar meson,
\begin{equation}
\gamma N \longrightarrow \mathcal{P} B \mathrm{,} \label{eq:PhotProdProcess}
\end{equation}
which is at the center of attention in this work. Here, $N = (p,n)$ denotes the nucleon, $\mathcal{P}$ is a pseudoscalar meson (e.g. $\pi^{0}$, $\pi^{+}$, $\pi^{-}$, $\eta$, $\eta^{\prime}$, $K^{+}$, $K^{-}$, $\ldots$) and the recoil baryon $B$ can be either a nucleon $N$ or a hyperon $(\Lambda, \Sigma)$. The photoproduction process allows for the extraction of $16$ non-redundant po\-la\-ri\-za\-tion observables at each point in phase space, i.e. center of mass energy and scattering angle $\left(W, \theta\right)$ (cf. Table \ref{tab:Observables} and \cite{Sandorfi:2010uv}). In addition to the unpolarized differential cross section,  the observables group into single-po\-la\-ri\-za\-tion measurements which comprise three additional quantities, as well as $12$ double-po\-la\-ri\-za\-tion observables that may be extracted by beam-target (BT), beam-recoil (BR) and target-recoil (TR) measurements. \newline
%
\begin{table*}[htb]
\RawFloats
\centering
\begin{tabular}{c|c|ccc|ccc|cccc}
\multicolumn{12}{c}{} \\
\hline
\hline
Beam &  & \multicolumn{3}{c|}{Target} & \multicolumn{3}{c|}{Recoil} & \multicolumn{4}{c}{Target + Recoil} \\
  & -& -& -& -& $ x' $ & $ y' $ & $ z' $ & $ x' $ & $ x' $ & $ z' $ & $ z' $  \\
  & -& $ x $ & $ y $ & $ z $ & -& -& -& $ x $ & $ z $ & $ x $ & $ z $ \\
\hline
  &  &  &  &  &  &  &  &  &   \\
unpolarized & $ \sigma_{0} $ &  & $ T $ &  &  & $ P $ &  & $T_{x'}$ & $L_{x'}$ & $T_{z'}$ & $L_{z'}$ \\
  &  &  &  &  &  &  &  &  &   \\
linear & $ \Sigma $ & $ H $ & $ P $ & $ G $ & $ O_{x'} $ & $ T $ & $ O_{z'} $ &  &  &  &  \\
  &  &  &  &  &  &  &  &  &   \\
circular &  & $ F $ &  & $ E $ & $ C_{x'} $ &  & $ C_{z'} $  &  &  &  &  \\
\hline
\hline
\end{tabular}

\caption{The unpolarized differential cross section $\sigma_{0}$ and 15 single- and double-po\-la\-ri\-za\-tion observables accessible in pseudoscalar meson photoproduction are listed. Unprimed coordinates refer to  CMS coordinates with the $\hat{z}$-axis chosen along the incident beam direction and $\hat{y}$-axis perpendicular to the reaction plane. Primed coordinates are rotated from the unprimed ones in such a way that the $\hat{z}^{\prime}$-axis points along the momentum of the pseudoscalar meson in the final state. The triple polarization observables have been omitted here. It is a fact that, for photoproduction of a single pseudoscalar meson, triple polarization measurements do not yield additional information that cannot be attained by measuring single and double polarization observables (see the discussion in ref. \cite{Sandorfi:2010uv}).}
\label{tab:Observables}
\end{table*}
%
The most recent experimental activities center around the facilities ELSA at Bonn 
\cite{Thiel:2012,Gottschall:2014,Hartmann:2014,Hartmann:2015,Mueller:2015}), MAMI at Mainz \cite{McNicoll:2010,Hornidge:2013,Akondi:2014,Adlarson:2015} and JLab at Newport News \cite{Dugger:2013}. With the modern technical developments like polarized beams and targets, the single spin polarization observables as well as BT observables have come into the experimental reach of investigation.
Using the self-analyzing power of hyperons, for example in $K \Lambda$ photoproduction \cite{Sandorfi:2010uv}, or, alternatively, the recently developed recoil polarimetry \cite{RecoilPol}, the remaining $8$ double-po\-la\-ri\-za\-tion observables of BR and TR have become in principle available as well. It has to be mentioned however, that especially for the latter method acceptable statistics is very hard to obtain. \newline
The new experimental information on single- and double-po\-la\-ri\-za\-tion observables is used to constrain the existing partial wave analyses (PWAs). Such approaches are for example the Bonn Gatchina PWA \cite{Anisovich:2011fc}, the SAID \cite{Arndt:2006bf} and MAID \cite{Drechsel:2007if} PWAs,
\begin{figure*}[htb]
\RawFloats
\centering
\begin{minipage}{0.48\linewidth}
\hspace*{-5pt}
\includegraphics[width=0.99\textwidth]{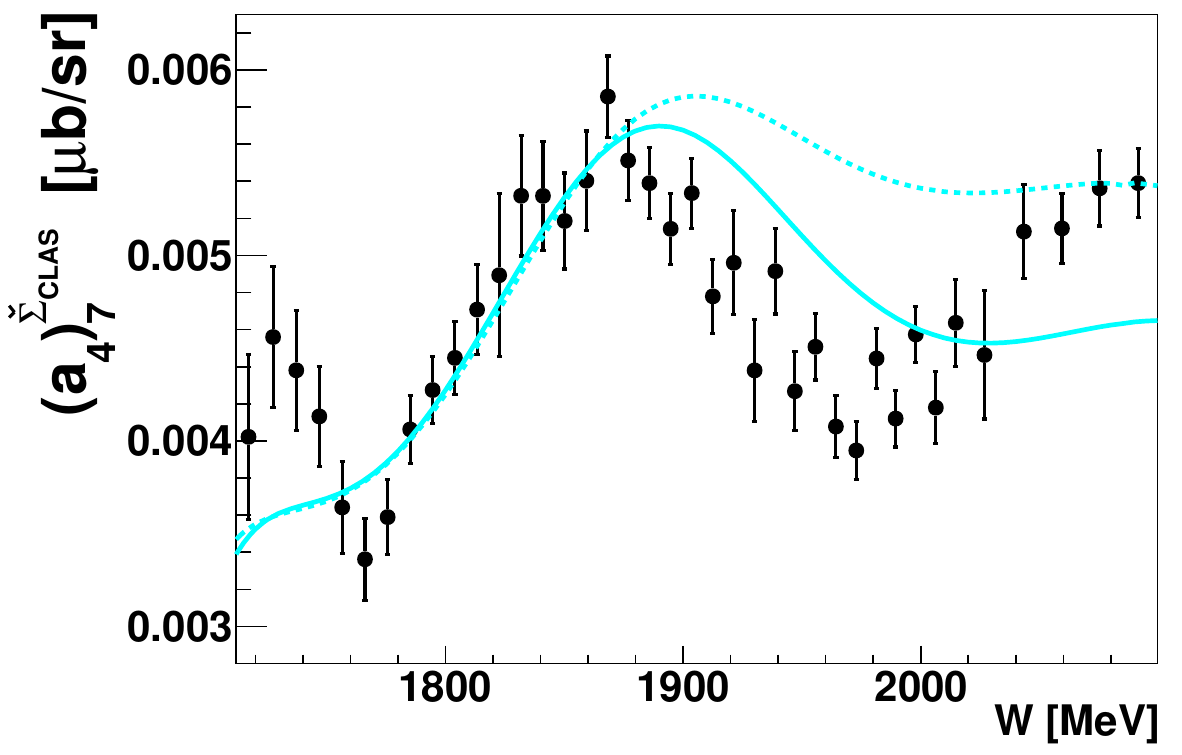}
\end{minipage}%
\begin{minipage}{.075\linewidth}
\vspace*{-25pt}
\hspace*{10pt}
\begin{equation}
\mathcal{C}_{7}^{\check{\Sigma}} \equiv \nonumber
\end{equation}
\end{minipage}%
\begin{minipage}{0.375\linewidth} \vspace*{2pt} \includegraphics[width=0.9\textwidth]{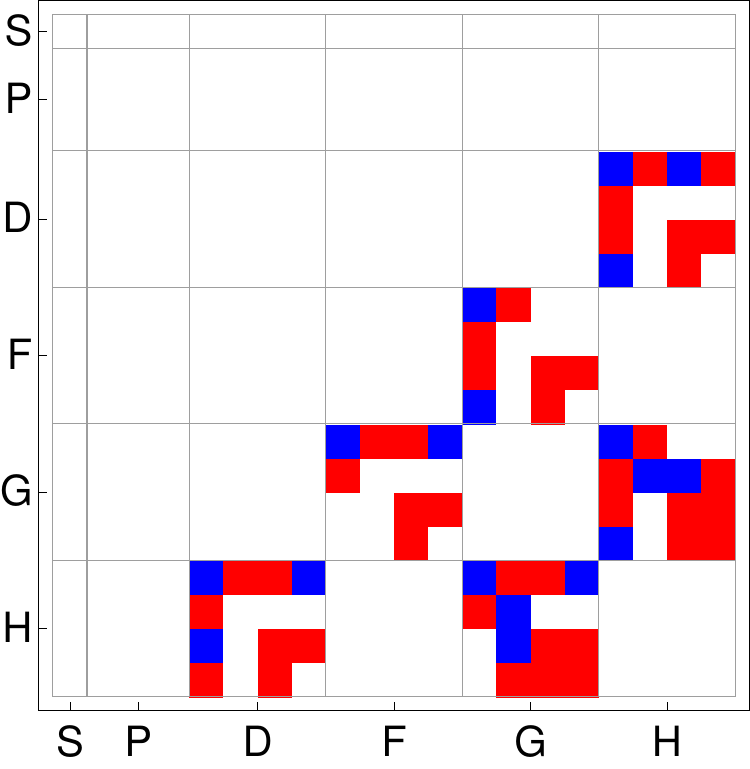} \end{minipage}
\caption{Left: Shown here is the Legendre coefficient $\left(a^{\check{\Sigma}}_{4}\right)_{7}$ (black dots), extracted from a Legendre fit to the recently published beam asymmetry data of the CLAS-collaboration \cite{Dugger:2013}. The dashed line represents the same coefficient as evaluated from the Bonn-Gatchina partial wave analysis solution BnGa2014-02, using multipoles contributing up to the $H$-waves $(E_{5\pm}. M_{5\pm})$. The solid line originates from a more recent fit by the Bonn-Gatchina group, where a new state in the $G$-waves, the $\Delta(2200) \frac{7}{2}^{-}$, was required in order to describe the data \cite{BnGaNewGResonance}. \newline Right: The partial wave content for the Legendre coefficient $\left(a^{\check{\Sigma}}_{4}\right)_{7}$ is shown up to $H$- waves. Only $D$-waves, $F$-waves, $G$-waves and $H$- waves are contributing. It is indicated by the colored boxes that only certain interferences of the multipoles can occur in $\left(a^{\check{\Sigma}}_{4}\right)_{7}$. Explicitly, only $\left<D,H\right>$-, $\left<F,G\right>$- and $\left<G,H\right>$- interferences are allowed.}
\label{fig:BnGaOldNewComparisonIntroduction}
\end{figure*}
the J\"ulich-Bonn dynamical coupled-channels model \cite{Ronchen:2012eg}, the approach of the ANL/Osaka group \cite{Kamano:2013iva} or the Giessen analysis \cite{Shklyar:2012js}. Analyses of this kind generally try to fit data on multiple reactions in an intricate unitary analysis scheme, which then yields resonance parameters directly in terms of positions and residues of the resonance poles in the scattering matrix. They are generally classified as energy-dependent (ED) fits. \newline
However, once only a single photoproduction channel is under consideration there exists also the possibility of directly fitting the truncated partial wave expansion of the full production amplitudes to the data \cite{Sandorfi:2010uv}, \cite{Grushin}, \cite{WBT}), thereby doing a Truncated Partial Wave Analysis (TPWA). This analysis scheme proceeds independently in each energy bin and therefore this method is also denoted as an energy-independent (EI) fit. Once partial wave amplitudes would be extracted uniquely including the phase, there are new methods proposed \cite{SvarcLplusP}) to determine the parameters of resonance poles from these single energy partial waves. A set of measured observables that facilitates the unique extraction of partial wave amplitudes up to an overall phase is called a complete experiment \cite{Sandorfi:2010uv}. \newline
This work is concerned neither with the above mentioned ED methods nor a full TPWA fit. Instead, a simple method is presented here which determines the dominant partial waves contributing to measured single- and double-po\-la\-ri\-za\-tion observables. As an example we focus on the reaction $\gamma p \to \pi^0 p$. The approach consists of fitting polynomials to angular distributions of observables and investigating the fit quality via the $\chi^{2}/\mathrm{ndf}$. The order and form of the polynomial is dictated by the truncation angular momentum $\text{L}_{\text{max}}$ and the TPWA formulas. The parameter $\text{L}_{\text{max}}$ then has to be increased until $\chi^{2}/\mathrm{ndf}$ is close to unity. These simple angular distribution fits to each single- and double-po\-la\-ri\-za\-tion observable will provide the information which $\text{L}_{\text{max}}$ is needed to describe the measured observable within the statical accuracy. \newline
 It has to be stressed here that in order to perform a unique partial wave analysis, the fitted observables not only have to be mathematically complete. Also, the statistical precision of the fitted data has to be good enough for them to be sensitive to high $\ell$ contributions. The method proposed in this work provides a quick means to check the precision and sensitivity of the data. Generally an observable can only give meaningful information on a resonant state of a certain $\ell$ if the precision of the measurement is good enough. \newline
 The method proposed in this paper allows to project out high spin partial wave contributions from any measured observable as long as the measurement has the necessary statistical accuracy. The strength of the method can be made explicit, once these fit coefficient are compared to different PWA solutions. Figure \ref{fig:BnGaOldNewComparisonIntroduction} shows for example the Legendre-coefficient $\left(a^{\check{\Sigma}}_{4}\right)_{7}$ extracted from the beam asymmetry $\Sigma$ from CLAS \cite{Dugger:2013}. \newline
 Several interesting facts can be said about the composition of this coefficient in terms of partial waves. The quantity $\left(a^{\check{\Sigma}}_{4}\right)_{7}$ is generally a bilinear hermitean form in the multipoles, which is illustrated in Figure \ref{fig:BnGaOldNewComparisonIntroduction} as well. In the colored plot on the right hand side, the coefficients of all partial wave interference-terms in a truncation at the $H$-waves ($\ell_{\mathrm{max}} = 5$) are represented by small colored boxes. Red boxes show entries with a positive sign, blue boxes those with a negative sign. More information on this particular scheme for representing partial wave compositions of the Legendre coefficients is provided in section \ref{sec:DescriptionCompositionLegCoeffs} of this work. \newline In case the only significant contribution to the coefficient would originate from $S$, $P$, $D$ and $F$ waves ($\ell$ = 0, 1, 2, and 3) and all higher partial waves would vanish, this coefficient would be exactly zero. The first none-zero contribution comes from an interference of $F$- with $G$- waves ($\ell$=4) and $D$- with $H$-waves ($\ell$=5).  \newline
 The $F$-wave multipoles $E_{3\pm}$ and $M_{3\pm}$ in the $p \pi^{0}$-channel are dominating the mass region of $1600$ - $2200$ $\mathrm{MeV}$ because of the two four-star resonances $N(1680)\frac{5}{2}^{+}$ \newline $\left(F_{15} (1680)\right)$ and $\Delta(1950)\frac{7}{2}^{+}$ $\left(F_{37} (1950)\right)$. All PWA approaches like Bonn-Gatchina, MAID, J\"ulich-Bonn or SAID show a very similar size of the magnitude and the energy dependence for the two $F$- multipoles \cite{CommonPaper} (see Figure \ref{fig:MultipoleComparisons}). Therefore, the major part of possible differences in PWA- or model-descriptions of the coefficient $\left(a^{\check{\Sigma}}_{4}\right)_{7}$ has to come from different $G$- and $H$-wave contributions. \newline
 Figure \ref{fig:BnGaOldNewComparisonIntroduction} also shows the solution of BnGa2014-02 (dotted line) and a most recent BnGa solution (solid line), the same as mentioned above, which includes an additional new state $\Delta(2200) \frac{7}{2}^{-}$ \cite{BnGaNewGResonance}. This state has the same quantum numbers as the $E_{4-}$ and $M_{4-}$ multipoles. The solid line describes the $\left(a^{\check{\Sigma}}_{4}\right)_{7}$ coefficient significantly better and supports the finding of a new resonance in the $G$-wave ($E_{4-}$ and $M_{4-}$ multipoles). This is, as mentioned above, only possible because of the interference of the $G$-waves with the already well determined $F$-wave multipoles. \newline
We proceed by first introducing all the necessary formalism. Then, recent measured po\-la\-ri\-za\-tion observables $\Sigma$, $T$, $P$, $H$, $E$, $F$ and $G$ in $\pi^{0}$-photoproduction are analyzed along with the unpolarized cross section $\sigma_0$ in regard of their dominant partial wave contributions. An appendix contains rather elaborate formulas and pictures that support the interpretation.
%
%

\section{Basic formalism for the deduction of dominant partial wave contributions} \label{sec:Formalism}
\begin{figure*}[ht]
\RawFloats
\centering
\includegraphics[width=0.9\textwidth]{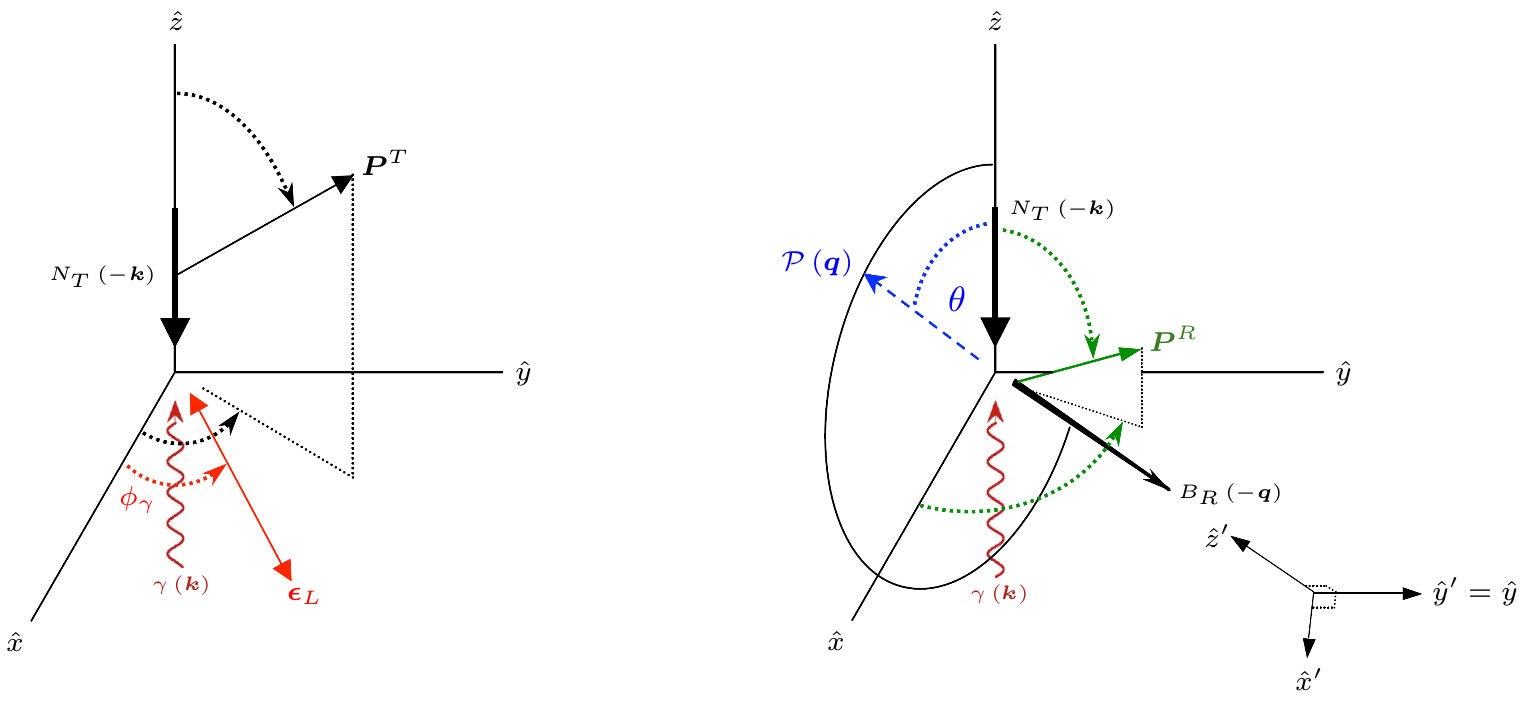}
\caption{The specifications of momenta and polarization vectors for pseudoscalar meson photoproduction are shown here in the CMS frame. \newline Left: The initial state contains a photon $\gamma$ with momentum  $\vec{k}$ and polarization vector $\vec{\epsilon}_{L}$ lying in the $\left< \hat{x} - \hat{y} \right>$ plane, which has an angle $\phi_{\gamma}$ towards the so called reaction plane, i.e. the $\left< \hat{x} - \hat{z} \right>$ plane. The target nucleon $N_{T}$ has momentum $-\vec{k}$ in the CMS and it's polarization is specified by the vector $\vec{P}^{T}$. Right: In the final state, the pseudoscalar meson $\mathcal{P}$ leaves the reaction with a CMS momentum vector $\vec{q}$, which is by convention chosen to lie in the reaction- (i.e. $\left< \hat{x} - \hat{z} \right>$) plane. The CMS scattering angle $\theta$ is defined versus the $\hat{z}$-axis. The recoil baryon $B_{R}$ leaves the process having CMS momentum $- \vec{q}$ and a polarization specified by the vector $\vec{P}^{R}$. \newline
The picture was taken from reference \cite{Sandorfi:2010uv}, although some relabelling of the occuring vectors and angles has been applied.}
\label{fig:coordinate}
\end{figure*}

The reaction amplitude for the photoproduction of pseudoscalar mesons is known to be decomposable in the center of mass system (CMS) \cite{CGLN}) as follows 
{\allowdisplaybreaks
\begin{align}
  F_{\mathrm{CGLN}} &= i \left( \vec{\sigma} \cdot \hat{\epsilon} \right) F_{1} + \left( \vec{\sigma} \cdot \hat{q} \right) 
  \Big[ \vec{\sigma} \cdot \left( \hat{k} \times \hat{\epsilon} \right) \Big] F_{2} \nonumber \\
  &\hspace*{8.5pt} + i \left( \vec{\sigma} \cdot \hat{k} \right) \left( \hat{q} \cdot \hat{\epsilon} \right) F_{3} 
  + i \left( 
   \vec{\sigma} \cdot \hat{q} \right) \left( \hat{q} \cdot \hat{\epsilon} \right) F_{4} \mathrm{,} \label{eq:DefCGLN} 
   \end{align}
   }
where $\hat{k}$ and $\hat{q}$ denote CMS momenta, $\vec{\sigma}$ is the Pauli spin operator and the complex CGLN amplitudes $F_{i} \left( W, \theta \right)$, once determined, completely specify the full amplitude. 
The expansions of CGLN amplitudes into multipoles are also well known \cite{CGLN}, \cite{Sandorfi:2010uv}).  For the sake of completeness, we list them again in eqs (\ref{eq:MultExpF1}) to (\ref{eq:MultExpF4}) (with $x = \cos \theta$). \newline
For certain photoproduction channels, for example for the $\pi^{0} p$ or $\eta p$ final states, it is a valid assumption to truncate the infinite multipole expansion at a finite angular momentum quantum number $\text{L}_{\text{max}}$, which close to the threshold should yield already a good approximation for the $F_{i}$. Resonances in the $s$-channel would then yield contributions to the finite number of energy-dependent complex multipoles according to the respective quantum numbers. \\
The po\-la\-ri\-za\-tion observables represent asymmetries among differential cross sections for measurements corresponding to different po\-la\-ri\-za\-tion states. Table \ref{tab:Observables} summarizes their definitions in terms of beam-, target- and recoil-po\-la\-ri\-za\-tion measurements. An illustration of the coordinates used to describe the photoproduction process is given in Figure \ref{fig:coordinate}, where center-of-mass coordinates are shown. \newline
Since the amplitude (\ref{eq:DefCGLN}) factors into the matrix element of polarized cross sections, it is possible to derive equations relating the $16$ po\-la\-ri\-za\-tion observables of pseudoscalar meson photoproduction to CGLN amplitudes. A consistent set of relations, corresponding up to signs to the formulas given in reference \cite{Sandorfi:2010uv}, is collected in Table \ref{tab:ObsInTermsOfCGLN}.

{\allowdisplaybreaks
\begin{align}
   F_{1} \left( W, \theta \right) &= \sum \limits_{\ell = 0}^{\infty} \Big\{ \left[ \ell M_{\ell+} \left( W \right) + E_{\ell+} \left( W \right) \right] P_{\ell+1}^{'} \left( x \right) \nonumber \\
 &  + \left[ \left( \ell+1 \right) M_{\ell-} \left( W \right) + E_{\ell-} \left( W \right) \right] P_{\ell-1}^{'} \left( x \right) \Big\} \mathrm{,} \label{eq:MultExpF1} \\
 F_{2} \left( W, \theta \right) &= \sum \limits_{\ell = 1}^{\infty} \left[ \left( \ell+1 \right) M_{\ell+} \left( W \right) + \ell M_{\ell-} \left( W \right) \right] \nonumber \\
 & \quad \quad \quad \times P_{\ell}^{'} \left( x \right) \mathrm{,} \label{eq:MultExpF2} \\
F_{3} \left( W, \theta \right) &= \sum \limits_{\ell = 1}^{\infty} \Big\{ \left[ E_{\ell+} \left( W \right) - M_{\ell+} \left( W \right) \right] P_{\ell+1}^{''} \left( x \right) \nonumber \\
 & \quad \quad \quad + \left[ E_{\ell-} \left( W \right) + M_{\ell-} \left( W \right) \right] P_{\ell-1}^{''} \left( x \right) \Big\} \mathrm{,} \label{eq:MultExpF3}  \\
F_{4} \left( W, \theta \right) &= \sum \limits_{\ell = 2}^{\infty} [ M_{\ell+} \left( W \right) - E_{\ell+} \left( W \right) \nonumber \\& - M_{\ell-} \left( W \right) - E_{\ell-} \left( W \right) ] P_{\ell}^{''} \left( x \right) \mathrm{.} \label{eq:MultExpF4}
 \end{align}
 }
%

%

A comparison of the sign conventions present in the literature can be found in \cite{Conventions}. The rather involved formulas in Table \ref{tab:ObsInTermsOfCGLN} can all be arranged in the shape of bilinear equations
\begin{align}
\check{\Omega}^{\alpha} \left( W, \theta \right) &= \frac{\rho}{2} \sum_{i,j=1}^{4} F_{i}^{\ast} \left( W, \theta \right) \hat{A}^{\alpha}_{ij} \left( \theta \right) F_{j} \left( W, \theta \right) \mathrm{,}\label{eq:DefObservable}\\  \alpha &= 1, \ldots, 16 \mathrm{,} \nonumber
\end{align}
where $\rho = \frac{q}{k}$ is the phase space factor and the matrices $\hat{A}^{\alpha}$ have to be hermitean in order for $\check{\Omega}^{\alpha}$ to be real. The quantities $\check{\Omega}^{\alpha}$ are denoted as the profile functions and the dimensionless po\-la\-ri\-za\-tion observables $\Omega^{\alpha}$ can be obtained via dividing by the unpolarized differential cross section $\Omega^{\alpha} = \frac{\check{\Omega}^{\alpha}}{\sigma_{0}}$. The notation with the index $\alpha$ labelling the observables originates from the paper by Chiang and Tabakin \cite{ChTab}, where the latter are written in the helicity basis using $16$ hermitean unitary $4 \times 4$ matrices. In this way, the TPWA can be brought into closed form for all $16$ observables as follows.\newline \newline
Once the multipole series (\ref{eq:MultExpF1}) to (\ref{eq:MultExpF4}) are truncated at a finite $\text{L}_{\text{max}}$, the maximal power of the resulting expansion in $\cos \theta$ can be read off the Legendre polynomials appearing in the expansion. For $F_{1}$ it is $\text{L}_{\text{max}}$, $F_{2}$ has highest power $(\text{L}_{\text{max}} - 1)$ as well as $F_{3}$ and for $F_{4}$ it is $(\text{L}_{\text{max}} - 2)$. By investigating the definitions of the profile functions in Table \ref{tab:ObsInTermsOfCGLN}, it is possible to infer the maximal power each observable has once the truncated partial wave expansion of the $F_{i}$ is inserted. \newline
This facilitates the expression of the $16$ profile functions in a TPWA as finite expansions in $\cos \theta$.
However, for practical fits it is advantageous to change the angular parametrization and expand the profile functions into associated Legendre Polynomials $P_{\ell}^{m} \left( \cos \theta \right)$ (cf. \cite{Abramowitz}). We have to state explicitly that we are using a definition of the associated Legendre polynomials \textit{without} the Condon-Shortley phase, i.e.  $P_{\ell}^{m} \left( x \right) = \left( 1 - x^{2} \right)^{\frac{m}{2}} \frac{d^{m}}{d x^{m}} P_{\ell} \left( x \right)$\footnote{The convention to write the associated Legendre polynomials $P_{\ell}^{m}$ without the Condon-Shortley phase
$(-1)^{m}$ corresponds to the way the latter are implemented in the ROOT-libraries. This can be seen under
the following link: \newline
\textit{https://root.cern/doc/master/group\_\_SpecFunc.html} \newline
All fits presented in this paper were performed with ROOT. Therefore, all definitions are kept consistent with the fits.}. In this form, the truncated partial wave expansion takes the shape
{\allowdisplaybreaks
\begin{align}
&\hspace*{-7pt}\check{\Omega}^{\alpha} \left( W, \theta \right) = \rho \hspace*{3pt} \sum \limits_{k = \beta_{\alpha}}^{2 \text{L}_{\text{max}} + \beta_{\alpha} + \gamma_{\alpha}} \left(a_{\text{L}_{\text{max}}}\right)_{k}^{\check{\Omega}^{\alpha}} \left( W \right) P^{\beta_{\alpha}}_{k} \left( \cos \theta \right) \mathrm{,}  \label{eq:LowEAssocLegParametrization1} \\
&\left(a_{\text{L}_{\text{max}}}\right)_{k}^{\check{\Omega}^{\alpha}} \left( W \right) = \left< \mathcal{M}_{\text{L}_{\text{max}}} \left( W \right) \right| \mathcal{C}_{k}^{\check{\Omega}^{\alpha}} \left| \mathcal{M}_{\text{L}_{\text{max}}} \left( W \right) \right> \mathrm{.} \label{eq:LowEAssocLegParametrization2}
\end{align}
}
Here, the energy-dependent real expansion coefficients \newline $\left(a_{\text{L}_{\text{max}}}\right)_{k}^{\alpha}$ are bilinear hermitean forms of the multipoles, defined via hermitean matrices $\mathcal{C}_{k}^{\check{\Omega}^\alpha}$, whose dimension is given by the number of multipoles in the respective truncation. For $\text{L}_{\text{max}}$ there are $4 \text{L}_{\text{max}}$ multipoles and the $\mathcal{C}_{k}^{\check{\Omega}^\alpha}$ are $(4 \text{L}_{\text{max}}) \times (4 \text{L}_{\text{max}})$ matrices. The parameters $\beta_{\alpha}$ and $\gamma_{\alpha}$ that define the angular expansions are collected in Table \ref{tab:AngularDistributions}. This notation is chosen in accord with reference \cite{LotharNStar}. \newline
Listings of particular Legendre coefficient matrices $ \mathcal{C}_{k}^{\check{\Omega}^{\alpha}}$ as well as an explanation of our notation for them can be found in section \ref{sec:DescriptionCompositionLegCoeffs} and Appendix \ref{sec:PWContentFormulas}. \newline
The angular parametrizations defined by an expansion into powers of $\cos \theta$ or associated Legendre polynomials are equally valid. This is true since the monomials $\cos^{n}\left(\theta\right)$ as well as the associated Legendre polynomials normalized to $\sin\left(\theta\right)$, i.e. $\frac{1}{\left(\sin\theta\right)^{\beta_{\alpha}}} P^{\beta_{\alpha}}_{k} \left( \cos \theta \right)$, are fully equivalent basis systems for the expansion of finite polynomials. The associated Legendre polynomials have the advantage of including the observable dependent $\sin \theta$ factors and are furthermore orthogonal on the full angular interval. This lowers the correlations among different angular fit parameters in practical a\-na\-ly\-ses. For this reason, we choose to work with equation (\ref{eq:LowEAssocLegParametrization1}) in the following. \newline
The approach for deducing the dominant partial wave contributions to po\-la\-ri\-za\-tion measurements consists of fitting finite expansions given by equation (\ref{eq:LowEAssocLegParametrization1}) to angular distributions of measured po\-la\-ri\-za\-tion observables in fixed energy bins. These fits minimize a standard error weighted $\chi^{2}$ which in the minimum then yields the parameter $\chi^{2}/\mathrm{ndf}$. The procedure starts usually at $\text{L}_{\text{max}} = 1$. In case $\chi^{2}/\mathrm{ndf}$ is significantly larger than 1, the truncation order $\text{L}_{\text{max}}$ has to be raised upon which again angular fits are performed. The assumption is that the order at which this procedure terminates gives a direct indication of the dominating partial wave contributions present in the measured observable at hand (cf. \cite{Grushin}). For all orders $\text{L}_{\text{max}}$ used in the procedure, the resulting $\chi^{2}/\mathrm{ndf}$ can be plotted versus energy, resulting in figures that allow an investigation of the energy dependence of certain partial wave contributions. \newline
Since the fit minimizes an error weighted functional, the order $\text{L}_{\text{max}}$ for which $\chi^{2}/\mathrm{ndf}$ approaches 1 also depends significantly on the precision and statistics of the measurement. Therefore one should distinguish at this point among partial wave contributions that are contained in the data theoretically and those that can be inferred directly from a measurement by this simple method. The partial wave series itself is an infinite series and the truncation of (\ref{eq:MultExpF1}) to (\ref{eq:MultExpF4}) already represents a source of error. In theory all partial waves contribute to the full amplitude at all energies. However, some partial waves may be so highly suppressed that a TPWA is a valid approximation. The error of the measurement then sets a limit on the order of contributing partial waves that can be seen in experiment, which is generally only a small subset of the infinity that is there theoretically. From a partial wave point of view this issue demands the increase of measurement precision. \newline
The $\chi^{2}$ criterion proposed here is no substitute for a full partial wave analysis, neither in an energy-dependent nor an energy independent fit. However it can serve as a means to obtain a quick first interpretation of measured po\-la\-ri\-za\-tion data. In addition, for the energy independent TPWA, which consists of the (numerical) solution of the equation system generated by equation (\ref{eq:LowEAssocLegParametrization2}), the criterion can serve as a first guide towards the selection of multipoles to be varied in this numerical procedure. As mentioned above, the truncation already represents a potential source of error. According to reference \cite{BowcockBurkhardt}, the effects of the truncation only show up in the highest partial waves varied in a TPWA. Therefore one could propose as a rule of thumb to vary one order above the $\text{L}_{\text{max}}$ that the $\chi^{2}$ criterion yields. One could also try to vary the multipoles up to this latter $\text{L}_{\text{max}}$ 
and study the differences to the $(\text{L}_{\text{max}} + 1)$ case. In addition, it has to be mentioned that the TPWA generally demands the complication of including the higher multipoles which are not varied as fixed parameters (cf. \cite{Sandorfi:2010uv}). These can be taken either from a model or some known physical background amplitude. \newline \newline
Since this work treats measurements of observables with beam and/or target po\-la\-ri\-za\-tion (Type S and BT), we close this section by specifying equation (\ref{eq:LowEAssocLegParametrization1}) to formulas that are valid for $\text{L}_{\text{max}} \geq 1$ in this particular case
{\allowdisplaybreaks
\begin{align}
\sigma_{0} \left( W, \theta \right) &= \rho \hspace*{3pt} \sum \limits_{k = 0}^{2 \text{L}_{\text{max}}} \left(a_{\text{L}_{\text{max}}}\right)_{k}^{\sigma_{0}} \left( W \right) P_{k} \left( \cos \theta \right) \mathrm{,}  \label{eq:LowEAssocLegParametrizationDCS}\\
\check{\Sigma} \left( W, \theta \right) &= \rho \hspace*{3pt} \sum \limits_{k = 2}^{2 \text{L}_{\text{max}}} \left(a_{\text{L}_{\text{max}}}\right)_{k}^{\check{\Sigma}} \left( W \right) P^{2}_{k} \left( \cos \theta \right) \mathrm{,}  \label{eq:LowEAssocLegParametrizationSigma} \\
\check{T} \left( W, \theta \right) &= \rho \hspace*{3pt} \sum \limits_{k = 1}^{2 \text{L}_{\text{max}}} \left(a_{\text{L}_{\text{max}}}\right)_{k}^{\check{T}} \left( W \right) P^{1}_{k} \left( \cos \theta \right) \mathrm{,}  \label{eq:LowEAssocLegParametrizationT}\\
\check{P} \left( W, \theta \right) &= \rho \hspace*{3pt} \sum \limits_{k = 1}^{2 \text{L}_{\text{max}}} \left(a_{\text{L}_{\text{max}}}\right)_{k}^{\check{P}} \left( W \right) P^{1}_{k} \left( \cos \theta \right) \mathrm{,}  \label{eq:LowEAssocLegParametrizationP}\\
\check{E} \left( W, \theta \right) &= \rho \hspace*{3pt} \sum \limits_{k = 0}^{2 \text{L}_{\text{max}}} \left(a_{\text{L}_{\text{max}}}\right)_{k}^{\check{E}} \left( W \right) P_{k} \left( \cos \theta \right) \mathrm{,}  \label{eq:LowEAssocLegParametrizationE}\\
\check{G} \left( W, \theta \right) &= \rho \hspace*{3pt} \sum \limits_{k = 2}^{2 \text{L}_{\text{max}}} \left(a_{\text{L}_{\text{max}}}\right)_{k}^{\check{G}} \left( W \right) P^{2}_{k} \left( \cos \theta \right) \mathrm{,}  \label{eq:LowEAssocLegParametrizationG}\\
\check{H} \left( W, \theta \right) &= \rho \hspace*{3pt} \sum \limits_{k = 1}^{2 \text{L}_{\text{max}}} \left(a_{\text{L}_{\text{max}}}\right)_{k}^{\check{H}} \left( W \right) P^{1}_{k} \left( \cos \theta \right) \mathrm{,}   \label{eq:LowEAssocLegParametrizationH} \\
\check{F} \left( W, \theta \right) &= \rho \hspace*{3pt} \sum \limits_{k = 1}^{2 \text{L}_{\text{max}}} \left(a_{\text{L}_{\text{max}}}\right)_{k}^{\check{F}} \left( W \right) P^{1}_{k} \left( \cos \theta \right) \mathrm{,}   \label{eq:LowEAssocLegParametrizationF}
\end{align}
}
where $P_{k} \left( \cos \theta \right) = P^{0}_{k} \left( \cos \theta \right)$. Finally, it is worth mentioning that the total cross section $\bar{\sigma} (W)$ is related to the zeroth Legendre coefficient of the differential cross section $\sigma_{0}$ in a very simple way. One can use the orthogonality of the Legendre polynomials to obtain the relation $\int_{-1}^{1} d x \hspace*{2pt} P_{\ell} (x) = \int_{-1}^{1} d x \hspace*{2pt} P_{\ell} (x) \hspace*{1.75pt} P_{0} (x) = 2 \delta_{\ell 0}$. Then, it is seen easily that the following equation holds
\begin{equation}
 \bar{\sigma} (W) = 4 \pi \frac{q}{k} \left(a_{\text{L}_{\text{max}}}\right)_{0}^{\sigma_{0}} \mathrm{.} \label{eq:RelationTCSZeroCoeff}
\end{equation}
%

%
\begin{table*}[htb]
\RawFloats
\centering
\begin{tabular}{cclcc}
\hline
\hline \\
Observable & $\alpha$ & CGLN-representation & Type \\
\hline \\
$ \sigma_{0} $ & $1$ & $ \rho \hspace*{3pt} \mathrm{Re} \left[ \left| F_{1} \right|^{2} + \left| F_{2} \right|^{2} - 2 \cos (\theta) F_{1}^{\ast} F_{2} + \frac{1}{2} \sin^{2} (\theta) \left\{ \left| F_{3} \right|^{2} + \left| F_{4} \right|^{2} + 2 F_{1}^{\ast} F_{4} + 2 F_{2}^{\ast} F_{3} + 2 \cos (\theta) F_{3}^{\ast} F_{4} \right\} \right] $ &  \\
$ \check{\Sigma} $ & $4$ & $ - \rho \hspace*{3pt} \frac{\sin^{2} (\theta)}{2} \hspace*{4pt} \mathrm{Re} \left[ \left| F_{3} \right|^{2} + \left| F_{4} \right|^{2} + 2 \left\{ F_{1}^{\ast} F_{4} + F_{2}^{\ast} F_{3} + \cos (\theta) F_{3}^{\ast} F_{4} \right\} \right] $ & S \\
$ \check{T} $ & $10$ & $ \rho \hspace*{3pt} \sin (\theta) \hspace*{4pt} \mathrm{Im} \Big[ F_{1}^{\ast} F_{3} - F_{2}^{\ast} F_{4} + \cos (\theta) \left\{ F_{1}^{\ast} F_{4} - F_{2}^{\ast} F_{3} \right\} - \sin^{2} (\theta) F_{3}^{\ast} F_{4} \Big] $ & \\
$ \check{P} $ & $12$ & $ \rho \hspace*{3pt} \sin(\theta) \hspace*{4pt} \mathrm{Im} \Big[ \left\{ 2 F_{2} + F_{3} + \cos (\theta) F_{4} \right\}^{\ast} F_{1} + F_{2}^{\ast} \left\{ \cos (\theta) F_{3} + F_{4} \right\} + \sin^{2} (\theta) F_{3}^{\ast} F_{4} \Big] $ & \\
 & & & & \\
$ \check{G} $ & $3$ & $ - \rho \hspace*{3pt} \sin^{2} (\theta) \hspace*{4pt} \mathrm{Im} \Big[ F_{4}^{\ast} F_{1} + F_{3}^{\ast} F_{2} \Big] $ & \\
$ \check{H} $ & $5$ & $ \rho \hspace*{3pt} \sin (\theta) \hspace*{4pt} \mathrm{Im} \Big[ \left\{ F_{2} + F_{3} + \cos (\theta) F_{4} \right\}^{\ast} F_{1} - \left\{ F_{1} + F_{4} + \cos (\theta) F_{3} \right\}^{\ast} F_{2} \Big] $ & BT \\
$ \check{E} $ & $9$ & $ \rho \hspace*{3pt} \mathrm{Re} \left[ \left| F_{1} \right|^{2} + \left| F_{2} \right|^{2} - 2 \cos (\theta) F_{1}^{\ast} F_{2} + \sin^{2} (\theta) \left\{ F_{4}^{\ast} F_{1} + F_{3}^{\ast} F_{2} \right\} \right] $ & \\
$ \check{F} $ & $11$ & $ \rho \hspace*{3pt} \sin (\theta) \hspace*{4pt} \mathrm{Re} \Big[ \left\{ F_{2} + F_{3} + \cos (\theta) F_{4} \right\}^{\ast} F_{1} - \left\{ F_{1} + F_{4} + \cos (\theta) F_{3} \right\}^{\ast} F_{2} \Big] $ & \\
 & & & & \\
$ \check{O}_{x'} $ & $14$ & $ \rho \hspace*{3pt} \sin (\theta) \hspace*{4pt} \mathrm{Im} \Big[ F_{3}^{\ast} F_{2} - F_{4}^{\ast} F_{1} + \cos (\theta) \left\{ F_{4}^{\ast} F_{2} - F_{3}^{\ast} F_{1} \right\} \Big] $ & \\
$ \check{O}_{z'} $ & $7$ & $ - \rho \hspace*{3pt} \sin^{2} (\theta) \hspace*{4pt} \mathrm{Im} \Big[ F_{3}^{\ast} F_{1} + F_{4}^{\ast} F_{2} \Big] $ & BR \\
$ \check{C}_{x'} $ & $16$ & $ \rho \hspace*{3pt} \sin (\theta) \hspace*{4pt} \mathrm{Re} \Big[ \left| F_{2} \right|^{2} - \left| F_{1} \right|^{2} + F_{3}^{\ast} F_{2} - F_{4}^{\ast} F_{1} + \cos (\theta) \left\{ F_{4}^{\ast} F_{2} - F_{3}^{\ast} F_{1} \right\} \Big] $ & \\
$ \check{C}_{z'} $ & $2$ & $ \rho \hspace*{3pt} \mathrm{Re} \Big[ - 2 F_{2}^{\ast} F_{1} + \cos (\theta) \left| F_{1} \right|^{2} + \cos (\theta) \left| F_{2} \right|^{2} - \sin^{2} (\theta) \left\{ F_{3}^{\ast} F_{1} + F_{4}^{\ast} F_{2} \right\} \Big] $ & \\
 & & & & \\
$ \check{T}_{x'} $ & $6$ & $ - \rho \hspace*{3pt} \frac{\sin^{2} (\theta)}{2} \hspace*{4pt} \mathrm{Re} \Big[ \cos (\theta) \left\{ \left| F_{3} \right|^{2} + \left| F_{4} \right|^{2} \right\} + 2 \left\{ F_{4}^{\ast} F_{3} + F_{3}^{\ast} F_{1} + F_{4}^{\ast} F_{2} \right\} \Big] $ & \\
$ \check{T}_{z'} $ & $13$ & $ \rho \hspace*{3pt} \sin (\theta) \hspace*{4pt} \mathrm{Re} \bigg[ \frac{\sin^{2} (\theta)}{2} \left\{ \left| F_{4} \right|^{2} - \left| F_{3} \right|^{2} \right\} +  F_{4}^{\ast} F_{1} - F_{3}^{\ast} F_{2} + \cos (\theta) \left\{F_{3}^{\ast} F_{1} - F_{4}^{\ast} F_{2} \right\} \bigg] $ & TR \\
$ \check{L}_{x'} $ & $8$ & $ \rho \hspace*{3pt} \sin (\theta) \hspace*{4pt} \mathrm{Re} \bigg[ \left| F_{1} \right|^{2} - \left| F_{2} \right|^{2} + F_{4}^{\ast} F_{1} - F_{3}^{\ast} F_{2} + \cos (\theta) \left\{ F_{3}^{\ast} F_{1} - F_{4}^{\ast} F_{2} \right\} + \frac{\sin^{2} (\theta)}{2} \left\{ \left| F_{4} \right|^{2} - \left| F_{3} \right|^{2} \right\} \bigg] $ & \\
$ \check{L}_{z'} $ & $15$ & $ \rho \hspace*{3pt} \mathrm{Re} \bigg[ 2 F_{2}^{\ast} F_{1} - \cos (\theta) \left\{ \left| F_{1} \right|^{2} + \left| F_{2} \right|^{2} \right\} + \sin^{2} (\theta) \left\{ \frac{\cos (\theta)}{2} \left( \left| F_{3} \right|^{2} + \left| F_{4} \right|^{2} \right) + F_{3}^{\ast} F_{1} + F_{4}^{\ast} F_{2} + F_{4}^{\ast} F_{3} \right\} \bigg] $ & \\
 & & & & \\
\hline
\hline
\end{tabular}
\caption{This table lists the definitions of profile functions $\check{\Omega}^{\alpha}$ in terms of CGLN amplitudes $F_{i}$. The sign conventions for observables used in this work correspond to the Bonn-Gatchina PWA, which again uses Fasano/Tabakin/Saghai conventions \cite{FTS}). For a useful comparison of sign conventions for photoproduction, see \cite{Conventions}. The phase space factor is $\rho = q/k$.}
\label{tab:ObsInTermsOfCGLN}
\end{table*}
\begin{table*}[htb]
\RawFloats
\centering
\begin{tabular}{ccc|cccccl||cccc|cccclllll}
\hline
\hline
  &  &  &  &  &  &  &  &  &  &  &  &  &  &  &  &  &  \\
Type & $ \check{\Omega}^{\alpha} $ & $\alpha$ &  & $ \beta_{\alpha} $ &  & $ \gamma_{\alpha} $ &  & $N_{a}$ & Type & $ \check{\Omega}^{\alpha} $ & $\alpha$ &  & $ \beta_{\alpha} $ &  & $ \gamma_{\alpha} $ &  &  $N_{a}$  \\
\hline
  &  &  &  &  &  &  &  &  &  &  &  &  &  &  &  &  &  \\
  & $ \sigma_{0} $ & $1$ &  & $ 0 $ &  & $ 0 $ &  & $2 \text{L}_{\text{max}} + 1$ &  & $ \check{O}_{x'} $ & $14$ &  & $ 1 $ &  & $ 0 $ &  &  $2 \text{L}_{\text{max}} + 1$ \\
 S & $ \check{\Sigma} $ & $4$ &  & $ 2 $ &  & $ -2 $ &  &  $2 \text{L}_{\text{max}} - 1$  & BR & $ \check{O}_{z'} $ & $7$ &  & $ 2 $ &  & $ -1 $ &  &  $2 \text{L}_{\text{max}}$  \\
  & $ \check{T} $ & $10$ &  & $ 1 $ &  & $ -1 $ &  & $2 \text{L}_{\text{max}}$ &  & $ \check{C}_{x'} $ & $16$ &  & $ 1 $ &  & $ 0 $ &   &   $2 \text{L}_{\text{max}} + 1$ \\
  & $ \check{P} $ & $12$ &  & $ 1 $ &  & $ -1 $ &  & $2 \text{L}_{\text{max}}$ &  & $ \check{C}_{z'} $ & $2$ &  & $ 0 $ &  & $ +1 $ &   &  $2 \text{L}_{\text{max}} + 2$ \\
\hline
  &  &  &  &  &  &  &  &  &  &  &  &  &  &  &  &  &  \\
  & $ \check{G} $ & $3$ &  & $ 2 $ &  & $ -2 $ &  & $2 \text{L}_{\text{max}} - 1$ &  & $ \check{T}_{x'} $ & $6$ &  & $ 2 $ &  & $ -1 $ &   &  $2 \text{L}_{\text{max}}$  \\
 BT & $ \check{H} $ & $5$ &  & $ 1 $ &  & $ -1 $ &  & $2 \text{L}_{\text{max}}$ & TR & $ \check{T}_{z'} $ & $13$ &  & $ 1 $ &  & $ 0 $ &  &  $2 \text{L}_{\text{max}} + 1$  \\
  & $ \check{E} $ & $9$ &  & $ 0 $ &  & $ 0 $ &  & $2 \text{L}_{\text{max}} + 1$ &  & $ \check{L}_{x'} $ & $8$ &  & $ 1 $ &  & $ 0 $ &   &  $2 \text{L}_{\text{max}} + 1$  \\
  & $ \check{F} $ & $11$ &  & $ 1 $ &  & $ -1 $ &  & $2 \text{L}_{\text{max}}$ &  & $ \check{L}_{z'} $ & $15$ &  & $ 0 $ &  & $ +1 $ &  &  $2 \text{L}_{\text{max}} + 2$ \\
\hline
\hline
\end{tabular}
\caption{The parameters given here define the angular parametrization (\ref{eq:LowEAssocLegParametrization1}) of the profile functions $\check{\Omega}^{\alpha}$ that arise in a truncated partial wave analysis. The notation is according to reference \cite{LotharNStar}. The parameter $\alpha$ labels the observables according to Chiang and Tabakin \cite{ChTab}.  The parameter $N_{a}$ counts the number of Legendre coefficients yielded by an observable in each truncation order $\text{L}_{\text{max}}$. It is given by $N_{a} = 2 \text{L}_{\text{max}} + \gamma_{\alpha} + 1$.}
\label{tab:AngularDistributions}
\end{table*}
%


\section{Application to recent po\-la\-ri\-za\-tion measurements} \label{sec:Application}\label{sec:DataBasis}
The measurements used for the TPWA fits were mostly obtained by the CBELSA/TAPS experiment. There is data available for the single-po\-la\-ri\-za\-tion observables $T$ and $P$ \cite{Hartmann:2014} and also for the double-po\-la\-ri\-za\-tion observables $E$ \cite{Gottschall:2014}, $G$ \cite{Thiel:2012} and $H$ \cite{Hartmann:2014}. An overview of the measurements and their energy ranges can be found in Tab. \ref{tab:DataBasis}. Since the observables $H$ and $P$ have been extracted utilizing linearly polarized photons, the available photon energy range is limited (compare e.g. \cite{Hartmann:2014}, \cite{Hartmann:2015}).\\
A very recent measurement \cite{AnnandEtAl:2016} of the BT-observable $F$ by the A2-collaboration was also considered. The reference \cite{AnnandEtAl:2016} also shows new data for the target asymmetry $T$. However, since this dataset has a large overlap in energy with the CBELSA/TAPS data for $T$ and is also in good agreement, we decided not to include it into the fits shown in this work. \newline
In a recent publication of the CLAS collaboration on the beam asymmetry $\Sigma$ \cite{Dugger:2013}, fits using associated Legendre polynomials were already performed and the resulting fit coefficients shown. For completeness, we performed the fits as well and we can confirm their results for the fit coefficients (see Fig. \ref{fig:Sclas_bins}). For the energy range of 551 MeV - 1475 MeV the beam asymmetry data of the GRAAL collaboration was used as well (see Fig. \ref{fig:Sgraal_bins}).

In order to extract the $\text{L}_{\text{max}}$, needed to describe the dimensionless po\-la\-ri\-za\-tion observables, the values for $\Omega^\alpha$ have to be multiplied by the unpolarized cross section $\sigma_{0}$. For each energy bin of the different observables, fits were performed according to equations (\ref{eq:LowEAssocLegParametrizationSigma}) to (\ref{eq:LowEAssocLegParametrizationH}). The fits were conducted with $\text{L}_{\text{max}}=1$ up to $\text{L}_{\text{max}}=4$, since no indications of resonances of higher order have been found in the given precision of the experimental data. Some specific datasets with a very high precision, namely the $\sigma_{0}$ and $\Sigma_{\mathrm{CLAS}}$-data, were also studied with $\text{L}_{\text{max}}=5$. To investigate the quality of the fits, the $\chi^{2}/\mathrm{ndf}$ of each fit are plotted against the energy of the incident photon and the CMS energy. Additionally, fits for selected energy bins and the resulting fit coefficients are shown for each observable. The results for the observable $\check{E}$ are given in Fig.~\ref{fig:e_bins}, for the spin dependent cross sections $\sigma^{\left(1/2\right)}$ and $\sigma^{\left(3/2\right)}$ derivable from $E$ (see sec. \ref{sec:SpinDependentDCS}) in Figures \ref{fig:s12_bins} and \ref{fig:s32_bins}, for $\check{G}$ in Fig.~\ref{fig:g_bins}, for $\check{H}$ in Fig.~\ref{fig:h_bins}, for $\check{P}$ in Fig.~\ref{fig:P_bins}, for $\check{T}$ in Fig.~\ref{fig:T_bins}, for $\check{\Sigma}_{\mathrm{GRAAL}}$ in Fig.~\ref{fig:Sgraal_bins} and for $\check{\Sigma}_{\mathrm{CLAS}}$ in Fig.~\ref{fig:Sclas_bins}.

\begin{table}
\RawFloats
\centering
\begin{tabular}{c|c|c|c}
\hline
\hline
Observable & Number of   & Energy Range & Reference(s) \\
 &  Energy Bins & $E_\gamma$ [MeV] &\\\hline
$\sigma_0$ & 266   & 218 - 1573 & \cite{Adlarson:2015}\\
$\Sigma$ & 70   & 551 - 1880 & \cite{GRAAL,Dugger:2013}\\
$T$ &  24  & 700 - 1900 & \cite{Hartmann:2014,Hartmann:2015}\\
$P$ &  8  & 650 - 950 & \cite{Hartmann:2014,Hartmann:2015}\\
$G$ & 19   & 630 - 1350 & \cite{Thiel:2012,Thiel:2015}\\
$E$ & 33   & 600 - 2300 & \cite{Gottschall:2014,Gottschall:2015}\\
$H$ &  8  & 650 - 950 & \cite{Hartmann:2014,Hartmann:2015}\\
$F$ &  34  & 440 - 1430 & \cite{AnnandEtAl:2016}\\
\hline\hline
\end{tabular} 
\caption{Recent new data for different observables in $\pi^0$ photoproduction from GRAAL, ELSA, JLab and MAMI.}
\label{tab:DataBasis}
\end{table}%

\subsection{Utilization of the Bonn-Gatchina cross section} \label{sec:BnGaDCSJustification}
\begin{figure}[htb]
\RawFloats
  \includegraphics[width=0.46\textwidth]{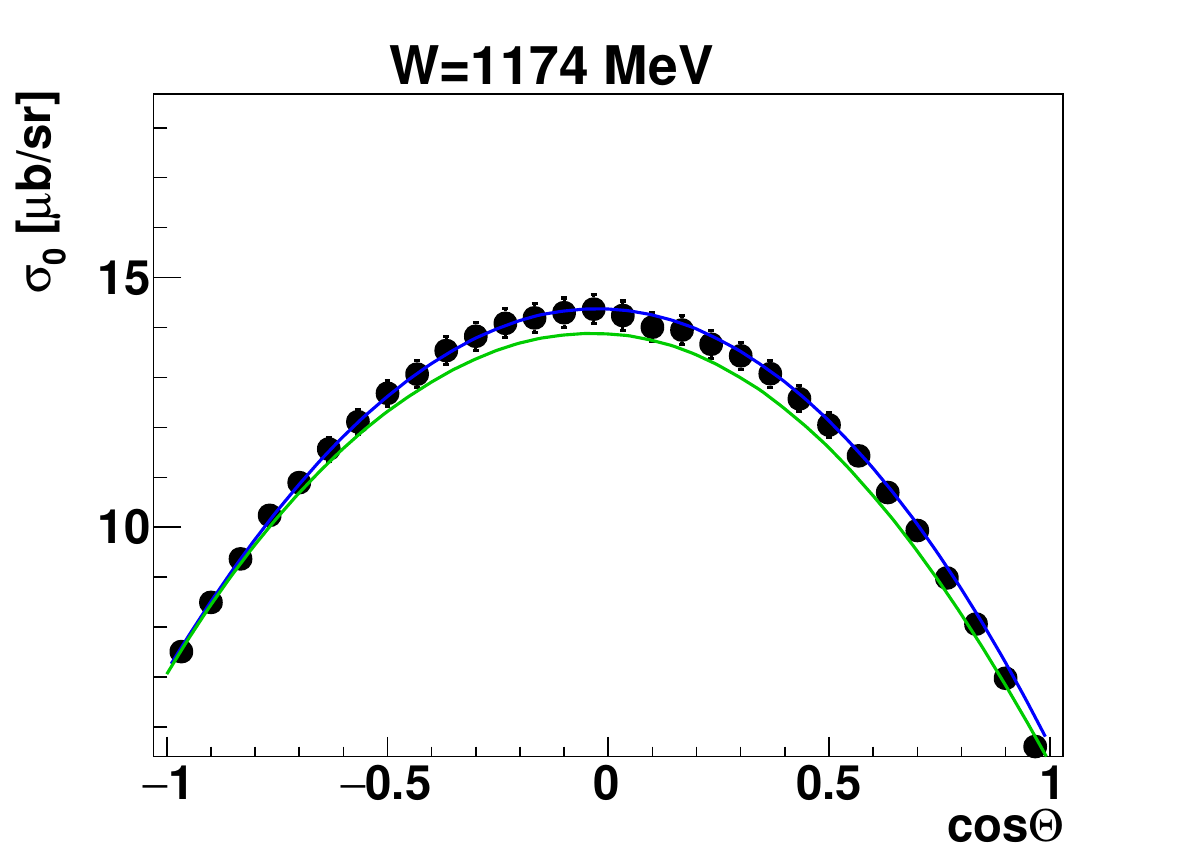}
  \includegraphics[width=0.46\textwidth]{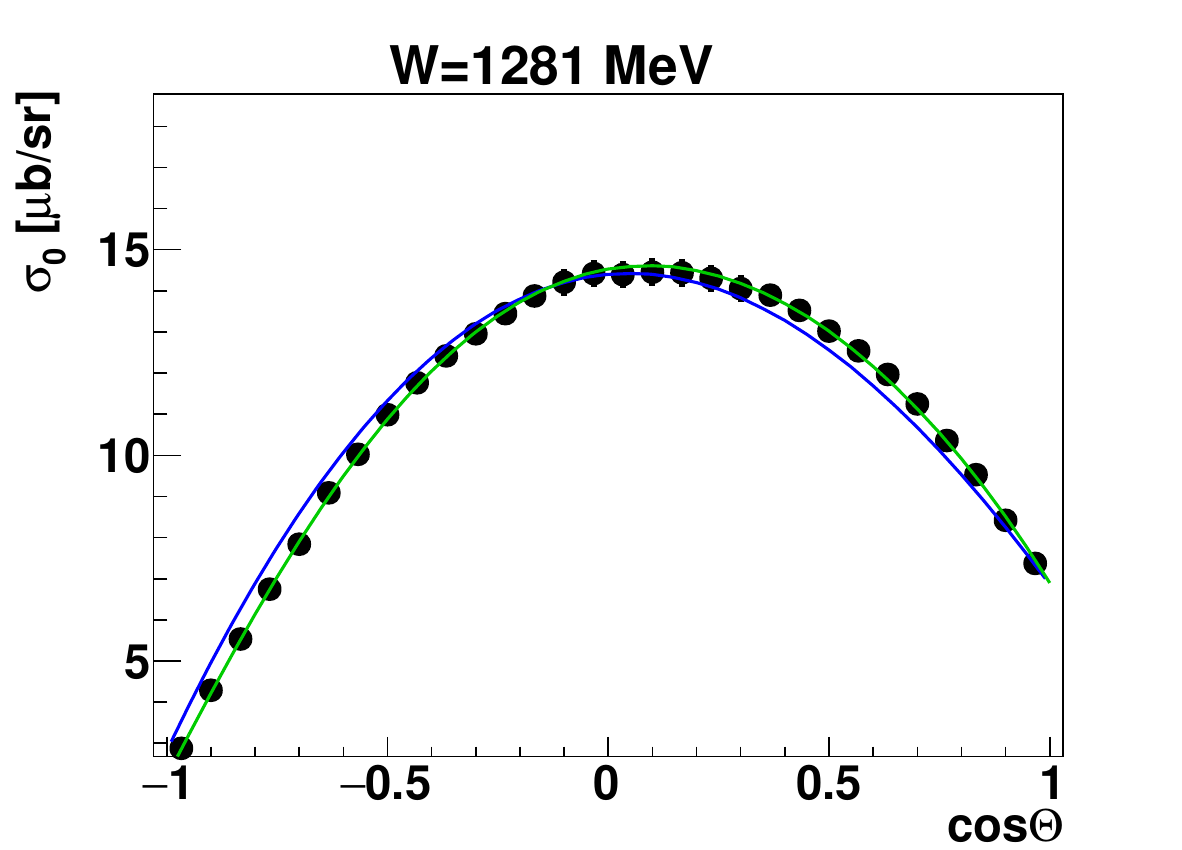}\\
  \includegraphics[width=0.46\textwidth]{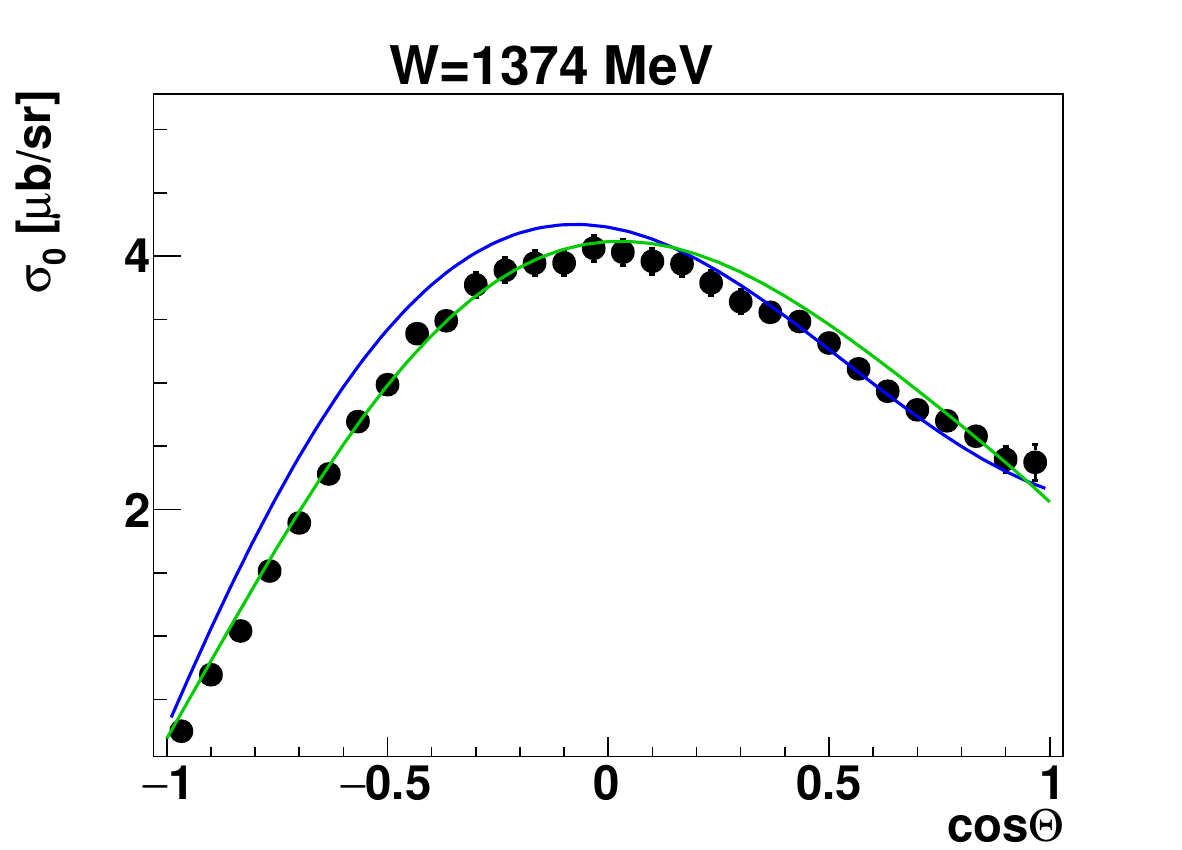}
  \includegraphics[width=0.46\textwidth]{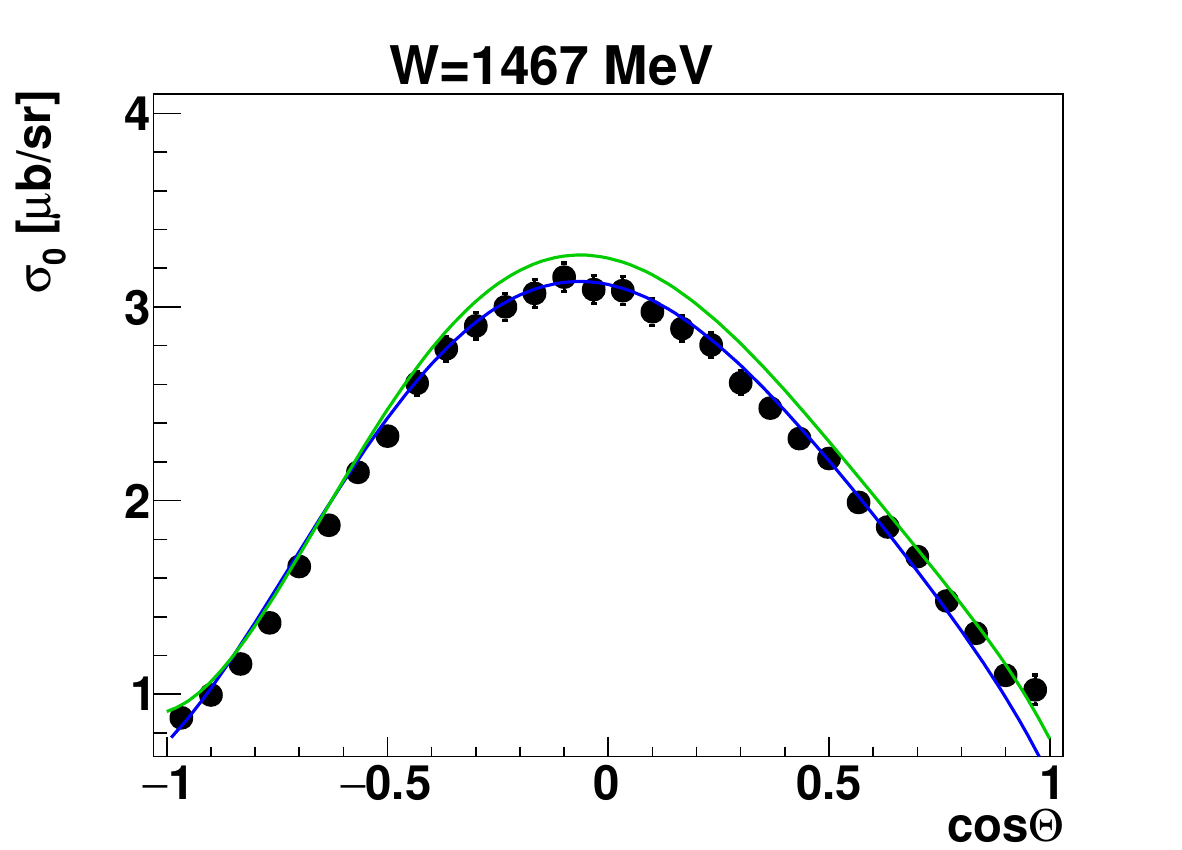}\\
  \includegraphics[width=0.46\textwidth]{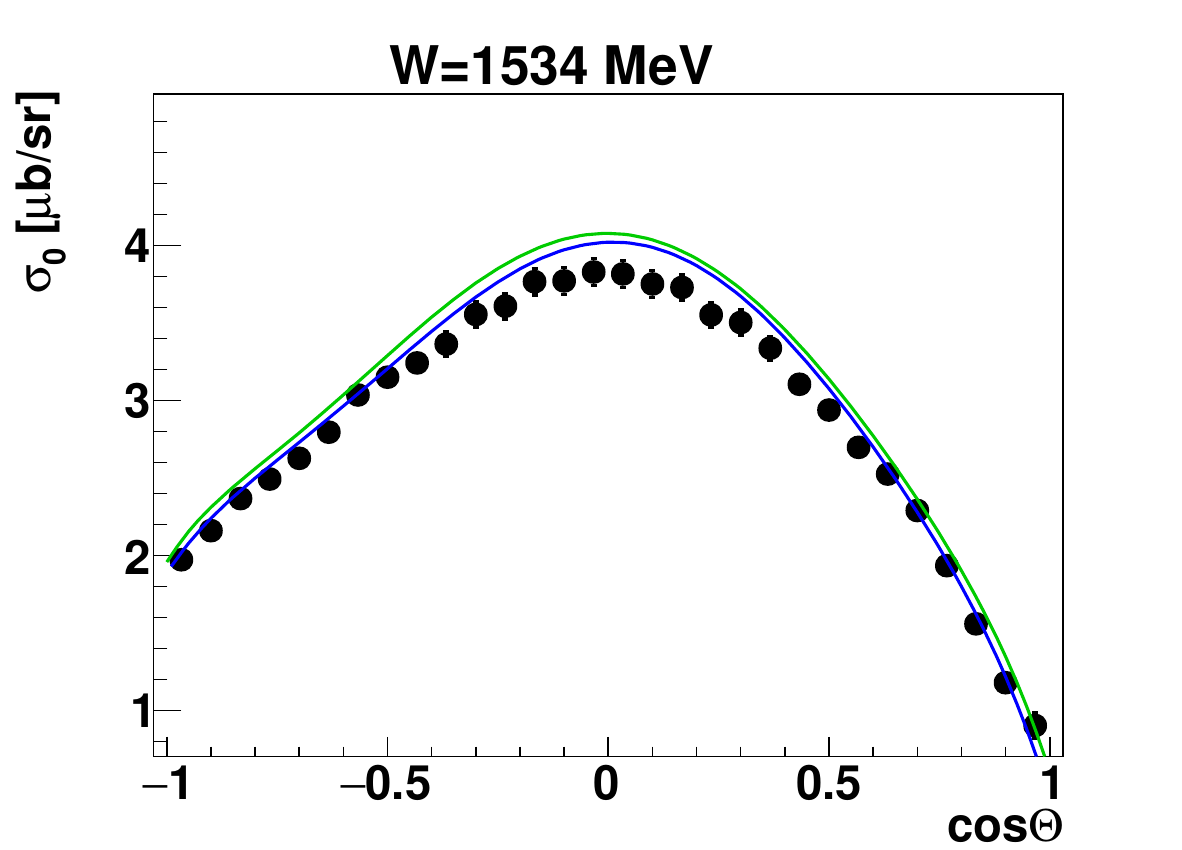}
  \includegraphics[width=0.46\textwidth]{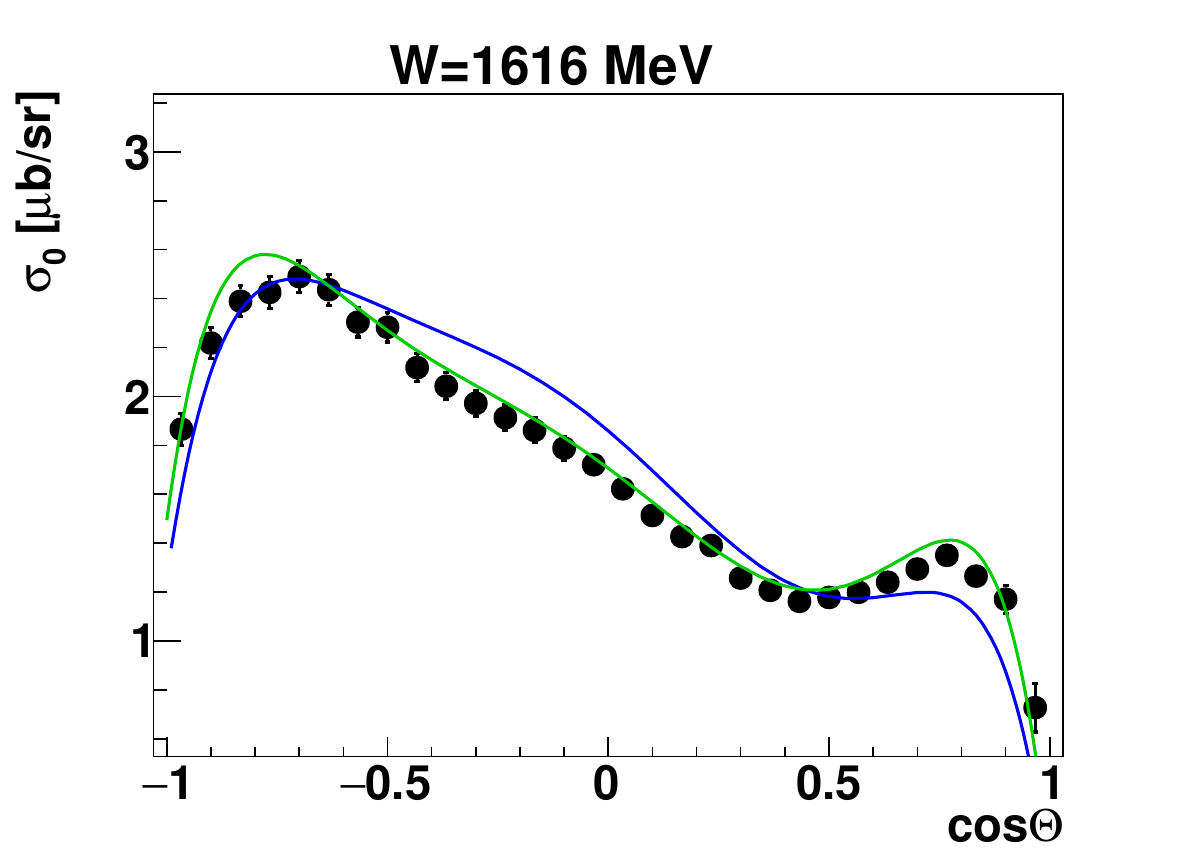}\\
  \includegraphics[width=0.46\textwidth]{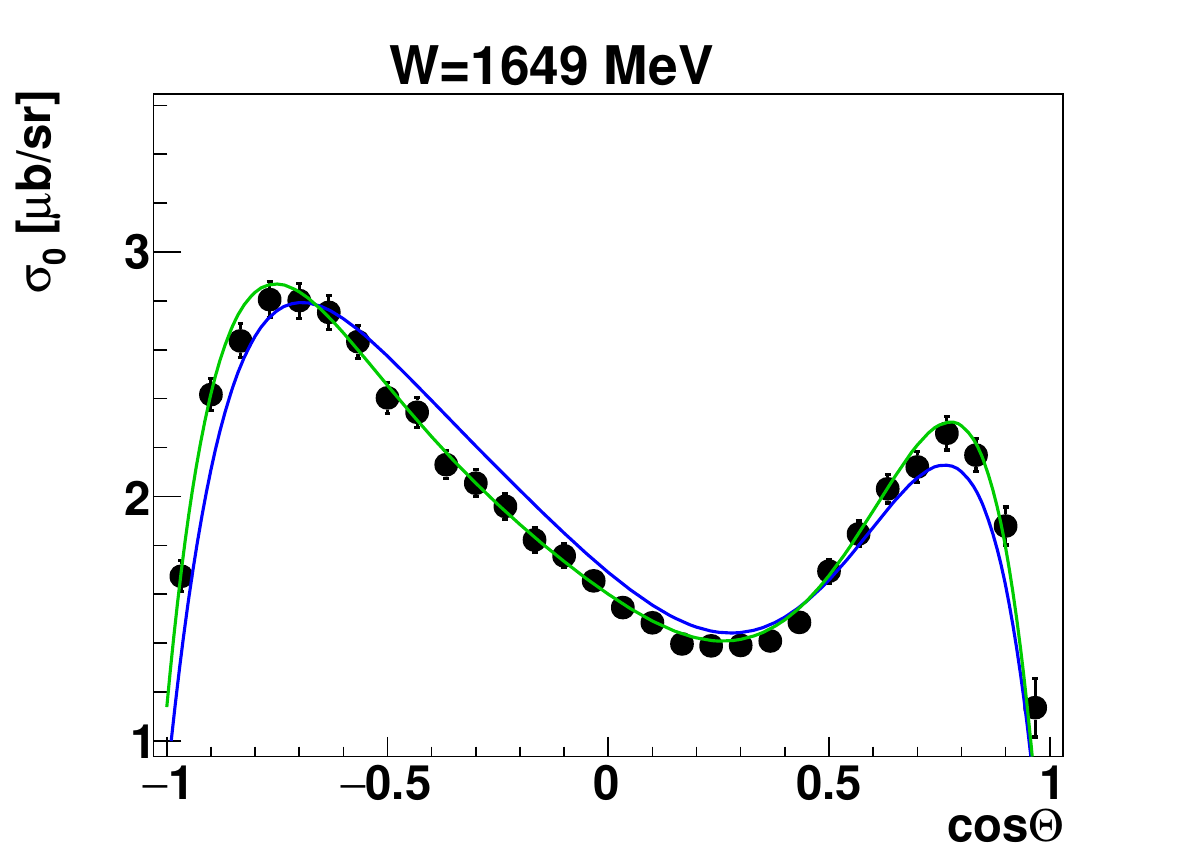}
  \includegraphics[width=0.46\textwidth]{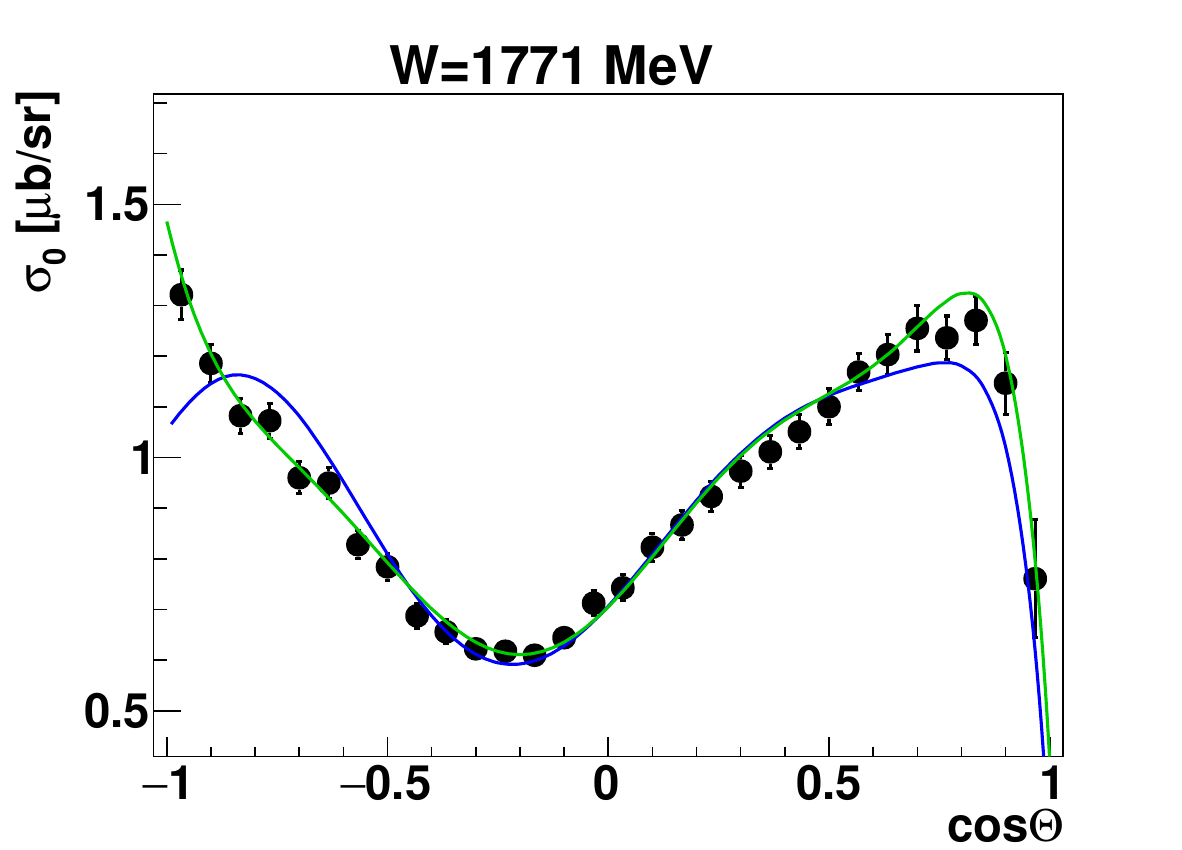}\\
  \includegraphics[width=0.46\textwidth]{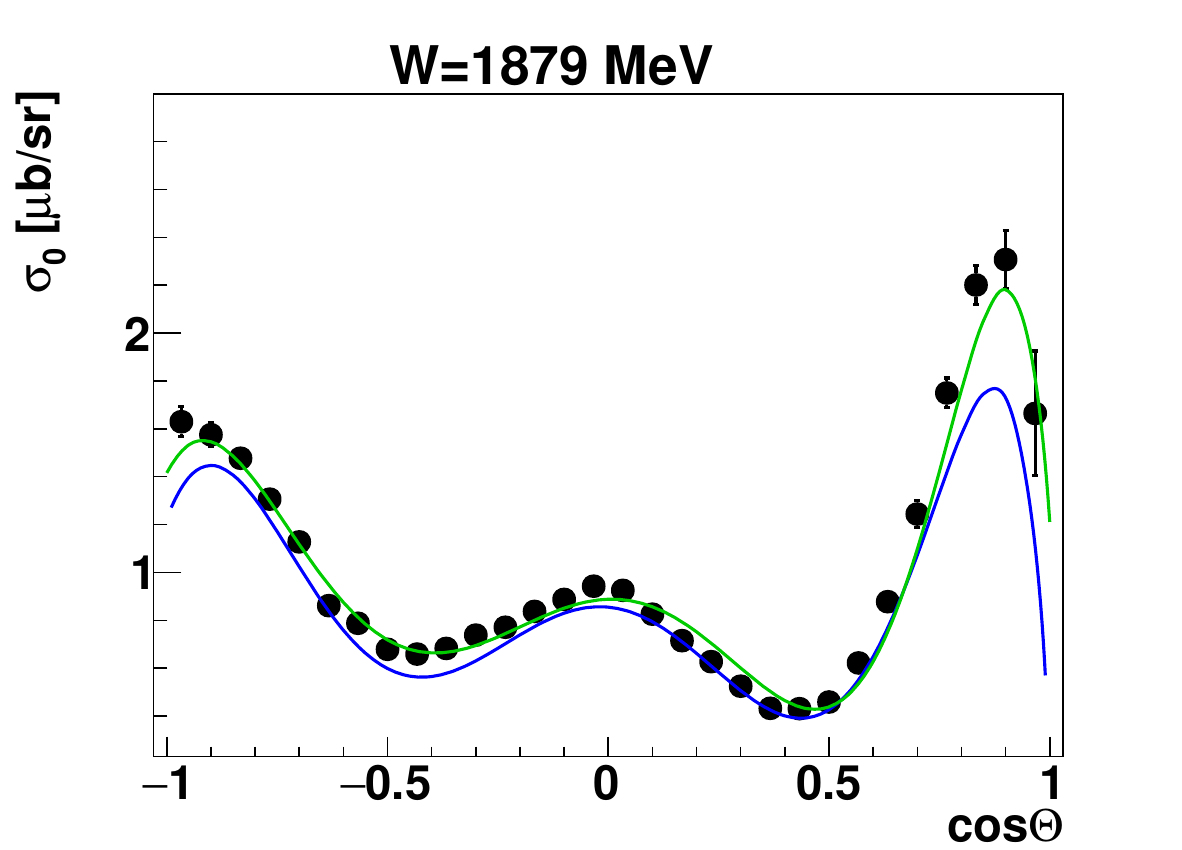}
  \includegraphics[width=0.46\textwidth]{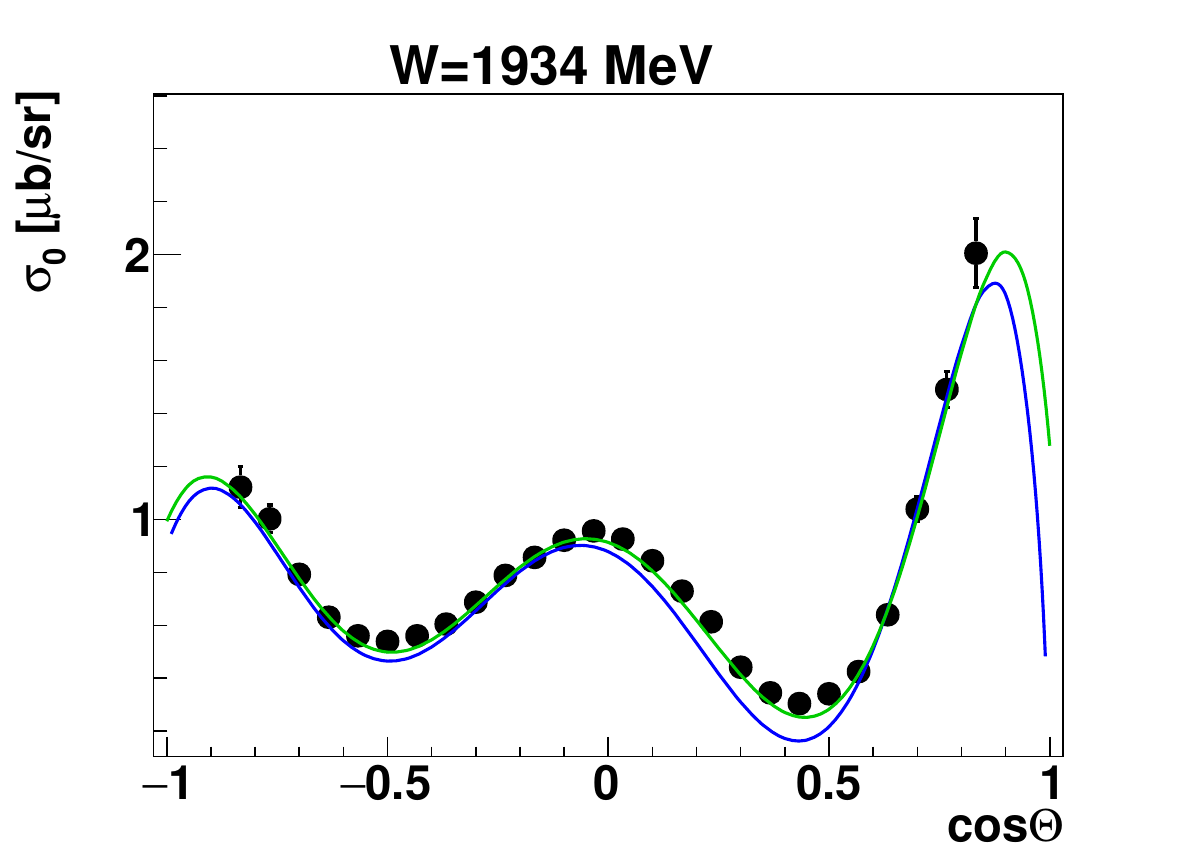}
\caption{Comparison of the BnGa2014-02 PWA solution (blue line) \cite{Bonn-Gatchina_2014} as well as the SAID PR15 solution \cite{Adlarson:2015} (green line) to the differential cross section $\sigma_0$ data measured by the A2 collaboration \cite{Adlarson:2015} for different energy bins. The new cross section data were included in the SAID PR15 fit but not in the BnGa2014-02 fit \cite{Bonn-Gatchina_2014}.}
\label{fig:wq0}       
\end{figure}

As already mentioned, the dimensionless observables $\Omega^{\alpha}$ need to be multiplied by the unpolarized cross section in order to determine the profile functions $\check{\Omega}^\alpha$.
The A2 collaboration recently published cross section data with a very high precision, covering a beam energy $\left(E_{\gamma}^{\mathrm{LAB}}\right)$ range of 218 MeV -1573 MeV in 4 MeV steps and the entire angular range with 30 data points in almost every energy bin \cite{Adlarson:2015}. This data set represents the most precise measurement of the $\gamma p \to \pi^{0} p $ cross section to date. In Fig. \ref{fig:wq0} a direct comparison of data and two PWAs for $\pi^{0}$ photoproduction is shown. The Bonn-Gatchina solution BnGa$2014$-$02$ \cite{Bonn-Gatchina_2014}, which has not used the A2 data for $\sigma_{0}$ as input, as well as a new fit from SAID, SAID PR15 \cite{Adlarson:2015}, which already fitted the A2 data can be seen. In Fig.\ref{fig:wq_bins} on the other hand, our fits to the A2 cross section are shown. \newline 

Upon inspection of Fig. \ref{fig:wq0}, it can be seen that some discrepancies between data and PWA descriptions are present. The SAID solution is closer to the data in most energy bins and especially in the higher region for $W \geq 1771 \hspace*{2pt} \mathrm{MeV}$, it is seen to correctly follow the structures of the data at the angular boundary regions. This however should come as no surprise since the SAID solution used the new A2 data as input, whereas Bonn-Gatchina has fitted older datasets to obtain the solution BnGa$2014$-$02$. On the contrary, for the latter solution the Bonn-Gatchina group has also utilized almost all of the po\-la\-ri\-za\-tion data investigated in this work (which the SAID fit has not). \newline
Therefore we would like to justify our choice of using the Bonn-Gatchina PWA cross section for the calculation of the profile functions $\check{\Omega}^{\alpha}$. One could of course choose either the original A2 data, or the SAID-cross section for this purpose. But then all discrepancies from Fig. \ref{fig:wq0} which would be corrected for $\sigma_{0}$ would turn up again in a ripple effect once the po\-la\-ri\-za\-tion data are investigated. \newline
Furthermore, the uncertainties published with the new A2 data cause inconsistencies once the results of fitted angular distributions are compared to PWAs (this is said in anticipation of section \ref{sec:Interpretation1stResRegion}).
Also, in the energy regions of all po\-la\-ri\-za\-tion observables investigated in this work except $\sigma_{0}$, i.e. the kinematic regime where the differential cross section is needed to evaluate profile functions, it can be seen that the Bonn-Gatchina cross section describes the data of the A2 collaboration rather well (cf. section \ref{sec:Interpretation1stResRegion}, especially the Legendre coefficients plotted in Figures \ref{fig:wq_bins} and \ref{fig:wq_bins_wsys}). Below the lowest energy point where a profile function is needed, $E_{\gamma}^{\mathrm{LAB}} = 551 \hspace*{2pt} \hspace*{2pt} \mathrm{MeV}$ ($W = 1385 \hspace*{2pt} \hspace*{2pt} \mathrm{MeV}$) from the $\Sigma$ measurements (Table \ref{tab:DataBasis}), there are discrepancies in the Legendre coefficient plots of Fig. \ref{fig:wq_bins} between the Bonn-Gatchina prediction and the A2 measurement. These are however irrelevant for the evaluation of the $\check{\Omega}^{\alpha}$. \newline
Therefore, at least as long as no new Bonn-Gatchina fit including the A2 cross section is avaliable, we still prefer the solution BnGa$2014$-$02$ for the evaluation of the profile functions. The effect of disregarding the $\sigma_{0}$ errors in the error propagations is expected to be negligible.

%
%

\subsection{Spin-dependent cross section} \label{sec:SpinDependentDCS}
The double po\-la\-ri\-za\-tion observable $E$ is the difference of two differential cross sections corresponding to states of the definite spin-z projections $1/2$ and $3/2$ in the CMS (see \cite{Gottschall:2014}). By using the unpolarized cross section $\sigma_0$, it is possible to extract the spin dependent cross sections $\sigma_{1/2}$ and $\sigma_{3/2}$ from the observable $E$ following 
\begin{align}
\sigma^{(1/2 | 3/2)}&=  \sigma_0 \cdot (1 \pm  E )\mathrm{~~~with}\\
\sigma_{1/2}+\sigma_{3/2} &= 2\sigma_0\mathrm{.}
\end{align}
Since only partial waves with total angular momentum quantum number equal or larger than $3/2$ can contribute to $\sigma_{3/2}$, additional information can be extracted by comparing the two different spin-z states. Similarly to the double-po\-la\-ri\-za\-tion observable $E$, the cross sections $\sigma_{1/2}$ and $\sigma_{3/2}$ can be written in a CGLN representation:
\begin{align}
 \sigma_{3/2} &= \rho \hspace*{3pt} \sin^2(\theta)\hspace*{4pt} \mathrm{Re} \Big[\frac{1}{2}\left(|F_3|^2+|F_4|^2\right)+\cos(\theta) F_3^\ast F_4 \Big]\\
\sigma_{1/2} &= \sigma_{3/2}+ 2 \rho \hspace*{4pt}\mathrm{Re} \Big[\left(|F_1|^2+|F_2|^2\right)-2\cos(\theta) F_1^\ast F_2 \nonumber\\
&+ \sin^2(\theta)(F_2^\ast F_3 + F_1^\ast F_4) \Big]\mathrm{.}
\end{align}
For a truncated PWA, the cross sections can be fitted following equation (\ref{eq:LowEAssocLegParametrization1}) by
\begin{align}
 \sigma_{3/2} &= \rho \hspace*{3pt} \sum \limits_{k = 2}^{2 \text{L}_{\text{max}}} \left(a_{\text{L}_{\text{max}}}\right)_{k}^{\sigma_{3/2}} \left( W \right) P^{2}_{k} \left( \cos \theta \right)\mathrm{,} \label{eq:LowEAssocLegParametrizationDCS1/2}\\
 \sigma_{1/2} &= \rho \hspace*{3pt} \sum \limits_{k = 0}^{2 \text{L}_{\text{max}}} \left(a_{\text{L}_{\text{max}}}\right)_{k}^{\sigma_{1/2}} \left( W \right) P_{k} \left( \cos \theta \right)\mathrm{.}\label{eq:LowEAssocLegParametrizationDCS3/2}
\end{align}
The results of the fits to $\sigma_{3/2}$ and $\sigma_{1/2}$ can be found in Fig. ~\ref{fig:s12_bins} and Fig. ~\ref{fig:s32_bins}. Since all resonant and background contributions from partial waves with $J=1/2$ are purged from $\sigma_{3/2}$, this observable will be particularly useful for detecting higher partial waves via interferences (cf. Sections \ref{sec:Interpretation2ndResRegion} to \ref{sec:Interpretation4thResRegion}).

\section{Legendre coefficients in terms of multipoles}\label{sec:DescriptionCompositionLegCoeffs}

Before commencing with the practical interpretation of fits to physical data, we would like to use this section in order to give more specifics on the partial wave contributions contained in the observables and to describe the composition of the Legendre coefficients in terms of multipoles. We give here a detailed description of $\check{E}$ for $\text{L}_{\text{max}} = 3$ ($F$-waves), since this order will also be encountered in the interpretation contained in the following sections. In appendix \ref{sec:PWContentFormulas}, the contributions to all considered observables are shown up to $\text{L}_{\text{max}} = 5$ ($H$-waves).
Truncating at the $F$-waves, the angular pa\-ra\-me\-tri\-za\-tion (\ref{eq:LowEAssocLegParametrizationE}) of $\check{E}$ reads
\begin{align}
 \check{E} &= \rho \hspace*{3pt} \Big( \left(a_{3}\right)^{\check{E}}_{0} \hspace*{1.5pt} P_{0} ( \cos \theta ) + \left(a_{3}\right)^{\check{E}}_{1} \hspace*{1.5pt} P_{1} ( \cos \theta )  \nonumber \\
 & \hspace*{22.5pt} + \left(a_{3}\right)^{\check{E}}_{2} \hspace*{1.5pt} P_{2} ( \cos \theta ) + \left(a_{3}\right)^{\check{E}}_{3} \hspace*{1.5pt} P_{3} ( \cos \theta ) \nonumber \\
 & \hspace*{22.5pt} + \left(a_{3}\right)^{\check{E}}_{4} \hspace*{1.5pt} P_{4} ( \cos \theta ) + \left(a_{3}\right)^{\check{E}}_{5} \hspace*{1.5pt} P_{5} ( \cos \theta ) \nonumber \\
 & \hspace*{22.5pt} + \left(a_{3}\right)^{\check{E}}_{6} \hspace*{1.5pt} P_{6} ( \cos \theta ) \Big) \mathrm{.} \label{eq:ExpEIntoLegLmax2}
\end{align}
The coefficient $\left(a_{3}\right)^{\check{E}}_{0} = \left< \mathcal{M}_{3} \right| \mathcal{C}_{0}^{\check{E}} \left| \mathcal{M}_{3} \right>$ is evaluated as an example in Table \ref{tab:CoeffsEright1}.

\begin{table*}[htb]
\RawFloats
\begin{small}
\begin{align}
\left(a_{2}\right)_{0}^{\check{E}} &= 
\left[ \begin{array}{ccccccccc} E_{0+}^{\ast} & E_{1+}^{\ast} & M_{1+}^{\ast} & M_{1-}^{\ast} & \ldots & E_{3+}^{\ast} & E_{3-}^{\ast} & M_{3+}^{\ast} & M_{3-}^{\ast} \end{array} \right]
\left[
\begin{array}{c|ccc|cccc|cccc}
 1 & 0 & 0 & 0 & 0 & 0 & 0 & 0 & 0 & 0 & 0 & 0 \\ \hline
 0 & 3 & 3 & 0 & 0 & 0 & 0 & 0 & 0 & 0 & 0 & 0 \\
 0 & 3 & -1 & 0 & 0 & 0 & 0 & 0 & 0 & 0 & 0 & 0 \\
 0 & 0 & 0 & 1 & 0 & 0 & 0 & 0 & 0 & 0 & 0 & 0 \\ \hline
 0 & 0 & 0 & 0 & 6 & 0 & 12 & 0 & 0 & 0 & 0 & 0 \\
 0 & 0 & 0 & 0 & 0 & -1 & 0 & -3 & 0 & 0 & 0 & 0 \\
 0 & 0 & 0 & 0 & 12 & 0 & -3 & 0 & 0 & 0 & 0 & 0 \\
 0 & 0 & 0 & 0 & 0 & -3 & 0 & 3 & 0 & 0 & 0 & 0 \\  \hline
 0 & 0 & 0 & 0 & 0 & 0 & 0 & 0 & 10 & 0 & 30 & 0 \\
 0 & 0 & 0 & 0 & 0 & 0 & 0 & 0 & 0 & -3 & 0 & -12 \\
 0 & 0 & 0 & 0 & 0 & 0 & 0 & 0 & 30 & 0 & -6 & 0 \\
 0 & 0 & 0 & 0 & 0 & 0 & 0 & 0 & 0 & -12 & 0 & 6
\end{array}
\right]
\left[ \begin{array}{c} E_{0+} \\ E_{1+} \\ M_{1+} \\ M_{1-} \\ E_{2+} \\ E_{2-} \\ M_{2+} \\ M_{2-} \\ E_{3+} \\ E_{3-} \\ M_{3+} \\ M_{3-}  \end{array} \right] \nonumber \\
 &= \left[ \begin{array}{ccccccccc} E_{0+}^{\ast} & E_{1+}^{\ast} & M_{1+}^{\ast} & M_{1-}^{\ast} & \ldots & E_{3+}^{\ast} & E_{3-}^{\ast} & M_{3+}^{\ast} & M_{3-}^{\ast} \end{array} \right]
\left[ \begin{array}{c} E_{0+} \\  3 E_{1+} + 3 M_{1+} \\  3 E_{1+} - M_{1+} \\  M_{1-} \\  6 E_{2+} + 12 M_{2+} \\ - E_{2-} - 3 M_{2-} \\  12 E_{2+} - 3 M_{2+} \\- 3 E_{2-} + 3 M_{2-} \\  10 E_{3+} + 30 M_{3+} \\ - 3 E_{3-} - 12 M_{3-} \\  30 E_{3+} - 6 M_{3+} \\- 12 E_{3-} + 6 M_{3-} \end{array} \right] \nonumber \\
 &= \left|E_{0+}\right|^2+3 \left| E_{1+} \right|^{2} + 3 E_{1+}^{\ast} M_{1+} + 3 M_{1+}^{\ast} E_{1+} -\left| M_{1+} \right|^{2}+\left|M_{1-}\right|^2+6 \left| E_{2+} \right|^{2} + 12 E_{2+}^{\ast} M_{2+} - \left| E_{2-} \right|^{2} - 3 E_{2-}^{\ast} M_{2-}  \nonumber \\
 & \hspace*{10pt} +12 M_{2+}^{\ast} E_{2+} -3 \left| M_{2+} \right|^{2} -3 M_{2-}^{\ast} E_{2-} + 3 \left| M_{2-} \right|^{2} + 10 \left| E_{3+} \right|^{2} + 30 E_{3+}^{\ast} M_{3+}  - 3 \left| E_{3-} \right|^{2} - 12 E_{3-}^{\ast} M_{3-}  \nonumber \\
 & \hspace*{10pt} + 30 M_{3+}^{\ast} E_{3+} - 6 \left| M_{3+}\right|^{2}  - 12 M_{3-}^{\ast} E_{3-} + 6 \left|M_{3-}\right|^{2}   \nonumber
\end{align}
\end{small}
\caption{The coefficient $\left(a_{3}\right)_{0}^{\check{E}}$ for an expansion of $\check{E}$ up to $\text{L}_{\text{max}} = 3$ is evaluated explicitly in terms of the matrix $\mathcal{C}_{0}^{\check{E}}$.}
\label{tab:CoeffsEright1}
\end{table*}
It is written first as defined by the symmetric matrix $\mathcal{C}_{0}^{\check{E}}$ and then evaluated as an expression in terms of bilinear products of multipoles. Expressions of the latter kind may be found at other places in the literature (see eg. \cite{FTS}). \newline
We also use this example to clarify our convention for sorting the complex multipoles into vectors. Table \ref{tab:CoeffsEright1} contains the example for $\text{L}_{\text{max}} = 3$, i.e.:
\begin{align}
\left| \mathcal{M}_{3} \right> &= \Big( E_{0+}, E_{1+}, M_{1+}, M_{1-}, E_{2+}, E_{2-}, M_{2+}, M_{2-}, \nonumber \\
 & \hspace*{20pt} E_{3+}, E_{3-}, M_{3+}, M_{3-} \Big)^{T} \mathrm{.} \label{eq:MultipoleVectorLmax2}
\end{align}
%
In order to save space, the partial wave compositions for all observables investigated in this work are given in a shorter graphical representation. All Legendre coefficients are included here for a truncation at the $F$-waves. \newline 
An explicit example, as well as a more detailed explanation of the color scheme, are provided in Table \ref{tab:CoeffsEright2}.  Each small colored block represents a matrix entry. Positive matrix entries have red color, negative ones are blue.

\begin{table*}[htb]
\RawFloats
\begin{minipage}{.5\linewidth}
\begin{equation}
\hspace*{30pt} \mathcal{C}_{0}^{\check{E}} = 
\left[
\begin{array}{c|ccc|cccc|cccc}
 1 & 0 & 0 & 0 & 0 & 0 & 0 & 0 & 0 & 0 & 0 & 0 \\ \hline
 0 & 3 & 3 & 0 & 0 & 0 & 0 & 0 & 0 & 0 & 0 & 0 \\
 0 & 3 & -1 & 0 & 0 & 0 & 0 & 0 & 0 & 0 & 0 & 0 \\
 0 & 0 & 0 & 1 & 0 & 0 & 0 & 0 & 0 & 0 & 0 & 0 \\ \hline
 0 & 0 & 0 & 0 & 6 & 0 & 12 & 0 & 0 & 0 & 0 & 0 \\
 0 & 0 & 0 & 0 & 0 & -1 & 0 & -3 & 0 & 0 & 0 & 0 \\
 0 & 0 & 0 & 0 & 12 & 0 & -3 & 0 & 0 & 0 & 0 & 0 \\
 0 & 0 & 0 & 0 & 0 & -3 & 0 & 3 & 0 & 0 & 0 & 0 \\  \hline
 0 & 0 & 0 & 0 & 0 & 0 & 0 & 0 & 10 & 0 & 30 & 0 \\
 0 & 0 & 0 & 0 & 0 & 0 & 0 & 0 & 0 & -3 & 0 & -12 \\
 0 & 0 & 0 & 0 & 0 & 0 & 0 & 0 & 30 & 0 & -6 & 0 \\
 0 & 0 & 0 & 0 & 0 & 0 & 0 & 0 & 0 & -12 & 0 & 6 \\
\end{array}
\right] \equiv \nonumber
\end{equation}
\end{minipage}
\begin{minipage}{.6\linewidth} \vspace*{0.7cm} \includegraphics[width=0.7\textwidth]{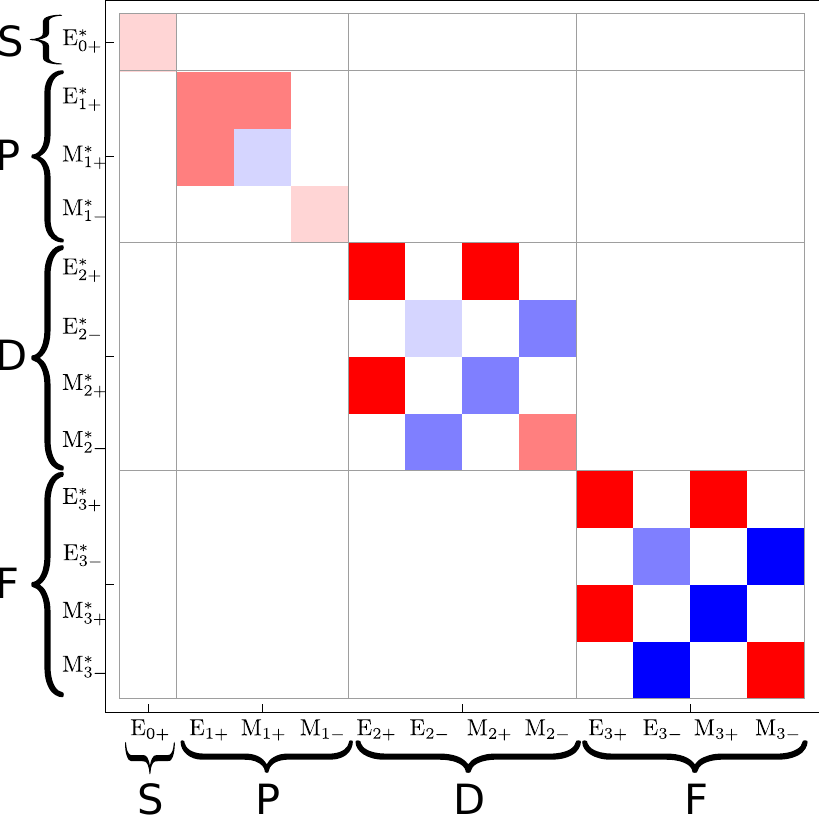} \end{minipage}
\caption{Matrix defining the coefficient $\left(a_{3}\right)_{0}^{\check{E}}$ for an expansion of $\check{E}$ up to $\text{L}_{\text{max}} = 3$. The color-scheme mentioned in the main text is exemplified here. In the color-plot on the right hand side, the individual multipoles contributing to every interference-term are indicated explicitly. Every matrix element corresponds to a particular interference term (see Table \ref{tab:CoeffsEright1}).}
\label{tab:CoeffsEright2}
\end{table*}
%
%

\begin{table*}[htb]
\RawFloats
\begin{minipage}{.075\linewidth}
\vspace*{-6.5pt}
\hspace*{5pt}
\begin{equation}
\mathcal{C}_{0}^{\check{E}} \equiv \nonumber
\end{equation}
\end{minipage}
\begin{minipage}{.3\linewidth} \vspace*{0.572cm} \includegraphics[width=0.875\textwidth]{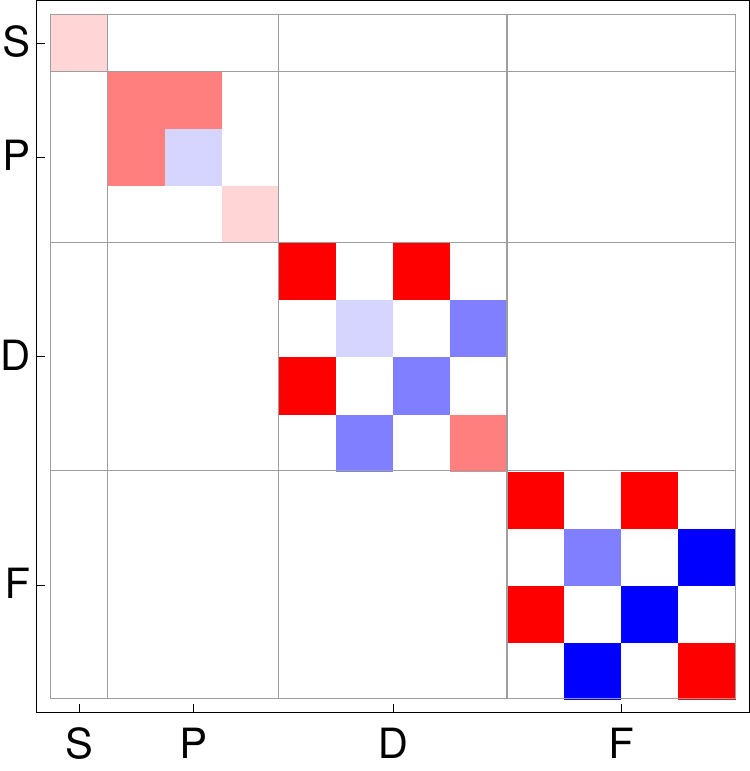} \end{minipage}
\begin{minipage}{.35\linewidth} \vspace*{0.500cm} \hspace*{-0.65cm}\includegraphics[width=1.15\textwidth]{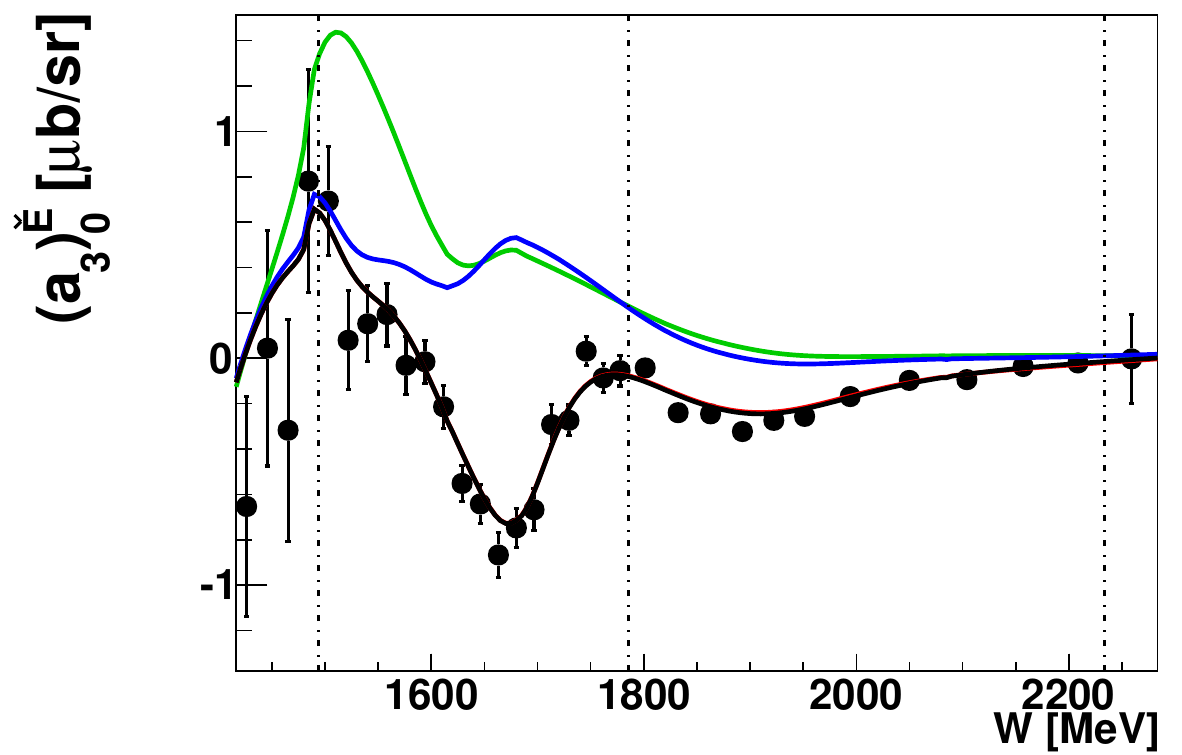}\end{minipage}
\begin{minipage}{.25\linewidth} \begin{align} \left(a_{3}\right)^{\check{E}}_{0} &= \left<S,S\right> + \left<P,P\right> \nonumber \\ & \hspace*{12.5pt} + \left<D,D\right> + \left<F,F\right> \nonumber \end{align} \end{minipage}
\caption{Left: Matrix $\mathcal{C}_{0}^{\check{E}}$, represented here in the color scheme, defines the coefficient $\left(a_{3}\right)_{0}^{\check{E}}$ for an expansion of $\check{E}$ up to $\text{L}_{\text{max}} = 3$, see also Table \ref{tab:CoeffsEright1}. Center: Coefficient $\left(a_{3}\right)_{0}^{\check{E}}$ obtained from a fit to the $E$-data (black points). For references to the data see Table \ref{tab:DataBasis}. Bonn Gatchina predictions, truncated at different $\text{L}_{\mathrm{max}}$ ($\text{L}_{\mathrm{max}} = 1$ is drawn in green, $\text{L}_{\mathrm{max}} = 2$ in blue, $\text{L}_{\mathrm{max}} = 3$ in red and $\text{L}_{\mathrm{max}} = 4$ in black) are drawn as well. Right: All partial wave interferences for $\text{L}_{\text{max}} = 3$ are indicated in the notation of equations (\ref{eq:EScalarProductCoeff0}) to (\ref{eq:EScalarProductCoeff6}).}
\label{tab:CoeffsEright3}
\end{table*}
\begin{table*}[htb]
\RawFloats
\begin{minipage}{.075\linewidth}
\vspace*{-6.5pt}
\hspace*{5pt}
\begin{equation}
\mathcal{C}_{1}^{\check{E}} \equiv \nonumber
\end{equation}
\end{minipage}
\begin{minipage}{.3\linewidth} \vspace*{0.572cm} \includegraphics[width=0.875\textwidth]{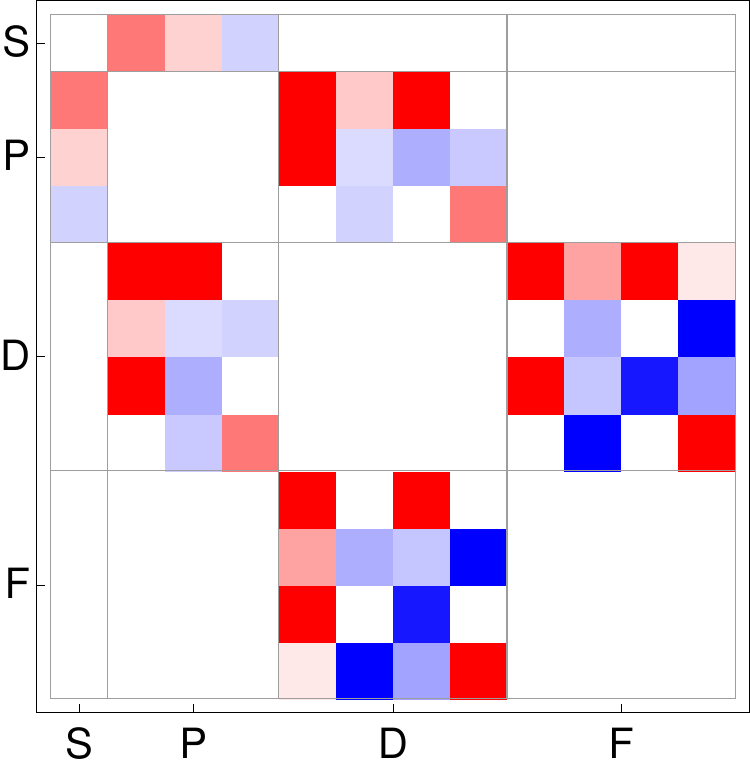} \end{minipage}
\begin{minipage}{.35\linewidth} \vspace*{0.500cm} \hspace*{-0.65cm}\includegraphics[width=1.15\textwidth]{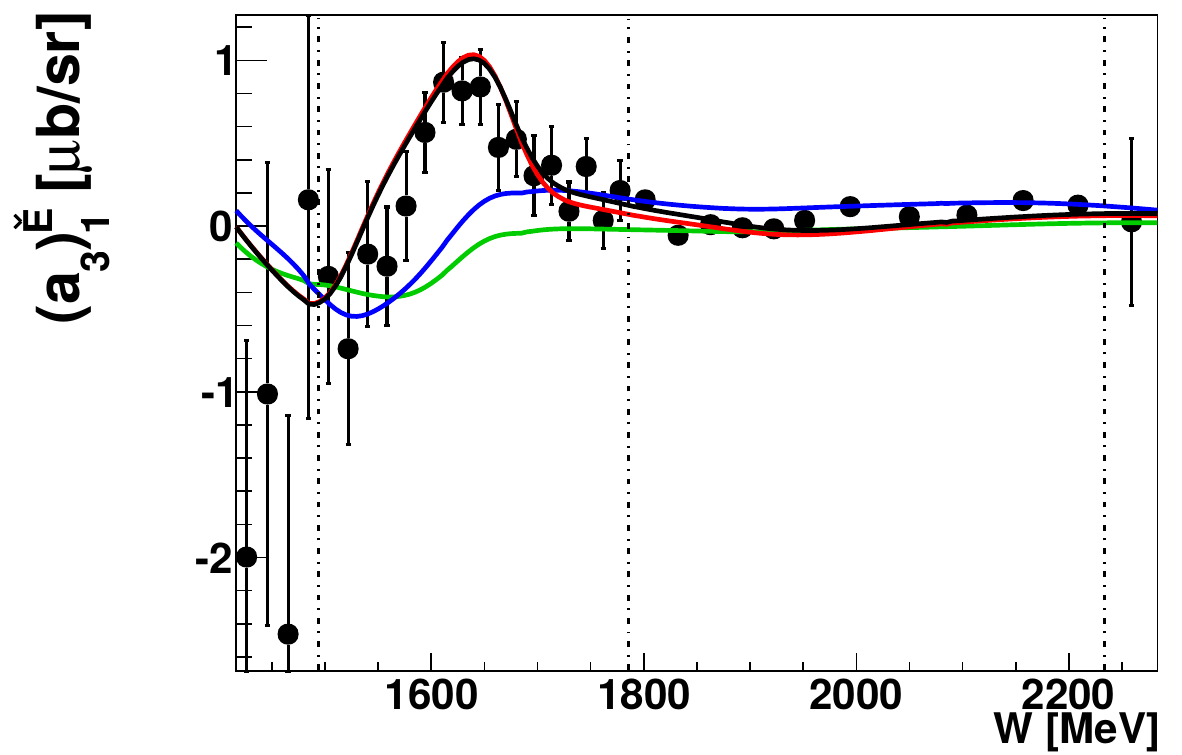}\end{minipage}
\begin{minipage}{.25\linewidth} \begin{align} \left(a_{3}\right)^{\check{E}}_{1} &= \left<S,P\right> + \left<P,D\right> \nonumber \\ & \hspace*{12.5pt} + \left<D,F\right> \nonumber \end{align} \end{minipage}
\caption{Left: Matrix $\mathcal{C}_{1}^{\check{E}}$, represented here in the color scheme, defines the coefficient $\left(a_{3}\right)_{1}^{\check{E}}$ for an expansion of $\check{E}$ up to $\text{L}_{\text{max}} = 3$, see also Table \ref{tab:CoeffsEright1}. Center: Coefficient $\left(a_{3}\right)_{1}^{\check{E}}$ obtained from a fit to the $E$-data (black points). For references to the data see Table \ref{tab:DataBasis}. Bonn Gatchina predictions, truncated at different $\text{L}_{\mathrm{max}}$ (same colors as in Table \ref{tab:CoeffsEright3}) are drawn as well. \newline Right: All partial wave interferences for $\text{L}_{\text{max}} = 3$ are indicated (in the notation of equations (\ref{eq:EScalarProductCoeff0}) to (\ref{eq:EScalarProductCoeff6})).}
\label{tab:CoeffsEright4}
\end{table*}
%

\begin{table*}[htb]
\RawFloats
\begin{minipage}{.075\linewidth}
\vspace*{-6.5pt}
\hspace*{5pt}
\begin{equation}
\mathcal{C}_{2}^{\check{E}} \equiv \nonumber
\end{equation}
\end{minipage}
\begin{minipage}{.3\linewidth} \vspace*{0.572cm} \includegraphics[width=0.875\textwidth]{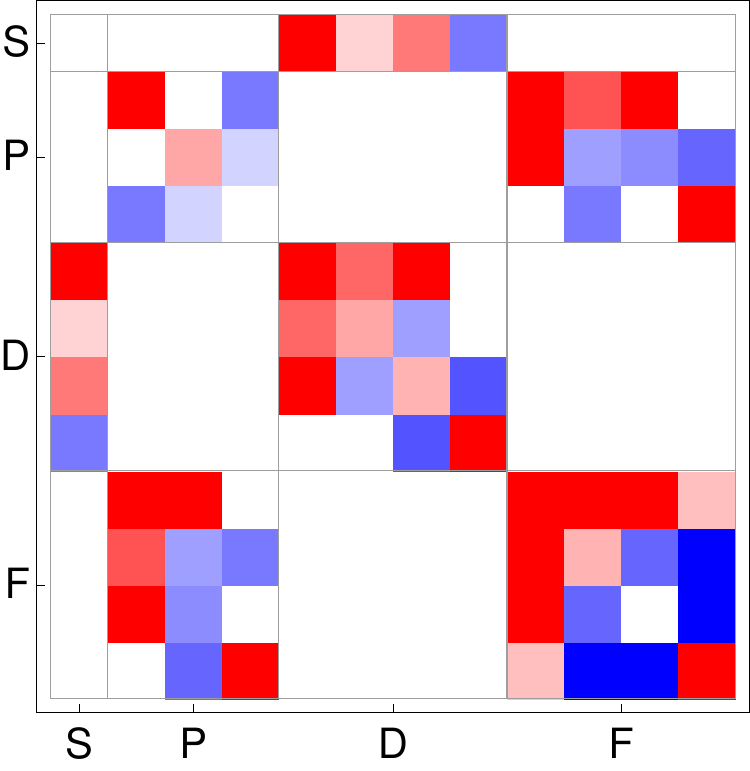} \end{minipage}
\begin{minipage}{.35\linewidth} \vspace*{0.500cm} \hspace*{-0.65cm}\includegraphics[width=1.15\textwidth]{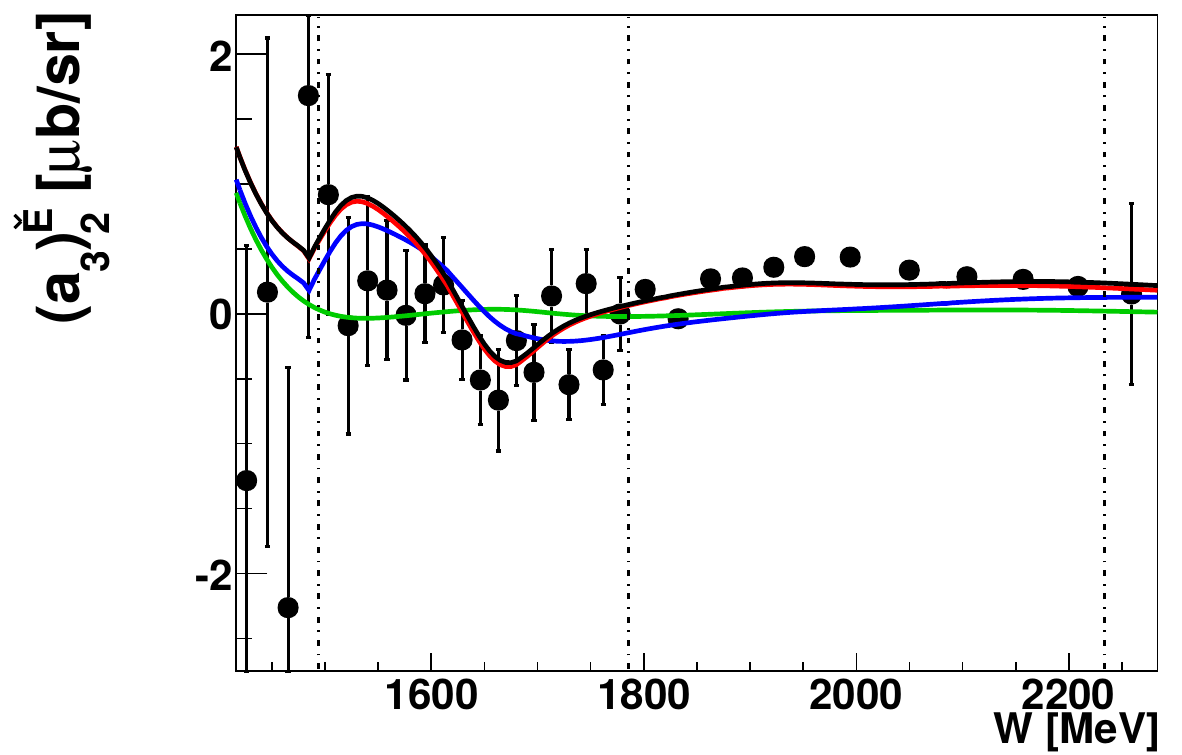}\end{minipage}
\begin{minipage}{.25\linewidth} \begin{align} \left(a_{3}\right)^{\check{E}}_{2} &= \left<P,P\right> + \left<S,D\right> \nonumber \\ & \hspace*{12.5pt} + \left<D,D\right> + \left<P,F\right> \nonumber \\ & \hspace*{12.5pt} + \left<F,F\right> \nonumber \end{align} \end{minipage}
\caption{Left: Matrix $\mathcal{C}_{2}^{\check{E}}$, represented here in the color scheme, defines the coefficient $\left(a_{3}\right)_{2}^{\check{E}}$ for an expansion of $\check{E}$ up to $\text{L}_{\text{max}} = 3$, see also Table \ref{tab:CoeffsEright1}. Center: Coefficient $\left(a_{3}\right)_{2}^{\check{E}}$ obtained from a fit to the $E$-data (black points). For references to the data see Table \ref{tab:DataBasis}. Bonn Gatchina predictions, truncated at different $\text{L}_{\mathrm{max}}$ (same colors as in Table \ref{tab:CoeffsEright3}) are drawn as well. \newline Right: All partial wave interferences for $\text{L}_{\text{max}} = 3$ are indicated (in the notation of equations (\ref{eq:EScalarProductCoeff0}) to (\ref{eq:EScalarProductCoeff6})).}
\label{tab:CoeffsEright5}
\end{table*}

%
The strength of the color shading corresponds to the magnitude of the entry. White entries are therefore zero. With the color scheme introduced here, the remaining Legendre coefficients $\left(a_{\text{L}_{\text{max}}}\right)^{\check{E}}_{(1,2,3,4,5,6)}$ defining $\check{E}$ can be represented as well. Their matrices are shown in Tables \ref{tab:CoeffsEright2} to \ref{tab:CoeffsEright6} and provide further examples for the color scheme. The graphical representation is shown for all investigated observables, for $\text{L}_{\text{max}} = 5$, in appendix \ref{sec:PWContentFormulas}. \newline
The thin gray lines separate definite partial wave interferences. These interference contributions can be rather obscured in the bilinear product expressions such as given in the last line of Table \ref{tab:CoeffsEright1}. \newline
Next we introduce a symbolic notation that clarifies the possible partial wave interferences contributing to a particular Legendre coefficient. We observe that each block separated by the thin lines in Tables \ref{tab:CoeffsEright1} to \ref{tab:CoeffsEright9} is exactly a contribution of bilinear products of multipoles $\mathcal{M}_{\ell_{1}}^{\ast} \mathcal{M}_{\ell_{2}}$ for two specific angular momenta $\ell_{1}$ and $\ell_{2}$. Each such block (or more precisely the contributions coming from two blocks, since all $\mathcal{C}_{k}^{\check{\Omega}^{\alpha}}$ are either symmetric or hermitean) is denoted by a 'scalar product' symbol $\left<-,-\right>$ indicating which partial waves contribute to the possible interferences in each coefficient (employing the standard spectrocopic notation for $S$-, $P$- $\ldots$ waves). In this way, the partial wave interferences in the coefficients defined by the matrices given in Tables \ref{tab:CoeffsEright2} to \ref{tab:CoeffsEright9} can be written compactly as:
{\allowdisplaybreaks
\begin{align}
 \left(a_{3}\right)^{\check{E}}_{0} &= \left<S,S\right> +\left<P,P\right> + \left<D,D\right> + \left<F,F\right> \mathrm{,} \label{eq:EScalarProductCoeff0} \\
 \left(a_{3}\right)^{\check{E}}_{1} &= \left<S,P\right> +\left<P,D\right> + \left<D,F\right> \mathrm{,} \label{eq:EScalarProductCoeff1} \\
 \left(a_{3}\right)^{\check{E}}_{2} &= \left<P,P\right> +\left<S,D\right> +\left<D,D\right> + \left<P,F\right> \nonumber \\
  & \hspace*{12.5pt}+ \left<F,F\right>  \mathrm{,} \label{eq:EScalarProductCoeff2} \\
 \left(a_{3}\right)^{\check{E}}_{3} &= \left<P,D\right> + \left<S,F\right> + \left<D,F\right> \mathrm{,} \label{eq:EScalarProductCoeff3} \\
 \left(a_{3}\right)^{\check{E}}_{4} &= \left<D,D\right> + \left<P,F\right> + \left<F,F\right> \mathrm{,} \label{eq:EScalarProductCoeff4} \\
  \left(a_{3}\right)^{\check{E}}_{5} &= \left<D,F\right> \mathrm{,} \label{eq:EScalarProductCoeff5} \\
 \left(a_{3}\right)^{\check{E}}_{6} &= \left<F,F\right> \mathrm{.} \label{eq:EScalarProductCoeff6}
\end{align}
}
For instance, the symbol $\left<D,D\right>$ appearing in the coefficient $\left(a_{2}\right)^{\check{E}}_{4}$ is just shorthand for a sum of bilinear products of multipoles multiplying only $D$- with $D$-waves, i.e.
\begin{equation}
\left< D,D \right> = \sum_{\mathcal{M},\mathcal{M}^{\prime}=\left\{E,M\right\}} \sum_{p,p^{\prime} = \left\{\pm\right\}} c_{p,p^{\prime}}^{\mathcal{M},\mathcal{M}^{\prime}} \mathcal{M}_{2 p}^{\ast} \mathcal{M}^{\prime}_{2 p^{\prime}} \mathrm{.} \label{eq:ExampleSumDD}
\end{equation}
The coefficients $c_{p,p^{\prime}}^{\mathcal{M},\mathcal{M}^{\prime}}$ appearing in this sum are stored in the corresponding matrix $\mathcal{C}_{4}^{\check{E}}$ (cf. Table \ref{tab:CoeffsEright7}). 

\begin{table*}[htb]
\RawFloats
\begin{minipage}{.075\linewidth}
\vspace*{-6.5pt}
\hspace*{5pt}
\begin{equation}
\mathcal{C}_{3}^{\check{E}} \equiv \nonumber
\end{equation}
\end{minipage}
\begin{minipage}{.3\linewidth} \vspace*{0.572cm} \includegraphics[width=0.875\textwidth]{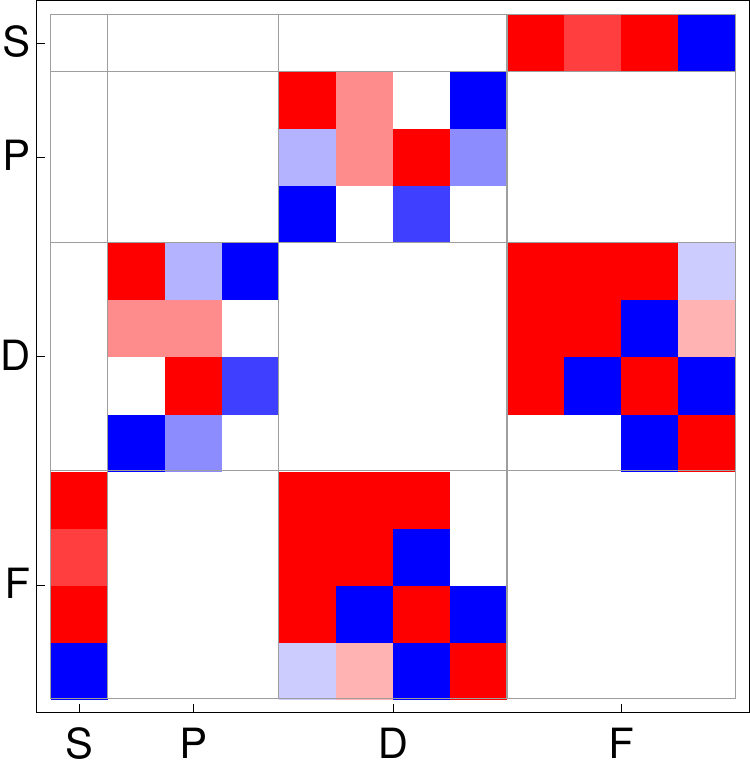} \end{minipage}
\begin{minipage}{.35\linewidth} \vspace*{0.500cm} \hspace*{-0.65cm}\includegraphics[width=1.15\textwidth]{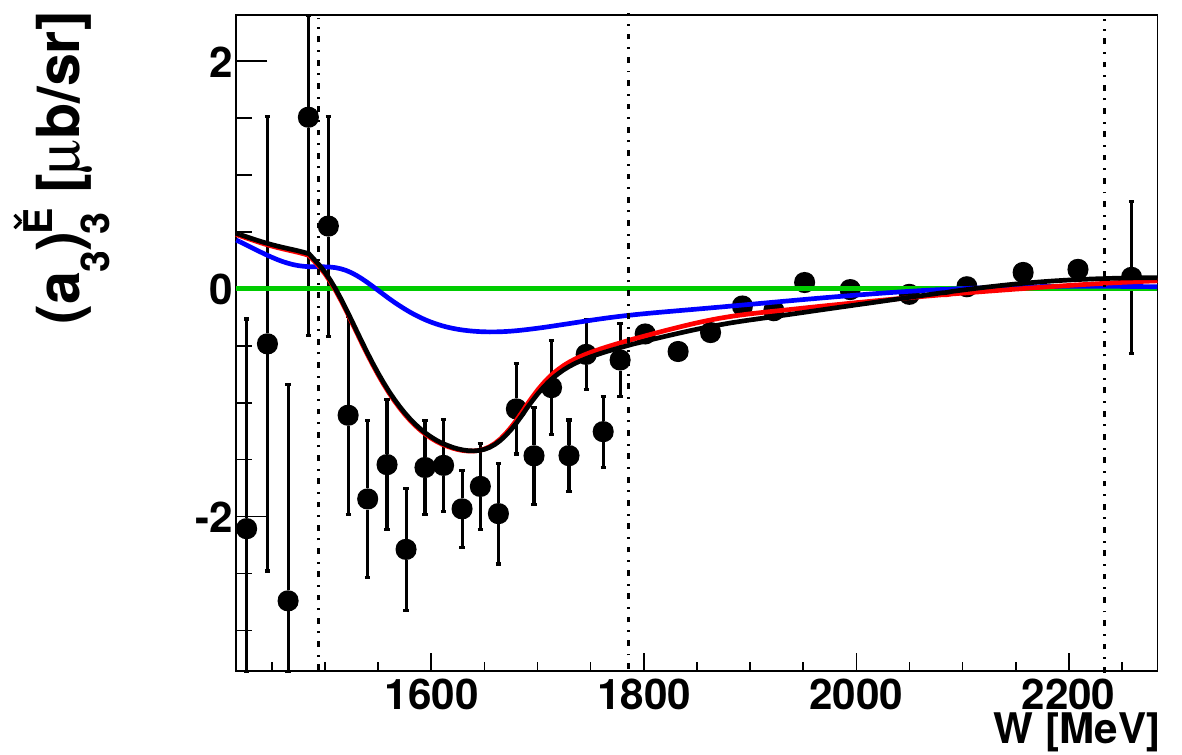}\end{minipage}
\begin{minipage}{.25\linewidth} \begin{align} \left(a_{3}\right)^{\check{E}}_{3} &= \left<P,D\right> + \left<S,F\right> \nonumber \\ & \hspace*{12.5pt} + \left<D,F\right>  \nonumber \end{align} \end{minipage}
\caption{Left: Matrix $\mathcal{C}_{3}^{\check{E}}$, represented here in the color scheme, defines the coefficient $\left(a_{3}\right)_{3}^{\check{E}}$ for an expansion of $\check{E}$ up to $\text{L}_{\text{max}} = 3$, see also Table \ref{tab:CoeffsEright1}. Center: Coefficient $\left(a_{3}\right)_{3}^{\check{E}}$ obtained from a fit to the $E$-data (black points). For references to the data see Table \ref{tab:DataBasis}. Bonn Gatchina predictions, truncated at different $\text{L}_{\mathrm{max}}$ (same colors as in Table \ref{tab:CoeffsEright3}) are drawn as well. \newline Right: All partial wave interferences for $\text{L}_{\text{max}} = 3$ are indicated (in the notation of equations (\ref{eq:EScalarProductCoeff0}) to (\ref{eq:EScalarProductCoeff6})).}
\label{tab:CoeffsEright6}
\end{table*}

%
\begin{table*}[htb]
\RawFloats
\begin{minipage}{.075\linewidth}
\vspace*{-6.5pt}
\hspace*{5pt}
\begin{equation}
\mathcal{C}_{4}^{\check{E}} \equiv \nonumber
\end{equation}
\end{minipage}
\begin{minipage}{.3\linewidth} \vspace*{0.572cm} \includegraphics[width=0.875\textwidth]{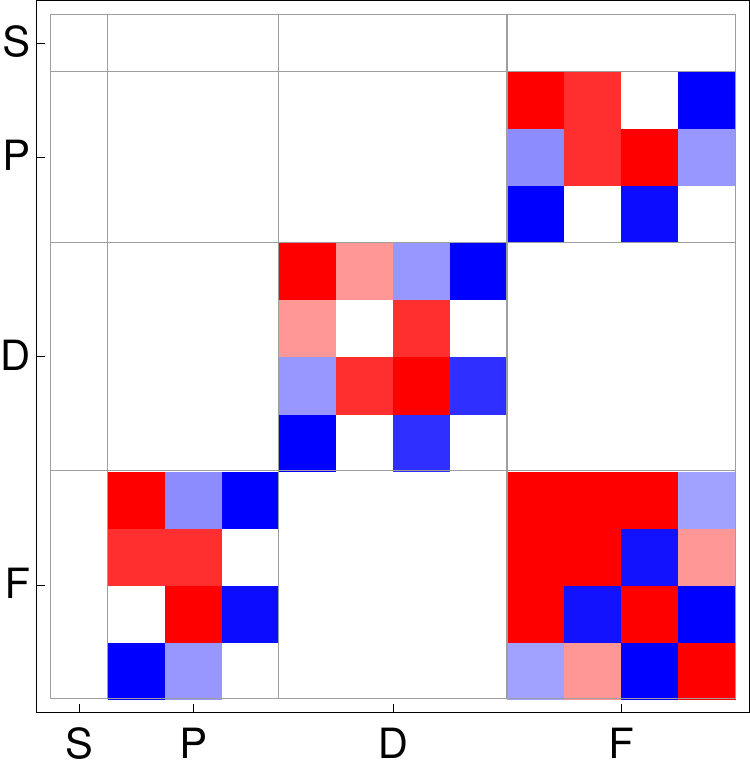} \end{minipage}
\begin{minipage}{.35\linewidth} \vspace*{0.500cm} \hspace*{-0.65cm}\includegraphics[width=1.15\textwidth]{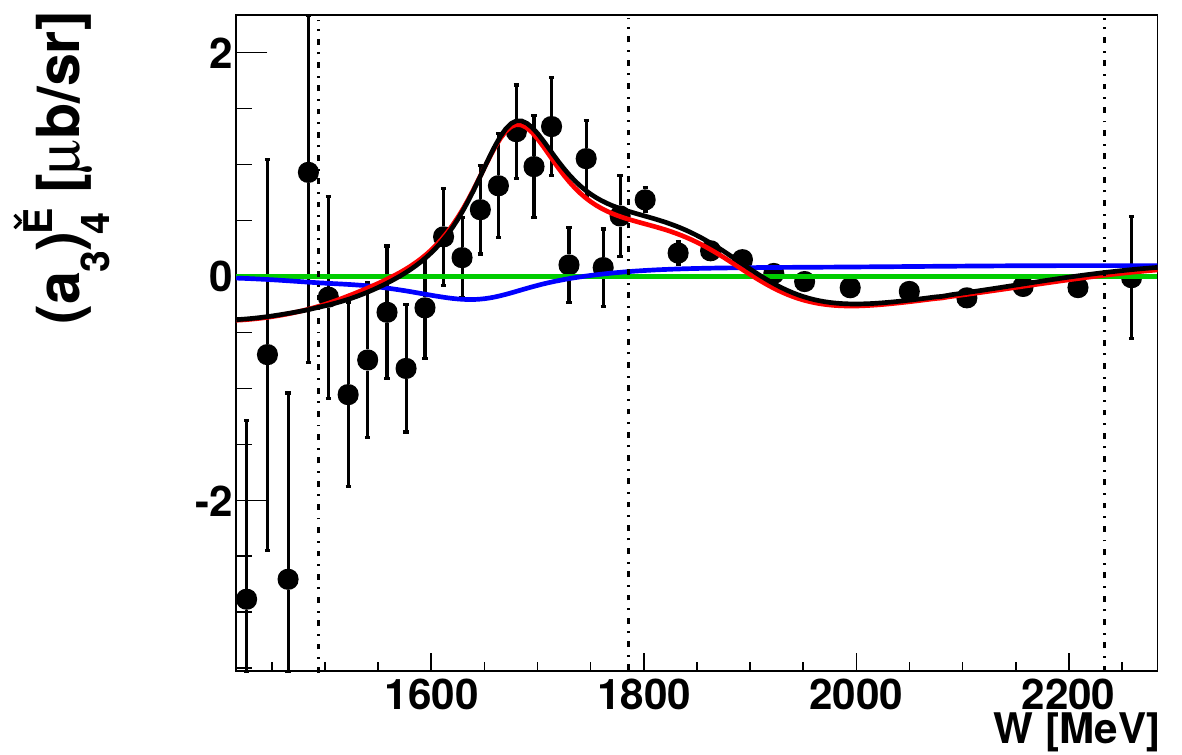}\end{minipage}
\begin{minipage}{.25\linewidth} \begin{align} \left(a_{3}\right)^{\check{E}}_{4} &= \left<D,D\right> + \left<P,F\right> \nonumber \\ & \hspace*{12.5pt} + \left<F,F\right>  \nonumber \end{align} \end{minipage}
\caption{Left: Matrix $\mathcal{C}_{4}^{\check{E}}$, represented here in the color scheme, defines the coefficient $\left(a_{3}\right)_{4}^{\check{E}}$ for an expansion of $\check{E}$ up to $\text{L}_{\text{max}} = 3$, see also Table \ref{tab:CoeffsEright1}. Center: Coefficient $\left(a_{3}\right)_{4}^{\check{E}}$ obtained from a fit to the $E$-data (black points). For references to the data see Table \ref{tab:DataBasis}. Bonn Gatchina predictions, truncated at different $\text{L}_{\mathrm{max}}$ (same colors as in Table \ref{tab:CoeffsEright3}) are drawn as well. \newline Right: All partial wave interferences for $\text{L}_{\text{max}} = 3$ are indicated (in the notation of equations (\ref{eq:EScalarProductCoeff0}) to (\ref{eq:EScalarProductCoeff6})).}
\label{tab:CoeffsEright7}
\end{table*}
\begin{table*}[htb]
\RawFloats
\begin{minipage}{.075\linewidth}
\vspace*{-6.5pt}
\hspace*{5pt}
\begin{equation}
\mathcal{C}_{5}^{\check{E}} \equiv \nonumber
\end{equation}
\end{minipage}
\begin{minipage}{.3\linewidth} \vspace*{0.572cm} \includegraphics[width=0.875\textwidth]{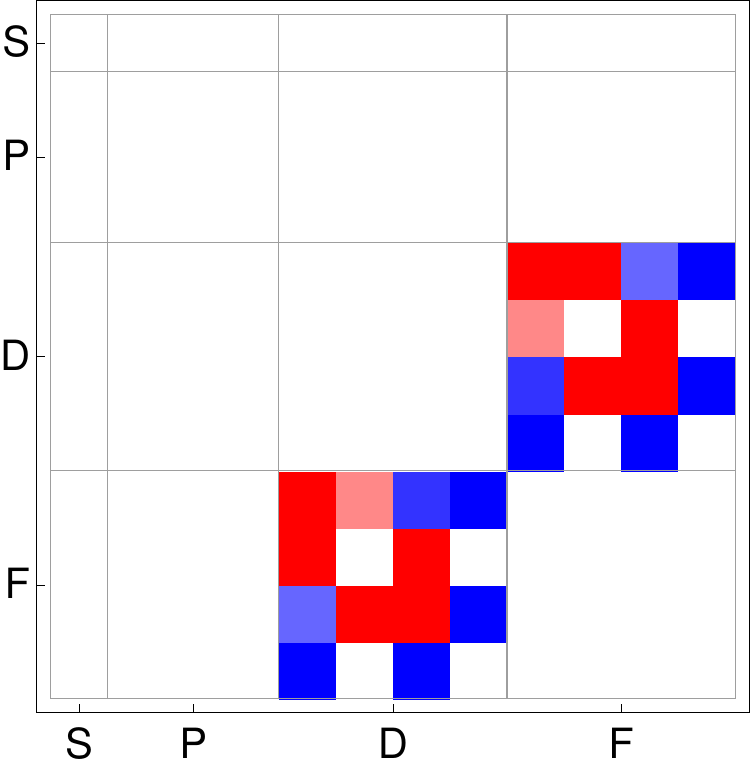} \end{minipage}
\begin{minipage}{.35\linewidth} \vspace*{0.500cm} \hspace*{-0.65cm}\includegraphics[width=1.15\textwidth]{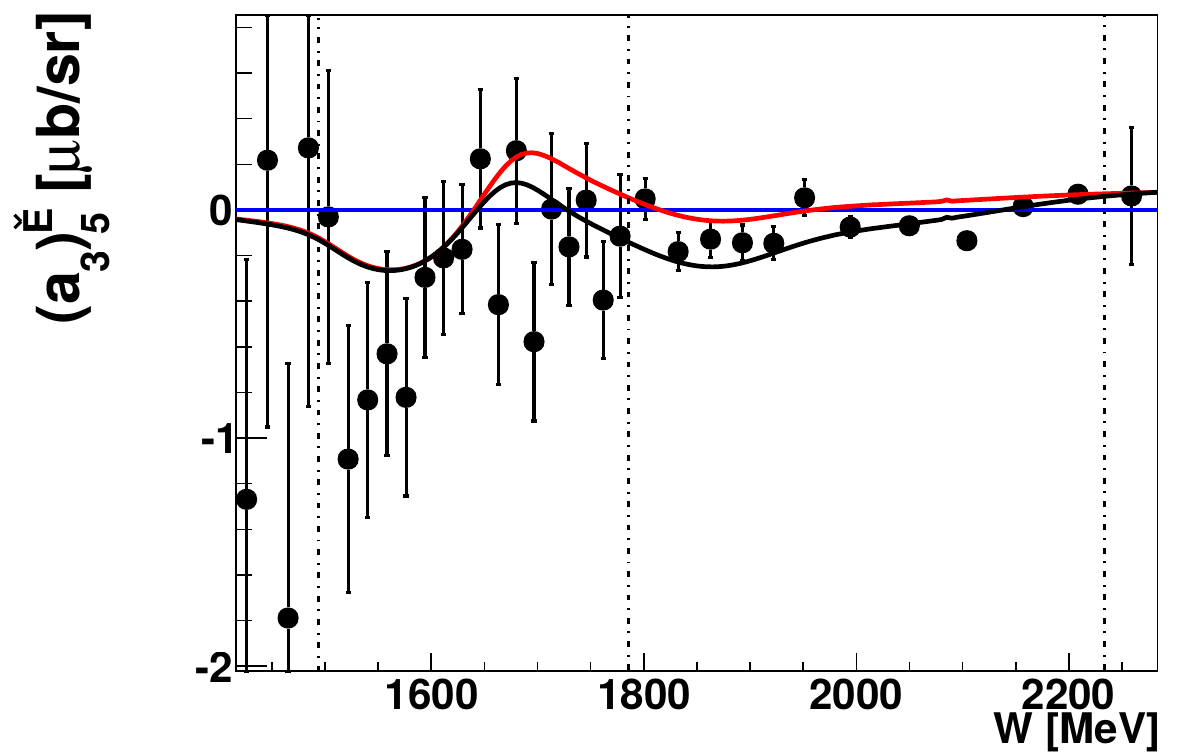}\end{minipage}
\begin{minipage}{.25\linewidth} \begin{align} \left(a_{3}\right)^{\check{E}}_{5} &= \left<D,F\right>  \nonumber \end{align} \end{minipage}
\caption{Left: Matrix $\mathcal{C}_{5}^{\check{E}}$, represented here in the color scheme, defines the coefficient $\left(a_{3}\right)_{5}^{\check{E}}$ for an expansion of $\check{E}$ up to $\text{L}_{\text{max}} = 3$, see also Table \ref{tab:CoeffsEright1}. Center: Coefficient $\left(a_{3}\right)_{5}^{\check{E}}$ obtained from a fit to the $E$-data (black points). For references to the data see Table \ref{tab:DataBasis}. Bonn Gatchina predictions, truncated at different $\text{L}_{\mathrm{max}}$ (same colors as in Table \ref{tab:CoeffsEright3}) are drawn as well. \newline Right: All partial wave interferences for $\text{L}_{\text{max}} = 3$ are indicated (in the notation of equations (\ref{eq:EScalarProductCoeff0}) to (\ref{eq:EScalarProductCoeff6})).}
\label{tab:CoeffsEright8}
\end{table*}
%
%
\begin{table*}[htb]
\RawFloats
\begin{minipage}{.075\linewidth}
\vspace*{-6.5pt}
\hspace*{5pt}
\begin{equation}
\mathcal{C}_{6}^{\check{E}} \equiv \nonumber
\end{equation}
\end{minipage}
\begin{minipage}{.3\linewidth} \vspace*{0.572cm} \includegraphics[width=0.875\textwidth]{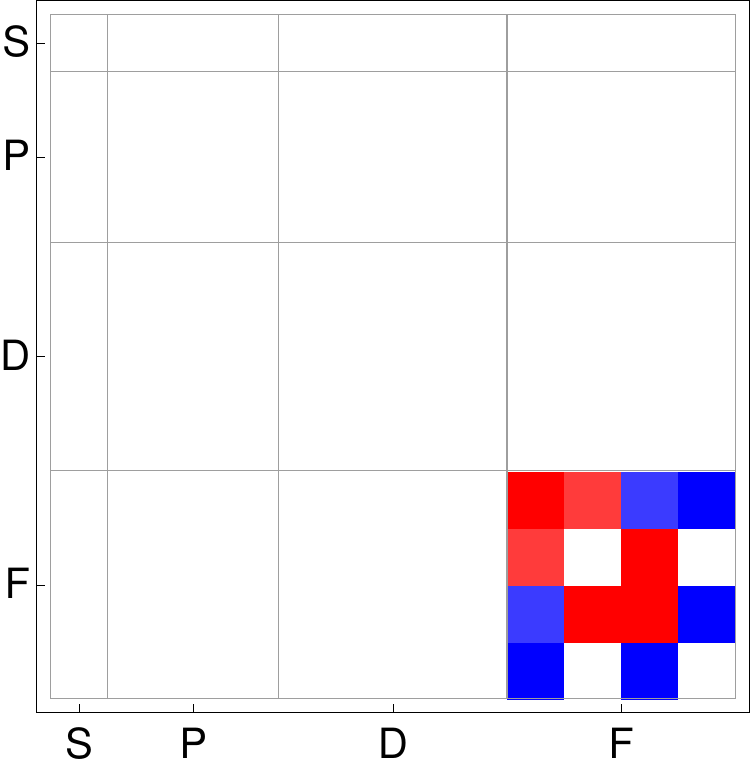} \end{minipage}
\begin{minipage}{.35\linewidth} \vspace*{0.500cm} \hspace*{-0.65cm}\includegraphics[width=1.15\textwidth]{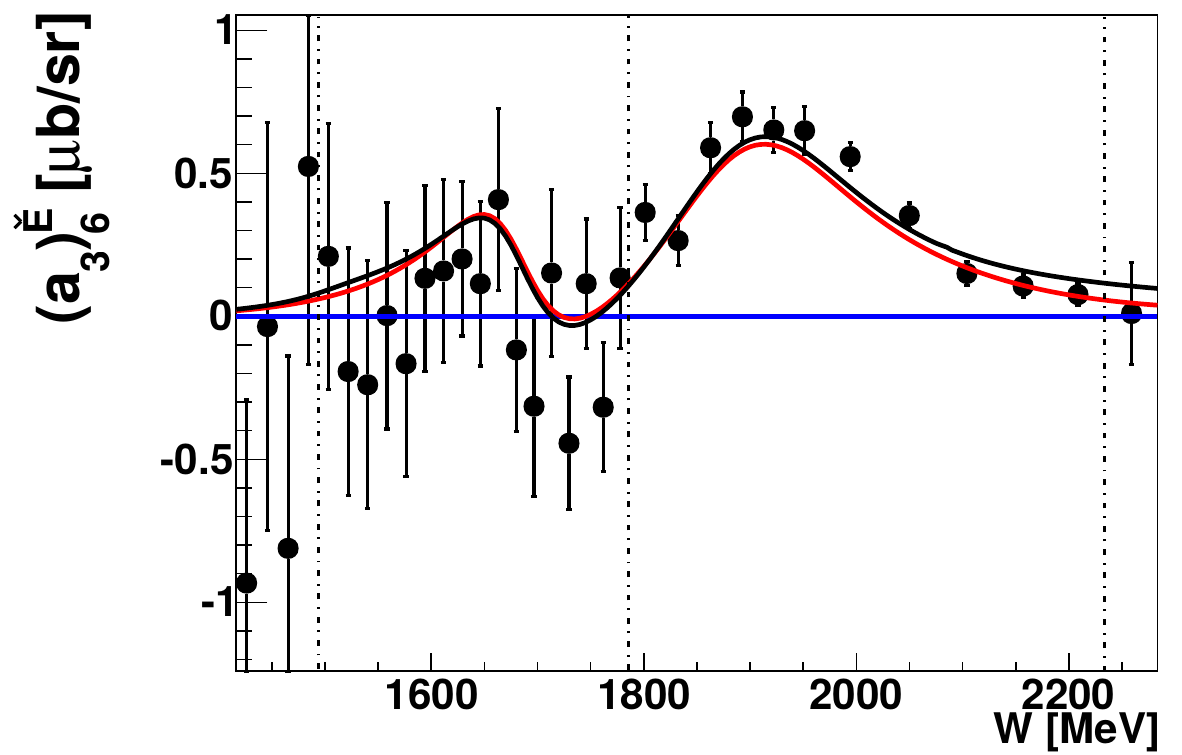}\end{minipage}
\begin{minipage}{.25\linewidth} \begin{align} \left(a_{3}\right)^{\check{E}}_{6} &= \left<F,F\right>  \nonumber \end{align} \end{minipage}
\caption{Left: Matrix $\mathcal{C}_{6}^{\check{E}}$, represented here in the color scheme, defines the coefficient $\left(a_{3}\right)_{6}^{\check{E}}$ for an expansion of $\check{E}$ up to $\text{L}_{\text{max}} = 3$, see also Table \ref{tab:CoeffsEright1}. Center: Coefficient $\left(a_{3}\right)_{6}^{\check{E}}$ obtained from a fit to the $E$-data (black points). For references to the data see Table \ref{tab:DataBasis}. Bonn Gatchina predictions, truncated at different $\text{L}_{\mathrm{max}}$ (same colors as in Table \ref{tab:CoeffsEright3}) are drawn as well. \newline Right: All partial wave interferences for $\text{L}_{\text{max}} = 3$ are indicated (in the notation of equations (\ref{eq:EScalarProductCoeff0}) to (\ref{eq:EScalarProductCoeff6})).}
\label{tab:CoeffsEright9}
\end{table*}

As a further example for this notation, the coefficient $\left(a_{2}\right)^{\check{E}}_{0}$ is given entirely by interferences among partial waves of the same order in $\ell$. Cross-interferences do not occur, as can be seen by inspection of the expression (\ref{eq:EScalarProductCoeff0}) or the matrix in Table \ref{tab:CoeffsEright2}. In the coefficient $\left(a_{2}\right)^{\check{E}}_{1}$ on the other hand, only cross-interference terms among $S$-, $P$-, and $D$-waves contribute (compare Table \ref{tab:CoeffsEright4} and equation (\ref{eq:EScalarProductCoeff1})). 
This brief way of representing partial wave interferences by scalar products is used repeatedly in the following discussion (cf. Sections \ref{sec:Interpretation1stResRegion} to \ref{sec:Interpretation4thResRegion}). It should however be reminded here that each $\left< -, - \right>$-term is in truth an elaborate bilinear form like (\ref{eq:ExampleSumDD}), depending on at least 4 different complex multipoles. \newline
Finally, in order to aid the readability of the ensuing section \ref{sec:Interpretation}, Tables \ref{tab:CoeffsEright3} to \ref{tab:CoeffsEright9} contain pictures that anticipate the results shown in the latter section. Here, we endow every color-scheme matrix with a picture of the associated fitted Legendre coefficient of $\check{E}$ in a truncation at $\text{L}_{\mathrm{max}} = 3$. The fitted values are plotted as black dots with errorbars. Furthermore, we plotted as a comparison continuous curves that represent the respective coefficient as calculated from Bonn-Gatchina multipoles up to a certain truncation order. Bonn-Gatchina curves are drawn for truncations at $\text{L}_{\mathrm{max}} = 1,2,3,4$, with corresponding colors explained in the picture in Table \ref{tab:CoeffsEright3}. \newline
We provide the plots here in order to motivate how the interpretative statements in section \ref{sec:Interpretation} were gained. One can look at the partial-wave interferences indicated by the color scheme plot on the left of each Table and compare with the corrections provided by each higher truncation order in the picture on the right. The interferences are also written again in the notation of equations (\ref{eq:EScalarProductCoeff0}) to (\ref{eq:EScalarProductCoeff6}) in the picture on the right. \newline
As an example, the coefficient $\left(a_{3}\right)_{0}^{\check{E}}$ (Table \ref{tab:CoeffsEright3}) shows appreciable corrections due to $F$-waves starting around $W = 1550 \hspace*{2pt} \mathrm{MeV}$. Judging from the matrix shown on the left, these corrections can only stem from $\left< F, F \right>$ interference terms. A correction due to $G$-waves is also drawn for comparison, and the influence of the latter is seen to be negligible. \newline
Another interesting coefficient is $\left(a_{3}\right)_{6}^{\check{E}}$, shown in Table \ref{tab:CoeffsEright9}. It is significantly non-zero in the energy-region starting at $W = 1800 \hspace*{2pt} \mathrm{MeV}$. There, it is also reproduced very well by the Bonn Gatchina curve truncated at the $F$-waves. The corresponding matrix shows that this correction is purely due to an $\left< F, F \right>$-term. The coefficient contains, for $\text{L}_{\mathrm{max}} = 3$ at least, no further interference-blocks. Therefore, it is already quite safe to interpret the correction to be associated with $F$-wave resonances in this specific energy region (see Table \ref{tab:MultipoleResonanceAssignments}).


\section{Interpretation}\label{sec:Interpretation}

\begin{figure*}
\begin{minipage}{\textwidth}
\floatbox[{\capbeside\thisfloatsetup{capbesideposition={right,top},capbesidewidth=7.8cm}}]{figure}[\FBwidth]
{\caption{The recent new differential cross section $\sigma_0$ data from MAMI \cite{Adlarson:2015} with only statistical error was fitted using associated Legendre polynomials according to eq. \ref{eq:LowEAssocLegParametrizationDCS} and truncating the partial wave expansion at $\text{L}_{\text{max}}=1\dots 5$. (a) The resulting $\chi^2/$ndf values of the different $\text{L}_{\text{max}}$-fits as a function of the center of mass energy W are shown. 
(b) 6 out of 265 selected angular distributions of $\sigma_0$ (black points) are plotted together with the different $\text{L}_{\text{max}}$ fits (solid lines) starting at W=1154 MeV up to 1855 MeV. (c) Comparison of the fit coefficients for $\text{L}_{\text{max}}=4$ (black points), $\left(a_{4}\right)^{\sigma_0}_{0\dots8}$ (see eq. \ref{eq:LowEAssocLegParametrizationDCS}), with the BnGa2014-02 solution truncated at different $\text{L}_{\text{max}}$ (solid lines). Colors same as in (a).}\label{fig:wq_bins}}
 {\includegraphics[width=0.49\textwidth, trim=0cm 0cm 1.8cm 0cm, clip]{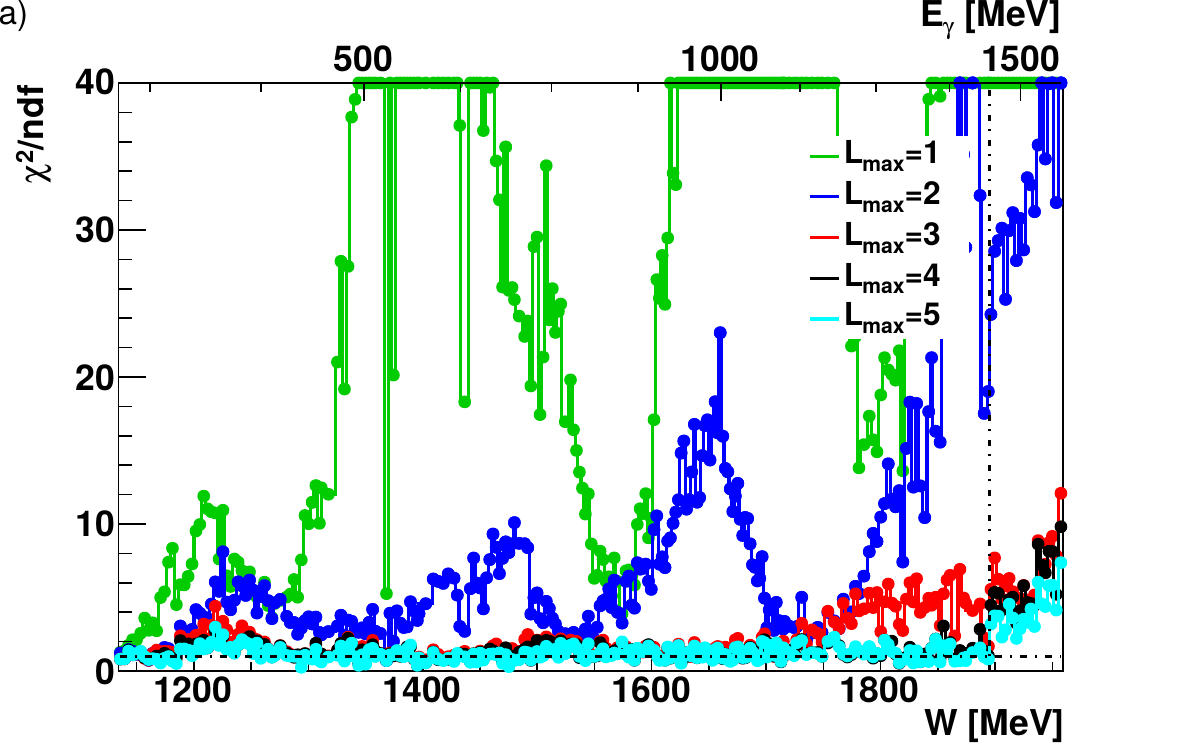}}
\end{minipage}\\

\begin{minipage}{\textwidth}
\centering
\hspace*{-0.45cm}
 \includegraphics[width=0.305\textwidth, trim=0cm 0cm 0.01cm 0.75cm, clip]{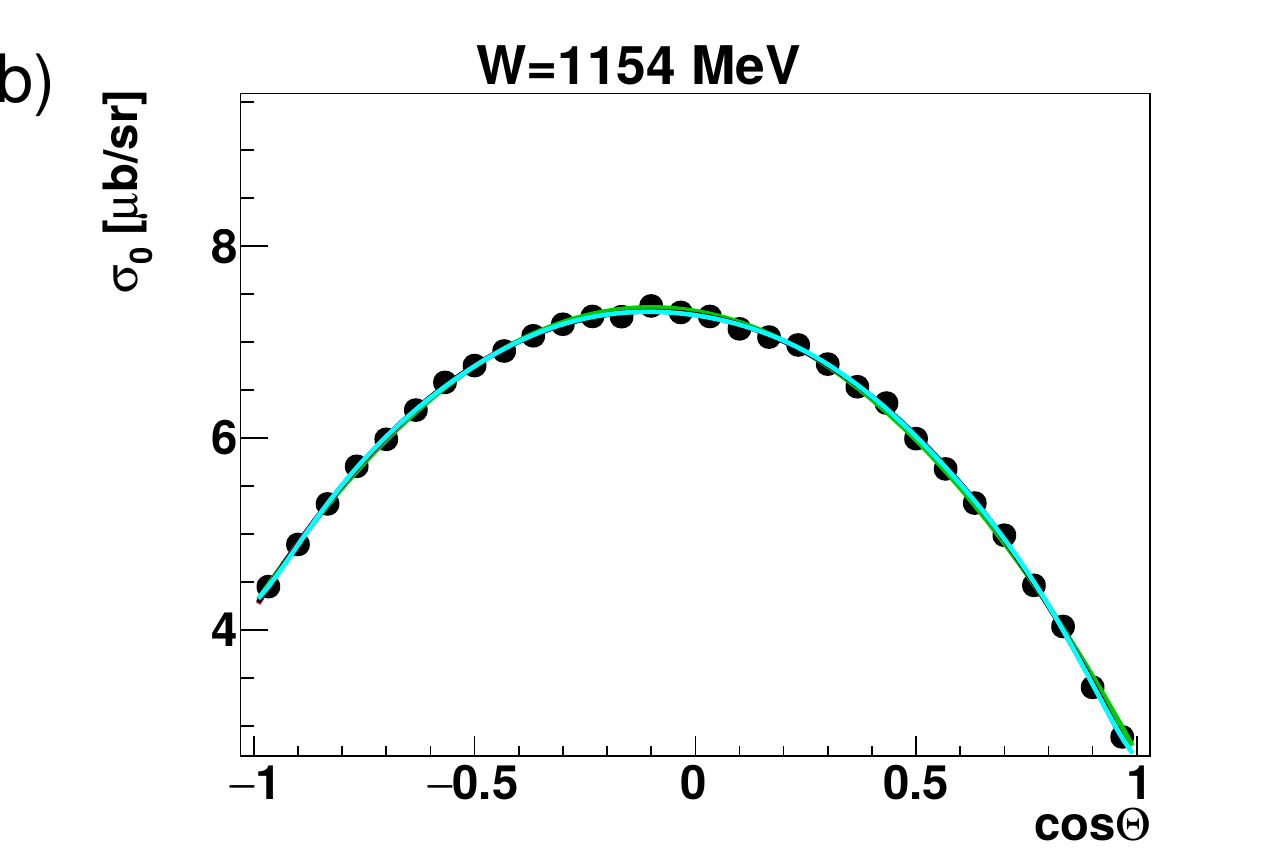}
  \includegraphics[width=0.285\textwidth, trim=0cm 0cm 0.01cm 0.75cm, clip]{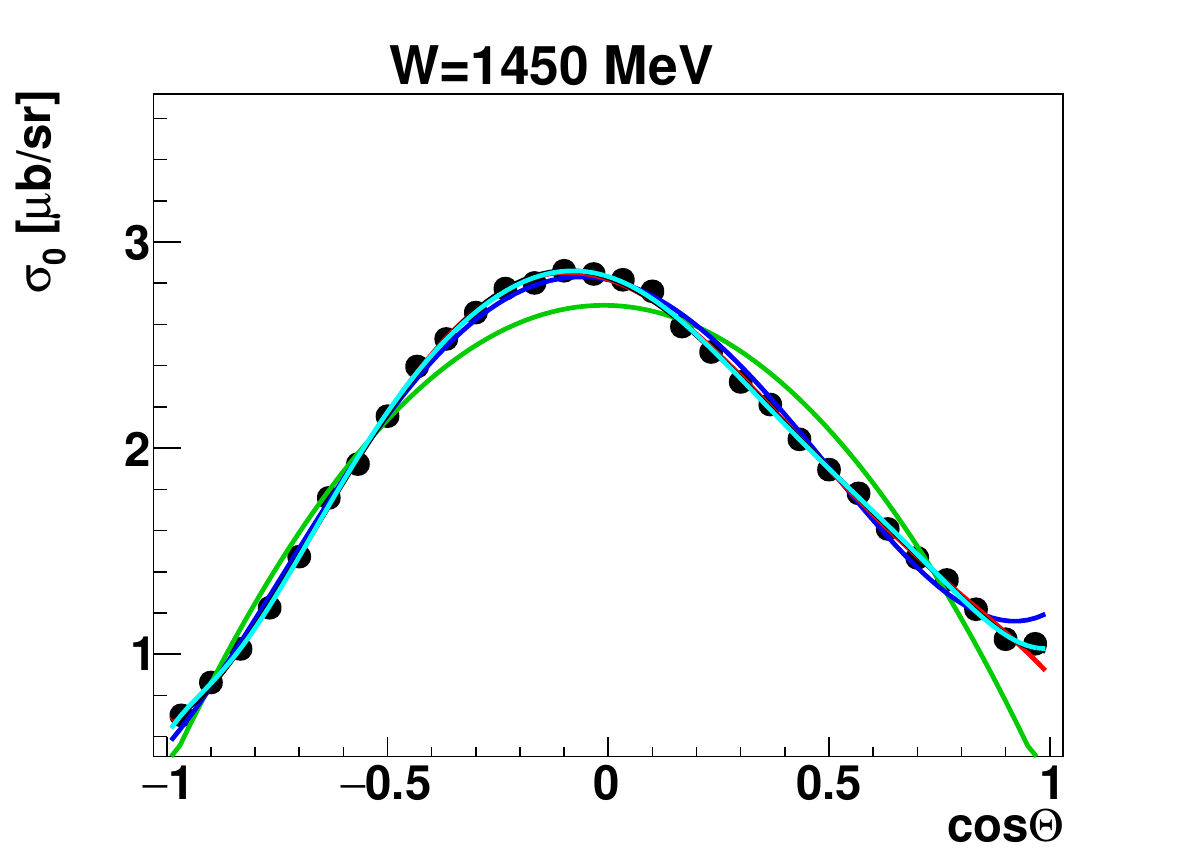}
  \includegraphics[width=0.285\textwidth, trim=0cm 0cm 0.01cm 0.75cm, clip]{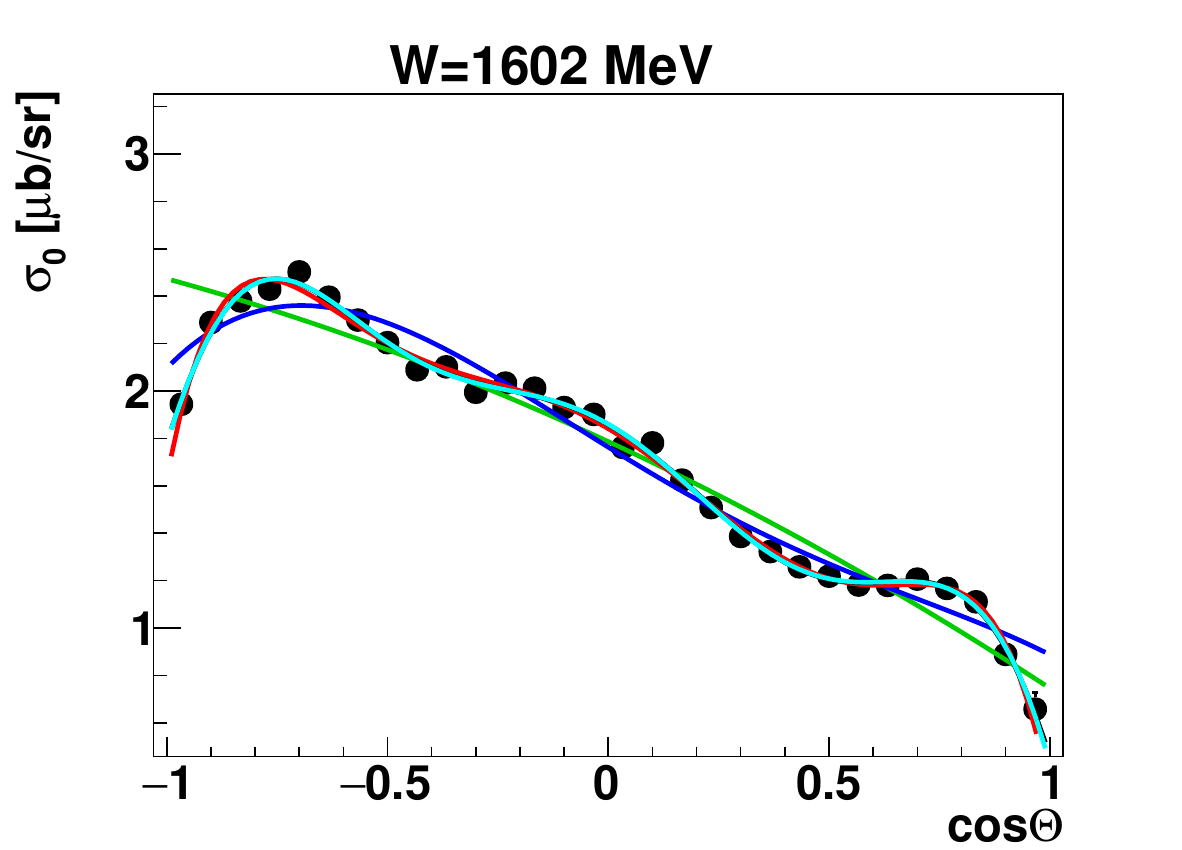}\\
  \includegraphics[width=0.285\textwidth, trim=0cm 0cm 0.01cm 0.75cm, clip]{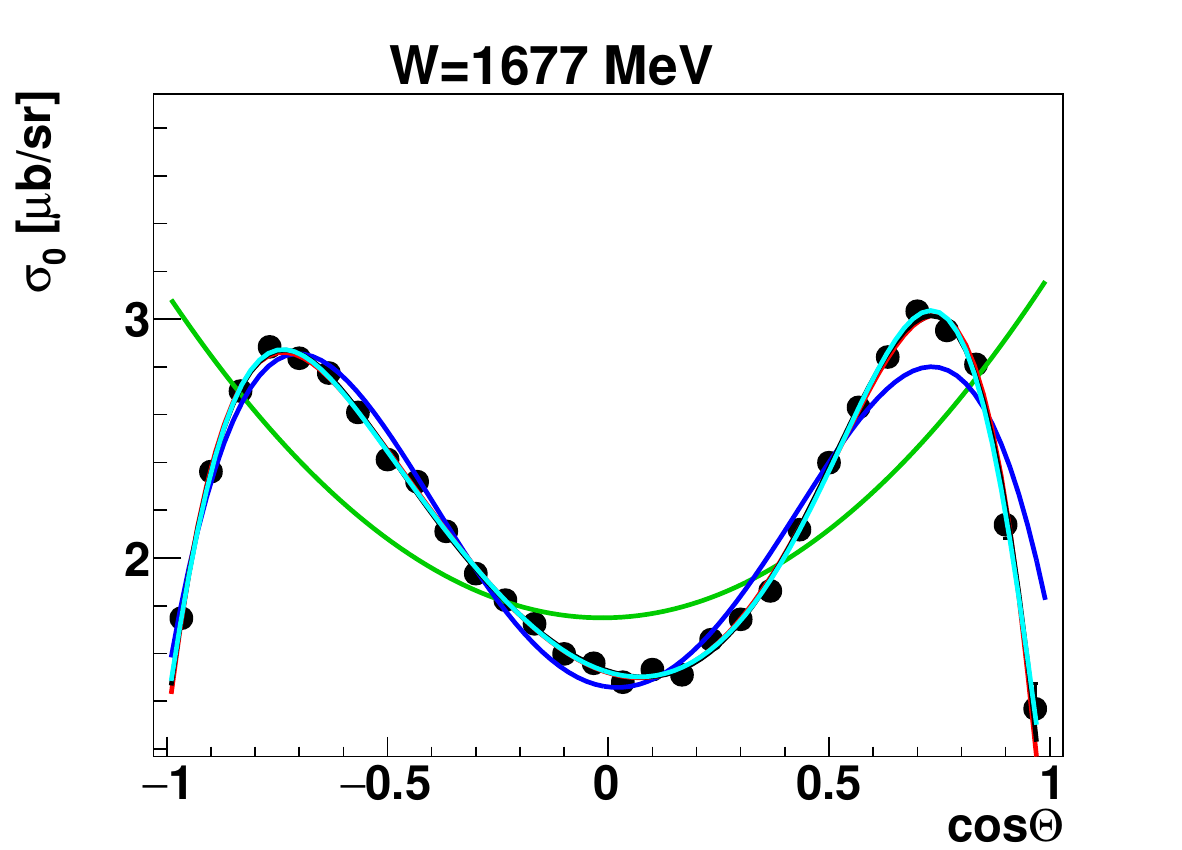}
  \includegraphics[width=0.285\textwidth, trim=0cm 0cm 0.01cm 0.75cm, clip]{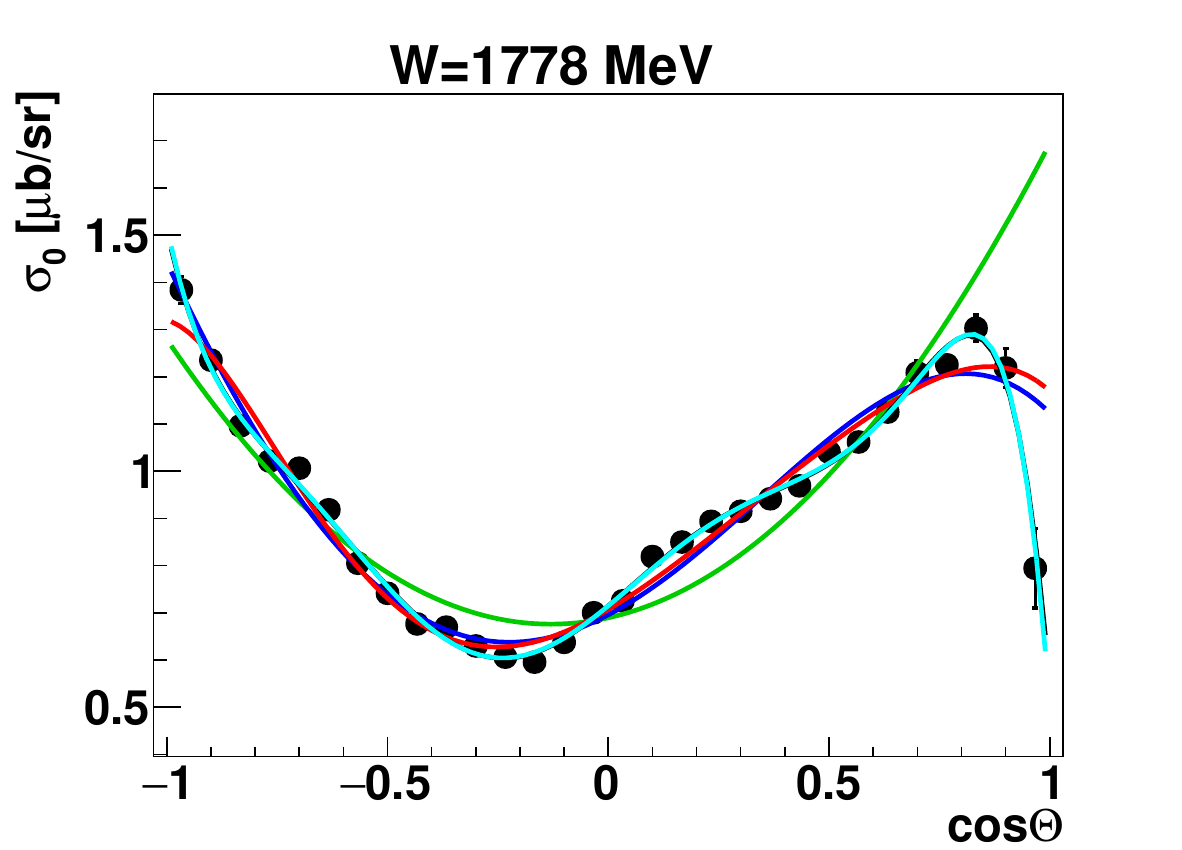}
  \includegraphics[width=0.285\textwidth, trim=0cm 0cm 0.01cm 0.75cm, clip]{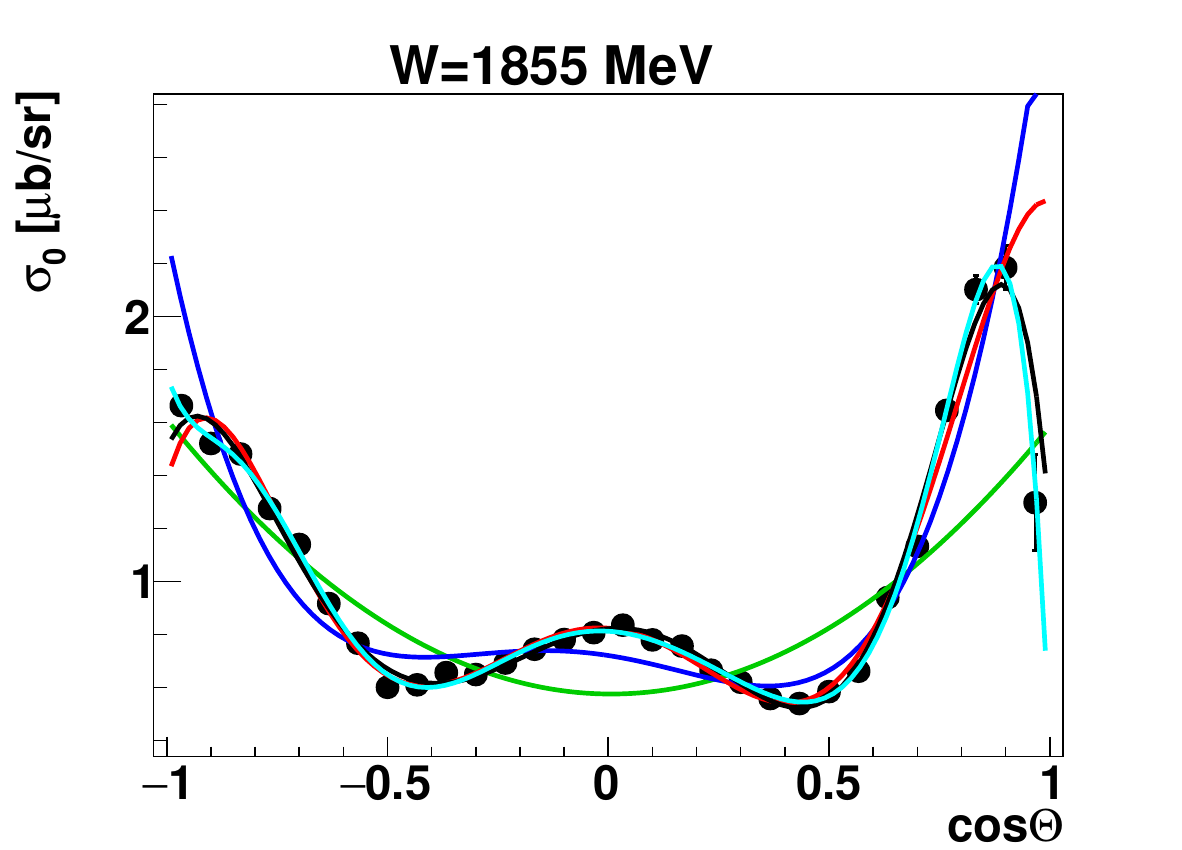}\\
 \vspace*{0.3cm}
  
  \hspace*{-23.5pt}\includegraphics[width=0.2905\textwidth]{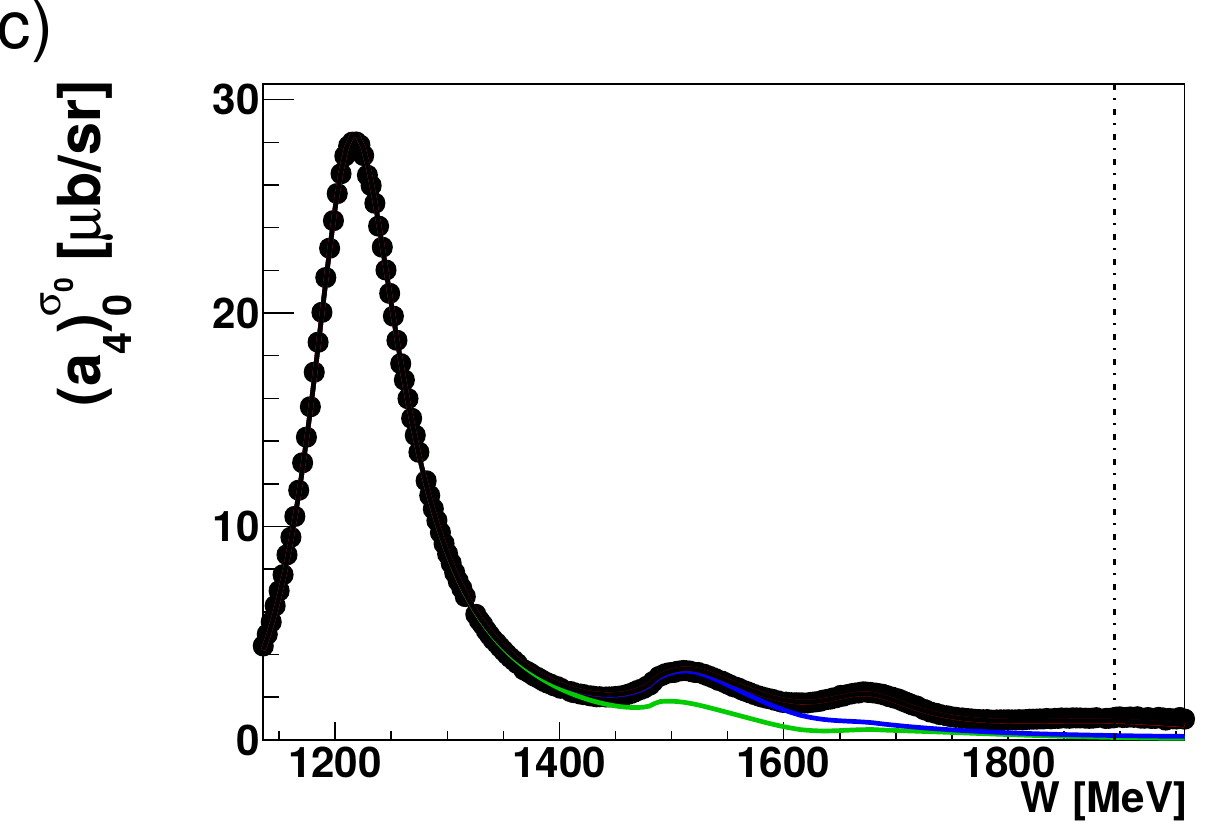}
  \includegraphics[width=0.285\textwidth]{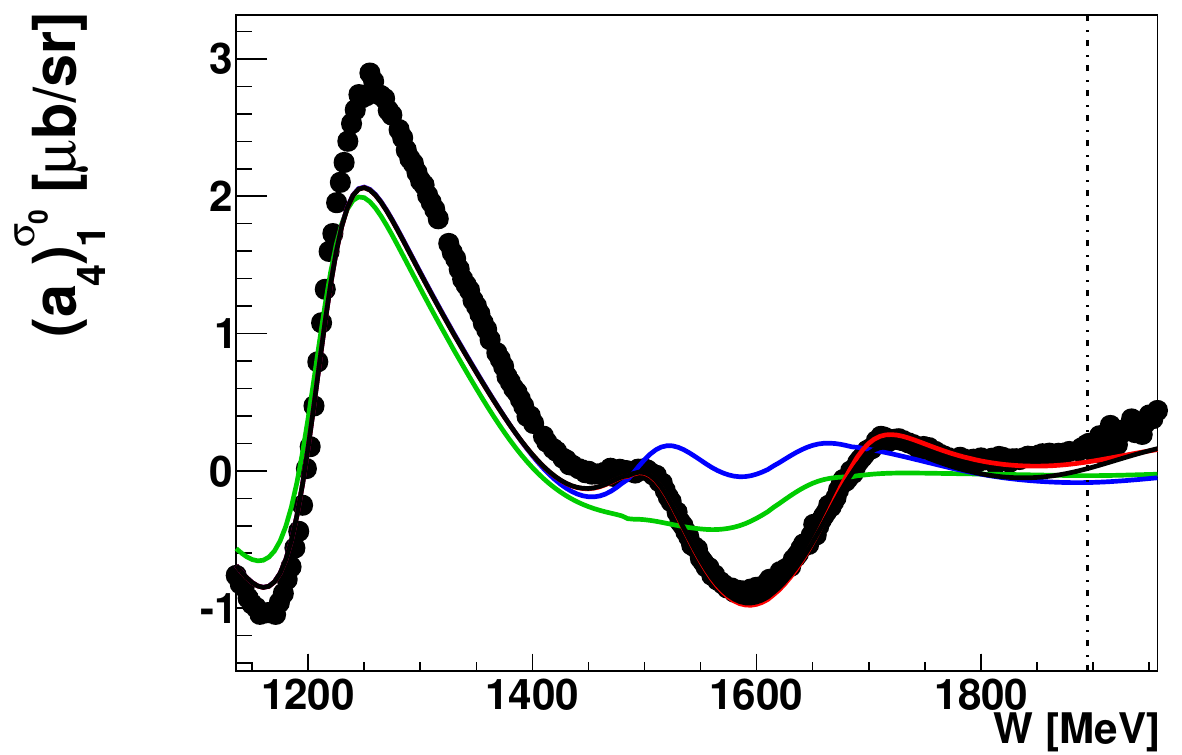}
  \includegraphics[width=0.285\textwidth]{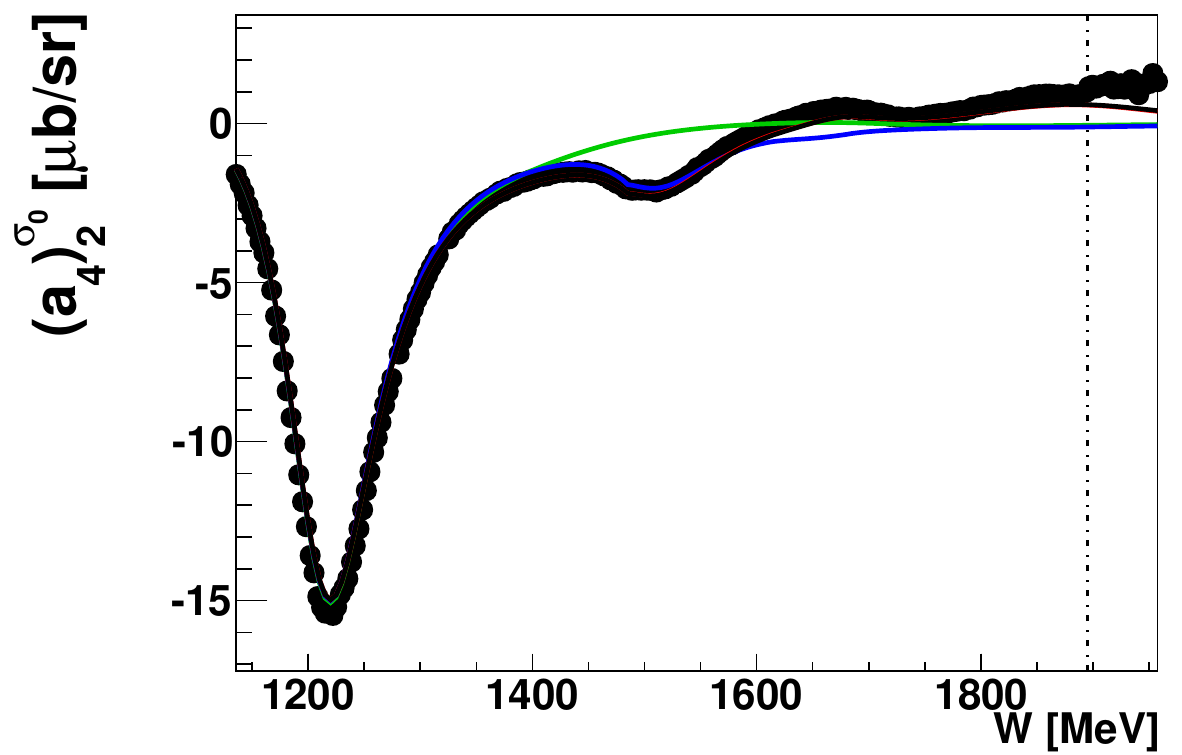}\\
  \hspace*{-19.5pt}\includegraphics[width=0.285\textwidth]{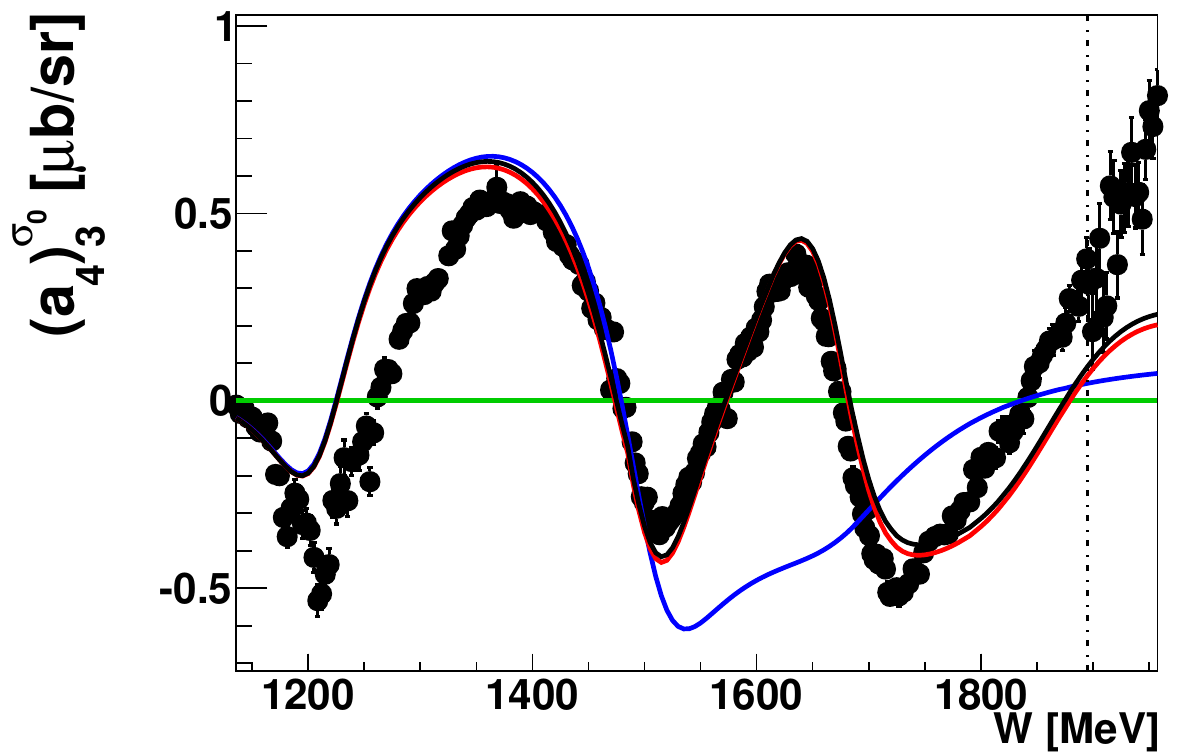}
  \includegraphics[width=0.285\textwidth]{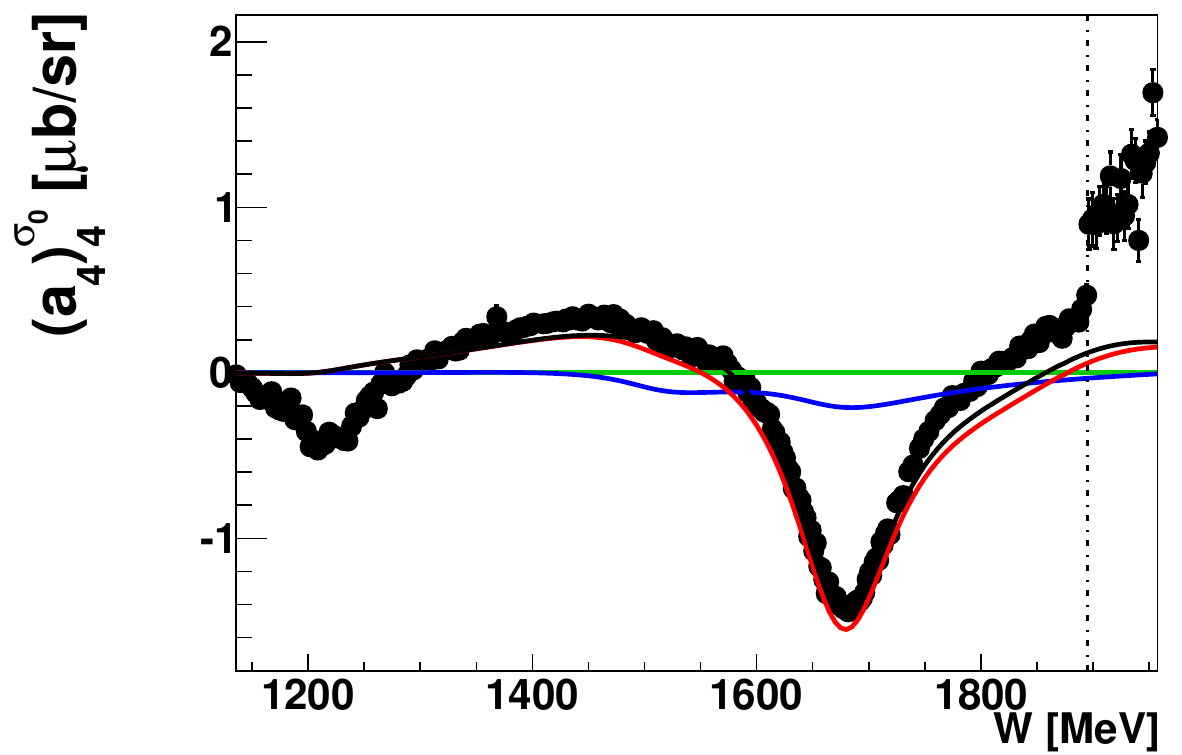}
  \includegraphics[width=0.285\textwidth]{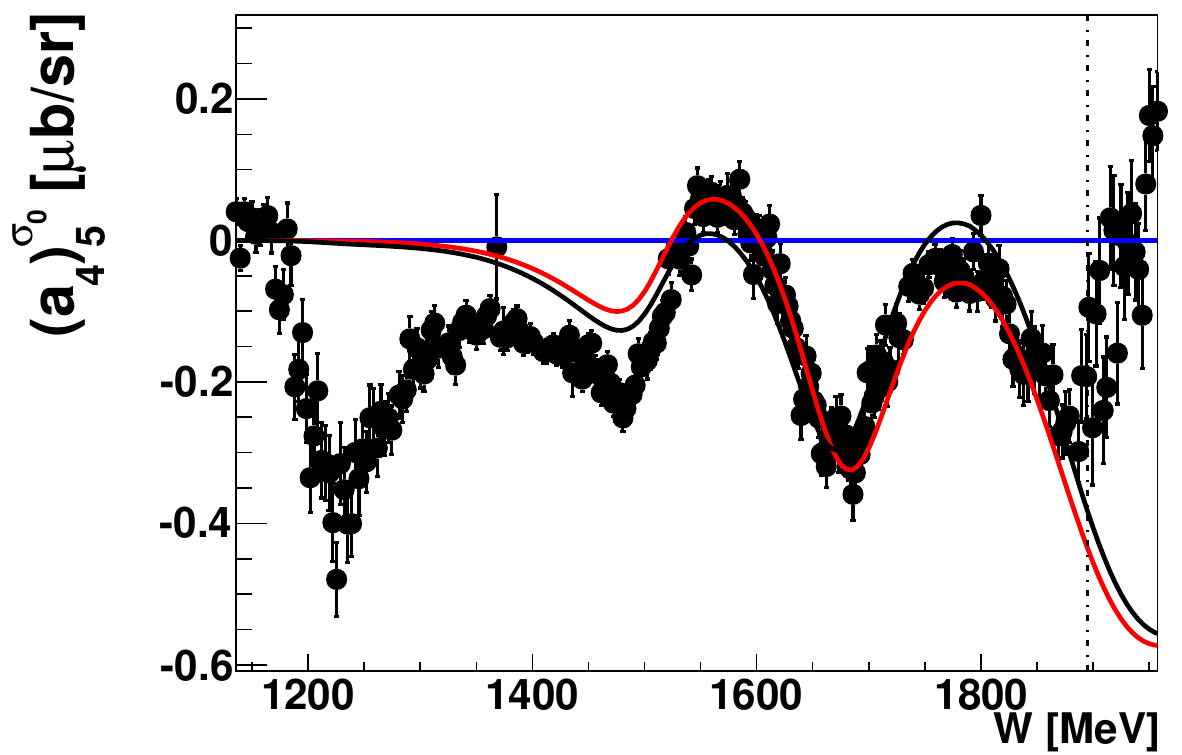}\\
  \hspace*{-19.5pt}\includegraphics[width=0.285\textwidth]{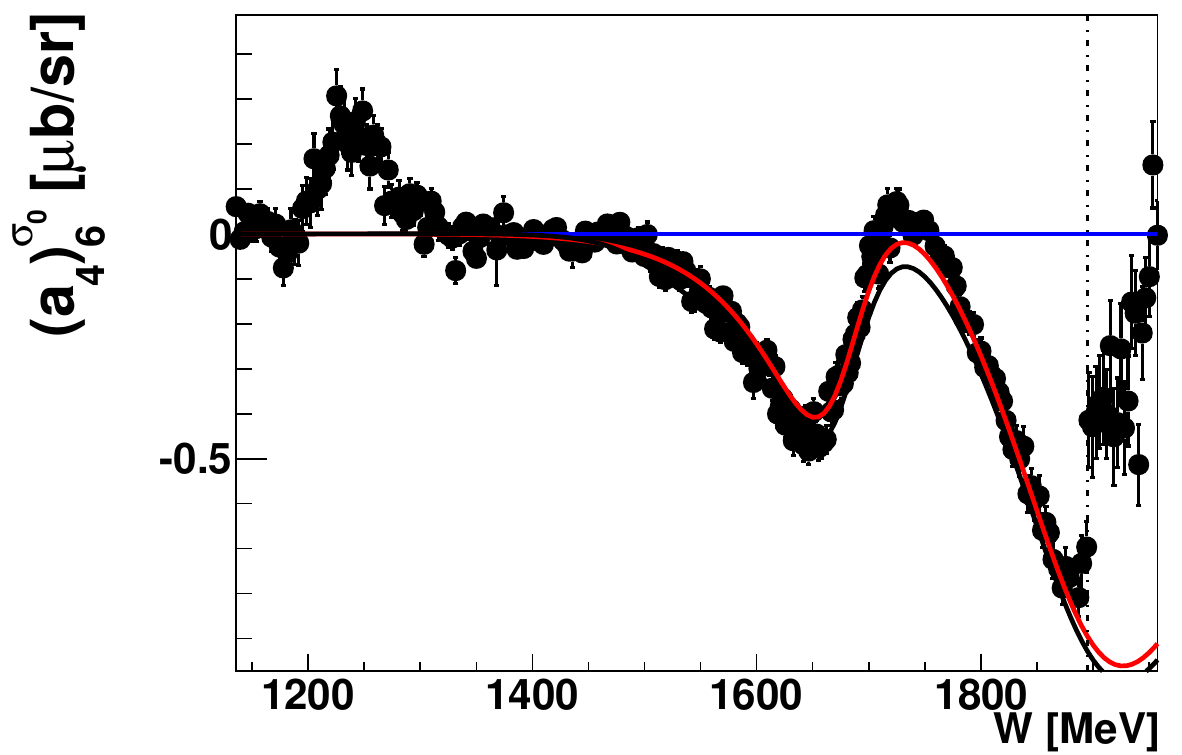}
  \includegraphics[width=0.285\textwidth]{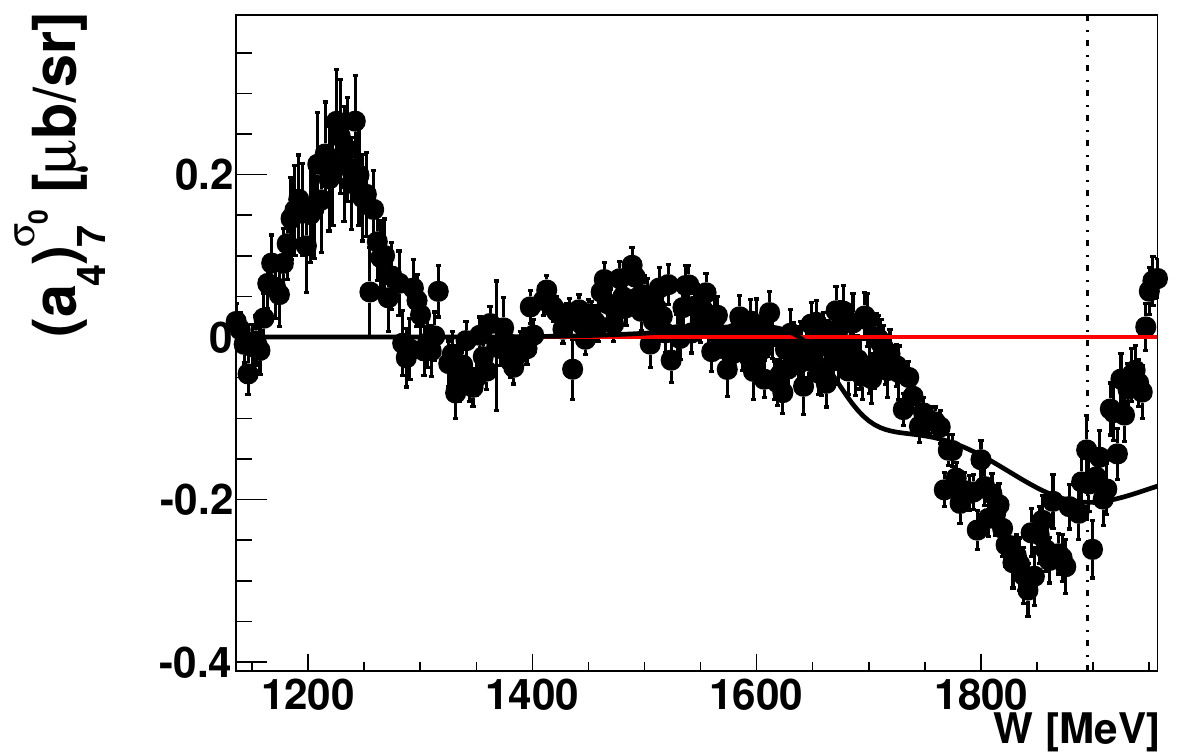}
  \includegraphics[width=0.285\textwidth]{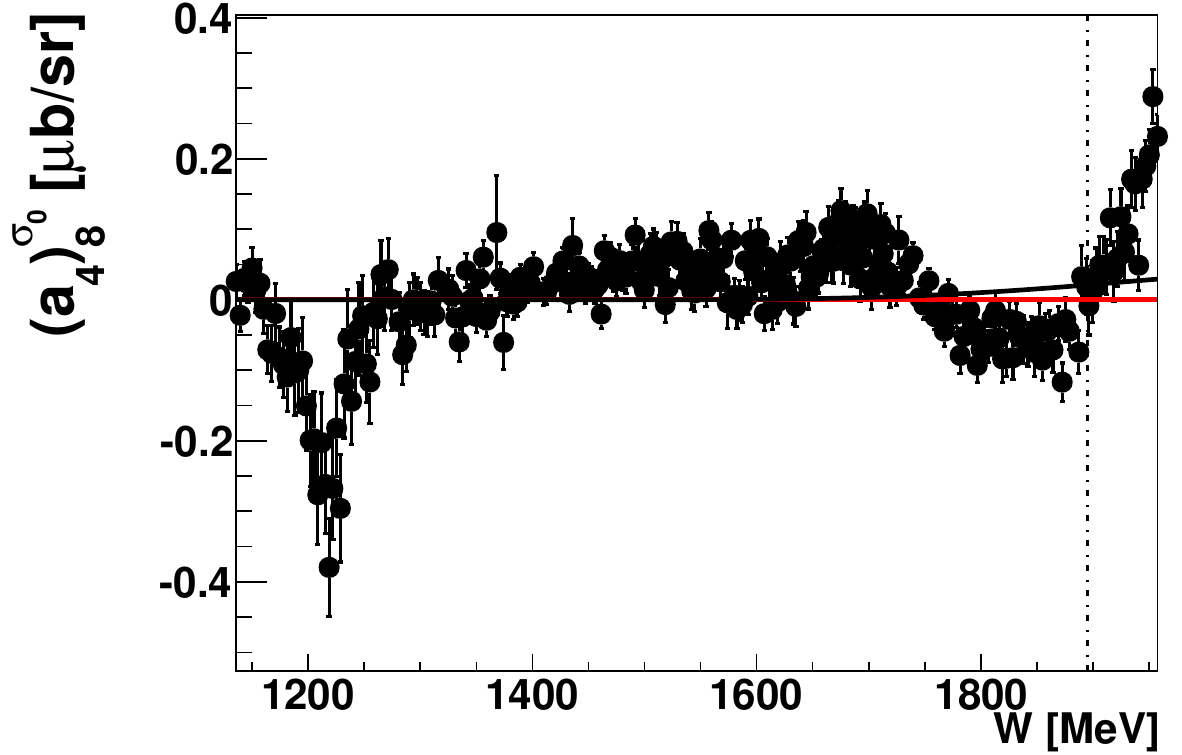}    
  \end{minipage}
\end{figure*}

\begin{figure*}
\begin{minipage}{\textwidth}
\floatbox[{\capbeside\thisfloatsetup{capbesideposition={right,top},capbesidewidth=7.8cm}}]{figure}[\FBwidth]
{\caption{The recent new differential cross section $\sigma_0$ data from MAMI \cite{Adlarson:2015} with statistical and systematical error was fitted using associated Legendre polynomials according to eq. \ref{eq:LowEAssocLegParametrizationDCS} and truncating the partial wave expansion at $\text{L}_{\text{max}}=1\dots 5$. (a) The resulting $\chi^2/$ndf values of the different $\text{L}_{\text{max}}$-fits as a function of the center of mass energy W are shown. (b) 6 out of 265 selected angular distributions of $\sigma_0$ (black points) are plotted together with the different $\text{L}_{\text{max}}$ fits (solid lines) starting at W=1154 MeV up to 1855 MeV. (c) Comparison of the fit coefficients for $\text{L}_{\text{max}}=4$ (black points), $\left(a_{4}\right)^{\sigma_0}_{0\dots8}$ (see eq. \ref{eq:LowEAssocLegParametrizationDCS}), with the BnGa2014-02 solution truncated at different $\text{L}_{\text{max}}$ (solid lines). Colors same as in (a).}\label{fig:wq_bins_wsys}}
{\includegraphics[width=0.49\textwidth, trim=0cm 0cm 1.8cm 0cm, clip]{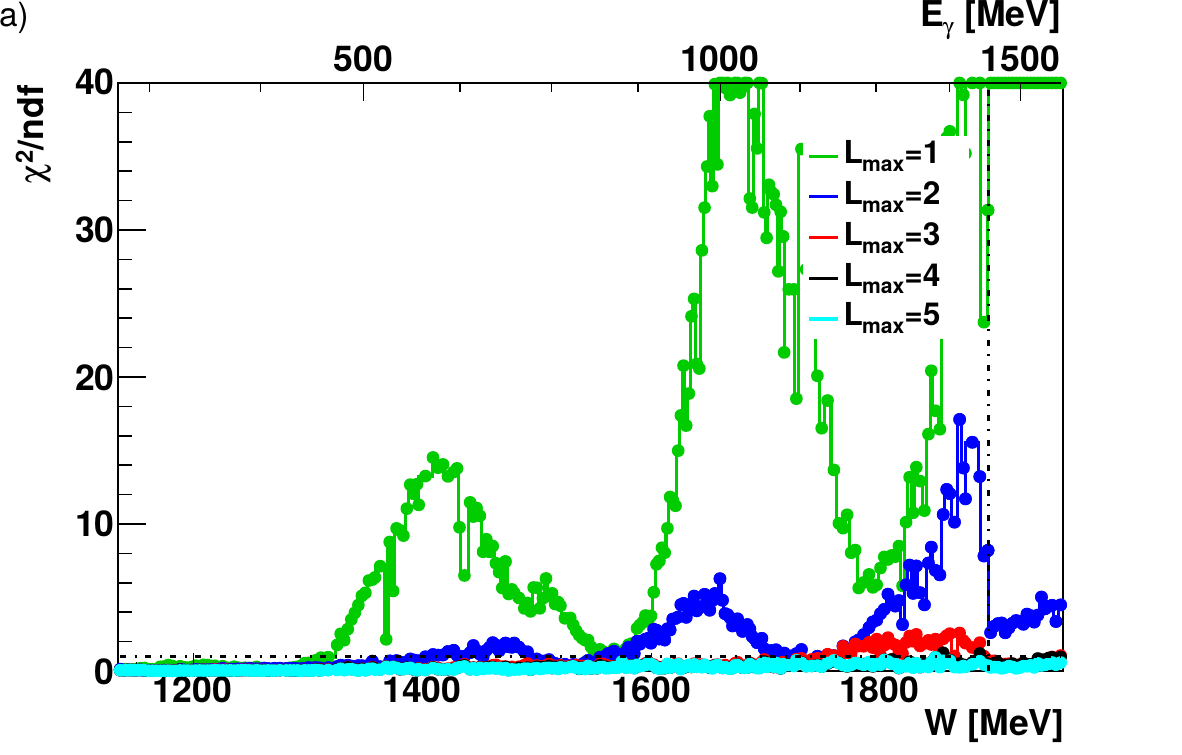}}
\end{minipage}\\

\begin{minipage}{\textwidth}
\centering
\hspace*{-0.45cm}
 \includegraphics[width=0.305\textwidth, trim=0cm 0cm 0.01cm 0.75cm, clip]{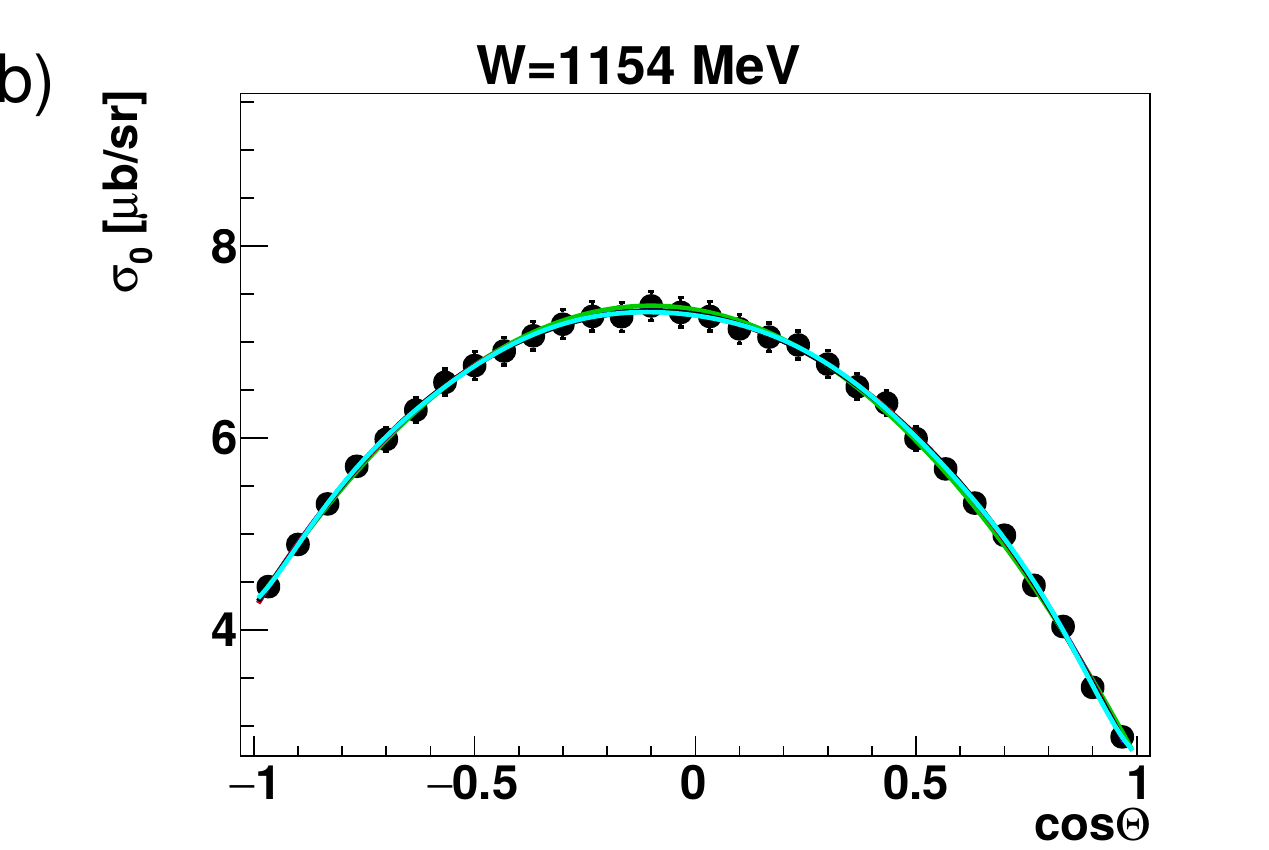}
  \includegraphics[width=0.285\textwidth, trim=0cm 0cm 0.01cm 0.75cm, clip]{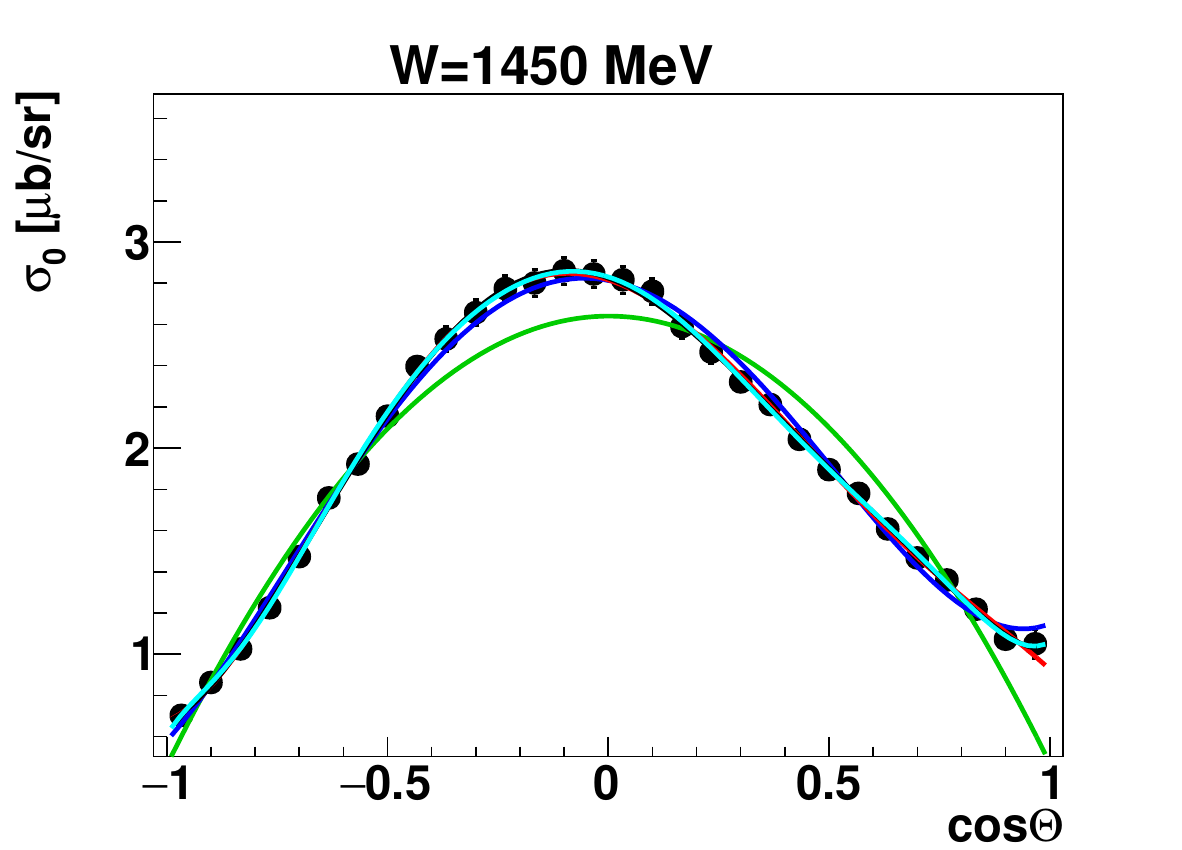}
  \includegraphics[width=0.285\textwidth, trim=0cm 0cm 0.01cm 0.75cm, clip]{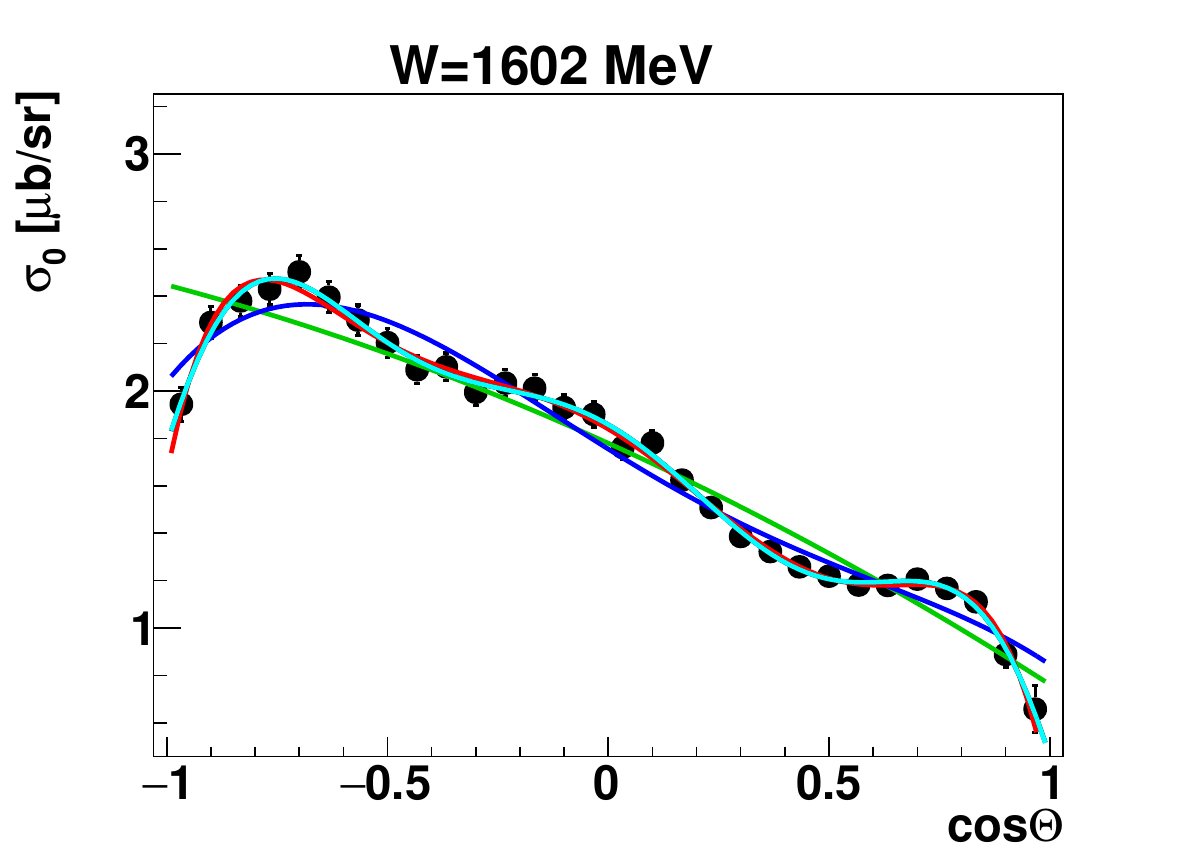}\\
  \includegraphics[width=0.285\textwidth, trim=0cm 0cm 0.01cm 0.75cm, clip]{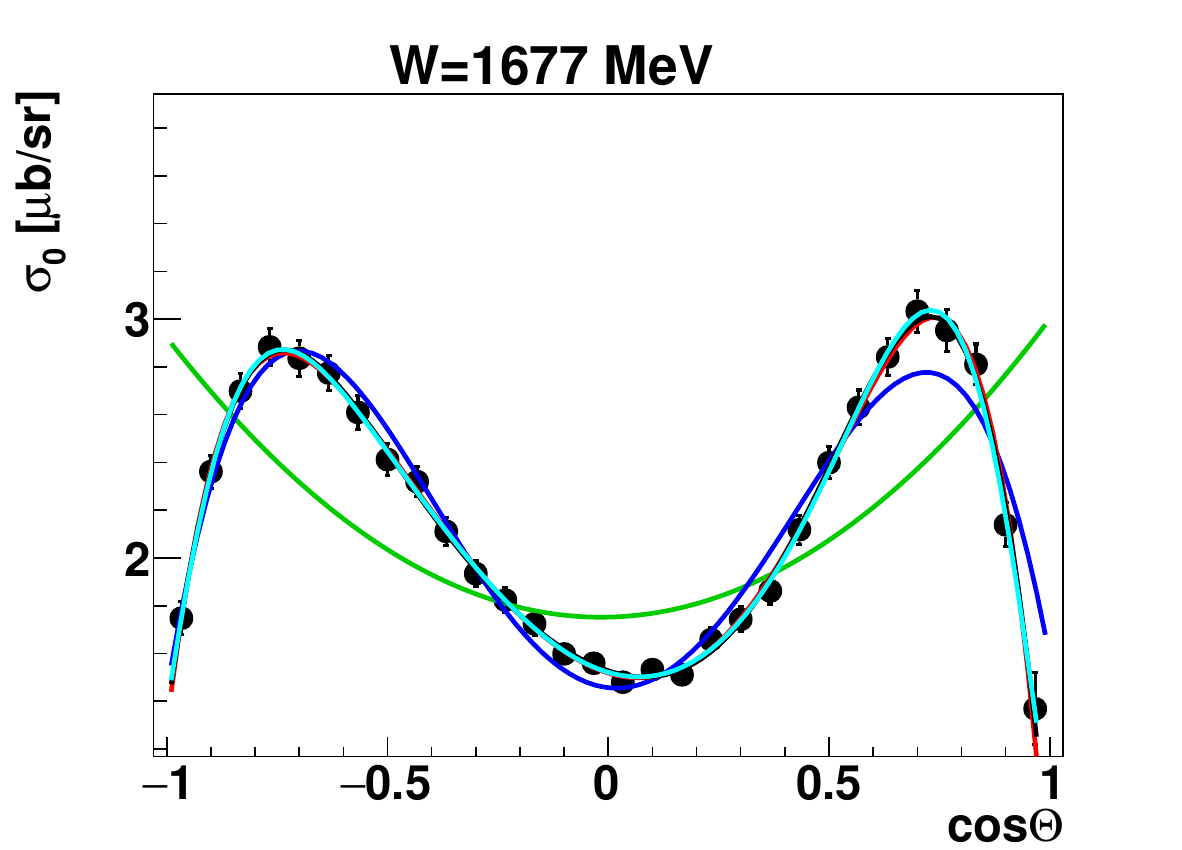}
  \includegraphics[width=0.285\textwidth, trim=0cm 0cm 0.01cm 0.75cm, clip]{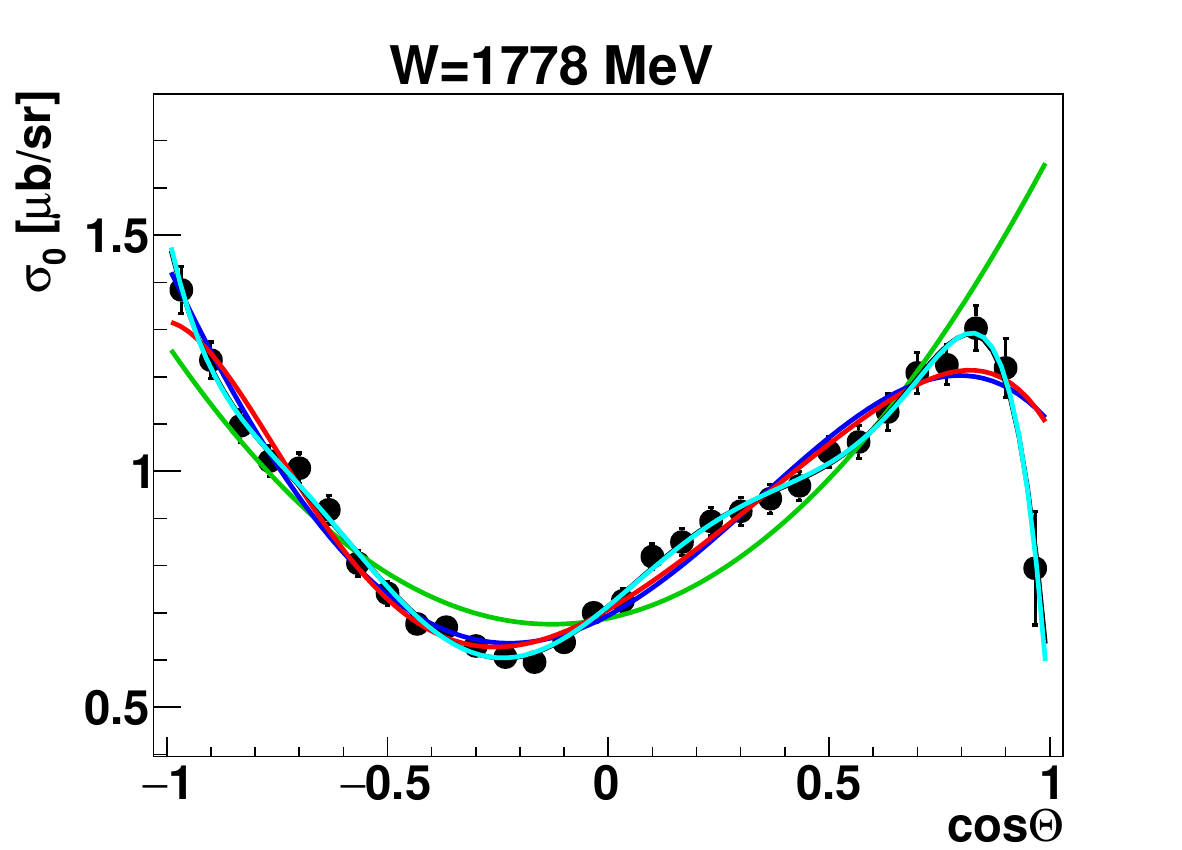}
  \includegraphics[width=0.285\textwidth, trim=0cm 0cm 0.01cm 0.75cm, clip]{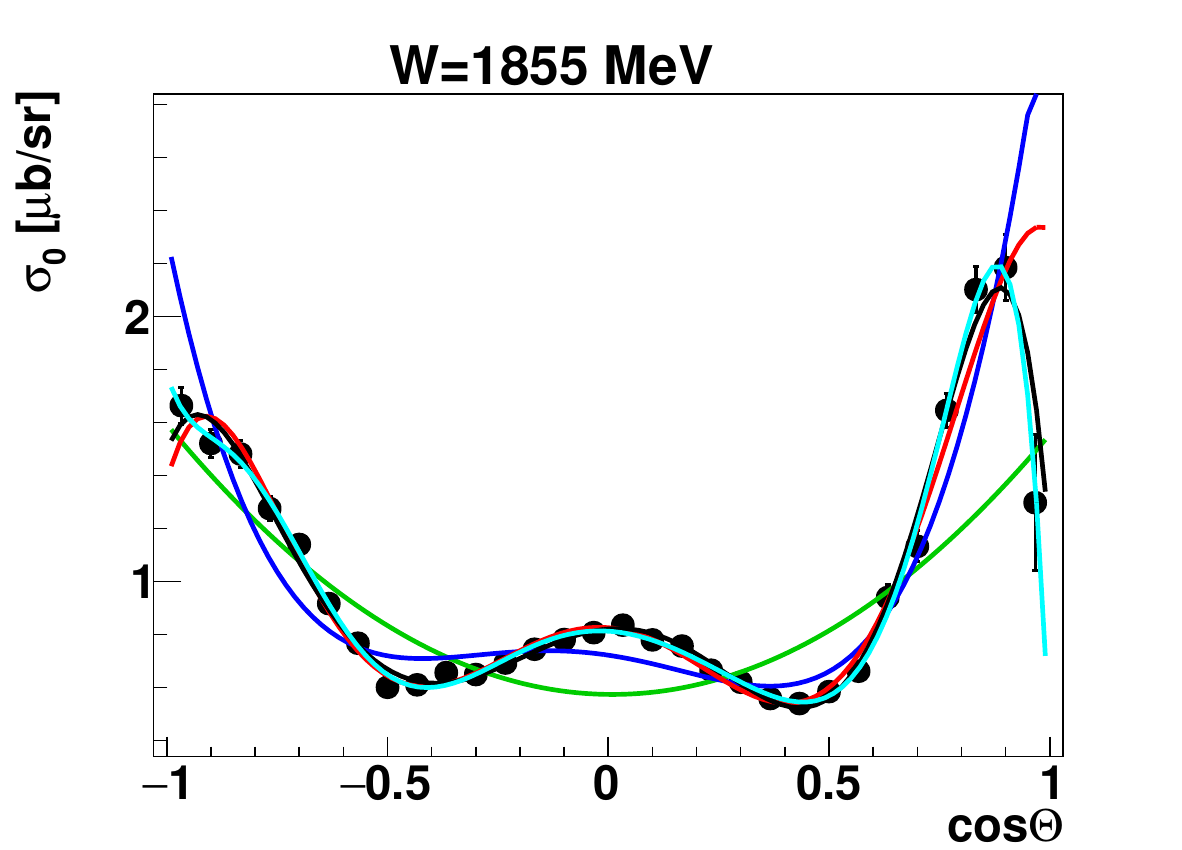}\\
 \vspace*{0.5cm}
  
  \hspace*{-23.5pt}\includegraphics[width=0.2905\textwidth]{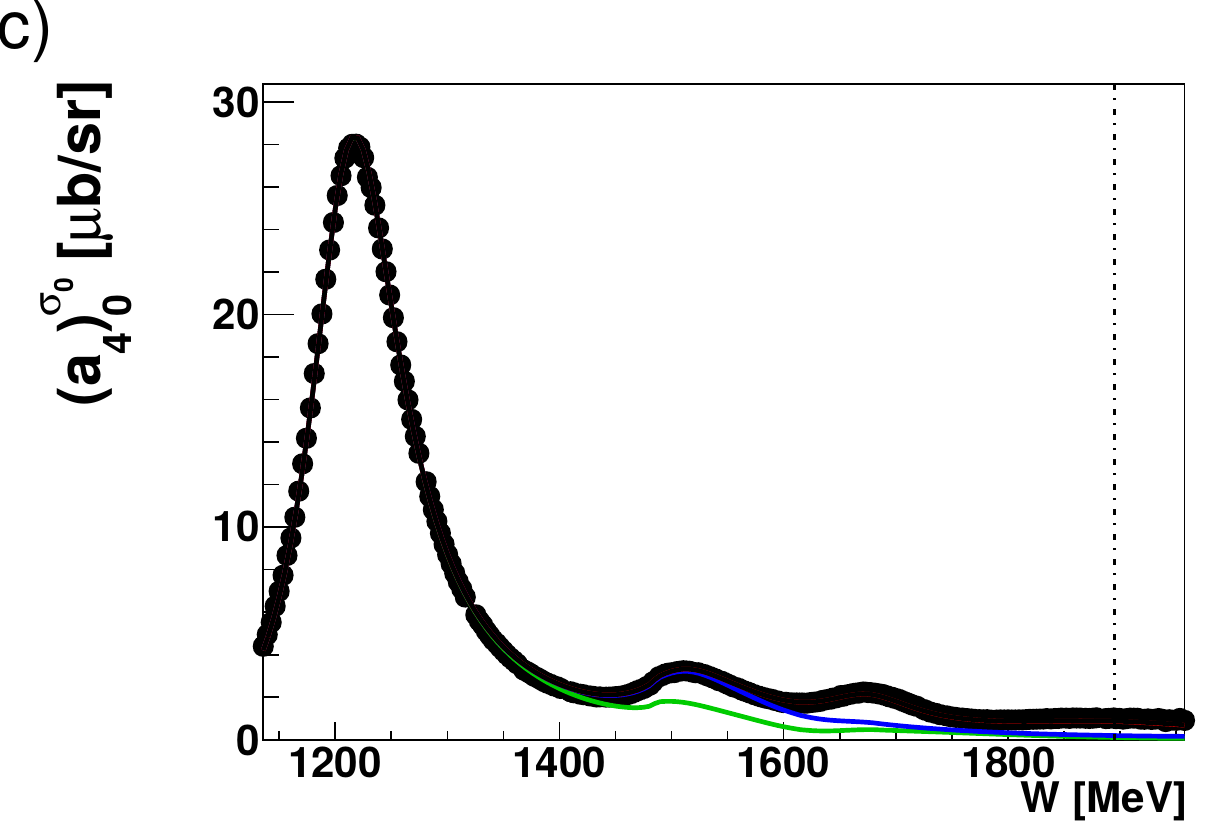}  
  \includegraphics[width=0.285\textwidth]{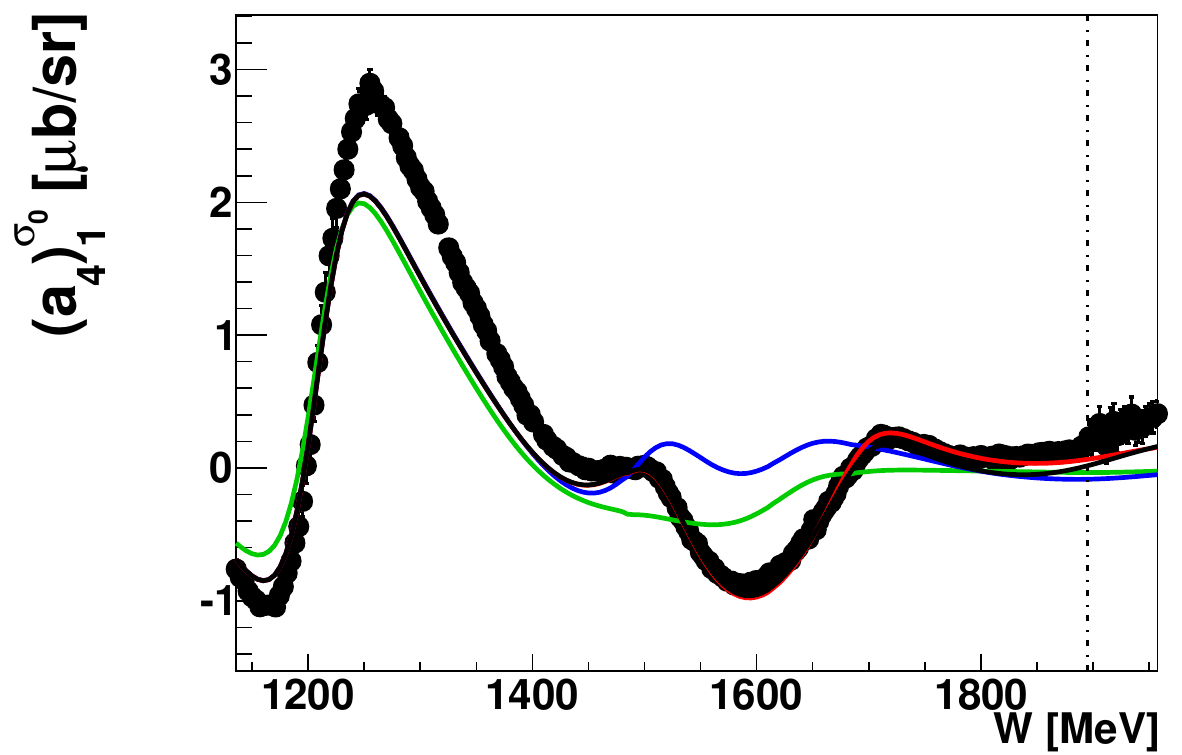}
  \includegraphics[width=0.285\textwidth]{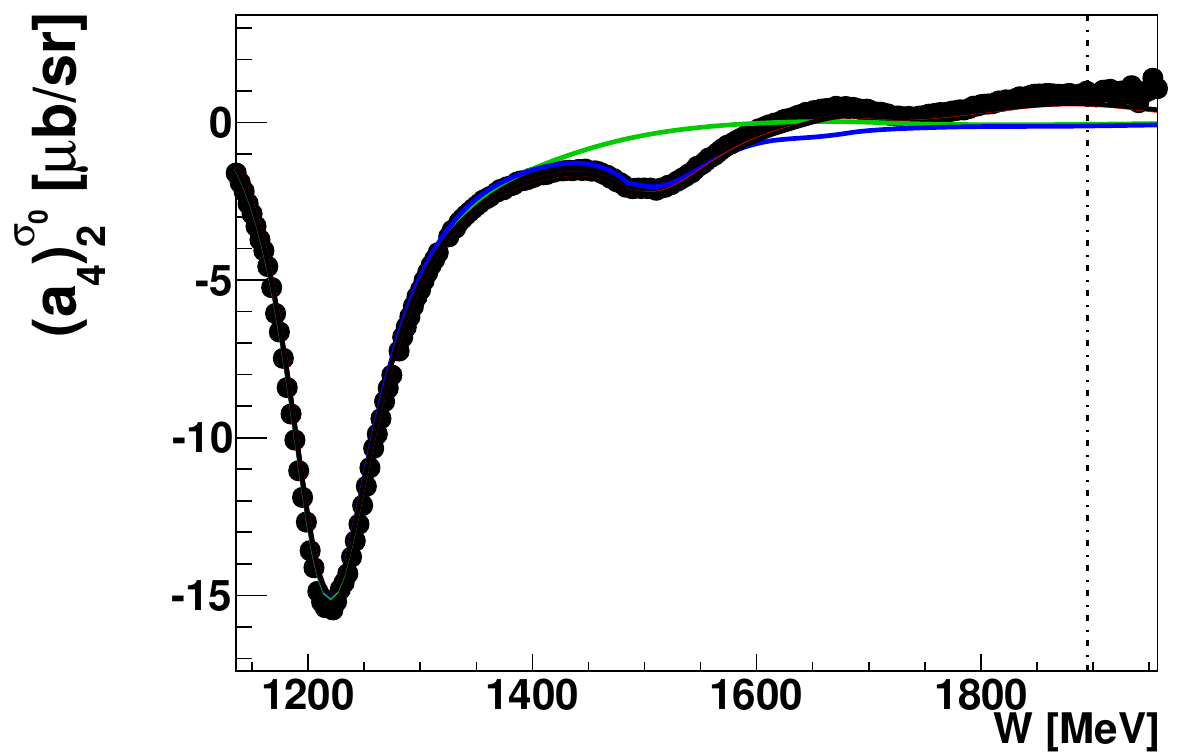}\\
  \hspace*{-19.5pt}\includegraphics[width=0.285\textwidth]{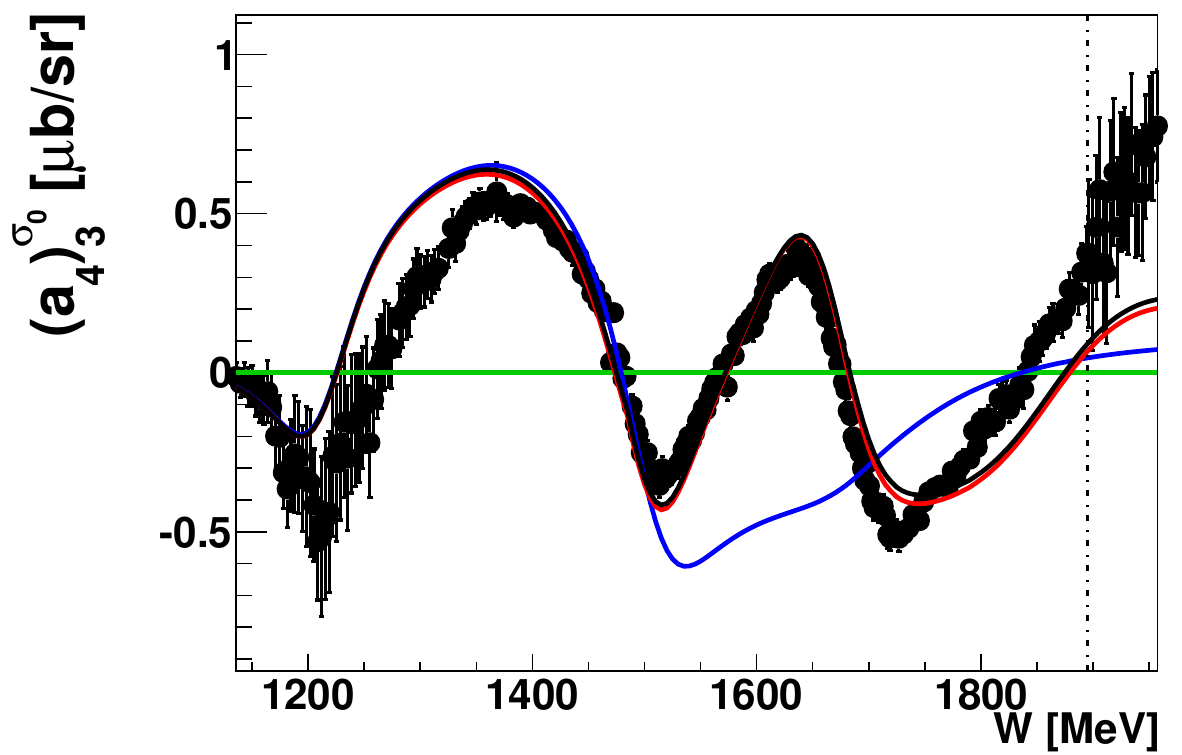}
  \includegraphics[width=0.285\textwidth]{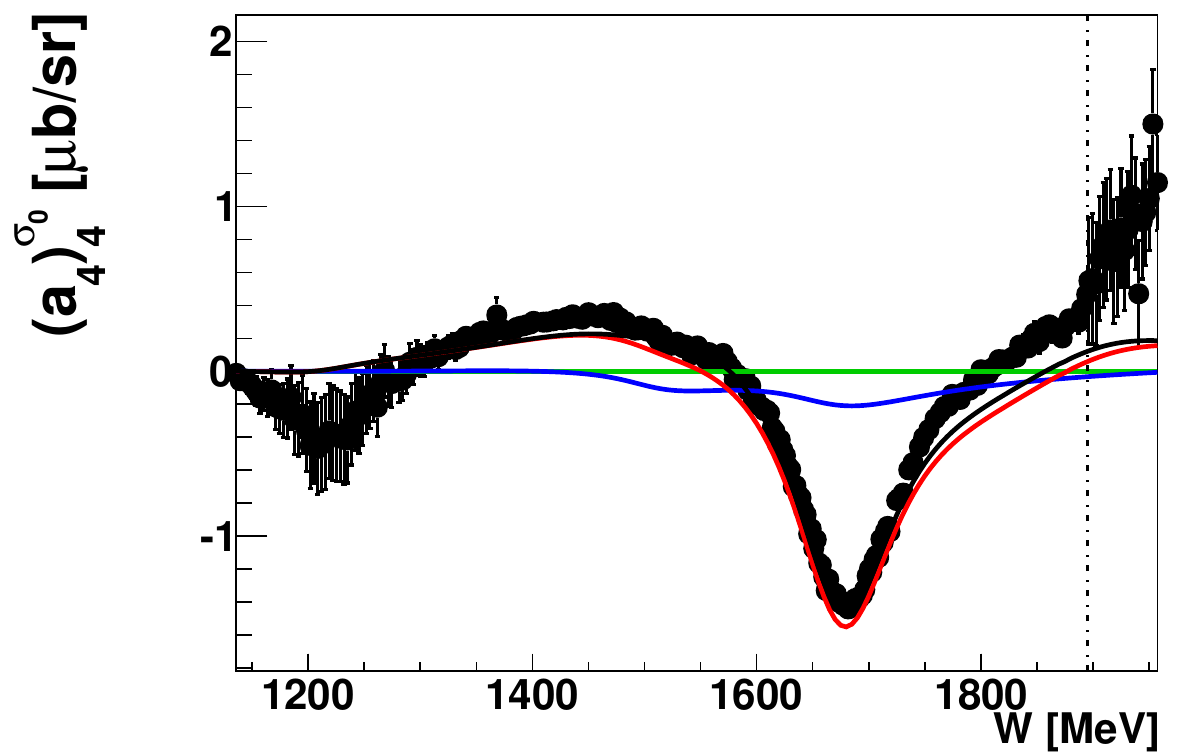}
  \includegraphics[width=0.285\textwidth]{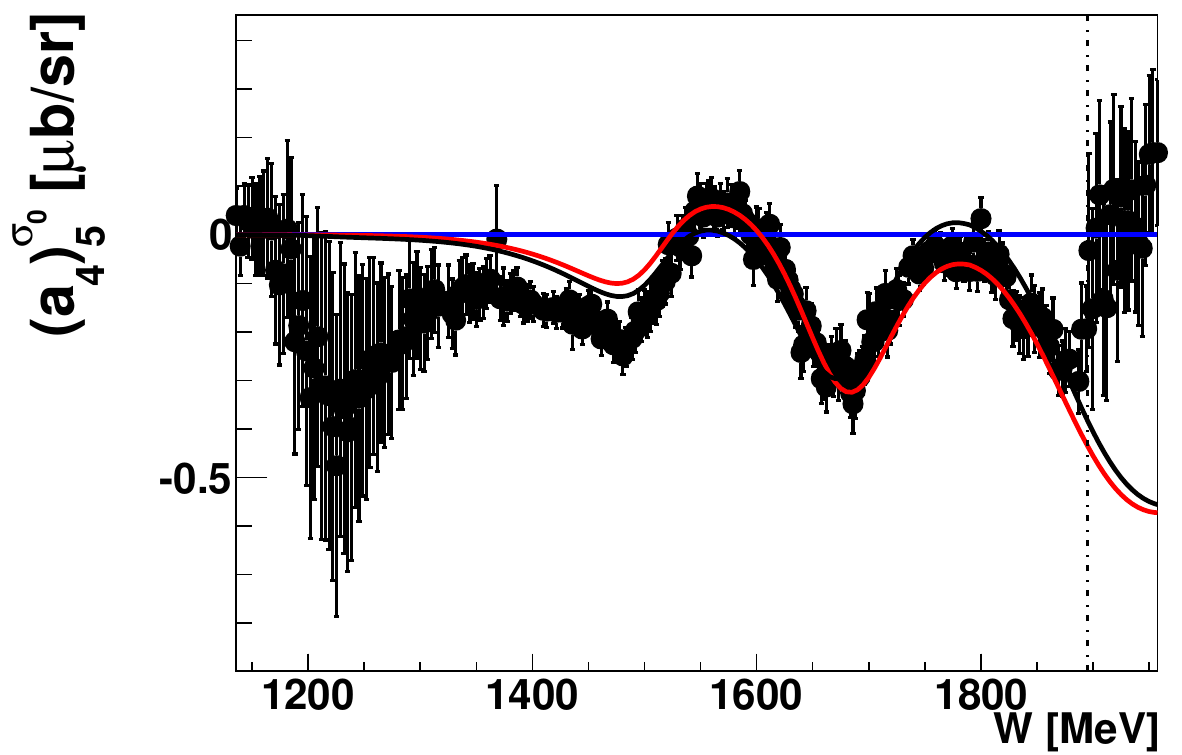}\\
  \hspace*{-19.5pt}\includegraphics[width=0.285\textwidth]{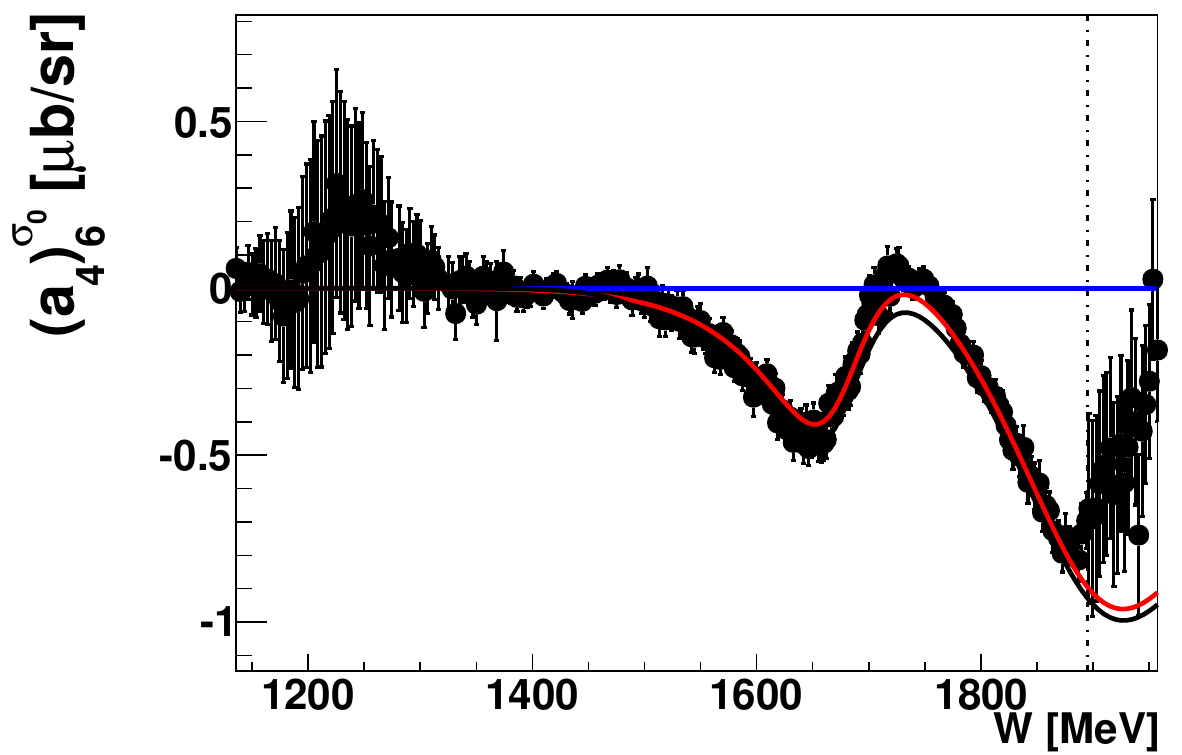}
  \includegraphics[width=0.285\textwidth]{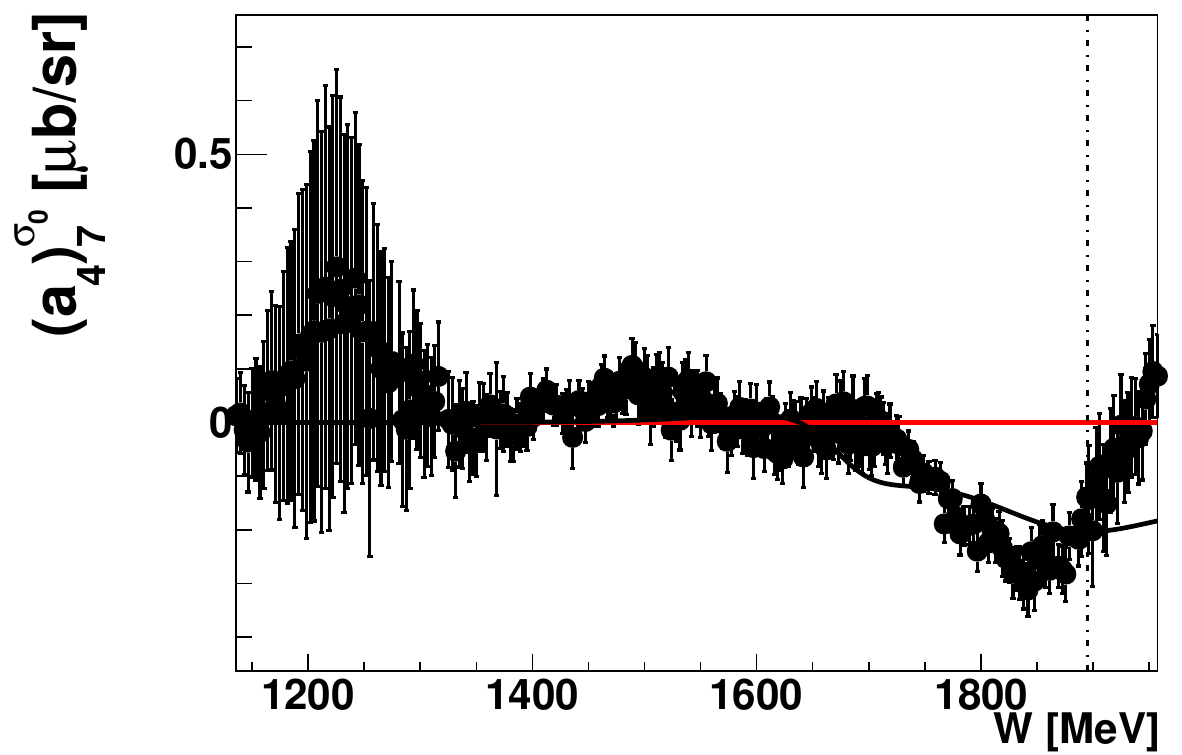}
  \includegraphics[width=0.285\textwidth]{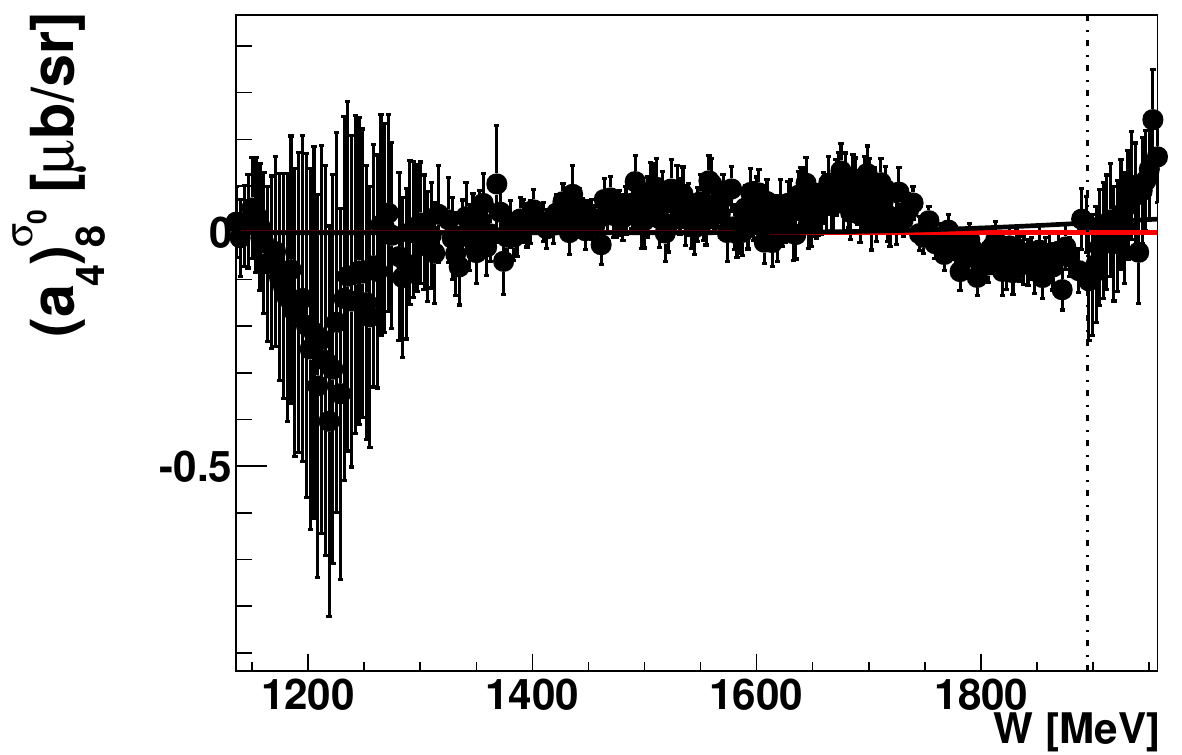}    
  \end{minipage}
\end{figure*}

\begin{figure*}[htb]
\centering
\includegraphics[width=0.285\textwidth]{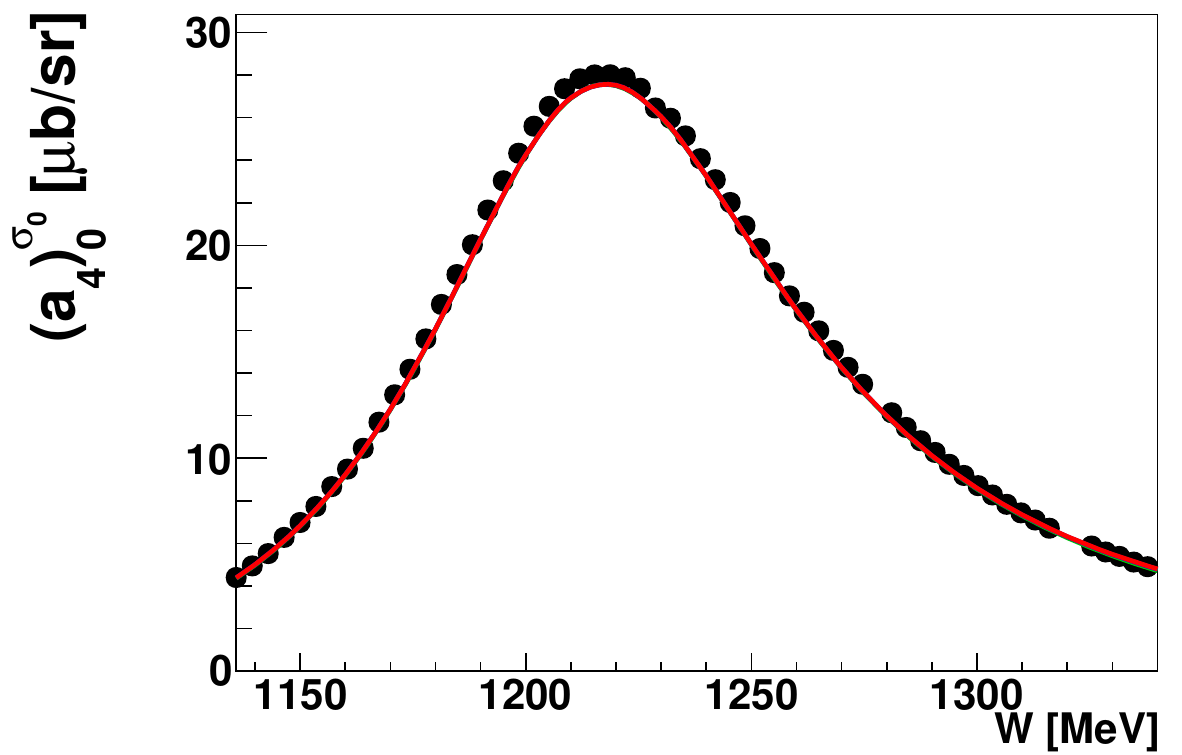}
  \includegraphics[width=0.285\textwidth]{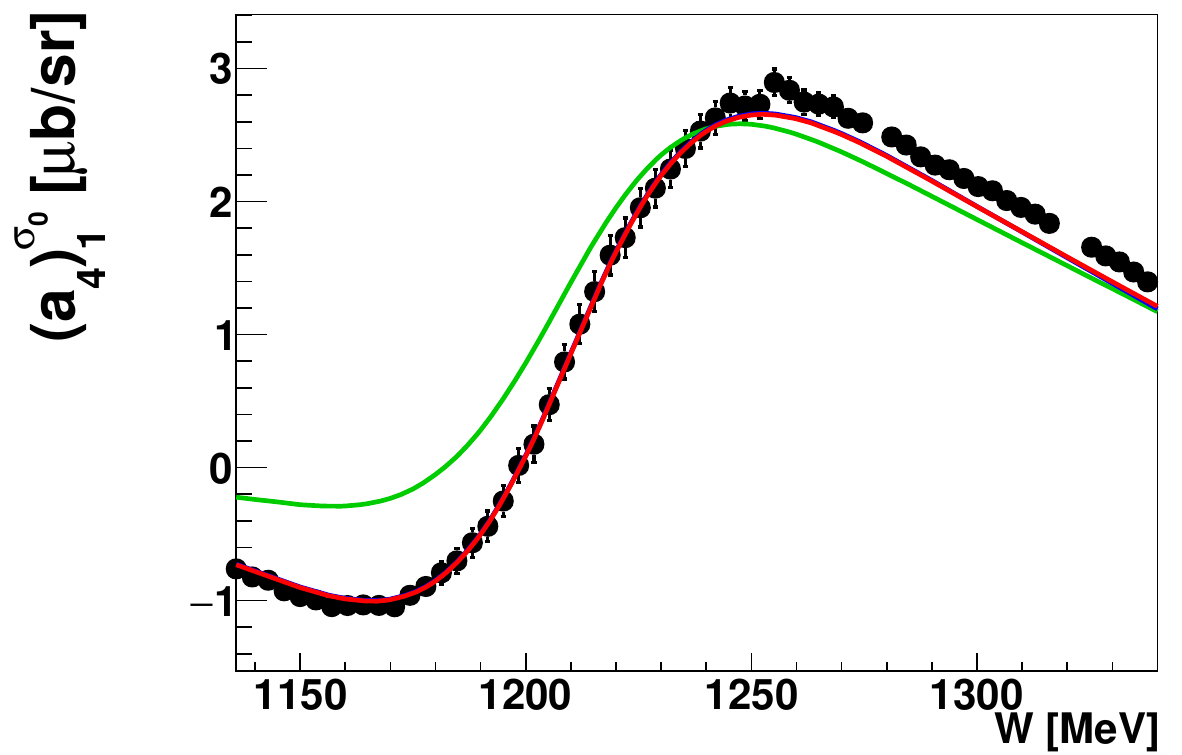}
  \includegraphics[width=0.285\textwidth]{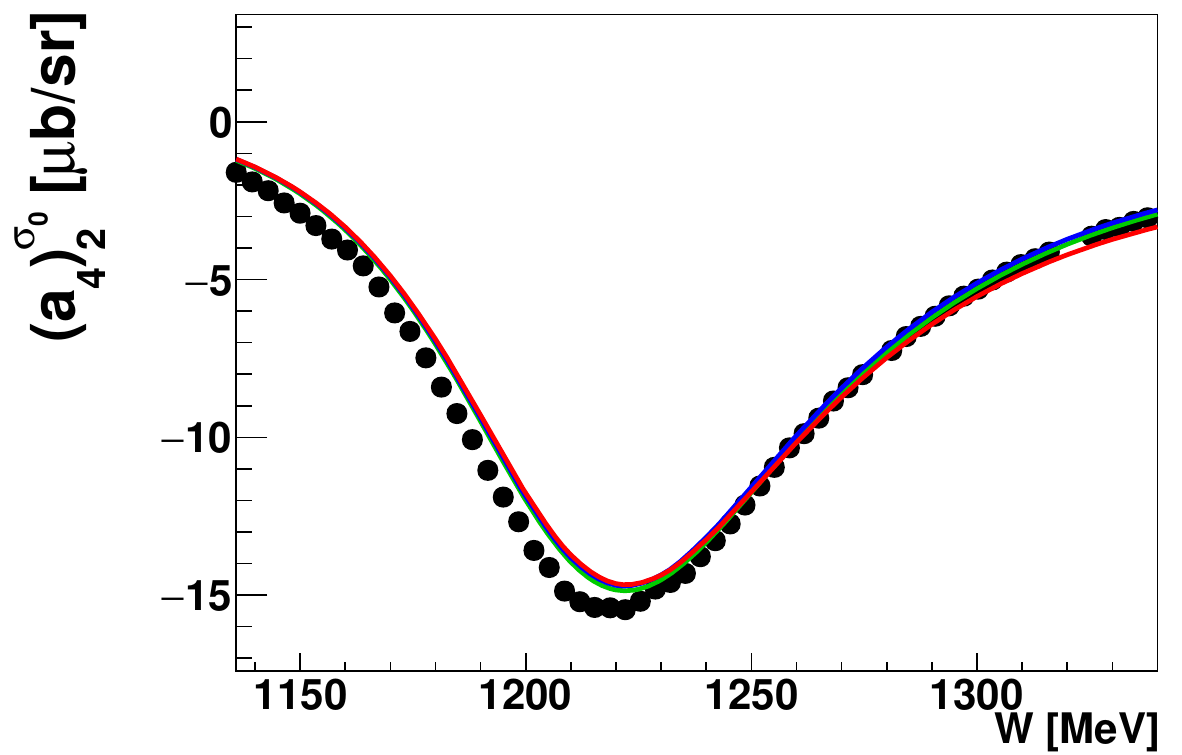}\\
  \hspace*{-19.5pt}\includegraphics[width=0.285\textwidth]{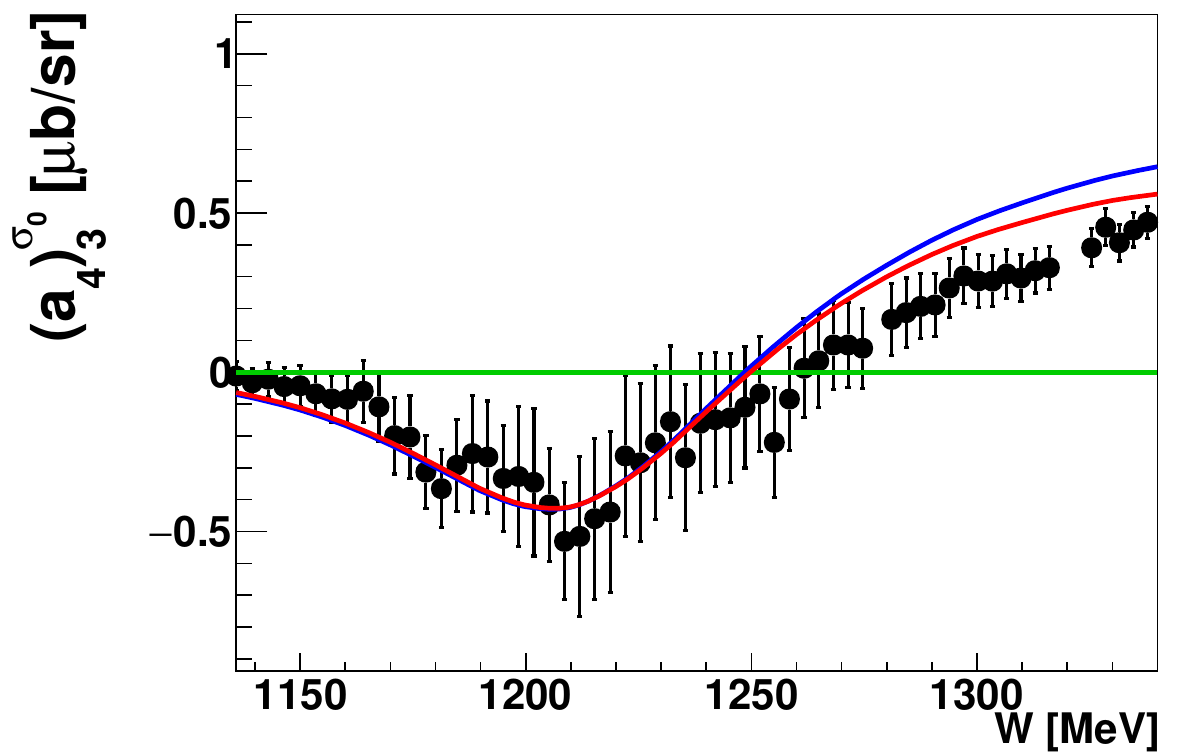}
  \includegraphics[width=0.285\textwidth]{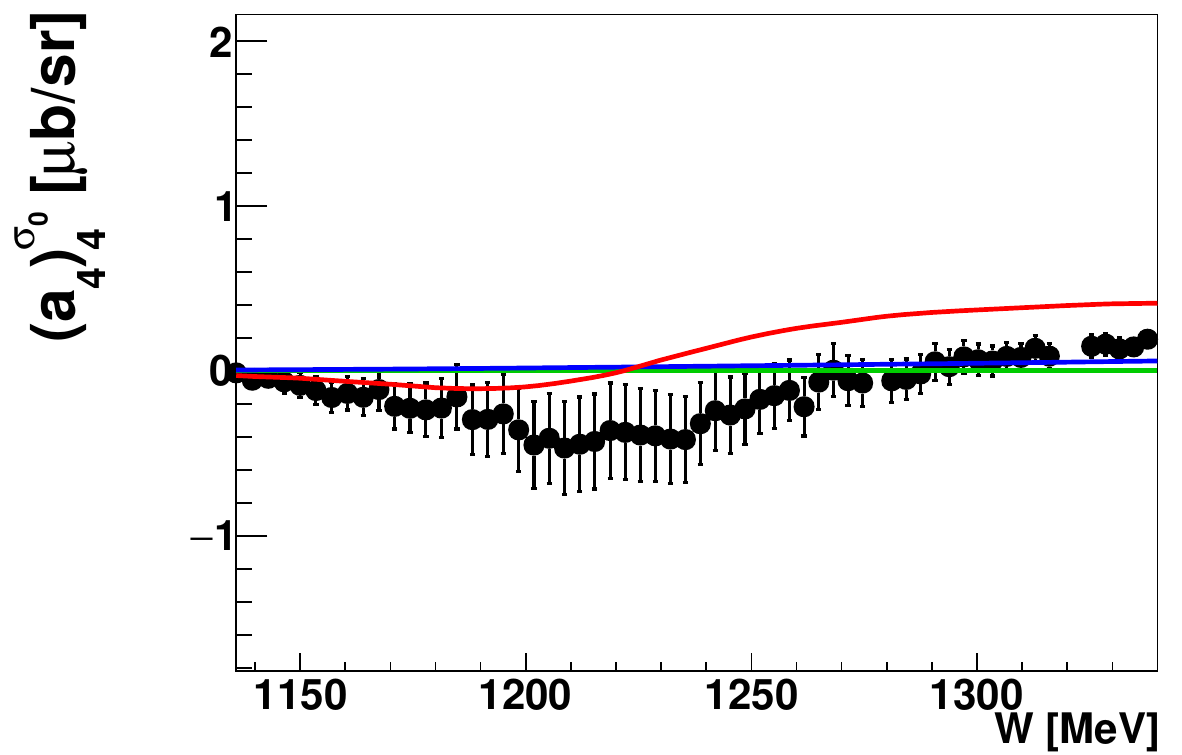}
\caption{The fitted Legendre coefficients $\left(a_{\text{L}_{\text{max}}}\right)^{\sigma_{0}}_{(0,\ldots,4)}$ are plotted in comparison to predictions from SAID-CM12 multipoles \cite{SAID} up to $\text{L}_{\text{max}} = 1$ (green), $\text{L}_{\text{max}} = 2$ (blue) and $\text{L}_{\text{max}} = 3$ (red).}
\label{fig:wq_Coeffs_SAID}       
\end{figure*}

\begin{figure*}[htb]
\centering
\includegraphics[width=0.285\textwidth]{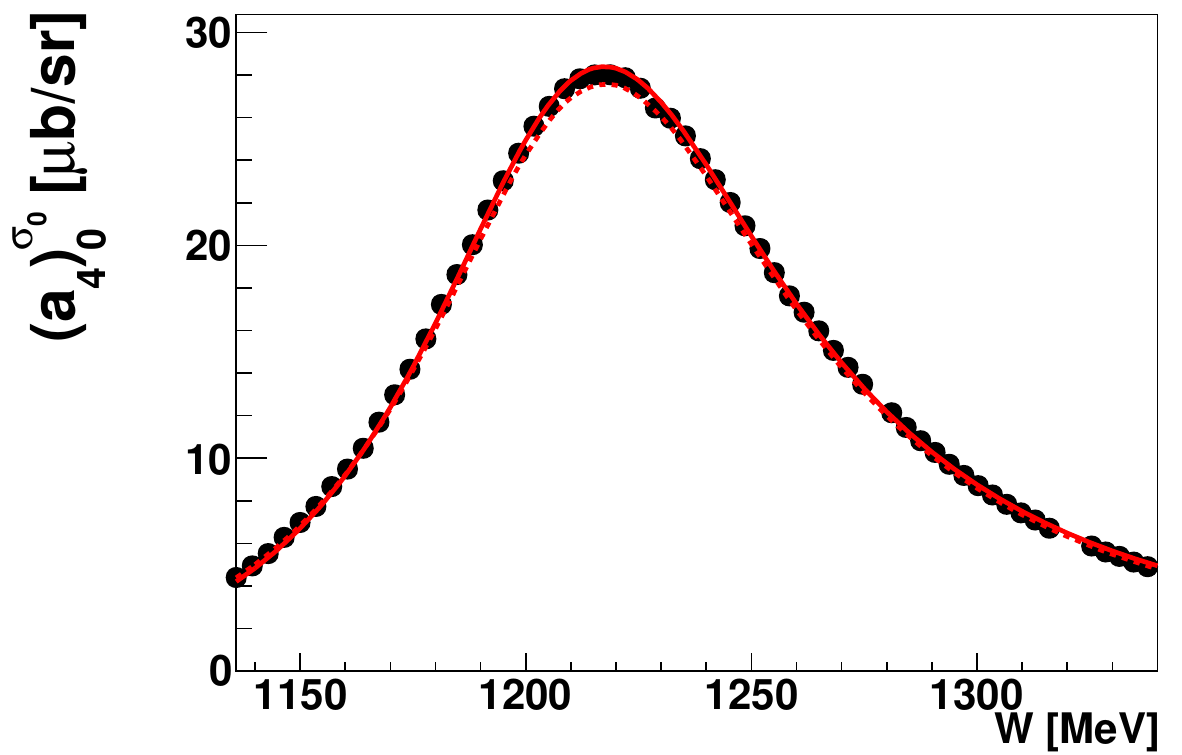}
  \includegraphics[width=0.285\textwidth]{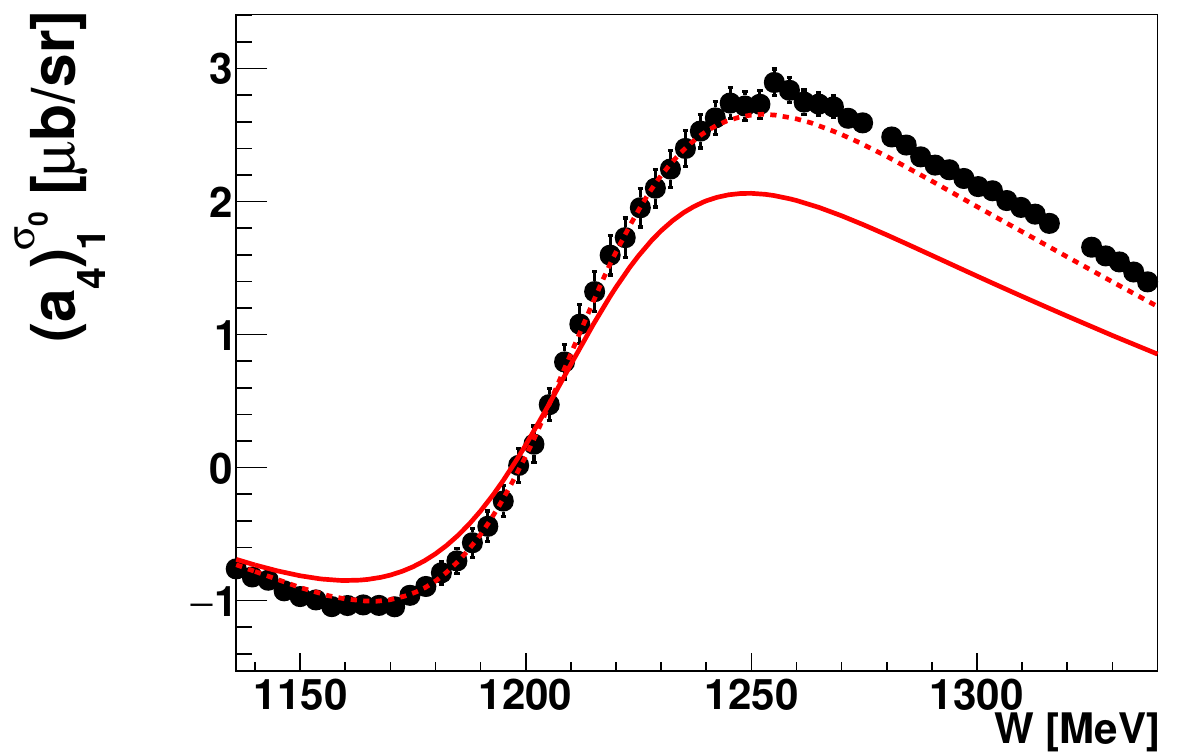}
  \includegraphics[width=0.285\textwidth]{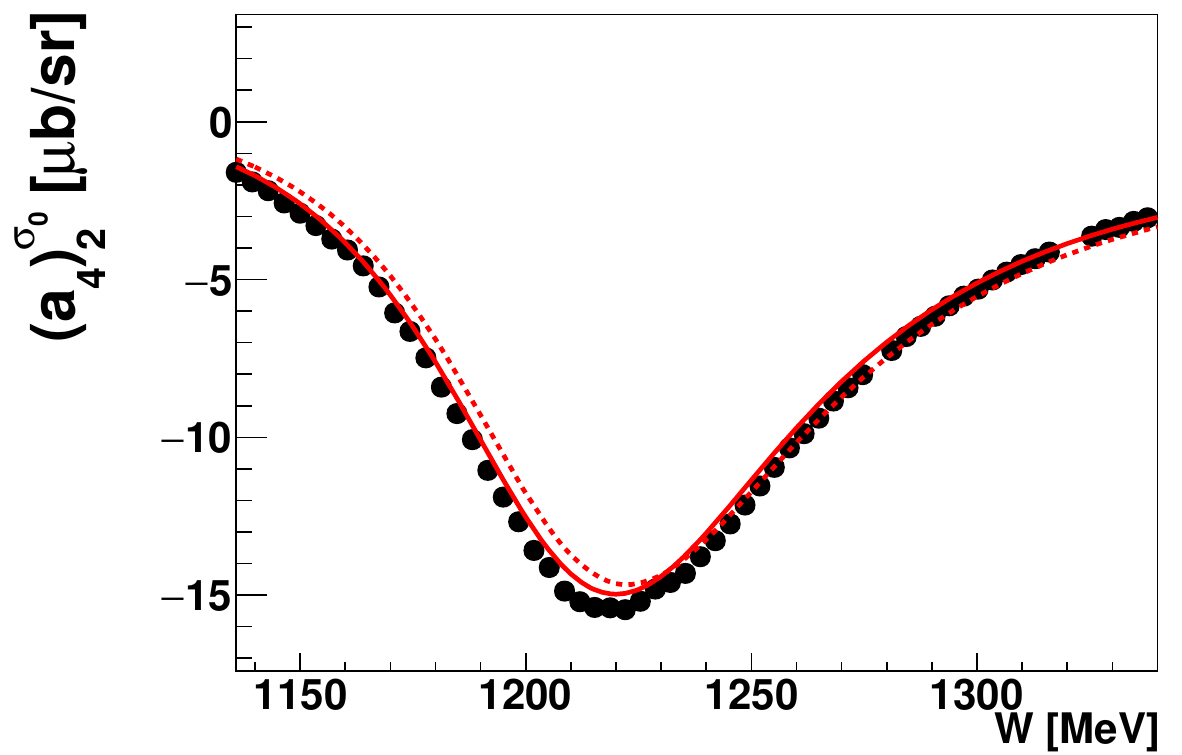}\\
  \hspace*{-19.5pt}\includegraphics[width=0.285\textwidth]{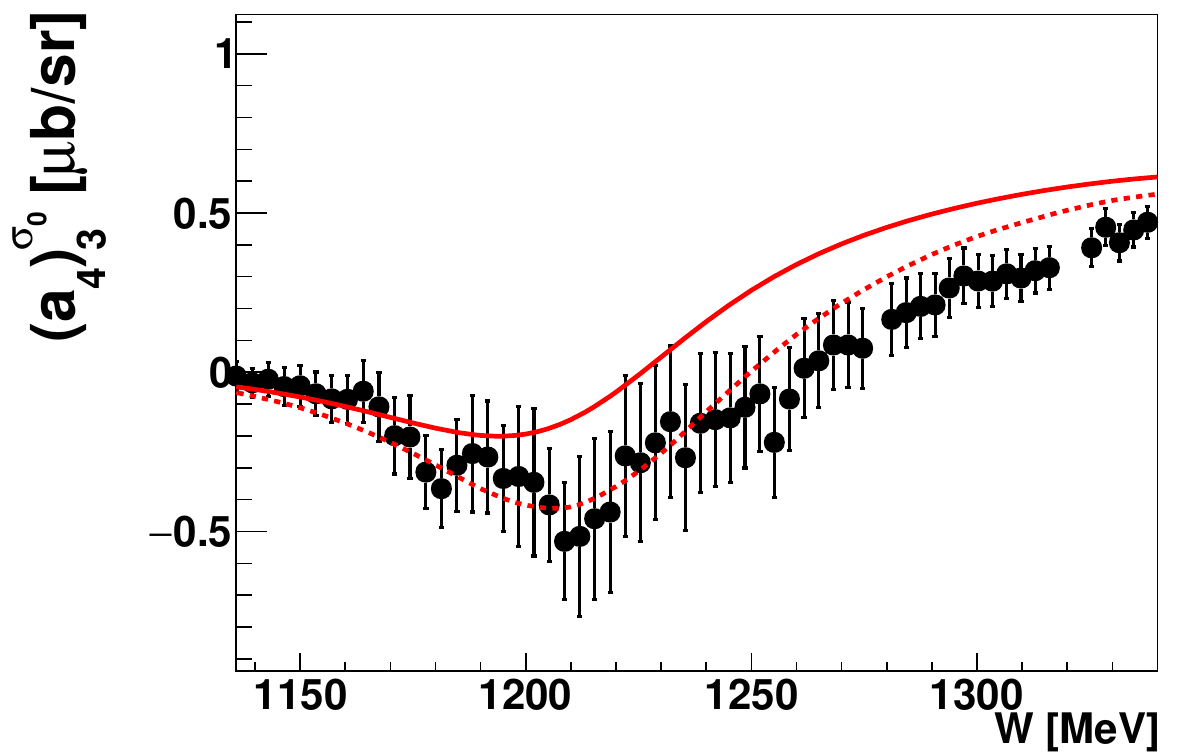}
  \includegraphics[width=0.285\textwidth]{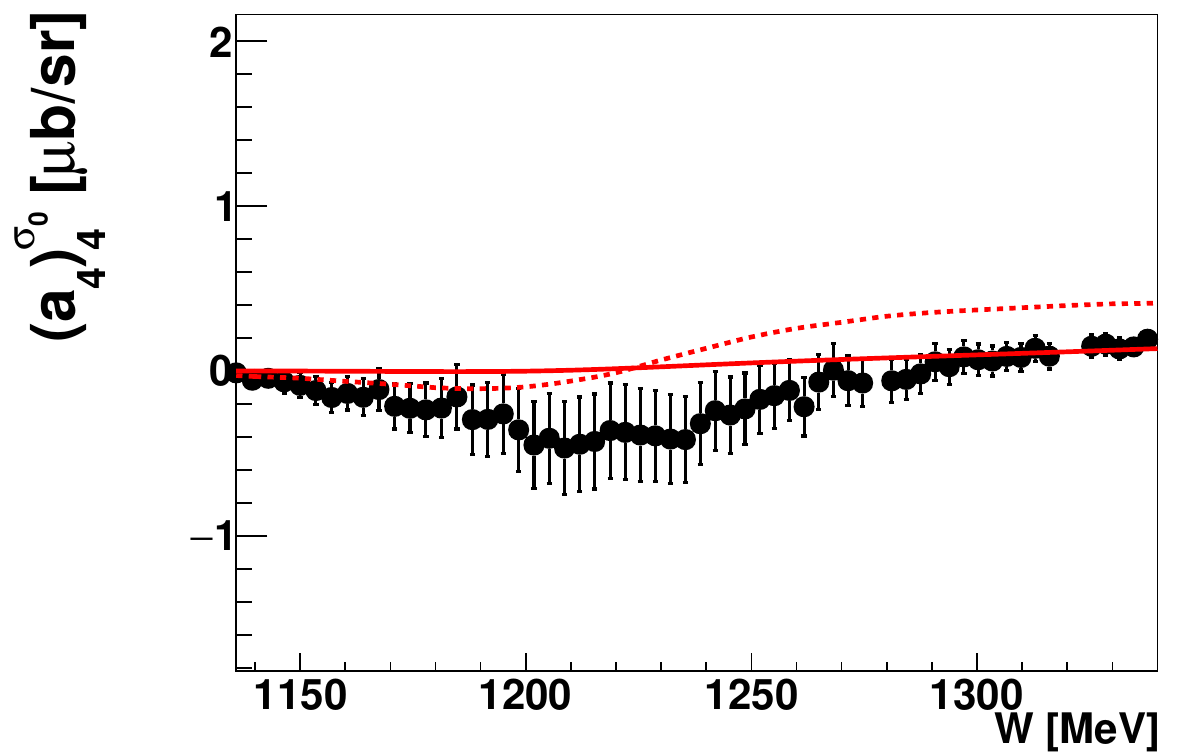}
\caption{The fitted Legendre coefficients $\left(a_{\text{L}_{\text{max}}}\right)^{\sigma_{0}}_{(0,\ldots,4)}$ are shown in comparison to predictions from BnGa2014-02 (red solid line) as well as SAID-CM12 \cite{SAID} multipoles (red dashed line) up to, in both cases, $\text{L}_{\text{max}} = 3$.}
\label{fig:wq_Coeffs_CompareSAIDBnGa}       
\end{figure*}

The following interpretation of the fits is organized ascending in energy, according to the four so-called 'resonance regions' present in the literature. The first resonance region denotes the energy range from threshold up to $E_{\gamma}^{\mathrm{LAB}} = 500 \hspace*{1.5pt} \hspace*{2pt} \mathrm{MeV}$ (or $W = 1350 \hspace*{1.5pt} \hspace*{2pt} \mathrm{MeV}$). The second region spans the energy interval $E_{\gamma}^{\mathrm{LAB}} \in \left[ 500, 900 \right] \hspace*{2pt} \mathrm{MeV}$ (or $W \in \left[ 1350, 1600 \right] \hspace*{2pt} \mathrm{MeV}$ respectively). We denote the energy range from $E_{\gamma}^{\mathrm{LAB}} = 900 \hspace*{1.5pt} \hspace*{2pt} \mathrm{MeV}$ ($W = 1600 \hspace*{1.5pt} \hspace*{2pt} \mathrm{MeV}$) up to $E_{\gamma}^{\mathrm{LAB}} = 1256 \hspace*{1.5pt} \hspace*{2pt} \mathrm{MeV}$ ($W = 1800 \hspace*{1.5pt} \hspace*{2pt} \mathrm{MeV}$) as the third 'resonance region'. Finally, we regard as the fourth resonance region the remaining energy regime from $W = 1800 \hspace*{1.5pt} \hspace*{2pt} \mathrm{MeV}$ up to the highest energy of the considered datasets, i.e. $E_{\gamma}^{\mathrm{LAB}} = 2300 \hspace*{1.5pt} \hspace*{2pt} \mathrm{MeV}$ ($W = 2250 \hspace*{1.5pt} \hspace*{2pt} \mathrm{MeV}$) from the $E$-dataset (cf. Table \ref{tab:DataBasis}). \newline
For each of these regions we consider the values of $\chi^{2}/\mathrm{ndf}$ vs. energy for specific truncation orders. Additionally we show examples of fitted angular distributions that illustrate the afore mentioned $\chi^{2}$-plots. Finally, and perhaps most importantly for the interpretation, the resulting Legendre coefficients are plotted for each observable and compared to predictions from the Bonn-Gatchina partial wave analysis. These comparisons can illustrate the sensitivity of certain observables to specific partial wave interferences rather well. Also, they can give a first hint of the dominating, possibly resonant, partial wave contributions in the data which may then be compared to
the current PDG resonances listed in Table \ref{tab:MultipoleResonanceAssignments}. A quicker reference relating multipoles to partial wave states and examples for well-established resonances is provided in Table \ref{tab:PartialWavesMultipoles}. \newline
In order to aid the interpretation, we have given the composition of the matrices defining the Legendre coefficients as bilinear hermitean forms of the multipoles (cf. Eq. (\ref{eq:LowEAssocLegParametrization2}) and section \ref{sec:DescriptionCompositionLegCoeffs}), for every observable up to $\text{L}_{\text{max}} = 5$ in App. \ref{sec:PWContentFormulas}. \newline
The $\chi^{2}/\mathrm{ndf}$ is strongly dependent on the number of data points as well as the covered angular range. Therefore, we have indicated the energy ranges of equivalent angular data points by dashed lines in Fig. \ref{fig:wq_bins} - \ref{fig:Sclas_bins} for all investigated datasets. By comparing the  $\chi^{2}/\mathrm{ndf}$ of the fits to the statistically equivalent data, found within two dashed lines respectively, it can be confirmed that the observed structures do not only depend on the angular coverage and number of data points, but really give information about the sensitivity on contributing resonances.

\begin{table}[htb]
\RawFloats
\centering
\begin{tabular}{l|c|l|l}
\hline\hline
$\ell_{\pi}$ & $P$ & Multipoles & Partial Wave States \\\hline
 & & & \\
$0$ $(S)$ & $-$ &$E_{0+}$  & $N \frac{1}{2}^{-}$, $\Delta \frac{1}{2}^{-}$ \\
 & & &  $N \left( 1535 \right) \frac{1}{2}^{-}$ \\
$1$ $(P)$ & $+$ &$E_{1 \pm}$, $M_{1-}$ & $N \frac{1}{2}^{+}$, $N \frac{3}{2}^{+}$, $\Delta \frac{1}{2}^{+}$, $\Delta \frac{3}{2}^{+}$\\
 & & &  $N \left( 1440 \right) \frac{1}{2}^{+}$, $\Delta \left( 1232 \right) \frac{3}{2}^{+}$ \\
$2$ $(D)$ & $-$ &$E_{2 \pm}$, $M_{2 \pm}$ & $N \frac{3}{2}^{-}$, $N \frac{5}{2}^{-}$, $\Delta \frac{3}{2}^{-}$, $\Delta \frac{5}{2}^{-}$ \\
 & & &  $N \left( 1520 \right) \frac{3}{2}^{-}$, $\Delta \left( 1700 \right) \frac{3}{2}^{-}$ \\
$3$ $(F)$ & $+$ &$E_{3 \pm}$, $M_{3 \pm}$ & $N \frac{5}{2}^{+}$, $N \frac{7}{2}^{+}$, $\Delta \frac{5}{2}^{+}$, $\Delta \frac{7}{2}^{+}$ \\
 & & &  $N \left( 1680 \right) \frac{5}{2}^{+}$, $\Delta \left( 1905 \right) \frac{5}{2}^{+}$ \\
$4$ $(G)$ & $-$ &$E_{4 \pm}$, $M_{4 \pm}$ & $N \frac{7}{2}^{-}$, $N \frac{9}{2}^{-}$, $\Delta \frac{7}{2}^{-}$, $\Delta \frac{9}{2}^{-}$ \\
 & & & $N \left( 2190 \right) \frac{7}{2}^{-}$ \\
$5$ $(H)$ & $+$ &$E_{5 \pm}$,  $M_{5 \pm}$ & $N \frac{9}{2}^{+}$, $N \frac{11}{2}^{+}$, $\Delta \frac{9}{2}^{+}$, $\Delta \frac{11}{2}^{+}$ \\
 & & & $N \left( 2220 \right) \frac{9}{2}^{+}$, $\Delta \left( 2420 \right) \frac{11}{2}^{+}$ \\
 & & & \\
\hline\hline
\end{tabular}\caption{List of the corresponding partial waves and multipoles to different values of $\ell_{\pi}$. The parity is given by $P=(-)^{\ell_{\pi}+1}$. The PDG notations for $N$ and $\Delta$ resonances are given. Examples of well established $N$- and $\Delta$-resonances (i.e. $\ast \ast \ast \ast$-resonances in the PDG) lowest in mass for a specific combination of quantum numbers are also given.}\label{tab:PartialWavesMultipoles}
\end{table} 

\begin{figure}[htb]
\RawFloats
\centering
\includegraphics[width=0.98\textwidth]{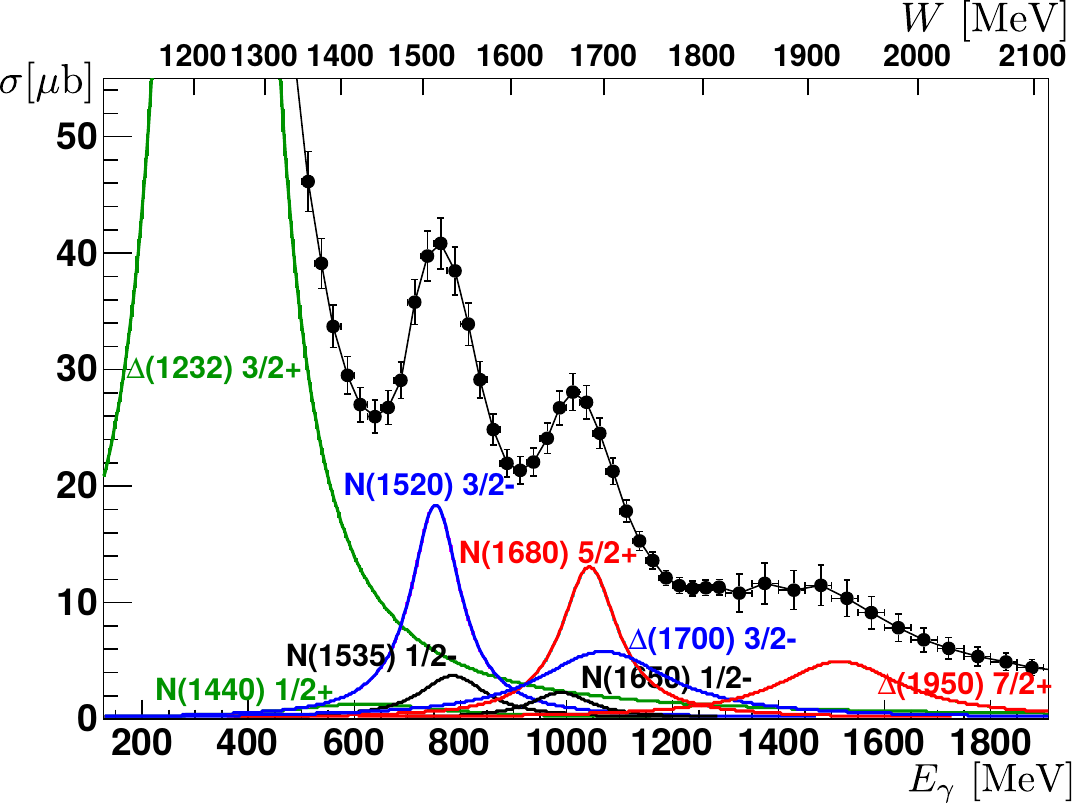}
\caption{The calculated Breit-Wigner amplitudes for different resonances in comparison to the measured total cross section for $\gamma p \rightarrow \pi^{0} p$ are shown \cite{Thiel:2015}. Remark: For any truncation order $\text{L}_{\text{max}}$, the total cross section generally stands in a very simple relation to the zeroth Legendre coefficient of the differential cross section $\sigma_{0}$, namely $\bar{\sigma} = 4 \pi \left( q / k \right) \left(a_{\text{L}_{\text{max}}}\right)_{0}^{\sigma_{0}}$.}
\label{fig:resonanzen}
\end{figure}


\subsection{First resonance region $\left( 1075 \hspace*{2pt} \mathrm{MeV} \lesssim W \lesssim 1350 \hspace*{2pt} \mathrm{MeV} \right)$} \label{sec:Interpretation1stResRegion}

In the first resonance region, the $\sigma_{0}$ measurement of the A2 collaboration has a significant overlap into this energy range. The new $F$-data from MAMI \cite{AnnandEtAl:2016} have some angular distributions in the first resonance region, but not enough to obtain good information on the energy-dependence of $\chi^{2}/\mathrm{ndf}$ or any partial wave.
The $\sigma_{0}$-data were investigated first by fitting each datapoint including just its published statistical error, cf. Fig. \ref{fig:wq_bins}.
Furthermore, a second fit of the whole dataset was done, endowing each datapoint with an error composed of the statistical ($\Delta \sigma_{0}^{\mathrm{stat.}}$) and systematic uncertainty ($\Delta \sigma_{0}^{\mathrm{sys.}}$) according to
\begin{equation}
\Delta \sigma_{0} = \sqrt{\left(\Delta \sigma_{0}^{\mathrm{stat.}}\right)^{2} + \left(\Delta \sigma_{0}^{\mathrm{sys.}}\right)^{2}} \mathrm{.} \label{eq:TotalErrorDCSA2}
\end{equation}
For comparison, the latter fit results are shown in Fig. \ref{fig:wq_bins_wsys}. \newline
It is immediately apparent that the fit using just the statistical error needs a truncation at $\text{L}_{\text{max}} = 4$ in order to describe the data in the first resonance region. This seems counter-intuitive, since the only well-established resonance in this region, the $\Delta(1232) \frac{3}{2}^{+}$ (cf. Fig. \ref{fig:resonanzen}), occurs in a $P$-wave and is long known to dominate all remaining partial waves by at least one order of magnitude. \newline 
However, once the systematic errors are included in the fit, it can be seen in Fig. \ref{fig:wq_bins_wsys} that a truncation at $\text{L}_{\text{max}} = 1$ is sufficient to describe the data, as may also be anticipated. These results already suggest that the systematic errors play an important role for the interpretation of the $\sigma_{0}$ data in the first resonace region, or more generally as soon as the statistical errors become smaller than the systematic errors. \newline
Also contained in Figures \ref{fig:wq_bins} and \ref{fig:wq_bins_wsys} are plots of the fitted Legendre coefficients (black points) compared to predictions of each coefficient (continuous lines), which are evaluated using Bonn-Gatchina multipoles only up to a specific truncation order. The color coding for the $\text{L}_{\text{max}}$ used in the Bonn-Gatchina prediction is kept in consistency with the one used in the $\chi^{2}$ plot at the top of Fig. \ref{fig:wq_bins} (the same colors are used as well in Figures \ref{fig:wq_bins_wsys} to \ref{fig:Sclas_bins}). The coefficients $\left(a_{4}\right)^{\sigma_{0}}_{(0,2)}$ are described well using Bonn-Gatchina multipoles up to $\text{L}_{\text{max}} = 1$. The remaining ones, $\left(a_{4}\right)^{\sigma_{0}}_{(1,3,\ldots,8)}$, show major discrepancies to the Bonn-Gatchina prediction even up to a truncation angular momentum of $4$. For $\left(a_{4}\right)^{\sigma_{0}}_{1}$ and $\left(a_{4}\right)^{\sigma_{0}}_{3}$ the difference shows a 'smooth' behaviour and the predictions for both coefficients are non-zero within the first resonance region. For $\left(a_{4}\right)^{\sigma_{0}}_{(4,\ldots,8)}$ however the differences are more 'peak-like' and the predictions even up to $\text{L}_{\text{max}} = 4$ vanish in the energy region considered here. These two observations can be interpreted consistently once the systematic error is incorporated into the fitting according to Eq. (\ref{eq:TotalErrorDCSA2}), the results of which are shown in Fig. \ref{fig:wq_bins_wsys}. It can be seen that the peak-like discrepancies in $\left(a_{4}\right)^{\sigma_{0}}_{(4,\ldots,8)}$ vanish within the range of their errors. \newline
 The results for the fitted coefficients are still clearly non-zero. Their errors now however support a scenario in which all of these coefficients would vanish. We therefore interpret the discrepancies in the higher coefficients to be systematic effects that are still present in the data. This interpretation is demanded by the fact that the dominant $S$- and $P$-waves can only occur in these coefficients as interferences with $F$-waves or higher (cf. App. \ref{sec:PWContentFormulas}). The latter is true especially for the dominant $P$-waves containing contribution from the $\Delta(1232) \frac{3}{2}^{+}$, which are not allowed to interfere with either $S$- or $D$-waves or with themselves. The coefficients $\left(a_{4}\right)^{\sigma_{0}}_{(7,8)}$ in particular do not get contributions from $S$- and $P$-waves at all (for $\text{L}_{\text{max}} \leq 5$), yet they show up as clearly non-zero once the data are fitted. The Bonn-Gatchina prediction, although containing a dominant $\Delta$-resonance contribution in the $P$-wave, determines all contributions of the above mentioned interferences to be practically zero. Therefore, we suggest that the peak-like discrepancies in the higher Legendre coefficients of $\sigma_{0}$ are very likely unrelated to any new physics information in the first resonance region, but are merely present due to the systematic uncertainties in the data. \newline
On the other hand, Legendre coefficients allowing a more interesting interpretation are $\left(a_{4}\right)^{\sigma_{0}}_{(1,3)}$. Here the disagreement is not corrected upon introducing the systematic errors into the fitting (cf. Fig. \ref{fig:wq_bins_wsys}). Rather the effect on the size of the errors is not so extreme in this case and the disagreement with Bonn-Gatchina is still significant. In order to clarify the situation, the fit results for $\left(a_{4}\right)^{\sigma_{0}}_{(0,\ldots,4)}$ are compared to predictions from SAID $S$-, $P$-, $D$- and $F$-waves in Fig. \ref{fig:wq_Coeffs_SAID}. Here, the agreement is a lot better. Therefore, the smooth discrepancies between Bonn-Gatchina and the fit results for $\left(a_{4}\right)^{\sigma_{0}}_{(1,3)}$ clearly show deficiencies of the PWA. Figure \ref{fig:wq_Coeffs_CompareSAIDBnGa}, where BnGa- and SAID-predictions truncated at $\text{L}_{\text{max}} = 3$ are compared in the $\Delta$-region, further serves to illustrate this point. \newline
As a physics interpretation of all data available in the first resonance region, it can be said that once the systematic uncertainties are included into the fitting, the influence of the $\Delta(1232) \frac{3}{2}^{+}$, which is long known to dominate here, is consistent with the $\chi^{2}$ plot shown in Fig. \ref{fig:wq_bins_wsys} (an $\text{L}_{\text{max}} = 1$-truncation is seen to clearly describe the data). The influence of all higher partial waves is small and can not be inferred by looking at the $\chi^{2}$. Their presence in the results of fits with only statistical errors (Fig. \ref{fig:wq_bins}) is highly unlikely to show any new physics. 

\subsection{Second resonance region $\left( 1350 \hspace*{2pt} \mathrm{MeV} \lesssim W \lesssim 1600 \hspace*{2pt} \mathrm{MeV} \right)$} \label{sec:Interpretation2ndResRegion}

\begin{figure*}
\begin{minipage}{\textwidth}
\floatbox[{\capbeside\thisfloatsetup{capbesideposition={right,top},capbesidewidth=7.8cm}}]{figure}[\FBwidth]
{\caption{The recent new double po\-la\-ri\-za\-tion observable $\check{E}$ data from ELSA \cite{Gottschall:2014,Gottschall:2015} with only statistical error was fitted using associated Legendre polynomials according to eq. \ref{eq:LowEAssocLegParametrizationE} and truncating the partial wave expansion at $\text{L}_{\text{max}}=1\dots 4$. (a) The resulting $\chi^2/$ndf values of the different $\text{L}_{\text{max}}$-fits as a function of the center of mass energy W are shown. (b) 6 out of 33 selected angular distributions of $\check{E}$ (black points) are plotted together with the different $\text{L}_{\text{max}}$ fits (solid lines) starting at W= 1522 MeV up to 2157 MeV. (c) Comparison of the fit coefficients for $\text{L}_{\text{max}}=3$ (black points), $\left(a_{3}\right)^{\check{E}}_{0\dots6}$ (see eq. \ref{eq:LowEAssocLegParametrizationE}), with the BnGa2014-02 solution truncated at different $\text{L}_{\text{max}}$ (solid lines). Colors same as in (a). \newline The fit coefficients $\left(a_{4}\right)^{\check{E}}_{7,8}$, for $\text{L}_{\text{max}} = 4$, are shown here as well.}\label{fig:e_bins}}
{\includegraphics[width=0.49\textwidth, trim=0cm 0cm 1.8cm 0cm, clip]{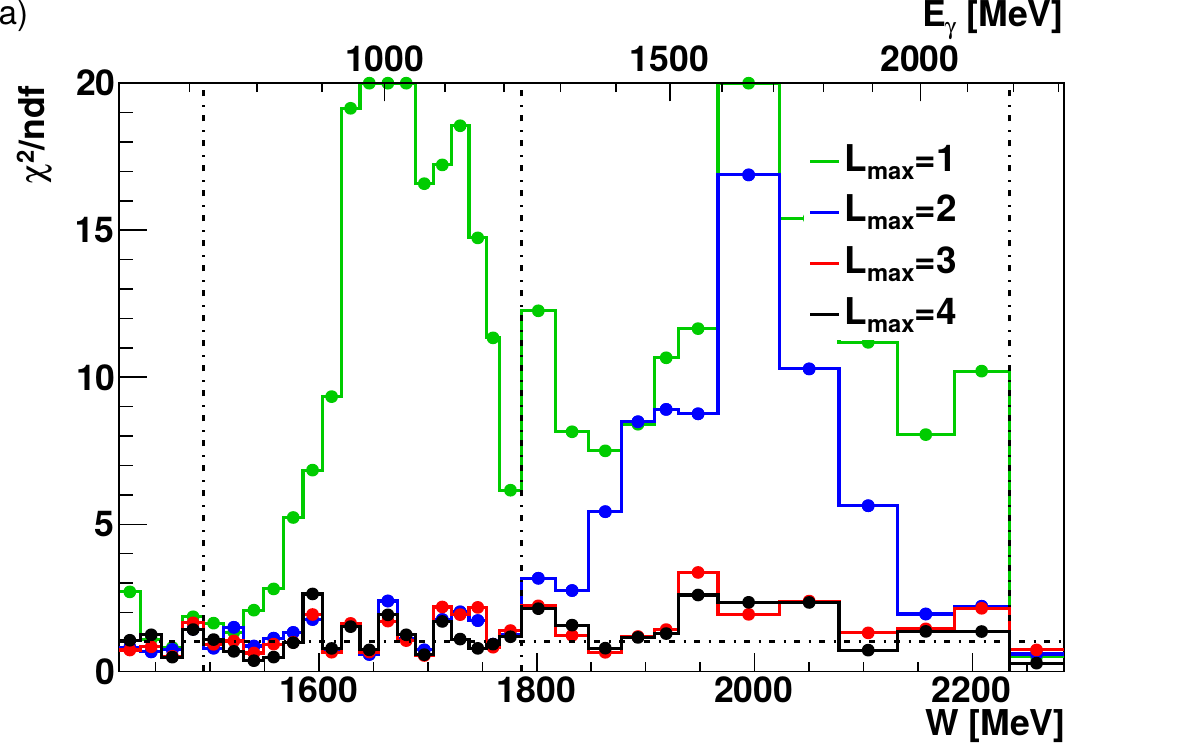}}
\end{minipage}\\

\begin{minipage}{\textwidth}
\centering
\hspace*{-0.45cm}
 \includegraphics[width=0.305\textwidth, trim=0cm 0cm 0.01cm 0.75cm, clip]{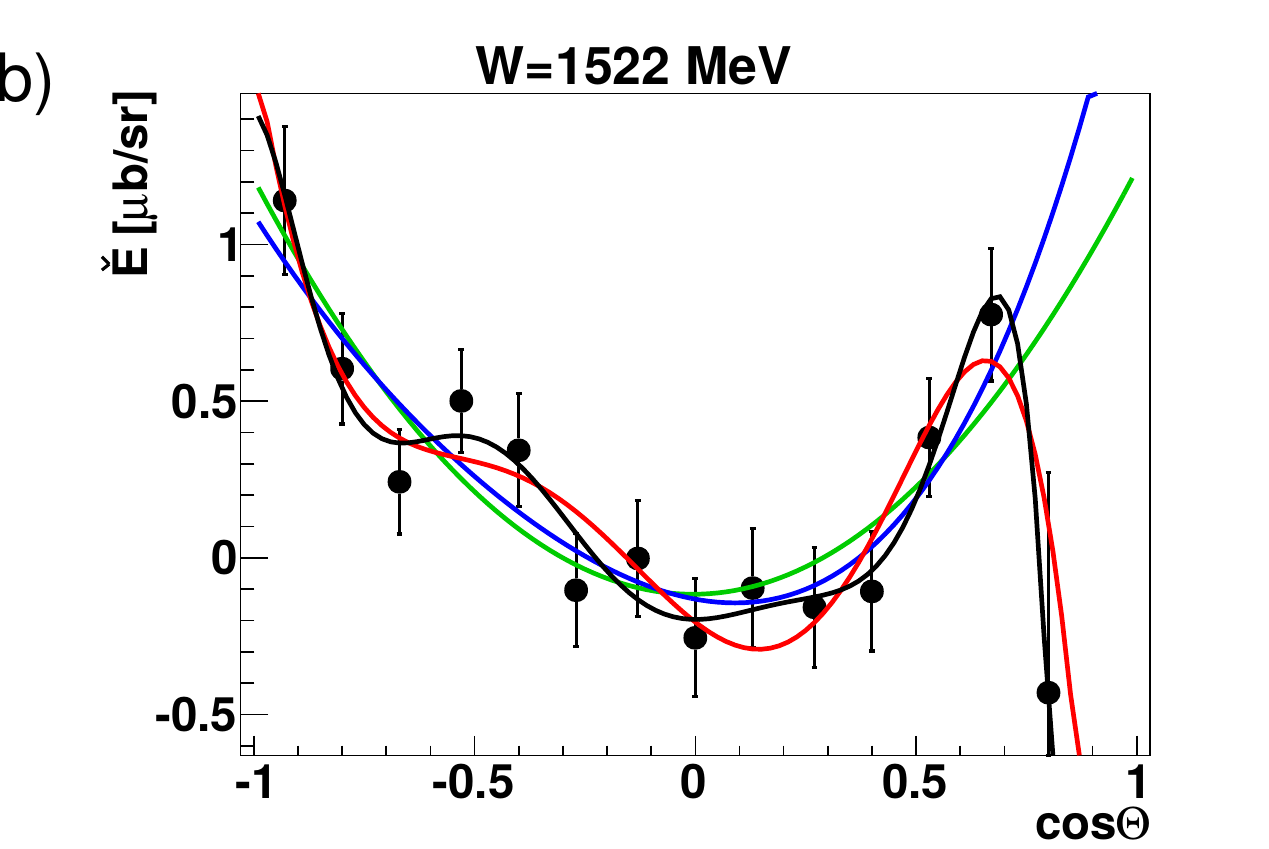}
  \includegraphics[width=0.285\textwidth, trim=0cm 0cm 0.01cm 0.75cm, clip]{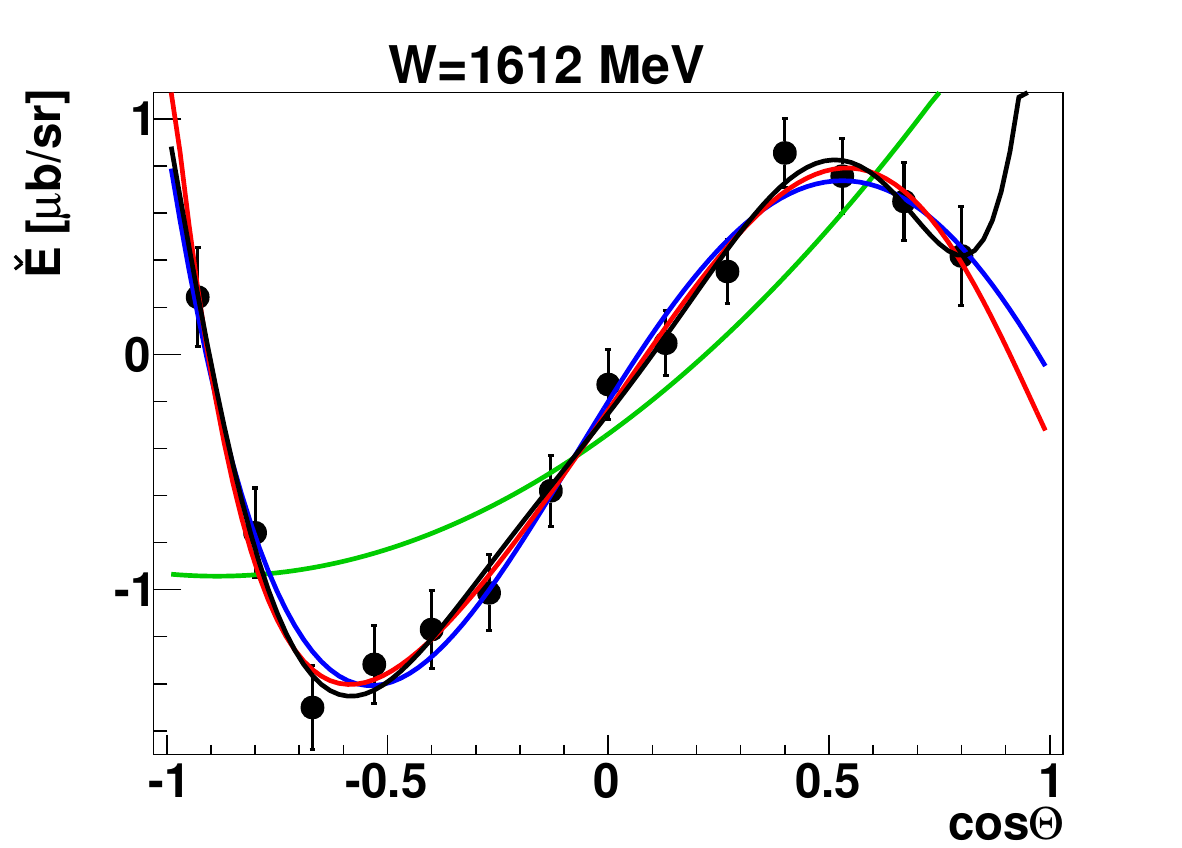}
  \includegraphics[width=0.285\textwidth, trim=0cm 0cm 0.01cm 0.75cm, clip]{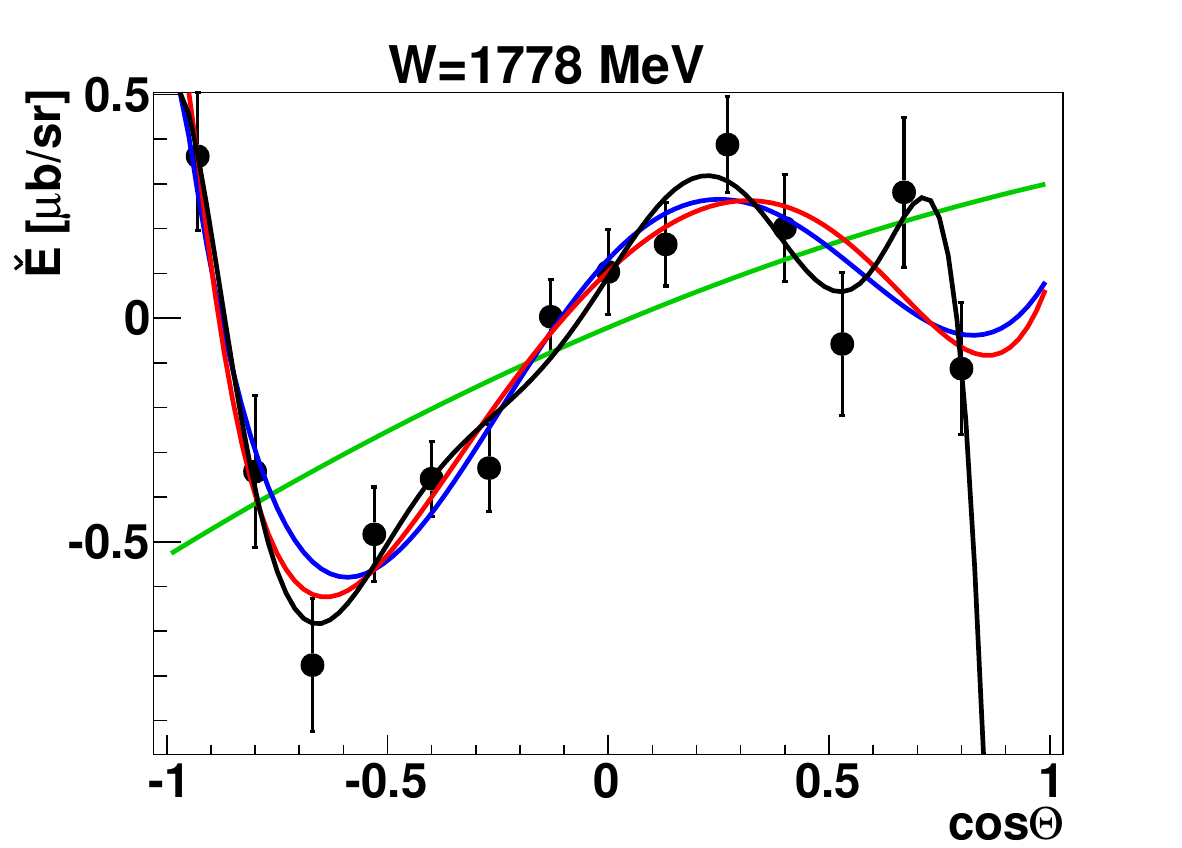}\\
  \includegraphics[width=0.285\textwidth, trim=0cm 0cm 0.01cm 0.75cm, clip]{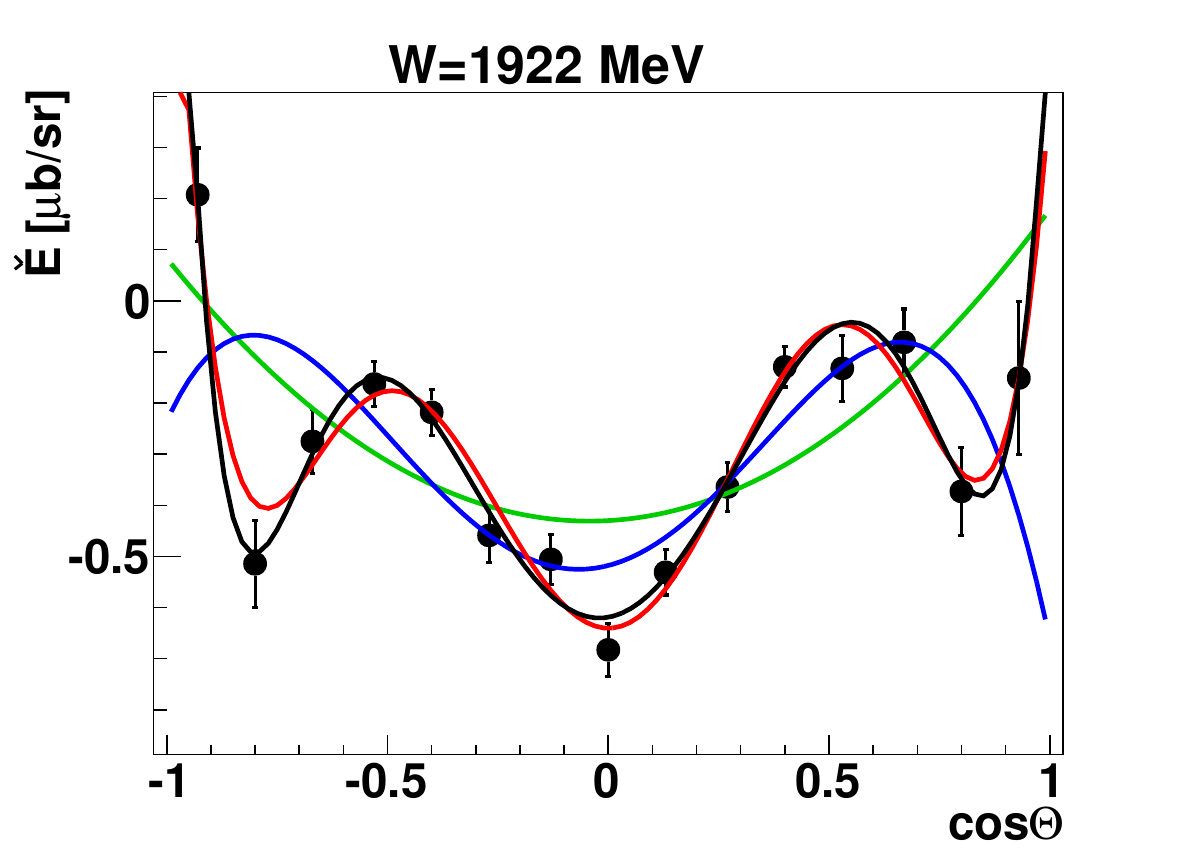}
  \includegraphics[width=0.285\textwidth, trim=0cm 0cm 0.01cm 0.75cm, clip]{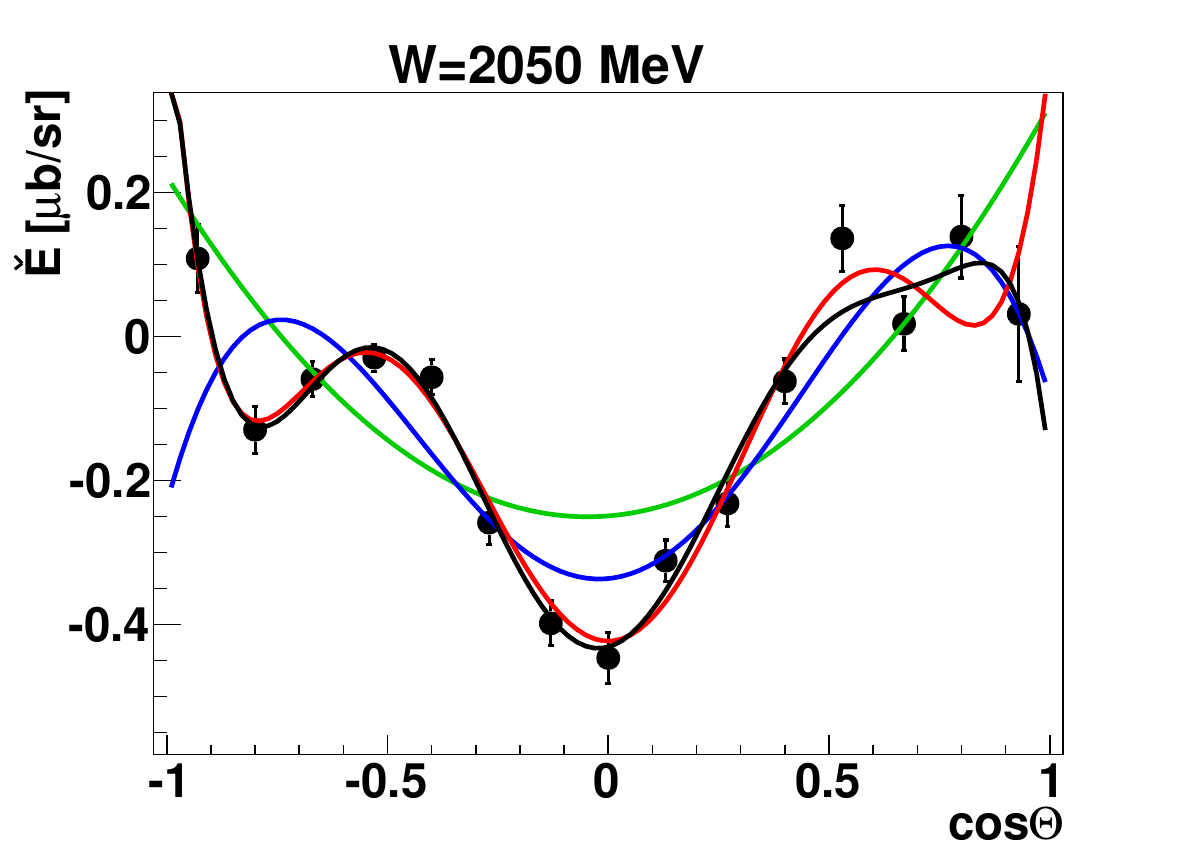}
  \includegraphics[width=0.285\textwidth, trim=0cm 0cm 0.01cm 0.75cm, clip]{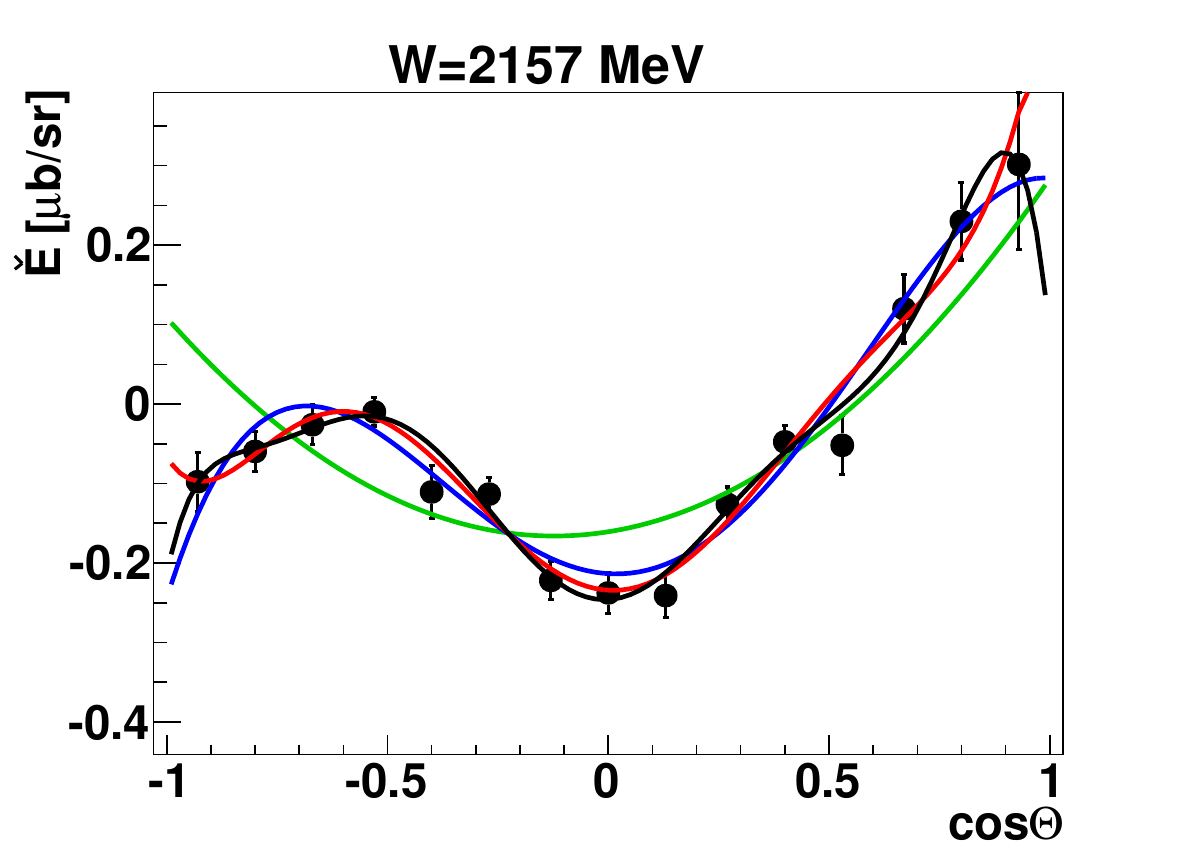}\\
 \vspace*{0.5cm}
  
  \hspace*{-23.5pt}\includegraphics[width=0.2905\textwidth]{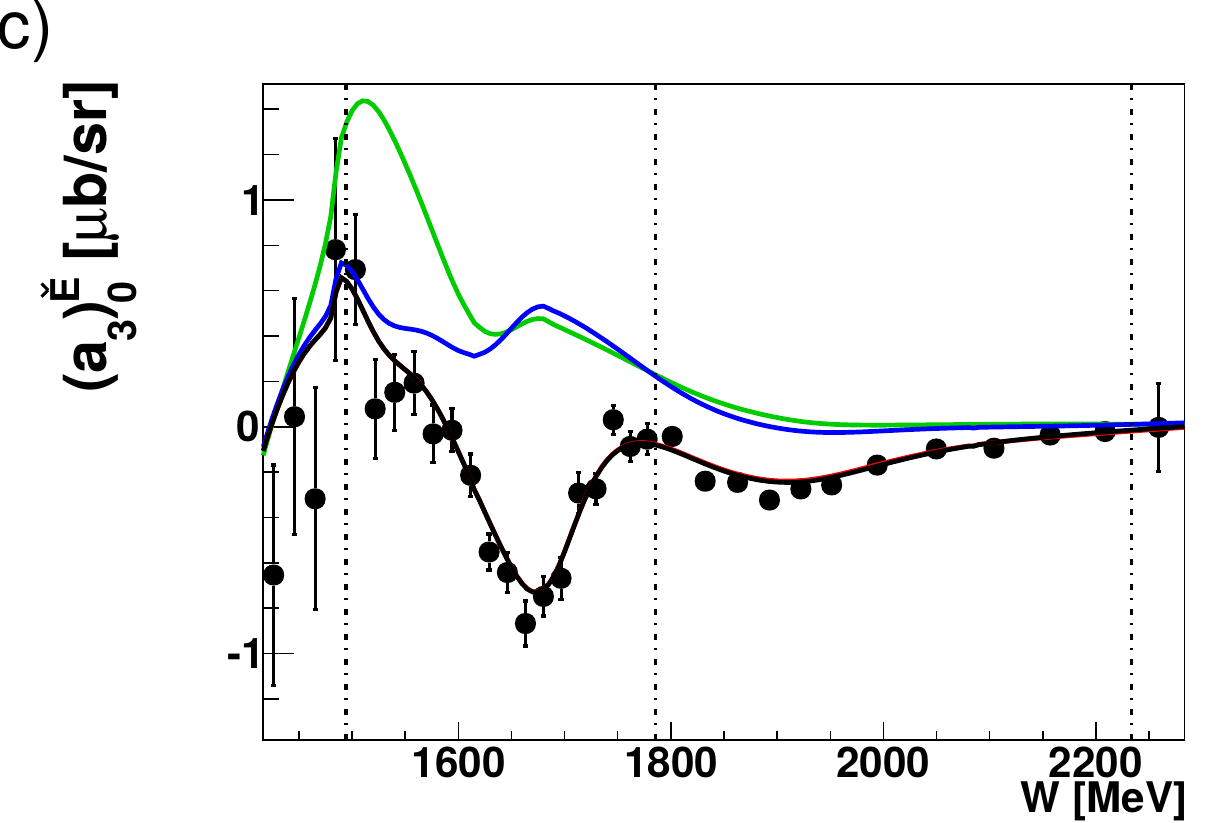}  
  \includegraphics[width=0.285\textwidth]{E_l3_coeff_1.pdf}
  \includegraphics[width=0.285\textwidth]{E_l3_coeff_2.pdf}\\
  \hspace*{-19.5pt}\includegraphics[width=0.285\textwidth]{E_l3_coeff_3.pdf}
  \includegraphics[width=0.285\textwidth]{E_l3_coeff_4.pdf}
  \includegraphics[width=0.285\textwidth]{E_l3_coeff_5.pdf}\\
  \hspace*{-19.5pt}\includegraphics[width=0.285\textwidth]{E_l3_coeff_6.pdf}
  \includegraphics[width=0.285\textwidth]{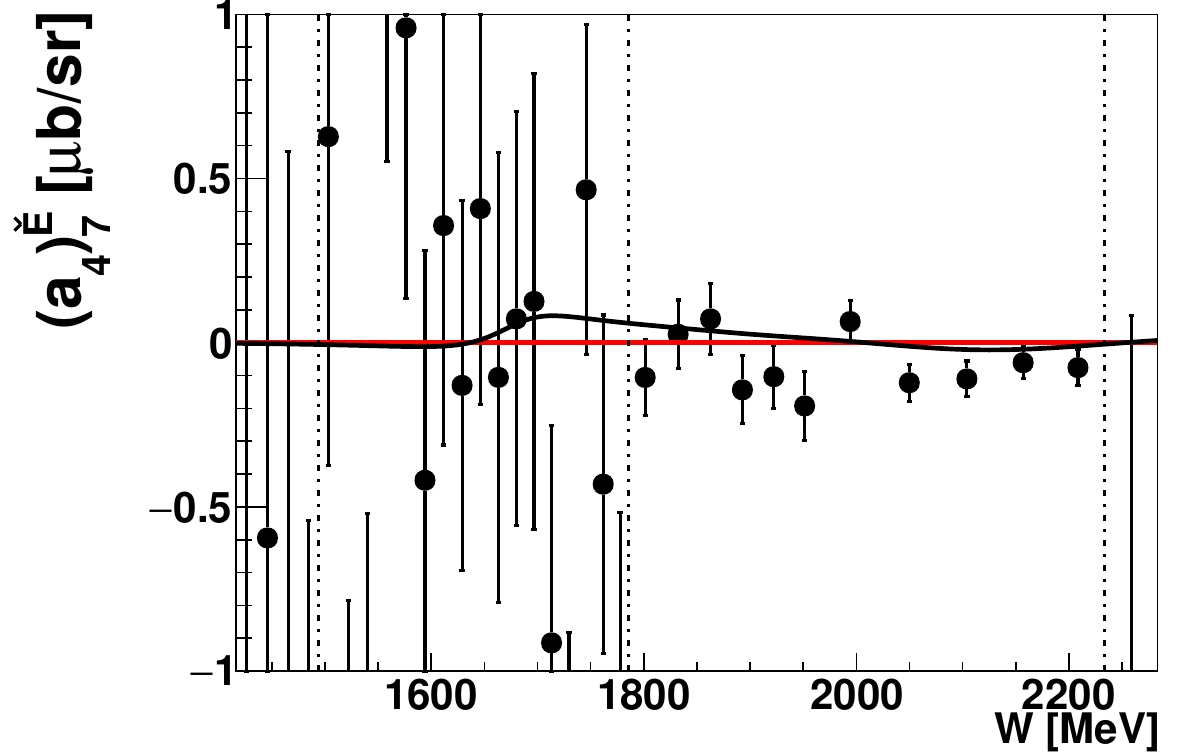}
  \includegraphics[width=0.285\textwidth]{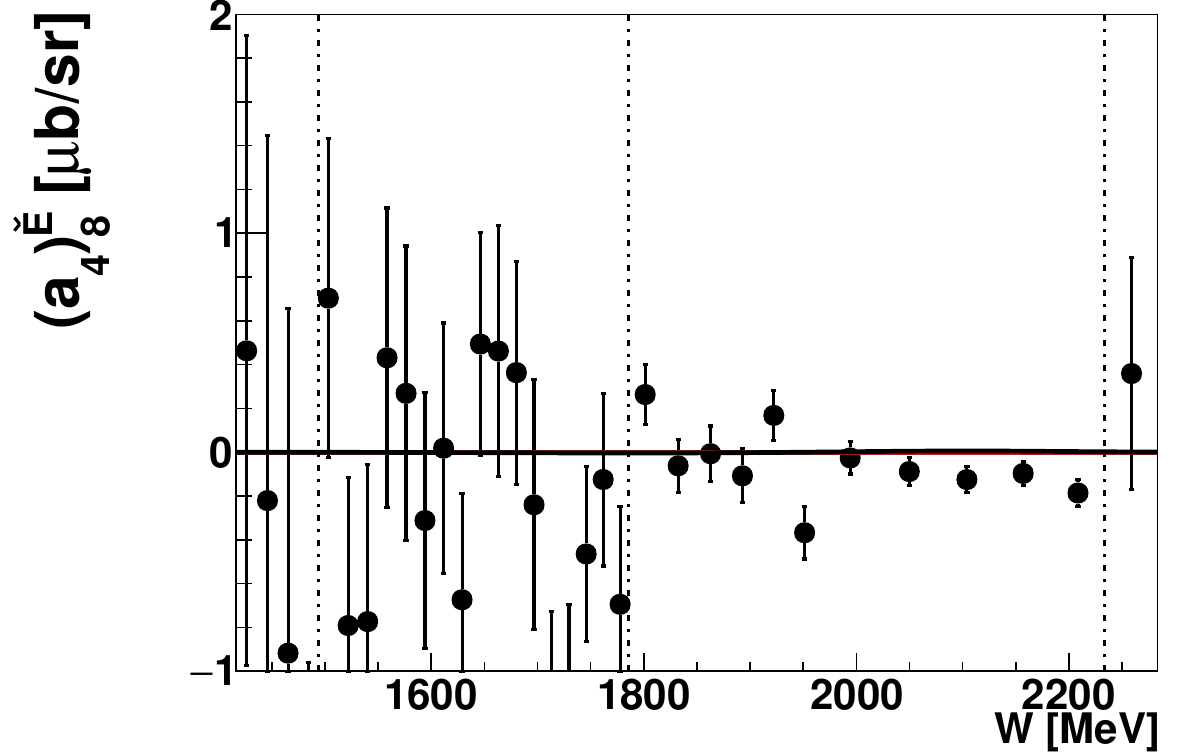}
  \end{minipage}
\end{figure*}

\begin{figure*}
\begin{minipage}{\textwidth}
\floatbox[{\capbeside\thisfloatsetup{capbesideposition={right,top},capbesidewidth=7.8cm}}]{figure}[\FBwidth]
{\caption{The recent new observable $\sigma_{1/2}$ data from ELSA \cite{Gottschall:2014,Gottschall:2015} with only statistical error was fitted using associated Legendre polynomials according to eq. \ref{eq:LowEAssocLegParametrizationDCS1/2} and truncating the partial wave expansion at $\text{L}_{\text{max}}=1\dots 4$. (a) The resulting $\chi^2/$ndf values of the different $\text{L}_{\text{max}}$-fits as a function of the center of mass energy W are shown. (b) 6 out of 33 selected angular distributions of $\sigma_{1/2}$ (black points) are plotted together with the different $\text{L}_{\text{max}}$ fits (solid lines) starting at W= 1522 MeV up to 2157 MeV. Only statistical errors were used. (c) Comparison of the fit coefficients for $\text{L}_{\text{max}}=3$ (black points), $\left(a_{3}\right)^{\sigma_{1/2}}_{0\dots6}$ (see eq. \ref{eq:LowEAssocLegParametrizationDCS1/2}), with the BnGa2014-02 solution truncated at different $\text{L}_{\text{max}}$ (solid lines). Colors same as in (a). The fit coefficients $\left(a_{4}\right)^{\sigma_{1/2}}_{7,8}$, for $\text{L}_{\text{max}} = 4$, are shown here as well.}\label{fig:s12_bins}}
{\includegraphics[width=0.49\textwidth, trim=0cm 0cm 1.8cm 0cm, clip]{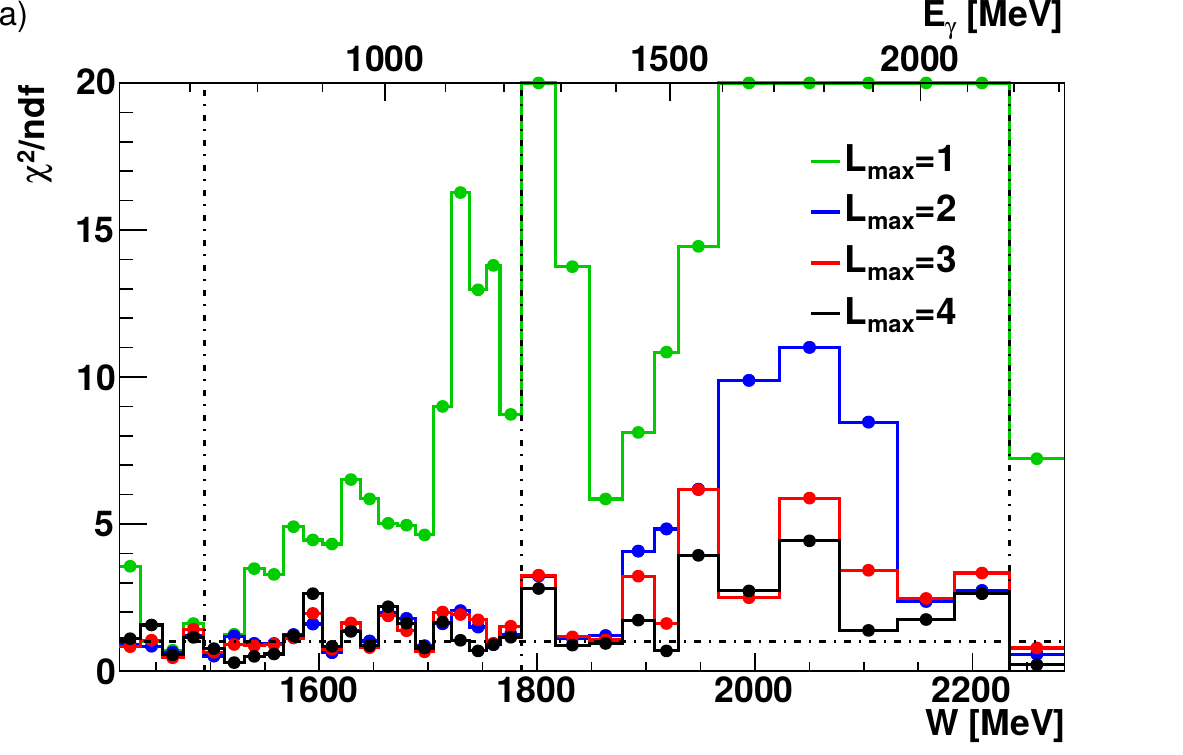}}
\end{minipage}\\

\begin{minipage}{\textwidth}
\centering
\hspace*{-0.45cm}
 \includegraphics[width=0.305\textwidth, trim=0cm 0cm 0.01cm 0.75cm, clip]{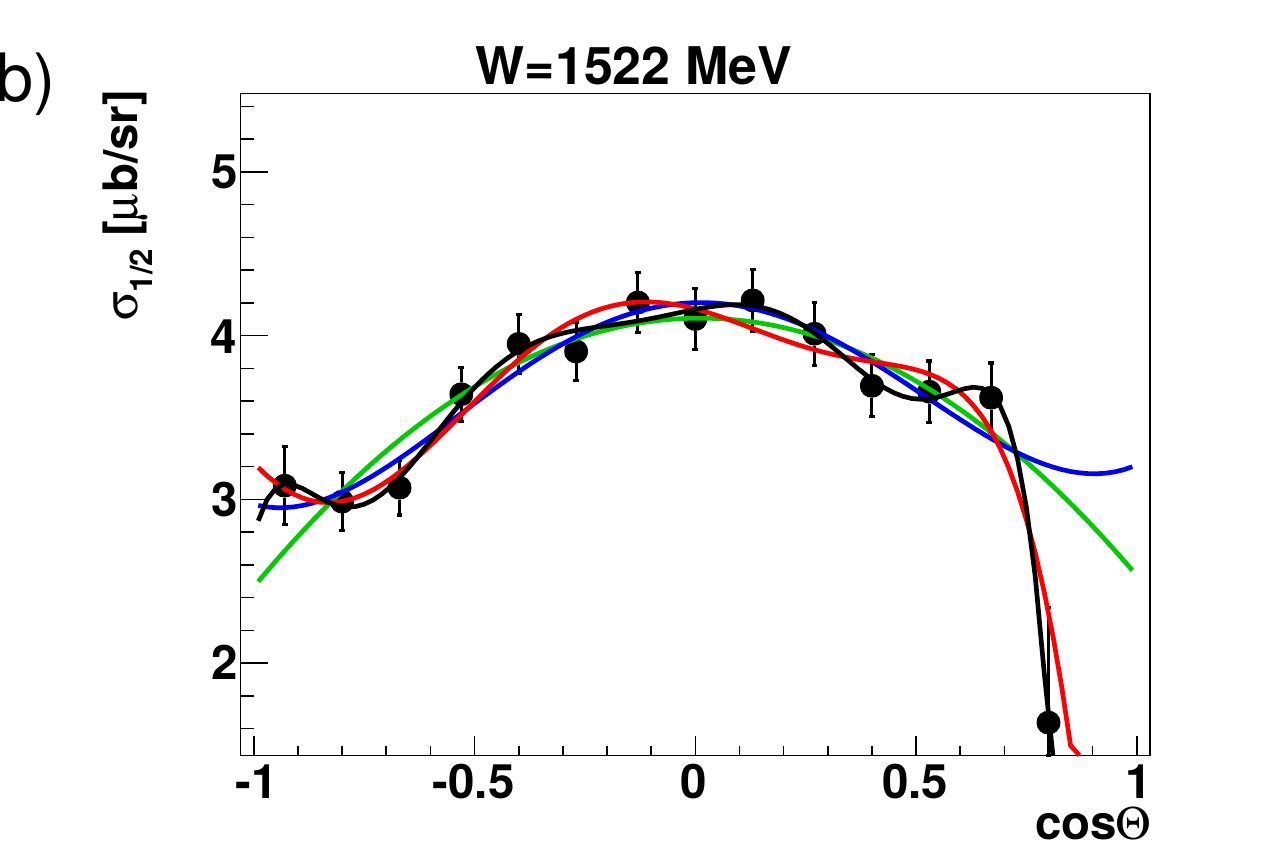}
  \includegraphics[width=0.285\textwidth, trim=0cm 0cm 0.01cm 0.75cm, clip]{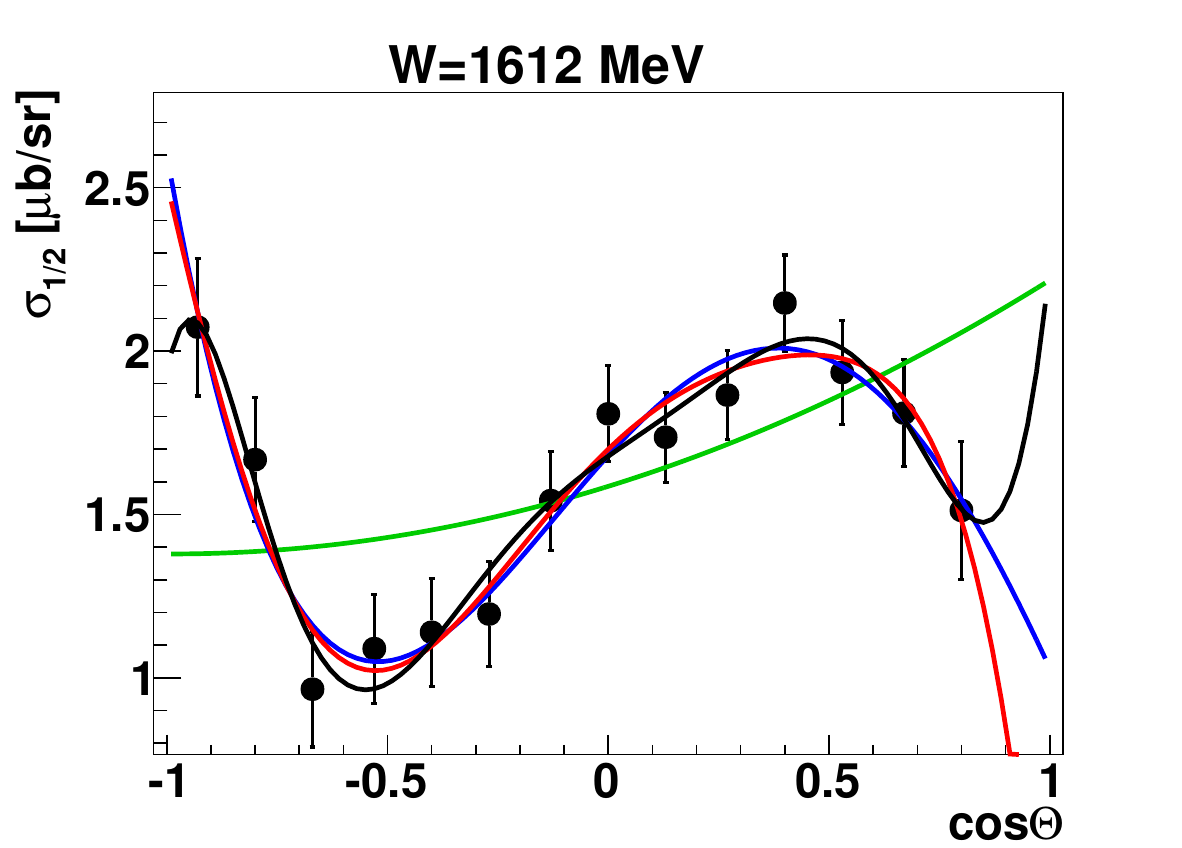}
  \includegraphics[width=0.285\textwidth, trim=0cm 0cm 0.01cm 0.75cm, clip]{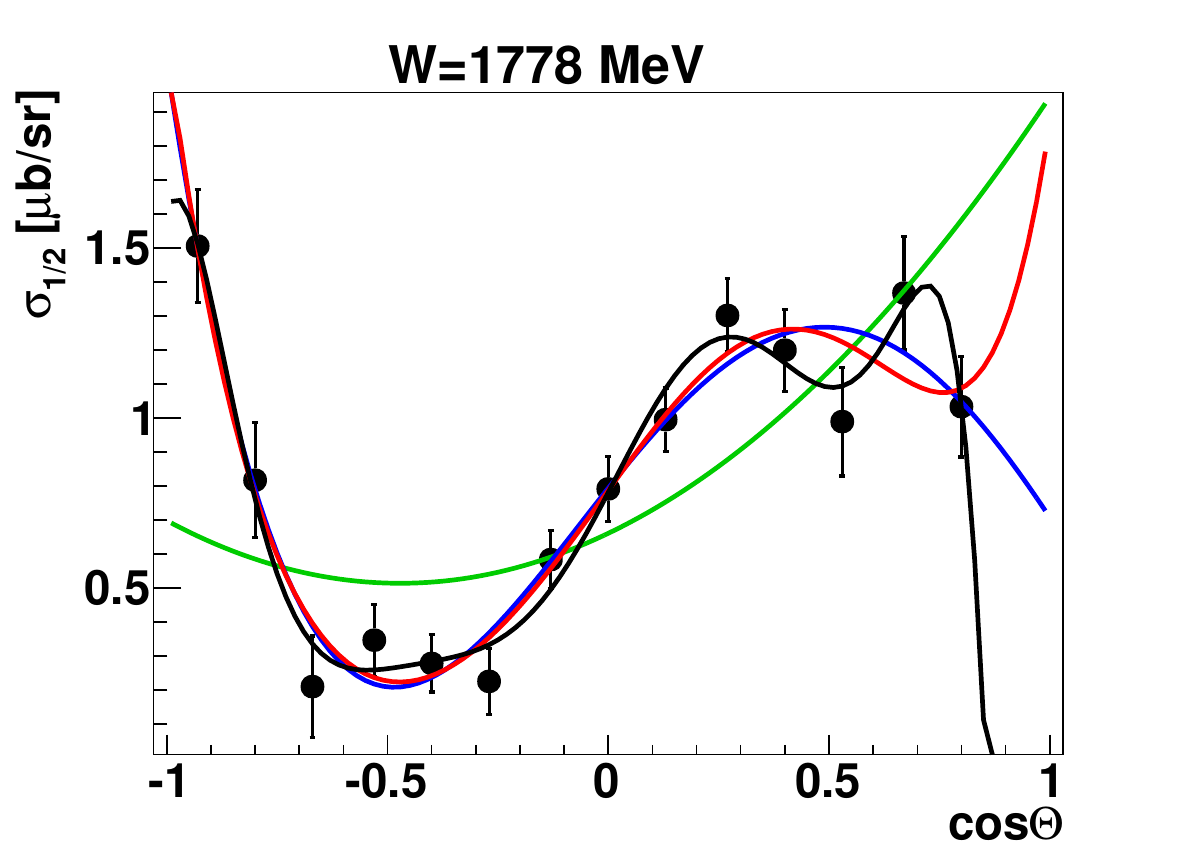}\\
  \includegraphics[width=0.285\textwidth, trim=0cm 0cm 0.01cm 0.75cm, clip]{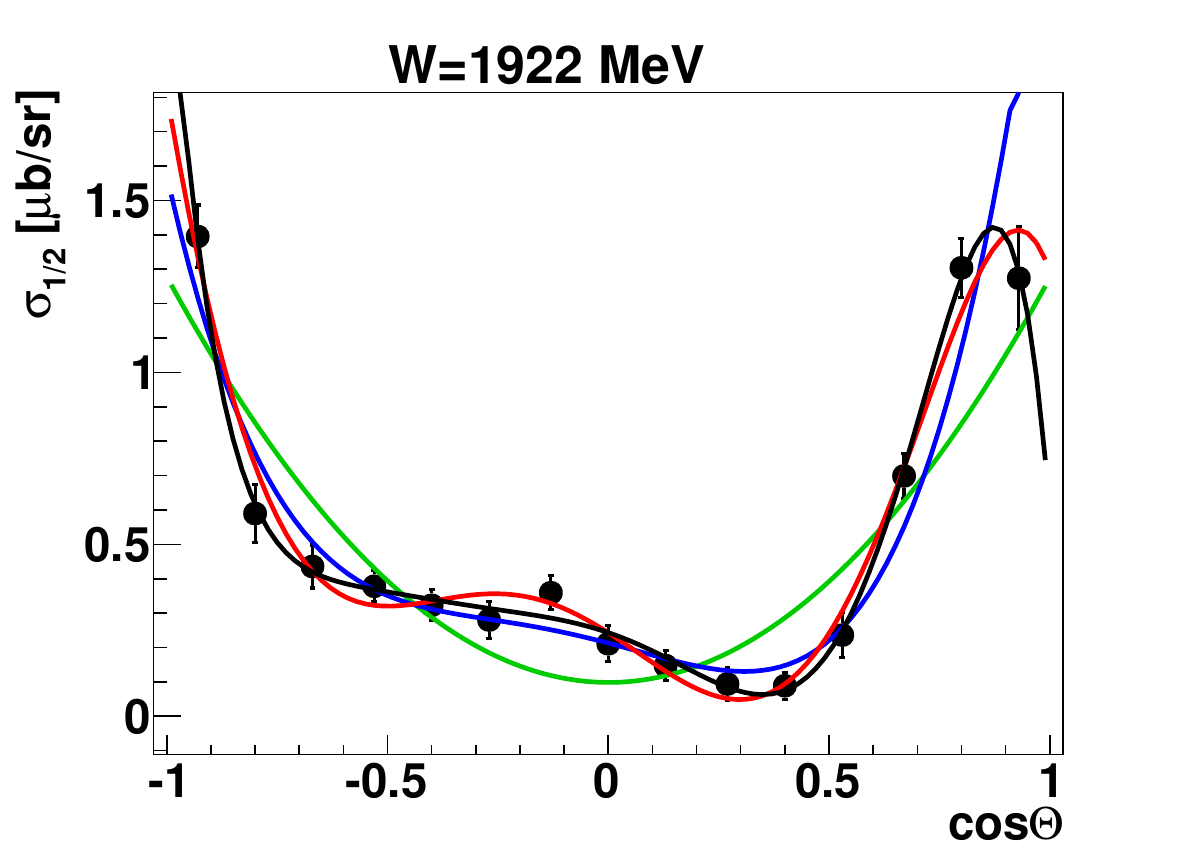}
  \includegraphics[width=0.285\textwidth, trim=0cm 0cm 0.01cm 0.75cm, clip]{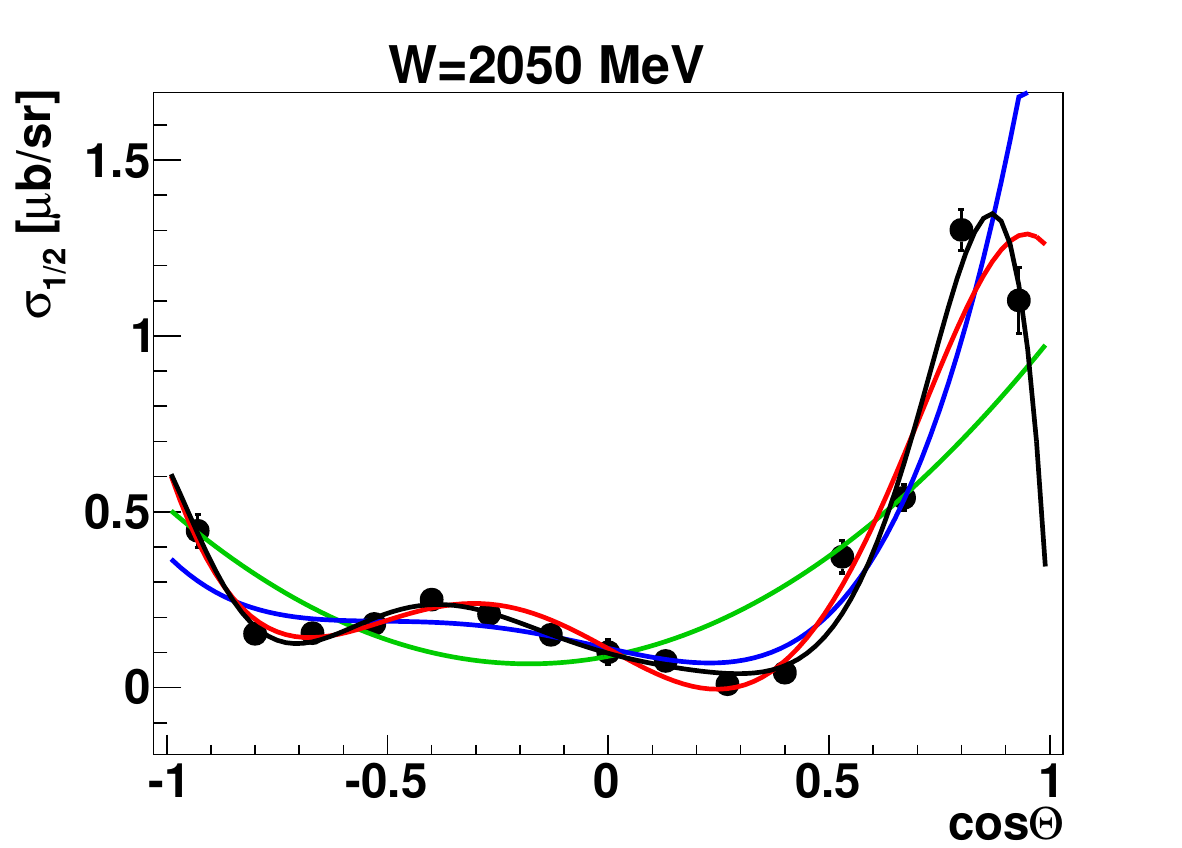}
  \includegraphics[width=0.285\textwidth, trim=0cm 0cm 0.01cm 0.75cm, clip]{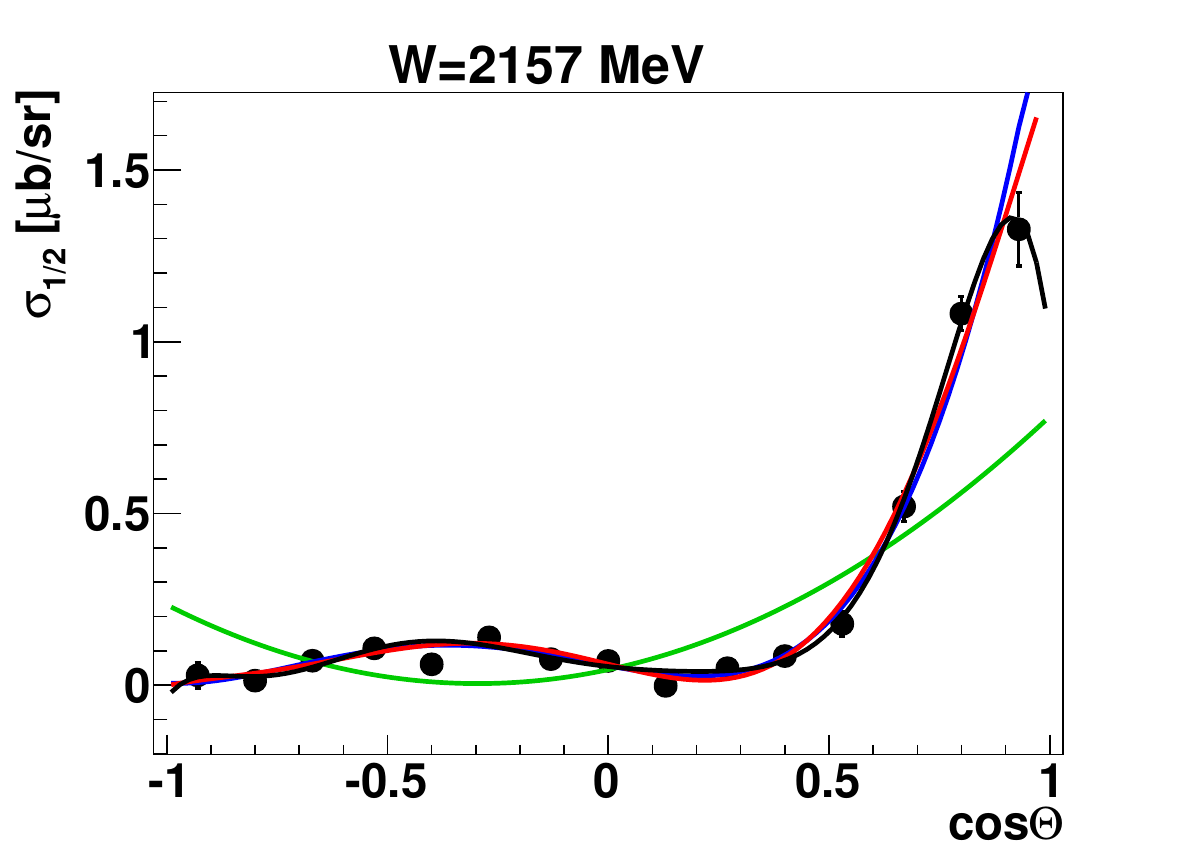}\\
 \vspace*{0.5cm}
  
  \hspace*{-23.5pt}\includegraphics[width=0.2905\textwidth]{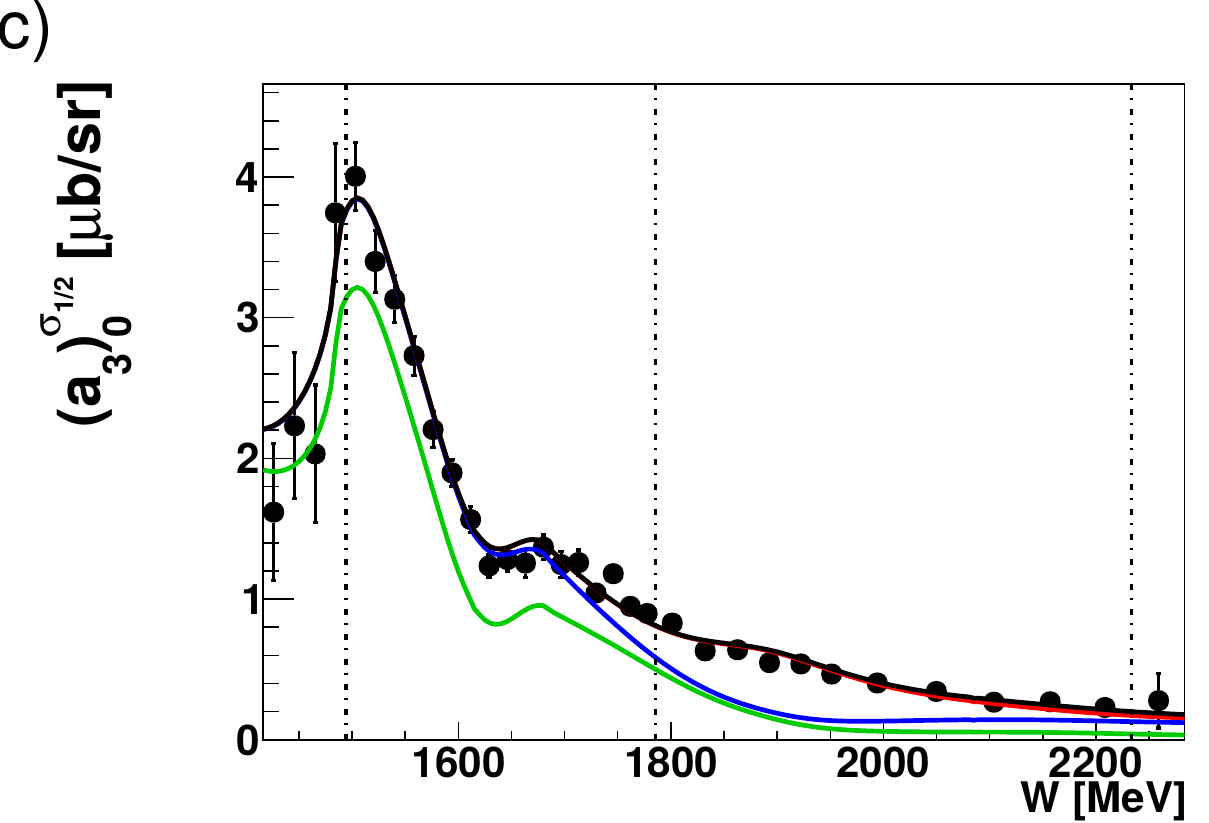}  
  \includegraphics[width=0.285\textwidth]{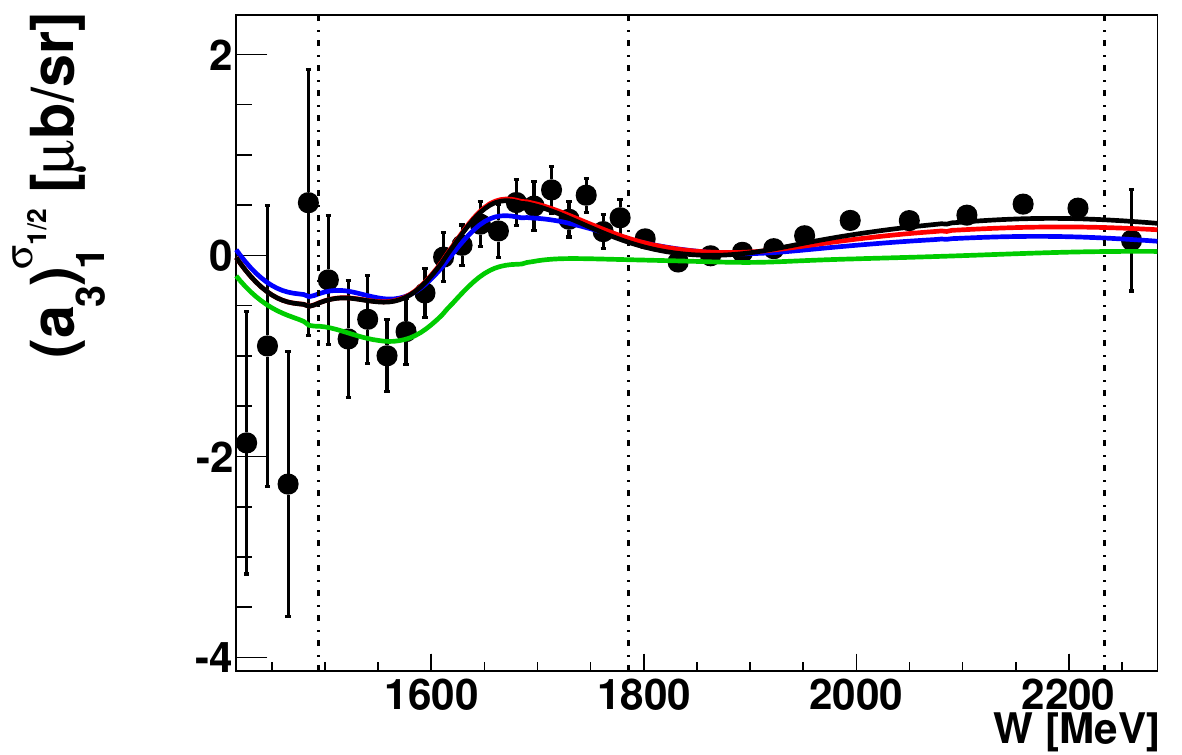}
  \includegraphics[width=0.285\textwidth]{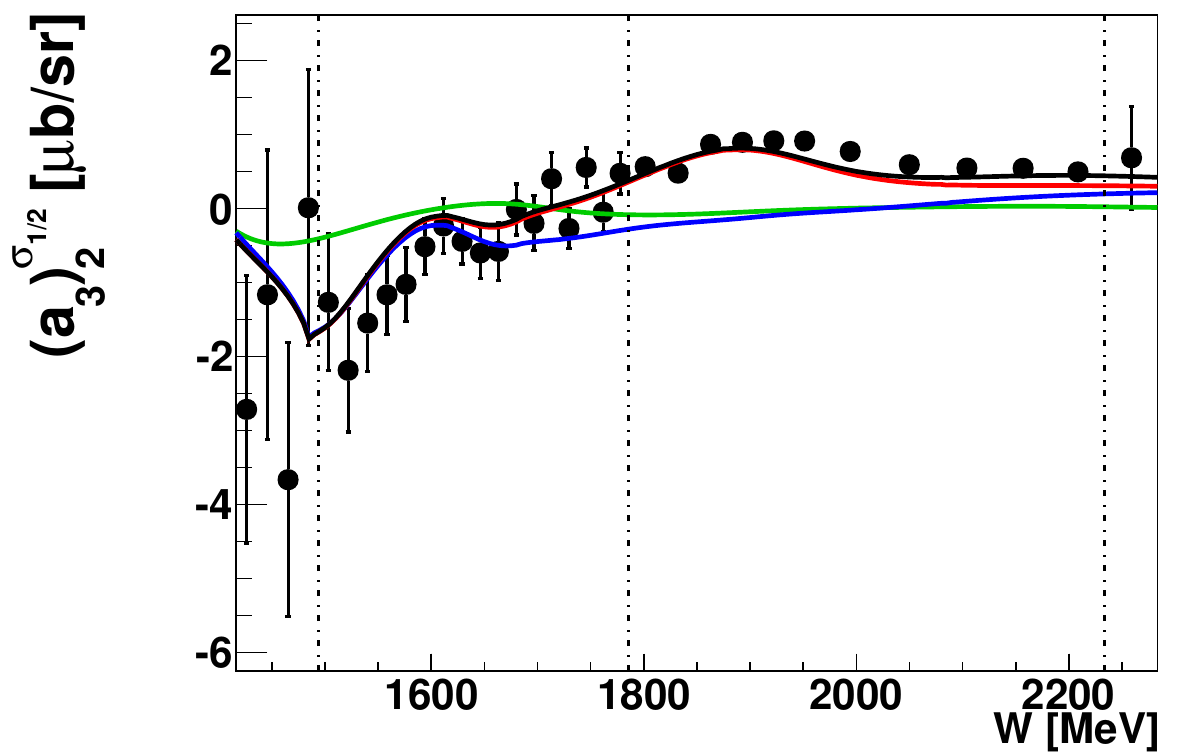}\\
  \hspace*{-19.5pt}\includegraphics[width=0.285\textwidth]{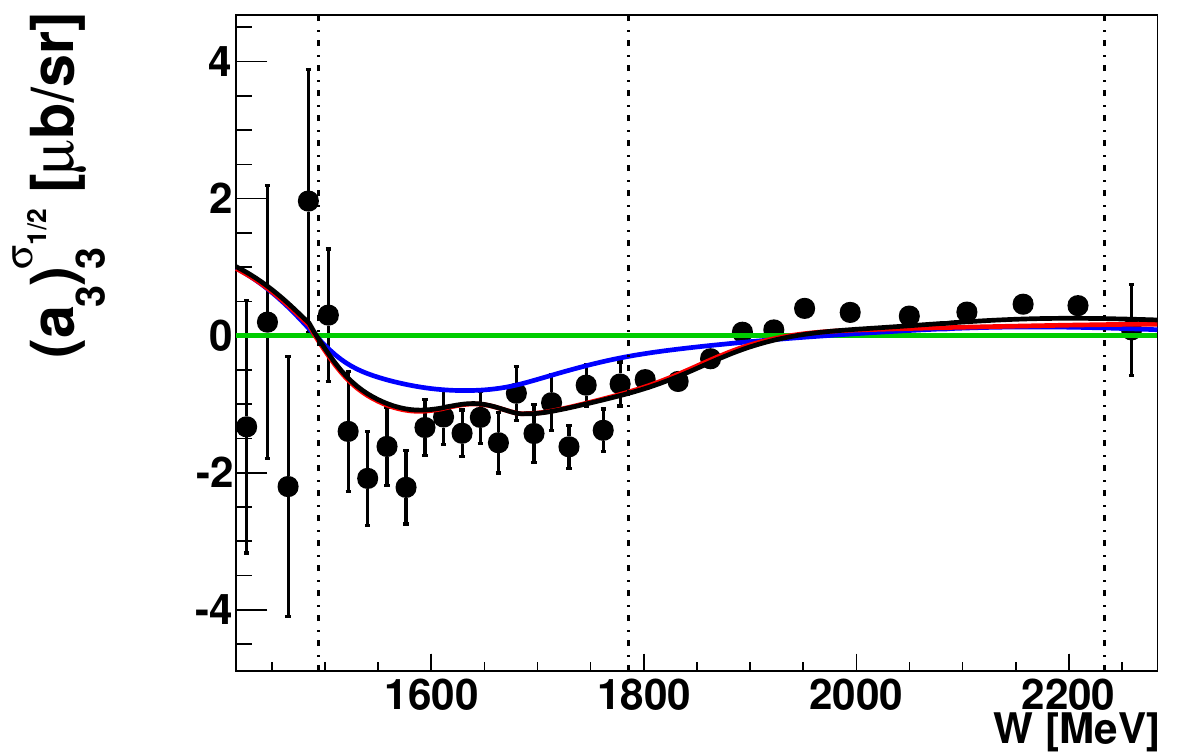}
  \includegraphics[width=0.285\textwidth]{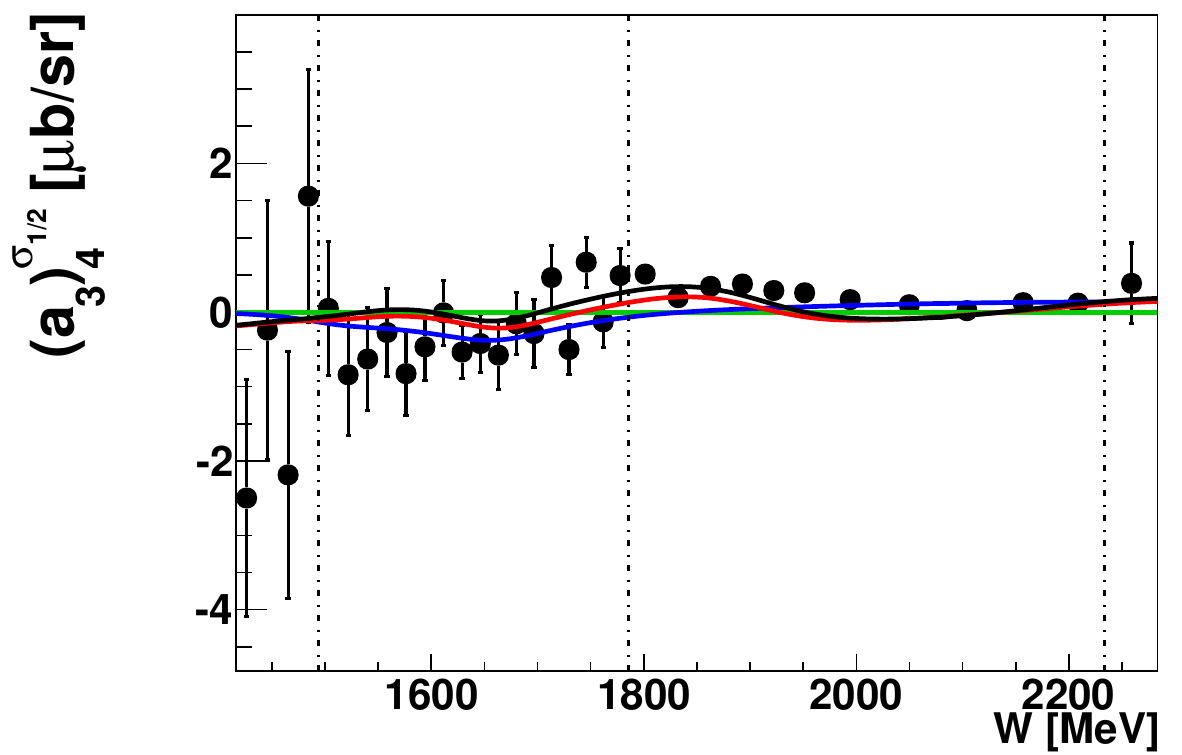}
  \includegraphics[width=0.285\textwidth]{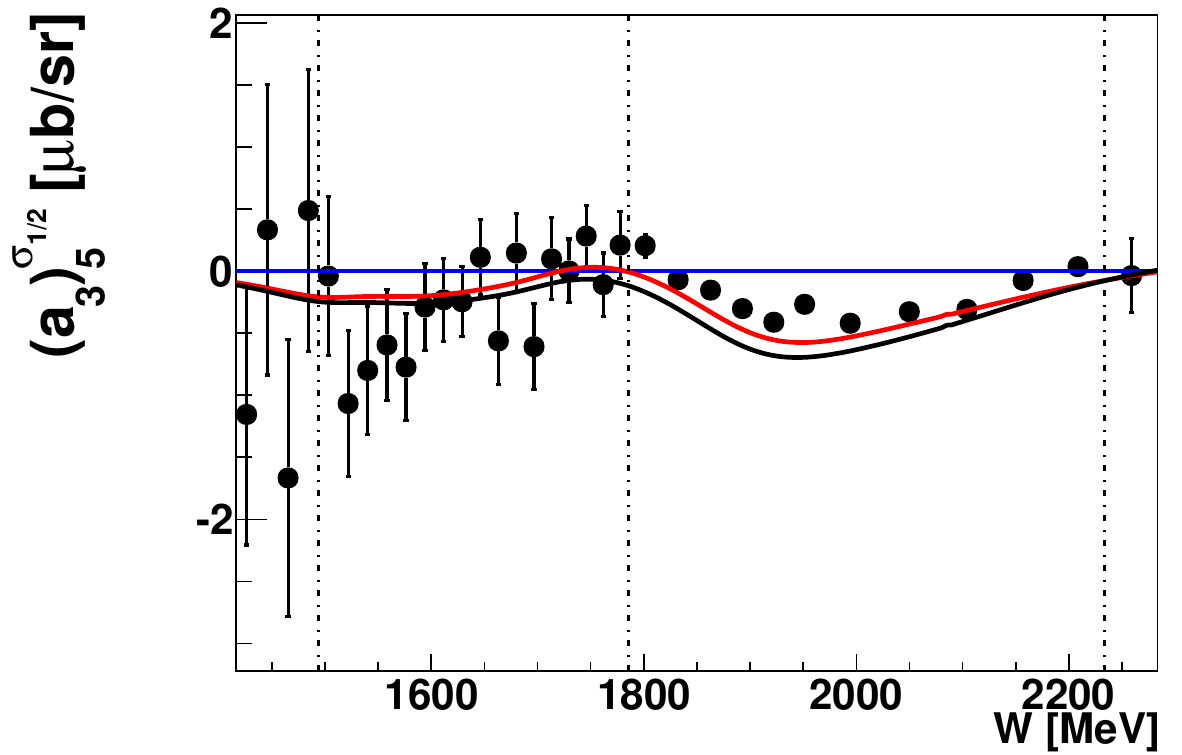}\\
  \hspace*{-19.5pt}\includegraphics[width=0.285\textwidth]{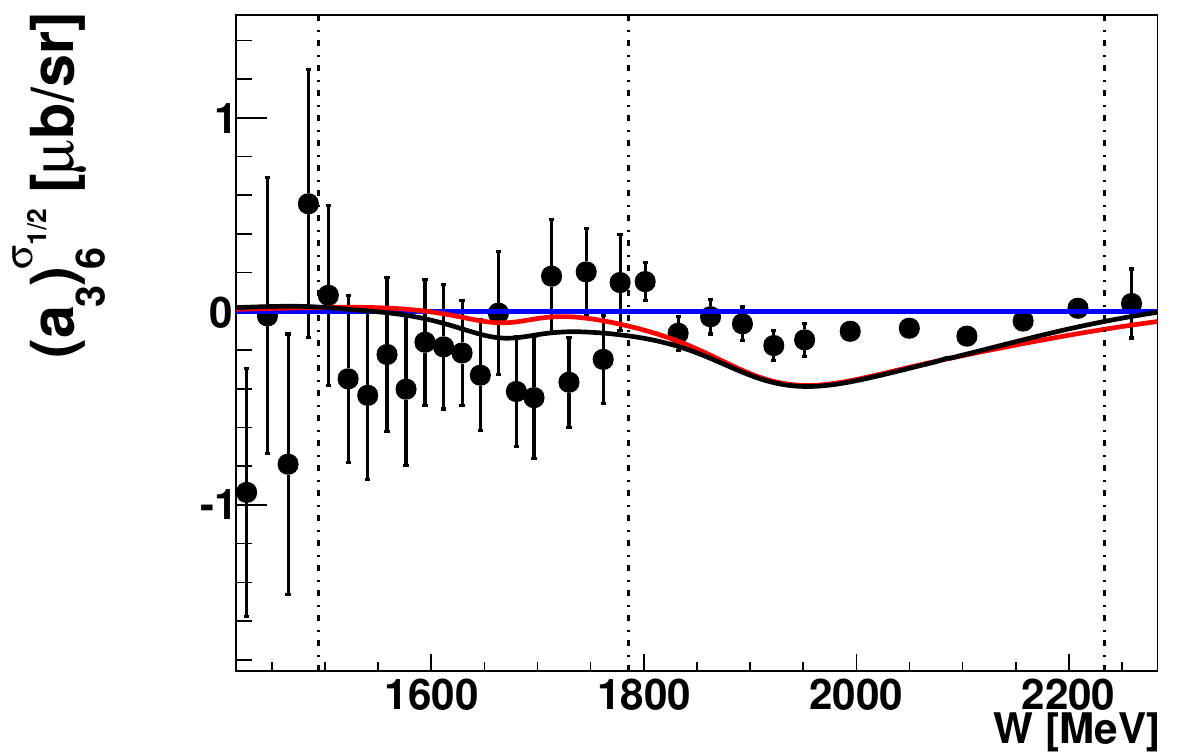}
  \includegraphics[width=0.285\textwidth]{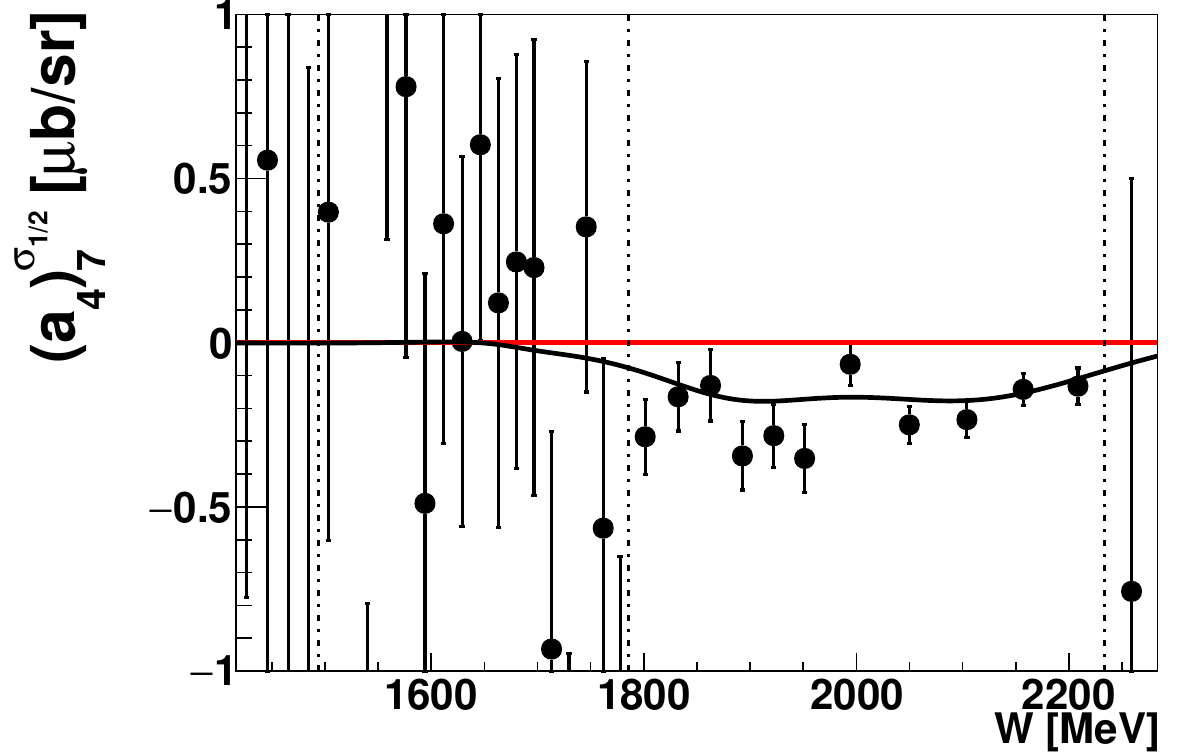}
  \includegraphics[width=0.285\textwidth]{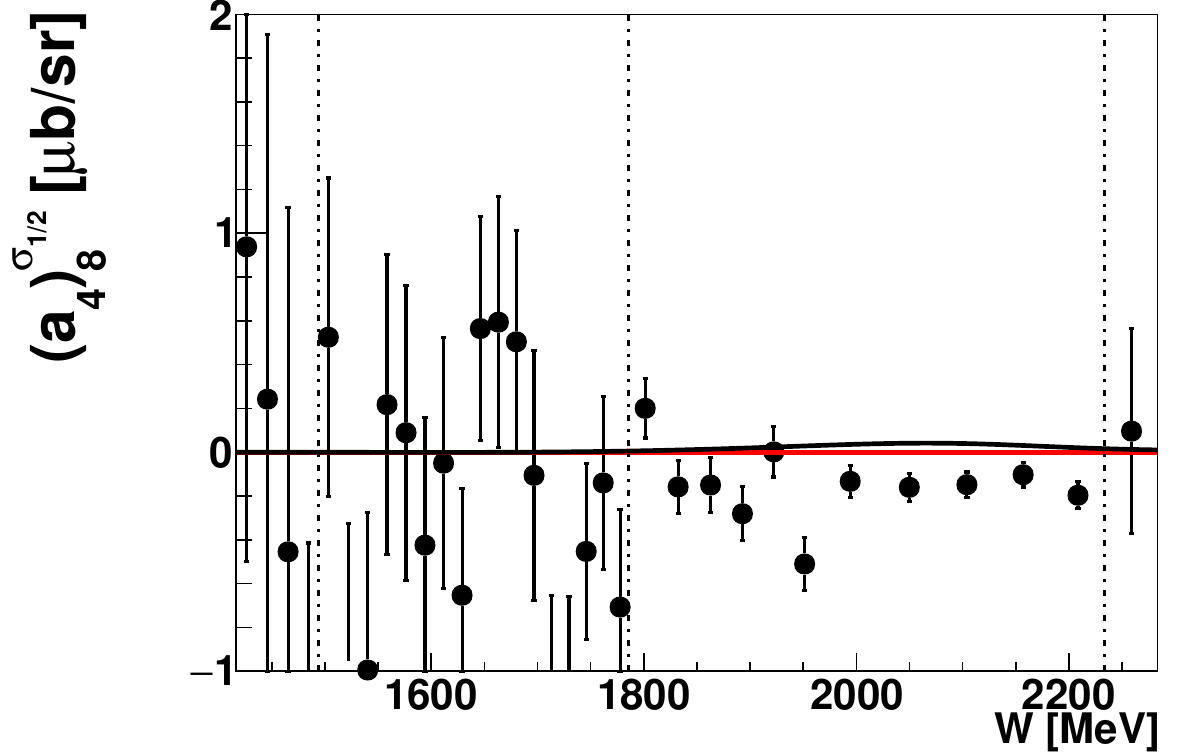}
  \end{minipage}
\end{figure*}

\begin{figure*}
\begin{minipage}{\textwidth}
\floatbox[{\capbeside\thisfloatsetup{capbesideposition={right,top},capbesidewidth=7.8cm}}]{figure}[\FBwidth]
{\caption{The recent new observable $\sigma_{3/2}$ data from ELSA \cite{Gottschall:2014,Gottschall:2015} with only statistical error was fitted using associated Legendre polynomials according to eq. \ref{eq:LowEAssocLegParametrizationDCS3/2} and truncating the partial wave expansion at $\text{L}_{\text{max}}=1\dots 4$. (a) The resulting $\chi^2/$ndf values of the different $\text{L}_{\text{max}}$-fits as a function of the center of mass energy W are shown. (b) 6 out of 33 selected angular distributions of $\sigma_{3/2}$ (black points) are plotted together with the different $\text{L}_{\text{max}}$ fits (solid lines) starting at W= 1522 MeV up to 2157 MeV. Only statistical errors were used. (c) Comparison of the fit coefficients for $\text{L}_{\text{max}}=4$ (black points), $\left(a_{4}\right)^{\sigma_{3/2}}_{2\dots8}$ (see eq. \ref{eq:LowEAssocLegParametrizationDCS3/2}), with the BnGa2014-02 solution truncated at different $\text{L}_{\text{max}}$ (solid lines). Colors same as in (a).}\label{fig:s32_bins}}
{\includegraphics[width=0.49\textwidth, trim=0cm 0cm 1.8cm 0cm, clip]{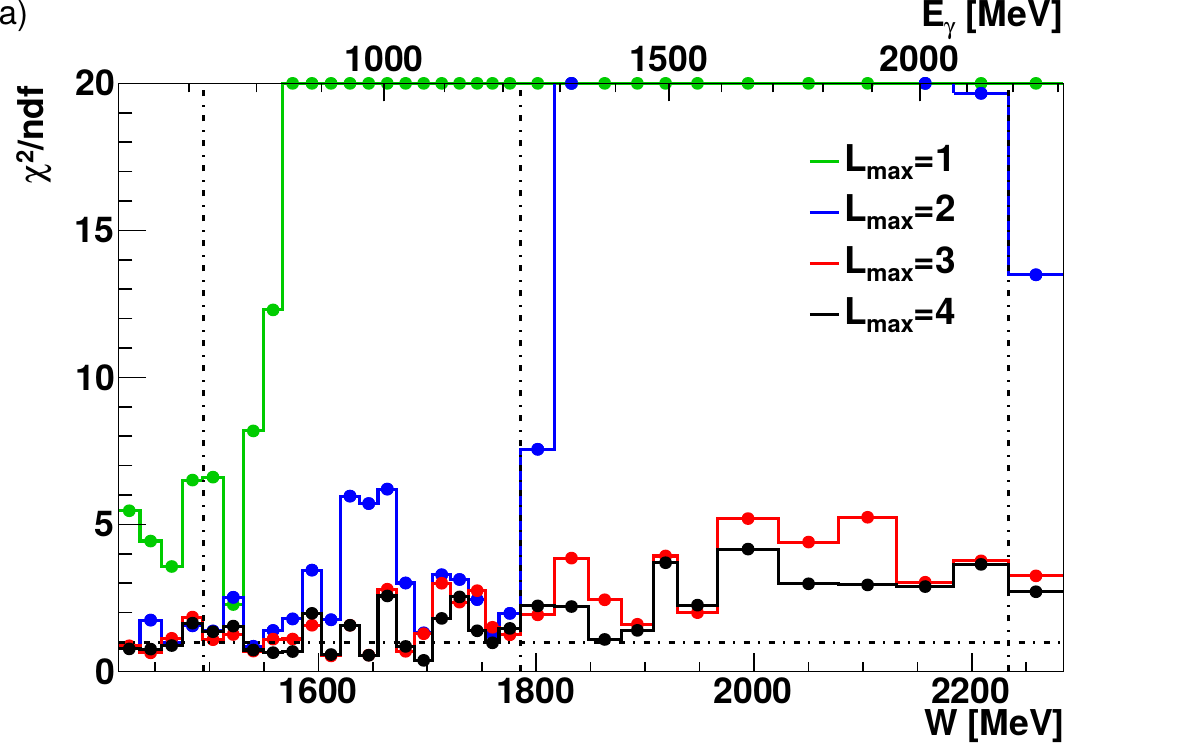}}
\end{minipage}\\

\begin{minipage}{\textwidth}
\centering
\hspace*{-0.45cm}
 \includegraphics[width=0.305\textwidth, trim=0cm 0cm 0.01cm 0.75cm, clip]{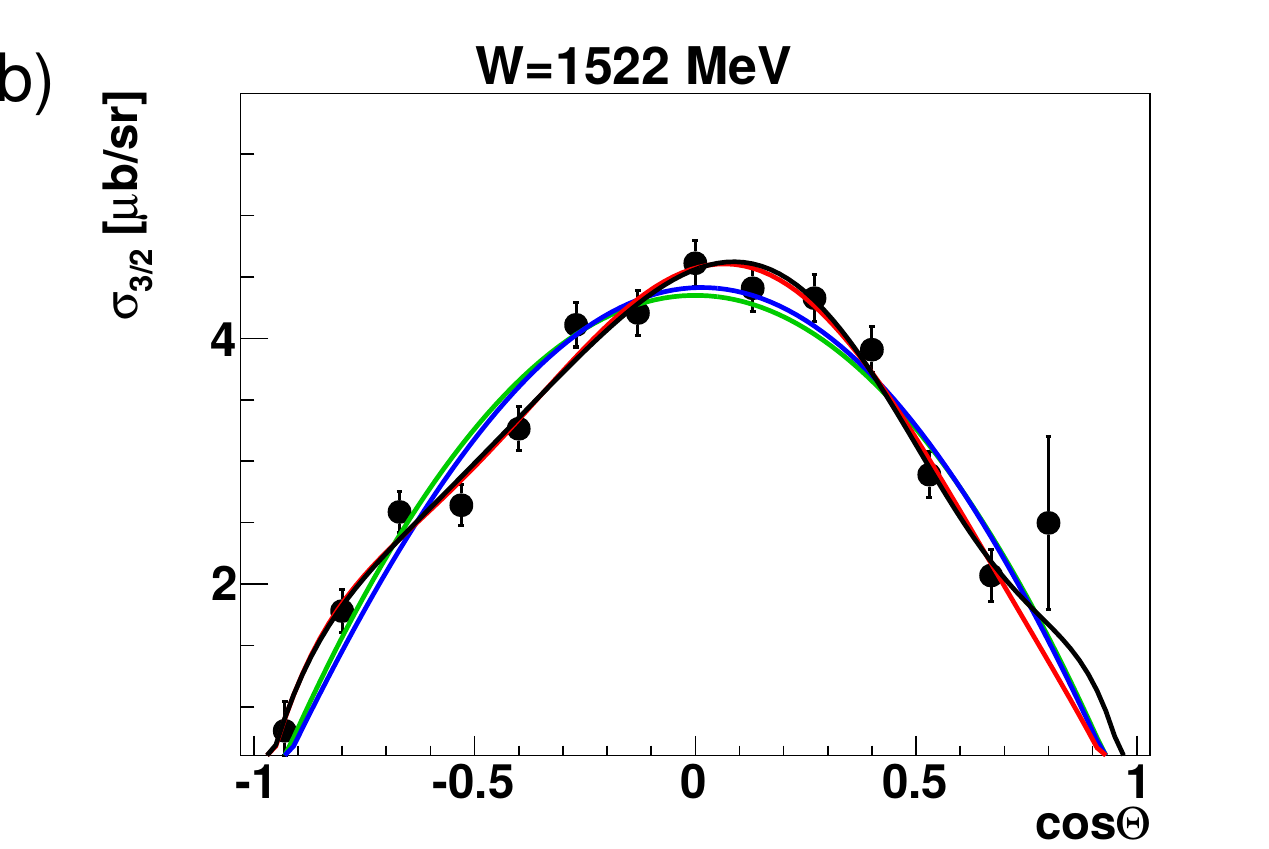}
  \includegraphics[width=0.285\textwidth, trim=0cm 0cm 0.01cm 0.75cm, clip]{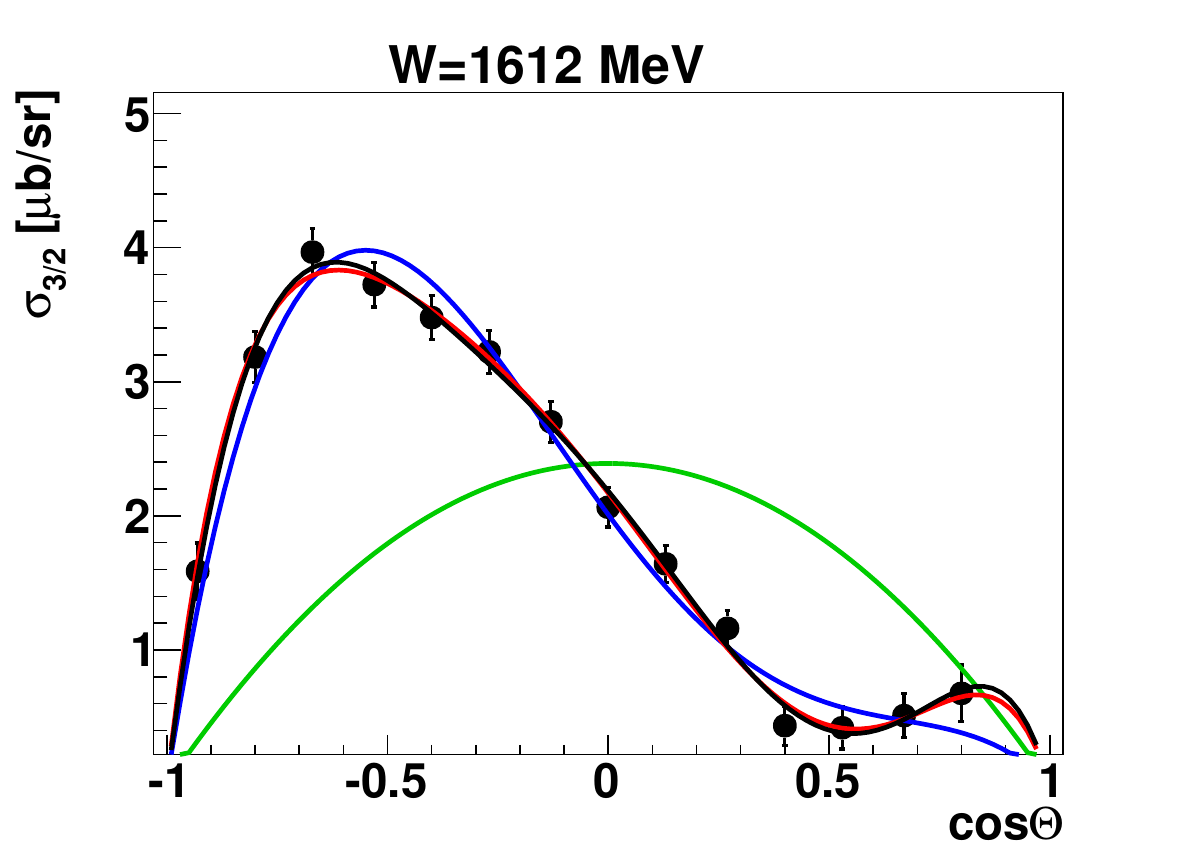}
  \includegraphics[width=0.285\textwidth, trim=0cm 0cm 0.01cm 0.75cm, clip]{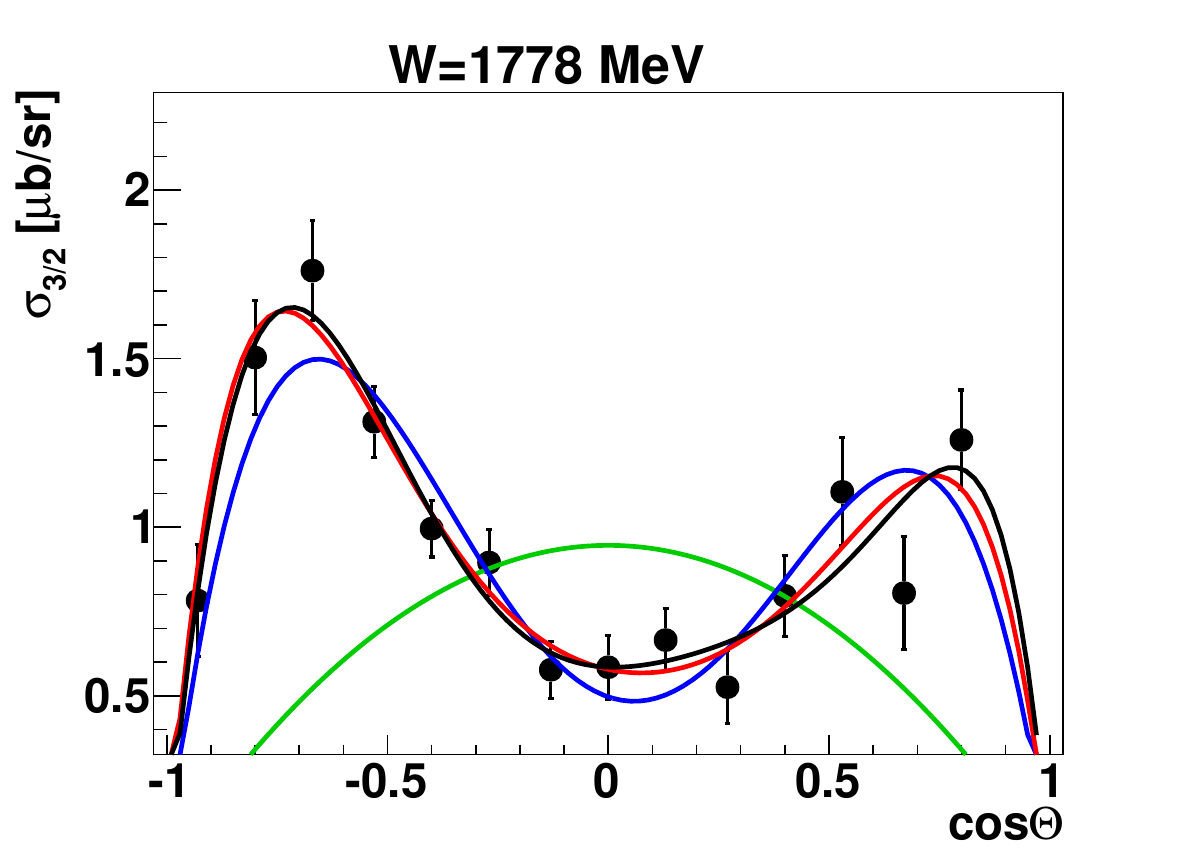}\\
  \includegraphics[width=0.285\textwidth, trim=0cm 0cm 0.01cm 0.75cm, clip]{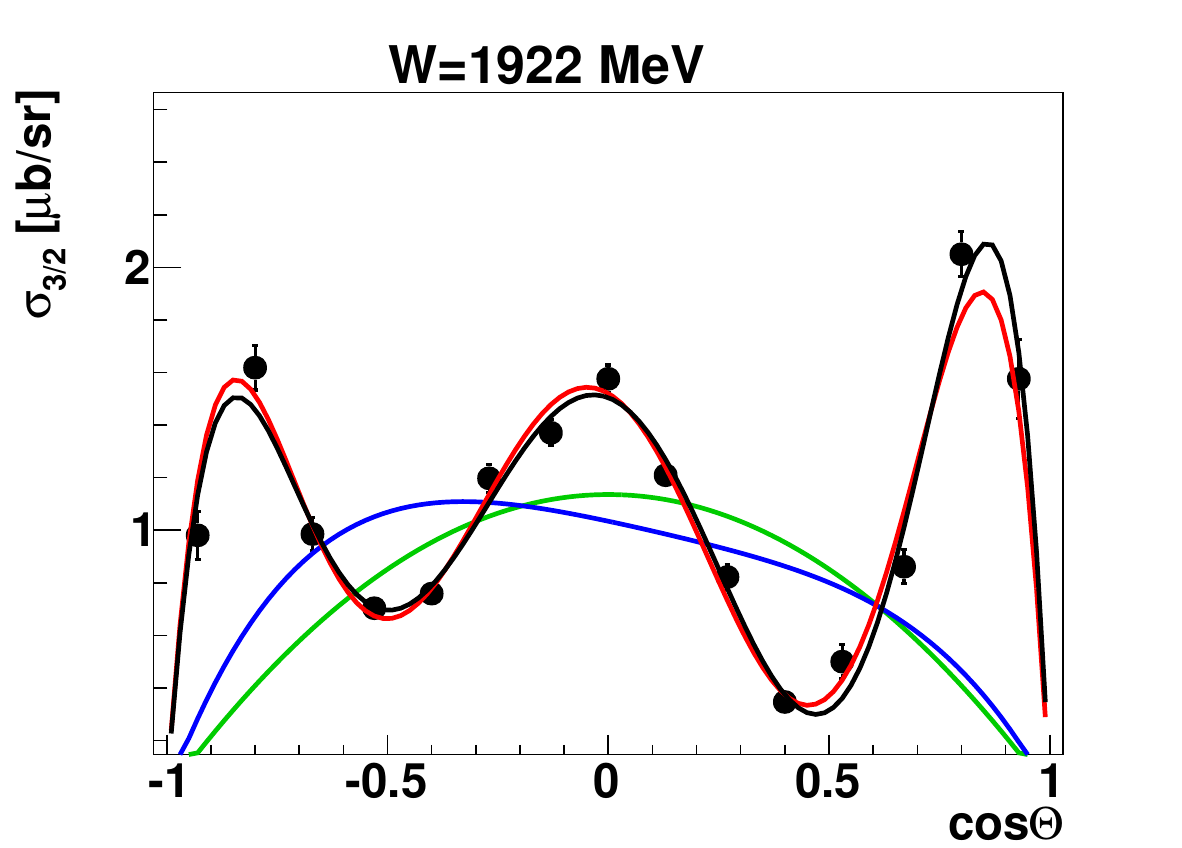}
  \includegraphics[width=0.285\textwidth, trim=0cm 0cm 0.01cm 0.75cm, clip]{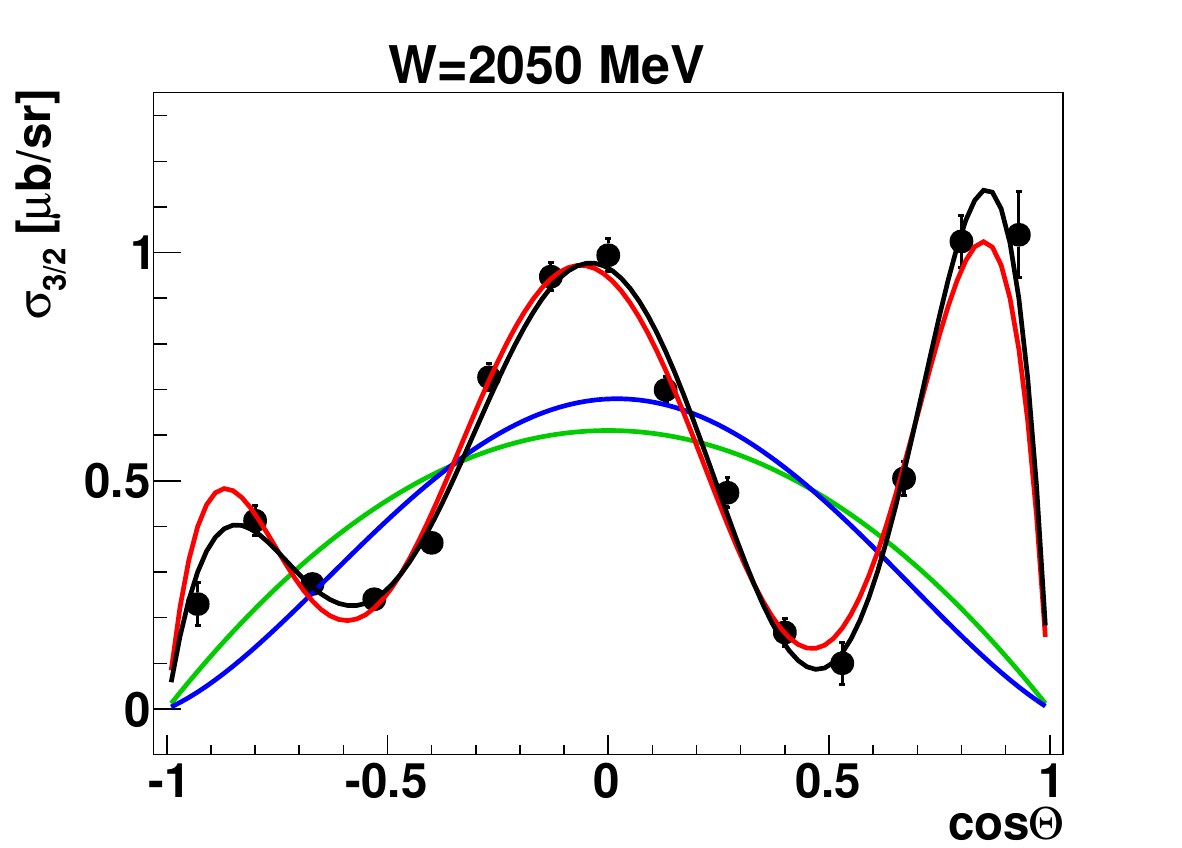}
  \includegraphics[width=0.285\textwidth, trim=0cm 0cm 0.01cm 0.75cm, clip]{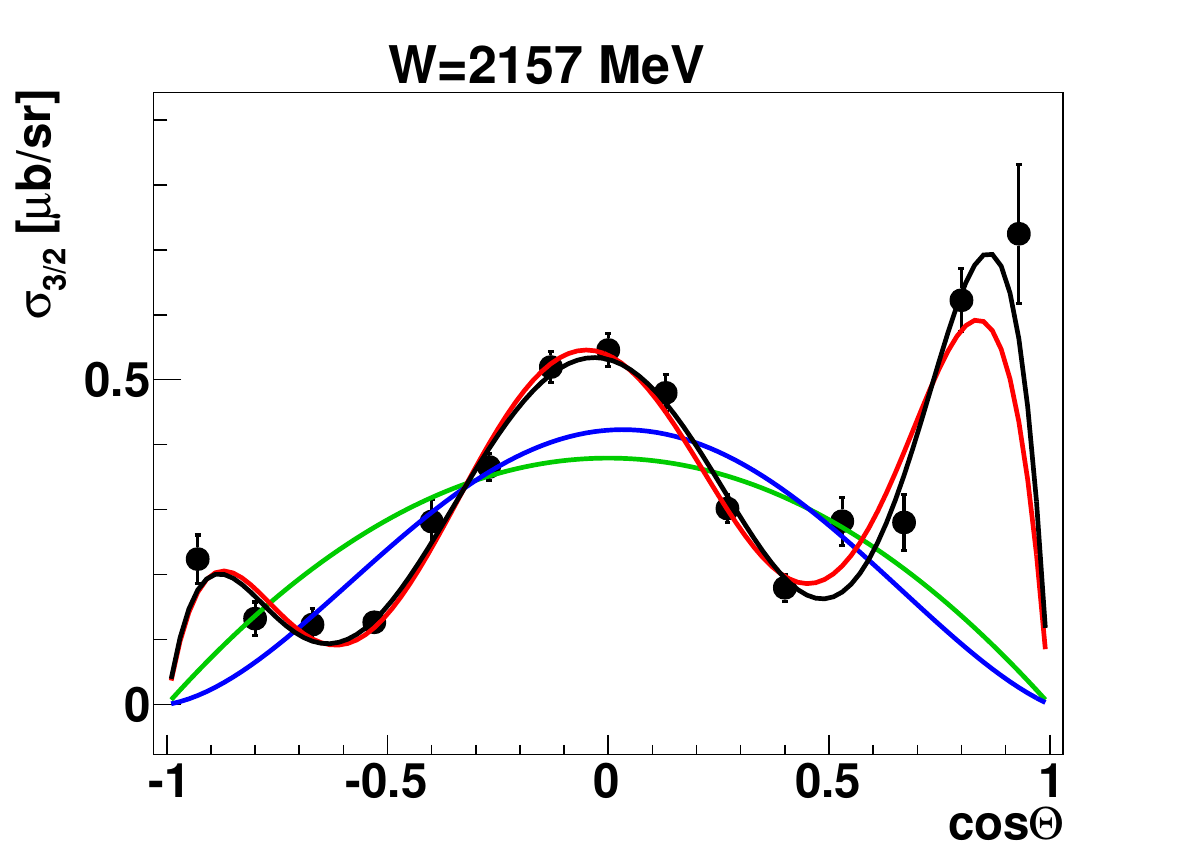}\\
 \vspace*{0.5cm}
  
  \hspace*{-23.5pt}\includegraphics[width=0.2905\textwidth]{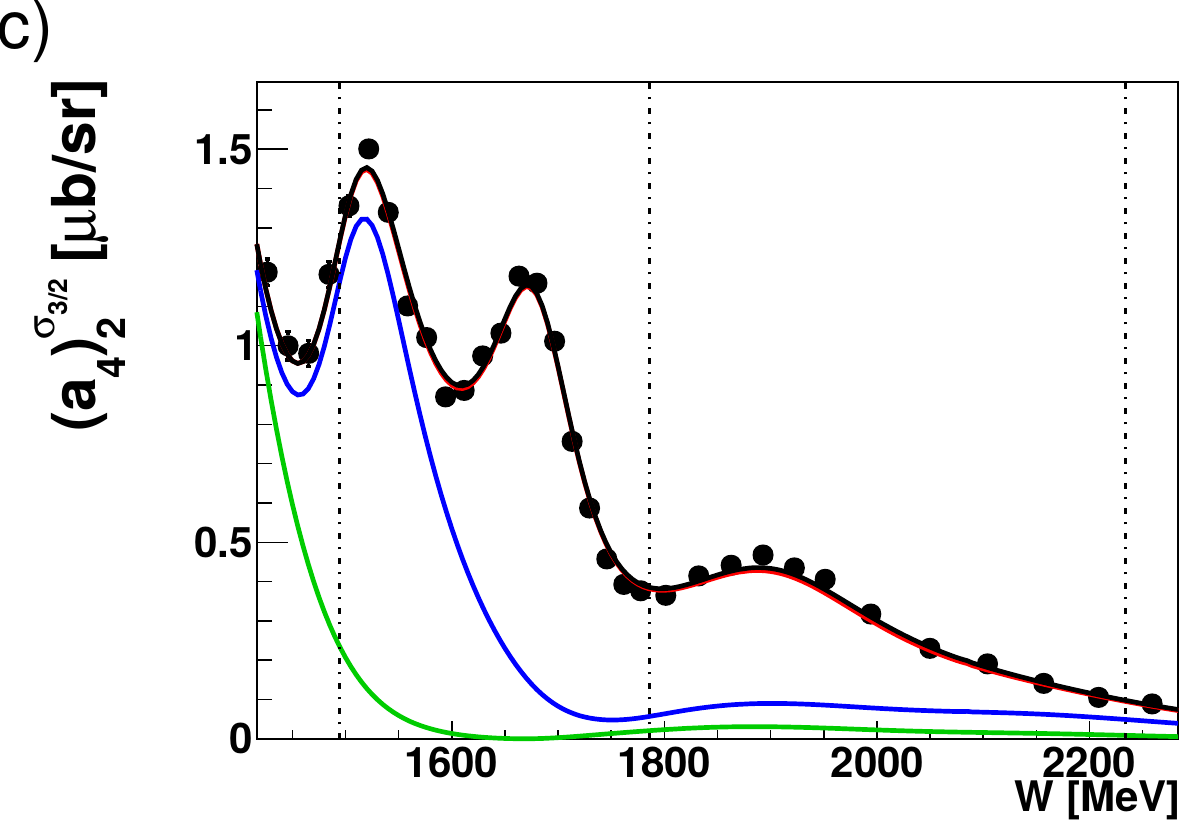}
  \includegraphics[width=0.285\textwidth]{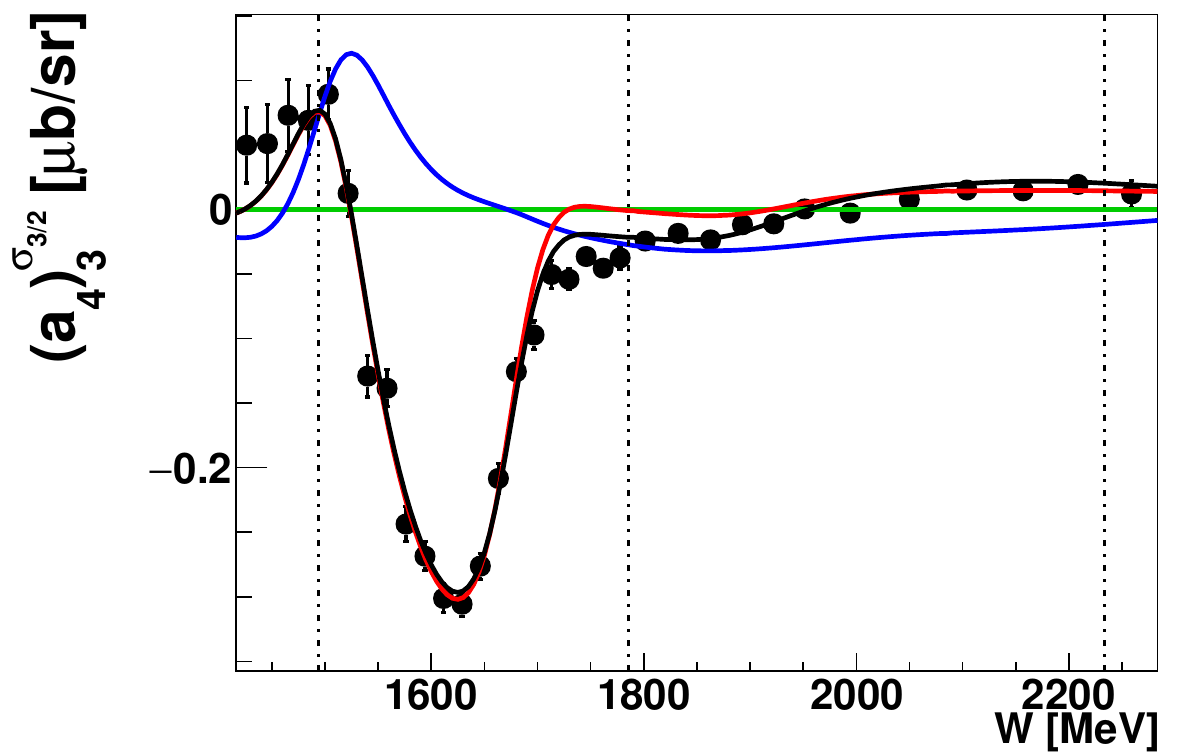}
  \includegraphics[width=0.285\textwidth]{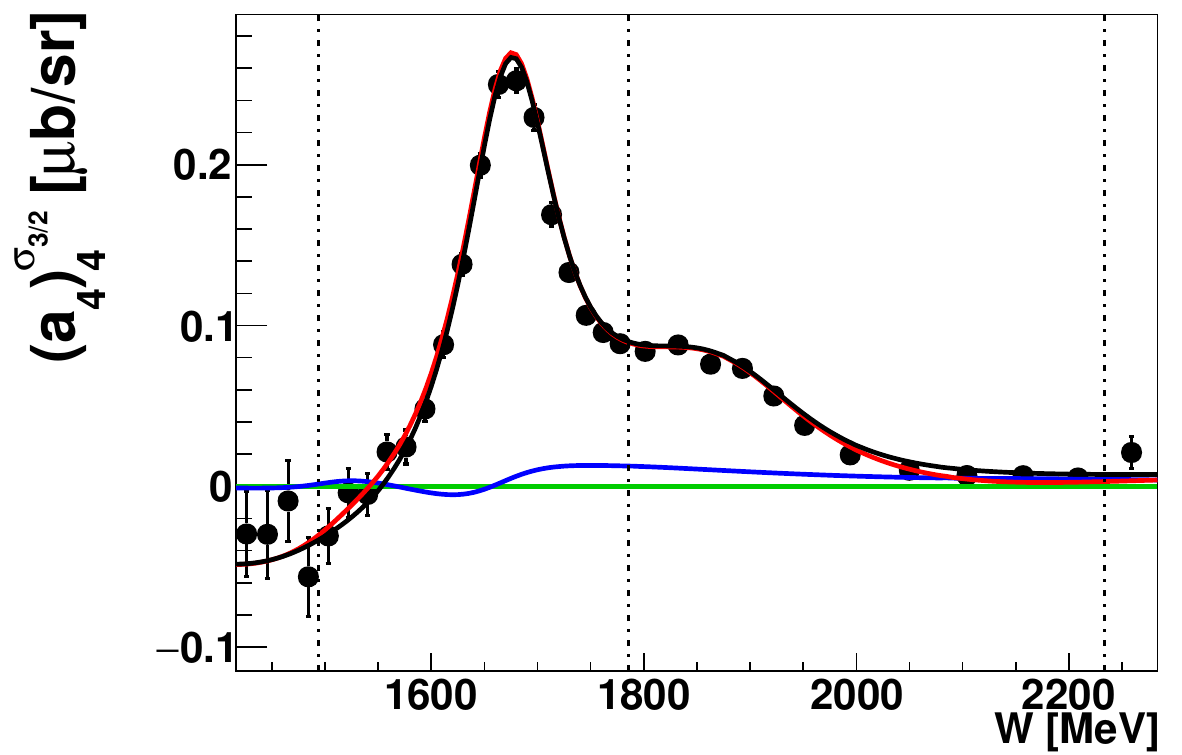}\\
  \hspace*{-19.5pt}\includegraphics[width=0.285\textwidth]{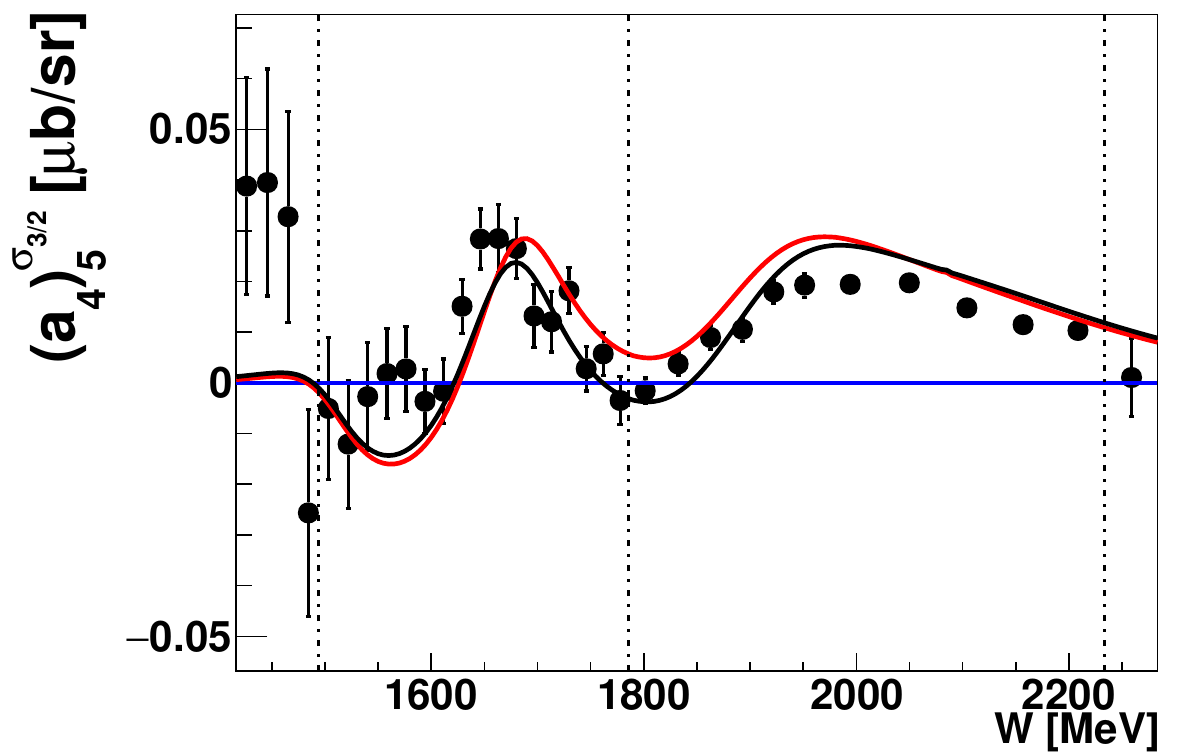}
  \includegraphics[width=0.285\textwidth]{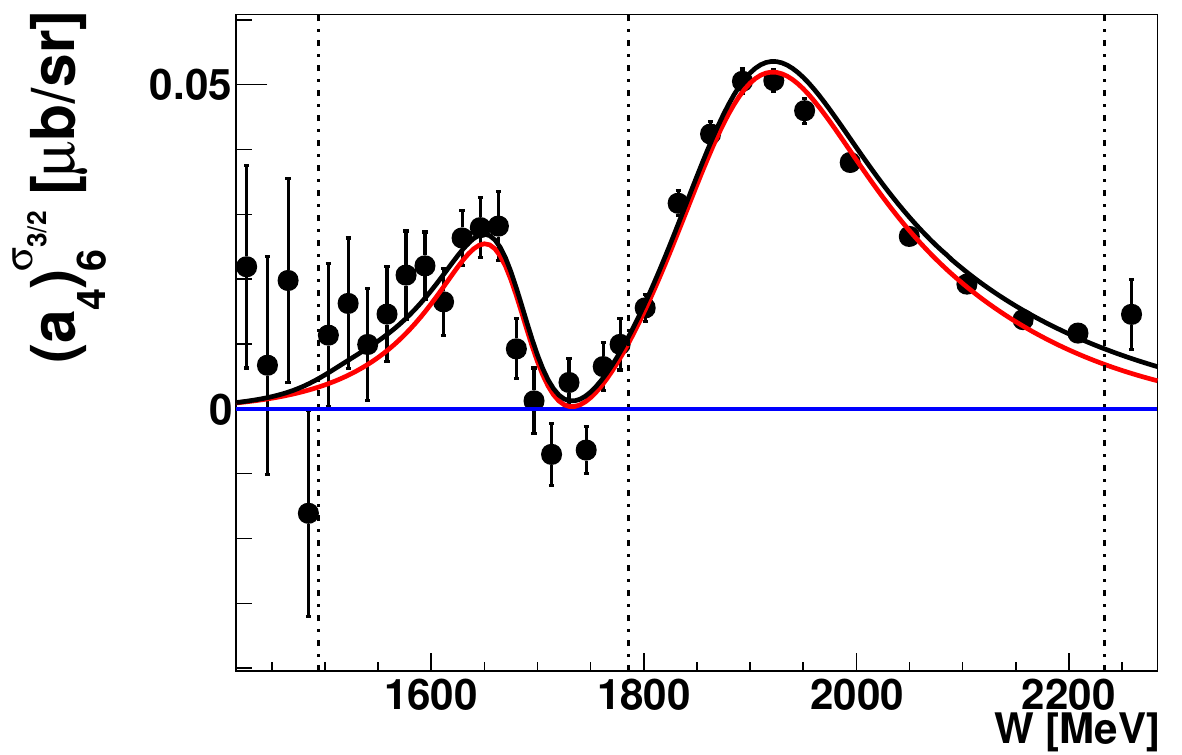}
  \includegraphics[width=0.285\textwidth]{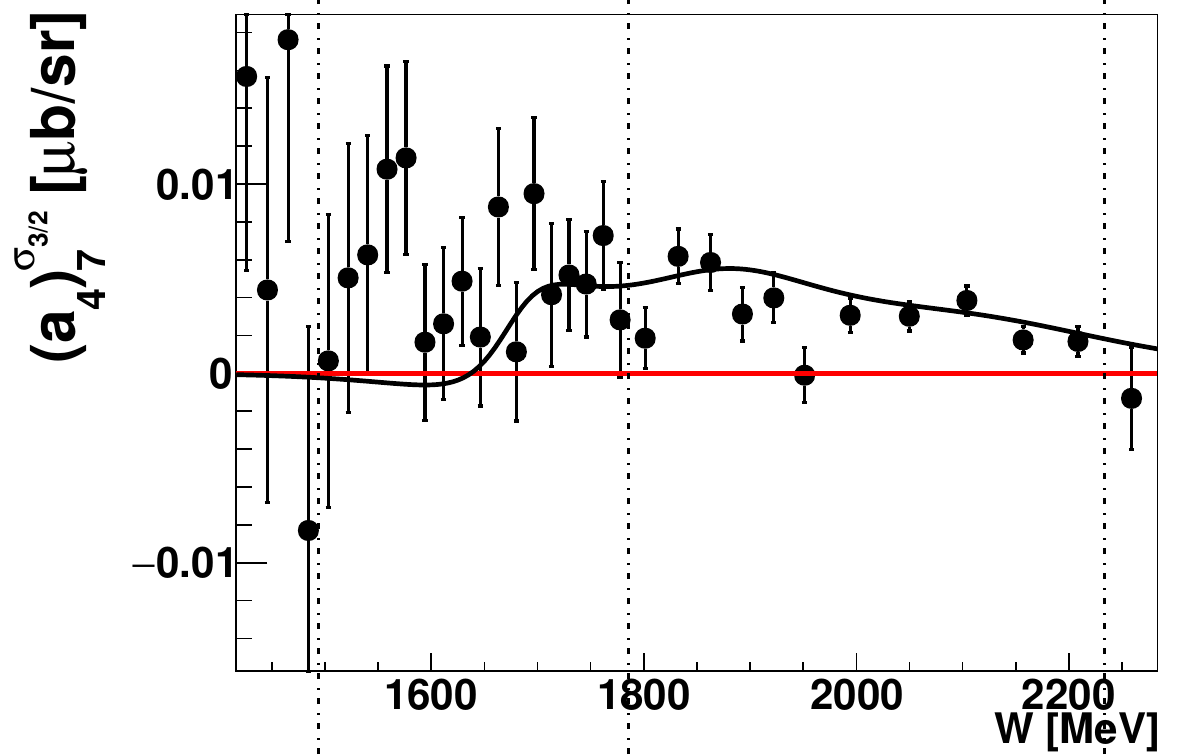} \\
  \hspace*{-19.5pt}\includegraphics[width=0.285\textwidth]{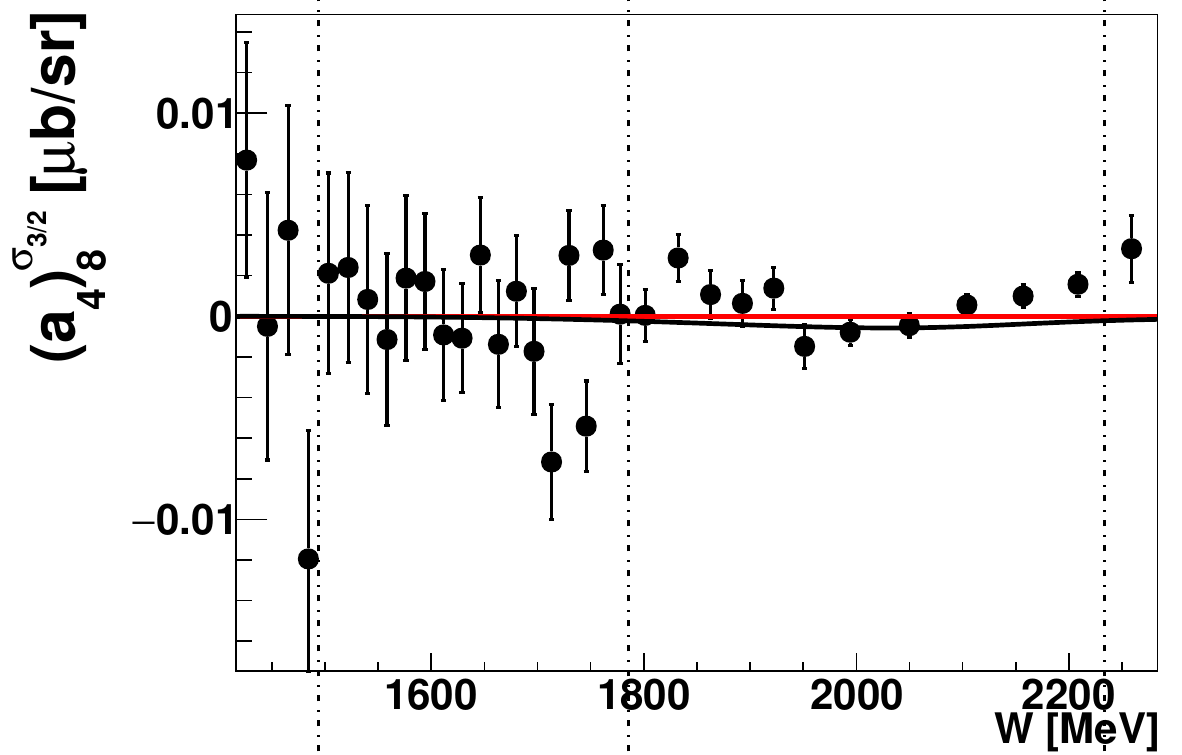}
  \end{minipage}
\end{figure*}

\begin{figure*}
\begin{minipage}{\textwidth}
\floatbox[{\capbeside\thisfloatsetup{capbesideposition={right,top},capbesidewidth=7.8cm}}]{figure}[\FBwidth]
{\caption{The recent new double po\-la\-ri\-za\-tion observable $\check{F}$ data from MAMI \cite{AnnandEtAl:2016} with only statistical error was fitted using associated Legendre polynomials according to eq. (\ref{eq:LowEAssocLegParametrizationF}) and truncating the partial wave expansion at $\text{L}_{\text{max}}=1\dots 4$. (a) The resulting $\chi^2/$ndf values of the different $\text{L}_{\text{max}}$-fits as a function of the center of mass energy W are shown. (b) 6 out of 34 selected angular distributions of $\check{F}$ (black points) are plotted together with the different $\text{L}_{\text{max}}$ fits (solid lines) starting at W= 1410 MeV up to 1842 MeV. (c) Comparison of the fit coefficients for $\text{L}_{\text{max}}=3$ (black points), $\left(a_{3}\right)^{\check{F}}_{1\dots6}$ (see eq. (\ref{eq:LowEAssocLegParametrizationF})), with the BnGa2014-02 solution truncated at different $\text{L}_{\text{max}}$ (solid lines). Colors same as in (a). A strong sensitivity to $\text{L}_{\text{max}}=2$ contributions and some signals of $\text{L}_{\text{max}}=3$-strength are visible for the whole energy range.}\label{fig:f_bins}}
{\includegraphics[width=0.49\textwidth, trim=0cm 0cm 1.8cm 0cm, clip]{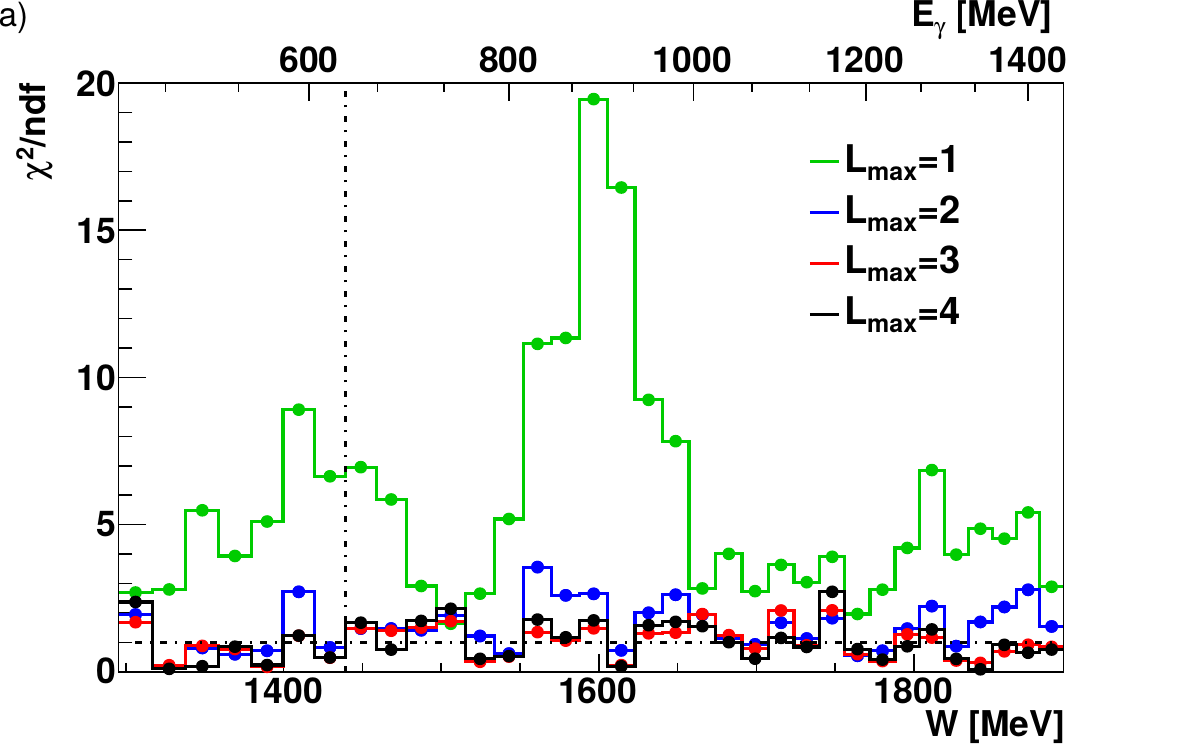}}
\end{minipage}\\

\begin{minipage}{\textwidth}
\centering
\hspace*{-0.45cm}
 \includegraphics[width=0.305\textwidth, trim=0cm 0cm 0.01cm 0.75cm, clip]{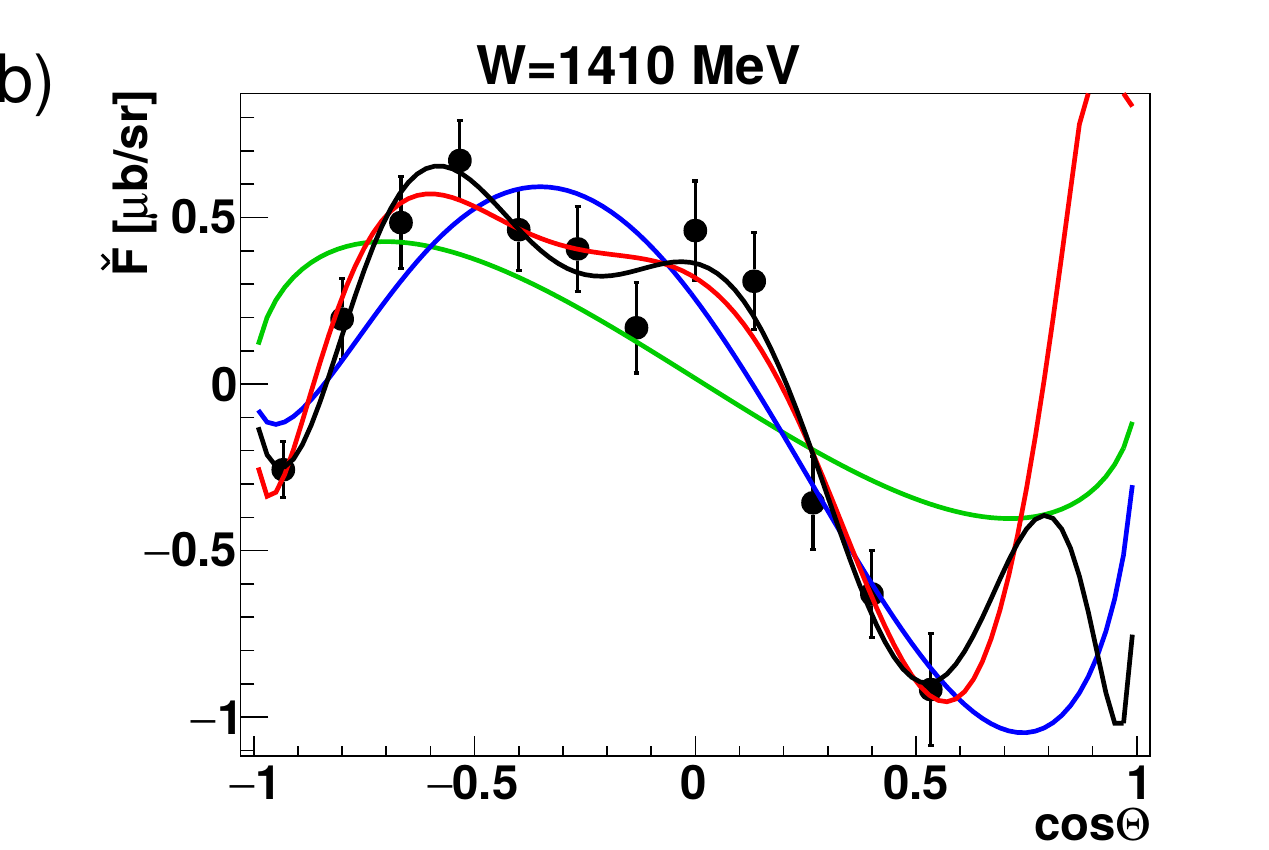}
  \includegraphics[width=0.285\textwidth, trim=0cm 0cm 0.01cm 0.75cm, clip]{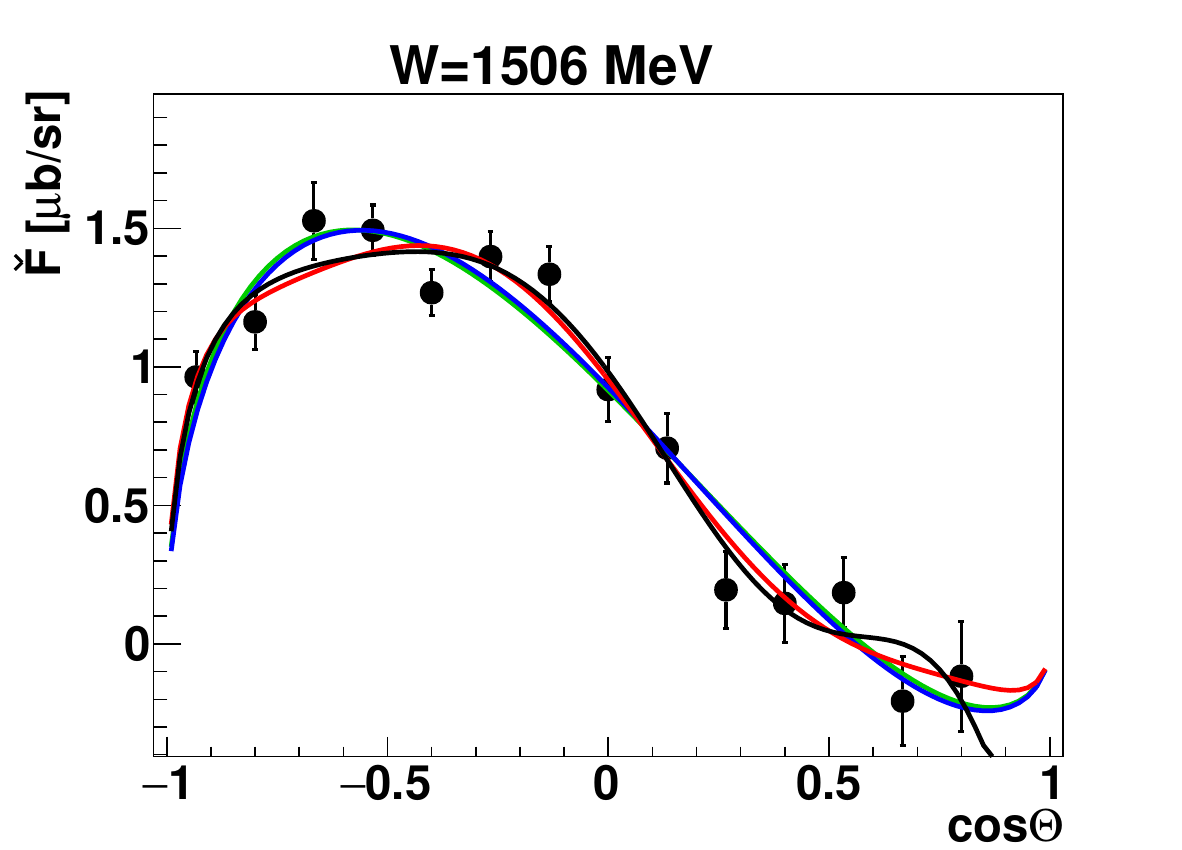}
  \includegraphics[width=0.285\textwidth, trim=0cm 0cm 0.01cm 0.75cm, clip]{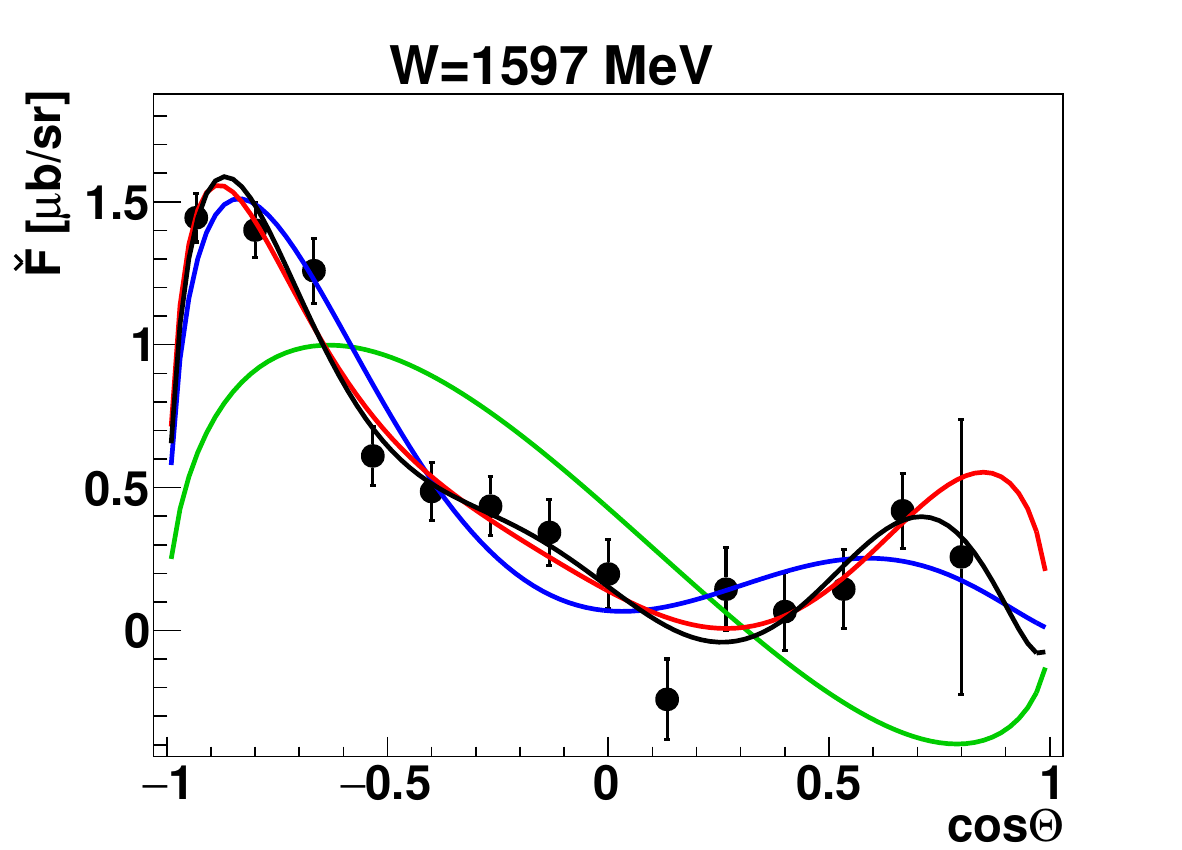}\\
  \includegraphics[width=0.285\textwidth, trim=0cm 0cm 0.01cm 0.75cm, clip]{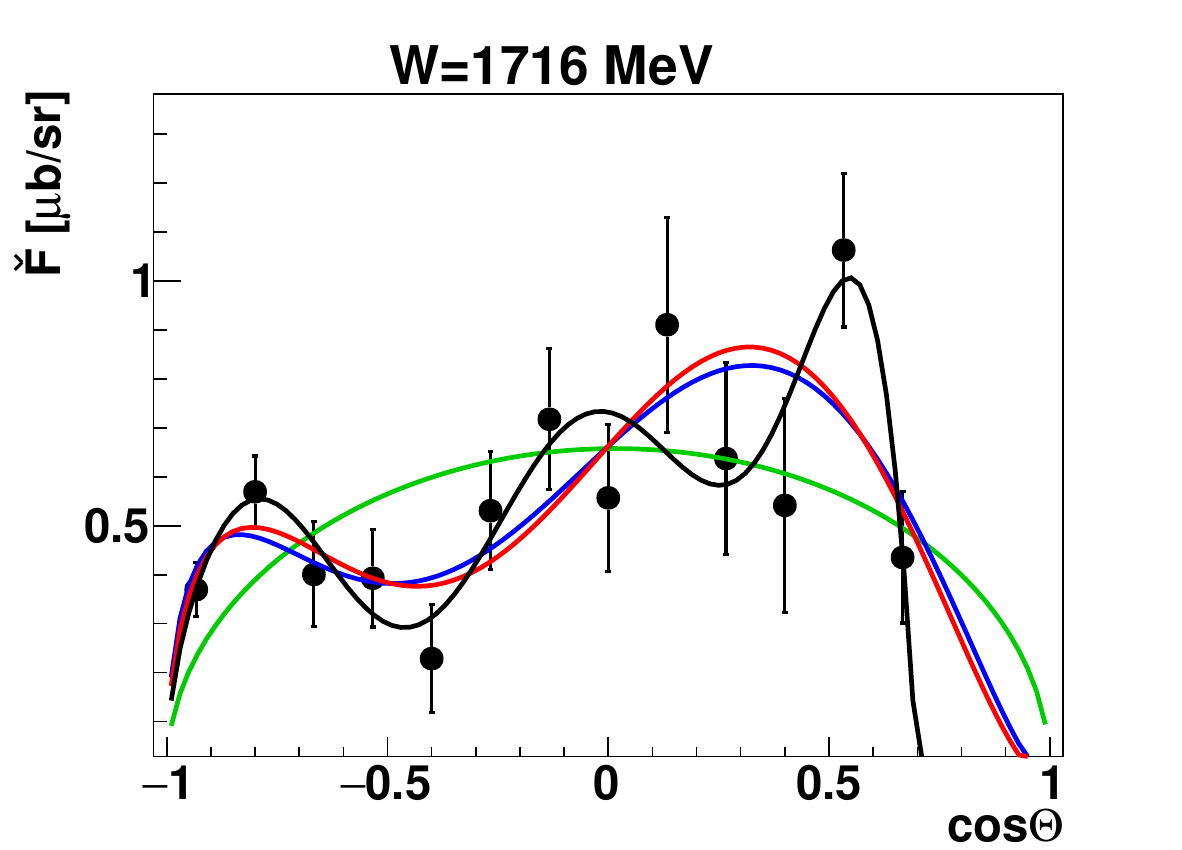}
  \includegraphics[width=0.285\textwidth, trim=0cm 0cm 0.01cm 0.75cm, clip]{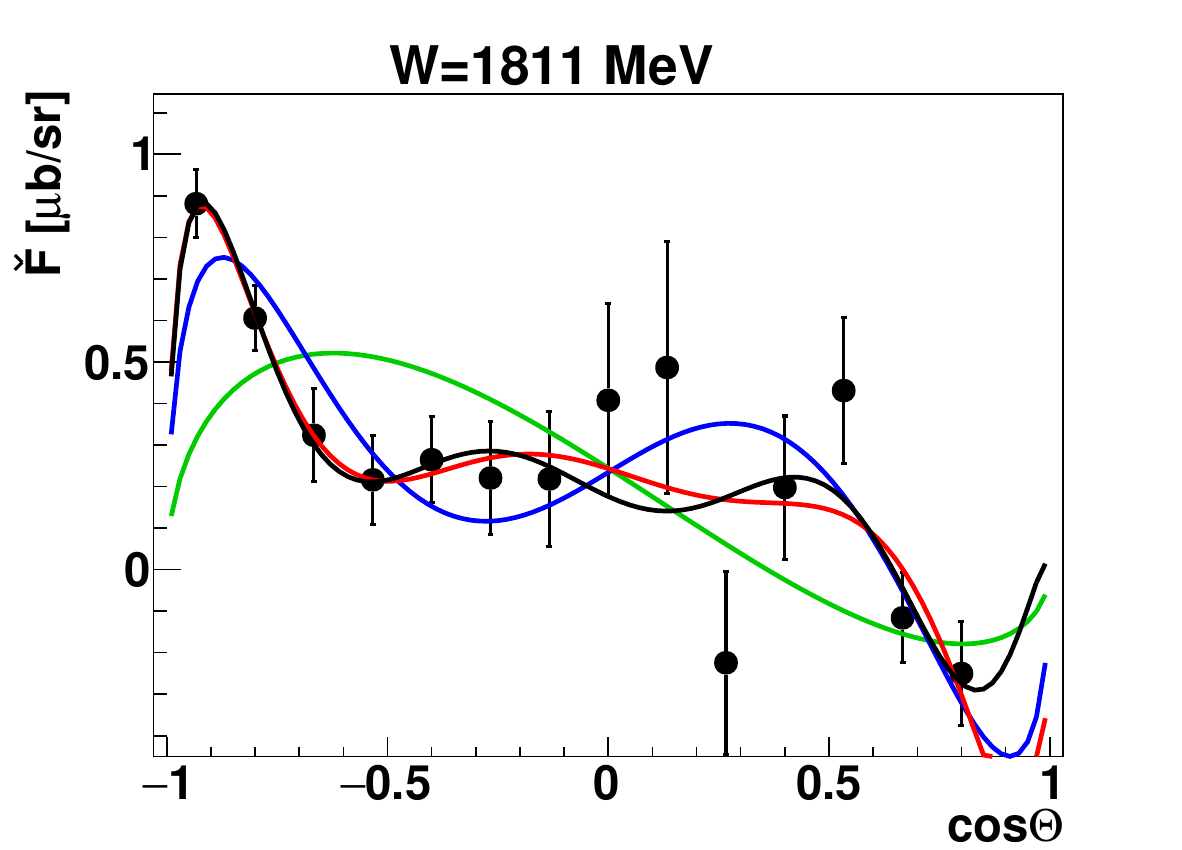}
  \includegraphics[width=0.285\textwidth, trim=0cm 0cm 0.01cm 0.75cm, clip]{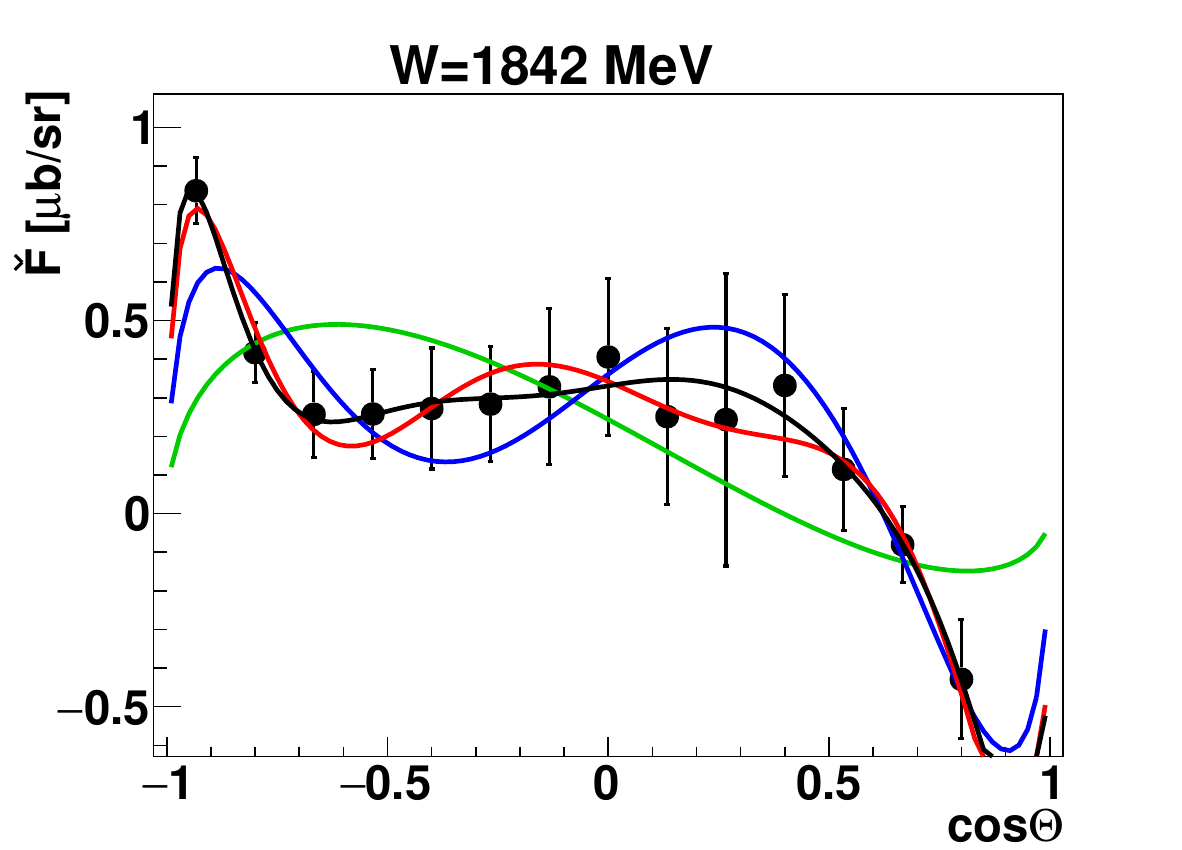}\\
 \vspace*{0.5cm}
  
  \hspace*{-23.5pt}\includegraphics[width=0.2905\textwidth]{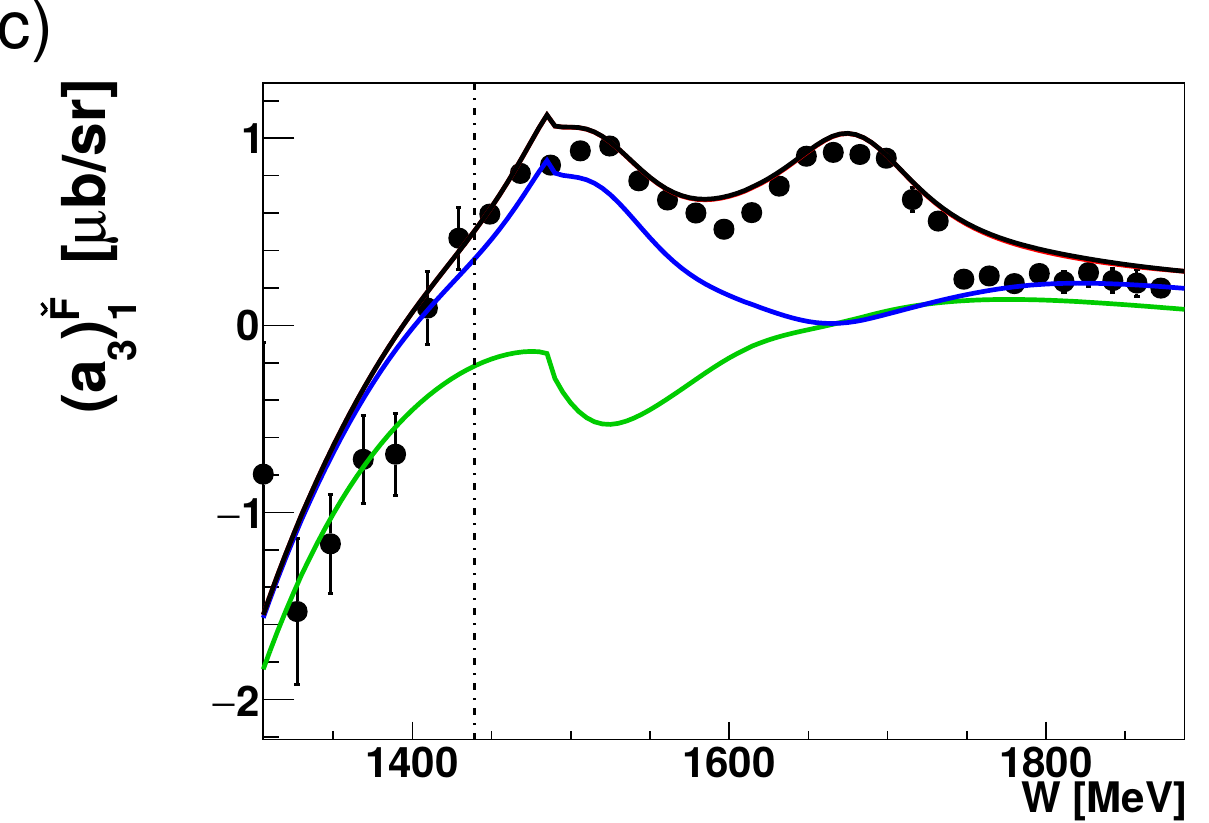}  
  \includegraphics[width=0.285\textwidth]{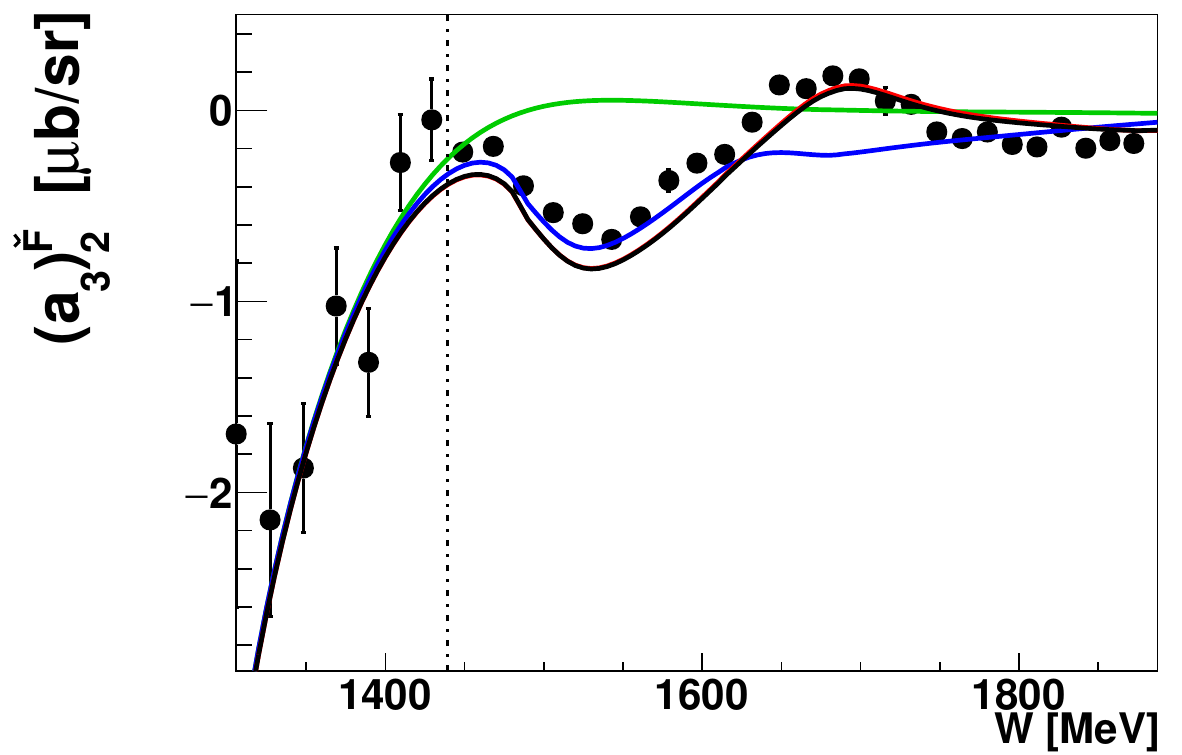}
  \includegraphics[width=0.285\textwidth]{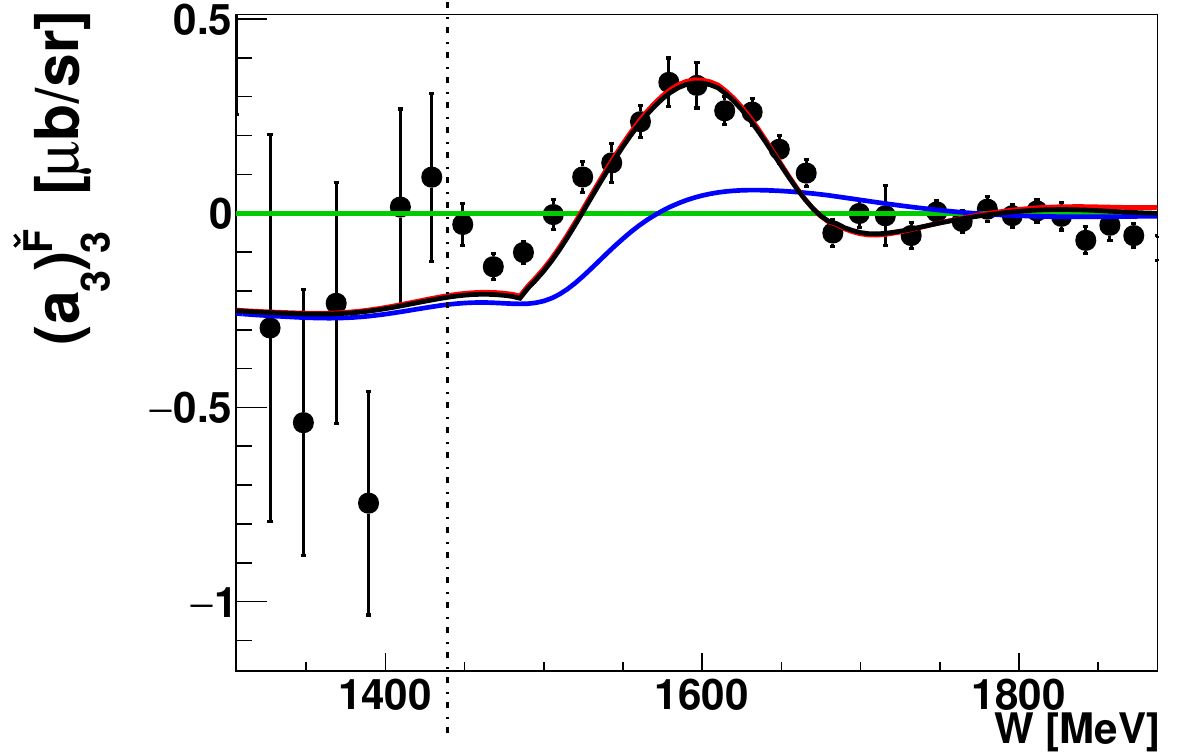}\\
  \hspace*{-19.5pt}\includegraphics[width=0.285\textwidth]{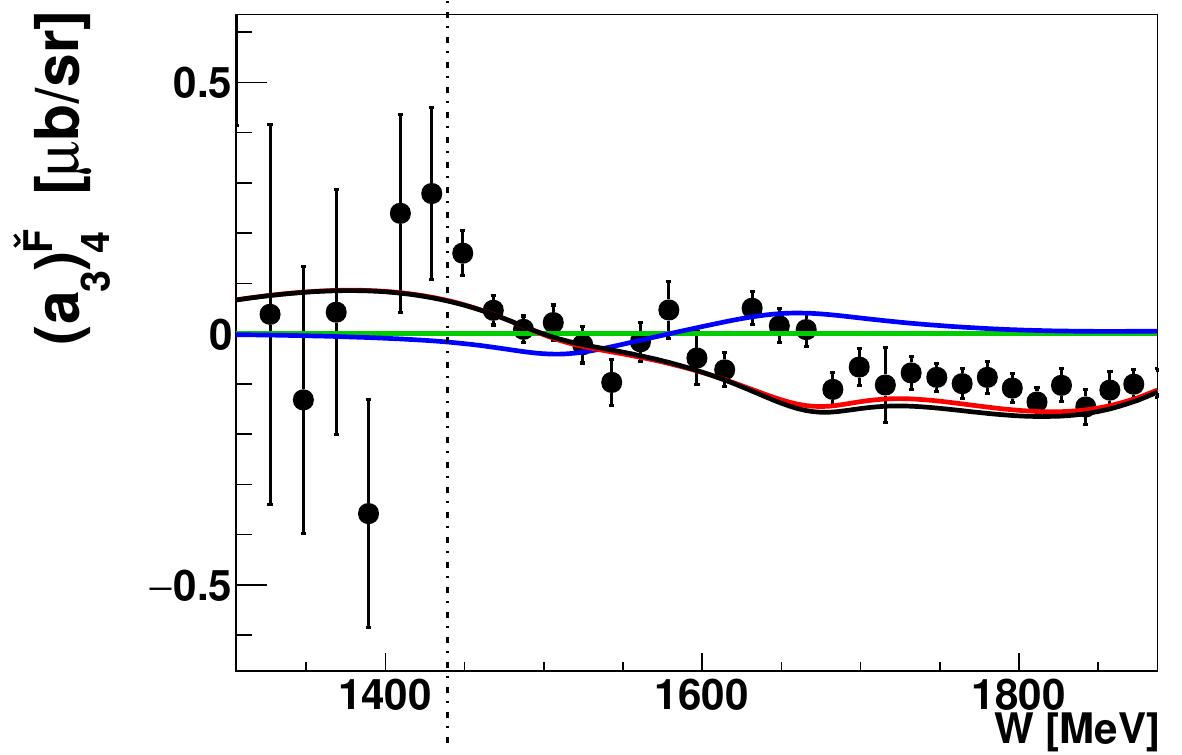}
  \includegraphics[width=0.285\textwidth]{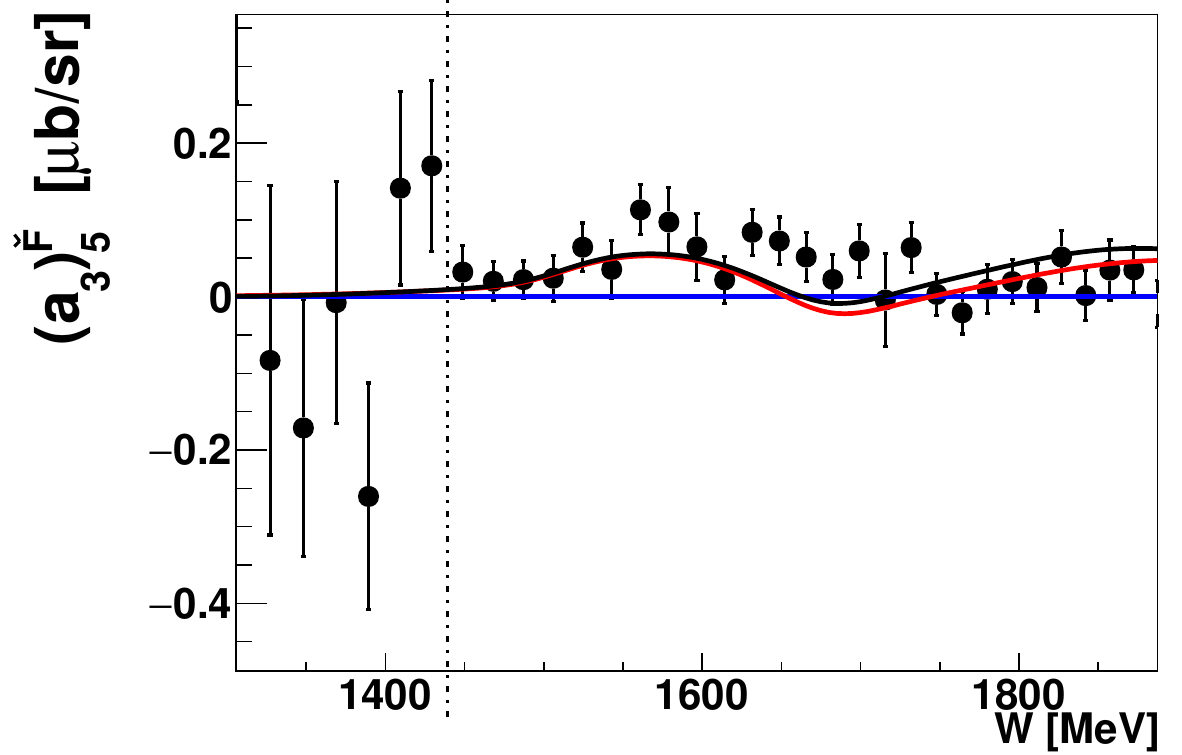}
  \includegraphics[width=0.285\textwidth]{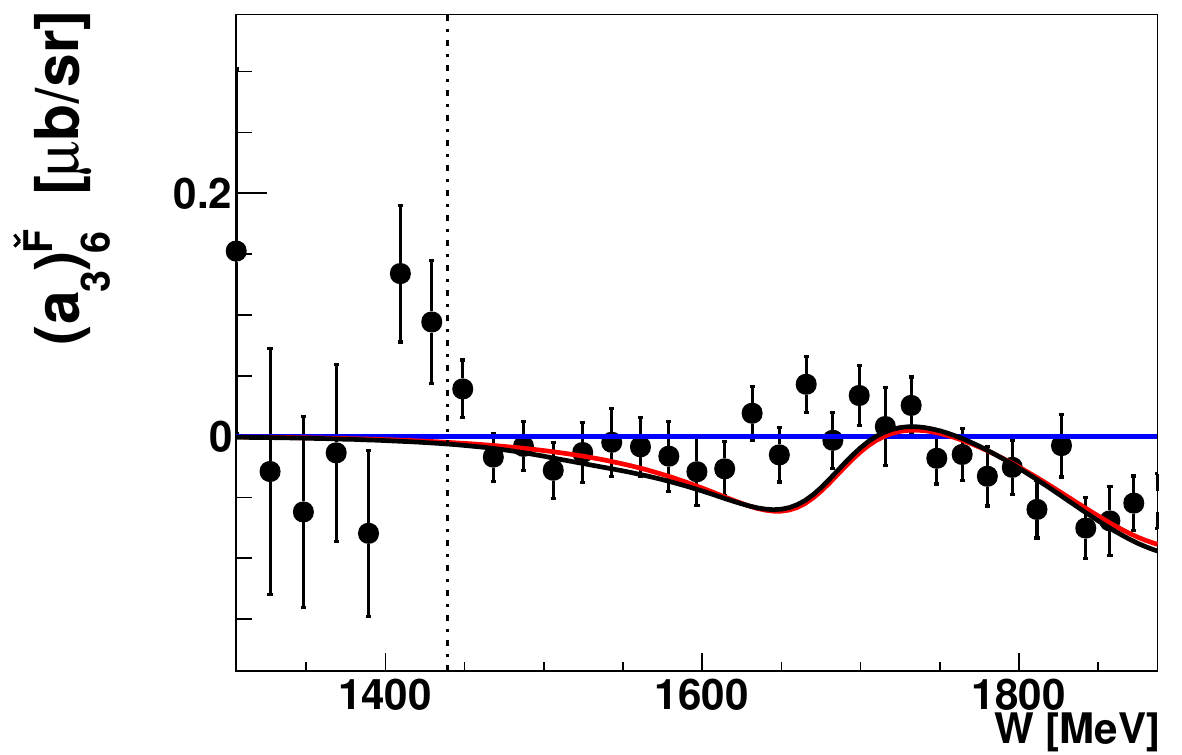}
  \end{minipage}
\end{figure*}

\begin{figure*}
\begin{minipage}{\textwidth}
\floatbox[{\capbeside\thisfloatsetup{capbesideposition={right,top},capbesidewidth=7.8cm}}]{figure}[\FBwidth]
{\caption{The recent new double po\-la\-ri\-za\-tion observable $\check{G}$ data from ELSA \cite{Thiel:2012,Thiel:2015} with only statistical error was fitted using associated Legendre polynomials according to eq. \ref{eq:LowEAssocLegParametrizationG} and truncating the partial wave expansion at $\text{L}_{\text{max}}=1\dots 4$. (a) The resulting $\chi^2/$ndf values of the different $\text{L}_{\text{max}}$-fits as a function of the center of mass energy W are shown. (b) 6 out of 19 selected angular distributions of $\check{G}$ (black points) are plotted together with the different $\text{L}_{\text{max}}$ fits (solid lines) starting at W= 1438 MeV up to 1822 MeV. (c) Comparison of the fit coefficients for $\text{L}_{\text{max}}=3$ (black points), $\left(a_{3}\right)^{\check{G}}_{2\dots6}$ (see eq. \ref{eq:LowEAssocLegParametrizationG}), with the BnGa2014-02 solution truncated at different $\text{L}_{\text{max}}$ (solid lines). Colors same as in (a).}\label{fig:g_bins}}
{\includegraphics[width=0.49\textwidth, trim=0cm 0cm 1.8cm 0cm, clip]{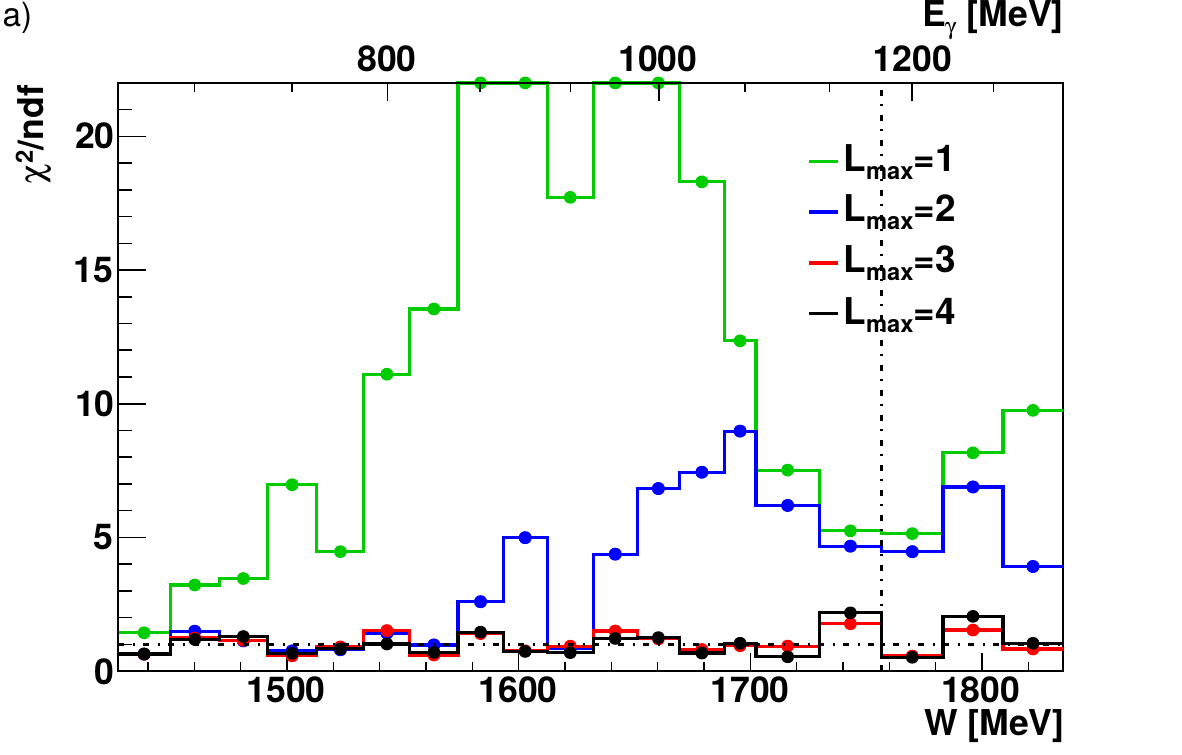}}
\end{minipage}\\

\begin{minipage}{\textwidth}
\centering
\hspace*{-0.45cm}
 \includegraphics[width=0.305\textwidth, trim=0cm 0cm 0.01cm 0.75cm, clip]{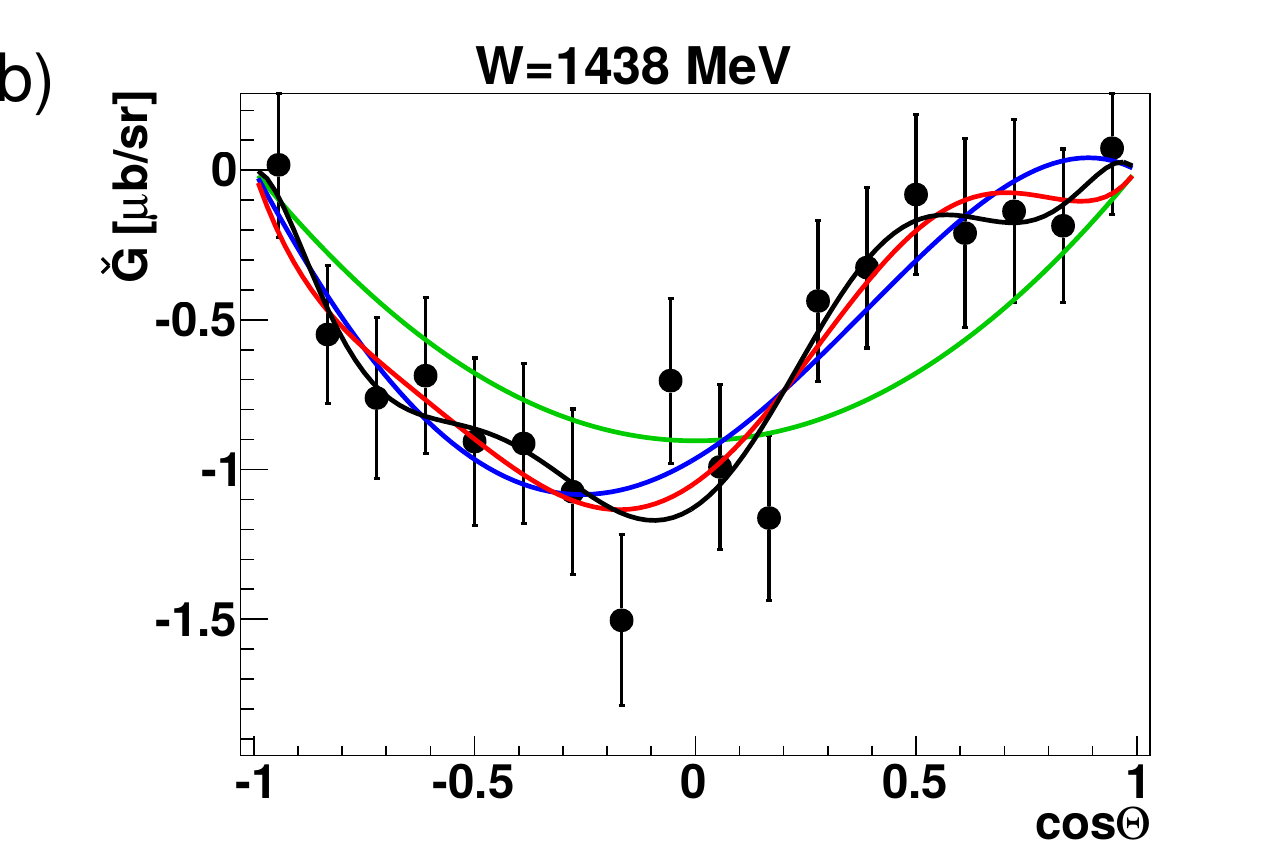}
  \includegraphics[width=0.285\textwidth, trim=0cm 0cm 0.01cm 0.75cm, clip]{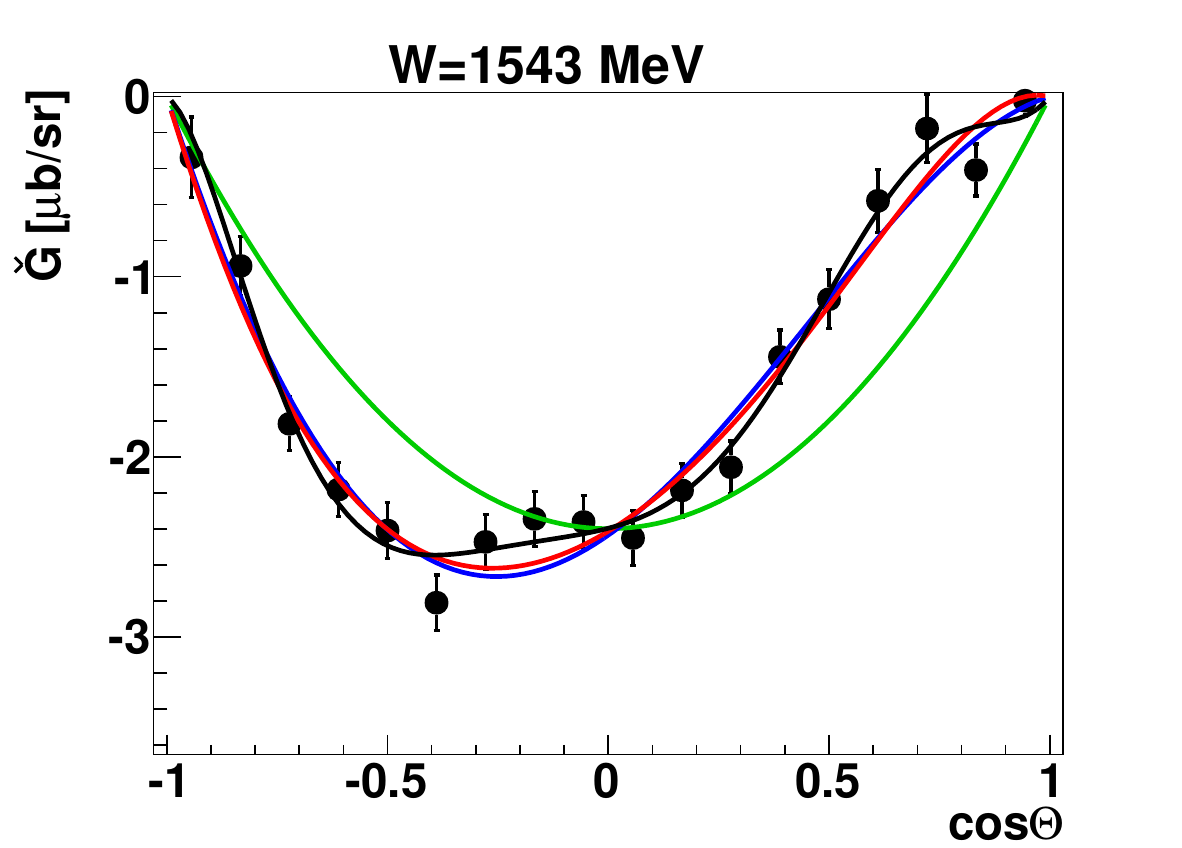}
  \includegraphics[width=0.285\textwidth, trim=0cm 0cm 0.01cm 0.75cm, clip]{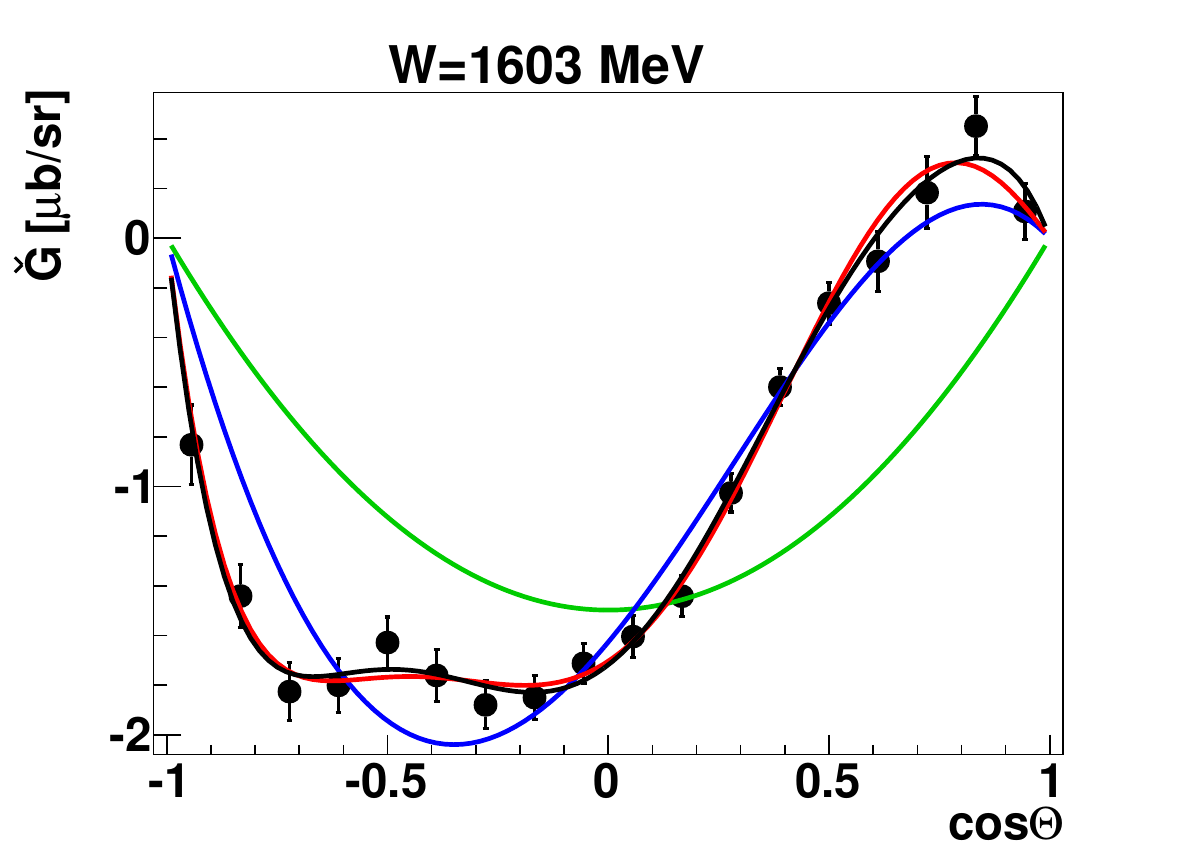}\\
  \includegraphics[width=0.285\textwidth, trim=0cm 0cm 0.01cm 0.75cm, clip]{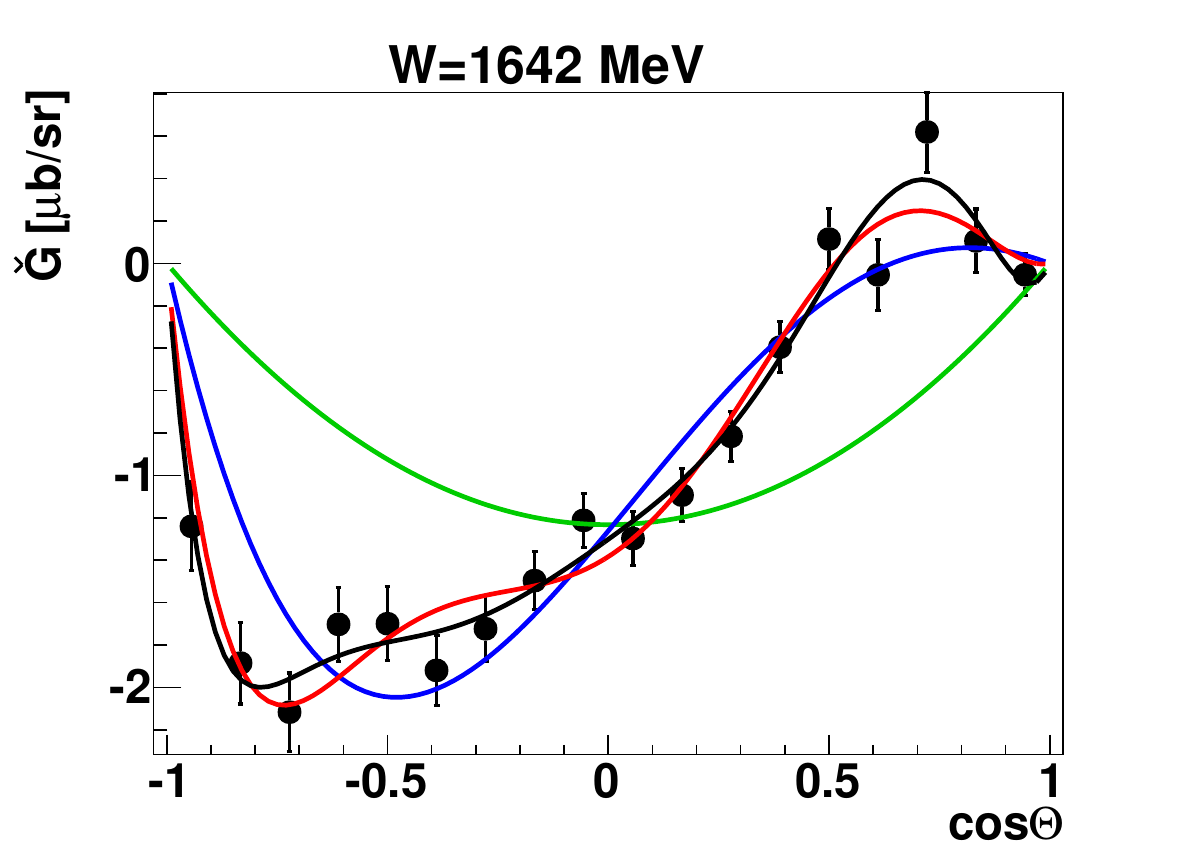}
  \includegraphics[width=0.285\textwidth, trim=0cm 0cm 0.01cm 0.75cm, clip]{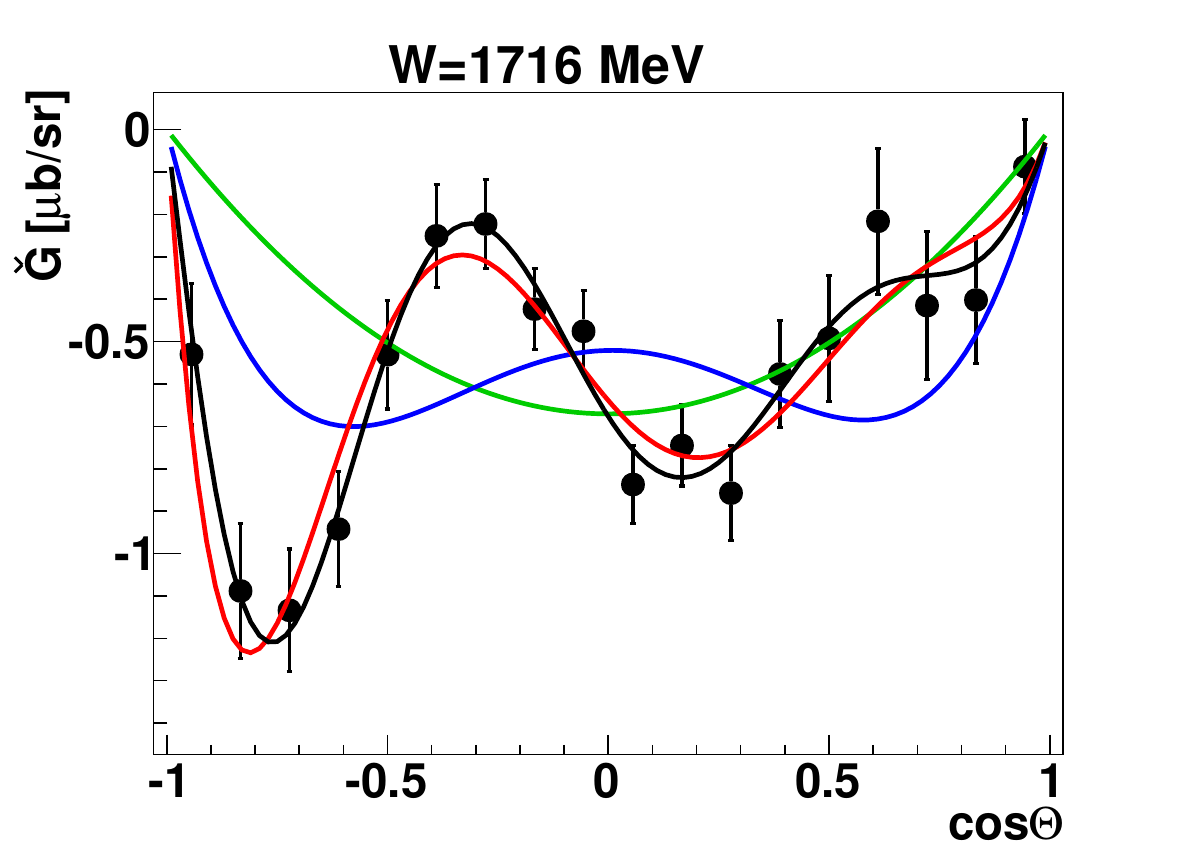}
  \includegraphics[width=0.285\textwidth, trim=0cm 0cm 0.01cm 0.75cm, clip]{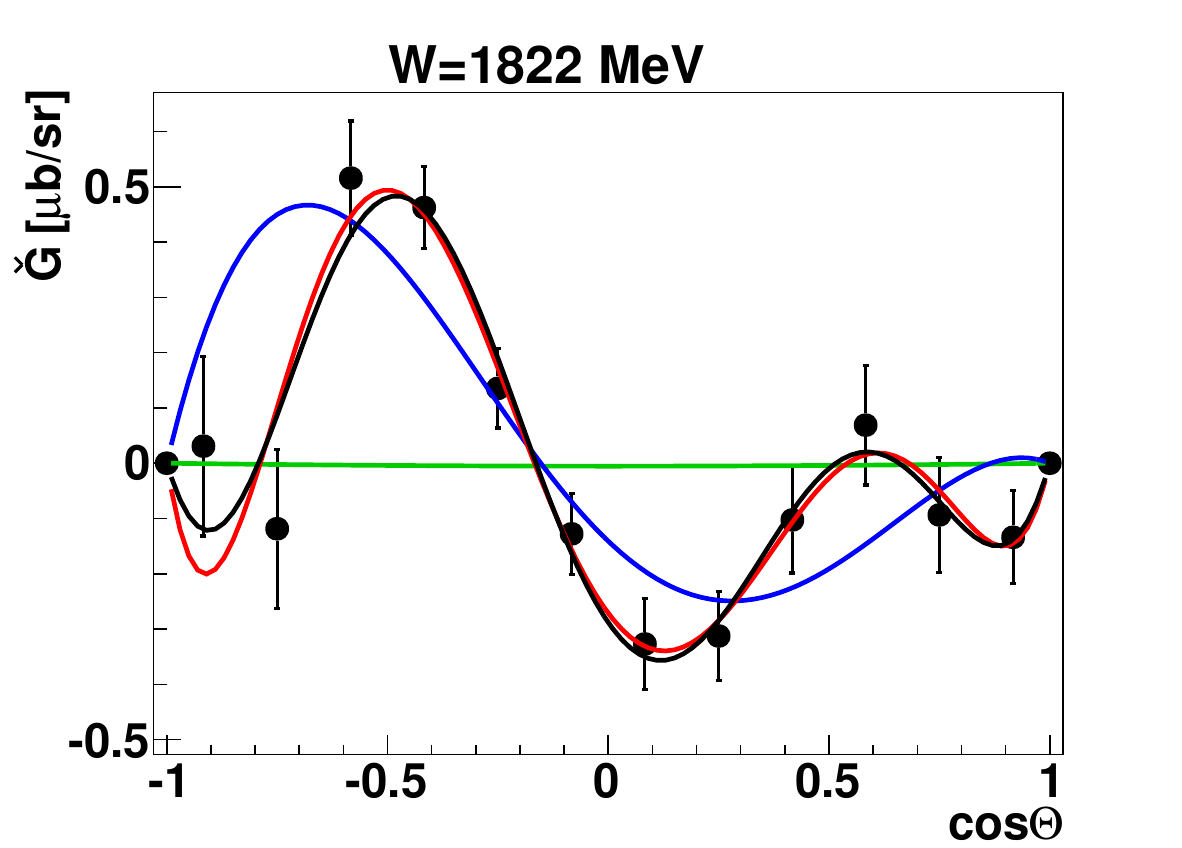}\\
 \vspace*{0.5cm}
  
  \hspace*{-23.5pt}\includegraphics[width=0.2905\textwidth]{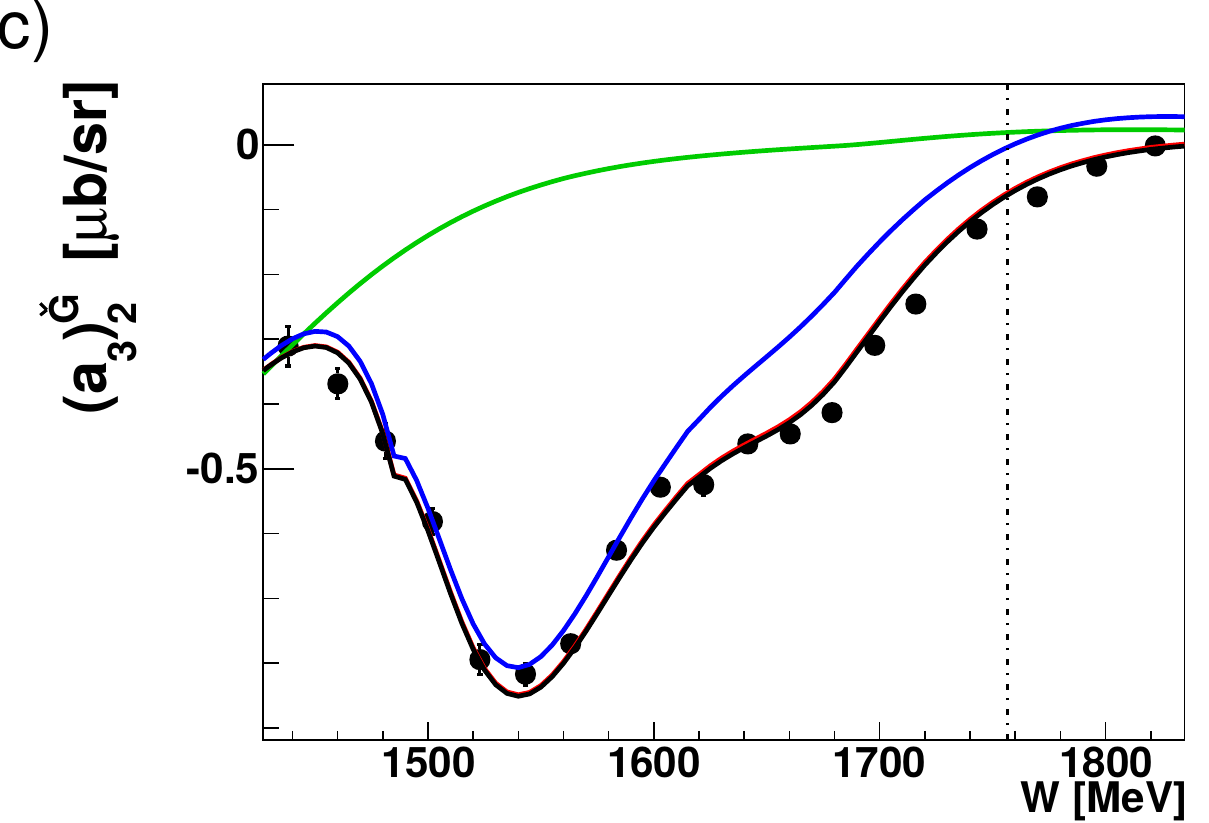}
  \includegraphics[width=0.285\textwidth]{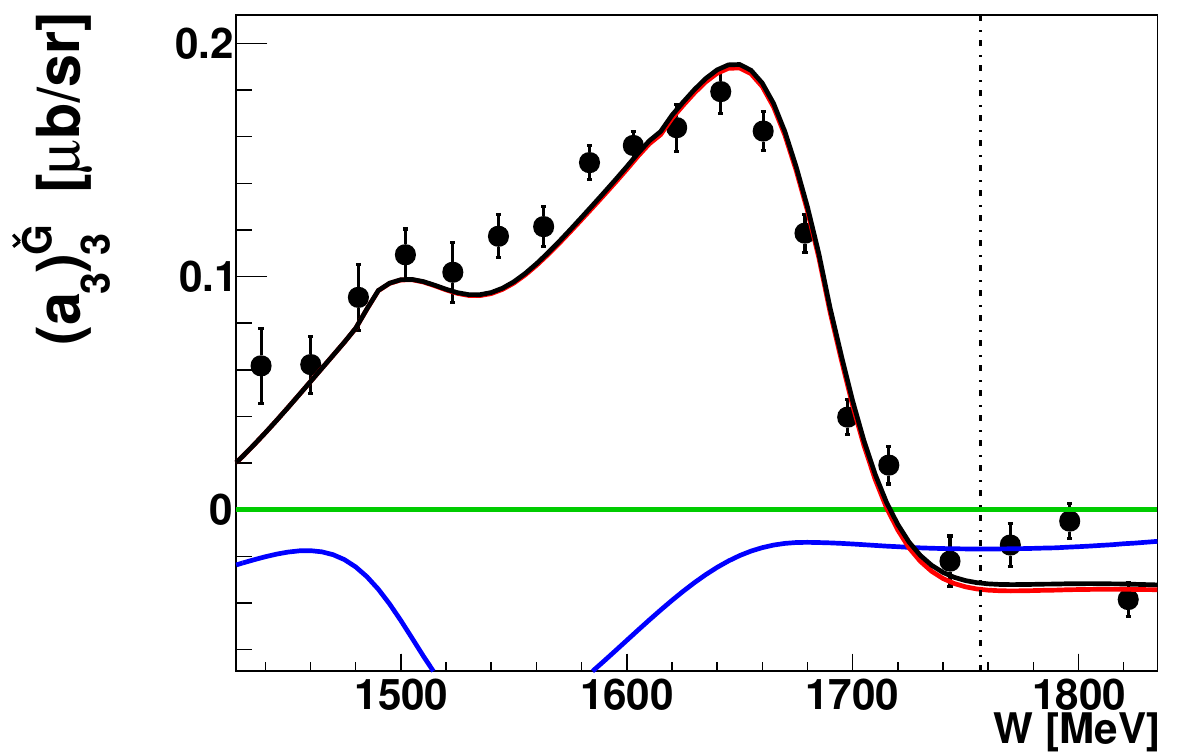}
  \includegraphics[width=0.285\textwidth]{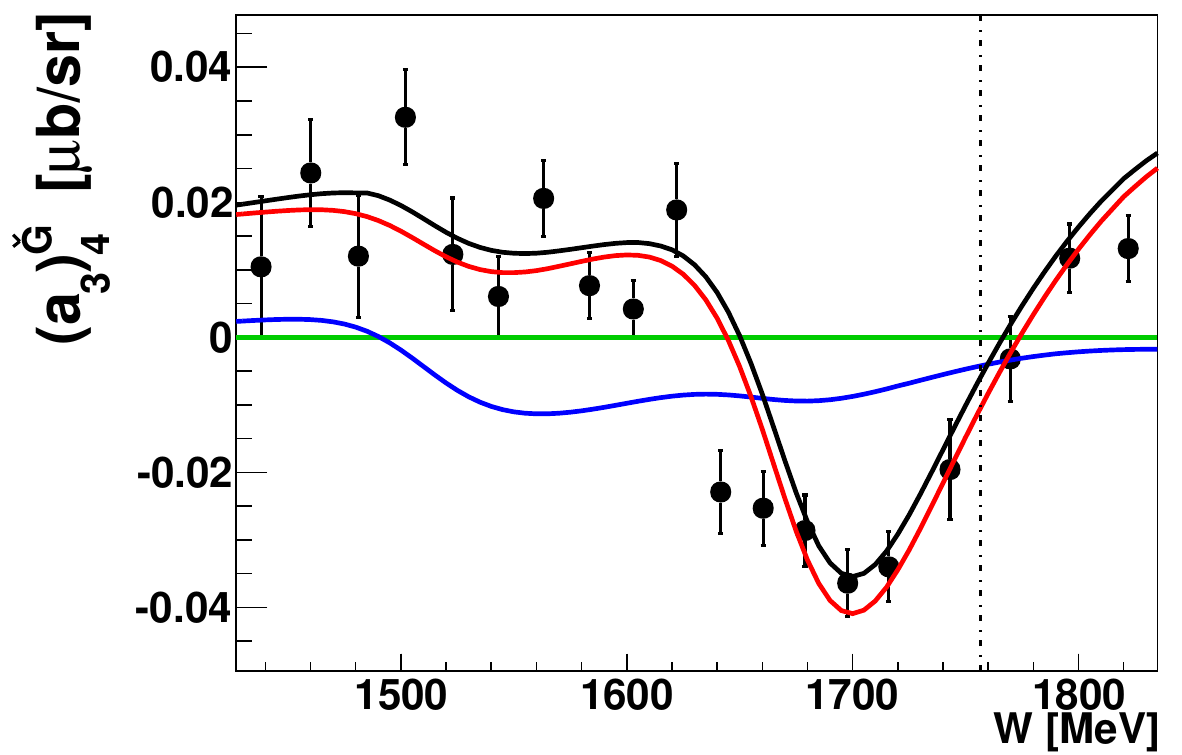}\\
  \hspace*{-19.5pt}\includegraphics[width=0.285\textwidth]{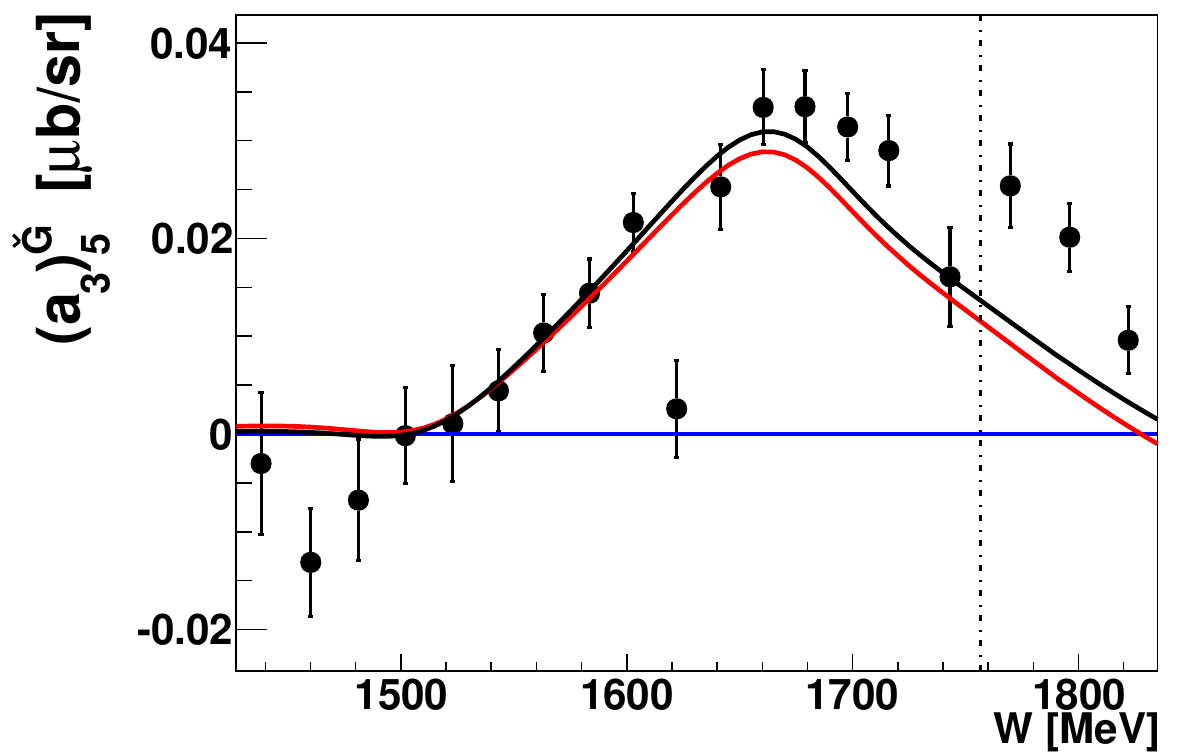}
  \includegraphics[width=0.285\textwidth]{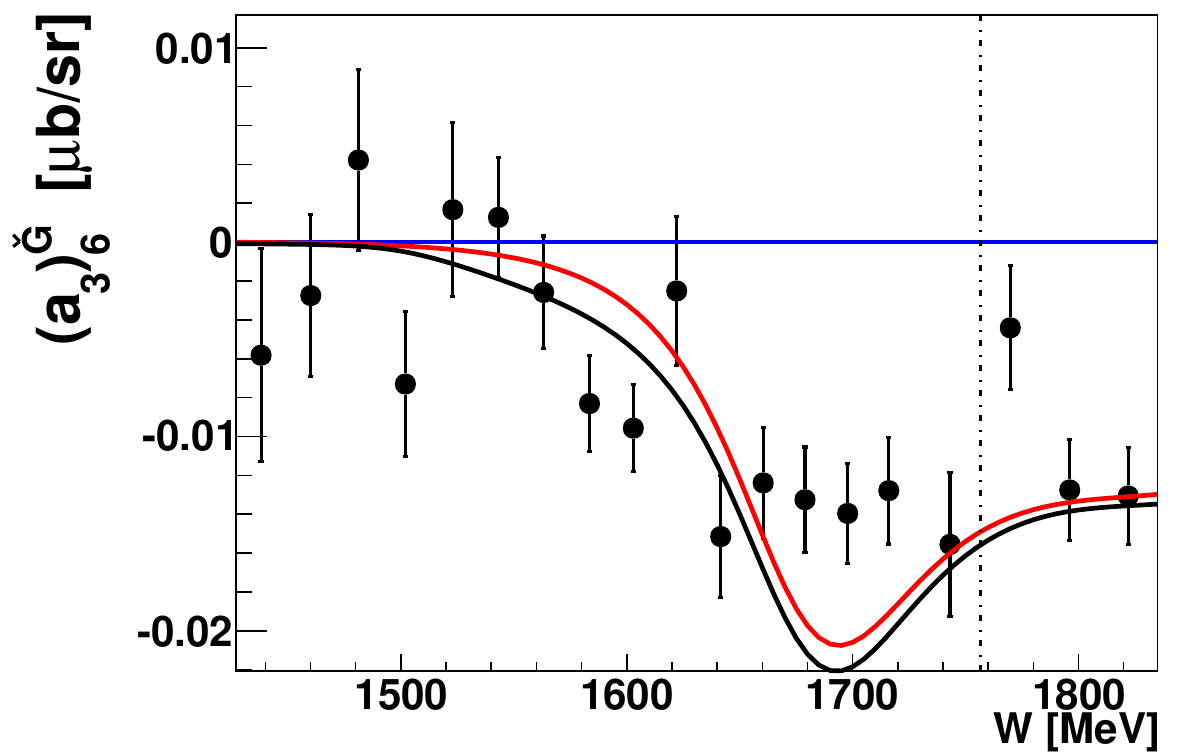}
  \end{minipage}
\end{figure*}

\begin{figure*}
\begin{minipage}{\textwidth}
\floatbox[{\capbeside\thisfloatsetup{capbesideposition={right,top},capbesidewidth=7.8cm}}]{figure}[\FBwidth]
{\caption{The recent new double po\-la\-ri\-za\-tion observable $\check{H}$ data from ELSA \cite{Hartmann:2014,Hartmann:2015} with only statistical error was fitted using associated Legendre polynomials according to eq. \ref{eq:LowEAssocLegParametrizationH} and truncating the partial wave expansion at $\text{L}_{\text{max}}=1\dots 4$. (a) The resulting $\chi^2/$ndf values of the different $\text{L}_{\text{max}}$-fits as a function of the center of mass energy W are shown. (b) 6 out of 8 selected angular distributions of $\check{H}$ (black points) are plotted together with the different $\text{L}_{\text{max}}$ fits (solid lines) starting at W= 1491 MeV up to 1593 MeV. (c) Comparison of the fit coefficients for $\text{L}_{\text{max}}=2$ (black points), $\left(a_{2}\right)^{\check{H}}_{1\ldots4}$ (see eq. \ref{eq:LowEAssocLegParametrizationH}), with the BnGa2014-02 solution truncated at different $\text{L}_{\text{max}}$ (solid lines). Colors same as in (a).}\label{fig:h_bins}}
{\includegraphics[width=0.49\textwidth, trim=0cm 0cm 1.8cm 0cm, clip]{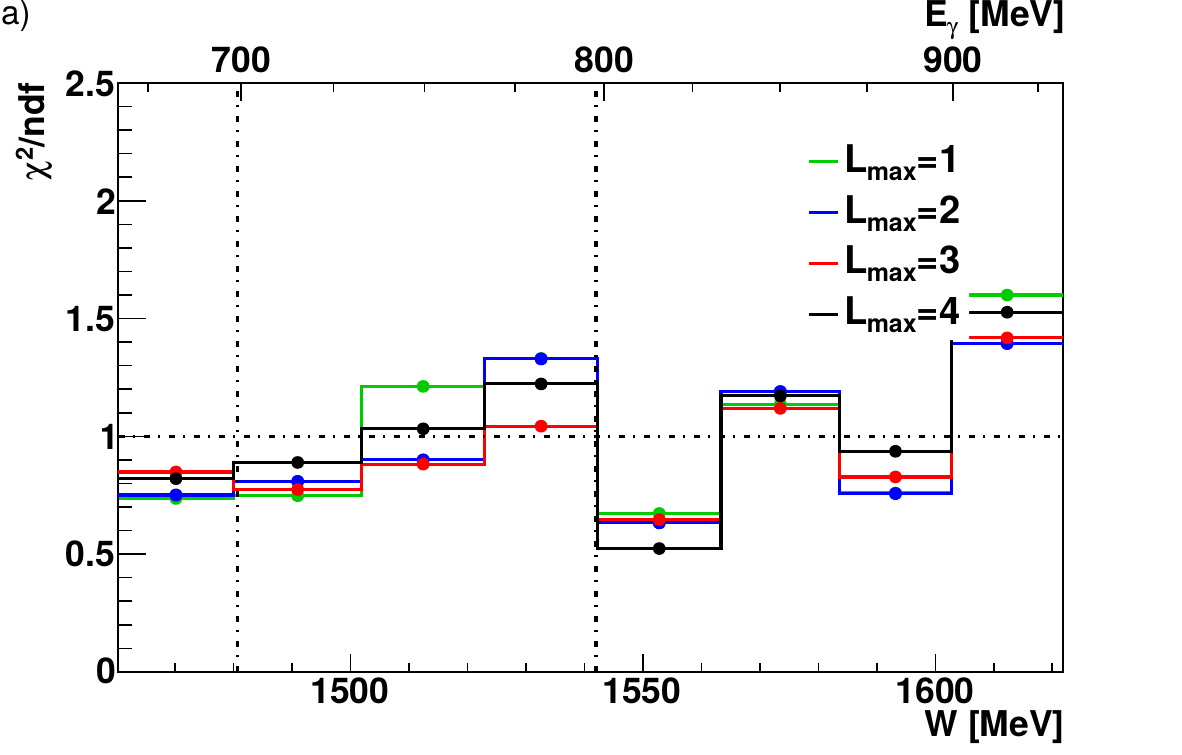}}
\end{minipage}\\

\begin{minipage}{\textwidth}
\centering
\hspace*{-0.45cm}
 \includegraphics[width=0.305\textwidth, trim=0cm 0cm 0.01cm 0.75cm, clip]{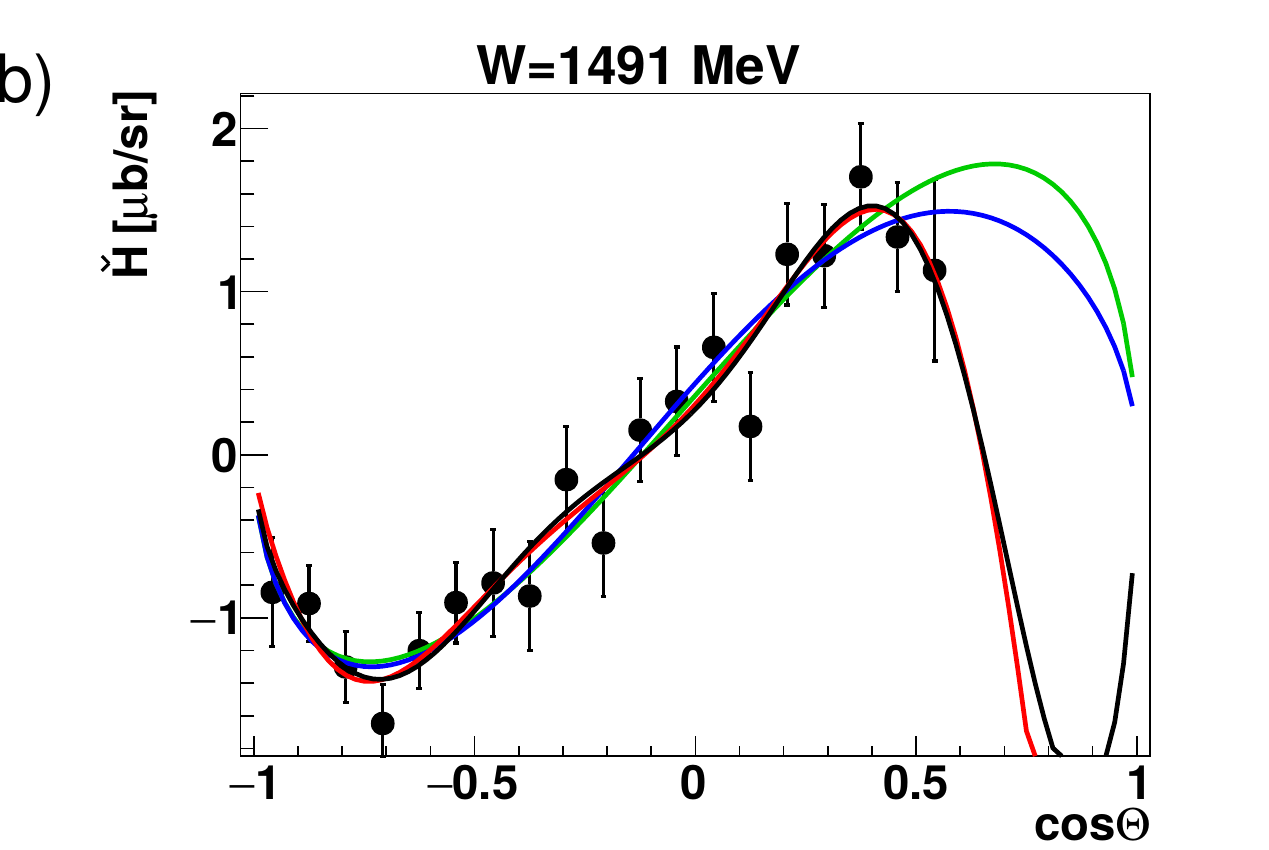}
  \includegraphics[width=0.285\textwidth, trim=0cm 0cm 0.01cm 0.75cm, clip]{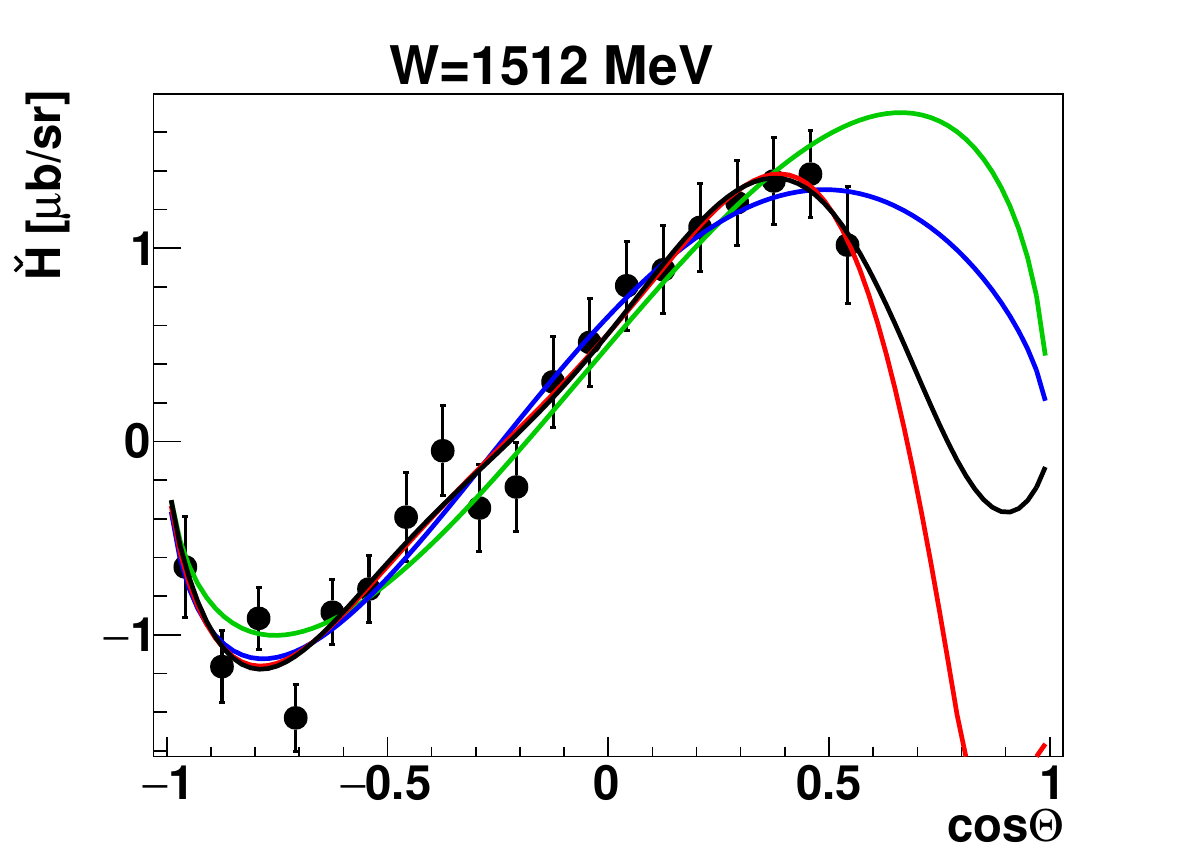}
  \includegraphics[width=0.285\textwidth, trim=0cm 0cm 0.01cm 0.75cm, clip]{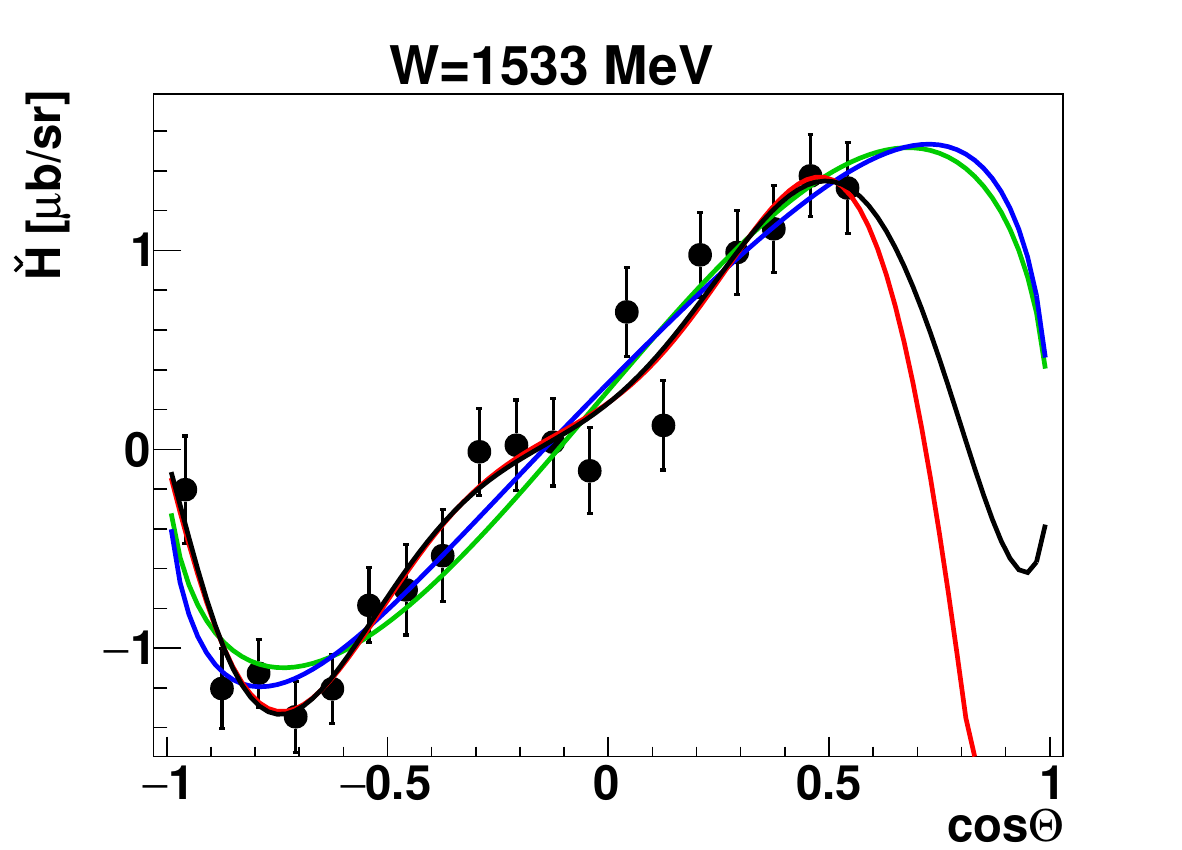}\\
  \includegraphics[width=0.285\textwidth, trim=0cm 0cm 0.01cm 0.75cm, clip]{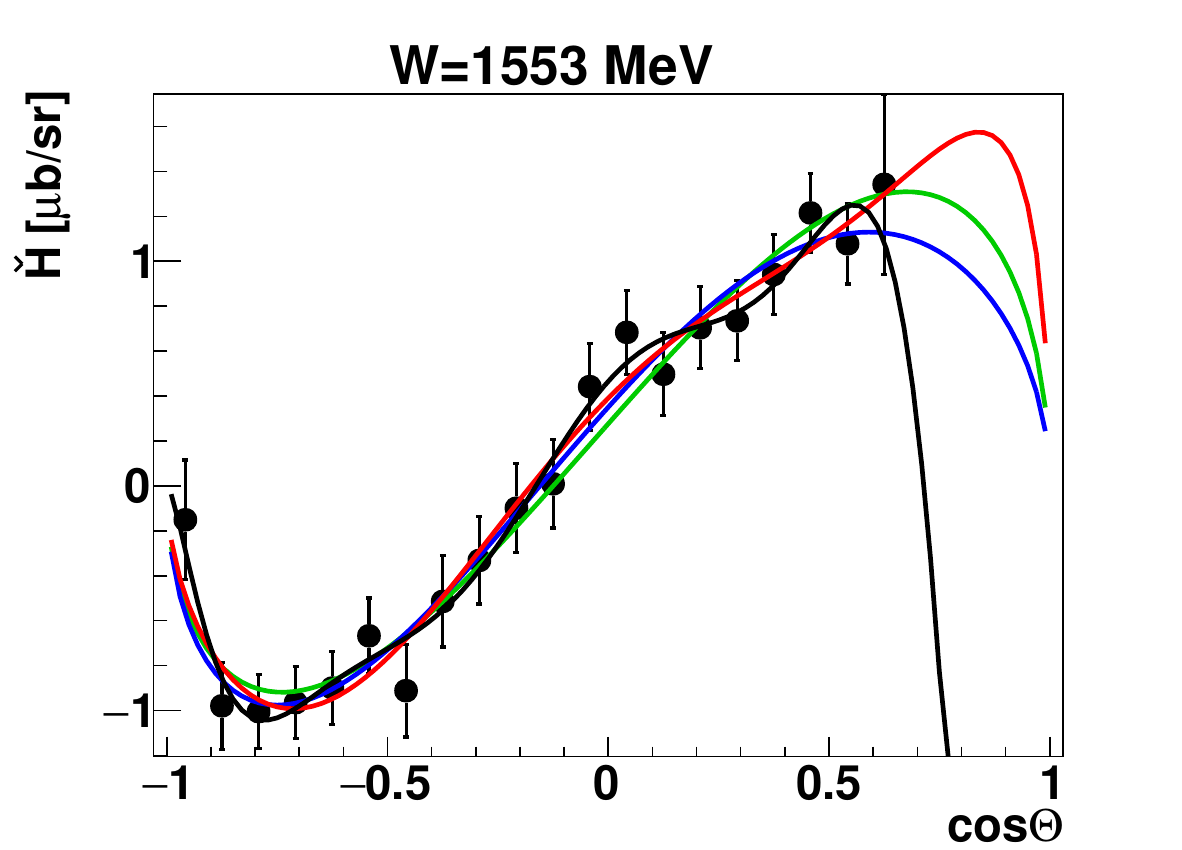}
  \includegraphics[width=0.285\textwidth, trim=0cm 0cm 0.01cm 0.75cm, clip]{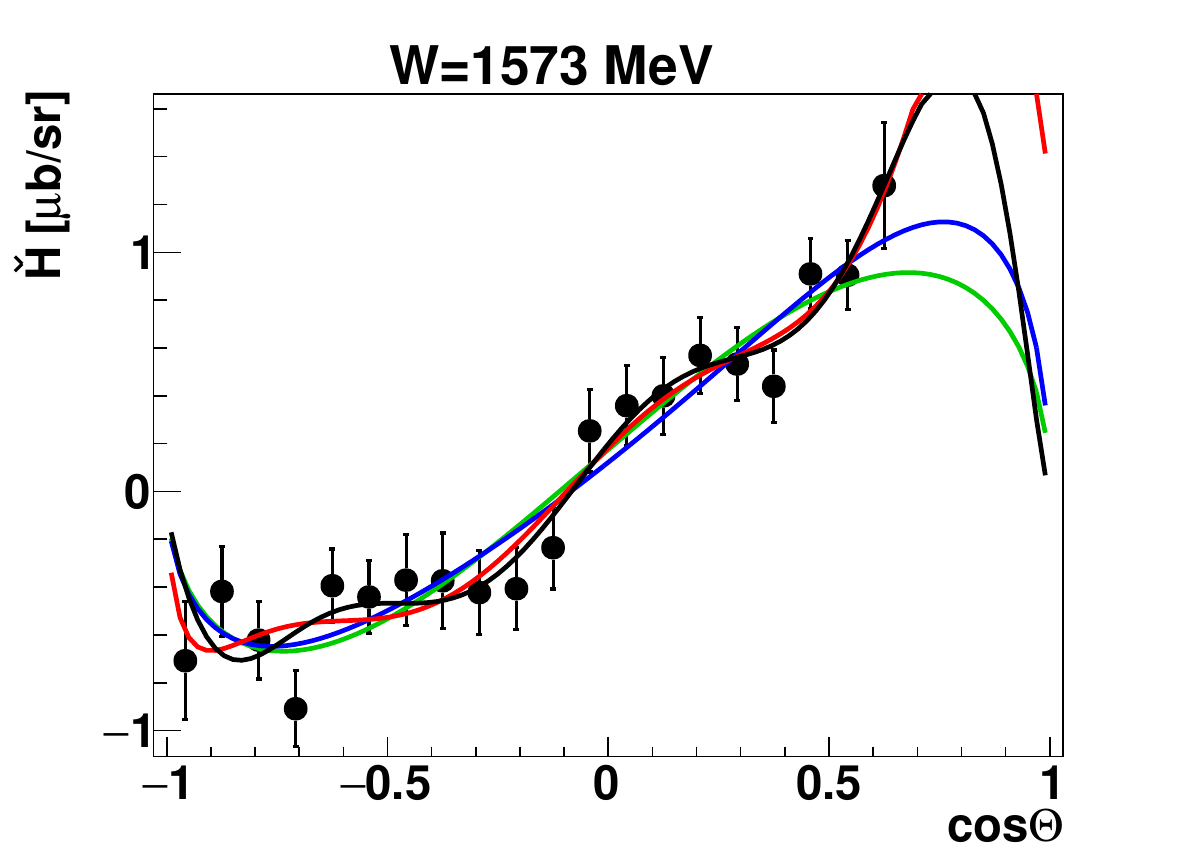}
  \includegraphics[width=0.285\textwidth, trim=0cm 0cm 0.01cm 0.75cm, clip]{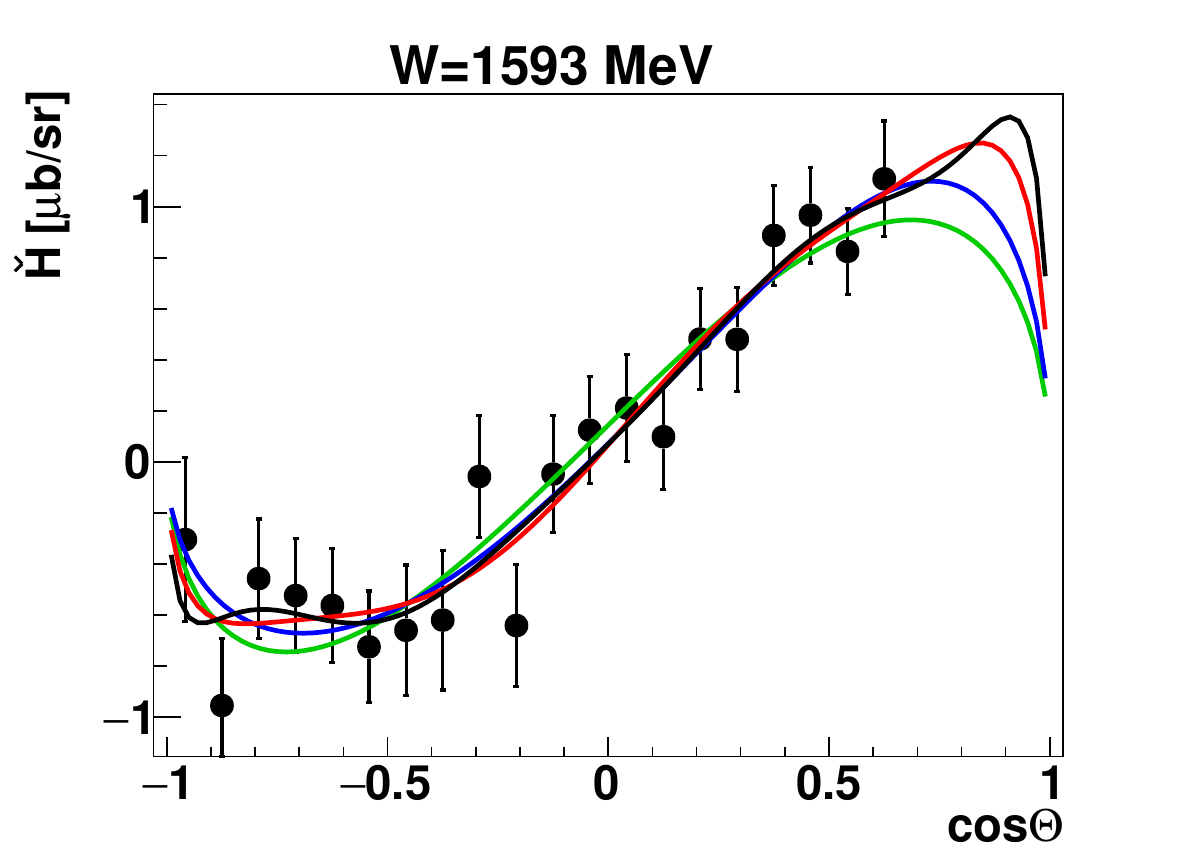}\\
 \vspace*{0.5cm}
  
  \hspace*{-23.5pt}\includegraphics[width=0.2905\textwidth]{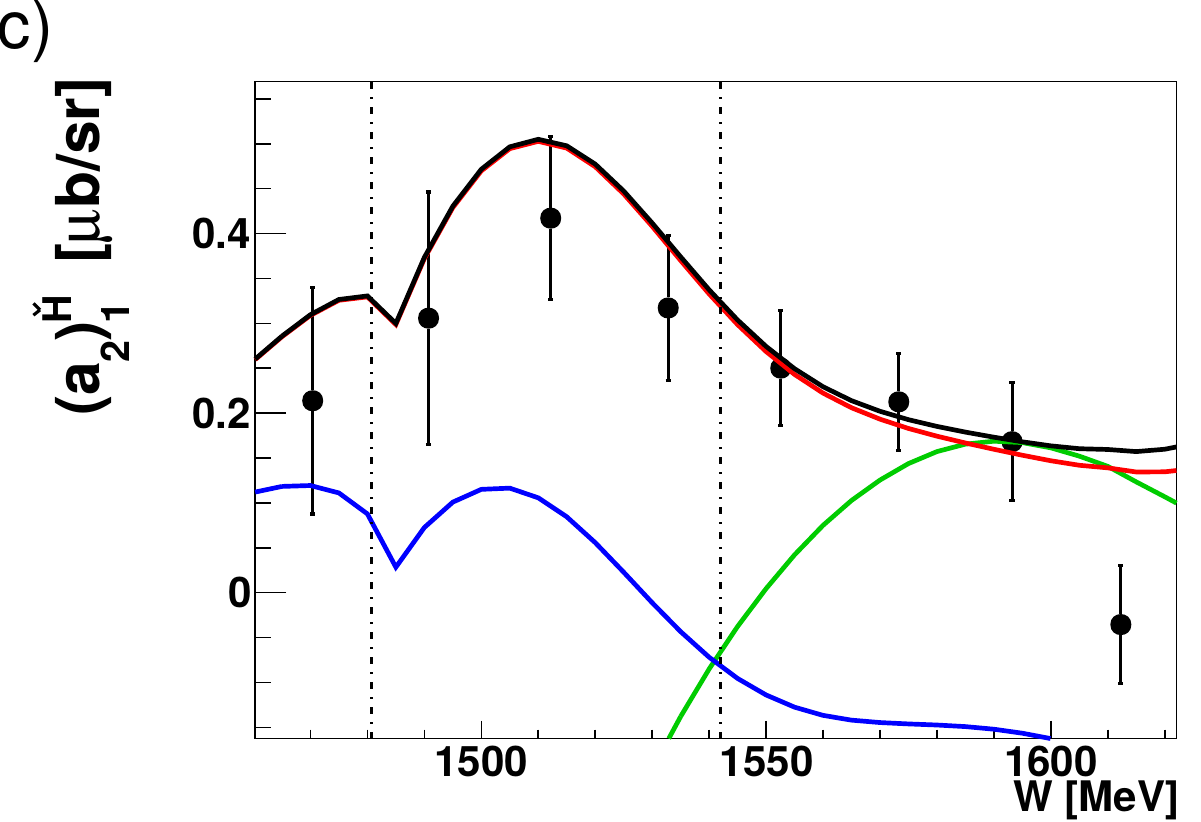}
  \includegraphics[width=0.285\textwidth]{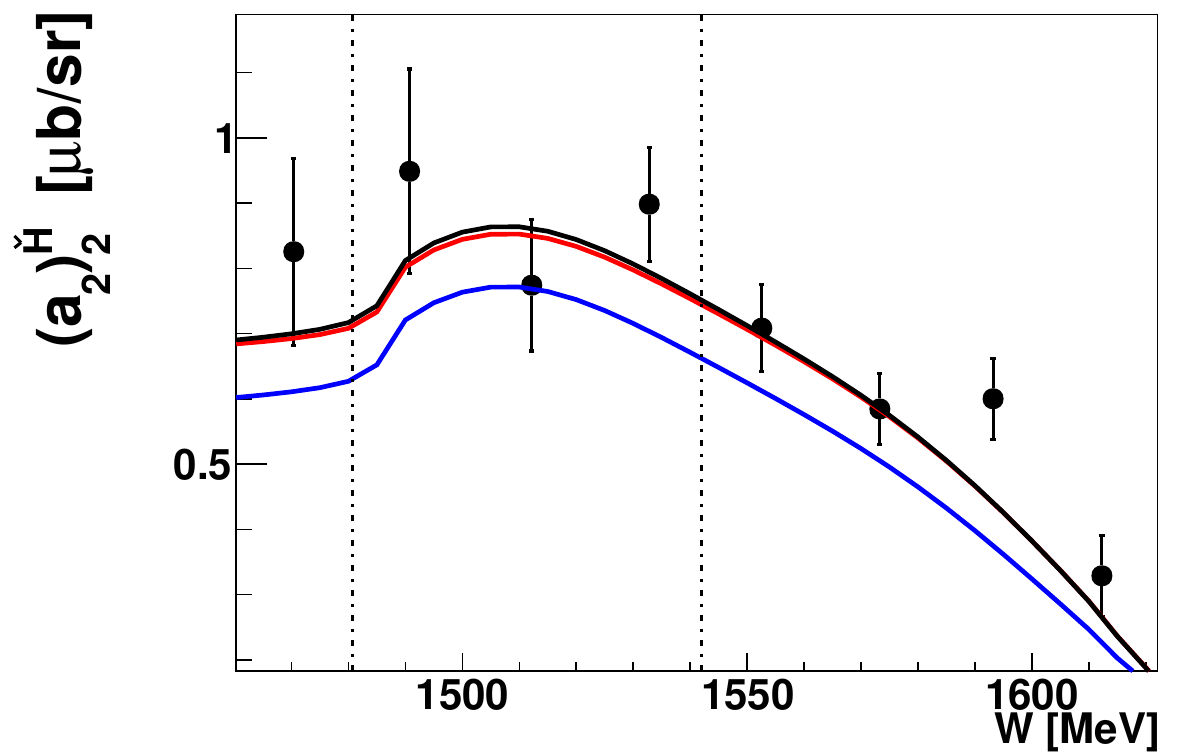}
  \includegraphics[width=0.285\textwidth]{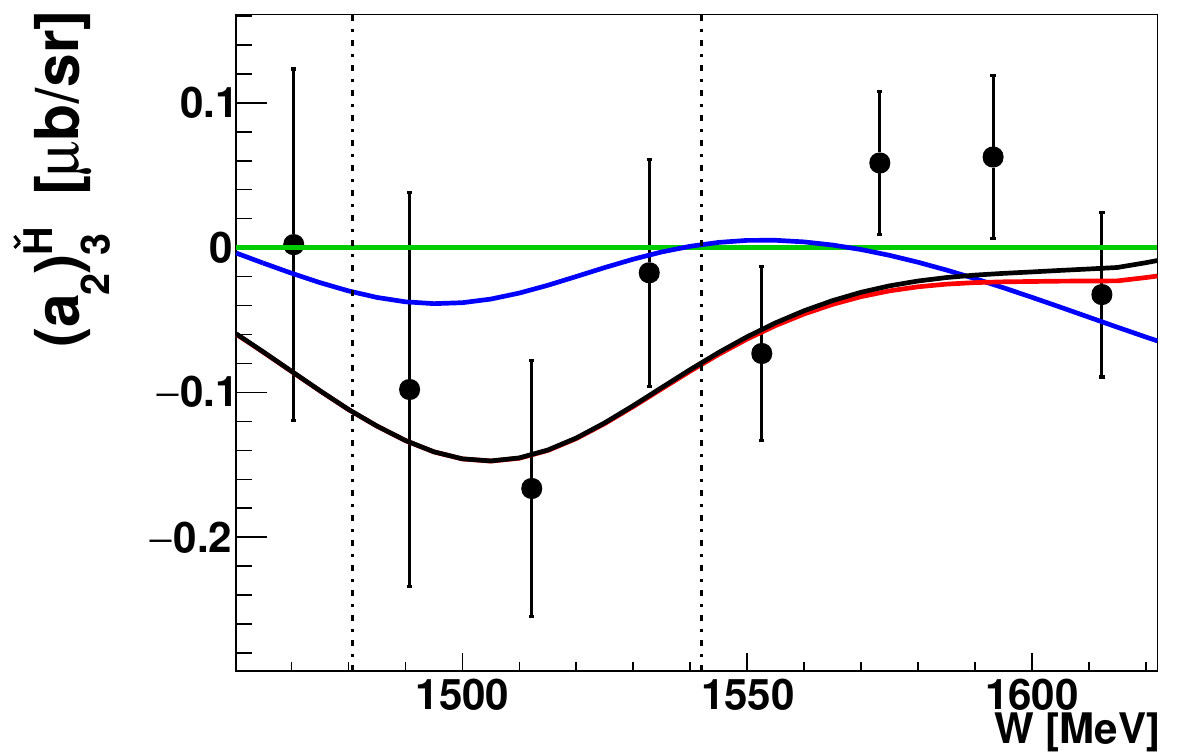} \\
  \hspace*{-19.5pt}\includegraphics[width=0.285\textwidth]{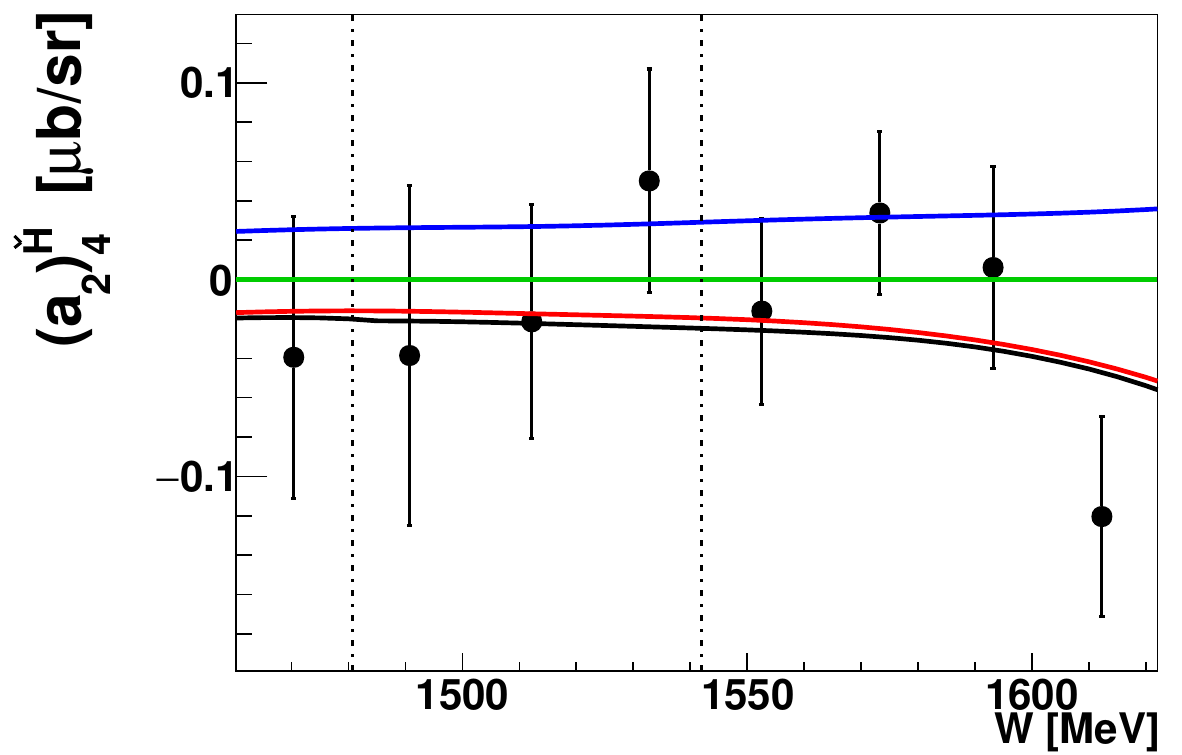}
  \end{minipage}
\end{figure*}

\begin{figure*}
\begin{minipage}{\textwidth}
\floatbox[{\capbeside\thisfloatsetup{capbesideposition={right,top},capbesidewidth=7.8cm}}]{figure}[\FBwidth]
{\caption{The recent new double po\-la\-ri\-za\-tion observable $\check{P}$ data from ELSA \cite{Hartmann:2014,Hartmann:2015} with only statistical error was fitted using associated Legendre polynomials according to eq. \ref{eq:LowEAssocLegParametrizationP} and truncating the partial wave expansion at $\text{L}_{\text{max}}=1\dots 4$. (a) The resulting $\chi^2/$ndf values of the different $\text{L}_{\text{max}}$-fits as a function of the center of mass energy W are shown. (b) 6 out of 8 selected angular distributions of $\check{P}$ (black points) are plotted together with the different $\text{L}_{\text{max}}$ fits (solid lines) starting at W= 1491 MeV up to 1612 MeV. (c) Comparison of the fit coefficients for $\text{L}_{\text{max}}=2$ (black points), $\left(a_{2}\right)^{\check{P}}_{1\dots 4}$ (see eq. \ref{eq:LowEAssocLegParametrizationP}), with the BnGa2014-02 solution truncated at different $\text{L}_{\text{max}}$ (solid lines). Colors same as in (a).}\label{fig:P_bins}}
{\includegraphics[width=0.49\textwidth, trim=0cm 0cm 1.8cm 0cm, clip]{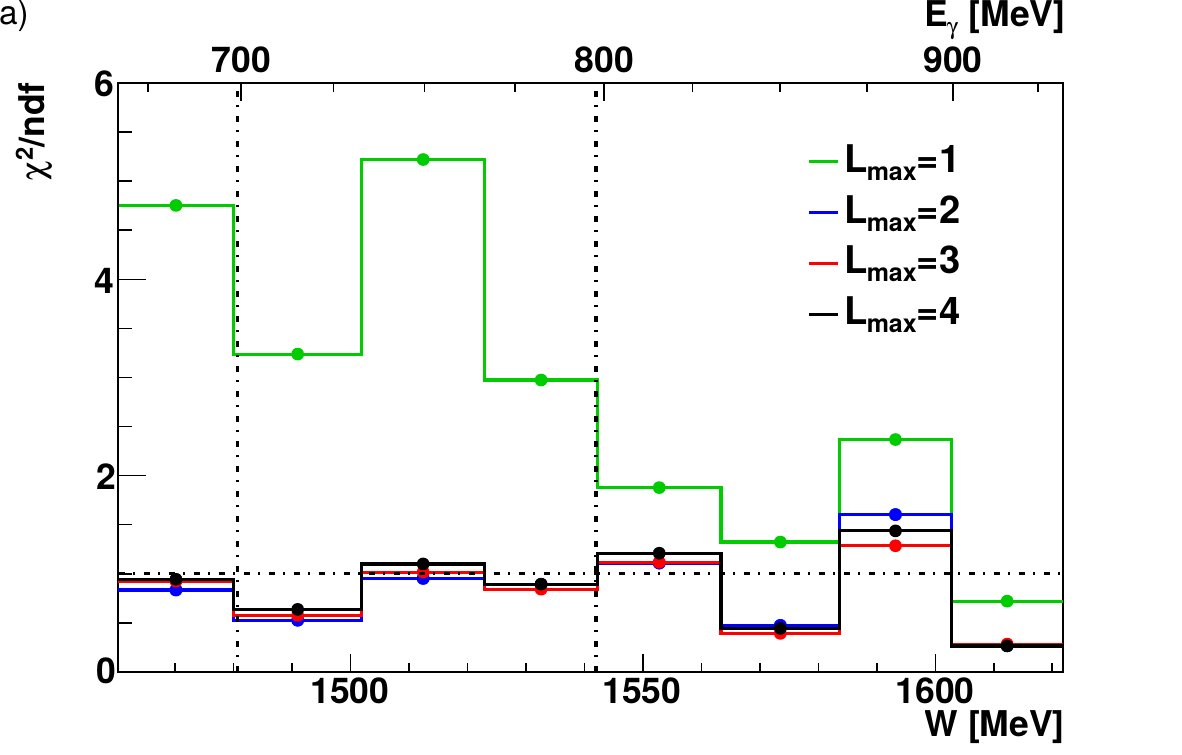}}
\end{minipage}\\

\begin{minipage}{\textwidth}
\centering
\hspace*{-0.45cm}
 \includegraphics[width=0.305\textwidth, trim=0cm 0cm 0.01cm 0.75cm, clip]{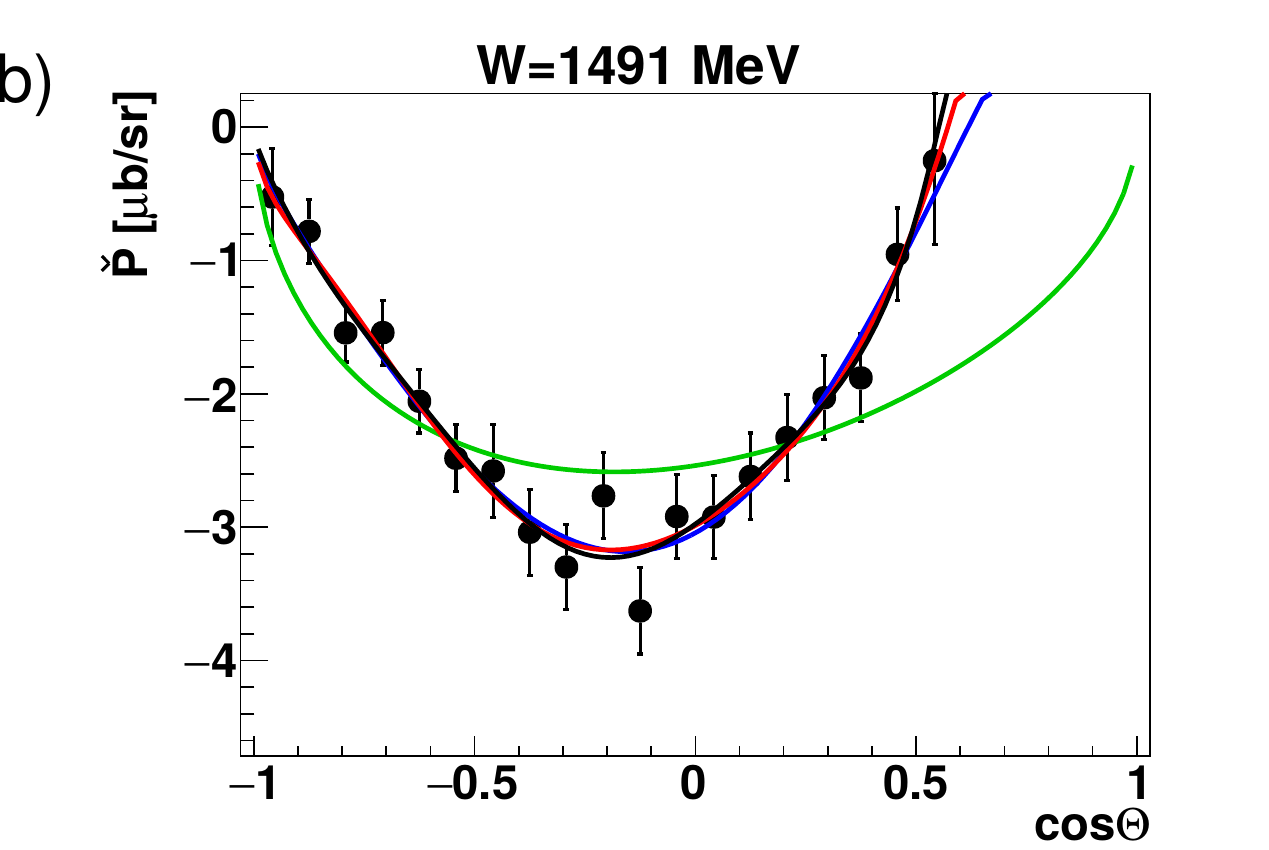}
  \includegraphics[width=0.285\textwidth, trim=0cm 0cm 0.01cm 0.75cm, clip]{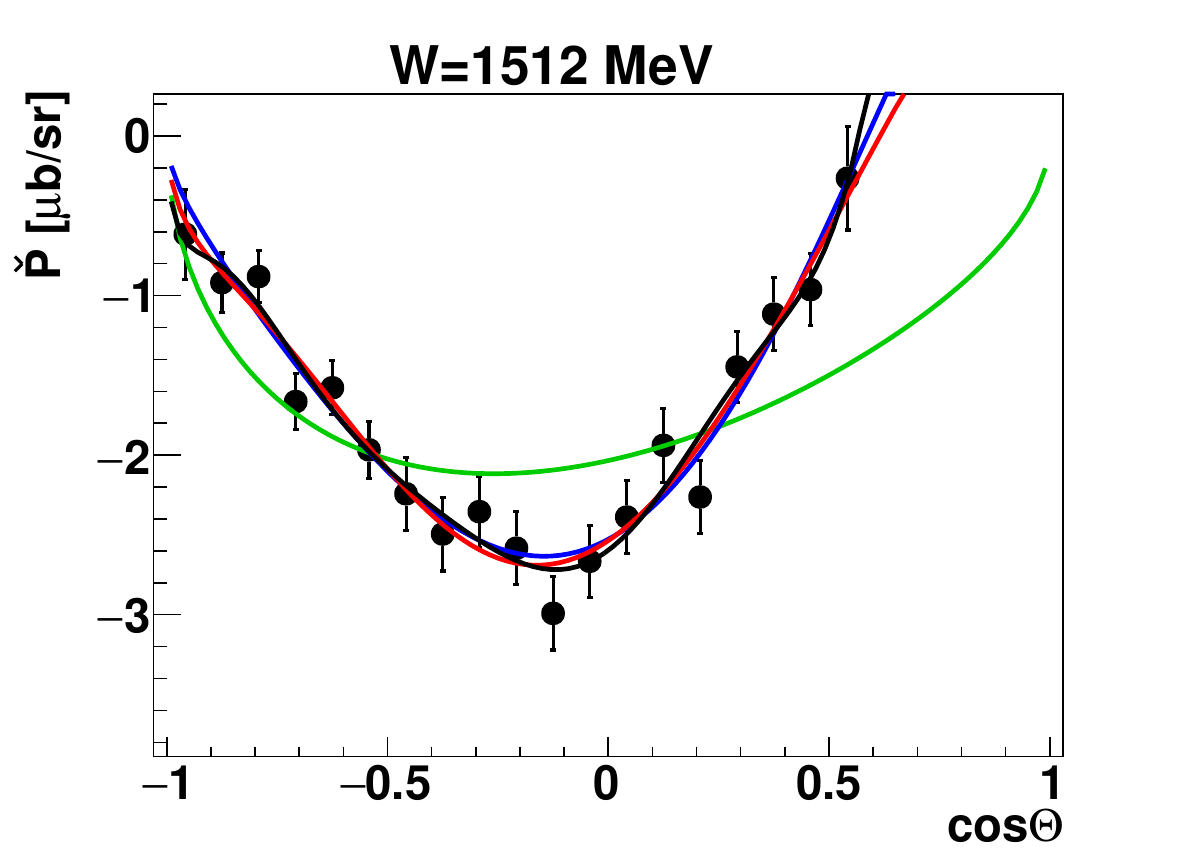}
  \includegraphics[width=0.285\textwidth, trim=0cm 0cm 0.01cm 0.75cm, clip]{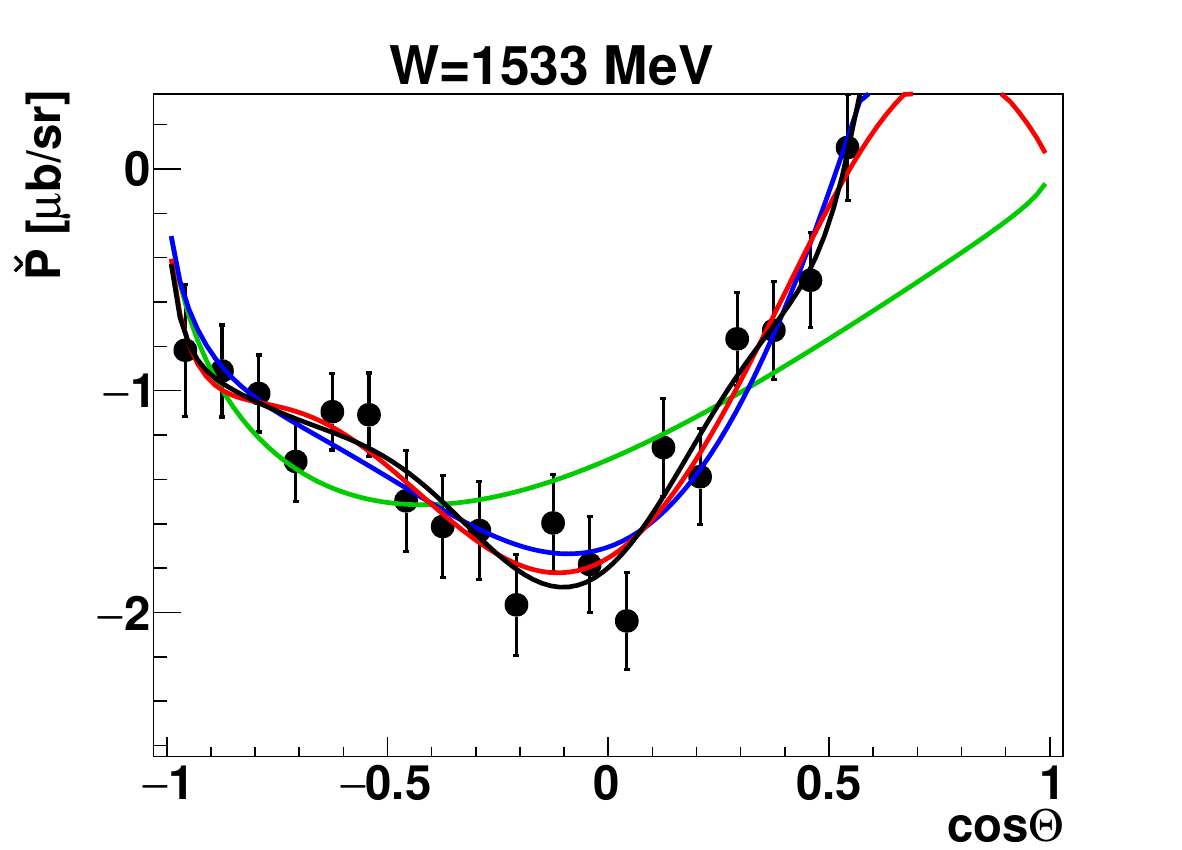}\\
  \includegraphics[width=0.285\textwidth, trim=0cm 0cm 0.01cm 0.75cm, clip]{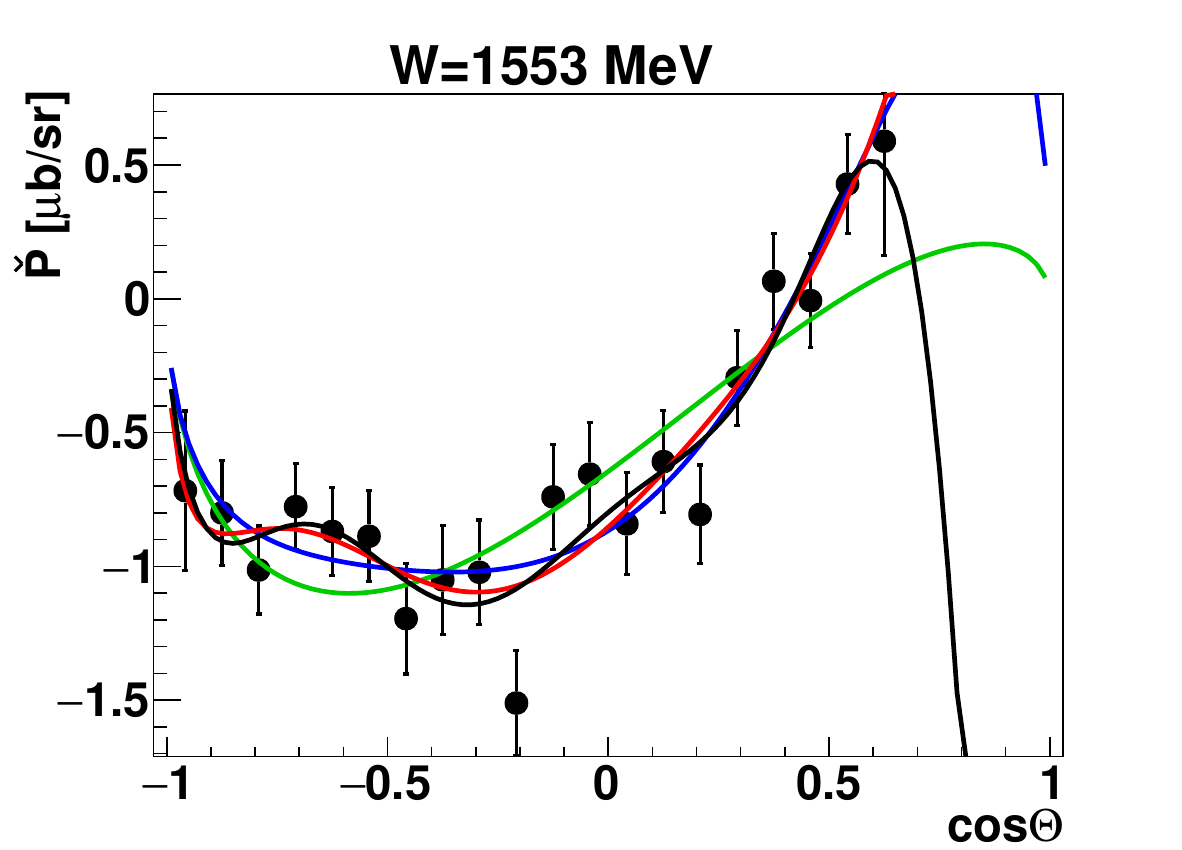}
  \includegraphics[width=0.285\textwidth, trim=0cm 0cm 0.01cm 0.75cm, clip]{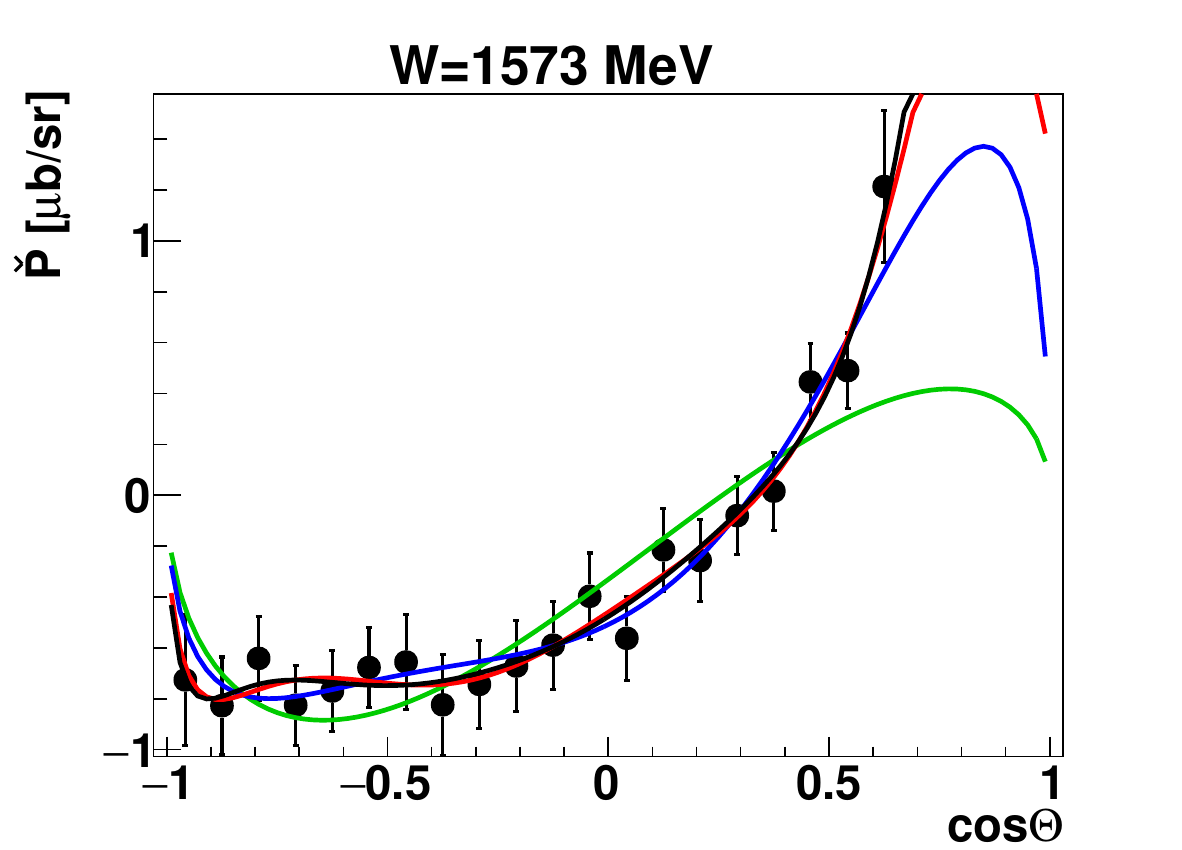}
  \includegraphics[width=0.285\textwidth, trim=0cm 0cm 0.01cm 0.75cm, clip]{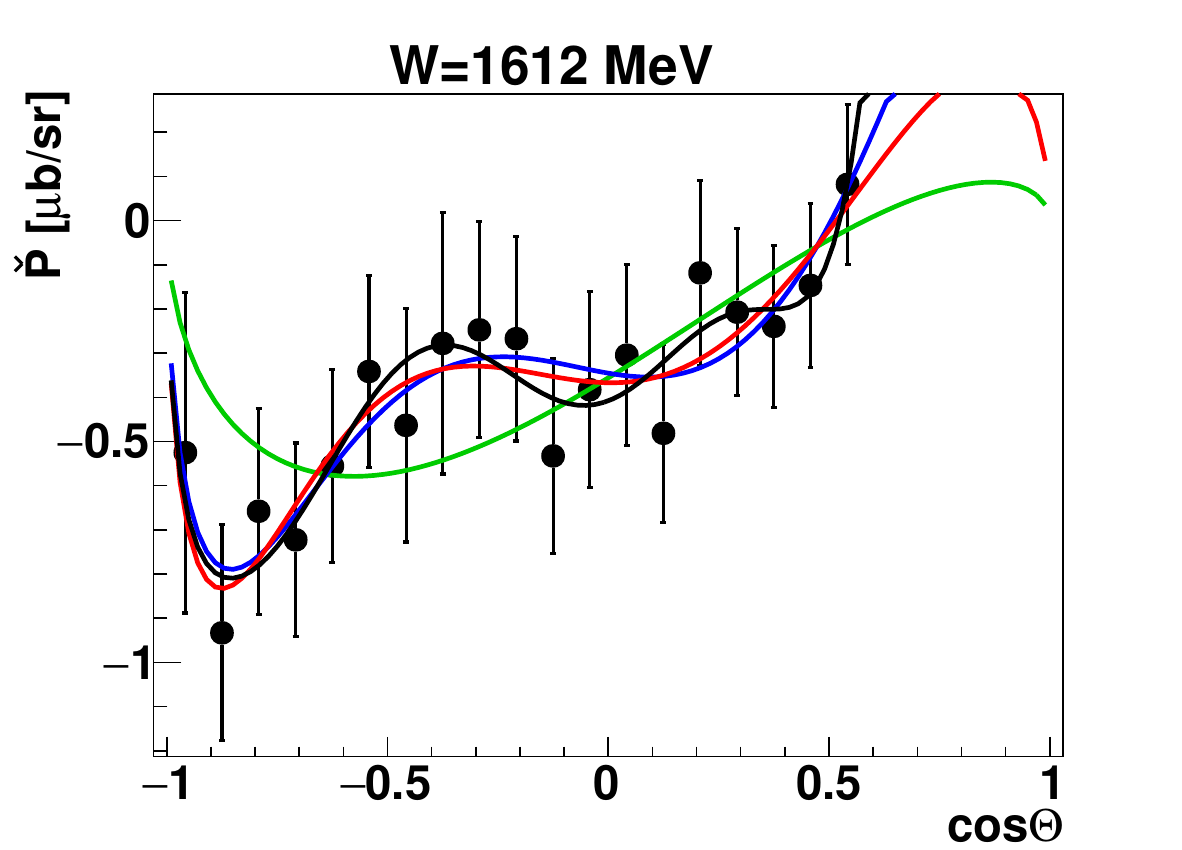}\\
 \vspace*{0.5cm}
  
  \hspace*{-23.5pt}\includegraphics[width=0.2905\textwidth]{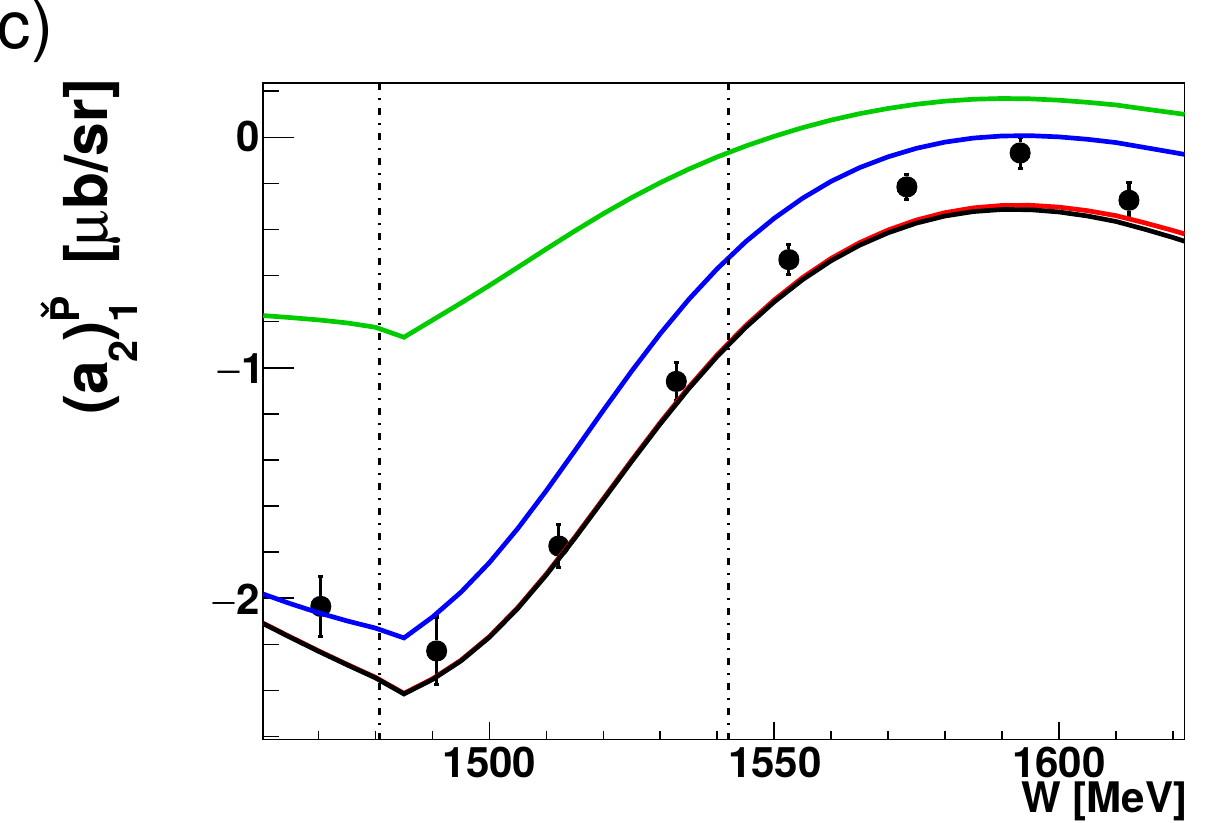}
  \includegraphics[width=0.285\textwidth]{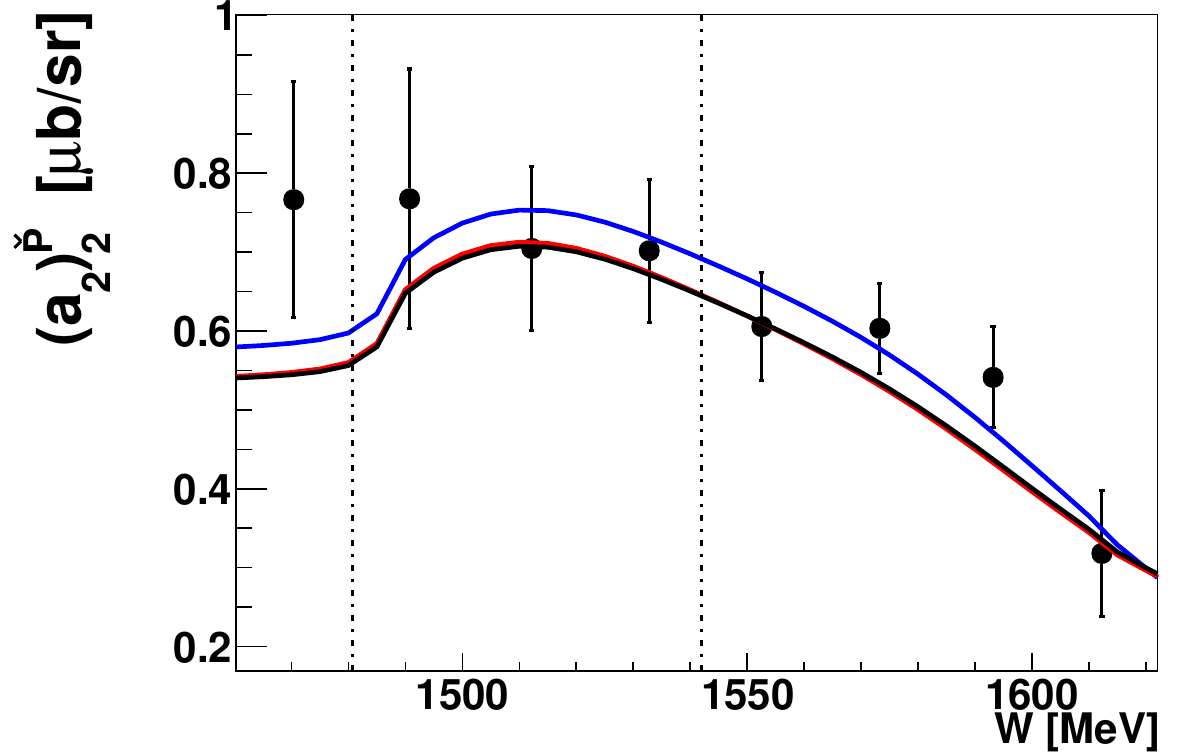}
  \includegraphics[width=0.285\textwidth]{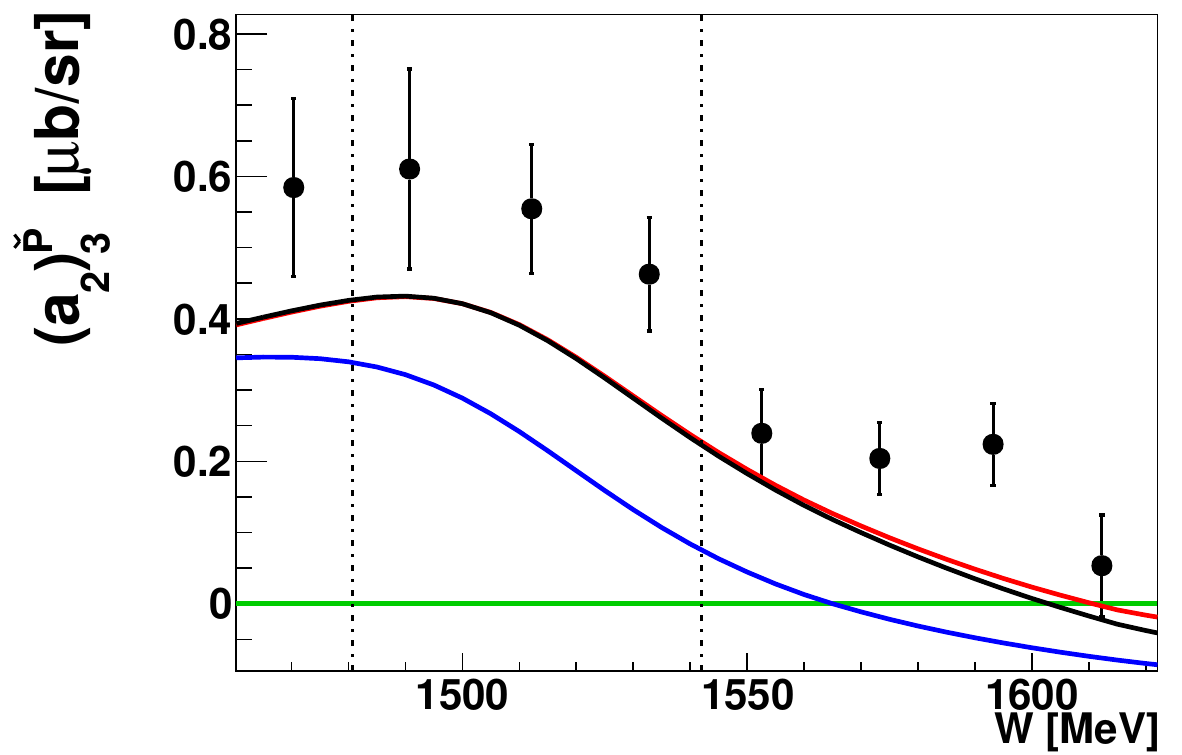}\\
  \hspace*{-19.5pt}\includegraphics[width=0.285\textwidth]{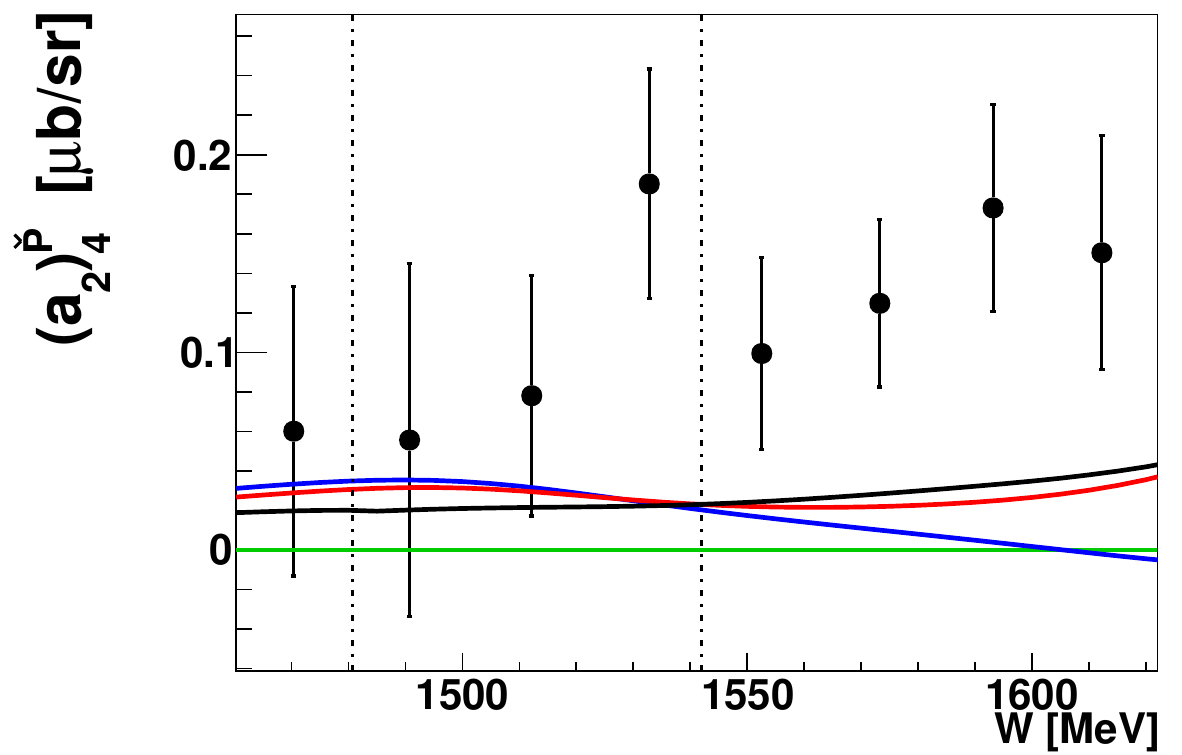}
  \end{minipage}
\end{figure*}

\begin{figure*}
\begin{minipage}{\textwidth}
\floatbox[{\capbeside\thisfloatsetup{capbesideposition={right,top},capbesidewidth=7.8cm}}]{figure}[\FBwidth]
{\caption{The recent new single po\-la\-ri\-za\-tion observable $\check{T}$ data from ELSA \cite{Hartmann:2014,Hartmann:2015} with only statistical error was fitted using associated Legendre polynomials according to eq. \ref{eq:LowEAssocLegParametrizationT} and truncating the partial wave expansion at $\text{L}_{\text{max}}=1\dots 4$. (a) The resulting $\chi^2/$ndf values of the different $\text{L}_{\text{max}}$-fits as a function of the center of mass energy W are shown. (b) 6 out of 24 selected angular distributions of $\check{T}$ (black points) are plotted together with the different $\text{L}_{\text{max}}$ fits (solid lines) starting at W= 1593 MeV up to 2085 MeV. (c) Comparison of the fit coefficients for $\text{L}_{\text{max}}=4$ (black points), $\left(a_{4}\right)^{\check{T}}_{1\dots 8}$ (see eq. \ref{eq:LowEAssocLegParametrizationT}), with the BnGa2014-02 solution truncated at different $\text{L}_{\text{max}}$ (solid lines). Colors same as in (a).}\label{fig:T_bins}}
{\includegraphics[width=0.49\textwidth, trim=0cm 0cm 1.8cm 0cm, clip]{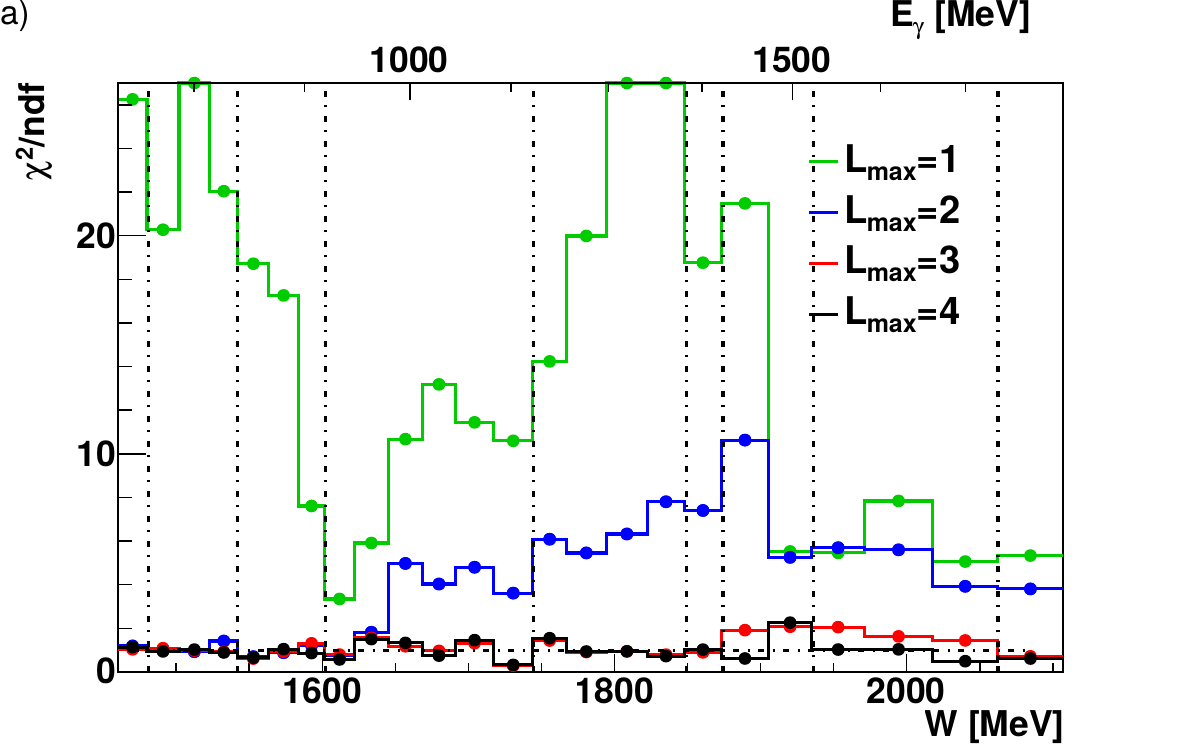}}
\end{minipage}\\

\begin{minipage}{\textwidth}
\centering
\hspace*{-0.45cm}
 \includegraphics[width=0.305\textwidth, trim=0cm 0cm 0.01cm 0.75cm, clip]{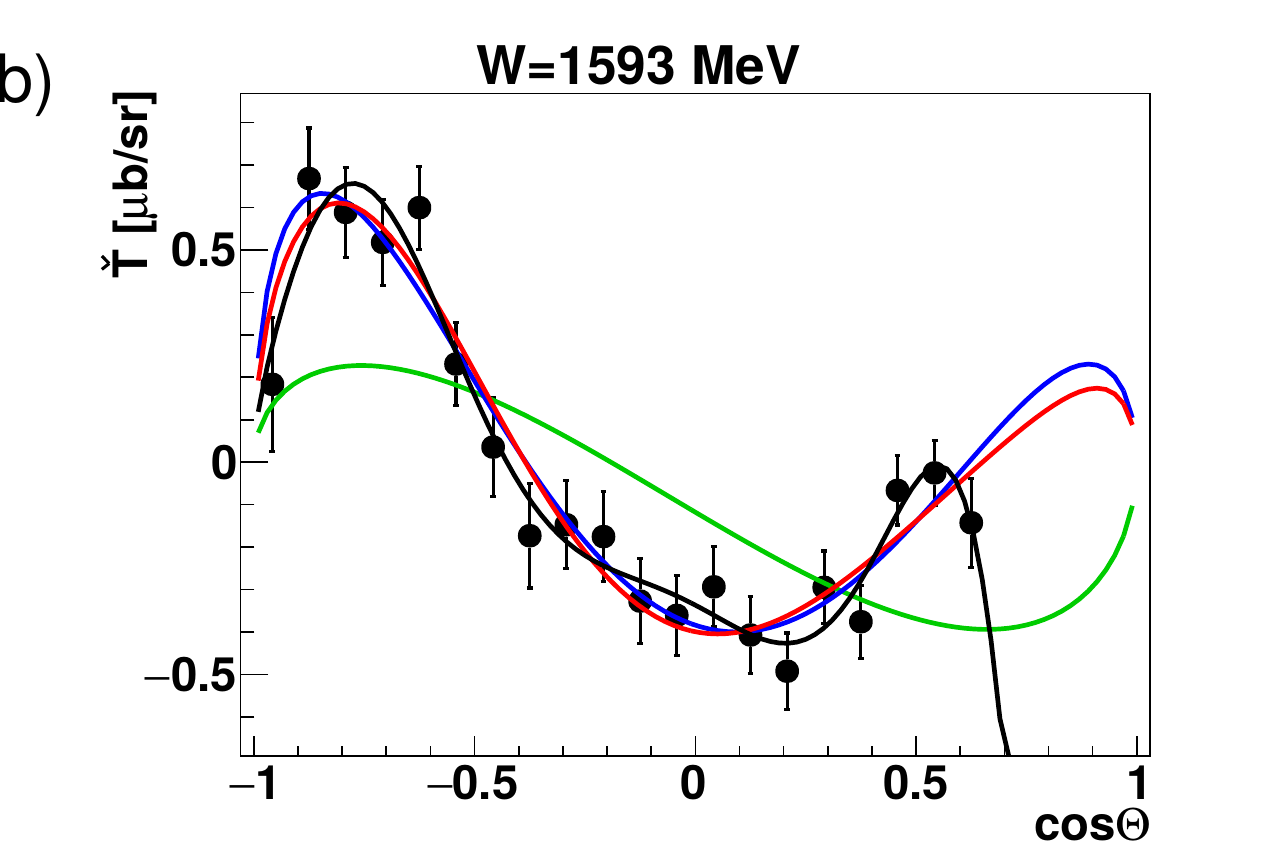}
  \includegraphics[width=0.285\textwidth, trim=0cm 0cm 0.01cm 0.75cm, clip]{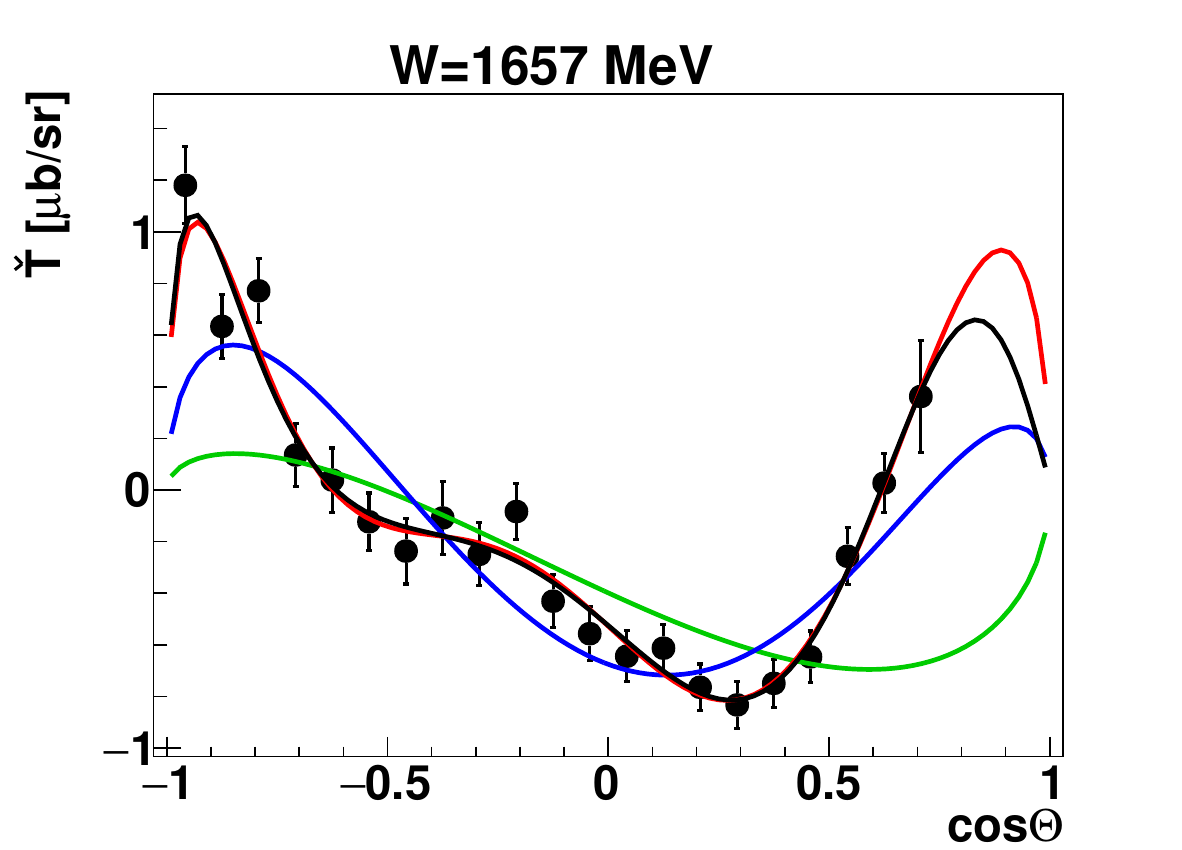}
  \includegraphics[width=0.285\textwidth, trim=0cm 0cm 0.01cm 0.75cm, clip]{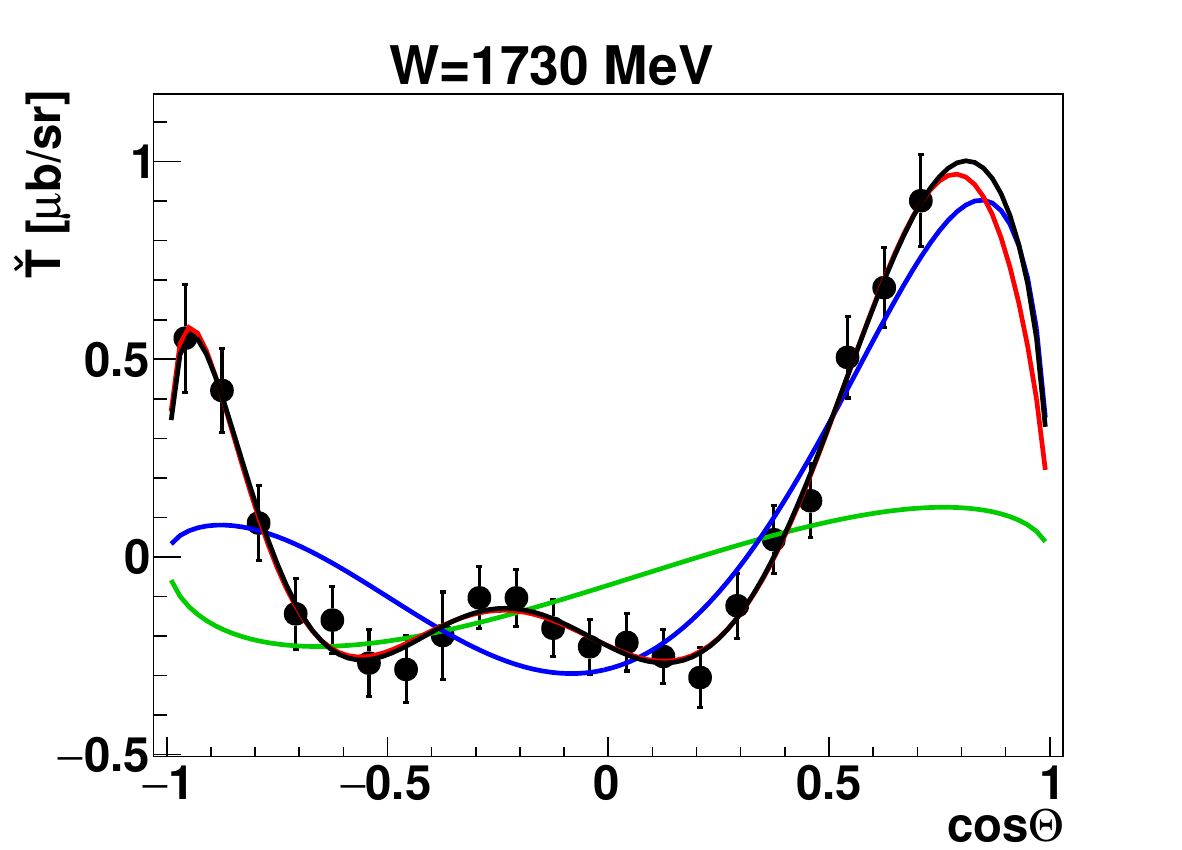}\\
  \includegraphics[width=0.285\textwidth, trim=0cm 0cm 0.01cm 0.75cm, clip]{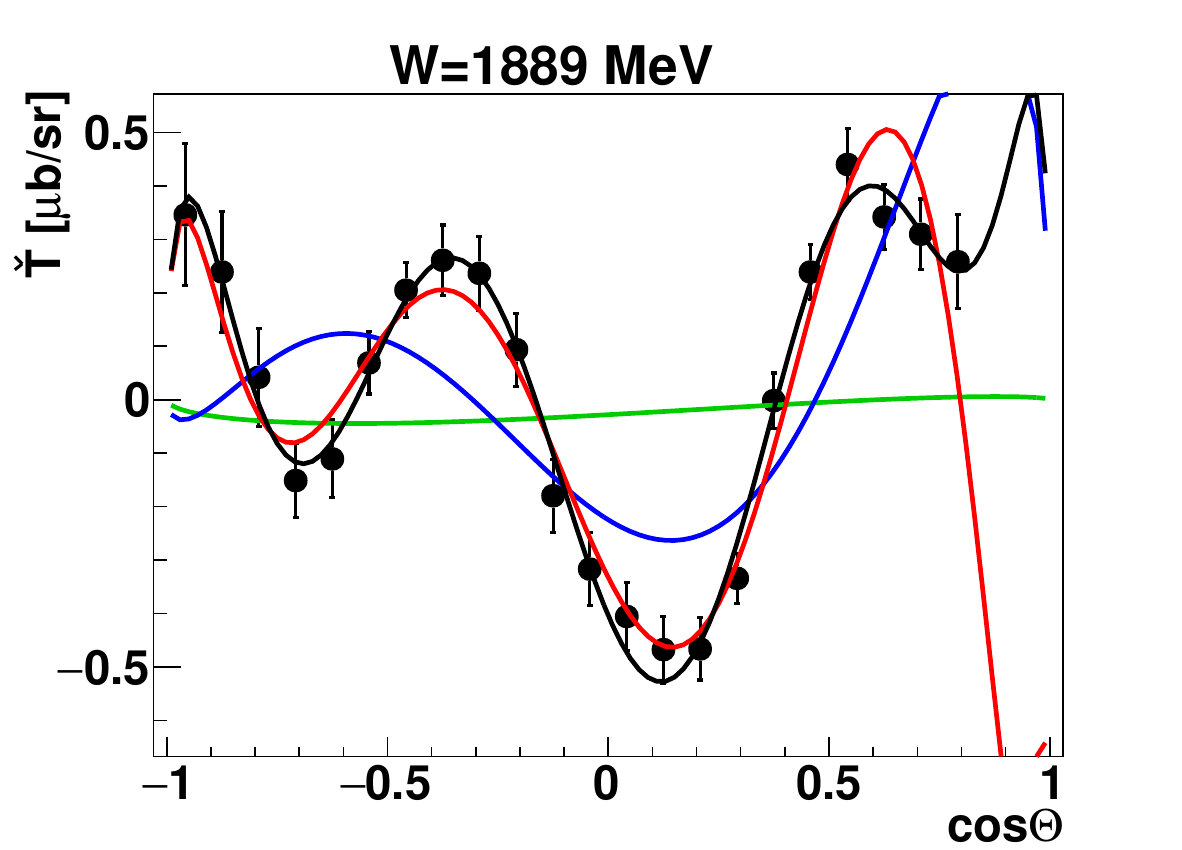}
  \includegraphics[width=0.285\textwidth, trim=0cm 0cm 0.01cm 0.75cm, clip]{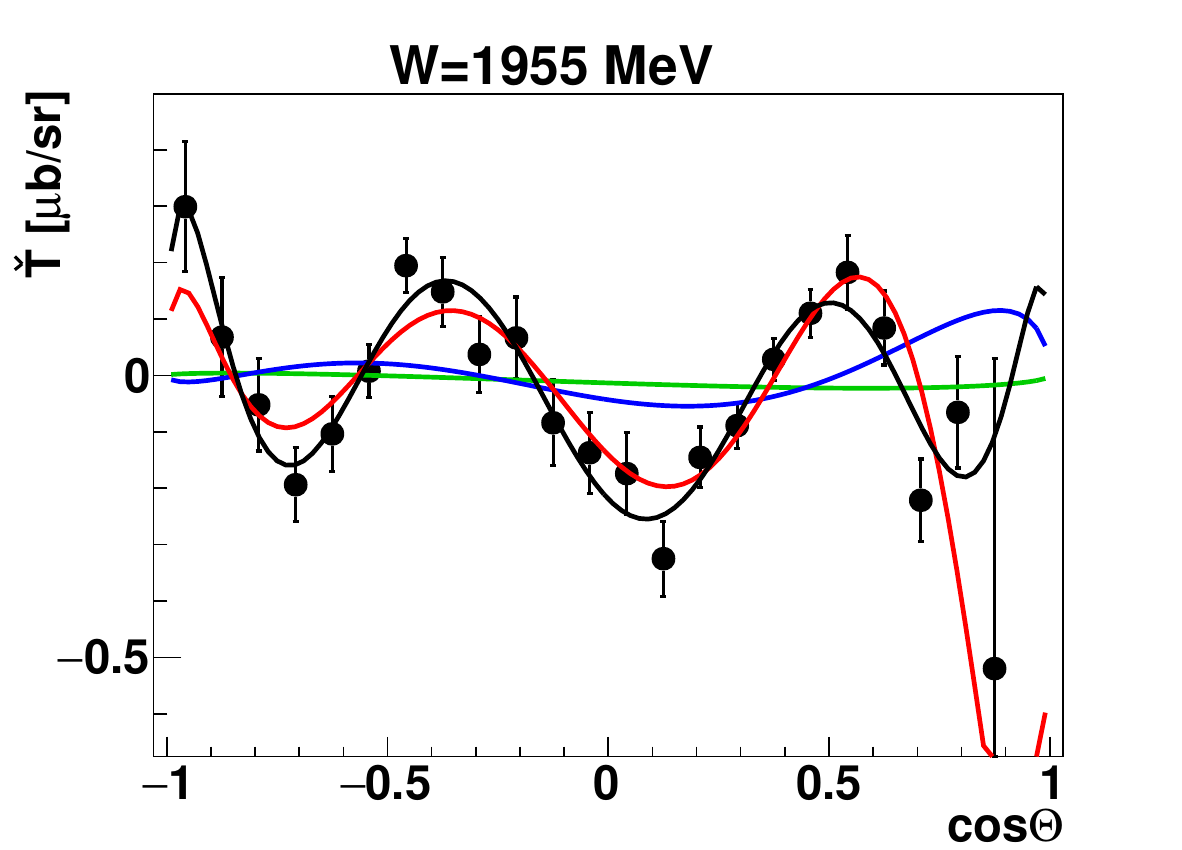}
  \includegraphics[width=0.285\textwidth, trim=0cm 0cm 0.01cm 0.75cm, clip]{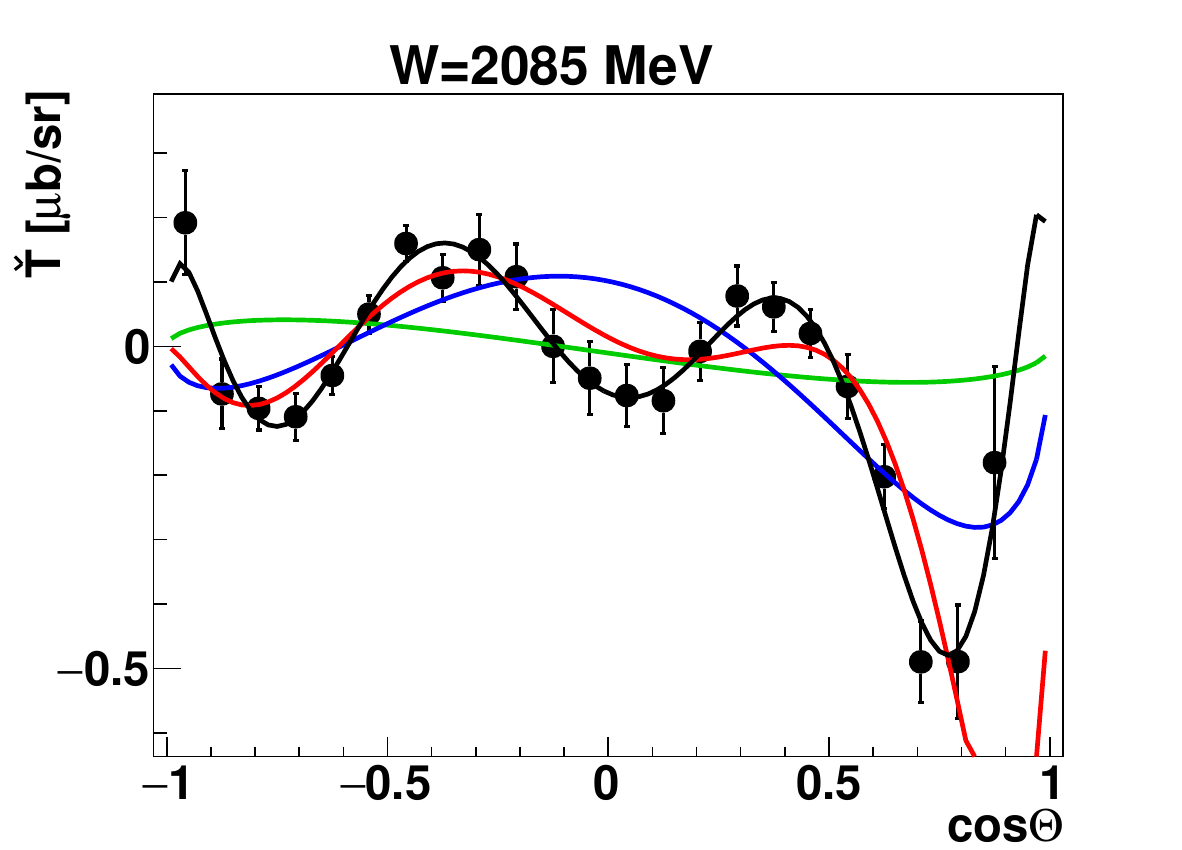}\\
 \vspace*{0.5cm}
  
  \hspace*{-23.5pt}\includegraphics[width=0.2905\textwidth]{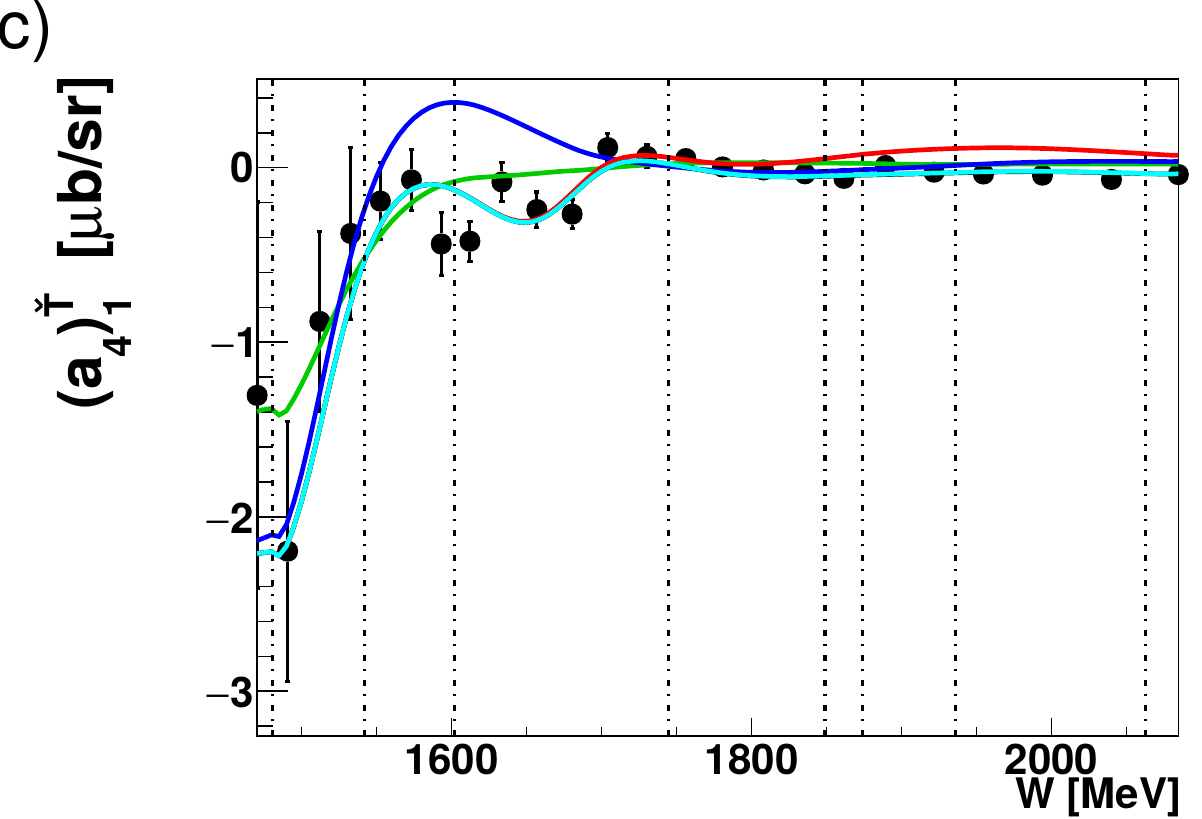}
  \includegraphics[width=0.285\textwidth]{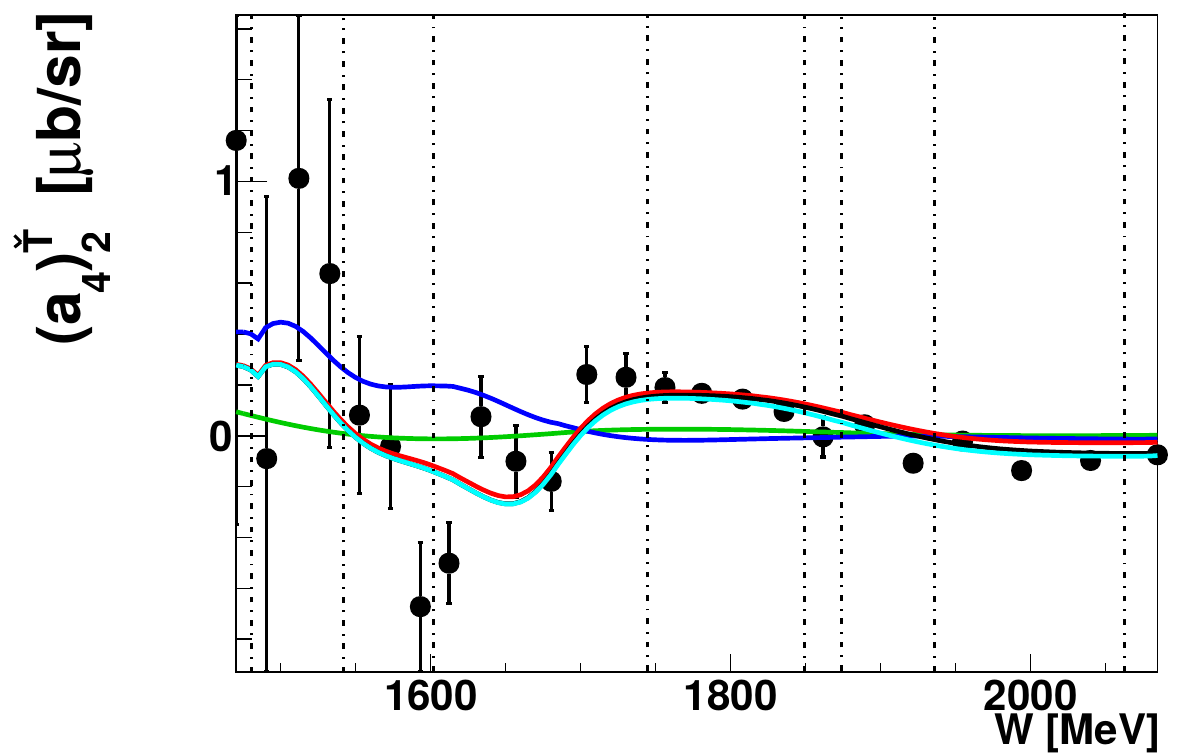}
  \includegraphics[width=0.285\textwidth]{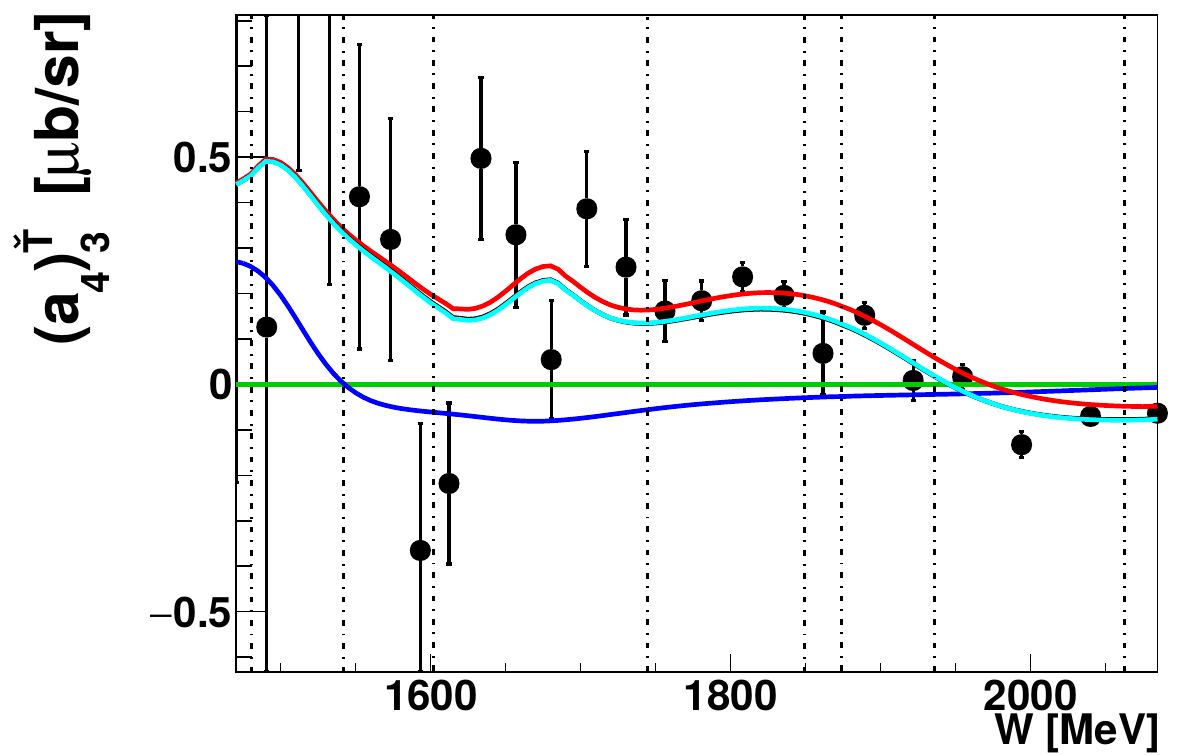}\\
  \hspace*{-19.5pt}\includegraphics[width=0.285\textwidth]{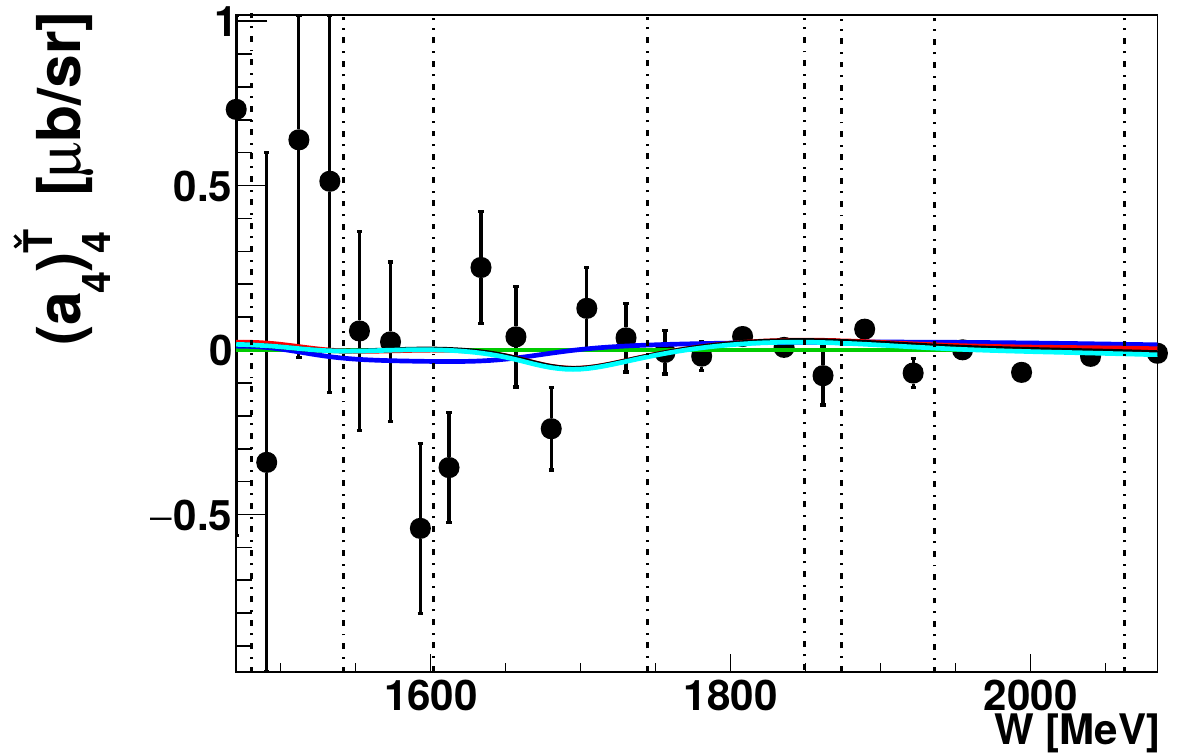}
  \includegraphics[width=0.285\textwidth]{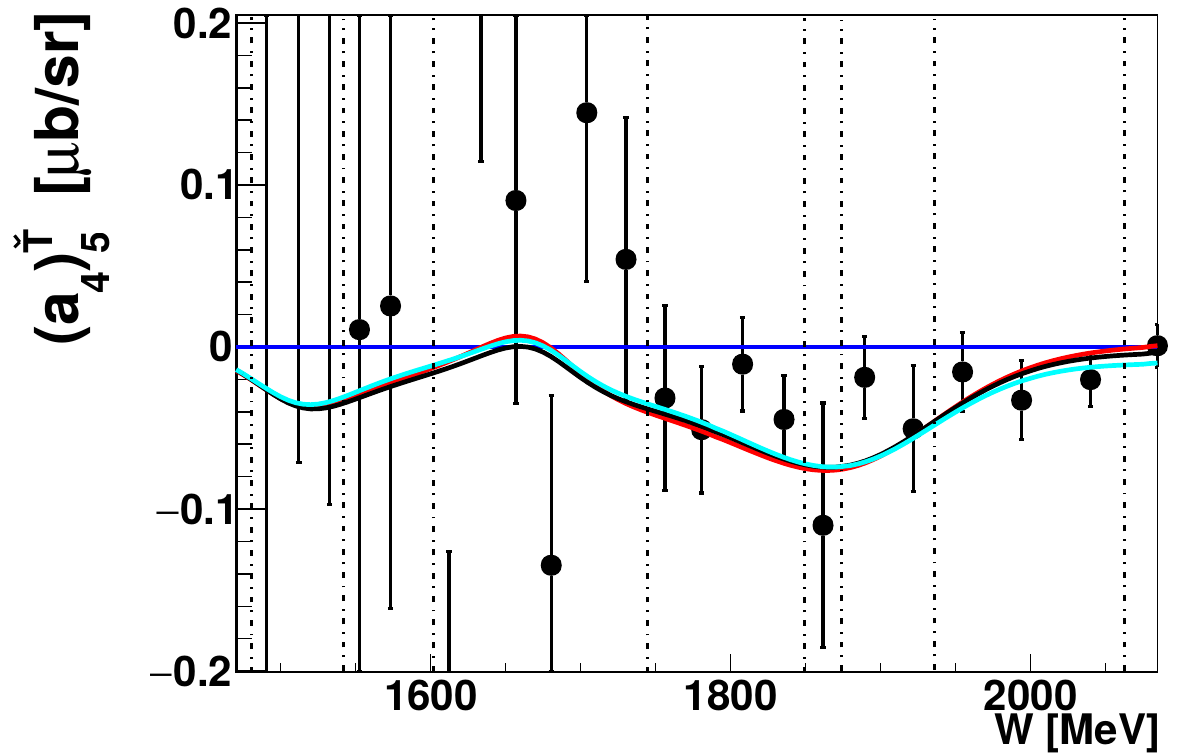}
  \includegraphics[width=0.285\textwidth]{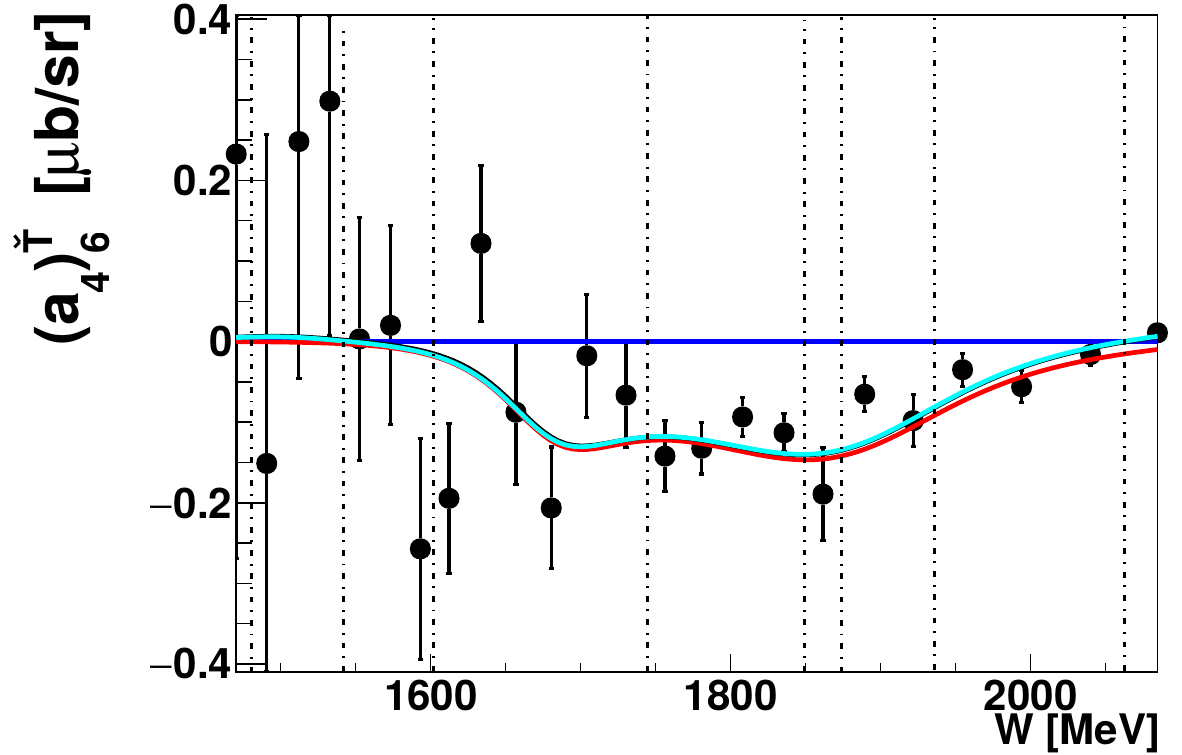} \\
  \hspace*{-19.5pt}\includegraphics[width=0.285\textwidth]{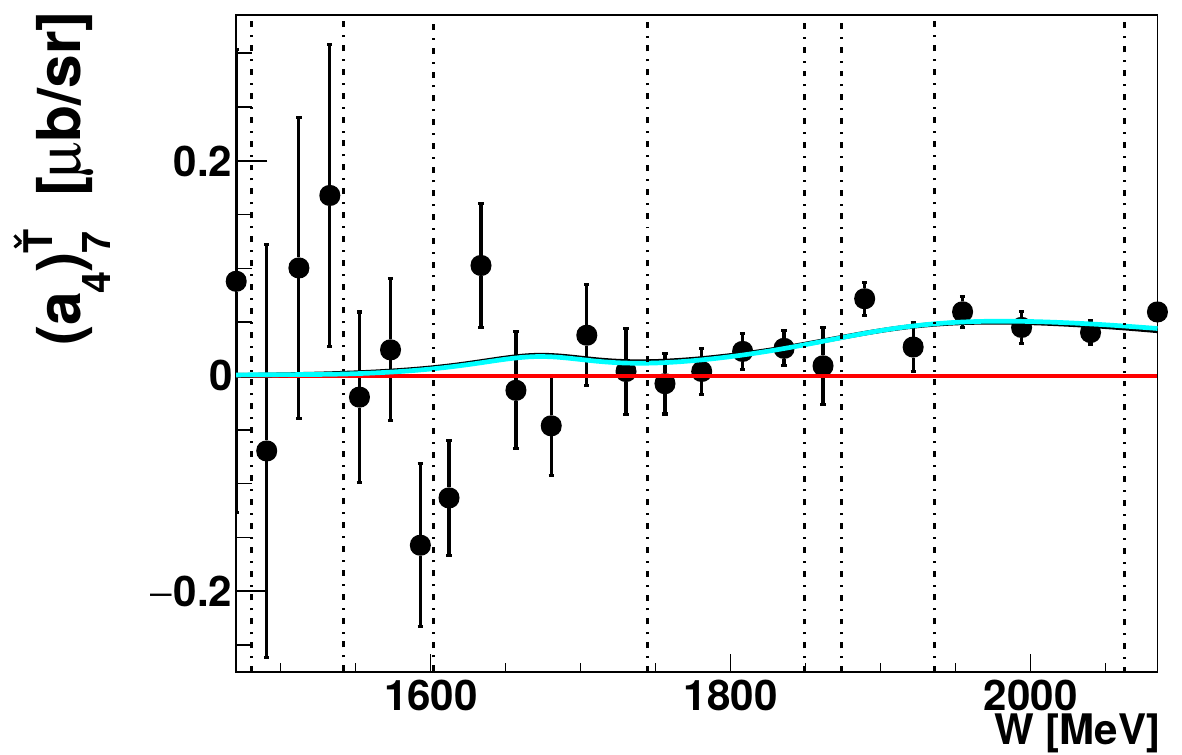}
  \includegraphics[width=0.285\textwidth]{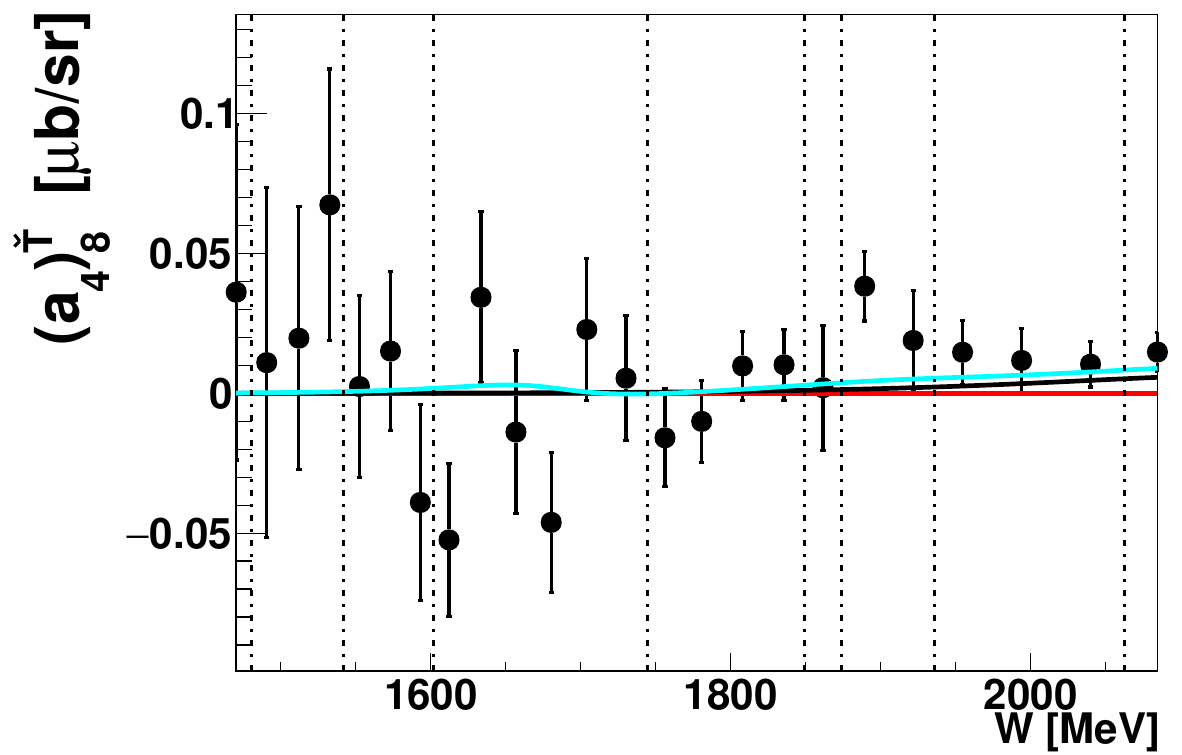}
  \end{minipage}
\end{figure*}

\begin{figure*}
\begin{minipage}{\textwidth}
\floatbox[{\capbeside\thisfloatsetup{capbesideposition={right,top},capbesidewidth=7.8cm}}]{figure}[\FBwidth]
{\caption{The beam asymmetry $\check{\Sigma}_{\mathrm{GRAAL}}$ data from GRAAL \cite{GRAAL} with only statistical error was fitted using associated Legendre polynomials according to eq. \ref{eq:LowEAssocLegParametrizationSigma} and truncating the partial wave expansion at $\text{L}_{\text{max}}=1\dots 4$. (a) The resulting $\chi^2/$ndf values of the different $\text{L}_{\text{max}}$-fits as a function of the center of mass energy W are shown. (b) 6 out of 31 selected angular distributions of $\check{\Sigma}_{\mathrm{GRAAL}}$ (black points) are plotted together with the different $\text{L}_{\text{max}}$ fits (solid lines) starting at W= 1504 MeV up to 1885 MeV. (c) Comparison of the fit coefficients for $\text{L}_{\text{max}}=4$ (black points), $\left(a_{4}\right)^{\check{\Sigma}_{\mathrm{GRAAL}}}_{2\dots 8}$ (see eq. \ref{eq:LowEAssocLegParametrizationSigma}), with the BnGa2014-02 solution truncated at different $\text{L}_{\text{max}}$ (solid lines). Colors same as in (a).}\label{fig:Sgraal_bins}}
{\includegraphics[width=0.49\textwidth, trim=0cm 0cm 1.8cm 0cm, clip]{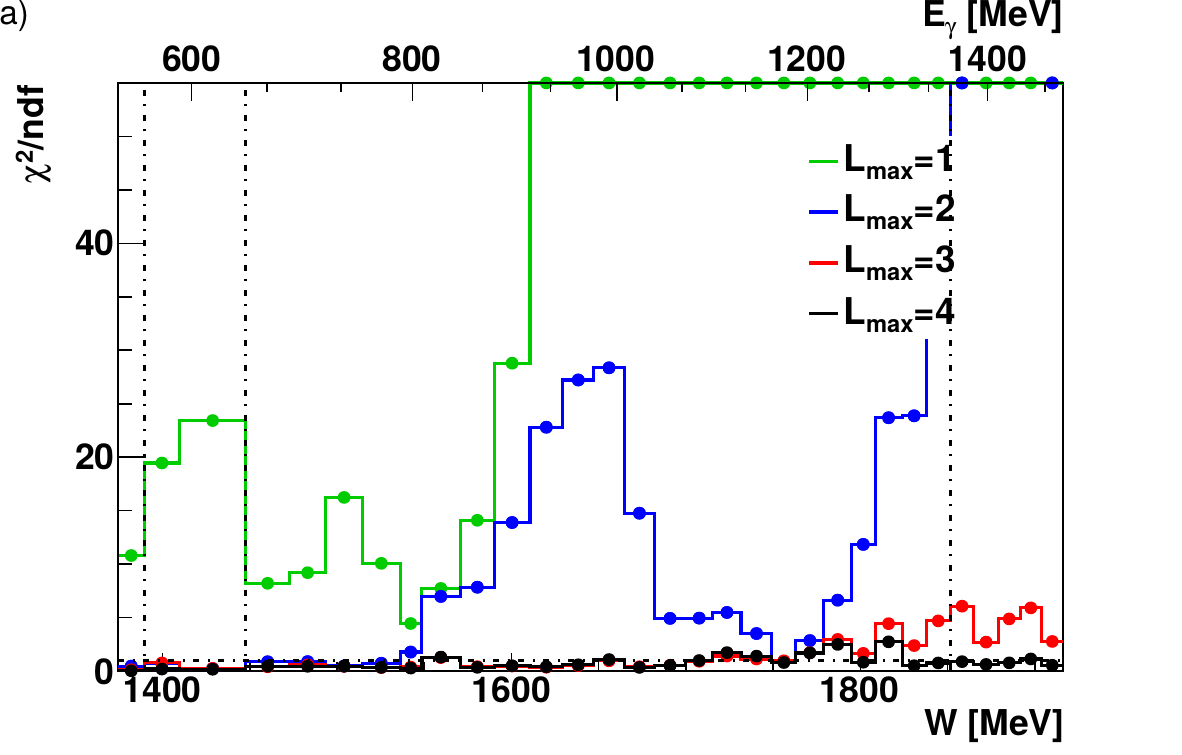}}
\end{minipage}\\

\begin{minipage}{\textwidth}
\centering
\hspace*{-0.45cm}
 \includegraphics[width=0.305\textwidth, trim=0cm 0cm 0.01cm 0.75cm, clip]{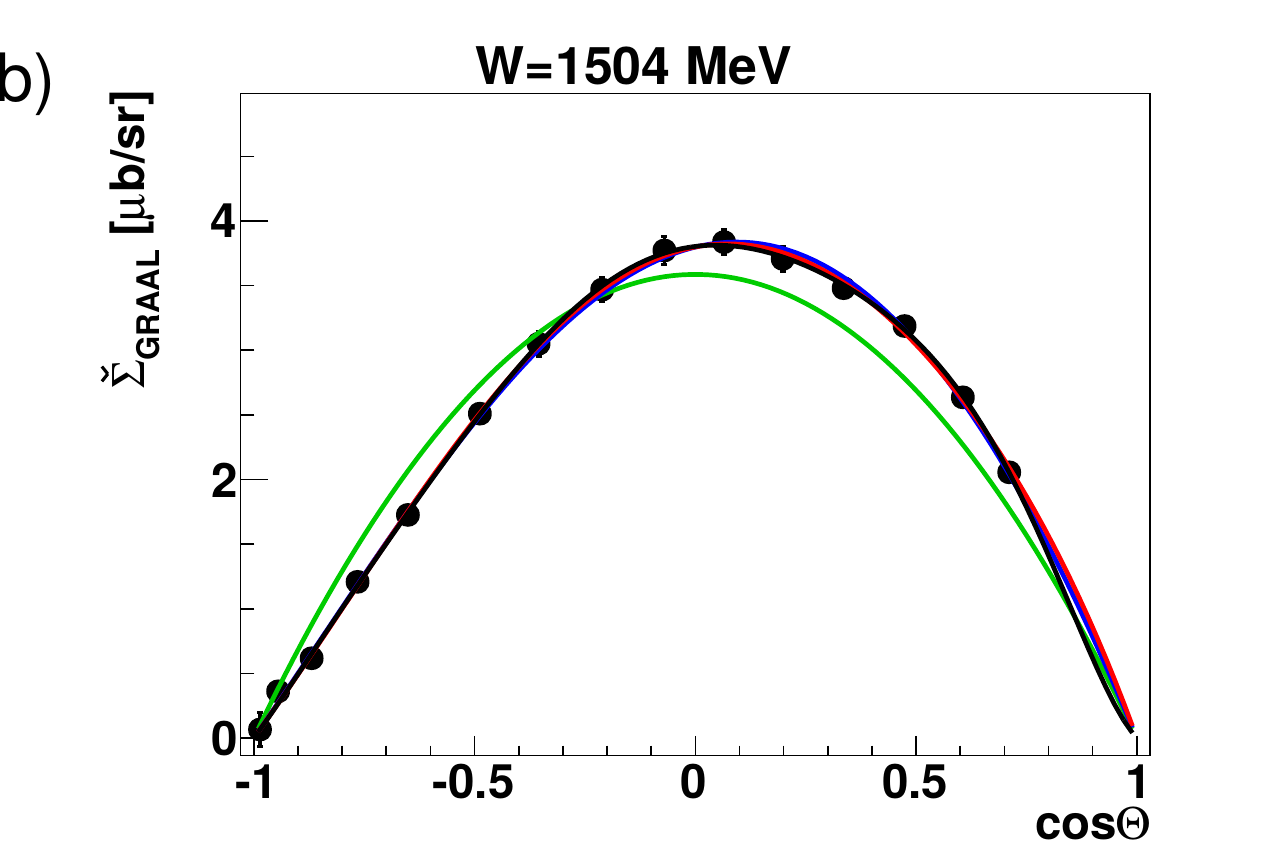}
  \includegraphics[width=0.285\textwidth, trim=0cm 0cm 0.01cm 0.75cm, clip]{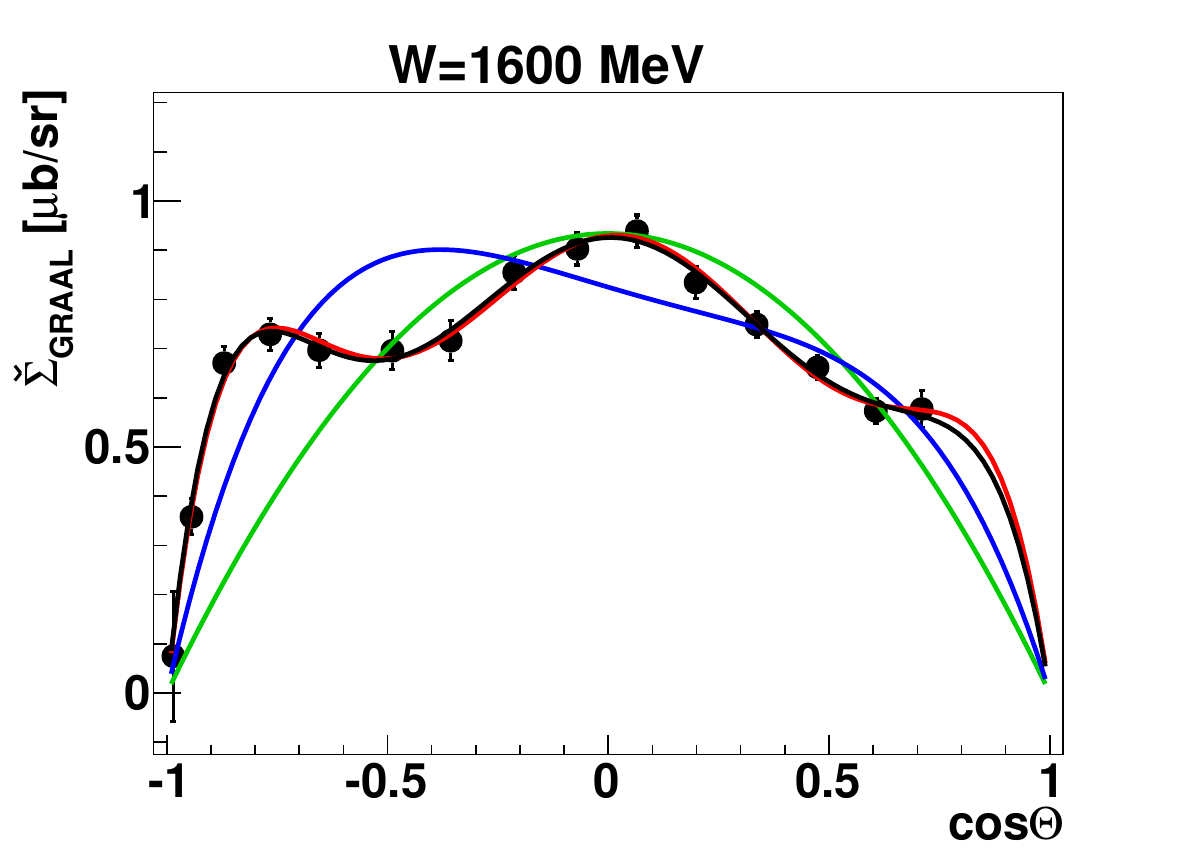}
  \includegraphics[width=0.285\textwidth, trim=0cm 0cm 0.01cm 0.75cm, clip]{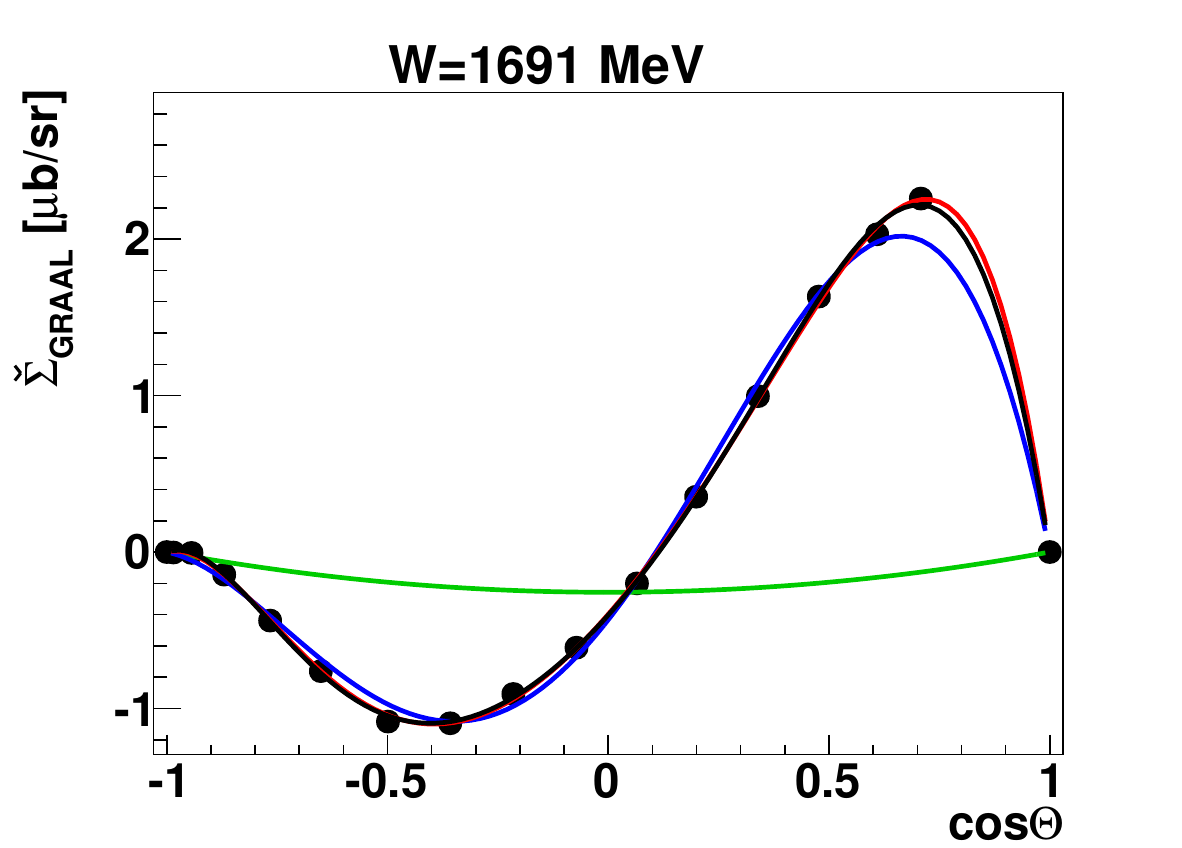}\\
  \includegraphics[width=0.285\textwidth, trim=0cm 0cm 0.01cm 0.75cm, clip]{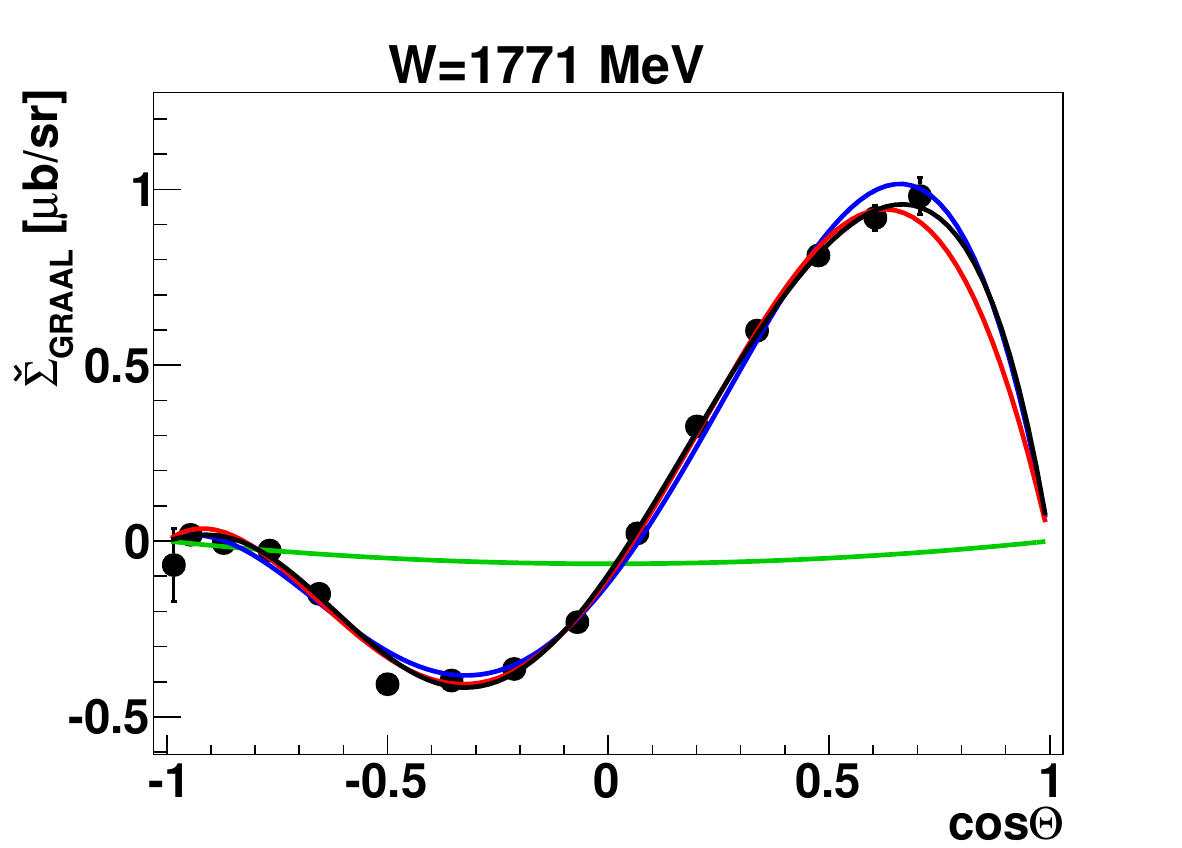}
  \includegraphics[width=0.285\textwidth, trim=0cm 0cm 0.01cm 0.75cm, clip]{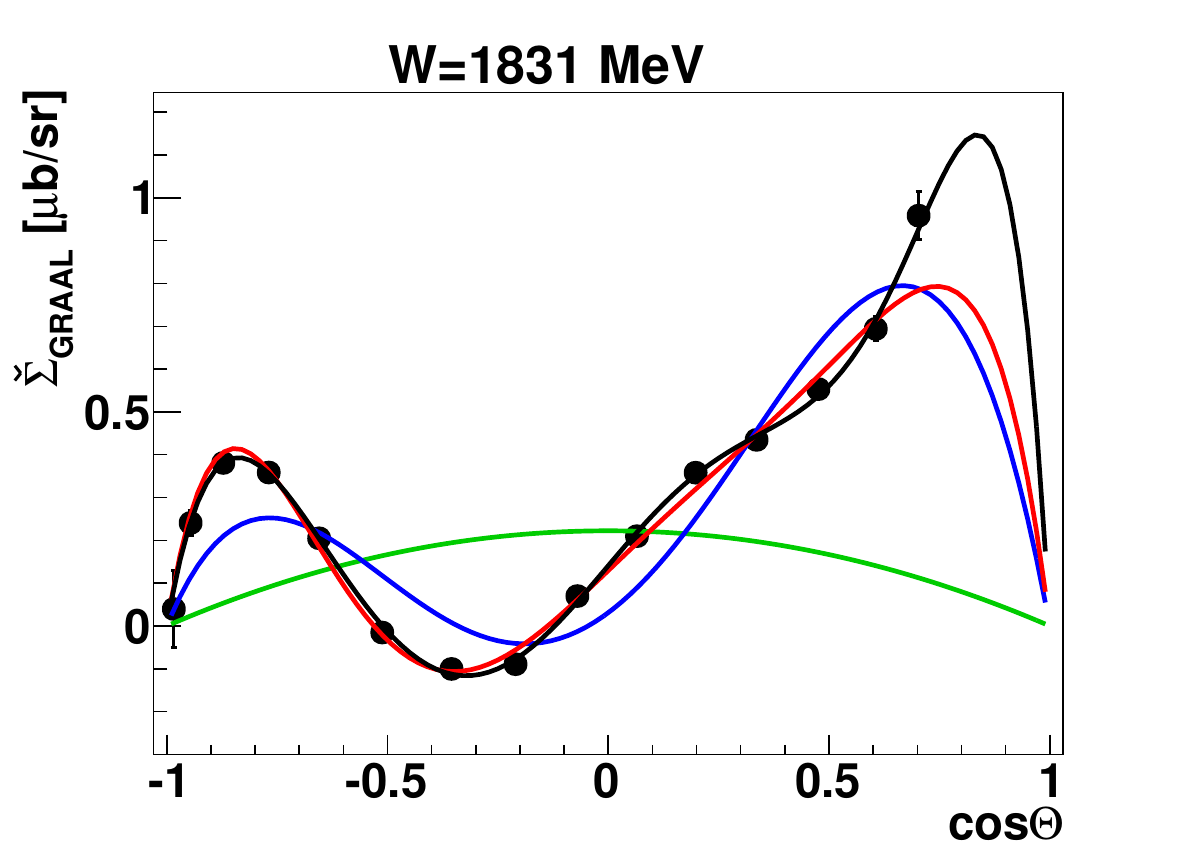}
  \includegraphics[width=0.285\textwidth, trim=0cm 0cm 0.01cm 0.75cm, clip]{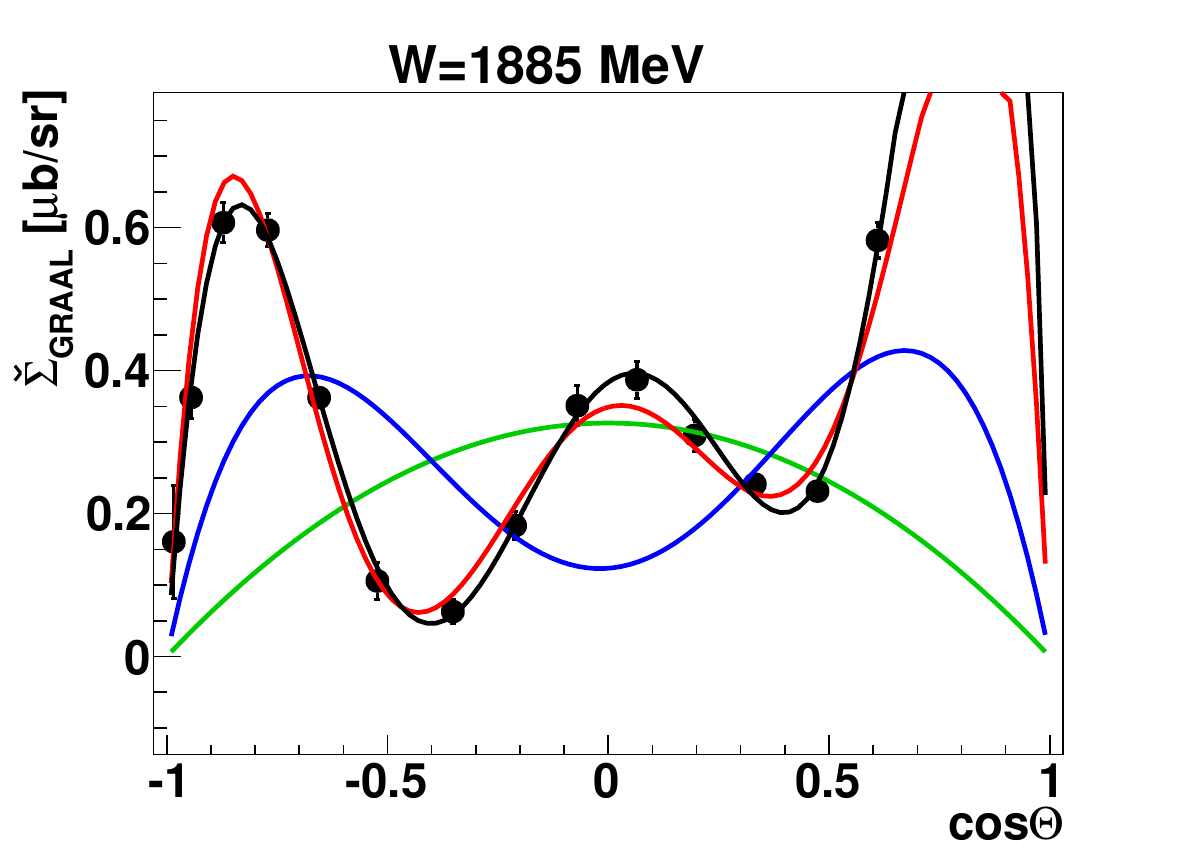}\\
 \vspace*{0.5cm}
  
  \hspace*{-23.5pt}\includegraphics[width=0.2905\textwidth]{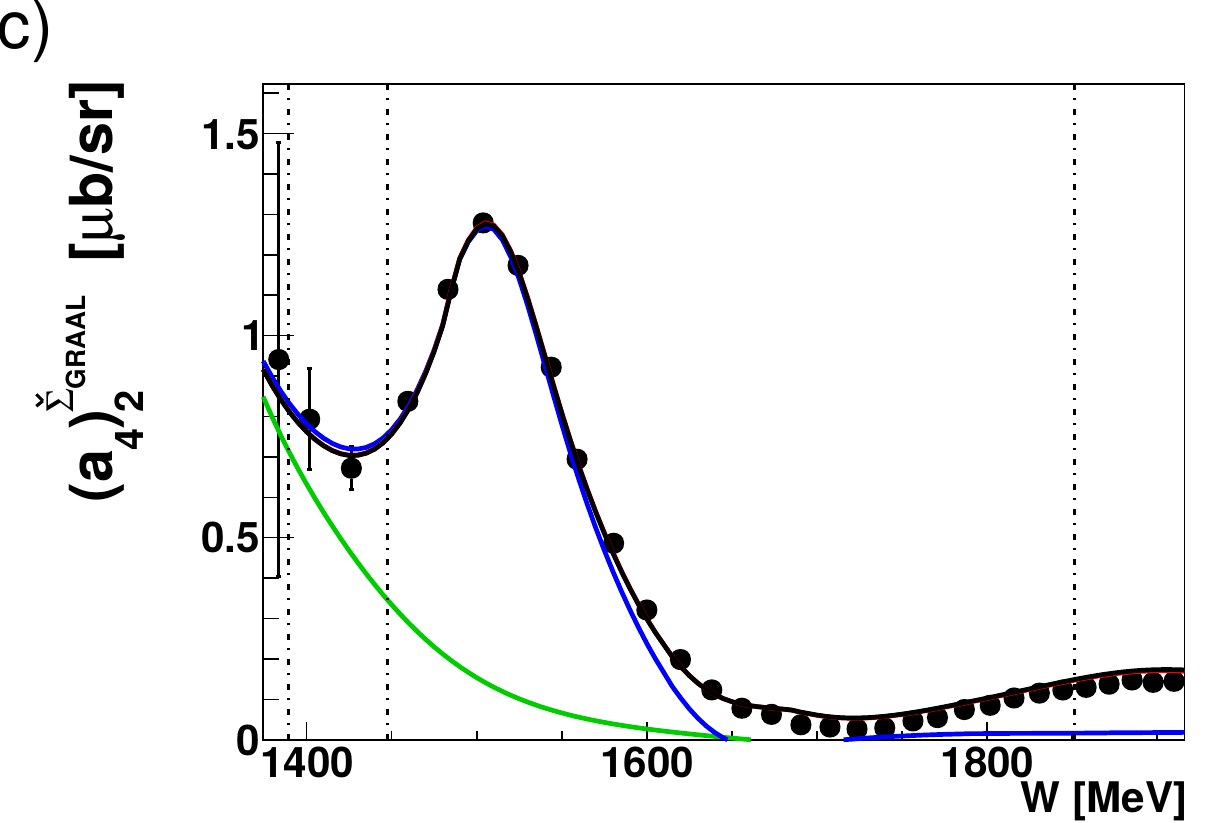}
  \includegraphics[width=0.285\textwidth]{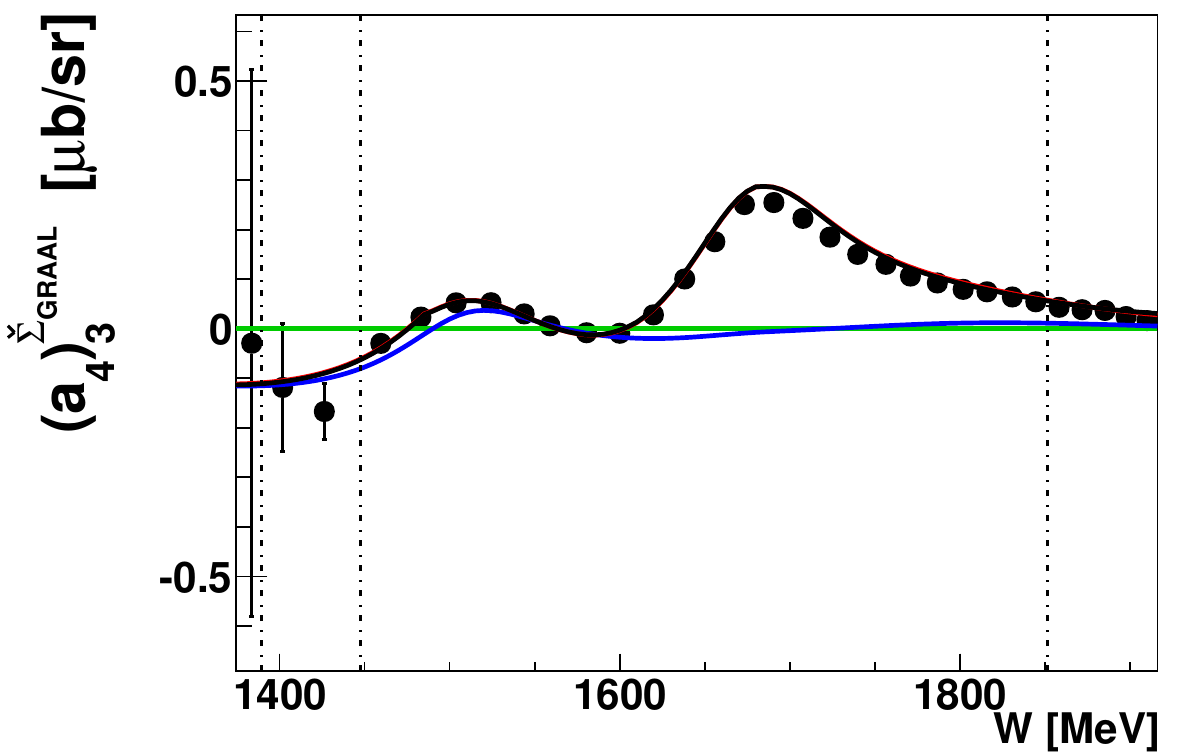}
  \includegraphics[width=0.285\textwidth]{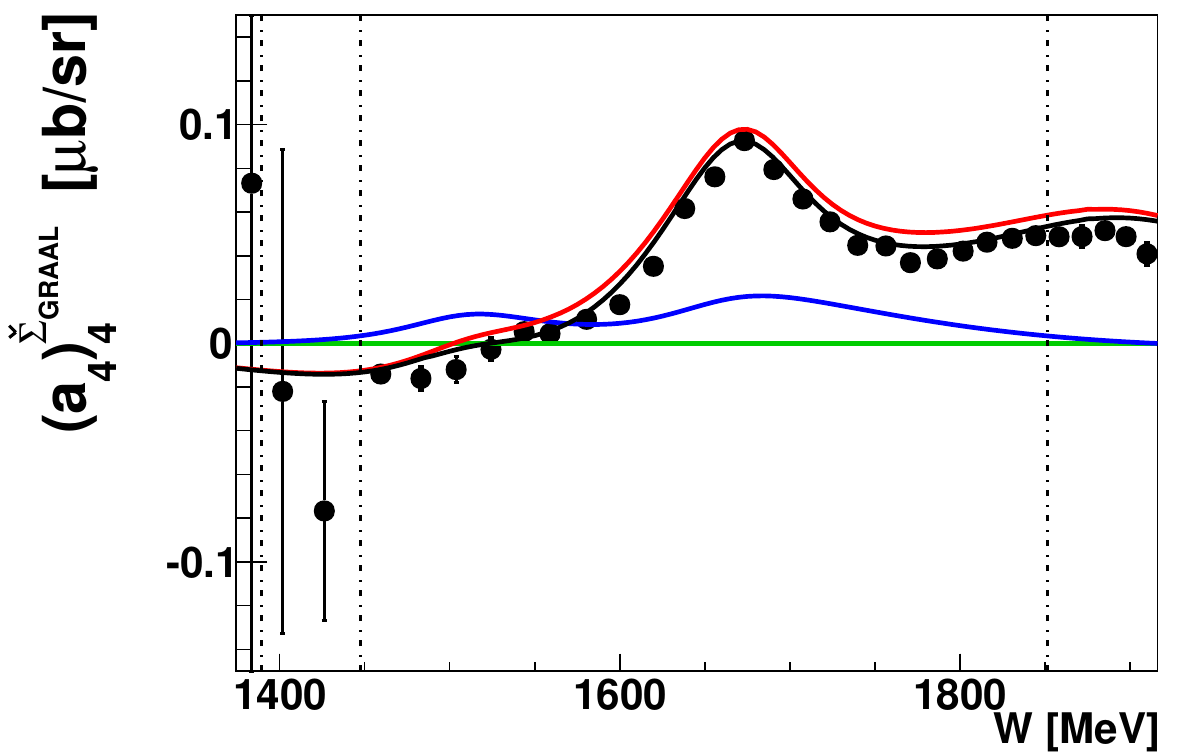}\\
  \hspace*{-19.5pt}\includegraphics[width=0.285\textwidth]{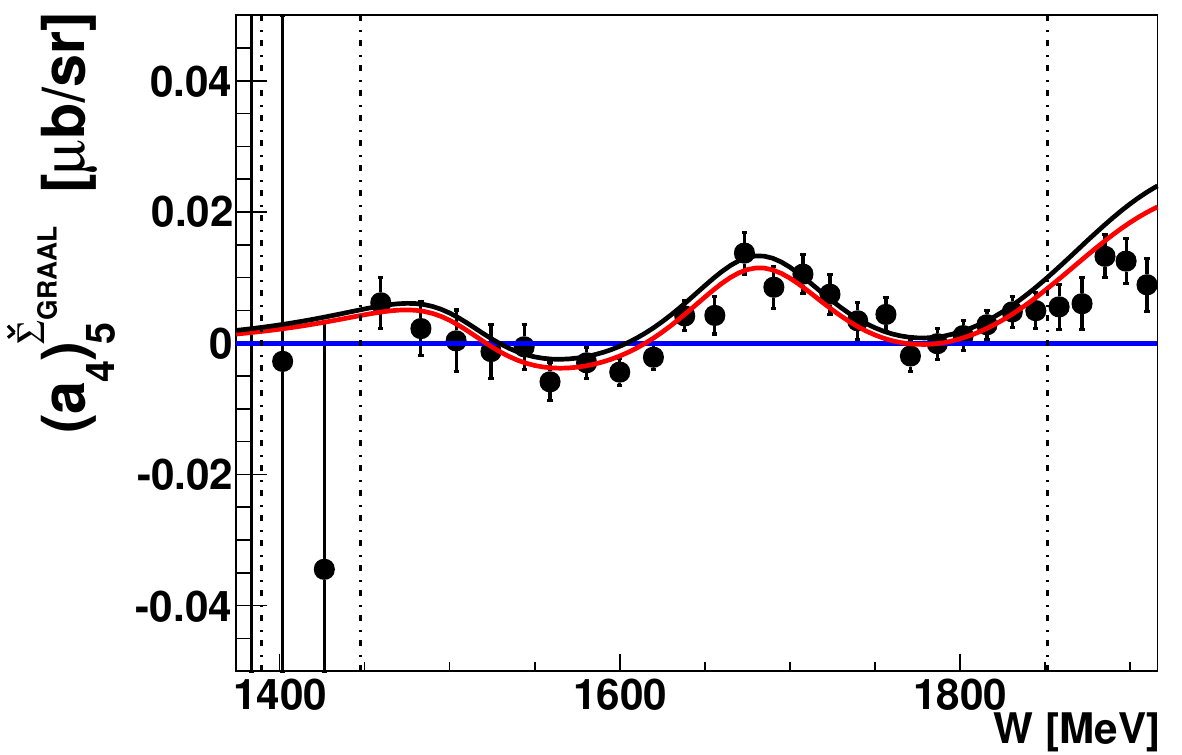}
  \includegraphics[width=0.285\textwidth]{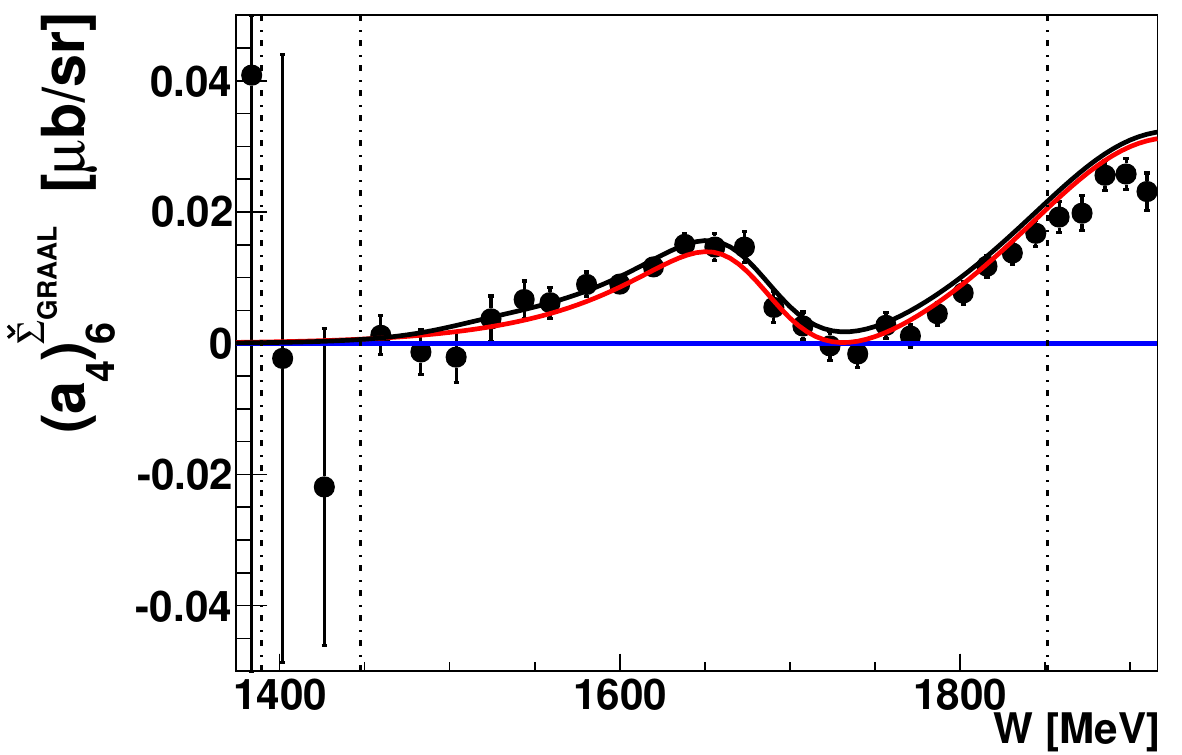}
  \includegraphics[width=0.285\textwidth]{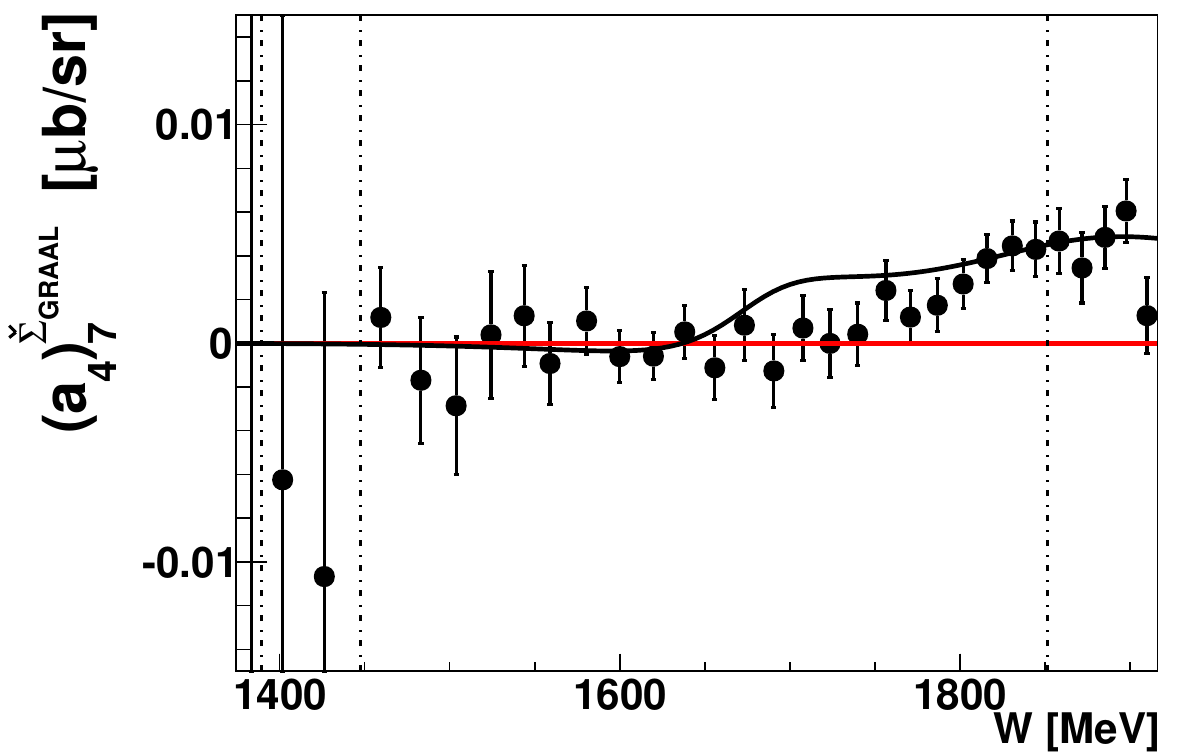}
  \hspace*{-19.5pt}\includegraphics[width=0.285\textwidth]{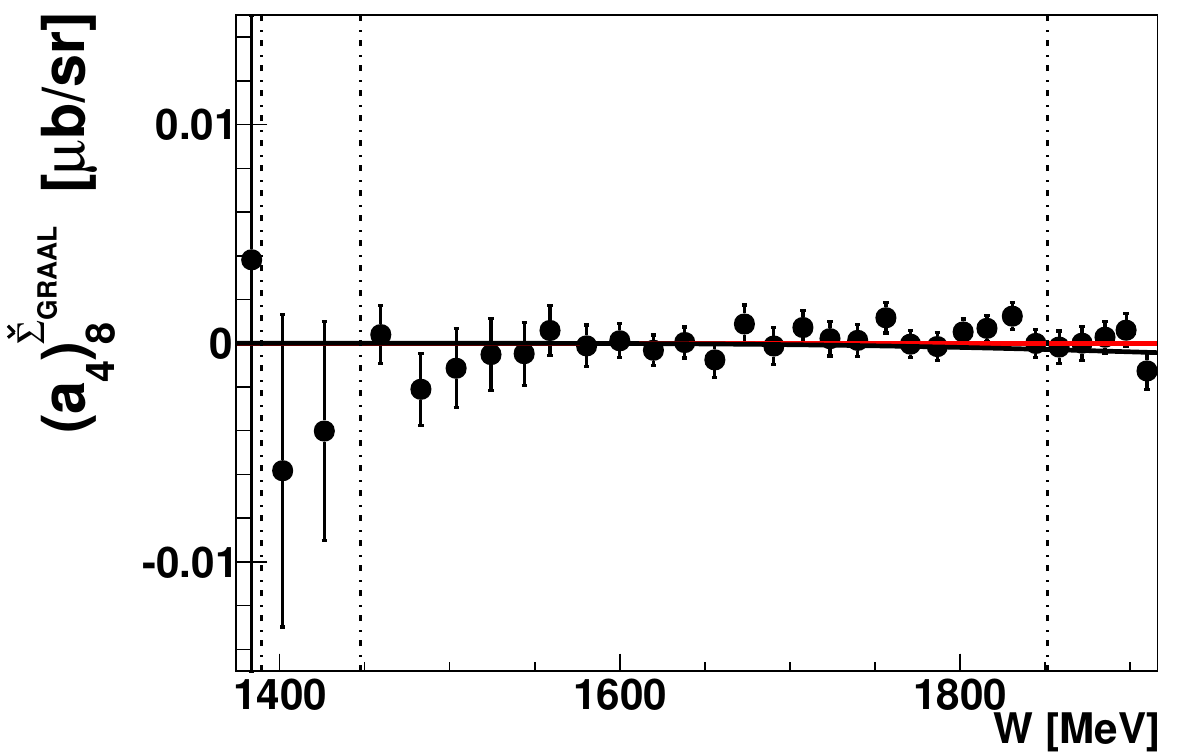}
  \end{minipage}
\end{figure*}

The data base for an interpretation of partial wave contributions in the second resonance region is a lot richer than that of the preceding section. In fact, all datasets except for the $\Sigma$ data from the CLAS collaboration have data here. The results of all fits can be seen in Figures \ref{fig:wq_bins}, \ref{fig:wq_bins_wsys}, \ref{fig:e_bins}, \ref{fig:s12_bins}, \ref{fig:s32_bins}, \ref{fig:f_bins}, \ref{fig:g_bins}, \ref{fig:h_bins}, \ref{fig:P_bins}, \ref{fig:T_bins} and \ref{fig:Sgraal_bins}. \newline
The plots of $\chi^{2}/\mathrm{ndf}$ vs. energy suggest that a description using up to $D$-waves is good for almost all observables under investigation. At the border to the third resonance region (around $W = 1600 \hspace*{2pt} \mathrm{MeV}$), the following datasets show small indications for $F$-waves in the $\chi^{2}/\mathrm{ndf}$:  $\sigma_{0}$ (Figures \ref{fig:wq_bins} and \ref{fig:wq_bins_wsys}), the spin dependent cross section $\sigma_{3/2}$ (Fig. \ref{fig:s32_bins}), the $G$ dataset from CBELSA/TAPS and the $\Sigma$ measurement from GRAAL (Figures \ref{fig:g_bins} and \ref{fig:Sgraal_bins}). The influence of $F$-waves shows up where the well confirmed $\ast\hspace*{0.75pt}\ast\hspace*{0.75pt}\ast\hspace*{0.75pt}\ast$-resonance $N (1680) \frac{5}{2}^{+}$ is known to exist (cf. Tables \ref{tab:MultipoleResonanceAssignments} and \ref{tab:PartialWavesMultipoles}). The second resonance region (mass range $ 1350 \hspace*{2pt} \mathrm{MeV} \lesssim W \lesssim 1600 \hspace*{2pt} \mathrm{MeV}$) contains no $F$-wave resonances according to the PDG (see as well Table \ref{tab:MultipoleResonanceAssignments}). Therefore, the data are in agreement with the already established resonances. \newline
The appearance of significant higher partial wave contributions in the A2 cross section remains questionable due to issues already discussed in Sec. \ref{sec:Interpretation1stResRegion}, especially since the fits to data including the systematic error show no such indications (cf. Fig. \ref{fig:wq_bins_wsys}). \newline
Once the fitted Legendre coefficients are compared to the BnGa2014-02 solution, many are well described with a truncation up to $D$-waves. Good examples are $\left(a_{4}\right)^{\sigma_{0}}_{(0,2)}$, $\left(a_{2}\right)^{\check{P}}_{2}$, $\left(a_{3}\right)^{\sigma_{1/2}}_{(0,\ldots,4)}$, $\left(a_{3}\right)^{\check{F}}_{2}$, $\left(a_{3}\right)^{\check{G}}_{2}$ and $\left(a_{4}\right)^{\check{\Sigma}_{\mathrm{GRAAL}}}_{2}$. 
Generally, it is seen that all those coefficients contain $\left< D, D \right>$-con\-tri\-bu\-tions. In regard of the well known $D$-wave resonance \newline $N\left(1520\right) \frac{3}{2}^{-}$, these facts seem not surprising. \newline 
However, some coefficients show structures that demand Bonn-Gat\-chi\-na $F$-waves in order to be reproduced. The Legendre coefficients $\left(a_{3}\right)^{\check{E}}_{(0,3)}$ for example both need slight corrections from $F$-waves approaching $W=1600 \hspace*{2pt} \mathrm{MeV}$. Further examples for small improvements due to $F$-wave contributions are $\left(a_{4}\right)^{\sigma_{3/2}}_{2}$, $\left(a_{3}\right)^{\check{F}}_{1}$, $\left(a_{2}\right)^{\check{H}}_{2}$, $\left(a_{2}\right)^{\check{P}}_{(1,2)}$ and \newline $\left(a_{4}\right)^{\check{\Sigma}}_{(4,6)}$ from fits to the GRAAL measurement. In all of these cases the Bonn-Gatchina description up to $D$-waves seems almost right, but the slight correction via the $F$-waves still gives an improved description of the fitted Legendre coefficients. This seems at first like a contradiction, since the $\chi^{2}$-tests above were suggesting in case of for instance the observables $\check{E}$, $\check{P}$ and $\check{H}$, that a truncation at lower $\text{L}_{\text{max}}$, i.e. $2$ or $1$, should be sufficient to describe the Legendre coefficients within the second resonance region. However, this apparent contradiction is not a real one. \newline
What can be observed here, is a phenomenon inherent to po\-la\-ri\-za\-tion observables, namely the possibility for partial waves of various orders to give interference contributions. Therefore, for each increased order in $\text{L}_{\text{max}}$, the multipoles corresponding to this increased order not only contribute to the higher Legendre coefficients according to Eqs (\ref{eq:LowEAssocLegParametrization1}) and (\ref{eq:LowEAssocLegParametrization2}), but also to the lower ones. Most crucially, higher partial waves who are themselves small can nonetheless yield nontrivial contributions via interference with dominant lower waves. For example, the coefficient $\left(a_{3}\right)^{\check{T}}_{3}$ is, in an $F$-wave truncation, made up entirely of $\left<P,D\right>$, $\left<S,F\right>$ and $\left<D,F\right>$ interference terms (for an explanation of this notation, see section \ref{sec:DescriptionCompositionLegCoeffs} and Appendix \ref{sec:PWContentFormulas}) and an effect is visible in Fig. \ref{fig:T_bins}. \newline
The above mentioned interference effect is even more pronounced in the coefficients $\left(a_{4}\right)^{\sigma_{3/2}}_{(3,5)}$, $\left(a_{4}\right)^{\sigma_{0}}_{(1,3)}$, $\left(a_{3}\right)^{\check{F}}_{3}$, \newline $\left(a_{3}\right)^{\check{G}}_{(3,5)}$, $\left(a_{2}\right)^{\check{H}}_{1}$ and $\left(a_{2}\right)^{\check{P}}_{3}$. The observable $\sigma_{3/2}$ deserves special mentioning here. For the coefficient $\left(a_{4}\right)^{\sigma_{3/2}}_{3}$ at around $W = 1500$ MeV, a highly nontrivial correction begins which is, in an $F$-wave truncation, given entirely by a $\left<D,F\right>$ interference term (see Figure \ref{fig:s32_bins} and Table \ref{tab:DCS32ColorPlots1}). The black curve, corresponding to a prediction including $G$-waves, shows an almost vanishing further correction and the blue $\text{L}_{\text{max}} = 2$ curve has no chance to correctly describe the Legendre coefficient. \newline
Apart from $\sigma_{3/2}$, the $H$ measurement from \newline CBELSA/TAPS deserves a special mentioning as well. It already has a good $\chi^{2}$ using a truncation at $\text{L}_{\text{max}} = 1$ (cf. Fig. \ref{fig:h_bins}), within the statistical precision. However, contributions from higher partial waves should not be disregarded for the $H$ measurement as well, especially in the first Legendre coefficient. \newline
Although the $\chi^{2}$ plot only shows indication of up to $P$-waves, contributions from Bonn-Gatchina $D$- and $F$-waves are needed in order to describe the coefficient. This is an extreme example of the above mentioned interference phenomenon. \newline
The Legendre coefficient $\left(a_{4}\right)^{\sigma_{3/2}}_{6}$ has small values but is also significantly non-zero. In an $F$-wave truncation, this quantity is given entirely by an $\left<F,F\right>$ term and the Bonn-Gatchina prediction for $\text{L}_{\text{max}}=3$ reproduces it rather well. Therefore, one can interpret the non-vanishing of $\left(a_{4}\right)^{\sigma_{3/2}}_{6}$ as a first hint of the $F$-wave resonances which dominate the third resonance region, reaching into the second one. A similar hint of a non-trivial $\left<F,F\right>$-contribution in the second resonance region can be obtained by looking at the coefficient $\left(a_{4}\right)^{\check{\Sigma}_{\mathrm{GRAAL}}}_{6}$. The quantities $\left(a_{4}\right)^{\sigma_{0}}_{(4,6)}$ get similar contributions as well. \newline
In order to summarize the physical results obtained for the second resonance region, it has to be said that most $\chi^{2}$ distributions are consistent with the dominance of well established $S$-, $P$- and $D$-wave resonances according to the PDG (cf. Table \ref{tab:PartialWavesMultipoles}). A particular example for these is given by the $N(1520) \frac{3}{2}^{-}$ (cf. Figure \ref{fig:resonanzen}). Only very few observables show first indications of $F$-waves in $\chi^{2}$. However, once the fit results are compared to the Bonn-Gatchina PWA, the influence of the lowest well-confirmed $F$-waves from the third resonance region can already be seen in a lot of Legendre coefficients. \newline
Therefore, po\-la\-ri\-za\-tion observables show their usefulness by being sensitive to a large variety of partial wave interferences. In particular, $\sigma_{3/2}$ shows up as an observable that is highly capable of detecting $F$-wave contributions.


\subsection{Third resonance region $\left( 1600 \hspace*{2pt} \mathrm{MeV} \lesssim W \lesssim 1800 \hspace*{2pt} \mathrm{MeV} \right)$} \label{sec:Interpretation3rdResRegion}

\begin{figure*}
\begin{minipage}{\textwidth}
\floatbox[{\capbeside\thisfloatsetup{capbesideposition={right,top},capbesidewidth=7.8cm}}]{figure}[\FBwidth]
{\caption{The beam asymmetry $\check{\Sigma}_{\mathrm{CLAS}}$ data from JLab \cite{Dugger:2013} with only statistical error was fitted using associated Legendre polynomials according to eq. \ref{eq:LowEAssocLegParametrizationSigma} and truncating the partial wave expansion at $\text{L}_{\text{max}}=1\dots 5$. (a) The resulting $\chi^2/$ndf values of the different $\text{L}_{\text{max}}$-fits as a function of the center of mass energy W are shown. (b) 6 out of 39 selected angular distributions of $\check{\Sigma}_{\mathrm{CLAS}}$ (black points) are plotted together with the different $\text{L}_{\text{max}}$ fits (solid lines) starting at W= 1776 MeV up to 2092 MeV. (c) Comparison of the fit coefficients for $\text{L}_{\text{max}}=4$ (black points), $\left(a_{4}\right)^{\check{\Sigma}_{\mathrm{CLAS}}}_{2\dots 8}$ (see eq. \ref{eq:LowEAssocLegParametrizationSigma}), with the BnGa2014-02 solution truncated at different $\text{L}_{\text{max}}$ (solid lines). Colors same as in (a).}\label{fig:Sclas_bins}}
{\includegraphics[width=0.49\textwidth, trim=0cm 0cm 1.8cm 0cm, clip]{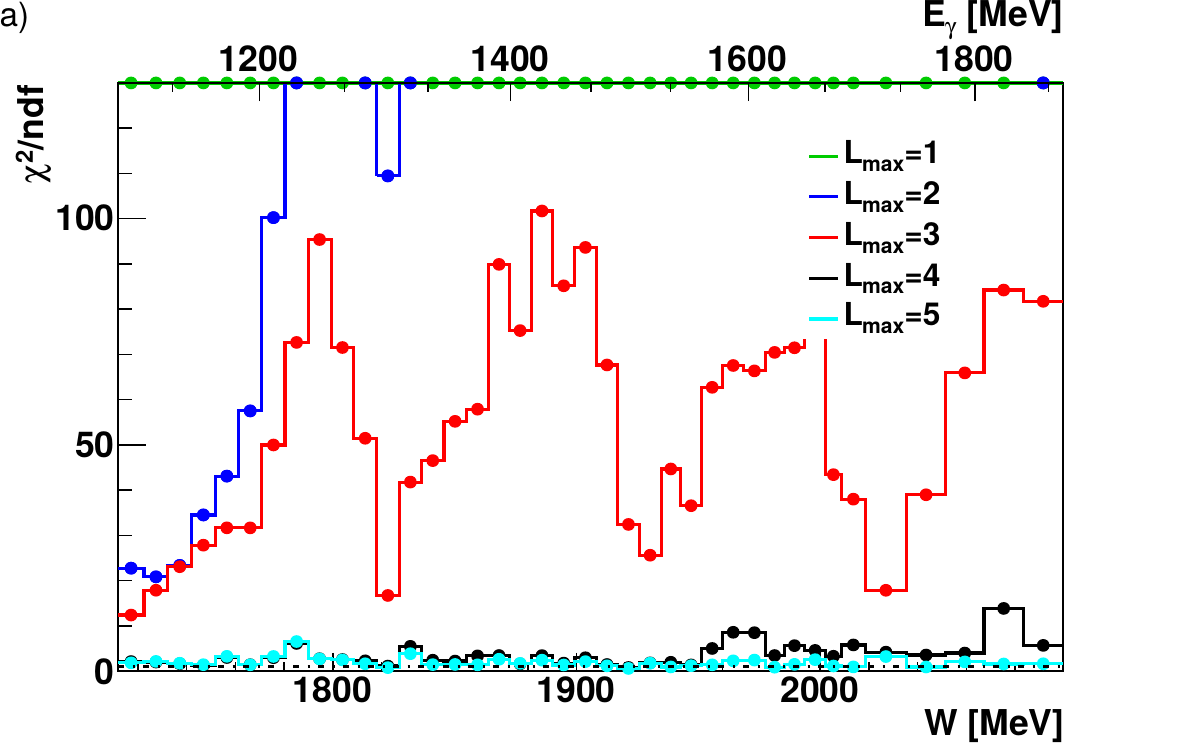}}
\end{minipage}\\

\begin{minipage}{\textwidth}
\centering
\hspace*{-0.45cm}
 \includegraphics[width=0.305\textwidth, trim=0cm 0cm 0.01cm 0.75cm, clip]{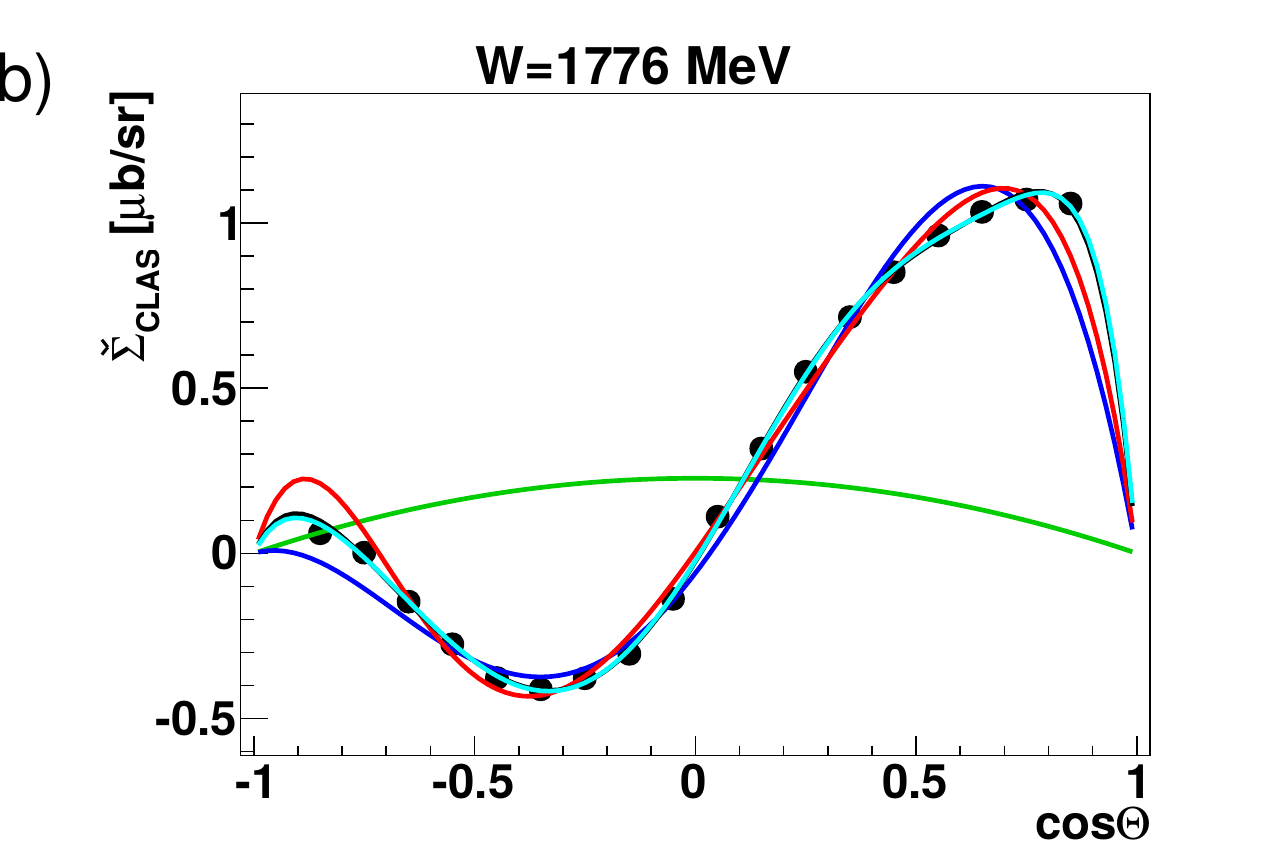}
  \includegraphics[width=0.285\textwidth, trim=0cm 0cm 0.01cm 0.75cm, clip]{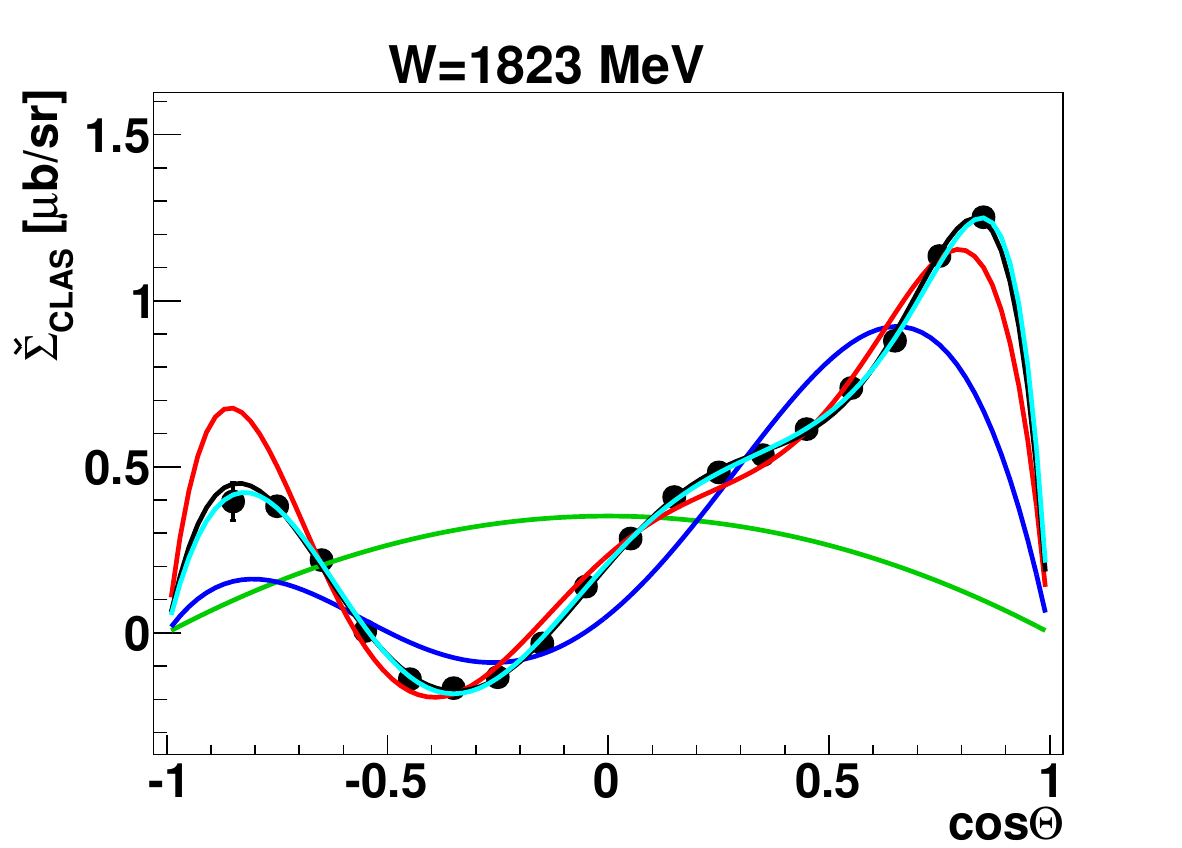}
  \includegraphics[width=0.285\textwidth, trim=0cm 0cm 0.01cm 0.75cm, clip]{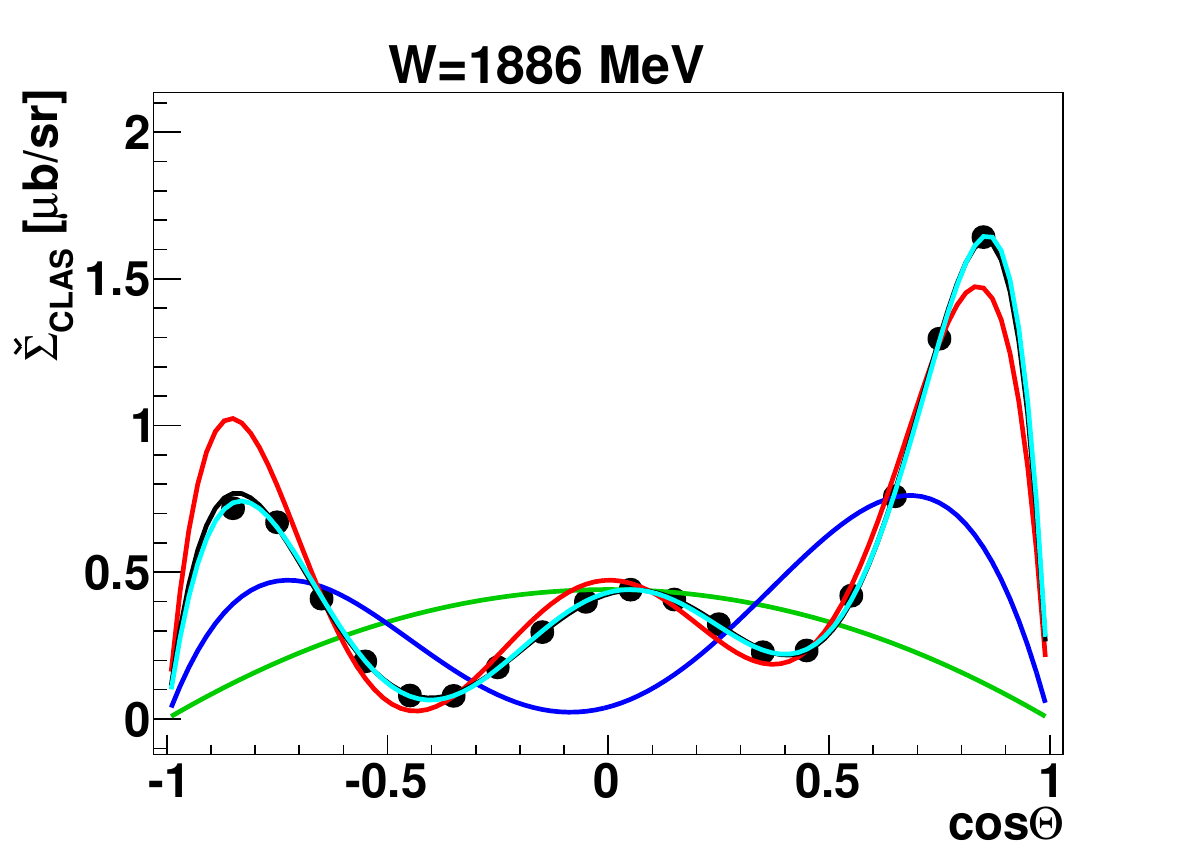}\\
  \includegraphics[width=0.285\textwidth, trim=0cm 0cm 0.01cm 0.75cm, clip]{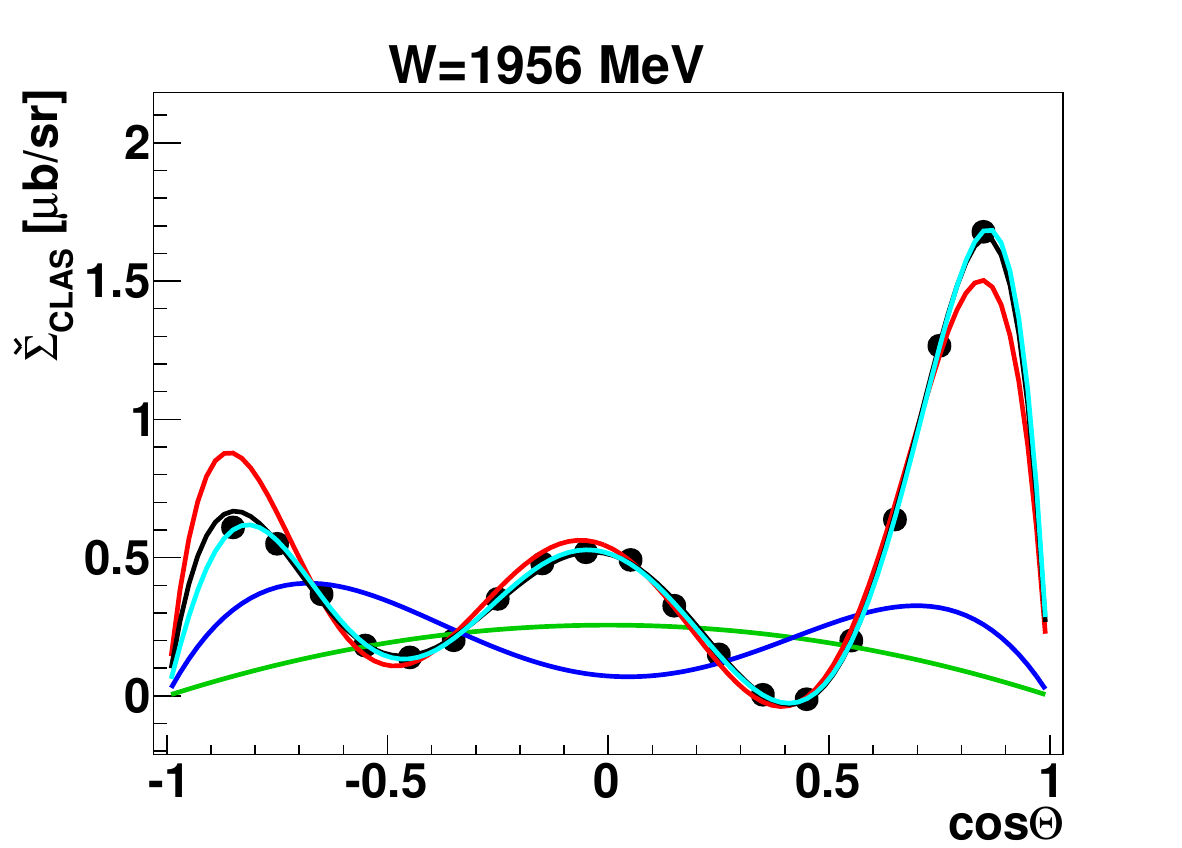}
  \includegraphics[width=0.285\textwidth, trim=0cm 0cm 0.01cm 0.75cm, clip]{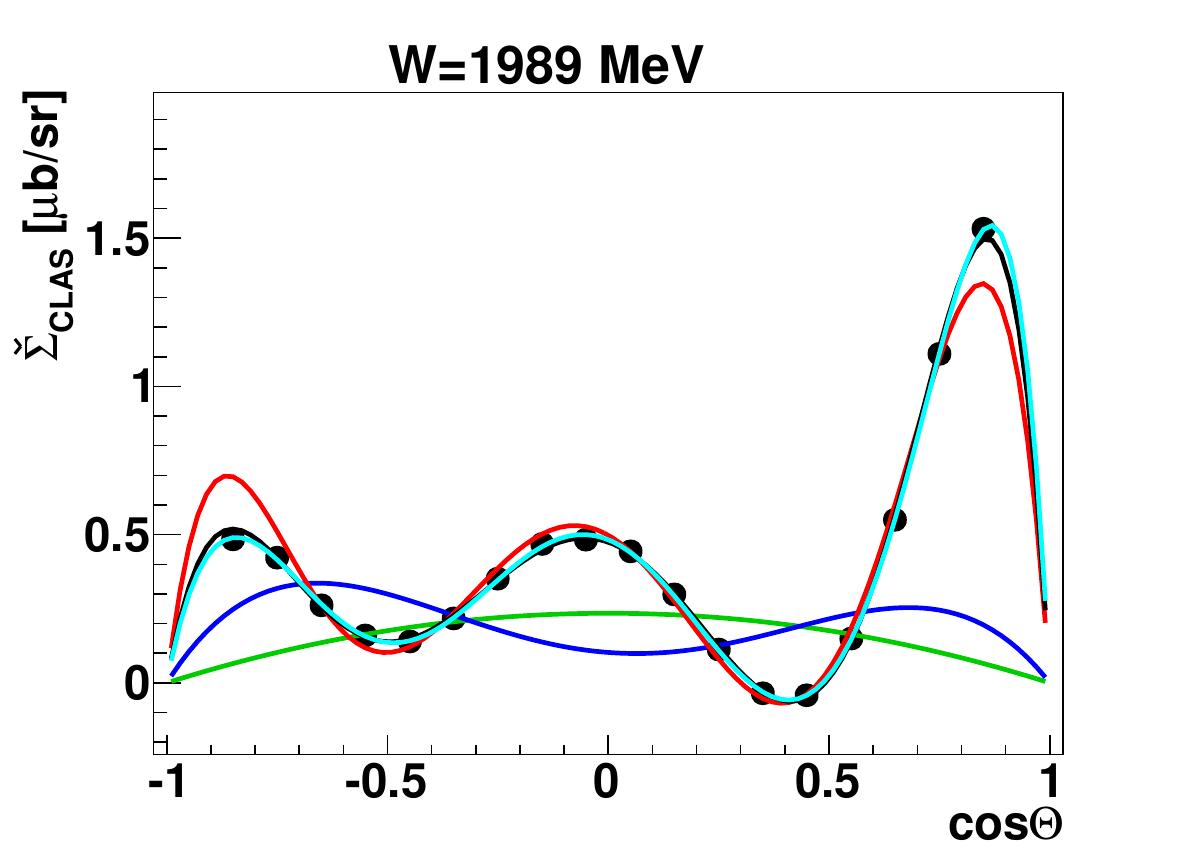}
  \includegraphics[width=0.285\textwidth, trim=0cm 0cm 0.01cm 0.75cm, clip]{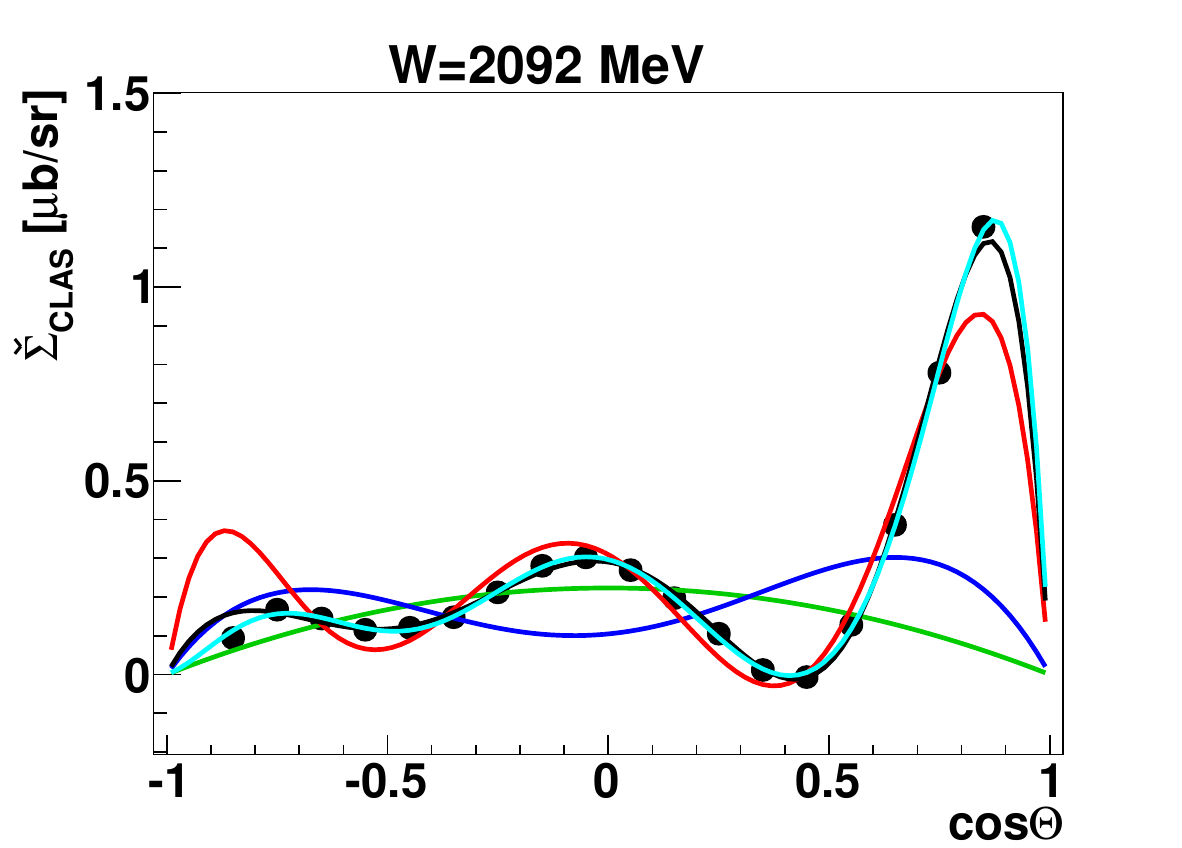}\\
 \vspace*{0.5cm}
  
  \hspace*{-23.5pt}\includegraphics[width=0.2905\textwidth]{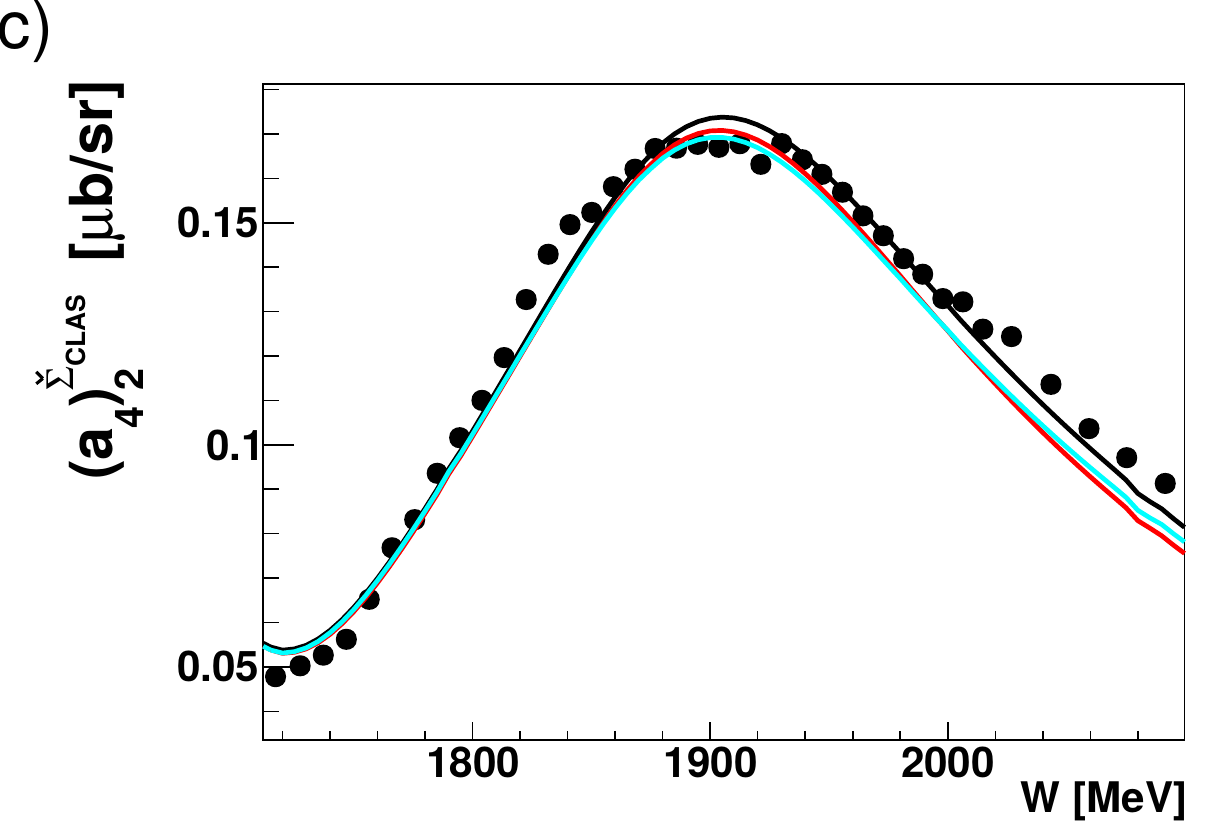}
  \includegraphics[width=0.285\textwidth]{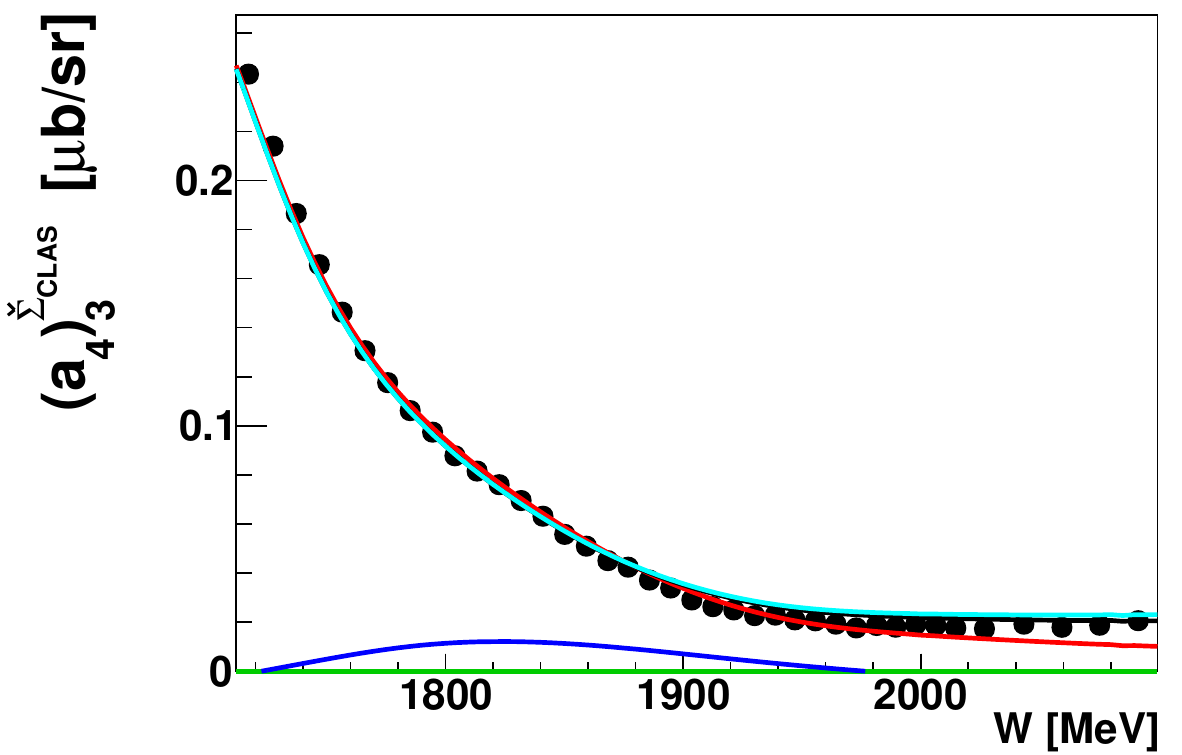}
  \includegraphics[width=0.285\textwidth]{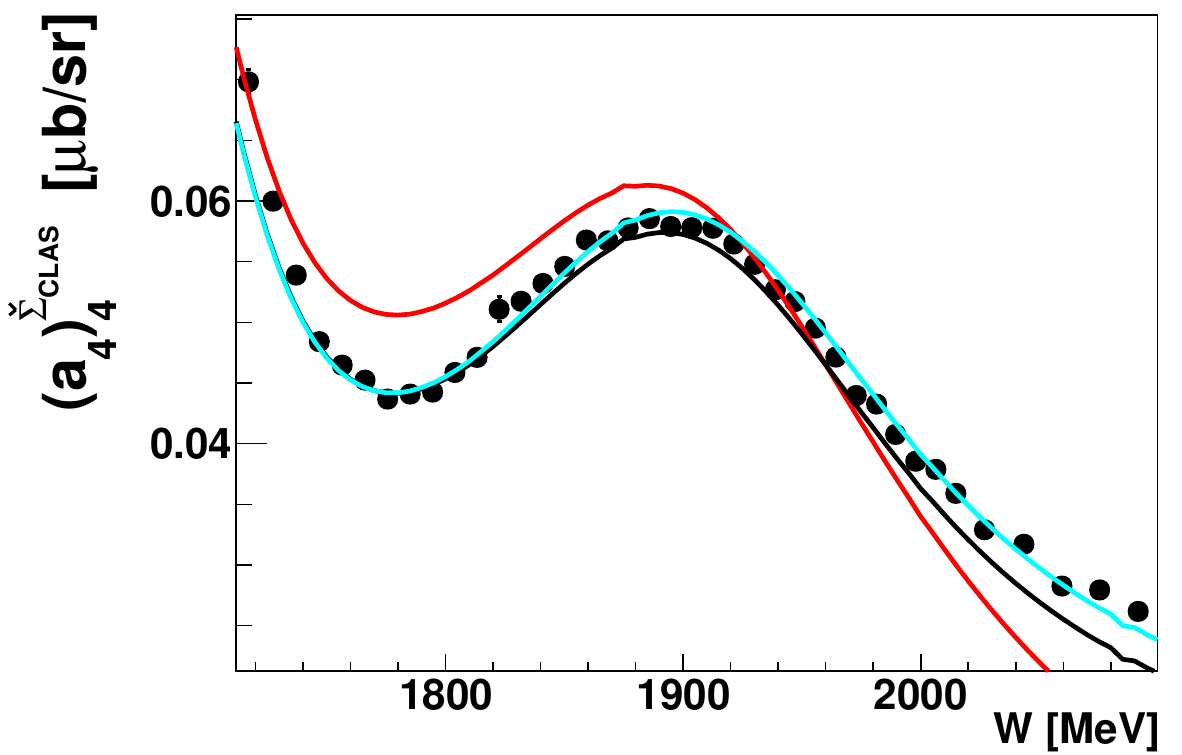}\\
  \hspace*{-19.5pt}\includegraphics[width=0.285\textwidth]{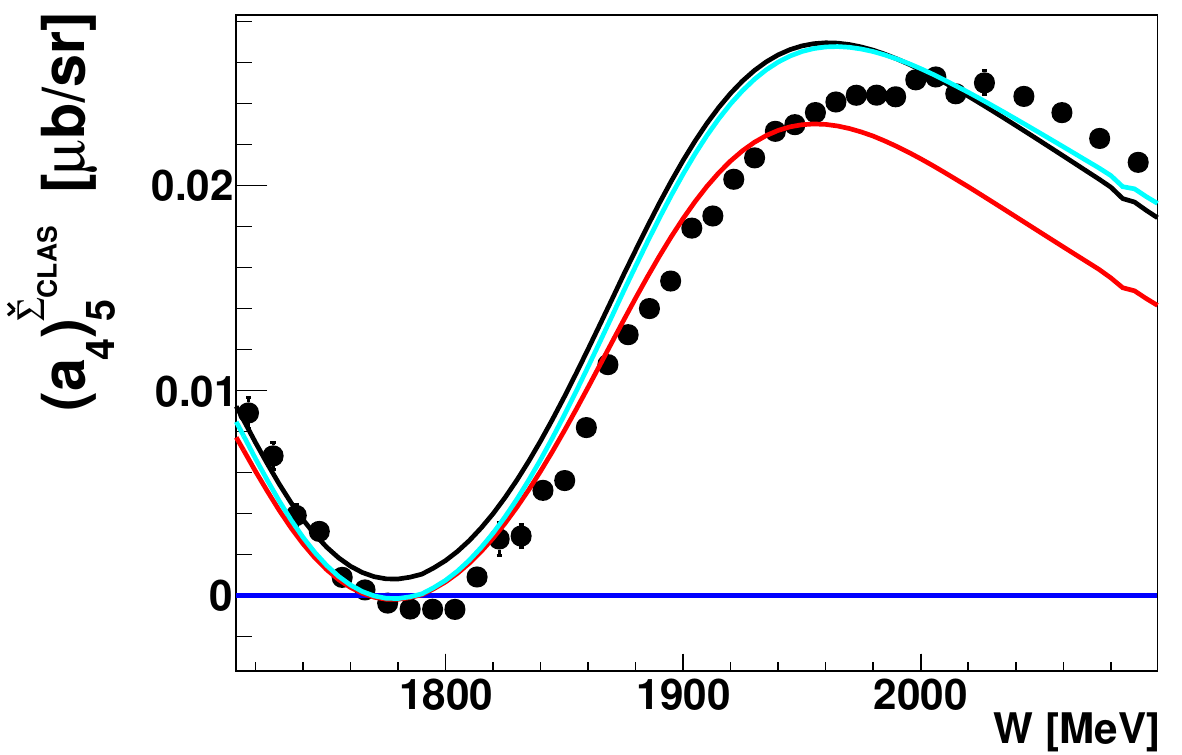}
  \includegraphics[width=0.285\textwidth]{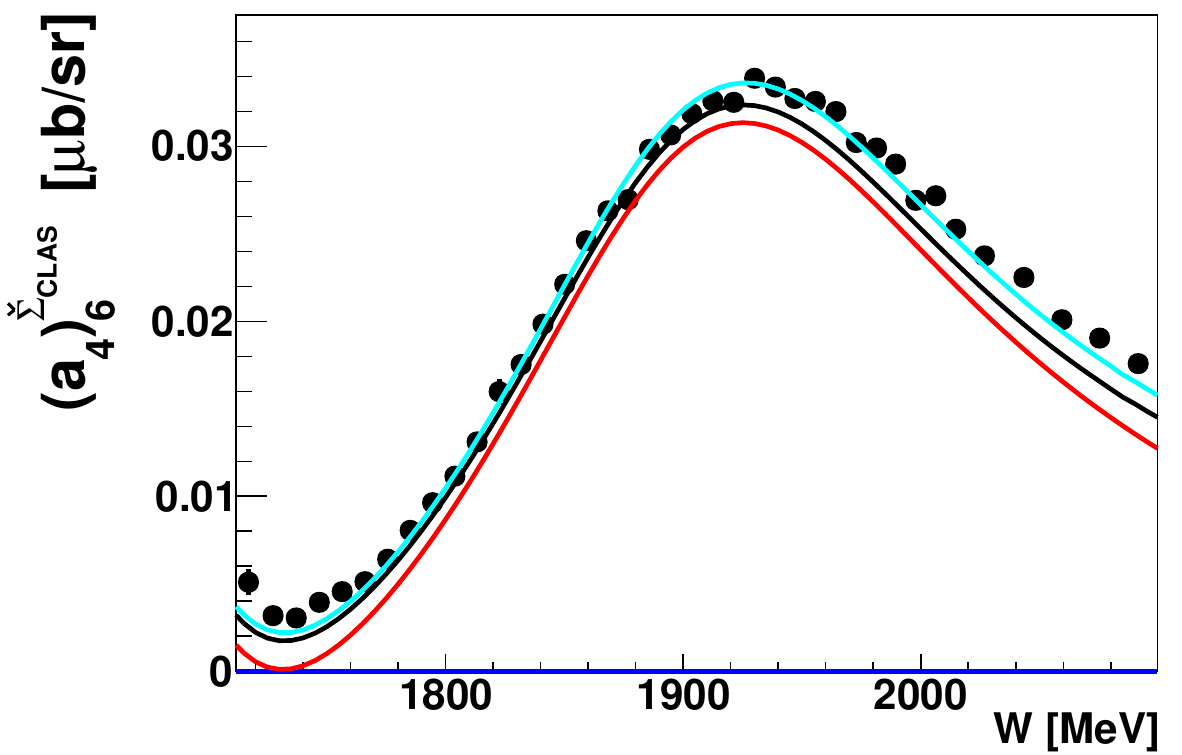}
  \includegraphics[width=0.285\textwidth]{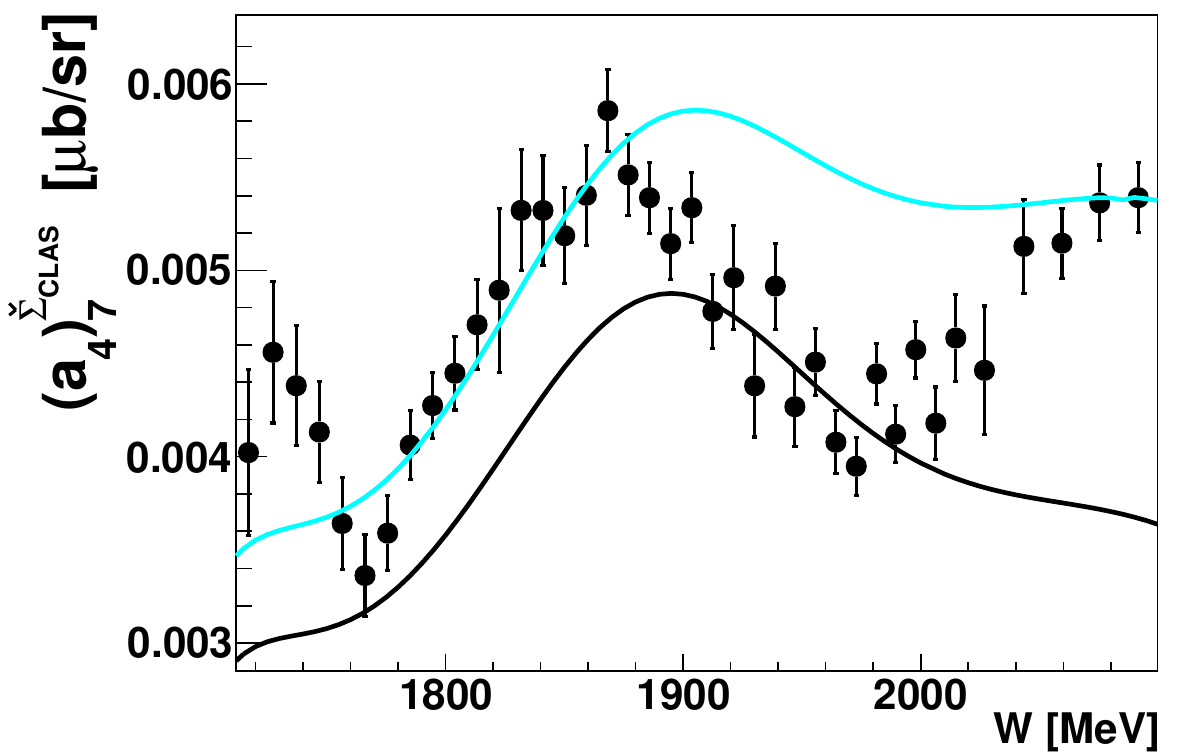} \\
  \hspace*{-19.5pt}\includegraphics[width=0.285\textwidth]{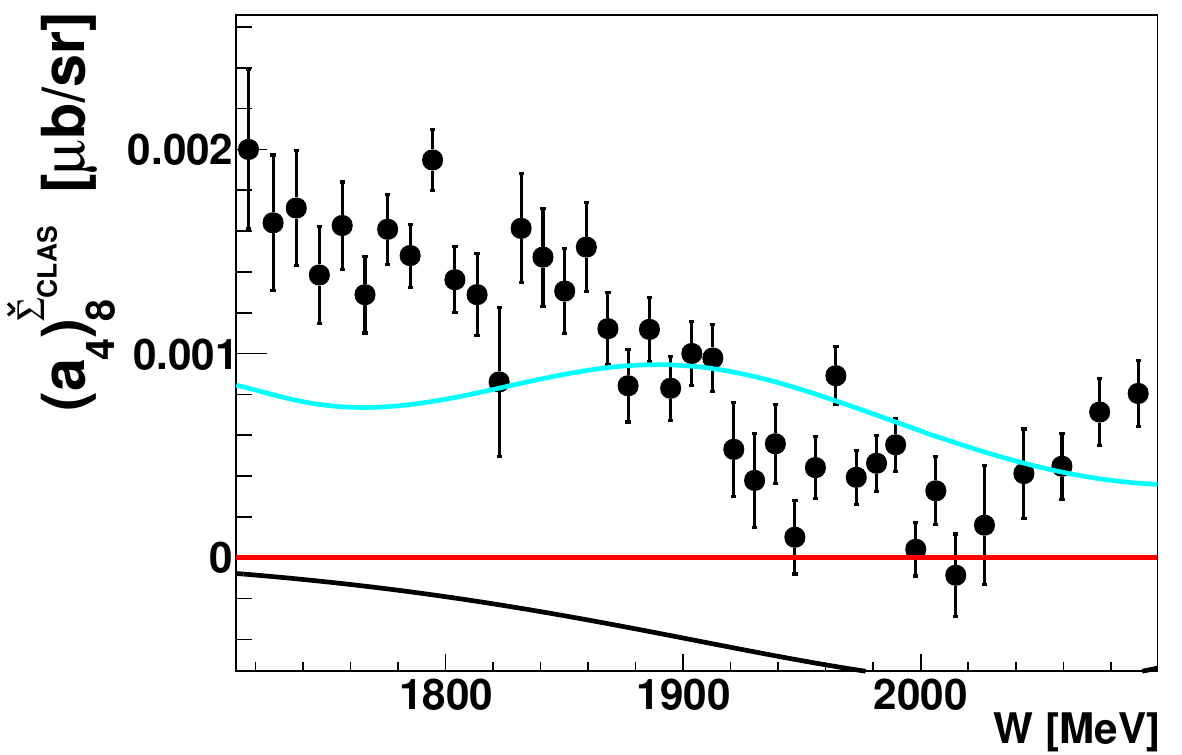}
  \end{minipage}
\end{figure*}

Almost all considered datasets have data in the third resonance region, except for the $P$ and $H$ measurements from the CBELSA/TAPS/TAPS collaboration (cf. Table \ref{tab:DataBasis}). \newline
The $\chi^{2}$ plots of almost all observables indicate the need for $F$-waves in order to describe the data. This is seen clearly in the rising flanks around $W = 1650 \hspace*{2pt} \mathrm{MeV}$ of the $\text{L}_{\text{max}} = 2$-curves in the $\chi^{2}$ plots of the $G$, $T$ and $\Sigma$ measurements (Figures \ref{fig:g_bins}, \ref{fig:T_bins}, \ref{fig:Sgraal_bins} and \ref{fig:Sclas_bins}). For $\sigma_{3/2}$ (Fig. \ref{fig:s32_bins}), this curve even has a pronounced peak around $W \approx 1650 \hspace*{2pt} \hspace*{2pt} \mathrm{MeV}$. Exceptions are the $E$-, $F$- and $\sigma_{1/2}$ observables which have already a good $\chi^{2}$ using a $D$-wave truncation (Figures \ref{fig:e_bins}, \ref{fig:f_bins} and \ref{fig:s12_bins}). The fits to $\sigma_{0}$ data both with and without inclusion of the systematic error (Figures \ref{fig:wq_bins} and \ref{fig:wq_bins_wsys}), as well as to the $\Sigma$ data from CLAS (Fig. \ref{fig:Sclas_bins}) even show indications for $G$-waves.
In this particular energy region, the possibility of this observation can be attributed to the high statistical precision of the latter datasets. \newline
Considering the Legendre coefficients in the third resonance region, it is seen in most cases that Bonn-Gatchina-predictions up to $F$-waves are needed for a good description. Exceptions that already agree with the $D$-wave predictions are $\left(a_{3}\right)^{\sigma_{1/2}}_{(0,\ldots,6)}$. Coefficients that even demand Bonn-Gatchina $G$-waves are $\left(a_{4}\right)^{\sigma_{0}}_{7}$, $\left(a_{4}\right)^{\sigma_{3/2}}_{5}$ and $\left(a_{4}\right)^{\Sigma_{\mathrm{CLAS}}}_{(4,7,8)}$. \newline
The latter coefficients allow an interpretation as influences of the dominant $G$-waves in the fourth resonance region (cf. Table \ref{tab:PartialWavesMultipoles}) down into the third region. The quantity $\left(a_{4}\right)^{\sigma_{0}}_{7}$ for example is a pure $\left<F,G\right>$-term in a $G$-wave truncation (see Table \ref{tab:DCSColorPlots2}). The predictions as well as fit results for this coefficient are clearly non-zero in the second half of the third resonance region. In $\left(a_{4}\right)^{\sigma_{3/2}}_{5}$, the $G$-wave correction due to $\left<P,G\right>$- and $\left<F,G\right>$ terms is admittedly small but necessary. The same is true for $\left(a_{4}\right)^{\Sigma_{\mathrm{CLAS}}}_{4}$ where the $G$-waves enter via $\left<S,G\right>$-, $\left<D,G\right>$- and $\left<G,G\right>$-terms. \newline
The observable $\sigma_{3/2}$ again facilitates the study of $F$-waves very well. The central peak of $\left(a_{4}\right)^{\sigma_{3/2}}_{2}$ around $W = 1700$ MeV in the third resonance region is completely missed by the Bonn-Gatchina prediction which is generated using up to $D$-waves. The inclusion of all partial waves with $\text{L}_{\text{max}} = 3$ makes the description perfect. In $\left(a_{4}\right)^{\sigma_{3/2}}_{3}$, the second half of the huge $\left<D,F\right>$-interference correction which was already mentioned in Sec. \ref{sec:Interpretation2ndResRegion} can be seen. The coefficient $\left(a_{4}\right)^{\sigma_{3/2}}_{4}$ also has a central peak in the third resonance region. Since it is composed solely of $\left<D,D\right>$-, $\left<P,F\right>$, and $\left<F,F\right>$-terms in an $F$-wave truncation and the $D$-wave prediction completely misses the fit result, it is tempting to associate this pronounced peak with the dominant $\ast \ast \ast \ast$-resonance $N(1680) \frac{5}{2}^{+}$. The maximum of the peak is even positioned very close to the resonance mass. The latter statement is also true with slightly less precision for the peak of the $\left(a_{4}\right)^{\sigma_{3/2}}_{6}$-coefficient. This quantity is given solely as an $\left<F,F\right>$-term for $\text{L}_{\text{max}}=3$. Its non-vanishing means again clear evidence for non-trivial $F$-wave contributions in the third resonance region. \newline
Further Legendre coefficients whose structure is given in terms of leading $\left<F,F\right>$-interferences are $\left(a_{3}\right)^{\check{E}}_{6}$, $\left(a_{3}\right)^{\check{G}}_{6}$ and $\left(a_{4}\right)^{\check{T}}_{6}$. While for the $E$-coefficient the errors are too large to confirm a signal of $F$-wave resonances, the $T$-coefficient, which is not zero, shows further good evidence for the dominant $F$-wave resonances. \newline
The physical interpretation of the results in the third resonance region is clearly that $F$-wave contributions can no longer be neglected and are important. These are stemming mainly from the well-established $F$-wave resonance with lowest mass, the $N(1680) \frac{5}{2}^{+}$ (cf. Fig. \ref{fig:resonanzen}). However there is also a $\ast \ast$-$F$-wave state, the $N(1860) \frac{5}{2}^{+}$ and a well-confirmed Delta state $\Delta(1905) \frac{5}{2}^{+}$ ($\ast \ast \ast \ast$) (for both cf. Table \ref{tab:PartialWavesMultipoles}) whose influence may be felt in the data here as well. Observables obtained by very precise measurements, i.e. the $\sigma_{0}$ and $\Sigma_{\mathrm{CLAS}}$ data, or the $\sigma_{3/2}$ cross section which by definition is expected to show a larger sensitivity to higher partial waves, show the influence of $G$-waves in the context of the comparison with Bonn-Gatchina predictions. Unsurprisingly, in almost all cases the $G$-waves enter via interferences with lower partial waves. One exception is the coefficient $\left(a_{4}\right)^{\check{\Sigma}_{\mathrm{CLAS}}}_{8}$, which also has $\left< G, G \right>$-contributions.

\subsection{Fourth resonance region $\left( 1800 \hspace*{2pt} \mathrm{MeV} \lesssim W \lesssim 2250 \hspace*{2pt} \mathrm{MeV} \right)$} \label{sec:Interpretation4thResRegion}

Within the region discussed here, all datasets that were discussed in Sec. \ref{sec:Interpretation3rdResRegion} are considered except for the $F$ and $G$ measurements, which do not have a very significant overlap. \newline
The $\chi^{2}$ plots mostly show clear evidence for $F$-waves and even small sensitivity to $G$-waves in selected bins. The latter statement is true for all observables. Datasets with a very large precision can even show indications of $H$-waves ($\text{L}_{\text{max}} = 5$). These can be seen for the fits of $\sigma_{0}$ (excluding the systematic error, Fig. \ref{fig:wq_bins}) and $\check{\Sigma}_{\mathrm{CLAS}}$ (Fig. \ref{fig:Sclas_bins}). The fit to $\sigma_{0}$ without systematic errors even suggests significant contributions beyond the $H$-waves for energies $W > 1900 \hspace*{2pt} \mathrm{MeV}$ (Fig. \ref{fig:wq_bins}). Once the systematic errors are included into the fitting (Fig. \ref{fig:wq_bins_wsys}), the $\text{L}_{\text{max}}$ implied by the $\chi^{2}$ plot is lowered significantly to $4$. In the highest mass region, for $W > 1900 \hspace*{2pt} \mathrm{MeV}$, it is even lowered to $3$. Therefore, in this highest energy region of the $\sigma_{0}$ measurement, the systematic errors are again important and their influence cannot be neglected. \newline
The latter fact can be confirmed by looking at the Legendre coefficients of $\sigma_{0}$. For $W < 1900 \hspace*{2pt} \mathrm{MeV}$, $\left(a_{4}\right)^{\sigma_{0}}_{(0,1,2,5,6)}$ are already well described using only up to Bonn-Gatchina $F$-waves. For $\left(a_{4}\right)^{\sigma_{0}}_{(3,4,7,8)}$, the Bonn-Gatchina description truncated even at $\text{L}_{\text{max}} = 4$ is slightly off. Those statements also hold for the fits including the systematic errors (Fig. \ref{fig:wq_bins_wsys}). The importance of the coefficient $\left(a_{4}\right)^{\sigma_{0}}_{7}$ shall be emphasised here, since it is a pure $\left<F,G\right>$-term in a $G$-wave truncation and clearly shows the non-trivial influence of $G$-waves in the fourth resonance region. This illustrates the potential within polarization measurements, namely to detect missing resonances by their interferences with well-established states, here for example the $\Delta (1905) \frac{5}{2}^{+}$, which couples to the $F_{35}$-wave. In the energy region beyond $W = 1900 \hspace*{2pt} \mathrm{MeV}$ however, all fit results for the higher Legendre coefficients $\left(a_{4}\right)^{\sigma_{0}}_{(3,\ldots,8)}$ from the pruely statistical fit show large discrepancies to the Bonn-Gatchina predictions (Fig. \ref{fig:wq_bins}). Taking into account the systematic errors substantially increases the errors of the Legendre coefficients in this highest energy regime (Fig. \ref{fig:wq_bins_wsys}). The disagreement with Bonn-Gatchina is still there, however due to the large errors it is less significant. \newline
The Legendre coefficients of the remaining observables are in most cases already well described using Bonn-Gatchina predictions up to $F$-waves. Exceptions that require a correction due to $G$-waves are $\left(a_{4}\right)^{\sigma_{3/2}}_{(3,5,7)}$, $\left(a_{4}\right)^{\sigma_{1/2}}_{7}$, $\left(a_{4}\right)^{\check{T}}_{7}$, \newline $\left(a_{4}\right)^{\check{\Sigma}_{\mathrm{GRAAL}}}_{7}$ and $\left(a_{4}\right)^{\check{\Sigma}_{\mathrm{CLAS}}}_{(4,5,6)}$. In both $\left(a_{4}\right)^{\sigma_{3/2}}_{(3,5)}$, $G$-waves enter as $\left<P,G\right>$- and $\left<F,G\right>$-interferences, the contribution of which cannot be neglected in the lower third of the fourth resonance region. For the higher energies, $\left(a_{4}\right)^{\sigma_{3/2}}_{3}$ is again well described by an $F$-wave truncation, while for $\left(a_{4}\right)^{\sigma_{3/2}}_{5}$ even truncations beyond $G$-waves were not found to be able to describe the coefficient well.
For $\left(a_{4}\right)^{\sigma_{3/2}}_{7}$, $\left(a_{4}\right)^{\sigma_{1/2}}_{7}$, $\left(a_{4}\right)^{\check{T}}_{7}$ and $\left(a_{4}\right)^{\check{\Sigma}_{\mathrm{GRAAL}}}_{7}$, $G$-waves enter purely via $\left<F,G\right>$-terms, which clearly cannot be neglected in the fourth resonance region. Nontrivial corrections can also be found for $\left(a_{4}\right)^{\check{\Sigma}_{\mathrm{CLAS}}}_{4}$ via $\left<S,G\right>$-, $\left<D,G\right>$- and $\left<G,G\right>$-terms. For $\left(a_{4}\right)^{\check{\Sigma}_{\mathrm{CLAS}}}_{5}$, $\left<P,G\right>$- and $\left<F,G\right>$-in\-ter\-fe\-ren\-ces become important. \newline
The coefficients $\left(a_{4}\right)^{\check{\Sigma}_{\mathrm{CLAS}}}_{(7,8)}$ are worth mentioning since they do not have a good description by the $G$-wave truncation, which is however improved drastically once $H$-waves are included. The $G$-wave prediction even gets the sign of $\left(a_{4}\right)^{\check{\Sigma}_{\mathrm{CLAS}}}_{8}$ wrong, an issue which is resolved by the $H$-waves. However, this discrepancy between the $G$-wave predictions and the data, as well as the apparent improvement upon including $H$-waves, may both very well be an artifact of our usage of the solution BnGa2014-02. In a more recent fit, the Bonn-Gatchina group was able to describe the $\Sigma_{\mathrm{CLAS}}$- and other data better by including a new $G$-wave state, namely $\Delta(2200) \frac{7}{2}^{-}$ \cite{BnGaNewGResonance}. \newline
As a physical conclusion, it can be said that the fourth resonance region is clearly dominated by $F$-waves. However, traces of small but non-negligible $G$-wave contributions can be found in the $\chi^{2}$ plots as well as the comparison to the Bonn-Gatchina PWA. Significant bumps that are produced by $\left<F,F\right>$-terms and may therefore be attributed to the resonances $N(1860) \frac{5}{2}^{+}$ ($\ast \ast$) and $\Delta(1905) \frac{5}{2}^{+}$ ($\ast \ast \ast \ast$) occur in the coefficients $\left(a_{3}\right)^{\check{E}}_{6}$, $\left(a_{4}\right)^{\sigma_{3/2}}_{6}$ and $\left(a_{4}\right)^{\check{T}}_{6}$.

\section{Conclusions and Outlook} \label{sec:Conclusions}

This work has given a detailed description of a method to infer the dominant partial wave contributions from po\-la\-ri\-za\-tion measurements, by fitting a Legendre parametrization to angular distributions and investigating the change in $\chi^{2}/\mathrm{ndf}$ for different truncation angular momenta $\text{L}_{\text{max}}$. \newline
For the proposed parametrization in terms of associated Legendre polynomials $P_{\ell}^{m} \left(\cos \theta\right)$, the detailed composition of every considered Legendre coefficient $\left(a_{\text{L}_{\text{max}}}\right)^{\check{\Omega}^{\alpha}}_{k}$ (belonging to a particular observable $\check{\Omega}^{\alpha}$) as a bilinear hermitean form in the multipoles was also provided. The analysis method was then applied in a survey of recent po\-la\-ri\-za\-tion measurements of single-spin and beam-target observables in $\gamma p \rightarrow \pi^{0} p$ \cite{Adlarson:2015}, \cite{GRAAL,Dugger:2013}, \cite{Hartmann:2014,Hartmann:2015}, \cite{Hartmann:2014,Hartmann:2015}, \cite{Thiel:2012,Thiel:2015}, \cite{Gottschall:2014,Gottschall:2015}, \cite{Hartmann:2014,Hartmann:2015} and \cite{AnnandEtAl:2016}). The results of the analysis were also compared to predictions from the Bonn-Gatchina PWA truncated at different $\text{L}_{\text{max}}$ (Lmax =1: S and P-waves only; Lmax =2: S, P and D-waves; Lmax =3: S, P, D and F-waves; Lmax =4: S, P, D, F and G-waves; Lmax =5: S, P, D, F, G and H-waves;). The interpretation of these comparisons also exposed the advantages and disadvantages of the method. \newline
On the one hand the method is simple and the analysis is quickly executed. Furthermore, the whole scheme is generalizable to $2$-body reactions that have either more or less involved spin-structures compared to photoproduction. Examples for such processes are $\pi N$-scattering, or the electroproduction of a single pseudoscalar meson. \newline
The procedure is furthermore directly sensitive to the precision as well as, primarily in the resulting Legendre coefficients, to the kinematic coverage of the data. Therefore it reflects the quality of the considered datasets in just a few steps. Furthermore, the method reliably determines the order $\text{L}_{\text{max}}$ beyond which the higher Legendre coefficients are supressed. \newline
The comparison with the Bonn-Gatchina PWA on the other hand has revealed that the $\chi^{2}$ criterion is in some cases not sensitive to partial wave interferences in the lower, non-supressed Legendre coefficients. Some of the considered datasets seemed to be very sensitive to interferences of small higher partial waves with dominant lower ones, which then were seen to still yield significant contributions in the lower $\left(a_{\text{L}_{\text{max}}}\right)^{\check{\Omega}^{\alpha}}_{k}$. The $H$ measurement is a particularly extreme example. \newline \newline
However, po\-la\-ri\-za\-tion observables have in the past been proposed as meaningful quantities precisely because of these interferences, and that the latter can give important information even on the supressed higher partial waves. Observables, at least in the context of this study, have shown to be sensitive to high-low partial wave interferences for two reasons: \newline
The first would be a very large precision in the measurement of the observable and correspondingly very small errors. The $\sigma_{0}$ data of the A2 collaboration and the $\Sigma$ measurement from CLAS have shown up to have this property. \newline
As a second reason, it may also be that the given observable is already sensitive by way of its physical definition. An example that has shown up in this work is the spin dependent cross section $\sigma_{3/2}$, which is purged from contributions of all multipoles with $J = 1/2$ and therefore also from all resonances with $J^{P} = \frac{1}{2}^{+}, \hspace*{1.5pt} \frac{1}{2}^{-}$ quantum numbers (or multipoles $E_{0+}$ and $M_{1-}$, respectively). \newline
The question now is: How does one proceed from the fin\-dings obtained by the proposed analysis method? At most, the energy-independent truncated partial wave analysis already mentioned in the introduction would suggest itself. It is then of interest how to choose the truncation angular momentum $\text{L}_{\text{max}}$ for such an energy independent multipole fit. Based on the results obtained in this work, the following procedure is recommended. \newline
In case a set of observables is fitted it should, for a truncated PWA be at least a mathematically complete set (cf. \cite{WBT}). Examples for such sets are $\left\{\sigma_{0}, \Sigma, T, P, G\right\}$ and $\left\{\sigma_{0}, \Sigma, T, P, F\right\}$. By this criterion, the data investigated in this work are already a mathematically overcomplete set.\newline \newline
Then one should investigate the $\chi^{2}$-plots of all those observables in the energy region where the datasets overlap. This yields the estimate $\text{L}_{\text{max}}^{\chi^{2}}$ from the observable that needs the highest truncation in order to get a good $\chi^{2}$. Once multipoles are fitted however, one should at least choose $\left(\text{L}_{\text{max}}^{\chi^{2}} + 1\right)$ for the multipole fit. The multipoles corresponding to the additional order may then be either fitted as well, or fixed to specific parameters from a model amplitude. \newline
For the datasets considered in this analysis, in case one makes a fit at the highest accessible energies, the considerations of the observable $\Sigma^{\mathrm{CLAS}}$ specifically suggest to analyze multipoles up to the $H$-waves. \newline
Whatever the further analysis steps are supposed to be, it is clear that an introductory survey according to the methods shown in this work is a useful first step. It can serve as a guideline for either a truncated single energy fit, or any other kind of fitting of the data. \newline 

\hspace*{1.5pt} \textbf{Acknowledgments} \newline

\hspace*{1.5pt} This work was supported by the Deutsche Forschungsgemeinschaft (SFB/TR16)
   and the European Community-Research Infrastructure Activity (FP7). \\
   The authors would like to thank Eberhard Klempt, Andrey Sarantsev, Victor Nikonov, Alexei V. Anisovich and Ulrike Thoma for fruitful discussions.

\appendix
%

\section{Partial wave contributions to angular fit coefficients} 
\label{sec:PWContentFormulas}
\vspace*{0cm}
In this appendix, we extend the notations and ideas introduced in section \ref{sec:DescriptionCompositionLegCoeffs} to the largest truncation angular momentum needed for all the interpretations in this work, i.e. $\text{L}_{\text{max}} = 5$. The observable $E$ shall again serve as an illustration. For the above mentioned truncation at the $H$-waves, the angular pa\-ra\-me\-tri\-za\-tion (\ref{eq:LowEAssocLegParametrizationE}) reads
\vspace*{0cm}
\begin{align}
 \check{E} &= \rho \hspace*{3pt} \Big( \left(a_{5}\right)^{\check{E}}_{0} \hspace*{1.5pt} P_{0} ( \cos \theta ) + \left(a_{5}\right)^{\check{E}}_{1} \hspace*{1.5pt} P_{1} ( \cos \theta )  \nonumber \\
  & \hspace*{22.5pt} + \left(a_{5}\right)^{\check{E}}_{2} \hspace*{1.5pt} P_{2} ( \cos \theta ) + \left(a_{5}\right)^{\check{E}}_{3} \hspace*{1.5pt} P_{3} ( \cos \theta )  \nonumber \\
  & \hspace*{22.5pt} + \left(a_{5}\right)^{\check{E}}_{4} \hspace*{1.5pt} P_{4} ( \cos \theta ) + \left(a_{5}\right)^{\check{E}}_{5} \hspace*{1.5pt} P_{5} ( \cos \theta )  \nonumber \\
  & \hspace*{22.5pt} + \left(a_{5}\right)^{\check{E}}_{6} \hspace*{1.5pt} P_{6} ( \cos \theta ) + \left(a_{5}\right)^{\check{E}}_{7} \hspace*{1.5pt} P_{7} ( \cos \theta )  \nonumber \\
  & \hspace*{22.5pt} + \left(a_{5}\right)^{\check{E}}_{8} \hspace*{1.5pt} P_{8} ( \cos \theta ) + \left(a_{5}\right)^{\check{E}}_{9} \hspace*{1.5pt} P_{9} ( \cos \theta )  \nonumber \\
  & \hspace*{22.5pt} + \left(a_{5}\right)^{\check{E}}_{10} \hspace*{1.5pt} P_{10} ( \cos \theta ) \Big) \mathrm{.} \label{eq:ExpEIntoLegLmax5}
\end{align}
Multipoles are now collected into a vector $\left| \mathcal{M}_{5} \right>$, which is defined in analogy to the vector $\left| \mathcal{M}_{2} \right>$ of equation (\ref{eq:MultipoleVectorLmax2}) in section \ref{sec:DescriptionCompositionLegCoeffs}, i.e.
\vspace*{0cm}
\begin{align}
\left| \mathcal{M}_{5} \right> &= \Big( E_{0+}, E_{1+}, M_{1+}, M_{1-}, E_{2+}, E_{2-}, M_{2+}, M_{2-}, \nonumber \\
 & \hspace*{20pt} E_{3+}, E_{3-}, M_{3+}, M_{3-}, E_{4+}, E_{4-}, M_{4+}, M_{4-}, \nonumber \\
& \hspace*{20pt} E_{5+}, E_{5-}, M_{5+}, M_{5-} \Big)^{T} \mathrm{.} \label{eq:MultipoleVectorLmax5}
\end{align}
Each higher $\ell$ was included in the pattern $\big( E_{\ell+}, E_{\ell-}, M_{\ell+},$ $M_{\ell-}\big)$. Since now 12 additional complex multipoles coming from the $F$-, $G$- and $H$-waves contribute as compared to the $D$-wave truncation discussed in the main text, the formulas defining the Legendre coefficients $\left(a_{5}\right)^{\check{E}}_{k}$ appearing in equation (\ref{eq:ExpEIntoLegLmax5}) become a lot more involved. The basic structure of bilinear hermitean forms has however not changed, of course. \newline
In the simplified scalar-product notation introduced in section \ref{sec:DescriptionCompositionLegCoeffs}, the partial-wave compositions of the Legendre-coefficients of $E$ can in an $H$-wave truncation be abbreviated as
\vspace*{0cm}
{\allowdisplaybreaks
\begin{align}
 \left(a_{5}\right)^{\check{E}}_{0} &= \left<S,S\right> +\left<P,P\right> + \left<D,D\right> \\ 
                   & \hspace*{11.25pt} + \left<F,F\right> + \left<G,G\right> + \left<H,H\right>\mathrm{,} \label{eq:EScalarProductLmax5Coeff0} \\
 \left(a_{5}\right)^{\check{E}}_{1} &= \left<S,P\right> +\left<P,D\right> + \left<D,F\right> \\ 
                   & \hspace*{11.25pt} + \left<F,G\right> +  \left<G,H\right>  \mathrm{,} \label{eq:EScalarProductLmax5Coeff1} \\
 \left(a_{5}\right)^{\check{E}}_{2} &= \left<P,P\right> +\left<S,D\right> + \left<D,D\right>  \nonumber \\
                   & \hspace*{11.25pt} + \left<P,F\right> + \left<F,F\right> + \left<D,G\right> \\
                   & \hspace*{11.25pt} + \left<G,G\right> + \left<F,H\right> + \left<H,H\right> \mathrm{,} \label{eq:EScalarProductLmax5Coeff2} \\
 \left(a_{5}\right)^{\check{E}}_{3} &= \left<P,D\right> +\left<S,F\right> + \left<D,F\right>  \nonumber \\
                   & \hspace*{11.25pt} + \left<P,G\right> + \left<F,G\right> + \left<D,H\right> \\
                   & \hspace*{11.25pt} + \left<G,H\right> \mathrm{,} \label{eq:EScalarProductLmax5Coeff3} \\
 \left(a_{5}\right)^{\check{E}}_{4} &= \left<D,D\right> +\left<P,F\right> + \left<F,F\right>  \nonumber \\
                   & \hspace*{11.25pt} + \left<S,G\right> + \left<D,G\right> + \left<G,G\right> \\
                   & \hspace*{11.25pt} + \left<P,H\right> + \left<F,H\right> + \left<H,H\right> \mathrm{,} \label{eq:EScalarProductLmax5Coeff4} \\
 \left(a_{5}\right)^{\check{E}}_{5} &= \left<D,F\right> +\left<P,G\right> + \left<F,G\right> \\
                   & \hspace*{11.25pt} + \left<S,H\right> + \left<D,H\right> + \left<G,H\right>  \mathrm{,} \label{eq:EScalarProductLmax5Coeff5} \\
 \left(a_{5}\right)^{\check{E}}_{6} &= \left<F,F\right> +\left<D,G\right> + \left<G,G\right>  \nonumber \\
                   & \hspace*{11.25pt} + \left<P,H\right> + \left<F,H\right> + \left<H,H\right> \mathrm{,} \label{eq:EScalarProductLmax5Coeff6} \\
 \left(a_{5}\right)^{\check{E}}_{7} &= \left<F,G\right> +\left<D,H\right> +\left<G,H\right> \mathrm{,} \label{eq:EScalarProductLmax5Coeff7} \\
 \left(a_{5}\right)^{\check{E}}_{8} &= \left<G,G\right> + \left<F,H\right> + \left<H,H\right> \mathrm{,} \label{eq:EScalarProductLmax5Coeff8} \\
 \left(a_{5}\right)^{\check{E}}_{9} &= \left<G,H\right> \mathrm{.} \label{eq:EScalarProductLmax5Coeff9} \\
 \left(a_{5}\right)^{\check{E}}_{10} &= \left<H,H\right> \mathrm{.} \label{eq:EScalarProductLmax5Coeff10}
\end{align}
}
The Legendre coefficients $\left(a_{5}\right)^{\check{E}}_{(0,\ldots,10)}$ of $\check{E}$ in an $H$-wave truncation are formed by composing bilinear forms out of the $\mathcal{C}_{(0,\ldots,10)}^{\check{E}}$ via the rules outlined in section \ref{sec:DescriptionCompositionLegCoeffs}. Tables \ref{tab:EColorPlots1} to \ref{tab:EColorPlots3} shows them plotted in the color scheme. The Tables \ref{tab:CoeffsEright1} to \ref{tab:CoeffsEright9} of section \ref{sec:DescriptionCompositionLegCoeffs}, defining $\check{E}$ in a truncation at $\text{L}_{\text{max}}=3$, can be found in Tables \ref{tab:EColorPlots1} to \ref{tab:EColorPlots3} in all blocks that define non-vanishing interferences among $S$-, $P$-, $D$- and $F$-waves (to be found just in the color plots corresponding to $\left(a_{3}\right)^{\check{E}}_{(0,\ldots,6)}$, of course). \newline
The remaining po\-la\-ri\-za\-tion observables investigated in this work are represented in the same color scheme in Tables \ref{tab:DCSColorPlots1} to \ref{tab:FColorPlots3}. Those pictures constitute a compact and quick reference showing the composition of Legendre coefficients in terms of multipoles. In particular, they show which kind of partial wave interferences are allowed in a particular Legendre coefficient, and which are not. Therefore, they are referenced repeatedly in the main text. \newline
One special property of all matrices representing Legendre coefficients of observables defined by a real part in Table \ref{tab:ObsInTermsOfCGLN} (i.e. $\left(\sigma_{0}, \hspace*{1.5pt} \Sigma, \hspace*{1.5pt} E, \hspace*{1.5pt} \sigma_{3/2}, \hspace*{1.5pt} \sigma_{1/2} \hspace*{1.5pt}\right)$ in this case here), is that they are symmetric. The have to have this property in each truncation order, such that the resulting observables are real. \newline
In case observables are defined by an imaginary part (i.e. $\left(T, \hspace*{1.5pt} P, \hspace*{1.5pt} G, \hspace*{1.5pt} H\right)$), the matrices corresponding to the Legendre coefficients are hermitean for the same reason. Therefore, each matrix entry for the latter case has a factor of $i$, which is however not shown explicitly in the color scheme.

\begin{table*}[htb]
\RawFloats
\begin{minipage}{.075\linewidth}
\vspace*{-6.5pt}
\hspace*{5pt}
\begin{equation}
\mathcal{C}_{0}^{\sigma_0} \equiv \nonumber
\end{equation}
\end{minipage}
\begin{minipage}{.3\linewidth} \vspace*{0.572cm} \includegraphics[width=0.875\textwidth]{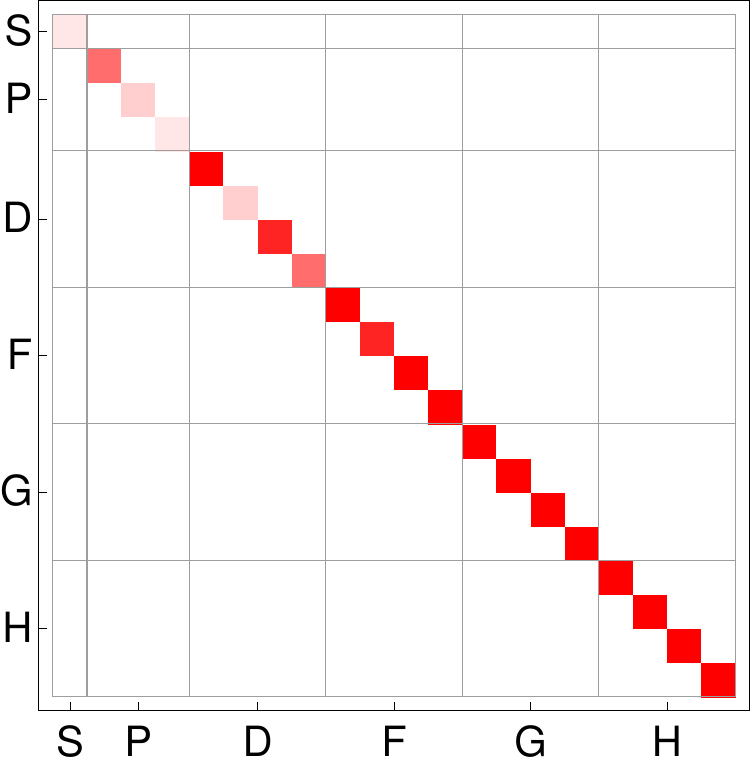} \end{minipage}
\begin{minipage}{.35\linewidth} \vspace*{0.500cm} \hspace*{-0.65cm}\includegraphics[width=1.15\textwidth]{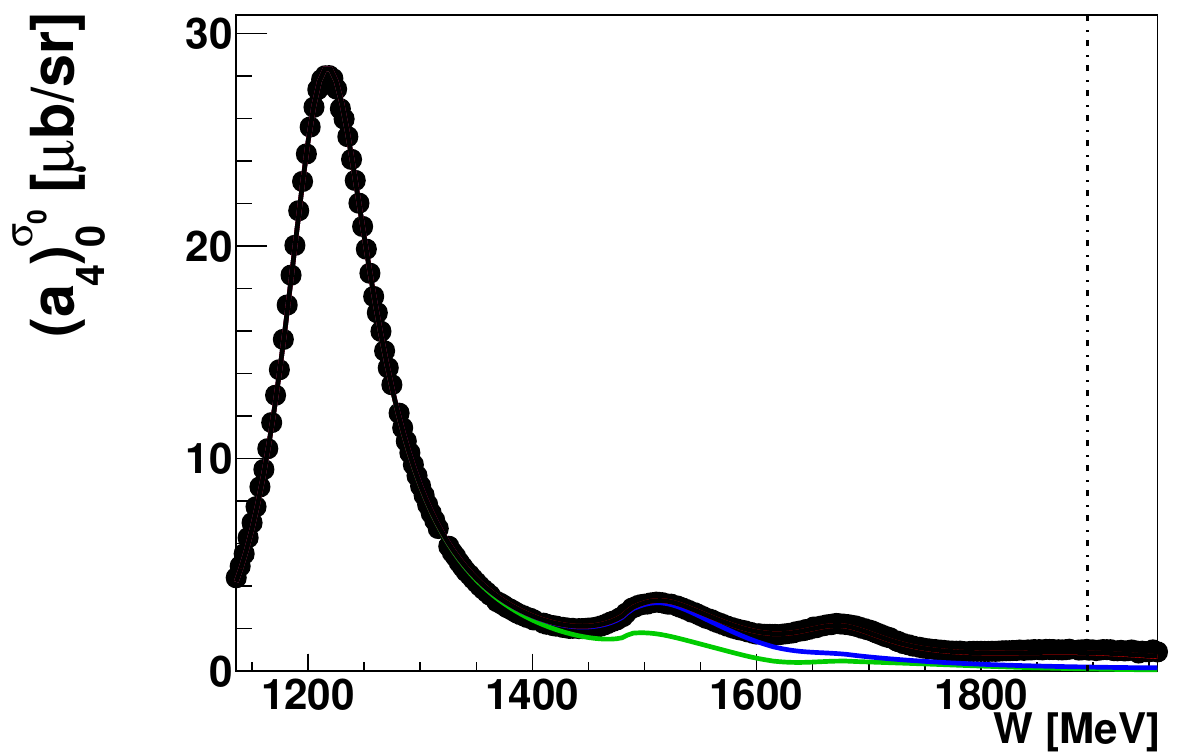}\end{minipage}
\begin{minipage}{.25\linewidth} \begin{align} \left(a_{5}\right)^{\sigma_0}_{0} &= \left<S,S\right> + \left<P,P\right> \nonumber \\ & \hspace*{12.5pt} + \left<D,D\right>  + \left<F,F\right>   \nonumber \\ & \hspace*{12.5pt} + \left<G,G\right>  + \left<H,H\right>  \nonumber \end{align} \end{minipage}

\begin{minipage}{.075\linewidth}
\vspace*{-6.5pt}
\hspace*{5pt}
\begin{equation}
\mathcal{C}_{1}^{\sigma_0} \equiv \nonumber
\end{equation}
\end{minipage}
\begin{minipage}{.3\linewidth} \vspace*{0.572cm} \includegraphics[width=0.875\textwidth]{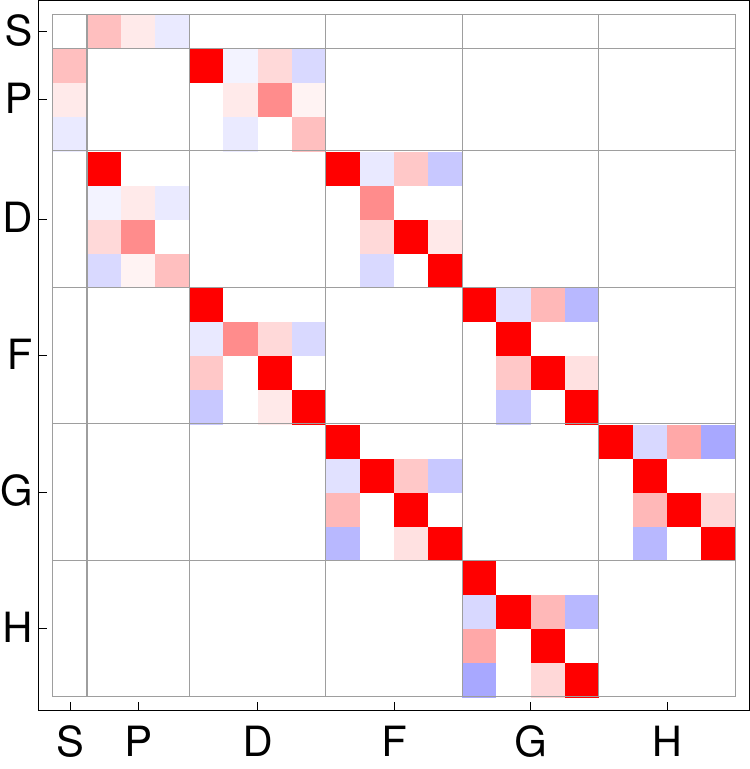} \end{minipage}
\begin{minipage}{.35\linewidth} \vspace*{0.500cm} \hspace*{-0.65cm}\includegraphics[width=1.15\textwidth]{WQ_l4_coeff_1_wsys.pdf}\end{minipage}
\begin{minipage}{.25\linewidth} \begin{align} \left(a_{5}\right)^{\sigma_0}_{1} &= \left<S,P\right> + \left<P,D\right> \nonumber \\ & \hspace*{12.5pt} + \left<D,F\right>  + \left<F,G\right> \nonumber \\ & \hspace*{12.5pt}  + \left<G,H\right>  \nonumber \end{align} \end{minipage}

\begin{minipage}{.075\linewidth}
\vspace*{-6.5pt}
\hspace*{5pt}
\begin{equation}
\mathcal{C}_{2}^{\sigma_0} \equiv \nonumber
\end{equation}
\end{minipage}
\begin{minipage}{.3\linewidth} \vspace*{0.572cm} \includegraphics[width=0.875\textwidth]{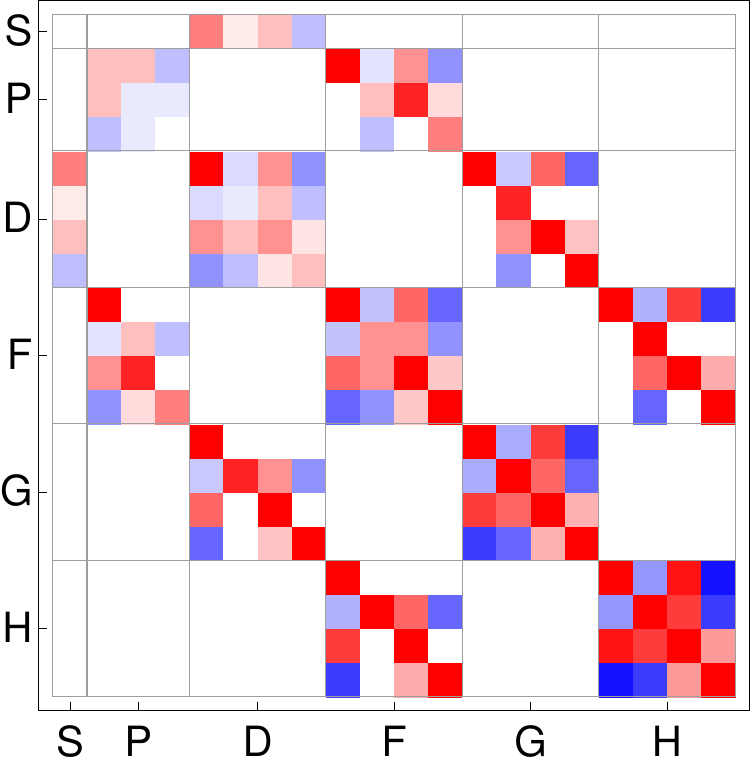} \end{minipage}
\begin{minipage}{.35\linewidth} \vspace*{0.500cm} \hspace*{-0.65cm}\includegraphics[width=1.15\textwidth]{WQ_l4_coeff_2_wsys.pdf}\end{minipage}
\begin{minipage}{.25\linewidth} \begin{align} \left(a_{5}\right)^{\sigma_0}_{2} &= \left<S,D\right> + \left<P,P\right> \nonumber \\ & \hspace*{12.5pt} + \left<P,F\right>  + \left<D,D\right> \nonumber \\ & \hspace*{12.5pt} + \left<D,G\right>  + \left<F,F\right>\nonumber \\ & \hspace*{12.5pt} + \left<F,H\right>  + \left<G,G\right> \nonumber \\ & \hspace*{12.5pt} + \left<H,H\right>\nonumber \end{align} \end{minipage}

\begin{minipage}{.075\linewidth}
\vspace*{-6.5pt}
\hspace*{5pt}
\begin{equation}
\mathcal{C}_{3}^{\sigma_0} \equiv \nonumber
\end{equation}
\end{minipage}
\begin{minipage}{.3\linewidth} \vspace*{0.572cm} \includegraphics[width=0.875\textwidth]{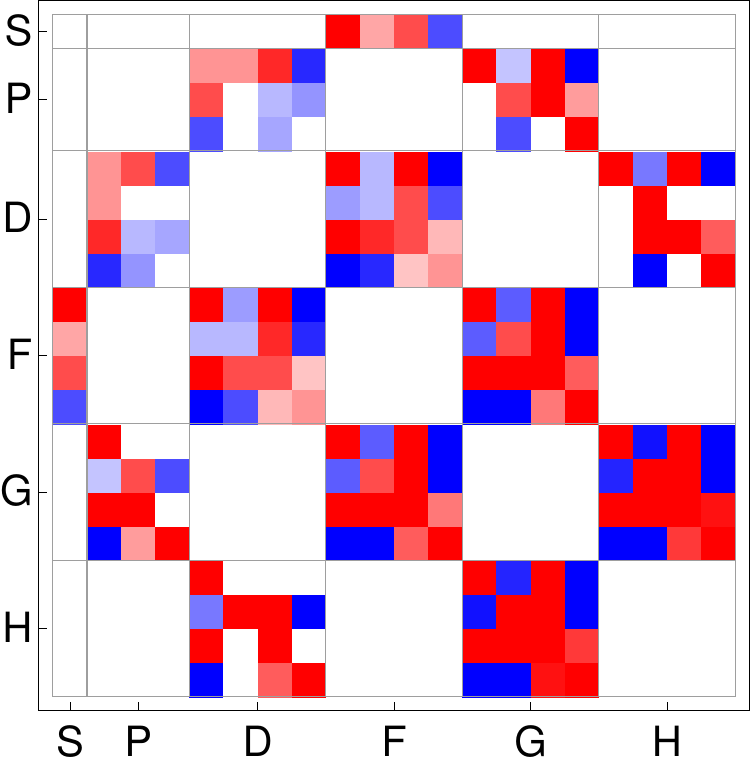} \end{minipage}
\begin{minipage}{.35\linewidth} \vspace*{0.500cm} \hspace*{-0.65cm}\includegraphics[width=1.15\textwidth]{WQ_l4_coeff_3_wsys.pdf}\end{minipage}
\begin{minipage}{.25\linewidth} \begin{align} \left(a_{5}\right)^{\sigma_0}_{3} &= \left<S,F\right> + \left<P,D\right> \nonumber \\ & \hspace*{12.5pt} + \left<D,F\right>  + \left<F,G\right> \nonumber \\ & \hspace*{12.5pt} + \left<P,G\right>  + \left<D,H\right>\nonumber \\ & \hspace*{12.5pt} + \left<G,H\right> \nonumber \end{align} \end{minipage}
\caption{%
Left: Matrices $\mathcal{C}_{0\cdots 3}^{\sigma_0}$, represented here in the color scheme, defines the coefficient $\left(a_{5}\right)_{0\cdots 3}^{\sigma_0}$ for an expansion of $\sigma_0$ up to $\text{L}_{\text{max}} = 5$. Center: Coefficients $\left(a_{4}\right)_{0\cdots 3}^{\sigma_0}$ obtained from a fit to the $\sigma_0$-data (black points). For references to the data see Table \ref{tab:DataBasis}. Bonn Gatchina predictions, truncated at different $\text{L}_{\mathrm{max}}$ ($\text{L}_{\mathrm{max}} = 1$ is drawn in green, $\text{L}_{\mathrm{max}} = 2$ in blue, $\text{L}_{\mathrm{max}} = 3$ in red and $\text{L}_{\mathrm{max}} = 4$ in black) are drawn as well. Right: All partial wave interferences for $\text{L}_{\text{max}} = 5$ are indicated.
}
\label{tab:DCSColorPlots1}
\end{table*}

\begin{table*}[htb]
\RawFloats
\begin{minipage}{.075\linewidth}
\vspace*{-6.5pt}
\hspace*{5pt}
\begin{equation}
\mathcal{C}_{4}^{\sigma_0} \equiv \nonumber
\end{equation}
\end{minipage}
\begin{minipage}{.3\linewidth} \vspace*{0.572cm} \includegraphics[width=0.875\textwidth]{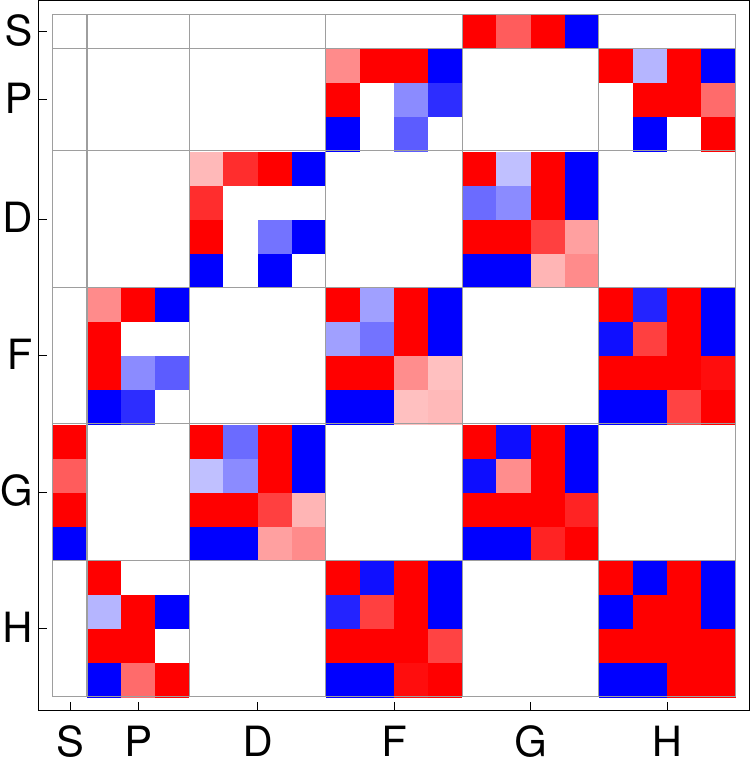} \end{minipage}
\begin{minipage}{.35\linewidth} \vspace*{0.500cm} \hspace*{-0.65cm}\includegraphics[width=1.15\textwidth]{WQ_l4_coeff_4_wsys.pdf}\end{minipage}
\begin{minipage}{.25\linewidth} \begin{align} \left(a_{5}\right)^{\sigma_0}_{4} &= \left<P,F\right> + \left<S,G\right> \nonumber \\ & \hspace*{12.5pt} + \left<P,H\right>  + \left<D,D\right>    \nonumber \\ & \hspace*{12.5pt} + \left<D,G\right>  + \left<F,F\right>    \nonumber \\ & \hspace*{12.5pt} + \left<F,H\right> + \left<G,G\right>  \nonumber \\ & \hspace*{12.5pt}   + \left<H,H\right>   \nonumber \end{align} \end{minipage}

\begin{minipage}{.075\linewidth}
\vspace*{-6.5pt}
\hspace*{5pt}
\begin{equation}
\mathcal{C}_{5}^{\sigma_0} \equiv \nonumber
\end{equation}
\end{minipage}
\begin{minipage}{.3\linewidth} \vspace*{0.572cm} \includegraphics[width=0.875\textwidth]{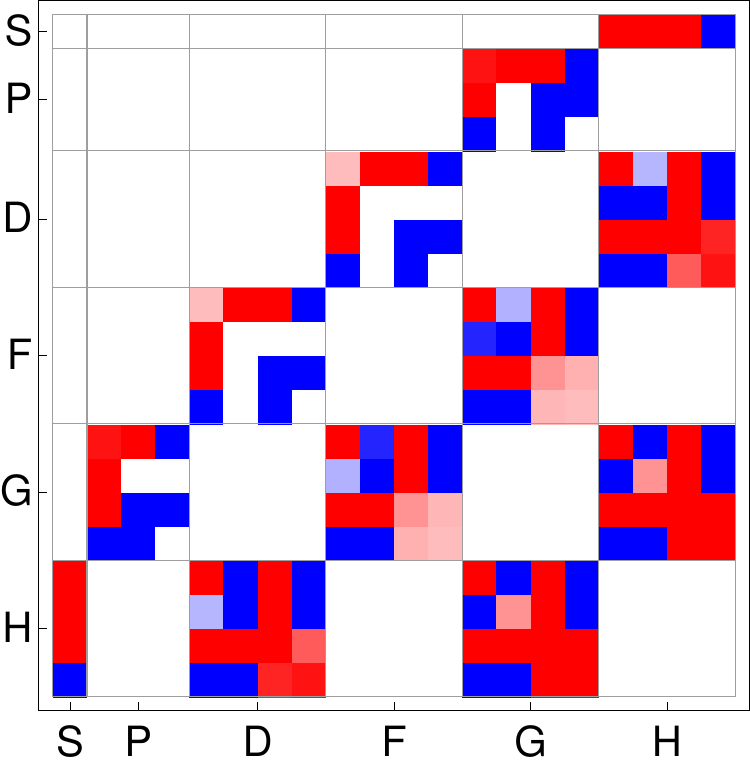} \end{minipage}
\begin{minipage}{.35\linewidth} \vspace*{0.500cm} \hspace*{-0.65cm}\includegraphics[width=1.15\textwidth]{WQ_l4_coeff_5_wsys.pdf}\end{minipage}
\begin{minipage}{.25\linewidth} \begin{align} \left(a_{5}\right)^{\sigma_0}_{5} &= \left<S,H\right> + \left<P,G\right> \nonumber \\ & \hspace*{12.5pt} + \left<D,F\right>  + \left<D,H\right>   \nonumber \\ & \hspace*{12.5pt} + \left<F,G\right>  + \left<G,H\right>  \nonumber  \end{align} \end{minipage}

\begin{minipage}{.075\linewidth}
\vspace*{-6.5pt}
\hspace*{5pt}
\begin{equation}
\mathcal{C}_{6}^{\sigma_0} \equiv \nonumber
\end{equation}
\end{minipage}
\begin{minipage}{.3\linewidth} \vspace*{0.572cm} \includegraphics[width=0.875\textwidth]{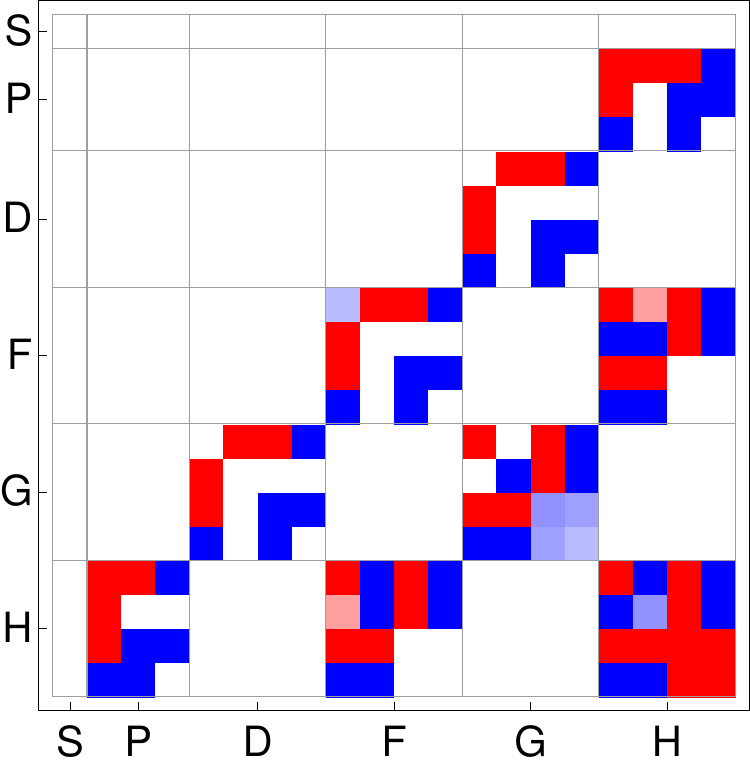} \end{minipage}
\begin{minipage}{.35\linewidth} \vspace*{0.500cm} \hspace*{-0.65cm}\includegraphics[width=1.15\textwidth]{WQ_l4_coeff_6_wsys.pdf}\end{minipage}
\begin{minipage}{.25\linewidth} \begin{align} \left(a_{5}\right)^{\sigma_0}_{6} &= \left<P,H\right> + \left<D,G\right> \nonumber \\ & \hspace*{12.5pt} + \left<F,F\right>  + \left<F,H\right> \nonumber \\ & \hspace*{12.5pt} + \left<G,G\right>  + \left<H,H\right>\nonumber \end{align} \end{minipage}

\begin{minipage}{.075\linewidth}
\vspace*{-6.5pt}
\hspace*{5pt}
\begin{equation}
\mathcal{C}_{7}^{\sigma_0} \equiv \nonumber
\end{equation}
\end{minipage}
\begin{minipage}{.3\linewidth} \vspace*{0.572cm} \includegraphics[width=0.875\textwidth]{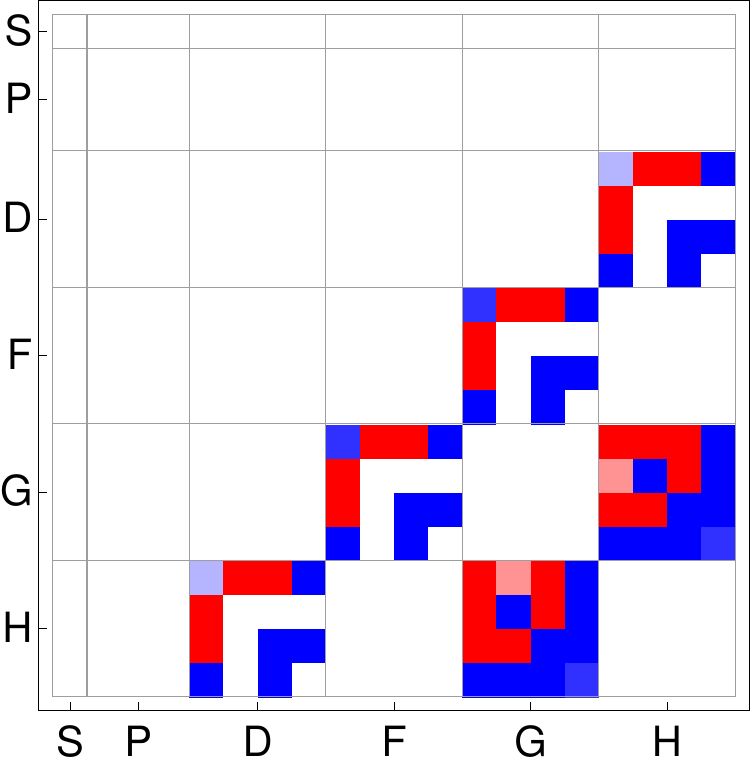} \end{minipage}
\begin{minipage}{.35\linewidth} \vspace*{0.500cm} \hspace*{-0.65cm}\includegraphics[width=1.15\textwidth]{WQ_l4_coeff_7_wsys.pdf}\end{minipage}
\begin{minipage}{.25\linewidth} \begin{align} \left(a_{5}\right)^{\sigma_0}_{7} &= \left<D,H\right> + \left<F,G\right> \nonumber \\ & \hspace*{12.5pt} + \left<G,H\right>   \nonumber \end{align} \end{minipage}
\caption{%
Left: Matrices $\mathcal{C}_{4\cdots 7}^{\sigma_0}$, represented here in the color scheme, defines the coefficient $\left(a_{5}\right)_{4\cdots 7}^{\sigma_0}$ for an expansion of $\sigma_0$ up to $\text{L}_{\text{max}} = 5$. Center: Coefficients $\left(a_{4}\right)_{4\cdots 7}^{\sigma_0}$ obtained from a fit to the $\sigma_0$-data (black points). For references to the data see Table \ref{tab:DataBasis}. Bonn Gatchina predictions, truncated at different $\text{L}_{\mathrm{max}}$ ($\text{L}_{\mathrm{max}} = 1$ is drawn in green, $\text{L}_{\mathrm{max}} = 2$ in blue, $\text{L}_{\mathrm{max}} = 3$ in red and $\text{L}_{\mathrm{max}} = 4$ in black) are drawn as well. Right: All partial wave interferences for $\text{L}_{\text{max}} = 5$ are indicated.
}
\label{tab:DCSColorPlots2}
\end{table*}

\begin{table*}[htb]
\RawFloats
\begin{minipage}{.075\linewidth}
\vspace*{-6.5pt}
\hspace*{5pt}
\begin{equation}
\mathcal{C}_{8}^{\sigma_0} \equiv \nonumber
\end{equation}
\end{minipage}
\begin{minipage}{.3\linewidth} \vspace*{0.572cm} \includegraphics[width=0.875\textwidth]{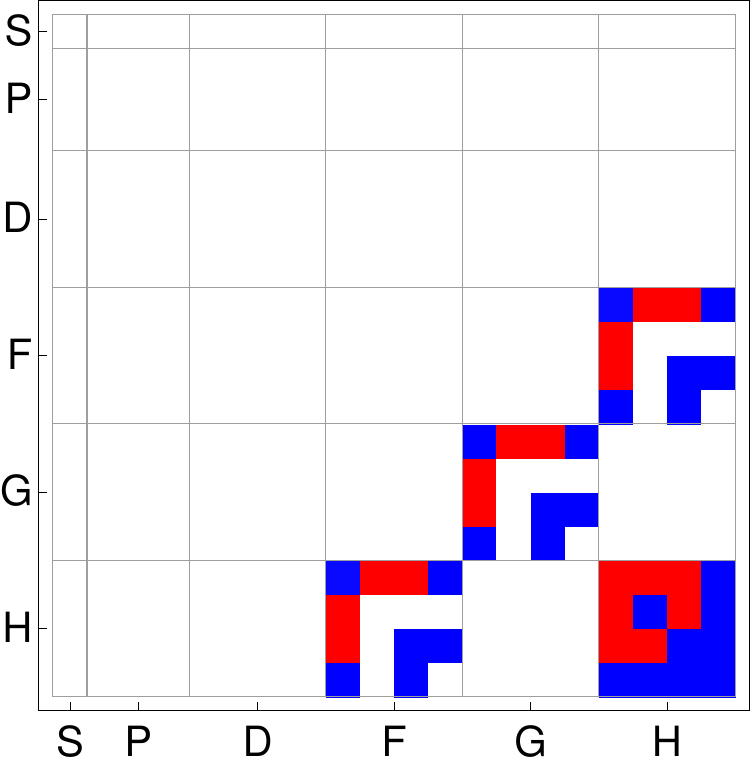} \end{minipage}
\begin{minipage}{.35\linewidth} \vspace*{0.500cm} \hspace*{-0.65cm}\includegraphics[width=1.15\textwidth]{WQ_l4_coeff_8_wsys.pdf}\end{minipage}
\begin{minipage}{.25\linewidth} \begin{align} \left(a_{5}\right)^{\sigma_0}_{8} &= \left<F,H\right> + \left<G,G\right> \nonumber \\ & \hspace*{12.5pt} + \left<H,H\right>    \nonumber \end{align} \end{minipage}

\begin{minipage}{.075\linewidth}
\vspace*{-6.5pt}
\hspace*{5pt}
\begin{equation}
\mathcal{C}_{9}^{\sigma_0} \equiv \nonumber
\end{equation}
\end{minipage}
\begin{minipage}{.3\linewidth} \vspace*{0.572cm} \includegraphics[width=0.875\textwidth]{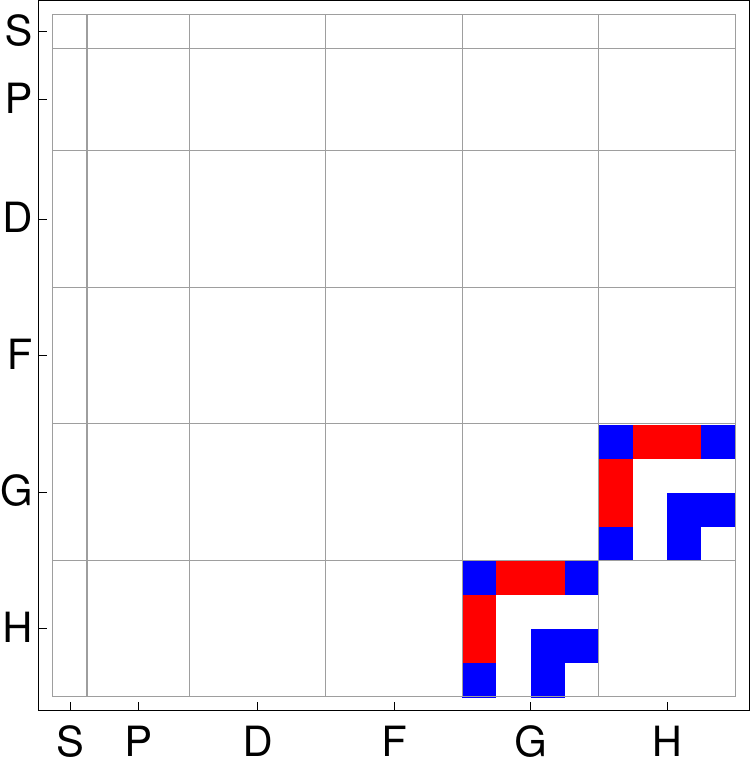} \end{minipage}
\begin{minipage}{.35\linewidth} \vspace*{0.500cm} \hspace*{-0.65cm}\includegraphics[width=1.15\textwidth]{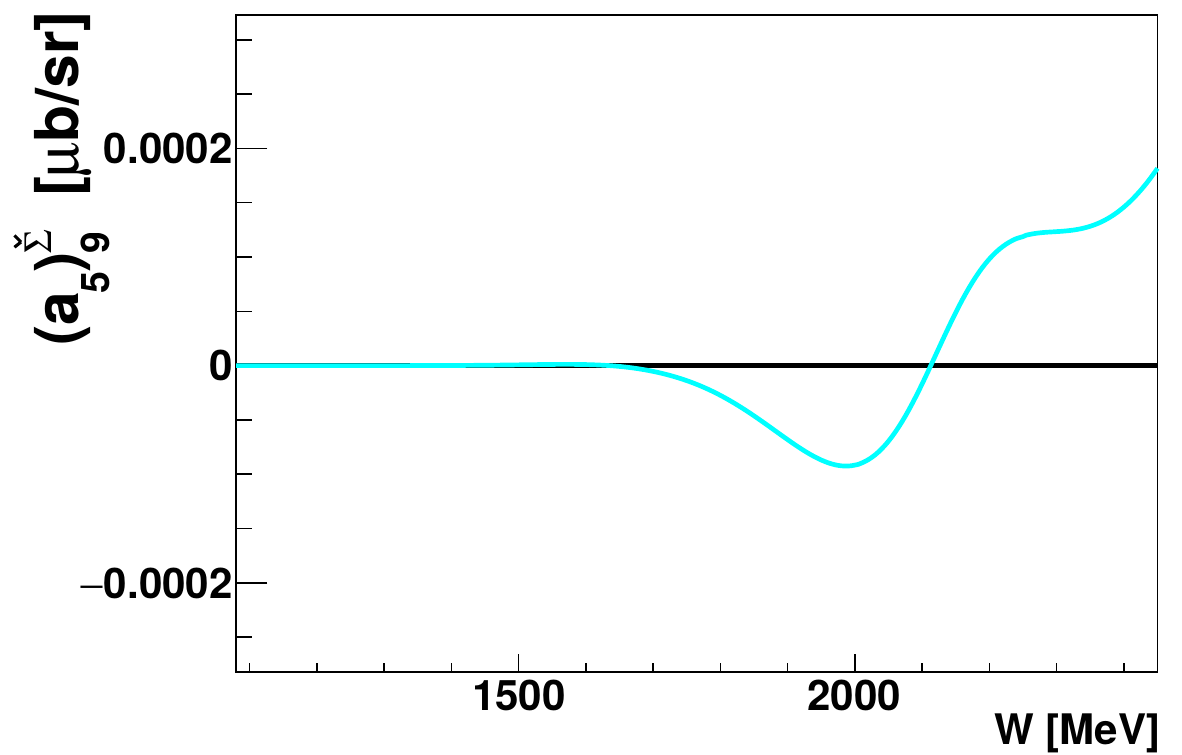}\end{minipage}
\begin{minipage}{.25\linewidth} \begin{align} \left(a_{5}\right)^{\sigma_0}_{9} &= \left<G,H\right>  \nonumber  \end{align} \end{minipage}

\begin{minipage}{.075\linewidth}
\vspace*{-6.5pt}
\hspace*{5pt}
\begin{equation}
\mathcal{C}_{10}^{\sigma_0} \equiv \nonumber
\end{equation}
\end{minipage}
\begin{minipage}{.3\linewidth} \vspace*{0.572cm} \includegraphics[width=0.875\textwidth]{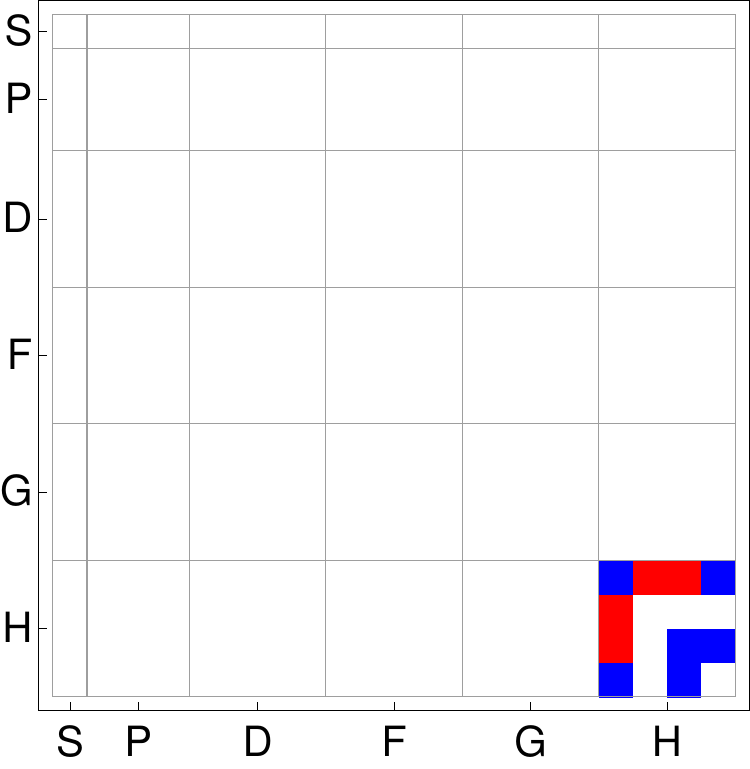} \end{minipage}
\begin{minipage}{.35\linewidth} \vspace*{0.500cm} \hspace*{-0.65cm}\includegraphics[width=1.15\textwidth]{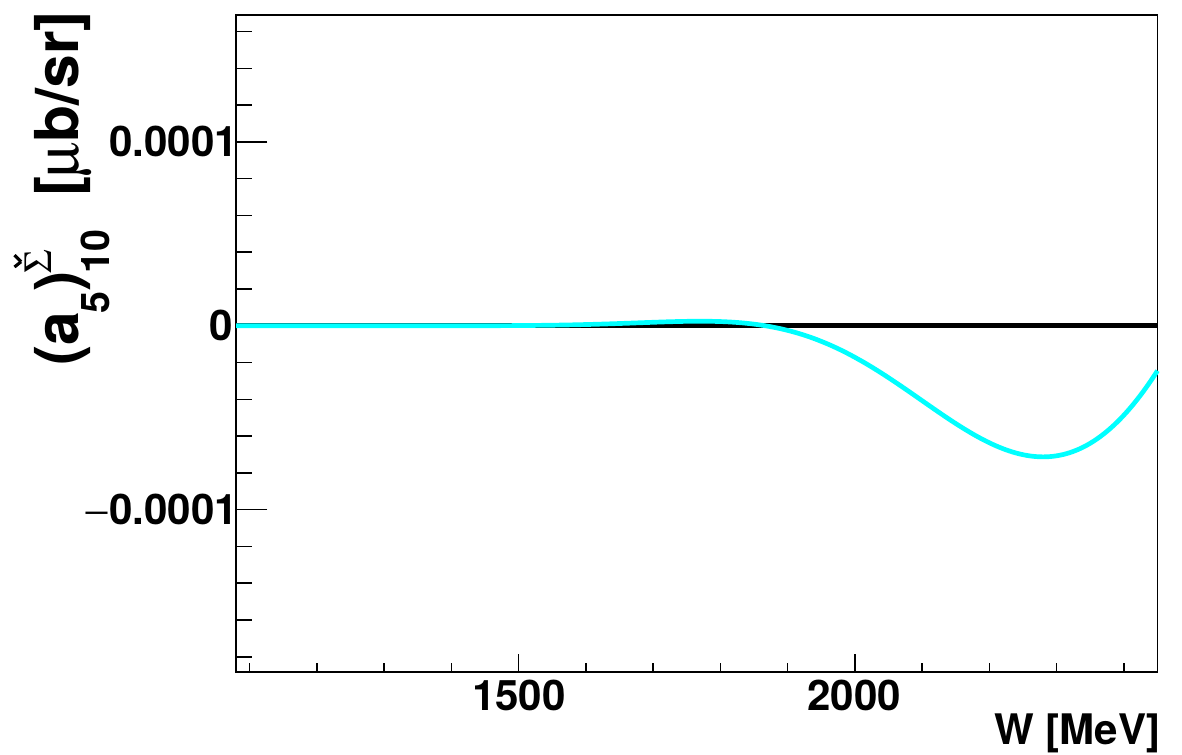}\end{minipage}
\begin{minipage}{.25\linewidth} \begin{align} \left(a_{5}\right)^{\sigma_0}_{10} &= \left<H,H\right>  \nonumber  \end{align} \end{minipage}

\caption{%
Left: Matrices $\mathcal{C}_{8\cdots 10}^{\sigma_0}$, represented here in the color scheme, defines the coefficient $\left(a_{5}\right)_{8\cdots 10}^{\sigma_0}$ for an expansion of $\sigma_0$ up to $\text{L}_{\text{max}} = 5$. Center: The coefficient $\left(a_{4}\right)_{8}^{\sigma_0}$ obtained from a fit to the $\sigma_0$-data (black points). For references to the data see Table \ref{tab:DataBasis}. Bonn Gatchina predictions, truncated at different $\text{L}_{\mathrm{max}}$ ($\text{L}_{\mathrm{max}} = 1$ is drawn in green, $\text{L}_{\mathrm{max}} = 2$ in blue, $\text{L}_{\mathrm{max}} = 3$ in red and $\text{L}_{\mathrm{max}} = 4$ in black) are drawn as well. For the highest non-fitted coefficients $\left(a_{5}\right)_{9, 10}^{\sigma_0}$, the Bonn Gatchina curves are shown (here, the truncation at $\text{L}_{\mathrm{max}} = 5$ is drawn in cyan). Right: All partial wave interferences for $\text{L}_{\text{max}} = 5$ are indicated.
}
\label{tab:DCSColorPlots3}
\end{table*}

\begin{table*}[htb]
\RawFloats
\begin{minipage}{.075\linewidth}
\vspace*{-6.5pt}
\hspace*{5pt}
\begin{equation}
\mathcal{C}_{2}^{\check{\Sigma}} \equiv \nonumber
\end{equation}
\end{minipage}
\begin{minipage}{.3\linewidth} \vspace*{0.572cm} \includegraphics[width=0.875\textwidth]{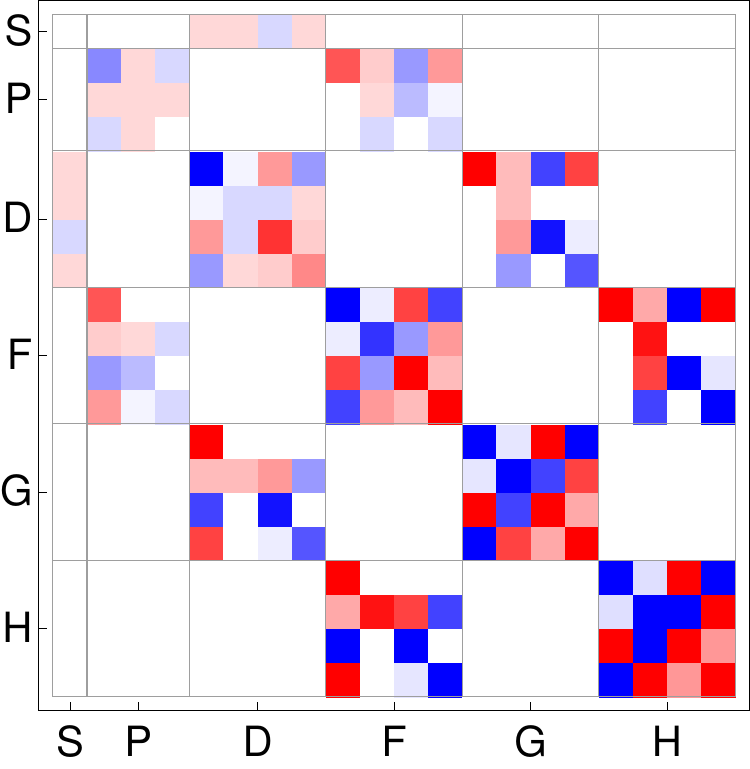} \end{minipage}
\begin{minipage}{.35\linewidth} \vspace*{0.500cm} \hspace*{-0.65cm}\includegraphics[width=1.15\textwidth]{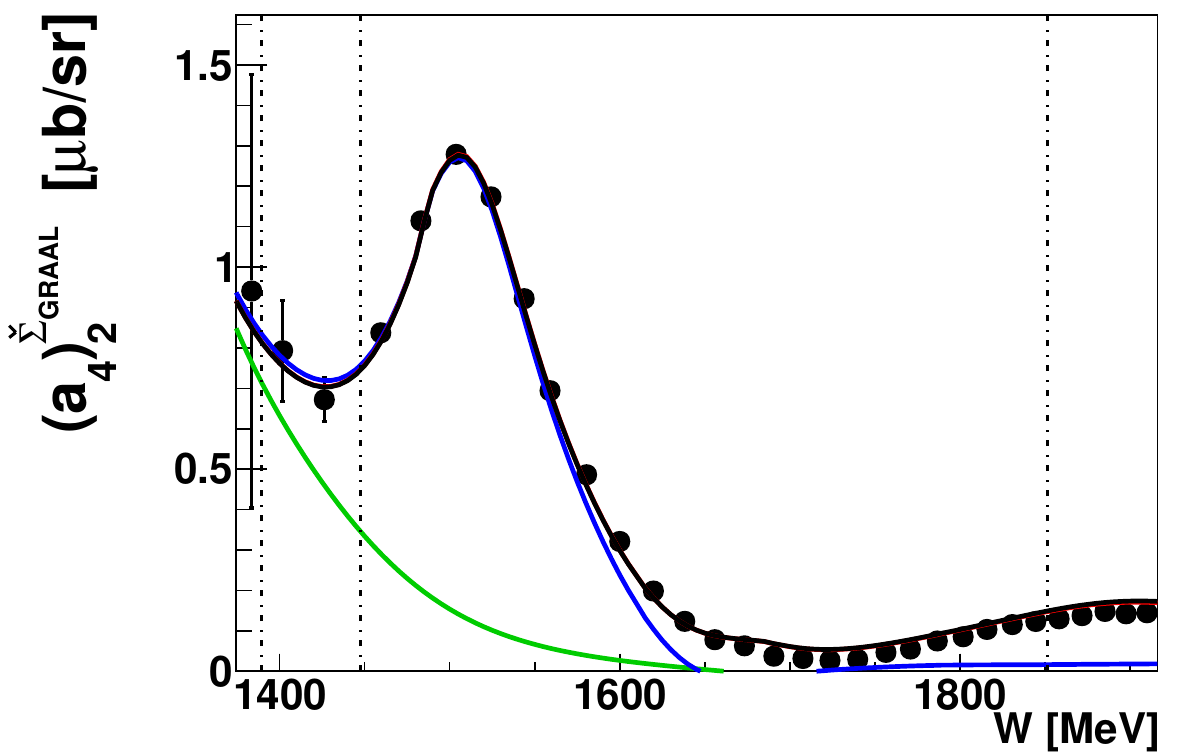}\end{minipage}
\begin{minipage}{.25\linewidth} \begin{align} \left(a_{5}\right)^{\check{\Sigma}}_{2} &= \left<S,D\right> + \left<P,P\right> \nonumber \\ & \hspace*{12.5pt} + \left<P,F\right> + \left<D,D\right>  \nonumber \\ & \hspace*{12.5pt}  + \left<D,G\right>   + \left<F,F\right> \nonumber \\ & \hspace*{12.5pt}   + \left<F,H\right>  + \left<G,G\right> \nonumber \\ & \hspace*{12.5pt} + \left<H,H\right>   \nonumber \end{align} \end{minipage}

\begin{minipage}{.075\linewidth}
\vspace*{-6.5pt}
\hspace*{5pt}
\begin{equation}
\mathcal{C}_{3}^{\check{\Sigma}} \equiv \nonumber
\end{equation}
\end{minipage}
\begin{minipage}{.3\linewidth} \vspace*{0.572cm} \includegraphics[width=0.875\textwidth]{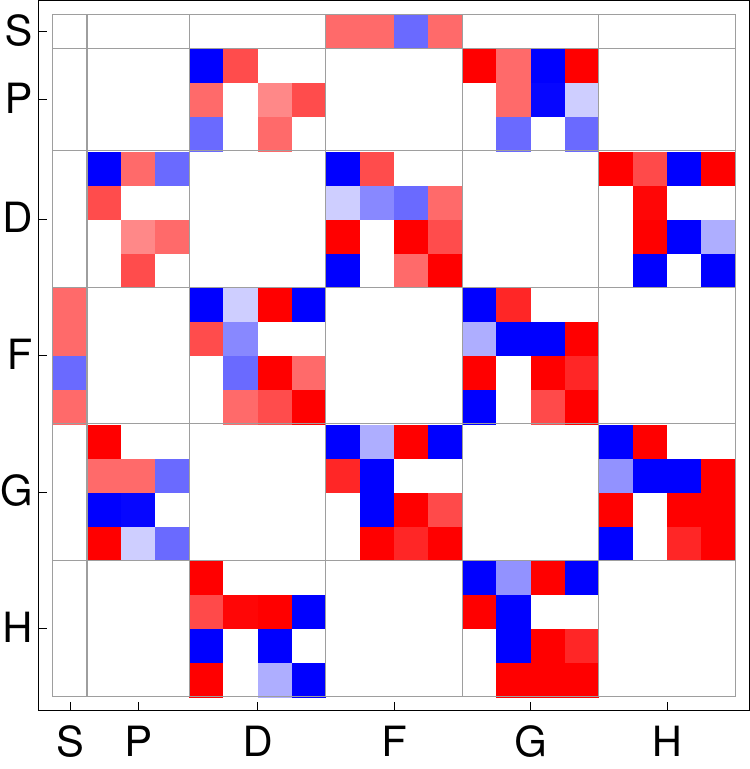} \end{minipage}
\begin{minipage}{.35\linewidth} \vspace*{0.500cm} \hspace*{-0.65cm}\includegraphics[width=1.15\textwidth]{Sgraal_l4_coeff_1.pdf}\end{minipage}
\begin{minipage}{.25\linewidth} \begin{align} \left(a_{5}\right)^{\check{\Sigma}}_{3} &= \left<S,F\right> + \left<P,D\right>  \nonumber \\ & \hspace*{12.5pt}+ \left<P,G\right> + \left<D,F\right>    \nonumber \\ & \hspace*{12.5pt} + \left<D,H\right> + \left<F,G\right>   \nonumber   \\ & \hspace*{12.5pt}   + \left<G,H\right> \nonumber \end{align} \end{minipage}

\begin{minipage}{.075\linewidth}
\vspace*{-6.5pt}
\hspace*{5pt}
\begin{equation}
\mathcal{C}_{4}^{\check{\Sigma}} \equiv \nonumber
\end{equation}
\end{minipage}
\begin{minipage}{.3\linewidth} \vspace*{0.572cm} \includegraphics[width=0.875\textwidth]{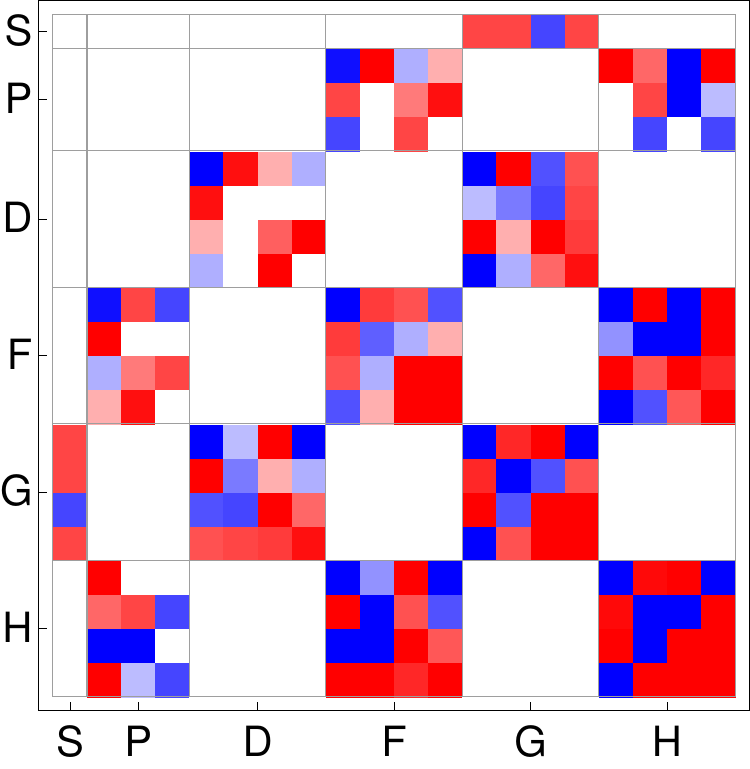} \end{minipage}
\begin{minipage}{.35\linewidth} \vspace*{0.500cm} \hspace*{-0.65cm}\includegraphics[width=1.15\textwidth]{Sgraal_l4_coeff_2.pdf}\end{minipage}
\begin{minipage}{.25\linewidth} \begin{align} \left(a_{5}\right)^{\check{\Sigma}}_{4} &= \left<S,G\right> + \left<P,F\right> \nonumber \\ & \hspace*{12.5pt} + \left<P,H\right>  + \left<D,D\right>   \nonumber \\ & \hspace*{12.5pt} + \left<D,G\right>  + \left<F,F\right>  \nonumber \\ & \hspace*{12.5pt} + \left<F,H\right>  + \left<G,G\right> \nonumber \\ & \hspace*{12.5pt} + \left<H,H\right>  \nonumber \end{align} \end{minipage}

\begin{minipage}{.075\linewidth}
\vspace*{-6.5pt}
\hspace*{5pt}
\begin{equation}
\mathcal{C}_{5}^{\check{\Sigma}} \equiv \nonumber
\end{equation}
\end{minipage}
\begin{minipage}{.3\linewidth} \vspace*{0.572cm} \includegraphics[width=0.875\textwidth]{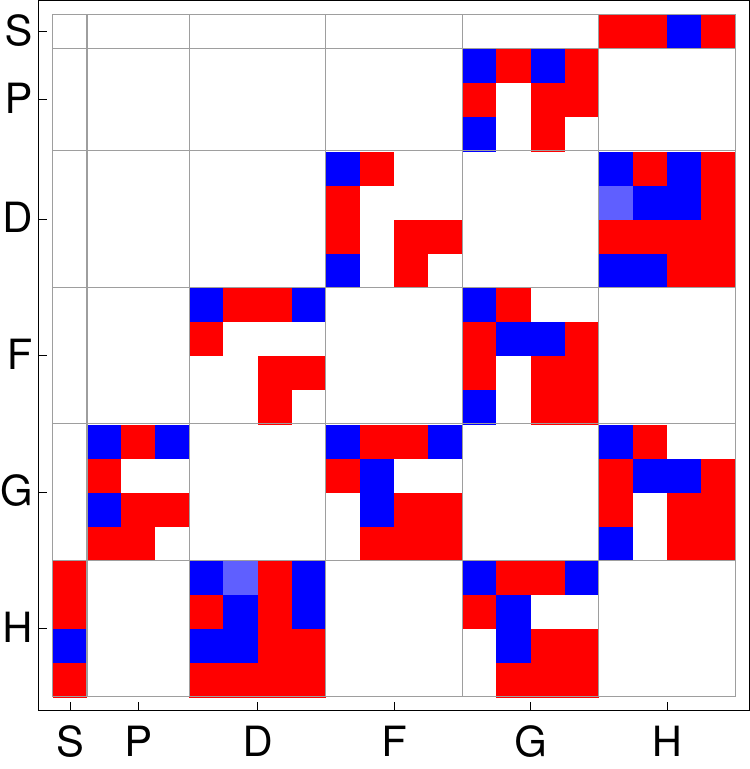} \end{minipage}
\begin{minipage}{.35\linewidth} \vspace*{0.500cm} \hspace*{-0.65cm}\includegraphics[width=1.15\textwidth]{Sgraal_l4_coeff_3.pdf}\end{minipage}
\begin{minipage}{.25\linewidth} \begin{align} \left(a_{5}\right)^{\check{\Sigma}}_{5} &= \left<S,H\right> + \left<P,G\right> \nonumber \\ & \hspace*{12.5pt} + \left<D,F\right>  + \left<D,H\right>   \nonumber \\ & \hspace*{12.5pt} + \left<F,G\right>  + \left<G,H\right>  \nonumber  \end{align} \end{minipage}
\caption{%
Left: Matrices $\mathcal{C}_{2\cdots 5}^{\check{\Sigma}}$, represented here in the color scheme, defines the coefficient $\left(a_{5}\right)_{2\cdots 5}^{\check{\Sigma}}$ for an expansion of $\check{\Sigma}$ up to $\text{L}_{\text{max}} = 5$. Center: Coefficients $\left(a_{4}\right)_{2\cdots 5}^{\check{\Sigma}_{\text{GRAAL}}}$ obtained from a fit to the $\check{\Sigma}_{\text{GRAAL}}$-data (black points). For references to the data see Table \ref{tab:DataBasis}. Bonn Gatchina predictions, truncated at different $\text{L}_{\mathrm{max}}$ ($\text{L}_{\mathrm{max}} = 1$ is drawn in green, $\text{L}_{\mathrm{max}} = 2$ in blue, $\text{L}_{\mathrm{max}} = 3$ in red and $\text{L}_{\mathrm{max}} = 4$ in black) are drawn as well. Right: All partial wave interferences for $\text{L}_{\text{max}} = 5$ are indicated.
}
\label{tab:SigmaColorPlots1}
\end{table*}

\begin{table*}[htb]
\RawFloats
\begin{minipage}{.075\linewidth}
\vspace*{-6.5pt}
\hspace*{5pt}
\begin{equation}
\mathcal{C}_{6}^{\check{\Sigma}} \equiv \nonumber
\end{equation}
\end{minipage}
\begin{minipage}{.3\linewidth} \vspace*{0.572cm} \includegraphics[width=0.875\textwidth]{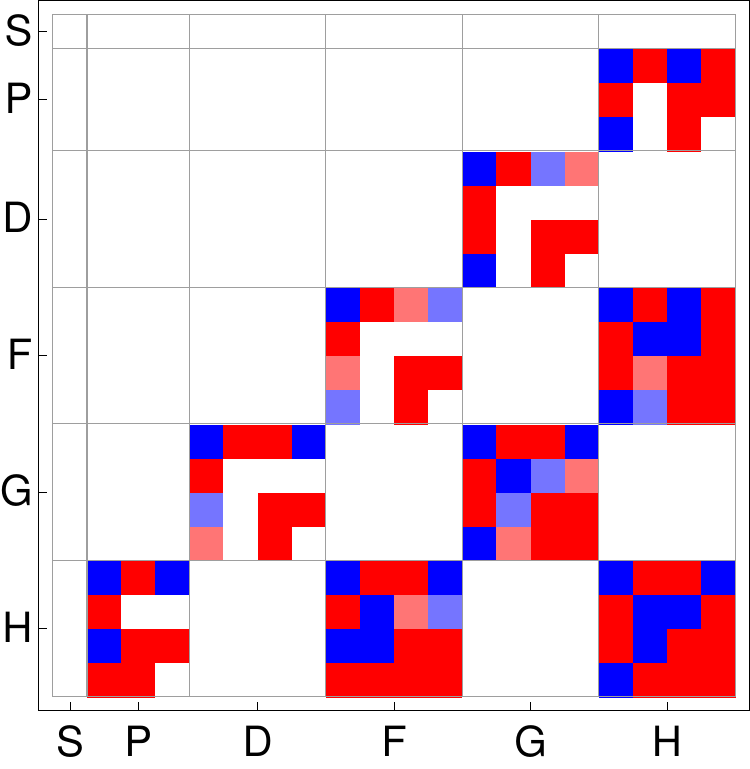} \end{minipage}
\begin{minipage}{.35\linewidth} \vspace*{0.500cm} \hspace*{-0.65cm}\includegraphics[width=1.15\textwidth]{Sgraal_l4_coeff_4.pdf}\end{minipage}
\begin{minipage}{.25\linewidth} \begin{align} \left(a_{5}\right)^{\check{\Sigma}}_{6} &= \left<P,H\right> + \left<D,G\right> \nonumber \\ & \hspace*{12.5pt} + \left<F,F\right>  + \left<F,H\right>   \nonumber \\ & \hspace*{12.5pt} + \left<G,G\right>  + \left<H,H\right> \nonumber \end{align} \end{minipage}

\begin{minipage}{.075\linewidth}
\vspace*{-6.5pt}
\hspace*{5pt}
\begin{equation}
\mathcal{C}_{7}^{\check{\Sigma}} \equiv \nonumber
\end{equation}
\end{minipage}
\begin{minipage}{.3\linewidth} \vspace*{0.572cm} \includegraphics[width=0.875\textwidth]{obs2_coeff6.pdf} \end{minipage}
\begin{minipage}{.35\linewidth} \vspace*{0.500cm} \hspace*{-0.65cm}\includegraphics[width=1.15\textwidth]{Sgraal_l4_coeff_5.pdf}\end{minipage}
\begin{minipage}{.25\linewidth} \begin{align} \left(a_{5}\right)^{\check{\Sigma}}_{7} &= \left<D,H\right> + \left<F,G\right> \nonumber \\ & \hspace*{12.5pt} + \left<G,H\right>   \nonumber \end{align} \end{minipage}

\begin{minipage}{.075\linewidth}
\vspace*{-6.5pt}
\hspace*{5pt}
\begin{equation}
\mathcal{C}_{8}^{\check{\Sigma}} \equiv \nonumber
\end{equation}
\end{minipage}
\begin{minipage}{.3\linewidth} \vspace*{0.572cm} \includegraphics[width=0.875\textwidth]{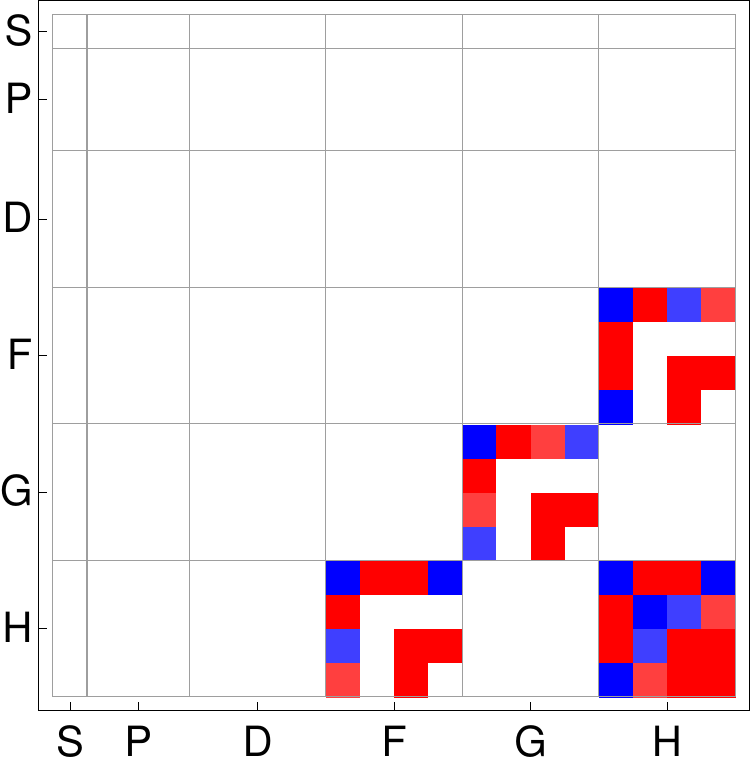} \end{minipage}
\begin{minipage}{.35\linewidth} \vspace*{0.500cm} \hspace*{-0.65cm}\includegraphics[width=1.15\textwidth]{Sgraal_l4_coeff_6.pdf}\end{minipage}
\begin{minipage}{.25\linewidth} \begin{align} \left(a_{5}\right)^{\check{\Sigma}}_{8} &= \left<F,H\right> + \left<G,G\right> \nonumber \\ & \hspace*{12.5pt} + \left<H,H\right>    \nonumber \end{align} \end{minipage}

\begin{minipage}{.075\linewidth}
\vspace*{-6.5pt}
\hspace*{5pt}
\begin{equation}
\mathcal{C}_{9}^{\check{\Sigma}} \equiv \nonumber
\end{equation}
\end{minipage}
\begin{minipage}{.3\linewidth} \vspace*{0.572cm} \includegraphics[width=0.875\textwidth]{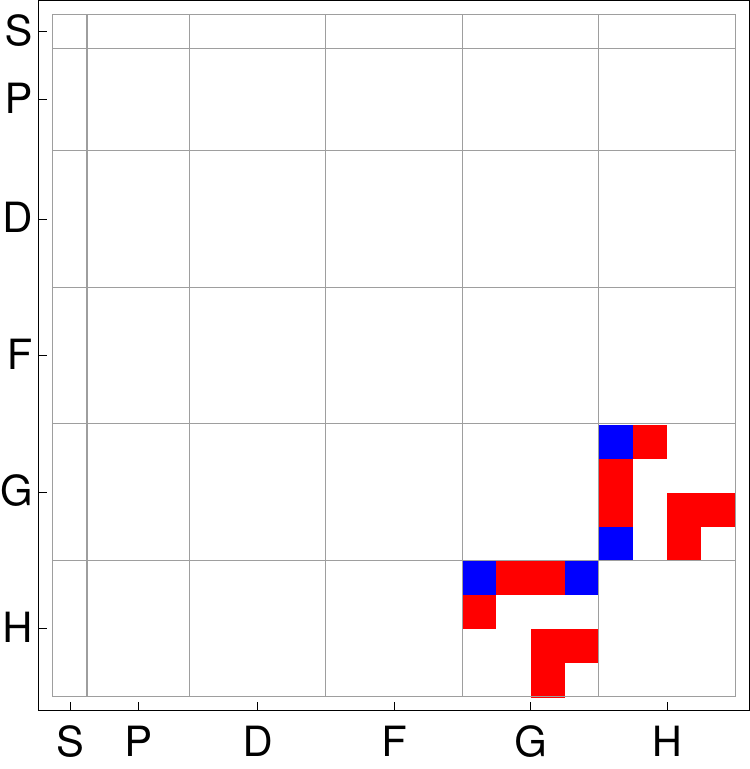} \end{minipage}
\begin{minipage}{.35\linewidth} \vspace*{0.500cm} \hspace*{-0.65cm}\includegraphics[width=1.15\textwidth]{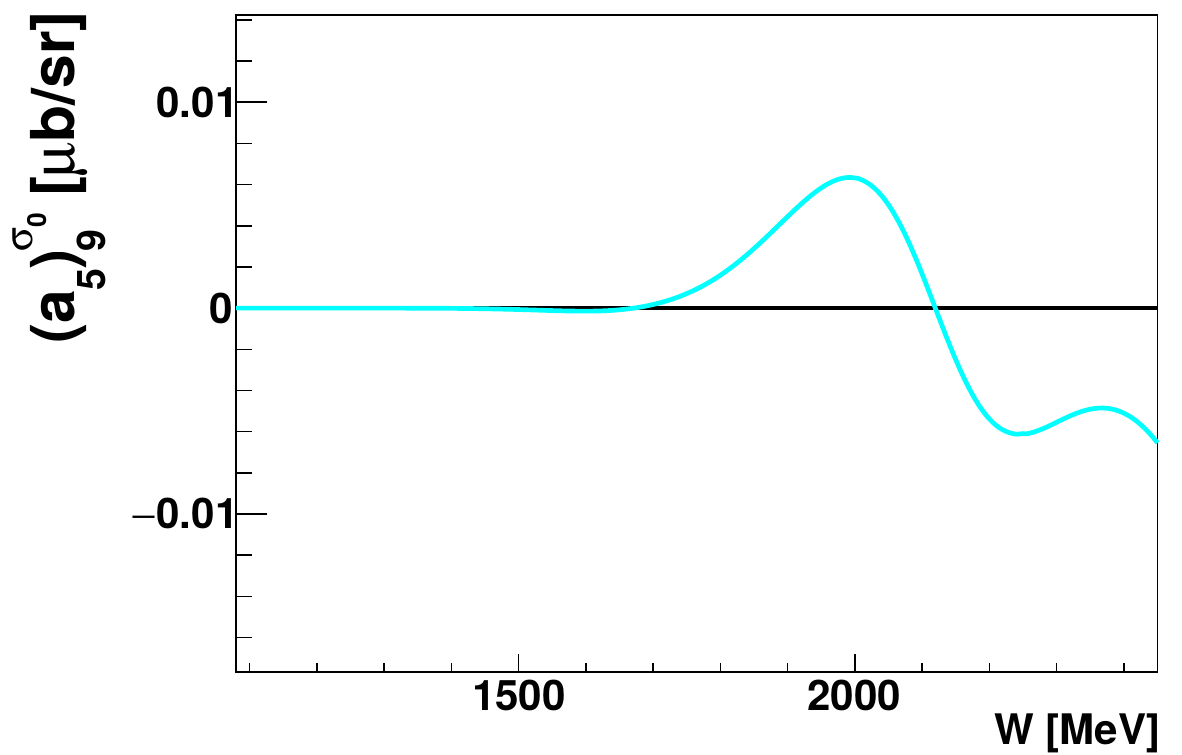}\end{minipage}
\begin{minipage}{.25\linewidth} \begin{align} \left(a_{5}\right)^{\check{\Sigma}}_{9} &= \left<G,H\right> \nonumber  \end{align} \end{minipage}
\caption{%
Left: Matrices $\mathcal{C}_{6\cdots 9}^{\check{\Sigma}}$, represented here in the color scheme, defines the coefficient $\left(a_{5}\right)_{6\cdots 9}^{\check{\Sigma}}$ for an expansion of $\check{\Sigma}$ up to $\text{L}_{\text{max}} = 5$. Center: Coefficients $\left(a_{4}\right)_{6\cdots 8}^{\check{\Sigma}_{\text{GRAAL}}}$ obtained from a fit to the $\check{\Sigma}_{\text{GRAAL}}$-data (black points). For references to the data see Table \ref{tab:DataBasis}. Bonn Gatchina predictions, truncated at different $\text{L}_{\mathrm{max}}$ ($\text{L}_{\mathrm{max}} = 1$ is drawn in green, $\text{L}_{\mathrm{max}} = 2$ in blue, $\text{L}_{\mathrm{max}} = 3$ in red and $\text{L}_{\mathrm{max}} = 4$ in black) are drawn as well. For the highest non-fitted coefficient $\left(a_{5}\right)_{9}^{\check{\Sigma}}$, the Bonn Gatchina curves are shown (here, the truncation at $\text{L}_{\mathrm{max}} = 5$ is drawn in cyan). Right: All partial wave interferences for $\text{L}_{\text{max}} = 5$ are indicated.
}
\label{tab:SigmaColorPlots2}
\end{table*}

 \begin{table*}[htb]
\RawFloats
\begin{minipage}{.075\linewidth}
\vspace*{-6.5pt}
\hspace*{5pt}
\begin{equation}
\mathcal{C}_{10}^{\check{\Sigma}} \equiv \nonumber
\end{equation}
\end{minipage}
\begin{minipage}{.3\linewidth} \vspace*{0.572cm} \includegraphics[width=0.875\textwidth]{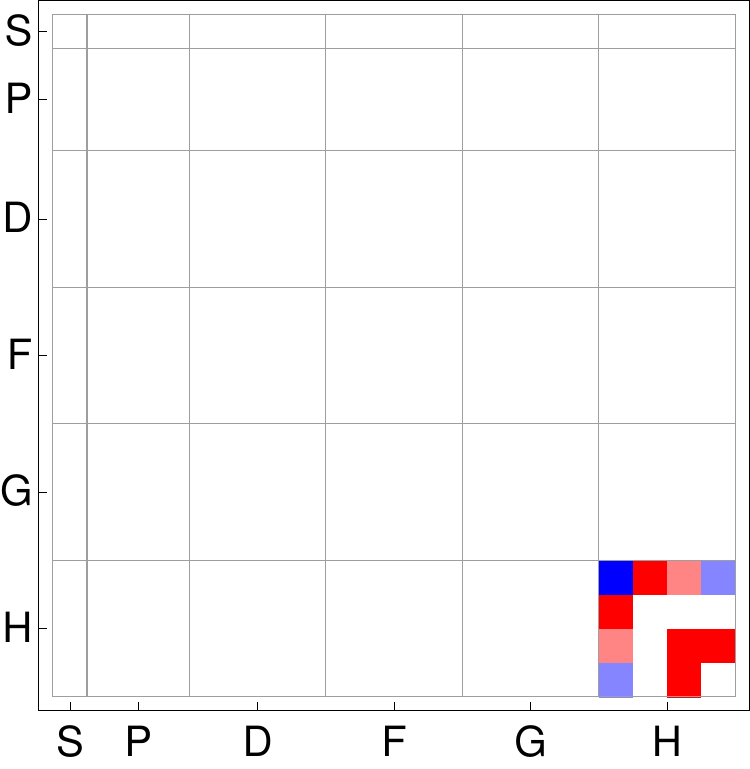} \end{minipage}
\begin{minipage}{.35\linewidth} \vspace*{0.500cm} \hspace*{-0.65cm}\includegraphics[width=1.15\textwidth]{sigma_5_10.pdf}\end{minipage}
\begin{minipage}{.25\linewidth} \begin{align} \left(a_{5}\right)^{\check{\Sigma}}_{10} &= \left<H,H\right>  \nonumber \end{align} \end{minipage}
\caption{%
Left: Matrix $\mathcal{C}_{10}^{\check{\Sigma}}$, represented here in the color scheme, defines the coefficient $\left(a_{5}\right)_{10}^{\check{\Sigma}}$ for an expansion of $\check{\Sigma}$ up to $\text{L}_{\text{max}} = 5$. Center: For the highest non-fitted coefficient $\left(a_{5}\right)_{10}^{\check{\Sigma}}$, the Bonn Gatchina curves are shown (here, the truncation at $\text{L}_{\mathrm{max}} = 5$ is drawn in cyan). Right: All partial wave interferences for $\text{L}_{\text{max}} = 5$ are indicated.
}
\label{tab:SigmaColorPlots3}
\end{table*}

\begin{table*}[htb]
\RawFloats
\begin{minipage}{.075\linewidth}
\vspace*{-6.5pt}
\hspace*{5pt}
\begin{equation}
\mathcal{C}_{2}^{\check{\Sigma}} \equiv \nonumber
\end{equation}
\end{minipage}
\begin{minipage}{.3\linewidth} \vspace*{0.572cm} \includegraphics[width=0.875\textwidth]{obs2_coeff1.pdf} \end{minipage}
\begin{minipage}{.35\linewidth} \vspace*{0.500cm} \hspace*{-0.65cm}\includegraphics[width=1.15\textwidth]{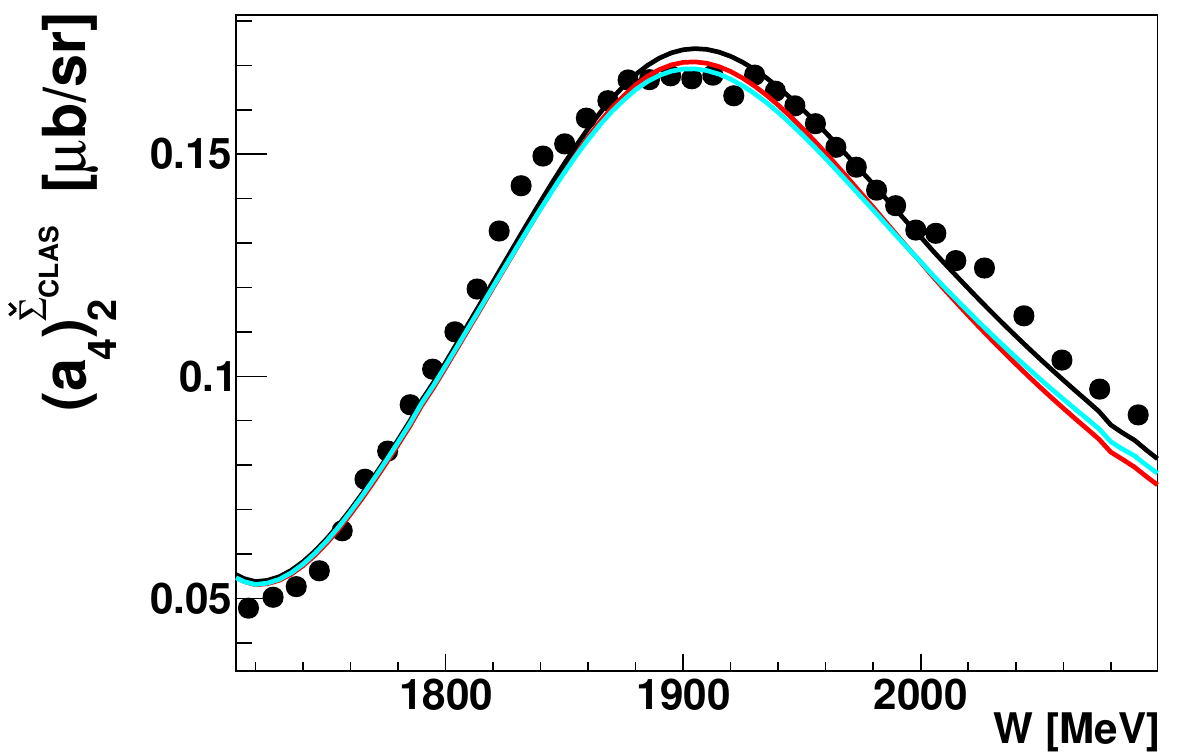}\end{minipage}
\begin{minipage}{.25\linewidth} \begin{align} \left(a_{5}\right)^{\check{\Sigma}}_{2} &= \left<S,D\right> + \left<P,P\right> \nonumber \\ & \hspace*{12.5pt} + \left<P,F\right> + \left<D,D\right>  \nonumber \\ & \hspace*{12.5pt}  + \left<D,G\right>   + \left<F,F\right> \nonumber \\ & \hspace*{12.5pt}   + \left<F,H\right>  + \left<G,G\right> \nonumber \\ & \hspace*{12.5pt} + \left<H,H\right>  \nonumber \end{align} \end{minipage}

\begin{minipage}{.075\linewidth}
\vspace*{-6.5pt}
\hspace*{5pt}
\begin{equation}
\mathcal{C}_{3}^{\check{\Sigma}} \equiv \nonumber
\end{equation}
\end{minipage}
\begin{minipage}{.3\linewidth} \vspace*{0.572cm} \includegraphics[width=0.875\textwidth]{obs2_coeff2.pdf} \end{minipage}
\begin{minipage}{.35\linewidth} \vspace*{0.500cm} \hspace*{-0.65cm}\includegraphics[width=1.15\textwidth]{Sclas_l4_coeff_1.pdf}\end{minipage}
\begin{minipage}{.25\linewidth} \begin{align} \left(a_{5}\right)^{\check{\Sigma}}_{3} &= \left<S,F\right> + \left<P,D\right>  \nonumber \\ & \hspace*{12.5pt}+ \left<P,G\right> + \left<D,F\right>    \nonumber \\ & \hspace*{12.5pt} + \left<D,H\right> + \left<F,G\right>   \nonumber   \\ & \hspace*{12.5pt}   + \left<G,H\right> \nonumber \end{align} \end{minipage}

\begin{minipage}{.075\linewidth}
\vspace*{-6.5pt}
\hspace*{5pt}
\begin{equation}
\mathcal{C}_{4}^{\check{\Sigma}} \equiv \nonumber
\end{equation}
\end{minipage}
\begin{minipage}{.3\linewidth} \vspace*{0.572cm} \includegraphics[width=0.875\textwidth]{obs2_coeff3.pdf} \end{minipage}
\begin{minipage}{.35\linewidth} \vspace*{0.500cm} \hspace*{-0.65cm}\includegraphics[width=1.15\textwidth]{Sclas_l4_coeff_2.pdf}\end{minipage}
\begin{minipage}{.25\linewidth} \begin{align} \left(a_{5}\right)^{\check{\Sigma}}_{4} &= \left<S,G\right> + \left<P,F\right> \nonumber \\ & \hspace*{12.5pt} + \left<P,H\right>  + \left<D,D\right>   \nonumber \\ & \hspace*{12.5pt} + \left<D,G\right>  + \left<F,F\right>  \nonumber \\ & \hspace*{12.5pt} + \left<F,H\right>  + \left<G,G\right> \nonumber \\ & \hspace*{12.5pt} + \left<H,H\right>  \nonumber \end{align} \end{minipage}

\begin{minipage}{.075\linewidth}
\vspace*{-6.5pt}
\hspace*{5pt}
\begin{equation}
\mathcal{C}_{5}^{\check{\Sigma}} \equiv \nonumber
\end{equation}
\end{minipage}
\begin{minipage}{.3\linewidth} \vspace*{0.572cm} \includegraphics[width=0.875\textwidth]{obs2_coeff4.pdf} \end{minipage}
\begin{minipage}{.35\linewidth} \vspace*{0.500cm} \hspace*{-0.65cm}\includegraphics[width=1.15\textwidth]{Sclas_l4_coeff_3.pdf}\end{minipage}
\begin{minipage}{.25\linewidth} \begin{align} \left(a_{5}\right)^{\check{\Sigma}}_{5} &= \left<S,H\right> + \left<P,G\right> \nonumber \\ & \hspace*{12.5pt} + \left<D,F\right>  + \left<D,H\right>   \nonumber \\ & \hspace*{12.5pt} + \left<F,G\right>  + \left<G,H\right>  \nonumber  \end{align} \end{minipage}
\caption{%
Left: Matrices $\mathcal{C}_{2\cdots 5}^{\check{\Sigma}}$, represented here in the color scheme, defines the coefficient $\left(a_{5}\right)_{2\cdots 5}^{\check{\Sigma}}$ for an expansion of $\check{\Sigma}$ up to $\text{L}_{\text{max}} = 5$. Center: Coefficients $\left(a_{4}\right)_{2\cdots 5}^{\check{\Sigma}_{\text{CLAS}}}$ obtained from a fit to the $\check{\Sigma}_{\text{CLAS}}$-data (black points). For references to the data see Table \ref{tab:DataBasis}. Bonn Gatchina predictions, truncated at different $\text{L}_{\mathrm{max}}$ ($\text{L}_{\mathrm{max}} = 1$ is drawn in green, $\text{L}_{\mathrm{max}} = 2$ in blue, $\text{L}_{\mathrm{max}} = 3$ in red, $\text{L}_{\mathrm{max}} = 4$ in black and $\text{L}_{\mathrm{max}} = 5$ in cyan) are drawn as well. Right: All partial wave interferences for $\text{L}_{\text{max}} = 5$ are indicated.
}
\label{tab:SigmaCLASColorPlots1}
\end{table*}

\begin{table*}[htb]
\RawFloats
\begin{minipage}{.075\linewidth}
\vspace*{-6.5pt}
\hspace*{5pt}
\begin{equation}
\mathcal{C}_{6}^{\check{\Sigma}} \equiv \nonumber
\end{equation}
\end{minipage}
\begin{minipage}{.3\linewidth} \vspace*{0.572cm} \includegraphics[width=0.875\textwidth]{obs2_coeff5.pdf} \end{minipage}
\begin{minipage}{.35\linewidth} \vspace*{0.500cm} \hspace*{-0.65cm}\includegraphics[width=1.15\textwidth]{Sclas_l4_coeff_4.pdf}\end{minipage}
\begin{minipage}{.25\linewidth} \begin{align} \left(a_{5}\right)^{\check{\Sigma}}_{6} &= \left<P,H\right> + \left<D,G\right> \nonumber \\ & \hspace*{12.5pt} + \left<F,F\right>  + \left<F,H\right>   \nonumber \\ & \hspace*{12.5pt} + \left<G,G\right>  + \left<H,H\right> \nonumber \end{align} \end{minipage}

\begin{minipage}{.075\linewidth}
\vspace*{-6.5pt}
\hspace*{5pt}
\begin{equation}
\mathcal{C}_{7}^{\check{\Sigma}} \equiv \nonumber
\end{equation}
\end{minipage}
\begin{minipage}{.3\linewidth} \vspace*{0.572cm} \includegraphics[width=0.875\textwidth]{obs2_coeff6.pdf} \end{minipage}
\begin{minipage}{.35\linewidth} \vspace*{0.500cm} \hspace*{-0.65cm}\includegraphics[width=1.15\textwidth]{Sclas_l4_coeff_5.pdf}\end{minipage}
\begin{minipage}{.25\linewidth} \begin{align} \left(a_{5}\right)^{\check{\Sigma}}_{7} &= \left<D,H\right> + \left<F,G\right> \nonumber \\ & \hspace*{12.5pt} + \left<G,H\right>   \nonumber \end{align} \end{minipage}

\begin{minipage}{.075\linewidth}
\vspace*{-6.5pt}
\hspace*{5pt}
\begin{equation}
\mathcal{C}_{8}^{\check{\Sigma}} \equiv \nonumber
\end{equation}
\end{minipage}
\begin{minipage}{.3\linewidth} \vspace*{0.572cm} \includegraphics[width=0.875\textwidth]{obs2_coeff7.pdf} \end{minipage}
\begin{minipage}{.35\linewidth} \vspace*{0.500cm} \hspace*{-0.65cm}\includegraphics[width=1.15\textwidth]{Sclas_l4_coeff_6.pdf}\end{minipage}
\begin{minipage}{.25\linewidth} \begin{align} \left(a_{5}\right)^{\check{\Sigma}}_{8} &= \left<F,H\right> + \left<G,G\right> \nonumber \\ & \hspace*{12.5pt} + \left<H,H\right>    \nonumber \end{align} \end{minipage}

\begin{minipage}{.075\linewidth}
\vspace*{-6.5pt}
\hspace*{5pt}
\begin{equation}
\mathcal{C}_{9}^{\check{\Sigma}} \equiv \nonumber
\end{equation}
\end{minipage}
\begin{minipage}{.3\linewidth} \vspace*{0.572cm} \includegraphics[width=0.875\textwidth]{obs2_coeff8.pdf} \end{minipage}
\begin{minipage}{.35\linewidth} \vspace*{0.500cm} \hspace*{-0.65cm}\includegraphics[width=1.15\textwidth]{Sigma_5_9.pdf}\end{minipage}
\begin{minipage}{.25\linewidth} \begin{align} \left(a_{5}\right)^{\check{\Sigma}}_{9} &= \left<G,H\right> \nonumber  \end{align} \end{minipage}
\caption{%
Left: Matrices $\mathcal{C}_{6\cdots 9}^{\check{\Sigma}}$, represented here in the color scheme, defines the coefficient $\left(a_{5}\right)_{6\cdots 9}^{\check{\Sigma}}$ for an expansion of $\check{\Sigma}$ up to $\text{L}_{\text{max}} = 5$. Center: Coefficients $\left(a_{4}\right)_{6\cdots 8}^{\check{\Sigma}_{\text{CLAS}}}$ obtained from a fit to the $\check{\Sigma}_{\text{CLAS}}$-data (black points). For references to the data see Table \ref{tab:DataBasis}. Bonn Gatchina predictions, truncated at different $\text{L}_{\mathrm{max}}$ ($\text{L}_{\mathrm{max}} = 1$ is drawn in green, $\text{L}_{\mathrm{max}} = 2$ in blue, $\text{L}_{\mathrm{max}} = 3$ in red, $\text{L}_{\mathrm{max}} = 4$ in black and $\text{L}_{\mathrm{max}} = 5$ in cyan) are drawn as well.  For the highest non-fitted coefficient $\left(a_{5}\right)_{9}^{\check{\Sigma}}$, the Bonn Gatchina curves are shown. Right: All partial wave interferences for $\text{L}_{\text{max}} = 5$ are indicated.
}
\label{tab:SigmaCLASColorPlots2}
\end{table*}


\begin{table*}[htb]
\RawFloats
\begin{minipage}{.075\linewidth}
\vspace*{-6.5pt}
\hspace*{5pt}
\begin{equation}
\mathcal{C}_{1}^{\check{T}} \equiv \nonumber
\end{equation}
\end{minipage}
\begin{minipage}{.3\linewidth} \vspace*{0.572cm} \includegraphics[width=0.875\textwidth]{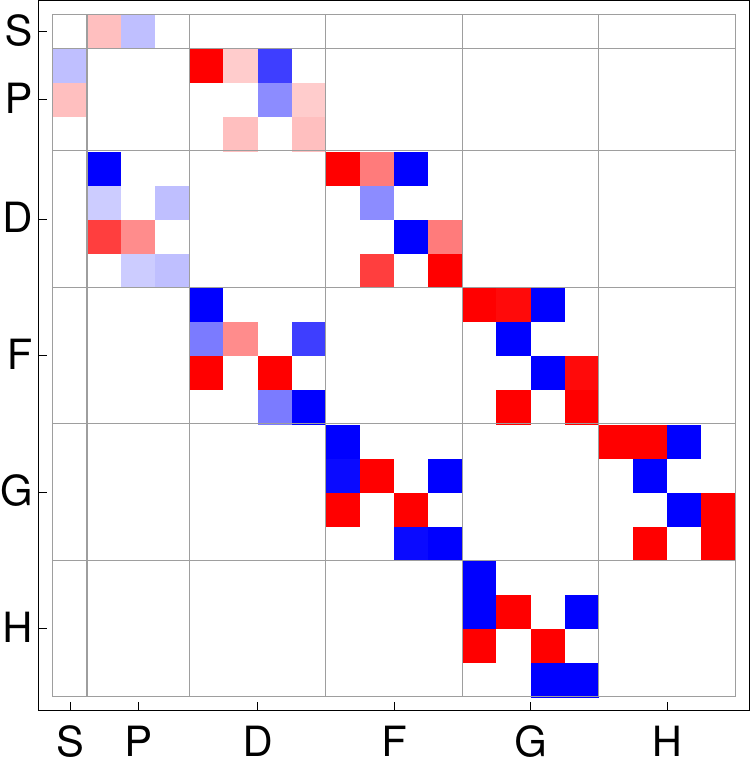} \end{minipage}
\begin{minipage}{.35\linewidth} \vspace*{0.500cm} \hspace*{-0.65cm}\includegraphics[width=1.15\textwidth]{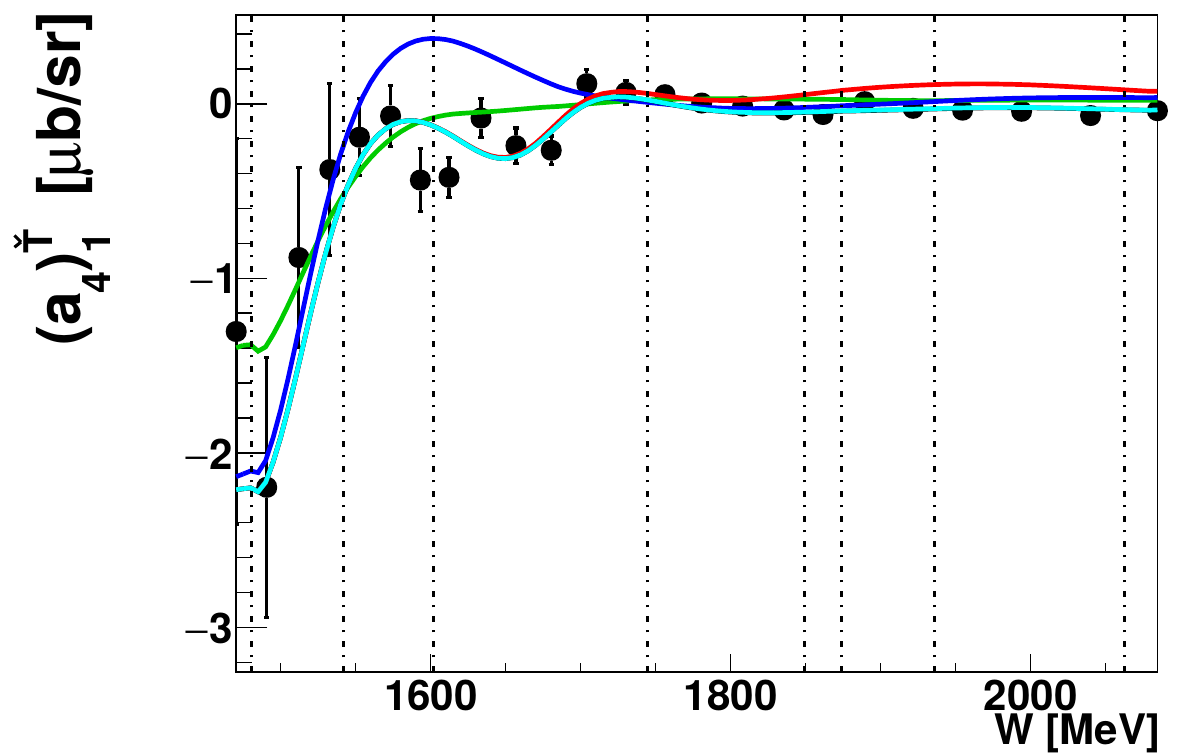}\end{minipage}
\begin{minipage}{.25\linewidth} \begin{align} \left(a_{5}\right)^{\check{T}}_{1} &= \left<S,P\right> + \left<P,D\right> \nonumber \\ & \hspace*{12.5pt} + \left<D,F\right> + \left<F,G\right>     \nonumber \\ & \hspace*{12.5pt}   + \left<G,H\right> \nonumber \end{align} \end{minipage}

\begin{minipage}{.075\linewidth}
\vspace*{-6.5pt}
\hspace*{5pt}
\begin{equation}
\mathcal{C}_{2}^{\check{T}} \equiv \nonumber
\end{equation}
\end{minipage}
\begin{minipage}{.3\linewidth} \vspace*{0.572cm} \includegraphics[width=0.875\textwidth]{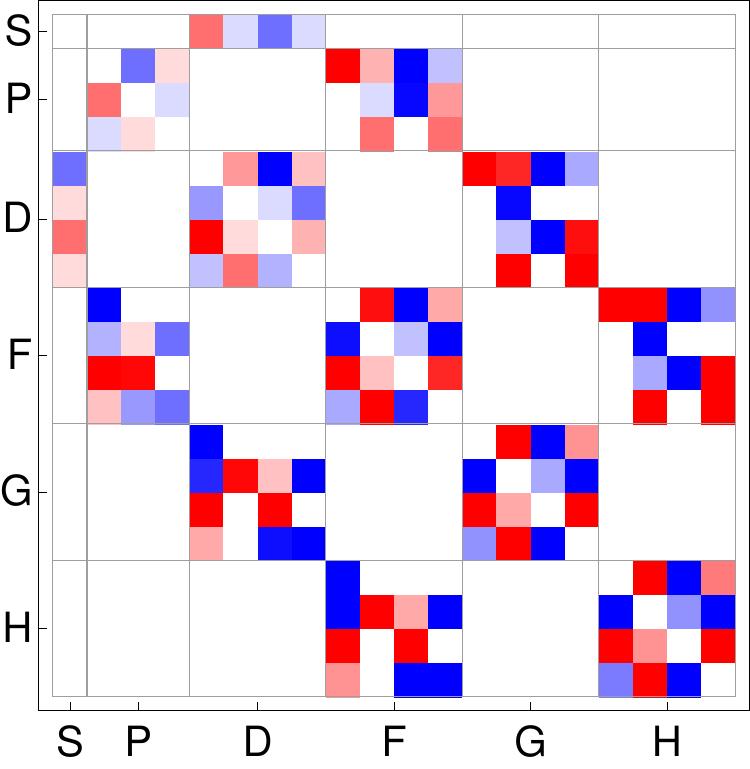} \end{minipage}
\begin{minipage}{.35\linewidth} \vspace*{0.500cm} \hspace*{-0.65cm}\includegraphics[width=1.15\textwidth]{T_l4_coeff_1.pdf}\end{minipage}
\begin{minipage}{.25\linewidth} \begin{align} \left(a_{5}\right)^{\check{T}}_{2} &= \left<S,D\right> + \left<P,P\right> \nonumber \\ & \hspace*{12.5pt} + \left<P,F\right>  + \left<D,D\right>   \nonumber \\ & \hspace*{12.5pt} + \left<D,G\right>  + \left<F,F\right>   \nonumber  \\ & \hspace*{12.5pt} + \left<F,H\right>  + \left<G,G\right>   \nonumber \\ & \hspace*{12.5pt} + \left<H,H\right>    \nonumber  \end{align} \end{minipage}

\begin{minipage}{.075\linewidth}
\vspace*{-6.5pt}
\hspace*{5pt}
\begin{equation}
\mathcal{C}_{3}^{\check{T}} \equiv \nonumber
\end{equation}
\end{minipage}
\begin{minipage}{.3\linewidth} \vspace*{0.572cm} \includegraphics[width=0.875\textwidth]{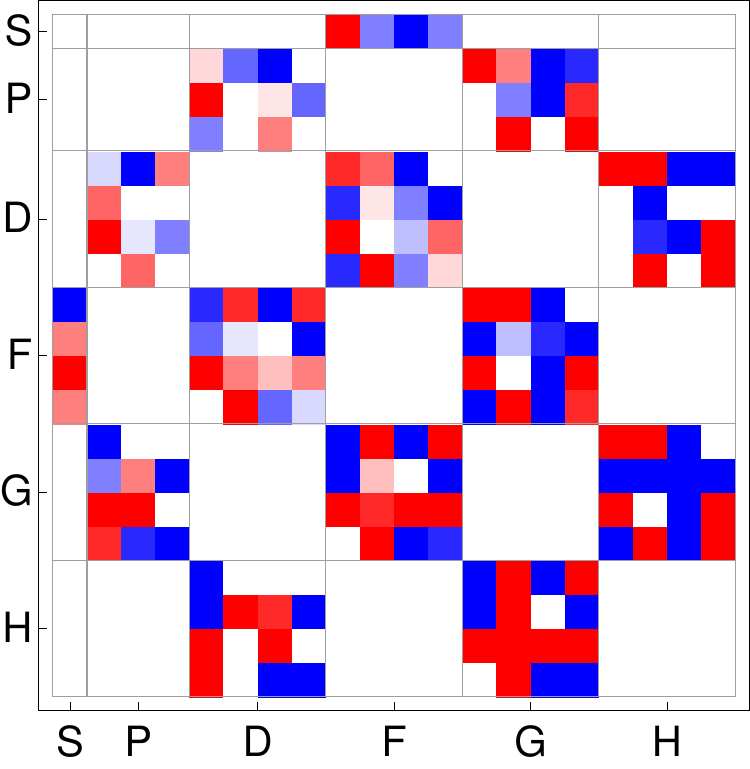} \end{minipage}
\begin{minipage}{.35\linewidth} \vspace*{0.500cm} \hspace*{-0.65cm}\includegraphics[width=1.15\textwidth]{T_l4_coeff_2.pdf}\end{minipage}
\begin{minipage}{.25\linewidth} \begin{align} \left(a_{5}\right)^{\check{T}}_{3} &= \left<S,F\right> + \left<P,D\right> \nonumber \\ & \hspace*{12.5pt} + \left<P,G\right>  + \left<D,F\right>   \nonumber  \\ & \hspace*{12.5pt} + \left<D,H\right>  + \left<F,G\right>   \nonumber \\ & \hspace*{12.5pt} + \left<G,H\right>   \nonumber   \end{align} \end{minipage}

\begin{minipage}{.075\linewidth}
\vspace*{-6.5pt}
\hspace*{5pt}
\begin{equation}
\mathcal{C}_{4}^{\check{T}} \equiv \nonumber
\end{equation}
\end{minipage}
\begin{minipage}{.3\linewidth} \vspace*{0.572cm} \includegraphics[width=0.875\textwidth]{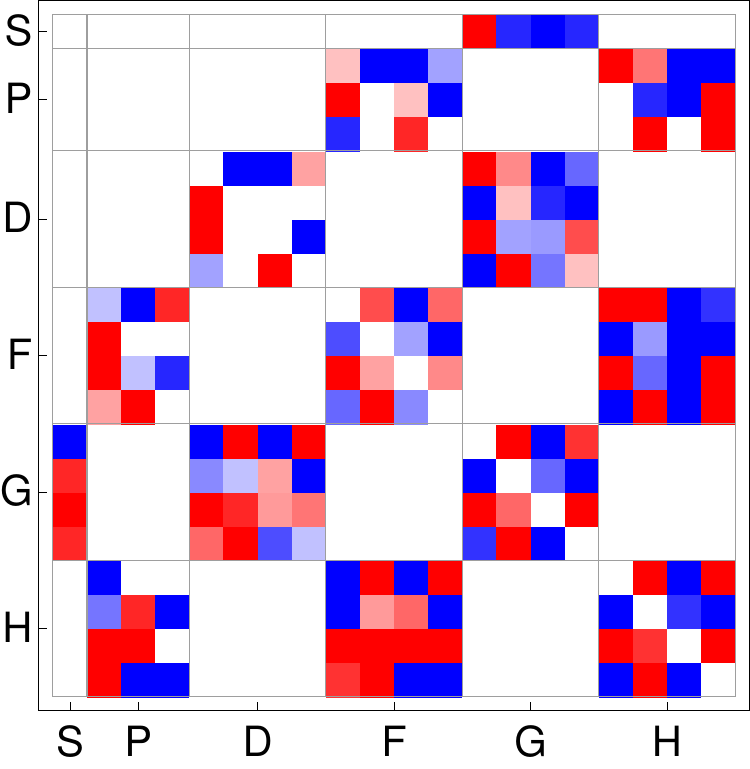} \end{minipage}
\begin{minipage}{.35\linewidth} \vspace*{0.500cm} \hspace*{-0.65cm}\includegraphics[width=1.15\textwidth]{T_l4_coeff_3.pdf}\end{minipage}
\begin{minipage}{.25\linewidth} \begin{align} \left(a_{5}\right)^{\check{T}}_{4} &= \left<S,G\right> + \left<P,F\right> \nonumber \\ & \hspace*{12.5pt} + \left<P,H\right>  + \left<D,D\right>   \nonumber   \\ & \hspace*{12.5pt} + \left<D,G\right>  + \left<F,F\right>   \nonumber \\ & \hspace*{12.5pt} + \left<F,H\right>  + \left<G,G\right>   \nonumber \\ & \hspace*{12.5pt} + \left<H,H\right>    \nonumber  \end{align} \end{minipage}
\caption{%
Left: Matrices $\mathcal{C}_{1\cdots 4}^{\check{T}}$, represented here in the color scheme, defines the coefficient $\left(a_{5}\right)_{1\cdots 4}^{\check{T}}$ for an expansion of $\check{T}$ up to $\text{L}_{\text{max}} = 5$. Center: Coefficients $\left(a_{4}\right)_{1\cdots 4}^{\check{T}}$ obtained from a fit to the $\check{T}$-data (black points). For references to the data see Table \ref{tab:DataBasis}. Bonn Gatchina predictions, truncated at different $\text{L}_{\mathrm{max}}$ ($\text{L}_{\mathrm{max}} = 1$ is drawn in green, $\text{L}_{\mathrm{max}} = 2$ in blue, $\text{L}_{\mathrm{max}} = 3$ in red, $\text{L}_{\mathrm{max}} = 4$ in black and $\text{L}_{\mathrm{max}} = 5$ in cyan) are drawn as well. Right: All partial wave interferences for $\text{L}_{\text{max}} = 5$ are indicated.
}
\label{tab:TColorPlots1}
\end{table*}

\begin{table*}[htb]
\RawFloats
\begin{minipage}{.075\linewidth}
\vspace*{-6.5pt}
\hspace*{5pt}
\begin{equation}
\mathcal{C}_{5}^{\check{T}} \equiv \nonumber
\end{equation}
\end{minipage}
\begin{minipage}{.3\linewidth} \vspace*{0.572cm} \includegraphics[width=0.875\textwidth]{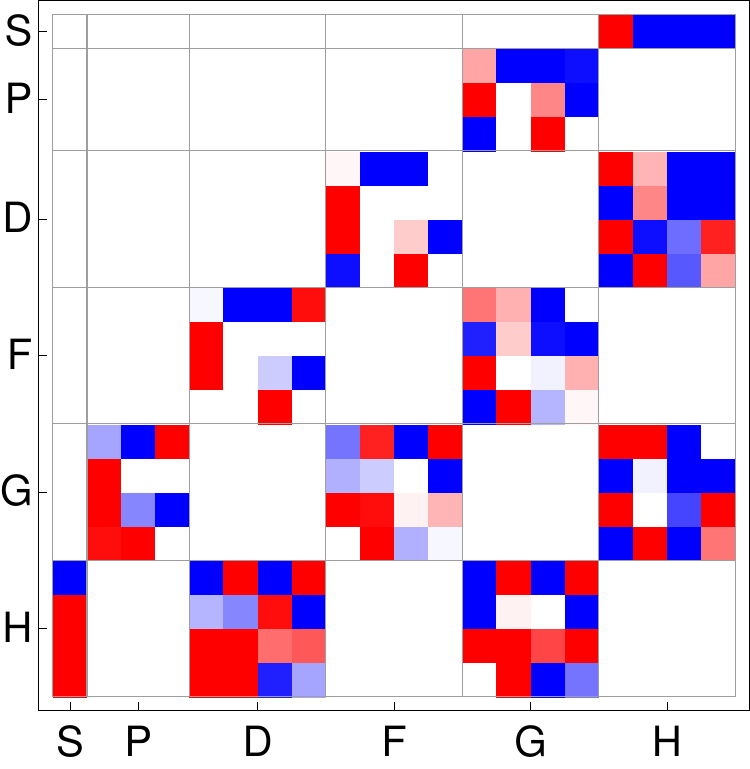} \end{minipage}
\begin{minipage}{.35\linewidth} \vspace*{0.500cm} \hspace*{-0.65cm}\includegraphics[width=1.15\textwidth]{T_l4_coeff_4.pdf}\end{minipage}
\begin{minipage}{.25\linewidth} \begin{align} \left(a_{5}\right)^{\check{T}}_{5} &= \left<S,H\right> + \left<P,G\right> \nonumber \\ & \hspace*{12.5pt} + \left<D,F\right>  + \left<D,H\right>   \nonumber  \\ & \hspace*{12.5pt} + \left<F,G\right>  + \left<G,H\right>   \nonumber\end{align} \end{minipage}

\begin{minipage}{.075\linewidth}
\vspace*{-6.5pt}
\hspace*{5pt}
\begin{equation}
\mathcal{C}_{6}^{\check{T}} \equiv \nonumber
\end{equation}
\end{minipage}
\begin{minipage}{.3\linewidth} \vspace*{0.572cm} \includegraphics[width=0.875\textwidth]{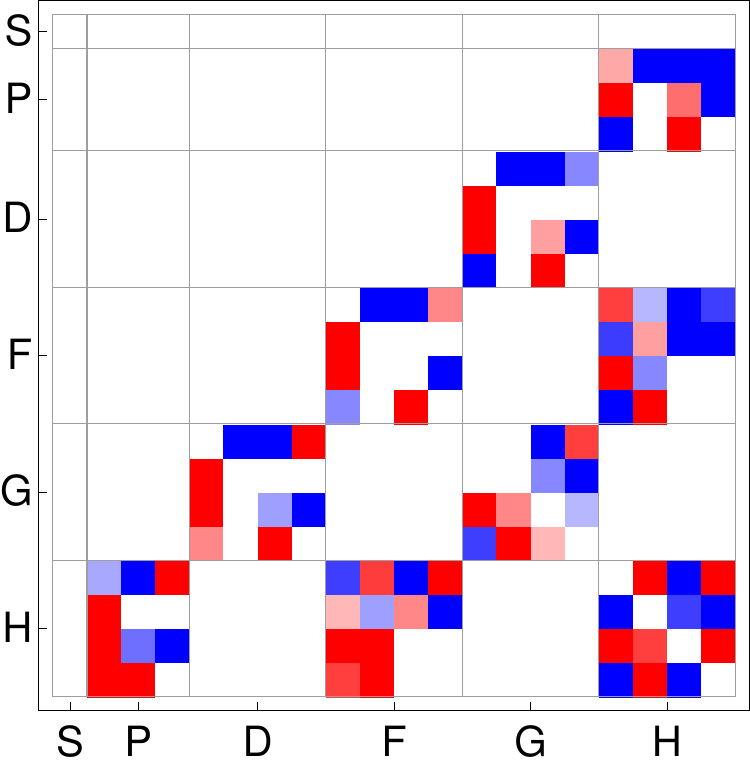} \end{minipage}
\begin{minipage}{.35\linewidth} \vspace*{0.500cm} \hspace*{-0.65cm}\includegraphics[width=1.15\textwidth]{T_l4_coeff_5.pdf}\end{minipage}
\begin{minipage}{.25\linewidth} \begin{align} \left(a_{5}\right)^{\check{T}}_{6} &= \left<P,H\right> + \left<D,G\right> \nonumber \\ & \hspace*{12.5pt} + \left<F,F\right>  + \left<F,H\right>   \nonumber  \\ & \hspace*{12.5pt} + \left<G,G\right>  + \left<H,H\right>   \nonumber   \end{align} \end{minipage}

\begin{minipage}{.075\linewidth}
\vspace*{-6.5pt}
\hspace*{5pt}
\begin{equation}
\mathcal{C}_{7}^{\check{T}} \equiv \nonumber
\end{equation}
\end{minipage}
\begin{minipage}{.3\linewidth} \vspace*{0.572cm} \includegraphics[width=0.875\textwidth]{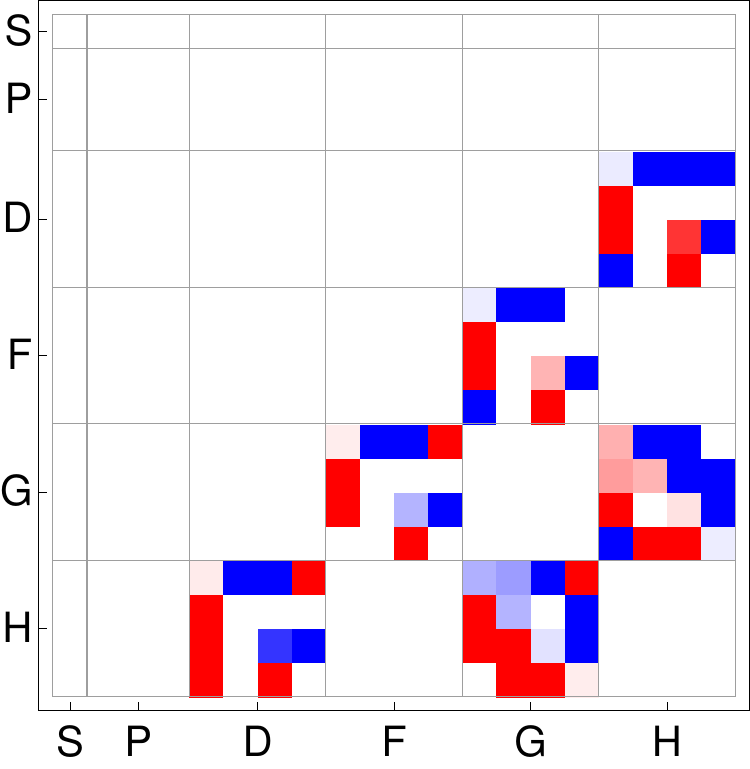} \end{minipage}
\begin{minipage}{.35\linewidth} \vspace*{0.500cm} \hspace*{-0.65cm}\includegraphics[width=1.15\textwidth]{T_l4_coeff_6.pdf}\end{minipage}
\begin{minipage}{.25\linewidth} \begin{align} \left(a_{5}\right)^{\check{T}}_{7} &= \left<D,H\right> + \left<F,G\right> \nonumber \\ & \hspace*{12.5pt} + \left<G,H\right>    \nonumber  \end{align} \end{minipage}

\begin{minipage}{.075\linewidth}
\vspace*{-6.5pt}
\hspace*{5pt}
\begin{equation}
\mathcal{C}_{8}^{\check{T}} \equiv \nonumber
\end{equation}
\end{minipage}
\begin{minipage}{.3\linewidth} \vspace*{0.572cm} \includegraphics[width=0.875\textwidth]{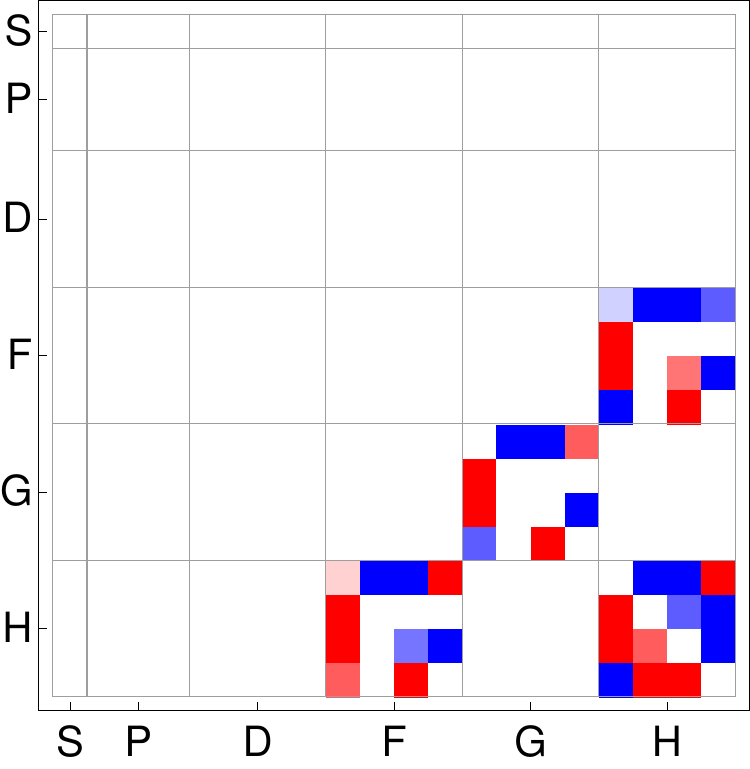} \end{minipage}
\begin{minipage}{.35\linewidth} \vspace*{0.500cm} \hspace*{-0.65cm}\includegraphics[width=1.15\textwidth]{T_l4_coeff_7.pdf}\end{minipage}
\begin{minipage}{.25\linewidth} \begin{align} \left(a_{5}\right)^{\check{T}}_{8} &= \left<F,H\right> + \left<G,G\right> \nonumber \\ & \hspace*{12.5pt} + \left<H,H\right>    \nonumber  \end{align} \end{minipage}
\caption{%
Left: Matrices $\mathcal{C}_{5\cdots 8}^{\check{T}}$, represented here in the color scheme, defines the coefficient $\left(a_{5}\right)_{5\cdots 8}^{\check{T}}$ for an expansion of $\check{T}$ up to $\text{L}_{\text{max}} = 5$. Center: Coefficients $\left(a_{4}\right)_{5\ldots 8}^{\check{T}}$ obtained from a fit to the $\check{T}$-data (black points). For references to the data see Table \ref{tab:DataBasis}. Bonn Gatchina predictions, truncated at different $\text{L}_{\mathrm{max}}$ ($\text{L}_{\mathrm{max}} = 1$ is drawn in green, $\text{L}_{\mathrm{max}} = 2$ in blue, $\text{L}_{\mathrm{max}} = 3$ in red, $\text{L}_{\mathrm{max}} = 4$ in black and $\text{L}_{\mathrm{max}} = 5$ in cyan) are drawn as well. Right: All partial wave interferences for $\text{L}_{\text{max}} = 5$ are indicated.
}
\label{tab:TColorPlots2}
\end{table*}
 
\begin{table*}[htb]
\RawFloats
\begin{minipage}{.075\linewidth}
\vspace*{-6.5pt}
\hspace*{5pt}
\begin{equation}
\mathcal{C}_{9}^{\check{T}} \equiv \nonumber
\end{equation}
\end{minipage}
\begin{minipage}{.3\linewidth} \vspace*{0.572cm} \includegraphics[width=0.875\textwidth]{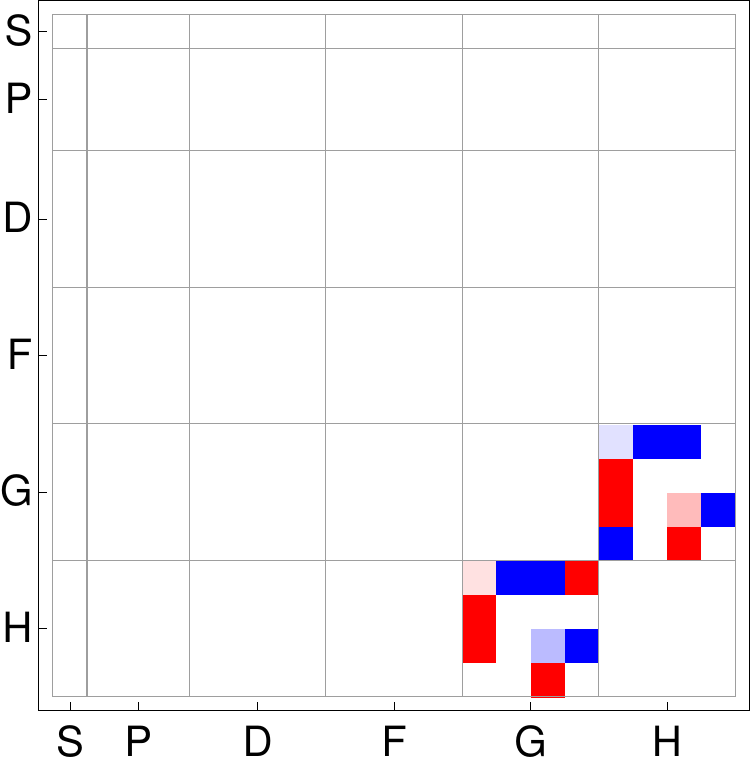} \end{minipage}
\begin{minipage}{.35\linewidth} \vspace*{0.500cm} \hspace*{-0.65cm}\includegraphics[width=1.15\textwidth]{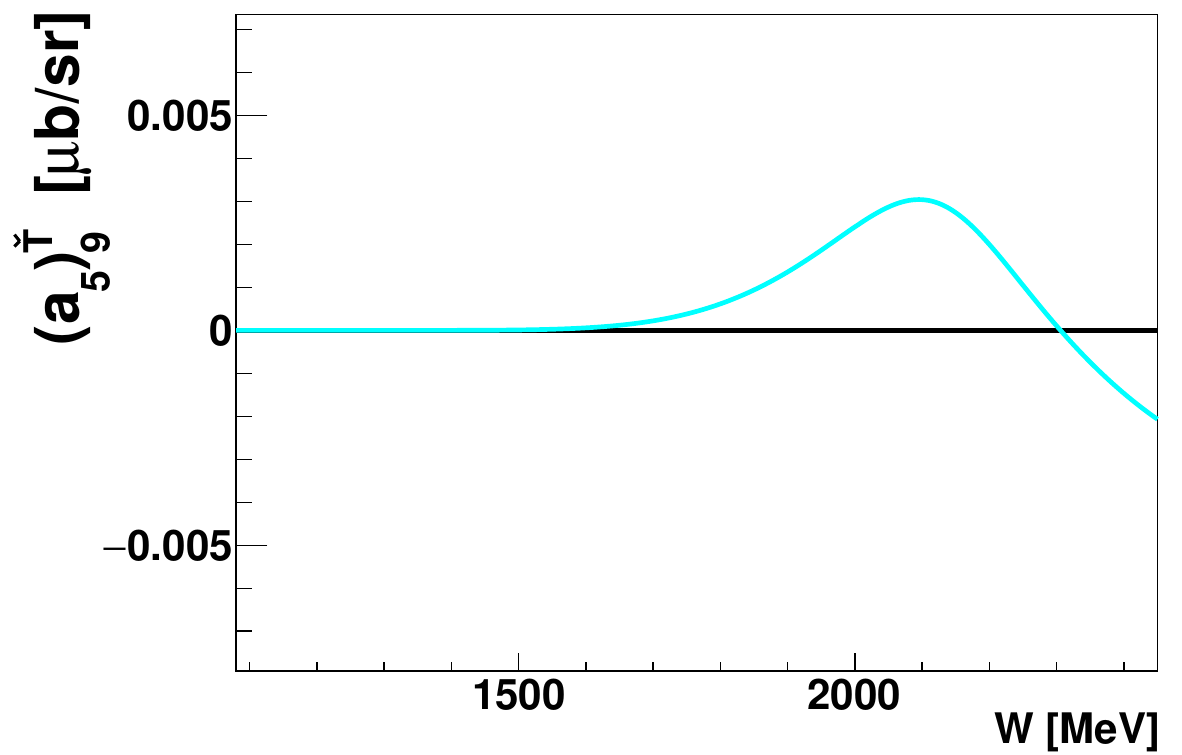}\end{minipage}
\begin{minipage}{.25\linewidth} \begin{align} \left(a_{5}\right)^{\check{T}}_{9} &= \left<G,H\right> \nonumber\end{align} \end{minipage}

\begin{minipage}{.075\linewidth}
\vspace*{-6.5pt}
\hspace*{5pt}
\begin{equation}
\mathcal{C}_{10}^{\check{T}} \equiv \nonumber
\end{equation}
\end{minipage}
\begin{minipage}{.3\linewidth} \vspace*{0.572cm} \includegraphics[width=0.875\textwidth]{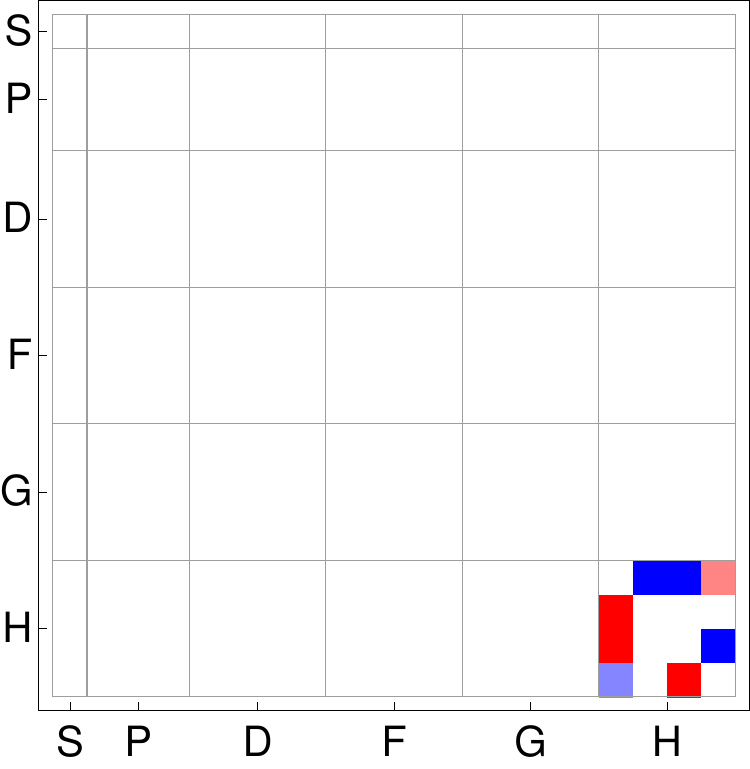} \end{minipage}
\begin{minipage}{.35\linewidth} \vspace*{0.500cm} \hspace*{-0.65cm}\includegraphics[width=1.15\textwidth]{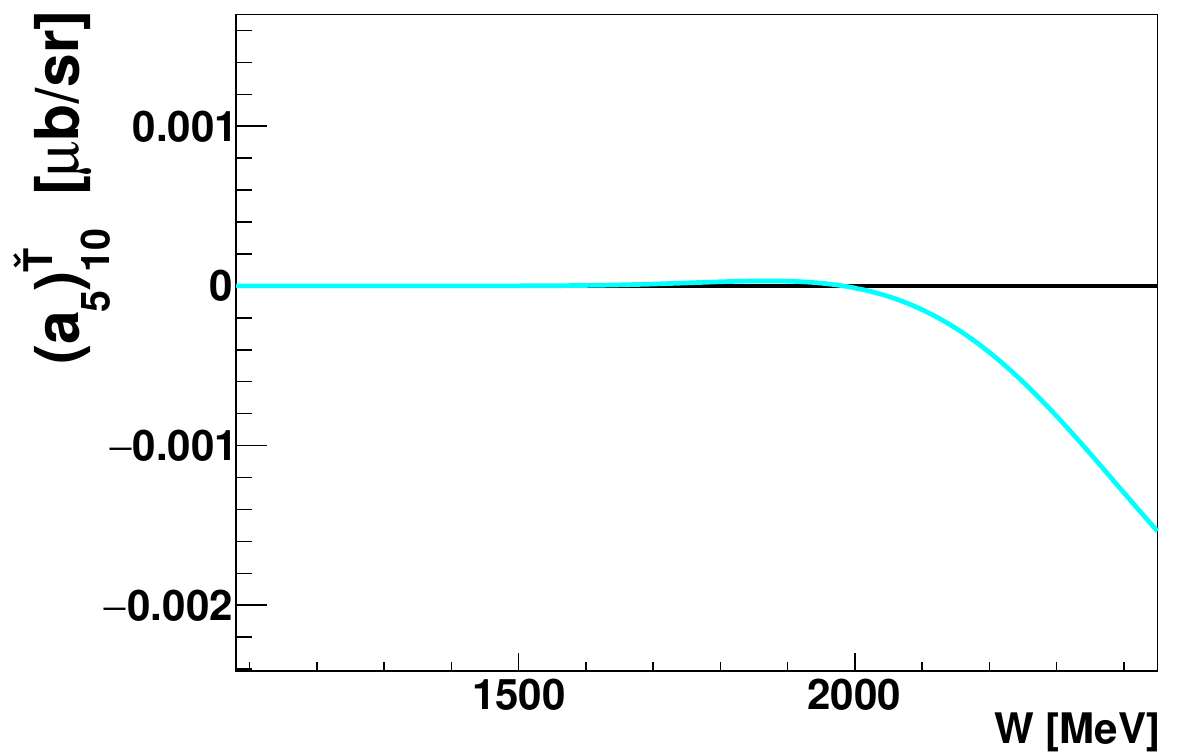}\end{minipage}
\begin{minipage}{.25\linewidth} \begin{align} \left(a_{5}\right)^{\check{T}}_{10} &=  \left<H,H\right>   \nonumber   \end{align} \end{minipage}
\caption{%
Left: Matrices $\mathcal{C}_{9,10}^{\check{T}}$, represented here in the color scheme, defines the coefficient $\left(a_{5}\right)_{9, 10}^{\check{T}}$ for an expansion of $\check{T}$ up to $\text{L}_{\text{max}} = 5$. Center: For the highest non-fitted coefficients $\left(a_{5}\right)_{9,10}^{\check{T}}$, the Bonn Gatchina curves are shown (here, the truncation at $\text{L}_{\mathrm{max}} = 5$ is drawn in cyan). Right: All partial wave interferences for $\text{L}_{\text{max}} = 5$ are indicated.
}
\label{tab:TColorPlots3}
\end{table*}

\begin{table*}[htb]
\RawFloats
\begin{minipage}{.075\linewidth}
\vspace*{-6.5pt}
\hspace*{5pt}
\begin{equation}
\mathcal{C}_{1}^{\check{P}} \equiv \nonumber
\end{equation}
\end{minipage}
\begin{minipage}{.3\linewidth} \vspace*{0.572cm} \includegraphics[width=0.875\textwidth]{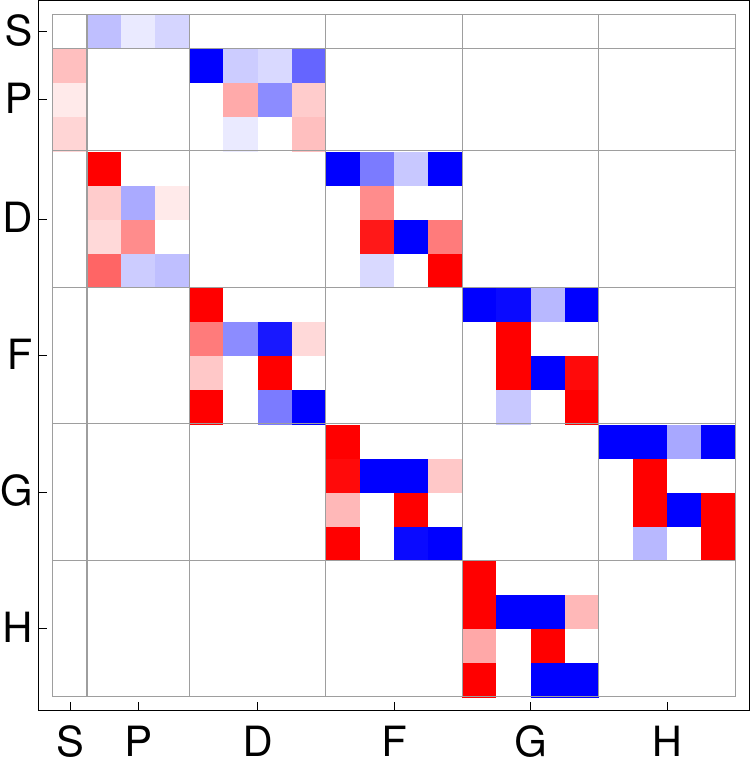} \end{minipage}
\begin{minipage}{.35\linewidth} \vspace*{0.500cm} \hspace*{-0.65cm}\includegraphics[width=1.15\textwidth]{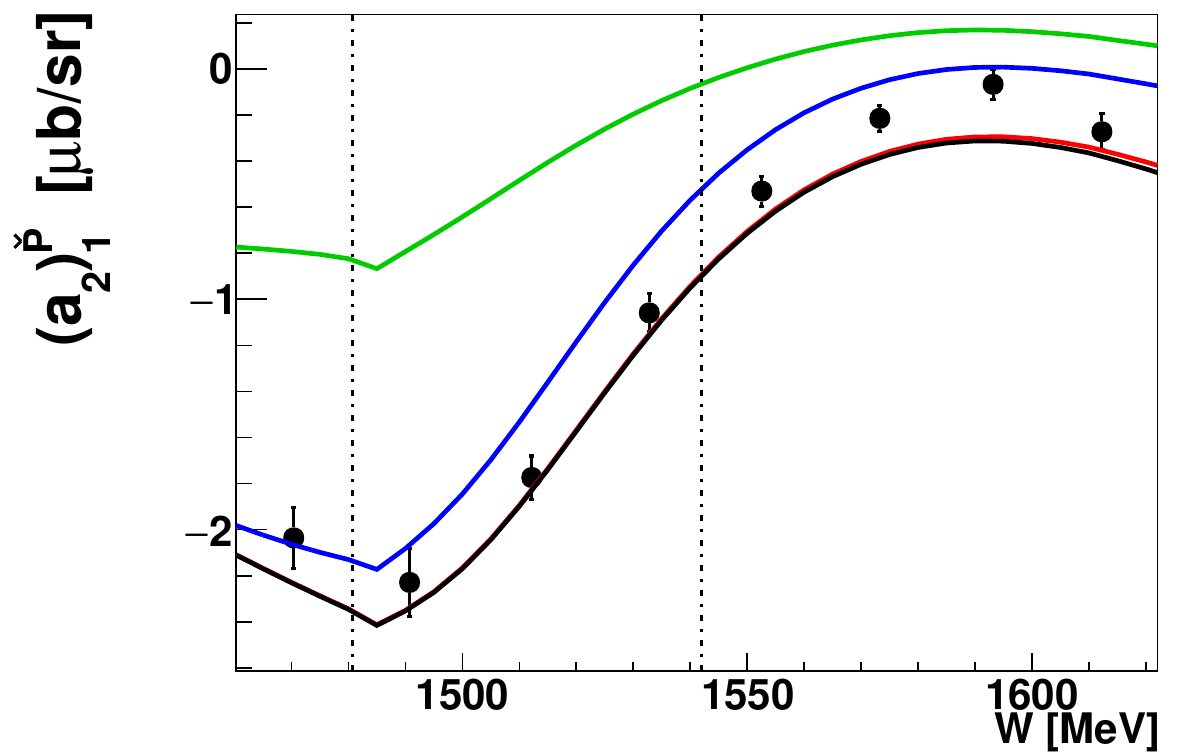}\end{minipage}
\begin{minipage}{.25\linewidth} \begin{align} \left(a_{5}\right)^{\check{P}}_{1} &= \left<S,P\right> + \left<P,D\right> \nonumber \\ & \hspace*{12.5pt} + \left<D,F\right> + \left<F,G\right> \nonumber \\ & \hspace*{12.5pt} + \left<G,H\right>   \nonumber  \end{align} \end{minipage}

\begin{minipage}{.075\linewidth}
\vspace*{-6.5pt}
\hspace*{5pt}
\begin{equation}
\mathcal{C}_{2}^{\check{P}} \equiv \nonumber
\end{equation}
\end{minipage}
\begin{minipage}{.3\linewidth} \vspace*{0.572cm} \includegraphics[width=0.875\textwidth]{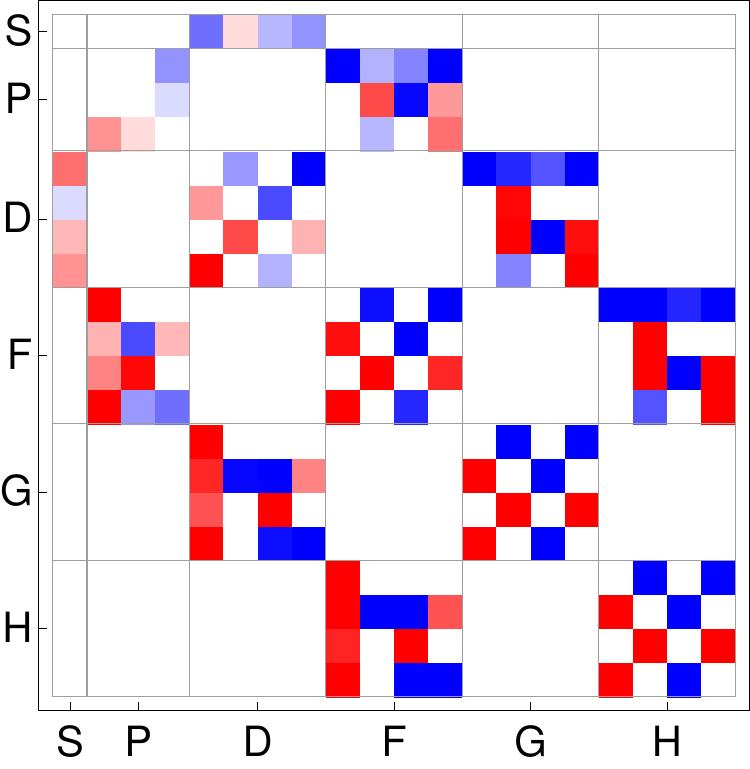} \end{minipage}
\begin{minipage}{.35\linewidth} \vspace*{0.500cm} \hspace*{-0.65cm}\includegraphics[width=1.15\textwidth]{P_l2_coeff_1.pdf}\end{minipage}
\begin{minipage}{.25\linewidth} \begin{align} \left(a_{5}\right)^{\check{P}}_{2} &= \left<S,D\right> + \left<P,P\right> \nonumber \\ & \hspace*{12.5pt} + \left<P,F\right>  + \left<D,D\right>   \nonumber \\ & \hspace*{12.5pt} + \left<D,G\right>  + \left<F,F\right>   \nonumber  \\ & \hspace*{12.5pt} + \left<F,H\right>  + \left<G,G\right>   \nonumber \\ & \hspace*{12.5pt} + \left<H,H\right>    \nonumber  \end{align} \end{minipage}

\begin{minipage}{.075\linewidth}
\vspace*{-6.5pt}
\hspace*{5pt}
\begin{equation}
\mathcal{C}_{3}^{\check{P}} \equiv \nonumber
\end{equation}
\end{minipage}
\begin{minipage}{.3\linewidth} \vspace*{0.572cm} \includegraphics[width=0.875\textwidth]{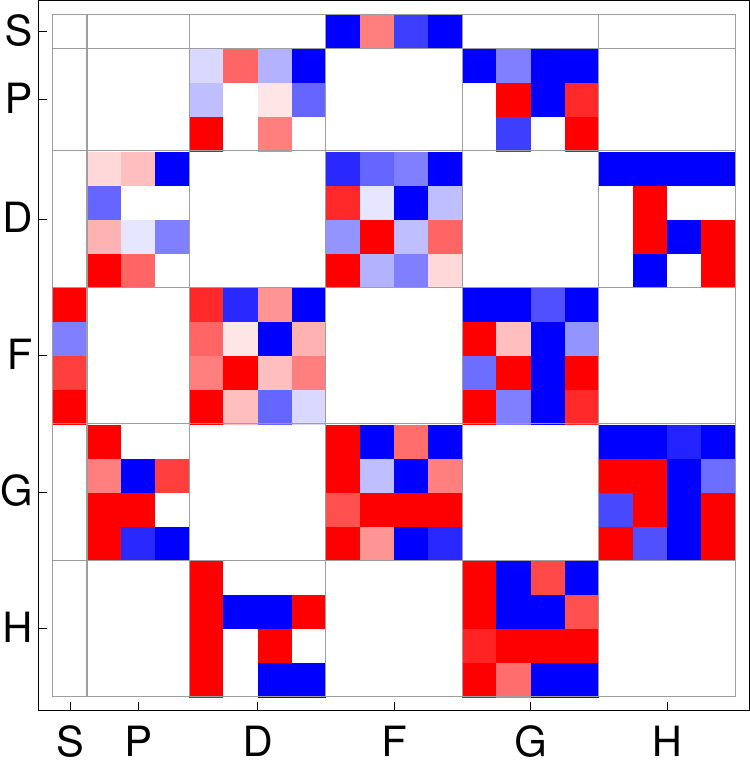} \end{minipage}
\begin{minipage}{.35\linewidth} \vspace*{0.500cm} \hspace*{-0.65cm}\includegraphics[width=1.15\textwidth]{P_l2_coeff_2.pdf}\end{minipage}
\begin{minipage}{.25\linewidth} \begin{align} \left(a_{5}\right)^{\check{P}}_{3} &= \left<S,F\right> + \left<P,D\right> \nonumber \\ & \hspace*{12.5pt} + \left<P,G\right>  + \left<D,F\right>   \nonumber  \\ & \hspace*{12.5pt} + \left<D,H\right>  + \left<F,G\right>   \nonumber \\ & \hspace*{12.5pt} + \left<G,H\right>   \nonumber   \end{align} \end{minipage}

\begin{minipage}{.075\linewidth}
\vspace*{-6.5pt}
\hspace*{5pt}
\begin{equation}
\mathcal{C}_{4}^{\check{P}} \equiv \nonumber
\end{equation}
\end{minipage}
\begin{minipage}{.3\linewidth} \vspace*{0.572cm} \includegraphics[width=0.875\textwidth]{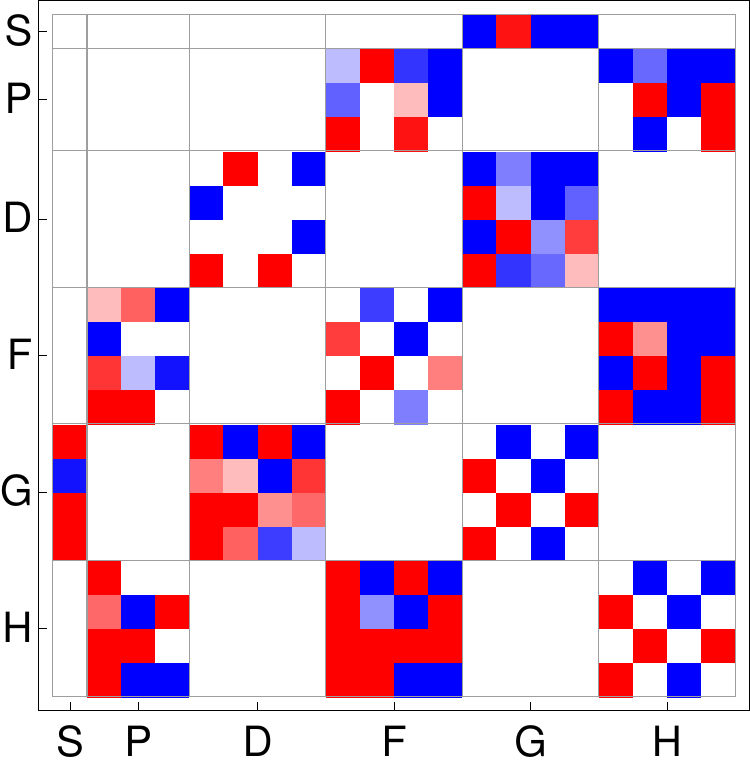} \end{minipage}
\begin{minipage}{.35\linewidth} \vspace*{0.500cm} \hspace*{-0.65cm}\includegraphics[width=1.15\textwidth]{P_l2_coeff_3.pdf}\end{minipage}
\begin{minipage}{.25\linewidth} \begin{align} \left(a_{5}\right)^{\check{P}}_{4} &= \left<S,G\right> + \left<P,F\right> \nonumber \\ & \hspace*{12.5pt} + \left<P,H\right>  + \left<D,D\right>   \nonumber   \\ & \hspace*{12.5pt} + \left<D,G\right>  + \left<F,F\right>   \nonumber \\ & \hspace*{12.5pt} + \left<F,H\right>  + \left<G,G\right>   \nonumber \\ & \hspace*{12.5pt} + \left<H,H\right>    \nonumber  \end{align} \end{minipage}
\caption{%
Left: Matrices $\mathcal{C}_{1\cdots 4}^{\check{P}}$, represented here in the color scheme, defines the coefficient $\left(a_{5}\right)_{1\cdots 4}^{\check{P}}$ for an expansion of $\check{P}$ up to $\text{L}_{\text{max}} = 5$. Center: Coefficients $\left(a_{2}\right)_{1\cdots 4}^{\check{P}}$ obtained from a fit to the $\check{P}$-data (black points). For references to the data see Table \ref{tab:DataBasis}. Bonn Gatchina predictions, truncated at different $\text{L}_{\mathrm{max}}$ ($\text{L}_{\mathrm{max}} = 1$ is drawn in green, $\text{L}_{\mathrm{max}} = 2$ in blue, $\text{L}_{\mathrm{max}} = 3$ in red and $\text{L}_{\mathrm{max}} = 4$ in black) are drawn as well. Right: All partial wave interferences for $\text{L}_{\text{max}} = 5$ are indicated.
}
\label{tab:PColorPlots1}
\end{table*}

\begin{table*}[htb]
\RawFloats
\begin{minipage}{.075\linewidth}
\vspace*{-6.5pt}
\hspace*{5pt}
\begin{equation}
\mathcal{C}_{5}^{\check{P}} \equiv \nonumber
\end{equation}
\end{minipage}
\begin{minipage}{.3\linewidth} \vspace*{0.572cm} \includegraphics[width=0.875\textwidth]{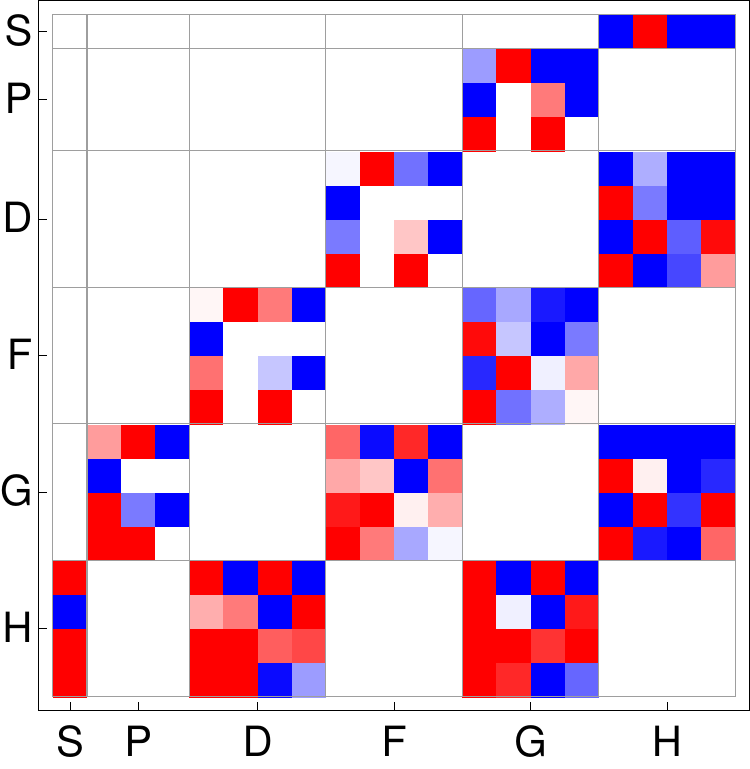} \end{minipage}
\begin{minipage}{.35\linewidth} \vspace*{0.500cm} \hspace*{-0.65cm}\includegraphics[width=1.15\textwidth]{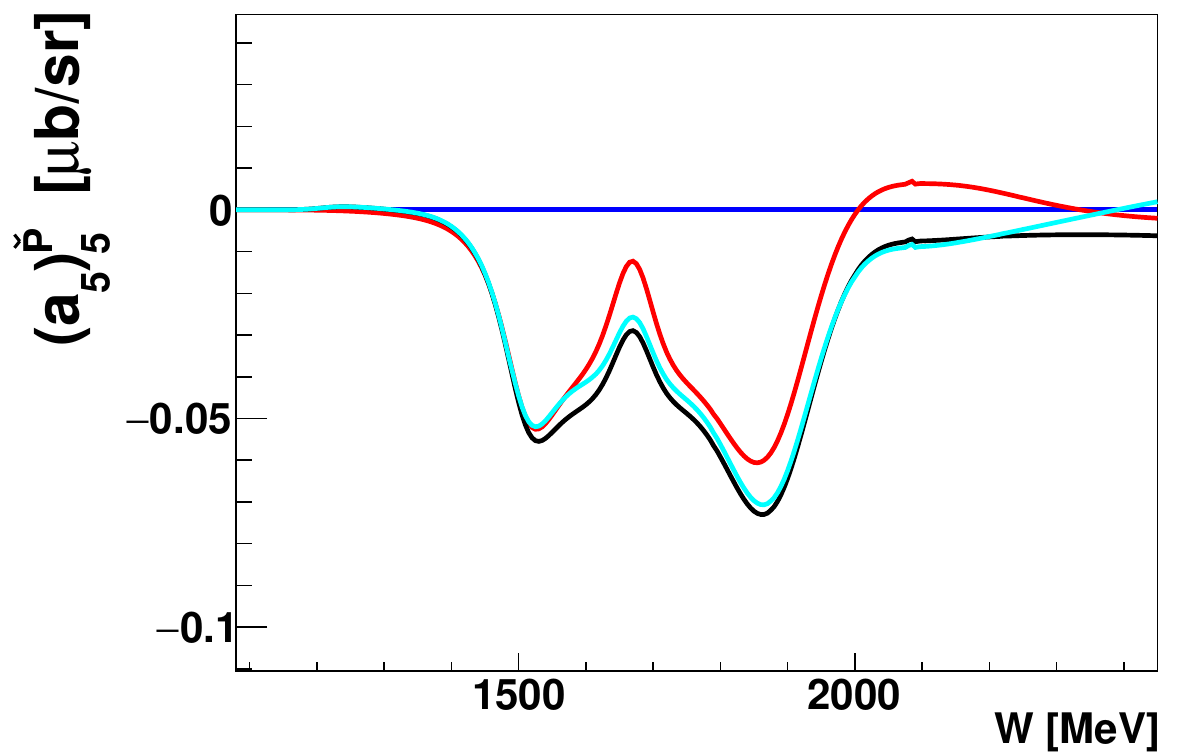}\end{minipage}
\begin{minipage}{.25\linewidth} \begin{align} \left(a_{5}\right)^{\check{P}}_{5} &= \left<S,H\right> + \left<P,G\right> \nonumber \\ & \hspace*{12.5pt} + \left<D,F\right>  + \left<D,H\right>   \nonumber  \\ & \hspace*{12.5pt} + \left<F,G\right>  + \left<G,H\right>   \nonumber\end{align} \end{minipage}

\begin{minipage}{.075\linewidth}
\vspace*{-6.5pt}
\hspace*{5pt}
\begin{equation}
\mathcal{C}_{6}^{\check{P}} \equiv \nonumber
\end{equation}
\end{minipage}
\begin{minipage}{.3\linewidth} \vspace*{0.572cm} \includegraphics[width=0.875\textwidth]{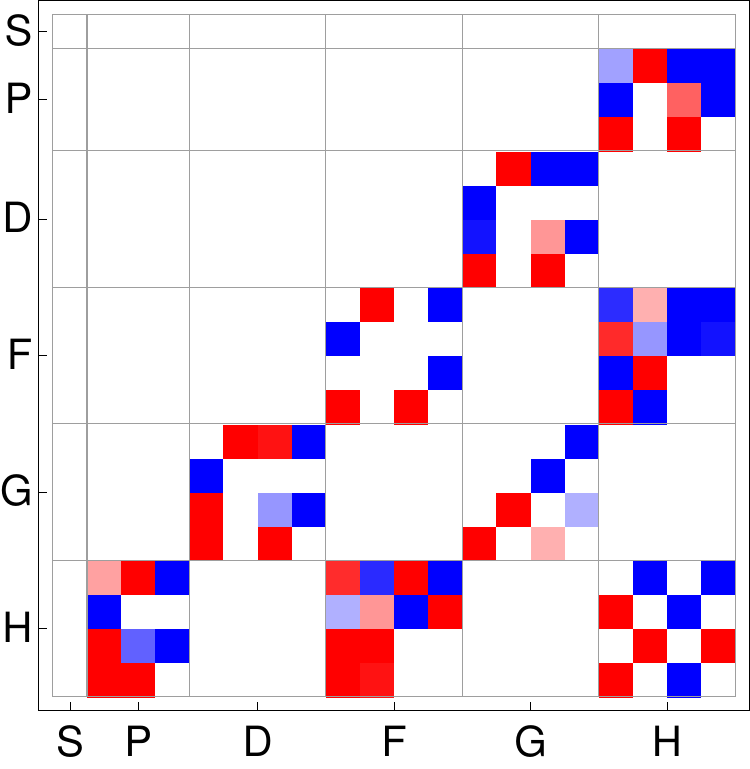} \end{minipage}
\begin{minipage}{.35\linewidth} \vspace*{0.500cm} \hspace*{-0.65cm}\includegraphics[width=1.15\textwidth]{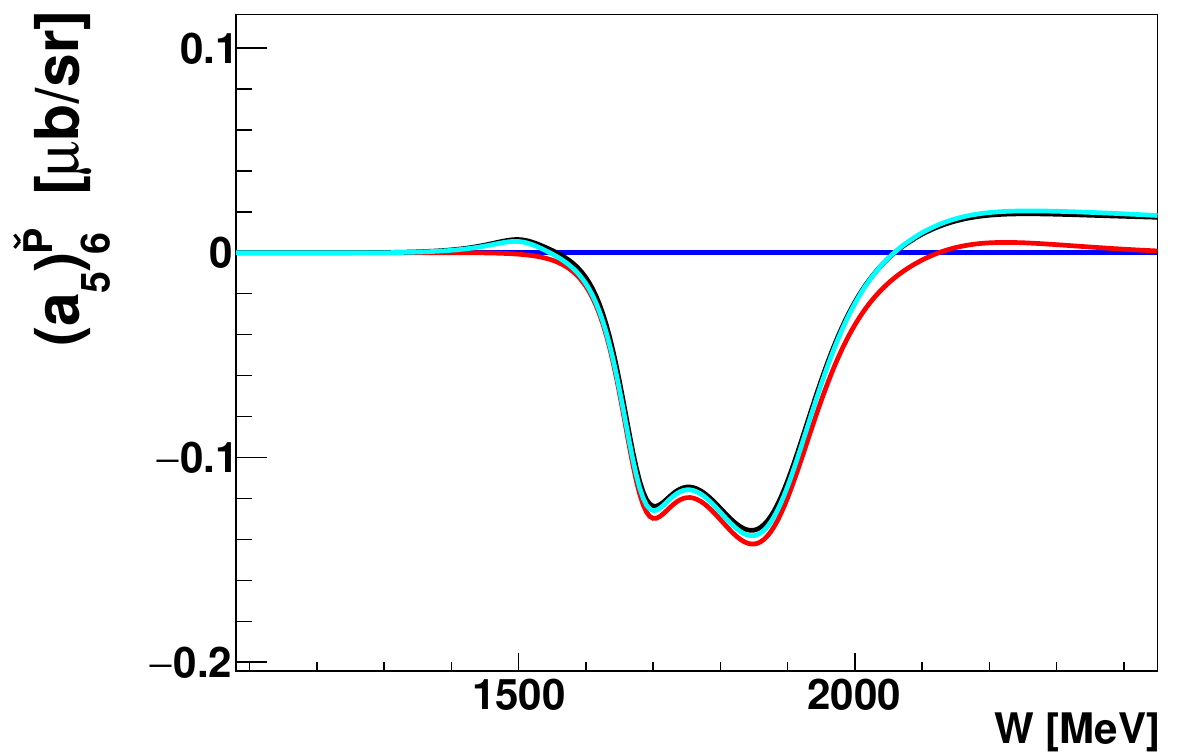}\end{minipage}
\begin{minipage}{.25\linewidth} \begin{align} \left(a_{5}\right)^{\check{P}}_{6} &= \left<P,H\right> + \left<D,G\right> \nonumber \\ & \hspace*{12.5pt} + \left<F,F\right>  + \left<F,H\right>   \nonumber  \\ & \hspace*{12.5pt} + \left<G,G\right>  + \left<H,H\right>   \nonumber   \end{align} \end{minipage}

\begin{minipage}{.075\linewidth}
\vspace*{-6.5pt}
\hspace*{5pt}
\begin{equation}
\mathcal{C}_{7}^{\check{P}} \equiv \nonumber
\end{equation}
\end{minipage}
\begin{minipage}{.3\linewidth} \vspace*{0.572cm} \includegraphics[width=0.875\textwidth]{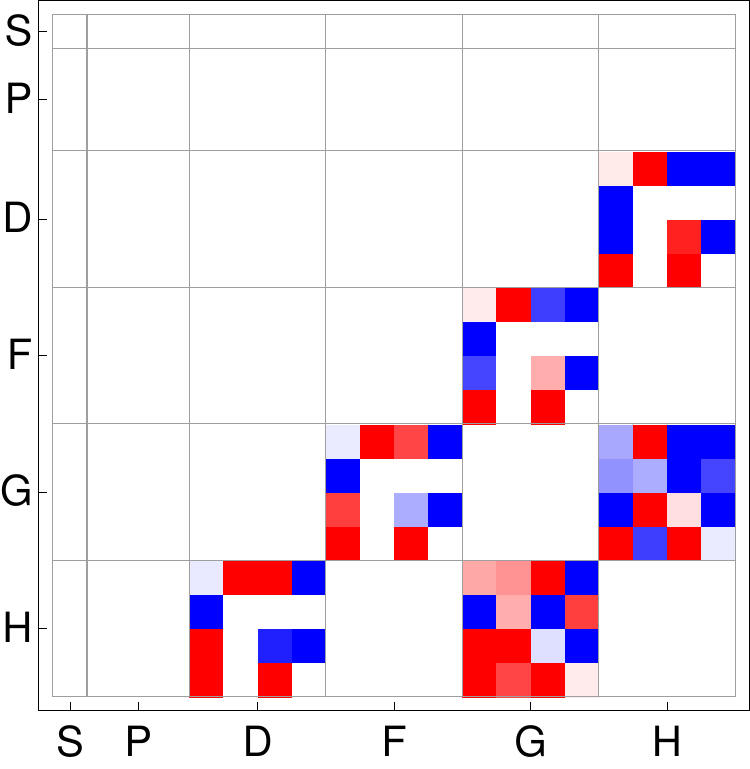} \end{minipage}
\begin{minipage}{.35\linewidth} \vspace*{0.500cm} \hspace*{-0.65cm}\includegraphics[width=1.15\textwidth]{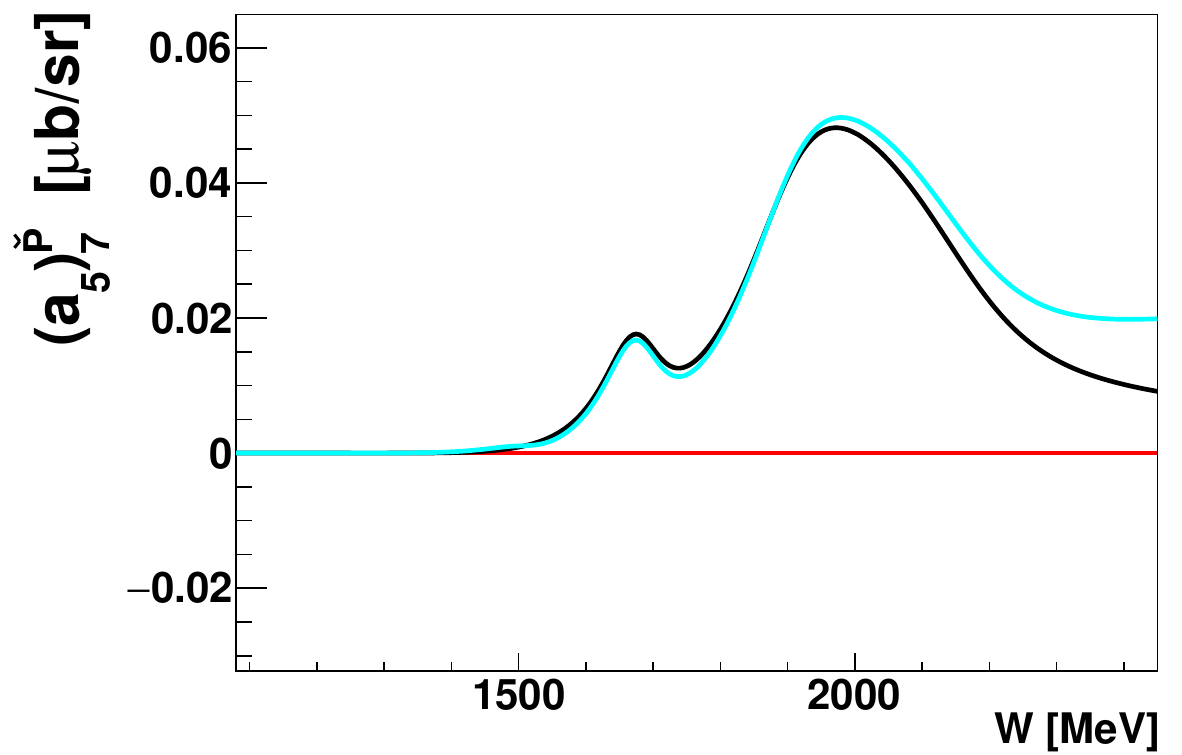}\end{minipage}
\begin{minipage}{.25\linewidth} \begin{align} \left(a_{5}\right)^{\check{P}}_{7} &= \left<D,H\right> + \left<F,G\right> \nonumber \\ & \hspace*{12.5pt} + \left<G,H\right>    \nonumber  \end{align} \end{minipage}

\begin{minipage}{.075\linewidth}
\vspace*{-6.5pt}
\hspace*{5pt}
\begin{equation}
\mathcal{C}_{8}^{\check{P}} \equiv \nonumber
\end{equation}
\end{minipage}
\begin{minipage}{.3\linewidth} \vspace*{0.572cm} \includegraphics[width=0.875\textwidth]{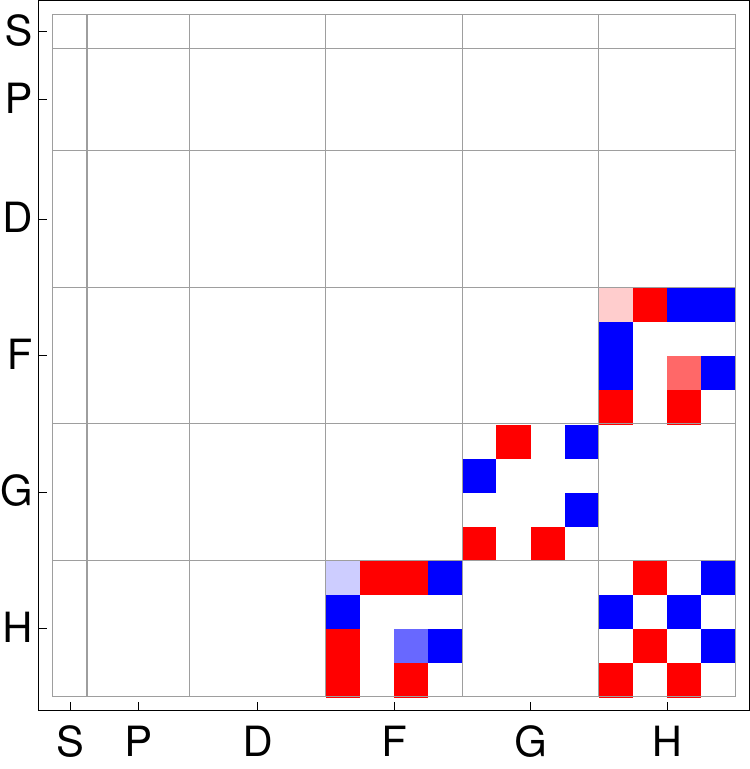} \end{minipage}
\begin{minipage}{.35\linewidth} \vspace*{0.500cm} \hspace*{-0.65cm}\includegraphics[width=1.15\textwidth]{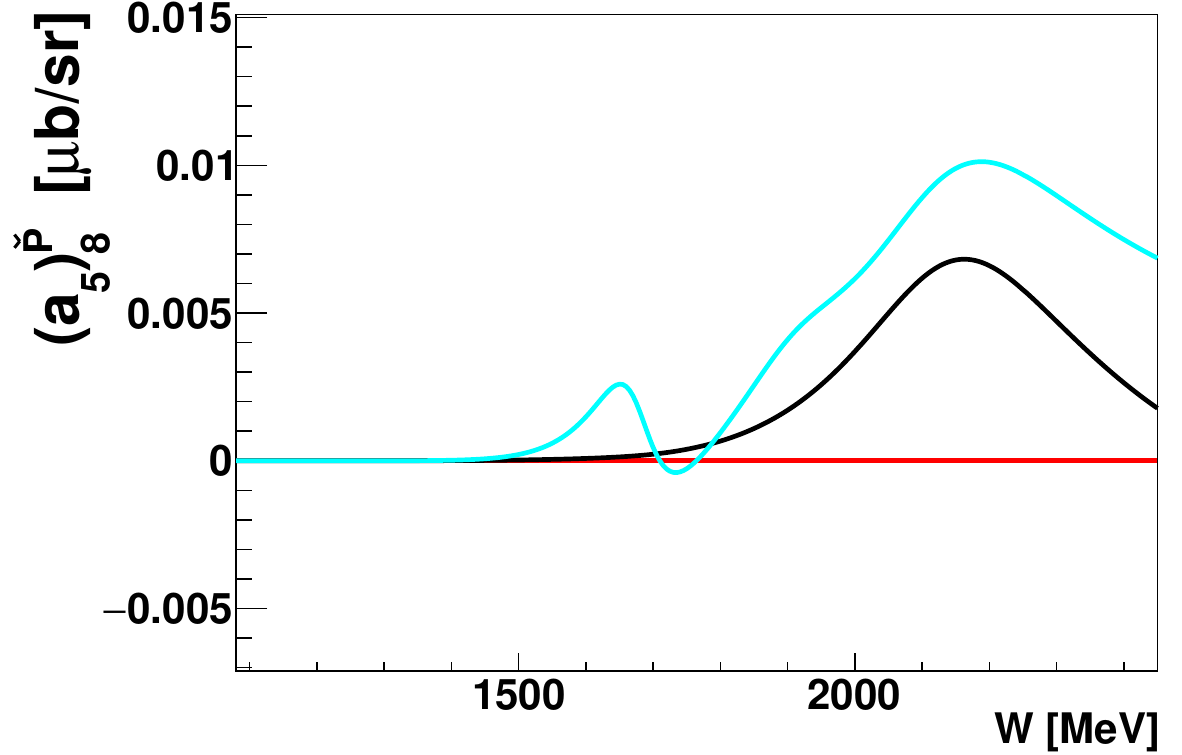}\end{minipage}
\begin{minipage}{.25\linewidth} \begin{align} \left(a_{5}\right)^{\check{P}}_{8} &= \left<F,H\right> + \left<G,G\right> \nonumber \\ & \hspace*{12.5pt} + \left<H,H\right>    \nonumber  \end{align} \end{minipage}
\caption{%
Left: Matrices $\mathcal{C}_{5\cdots 8}^{\check{P}}$, represented here in the color scheme, defines the coefficient $\left(a_{5}\right)_{5\cdots 8}^{\check{P}}$ for an expansion of $\check{P}$ up to $\text{L}_{\text{max}} = 5$. Center: For the higher non-fitted coefficients $\left(a_{5}\right)_{5 \ldots 8}^{\check{P}}$, the Bonn Gatchina curves are shown (here, the truncation at $\text{L}_{\mathrm{max}} = 5$ is drawn in cyan). Right: All partial wave interferences for $\text{L}_{\text{max}} = 5$ are indicated.
}
\label{tab:PColorPlots2}
\end{table*}
 
\begin{table*}[htb]
\RawFloats
\begin{minipage}{.075\linewidth}
\vspace*{-6.5pt}
\hspace*{5pt}
\begin{equation}
\mathcal{C}_{9}^{\check{P}} \equiv \nonumber
\end{equation}
\end{minipage}
\begin{minipage}{.3\linewidth} \vspace*{0.572cm} \includegraphics[width=0.875\textwidth]{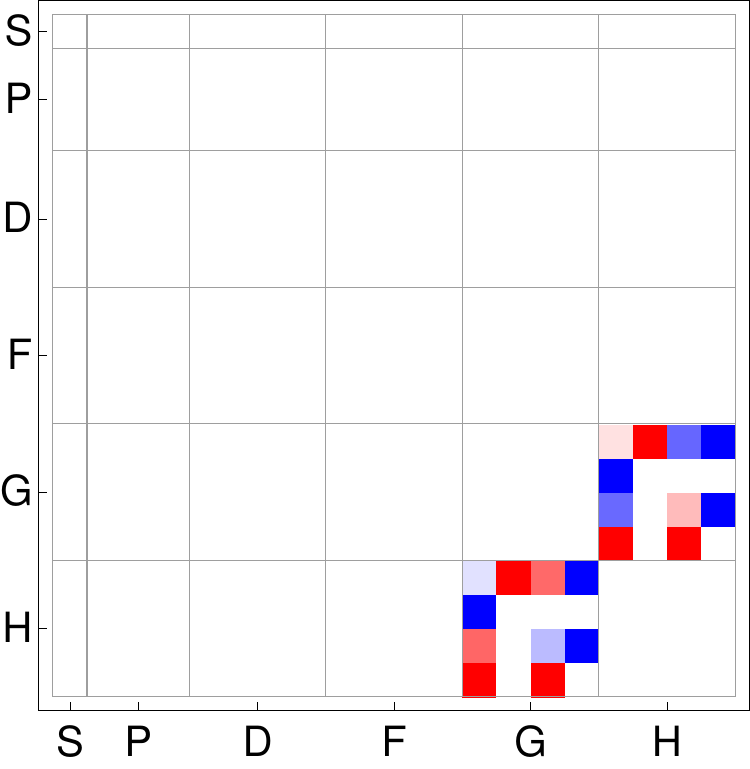} \end{minipage}
\begin{minipage}{.35\linewidth} \vspace*{0.500cm} \hspace*{-0.65cm}\includegraphics[width=1.15\textwidth]{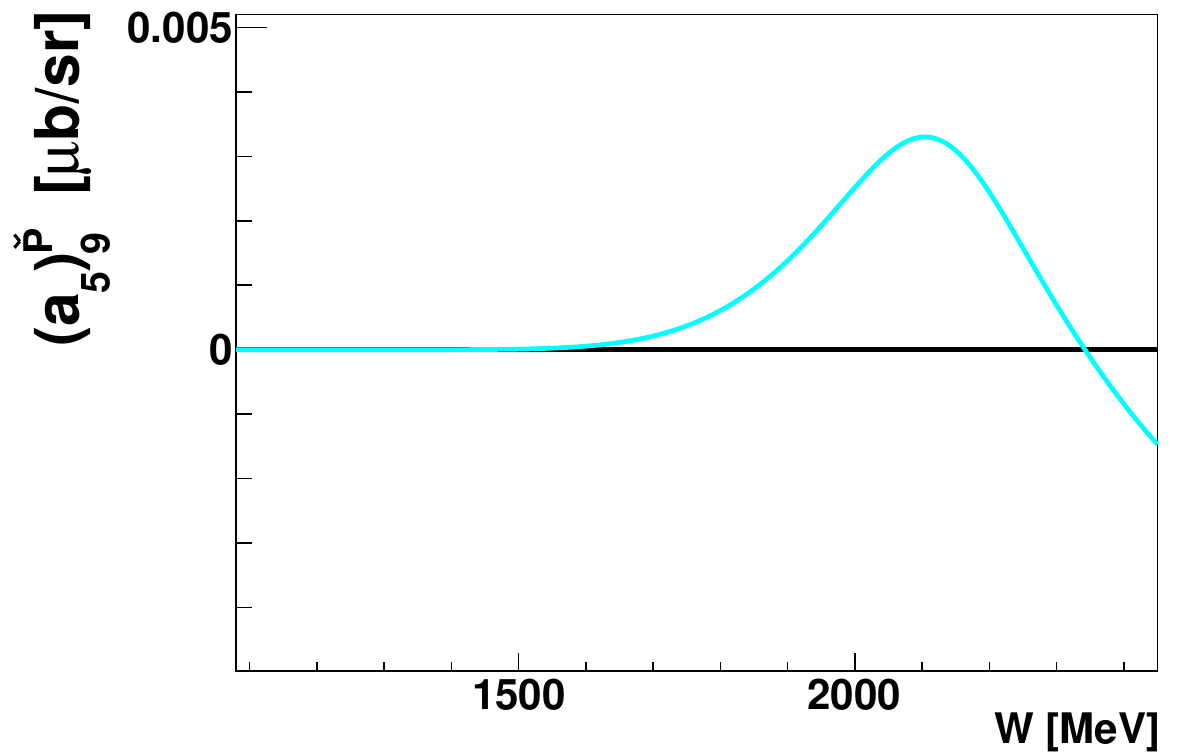}\end{minipage}
\begin{minipage}{.25\linewidth} \begin{align} \left(a_{5}\right)^{\check{P}}_{9} &= \left<G,H\right> \nonumber\end{align} \end{minipage}

\begin{minipage}{.075\linewidth}
\vspace*{-6.5pt}
\hspace*{5pt}
\begin{equation}
\mathcal{C}_{10}^{\check{P}} \equiv \nonumber
\end{equation}
\end{minipage}
\begin{minipage}{.3\linewidth} \vspace*{0.572cm} \includegraphics[width=0.875\textwidth]{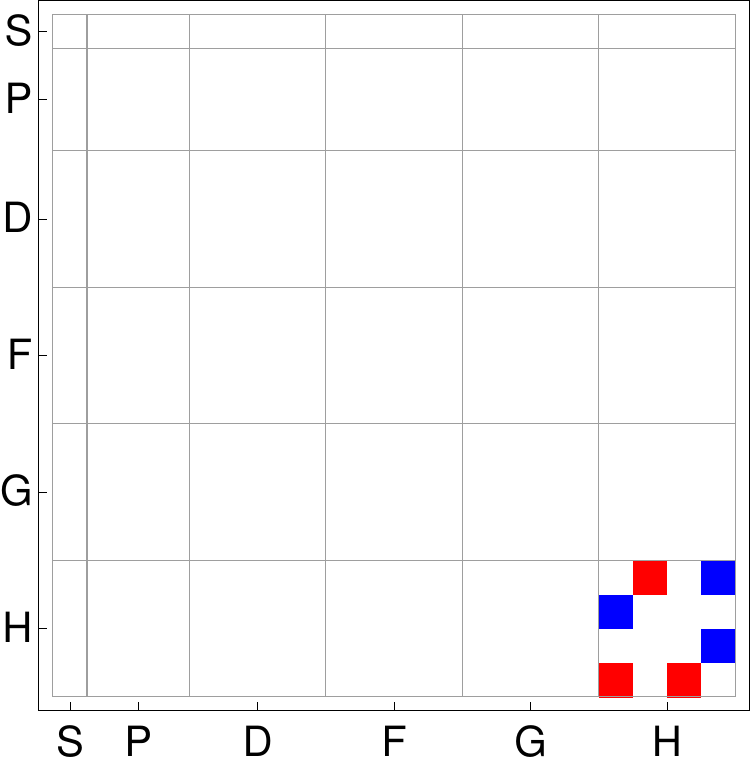} \end{minipage}
\begin{minipage}{.35\linewidth} \vspace*{0.500cm} \hspace*{-0.65cm}\includegraphics[width=1.15\textwidth]{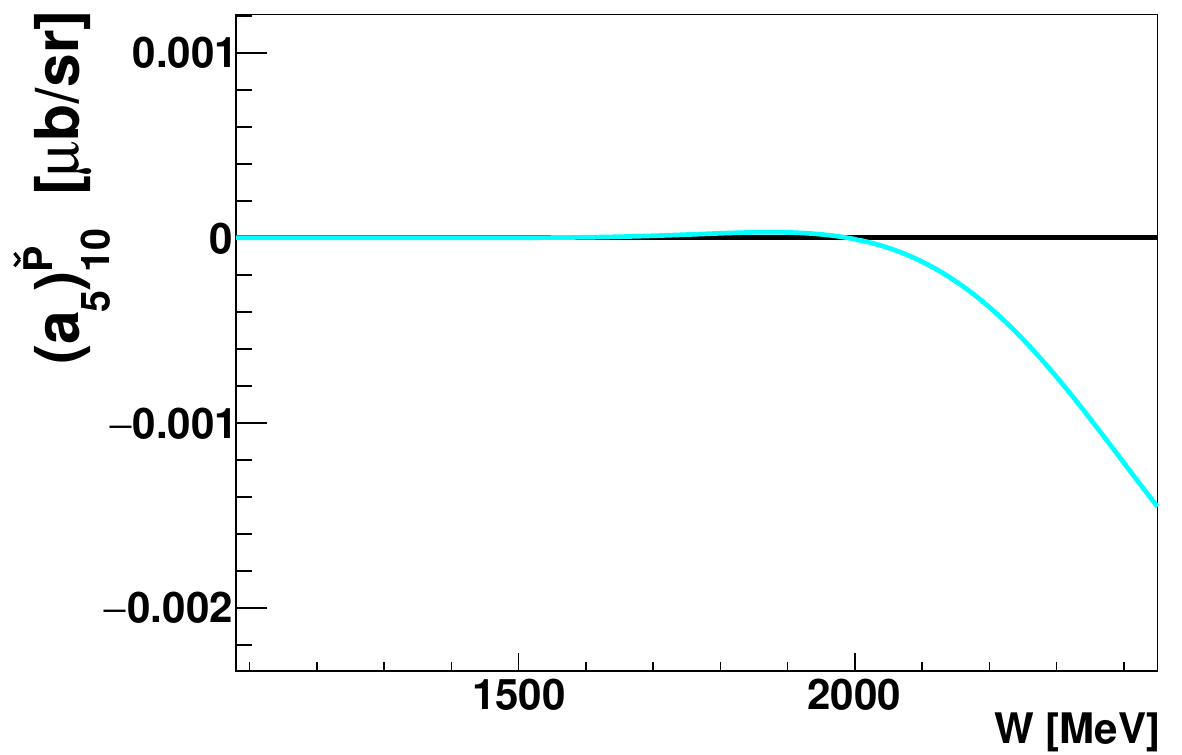}\end{minipage}
\begin{minipage}{.25\linewidth} \begin{align} \left(a_{5}\right)^{\check{P}}_{10} &=  \left<H,H\right>   \nonumber   \end{align} \end{minipage}
\caption{%
Left: Matrices $\mathcal{C}_{9, 10}^{\check{P}}$, represented here in the color scheme, defines the coefficient $\left(a_{5}\right)_{9, 10}^{\check{P}}$ for an expansion of $\check{P}$ up to $\text{L}_{\text{max}} = 5$. Center: For the highest non-fitted coefficient $\left(a_{5}\right)_{9, 10}^{\check{P}}$, the Bonn Gatchina curves are shown (here, the truncation at $\text{L}_{\mathrm{max}} = 5$ is drawn in cyan). Right: All partial wave interferences for $\text{L}_{\text{max}} = 5$ are indicated.
}
\label{tab:PColorPlots3}
\end{table*}


%
\begin{table*}[htb]
\RawFloats
\begin{minipage}{.075\linewidth}
\vspace*{-6.5pt}
\hspace*{5pt}
\begin{equation}
\mathcal{C}_{0}^{\check{E}} \equiv \nonumber
\end{equation}
\end{minipage}
\begin{minipage}{.3\linewidth} \vspace*{0.572cm} \includegraphics[width=0.875\textwidth]{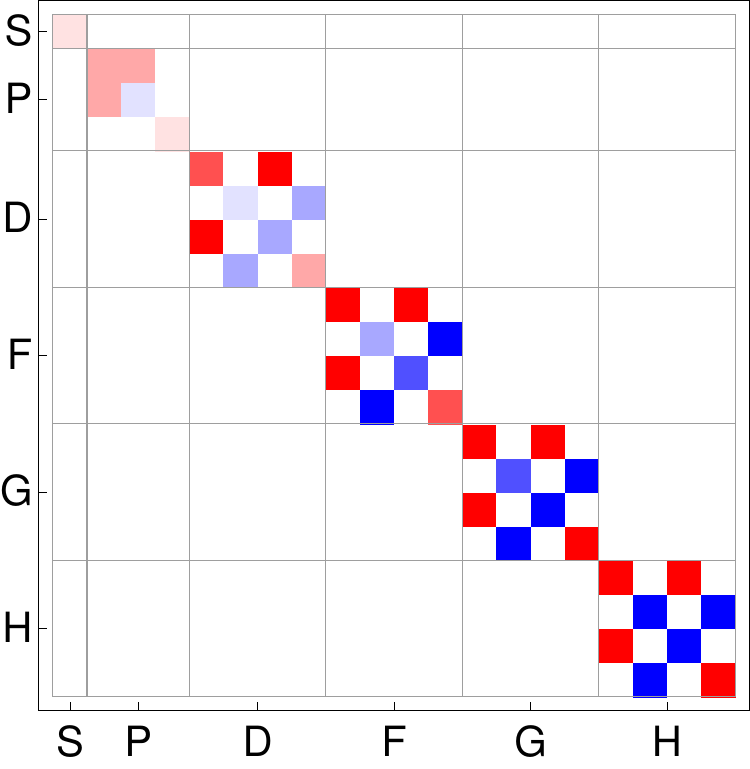} \end{minipage}
\begin{minipage}{.35\linewidth} \vspace*{0.500cm} \hspace*{-0.65cm}\includegraphics[width=1.15\textwidth]{E_l3_coeff_0_NoLetter.pdf}\end{minipage}
\begin{minipage}{.25\linewidth} \begin{align} \left(a_{5}\right)^{\check{E}}_{0} &= \left<S,S\right> + \left<P,P\right> \nonumber \\ & \hspace*{12.5pt} + \left<D,D\right>  + \left<F,F\right>   \nonumber \\ & \hspace*{12.5pt} + \left<G,G\right>  + \left<H,H\right>   \nonumber  \end{align} \end{minipage}

\begin{minipage}{.075\linewidth}
\vspace*{-6.5pt}
\hspace*{5pt}
\begin{equation}
\mathcal{C}_{1}^{\check{E}} \equiv \nonumber
\end{equation}
\end{minipage}
\begin{minipage}{.3\linewidth} \vspace*{0.572cm} \includegraphics[width=0.875\textwidth]{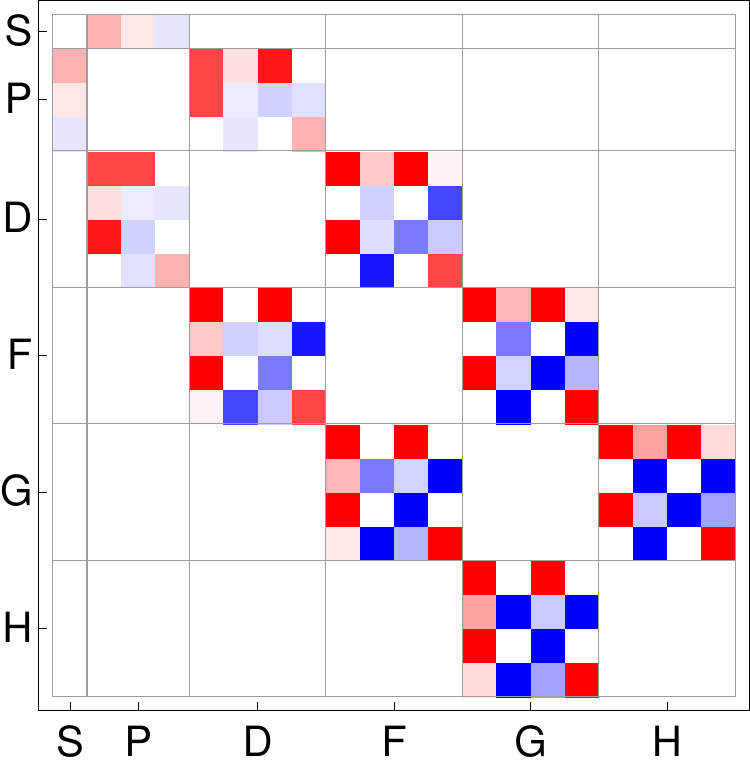} \end{minipage}
\begin{minipage}{.35\linewidth} \vspace*{0.500cm} \hspace*{-0.65cm}\includegraphics[width=1.15\textwidth]{E_l3_coeff_1.pdf}\end{minipage}
\begin{minipage}{.25\linewidth} \begin{align} \left(a_{5}\right)^{\check{E}}_{1} &= \left<S,P\right> + \left<P,D\right> \nonumber \\ & \hspace*{12.5pt} + \left<D,F\right>  + \left<F,G\right>   \nonumber \\ & \hspace*{12.5pt} + \left<G,H\right>    \nonumber   \end{align} \end{minipage}

\begin{minipage}{.075\linewidth}
\vspace*{-6.5pt}
\hspace*{5pt}
\begin{equation}
\mathcal{C}_{2}^{\check{E}} \equiv \nonumber
\end{equation}
\end{minipage}
\begin{minipage}{.3\linewidth} \vspace*{0.572cm} \includegraphics[width=0.875\textwidth]{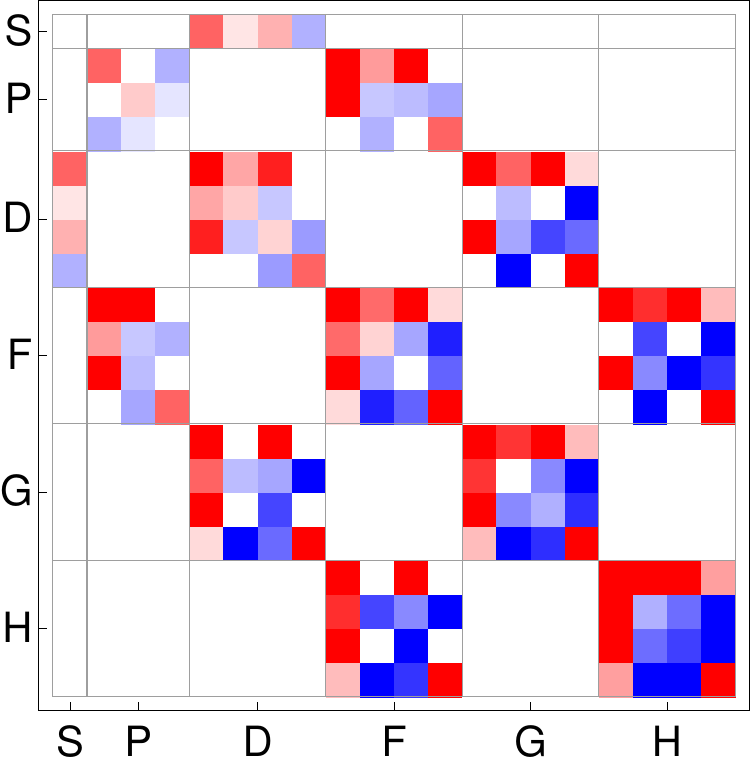} \end{minipage}
\begin{minipage}{.35\linewidth} \vspace*{0.500cm} \hspace*{-0.65cm}\includegraphics[width=1.15\textwidth]{E_l3_coeff_2.pdf}\end{minipage}
\begin{minipage}{.25\linewidth} \begin{align} \left(a_{5}\right)^{\check{E}}_{2} &= \left<P,P\right> + \left<S,D\right> \nonumber \\ & \hspace*{12.5pt} + \left<D,D\right>  + \left<P,F\right>   \nonumber  \\ & \hspace*{12.5pt} + \left<F,F\right>  + \left<D,G\right>   \nonumber \\ & \hspace*{12.5pt} + \left<G,G\right> + \left<F,H\right>  \nonumber \\ & \hspace*{12.5pt} + \left<H,H\right> \nonumber  \end{align} \end{minipage}

\begin{minipage}{.075\linewidth}
\vspace*{-6.5pt}
\hspace*{5pt}
\begin{equation}
\mathcal{C}_{3}^{\check{E}} \equiv \nonumber
\end{equation}
\end{minipage}
\begin{minipage}{.3\linewidth} \vspace*{0.572cm} \includegraphics[width=0.875\textwidth]{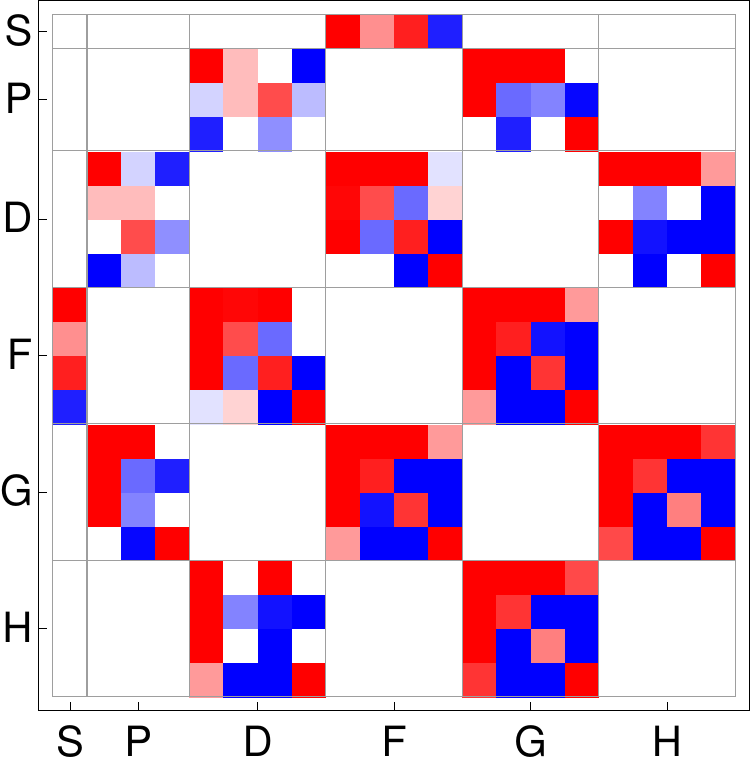} \end{minipage}
\begin{minipage}{.35\linewidth} \vspace*{0.500cm} \hspace*{-0.65cm}\includegraphics[width=1.15\textwidth]{E_l3_coeff_3.pdf}\end{minipage}
\begin{minipage}{.25\linewidth} \begin{align} \left(a_{5}\right)^{\check{E}}_{3} &= \left<P,D\right> + \left<S,F\right> \nonumber \\ & \hspace*{12.5pt} + \left<D,F\right>  + \left<P,G\right>   \nonumber   \\ & \hspace*{12.5pt} + \left<F,G\right>  + \left<D,H\right>   \nonumber \\ & \hspace*{12.5pt} + \left<G,H\right>    \nonumber  \end{align} \end{minipage}
\caption{%
Left: Matrices $\mathcal{C}_{0\cdots 3}^{\check{E}}$, represented here in the color scheme, defines the coefficient $\left(a_{5}\right)_{0\cdots 3}^{\check{E}}$ for an expansion of $\check{E}$ up to $\text{L}_{\text{max}} = 5$. Center: Coefficients $\left(a_{3}\right)_{0\cdots 3}^{\check{E}}$ obtained from a fit to the $\check{E}$-data (black points). For references to the data see Table \ref{tab:DataBasis}. Bonn Gatchina predictions, truncated at different $\text{L}_{\mathrm{max}}$ ($\text{L}_{\mathrm{max}} = 1$ is drawn in green, $\text{L}_{\mathrm{max}} = 2$ in blue, $\text{L}_{\mathrm{max}} = 3$ in red and $\text{L}_{\mathrm{max}} = 4$ in black) are drawn as well. Right: All partial wave interferences for $\text{L}_{\text{max}} = 5$ are indicated.
}
\label{tab:EColorPlots1}
\end{table*}
\begin{table*}[htb]
\RawFloats
\begin{minipage}{.075\linewidth}
\vspace*{-6.5pt}
\hspace*{5pt}
\begin{equation}
\mathcal{C}_{4}^{\check{E}} \equiv \nonumber
\end{equation}
\end{minipage}
\begin{minipage}{.3\linewidth} \vspace*{0.572cm} \includegraphics[width=0.875\textwidth]{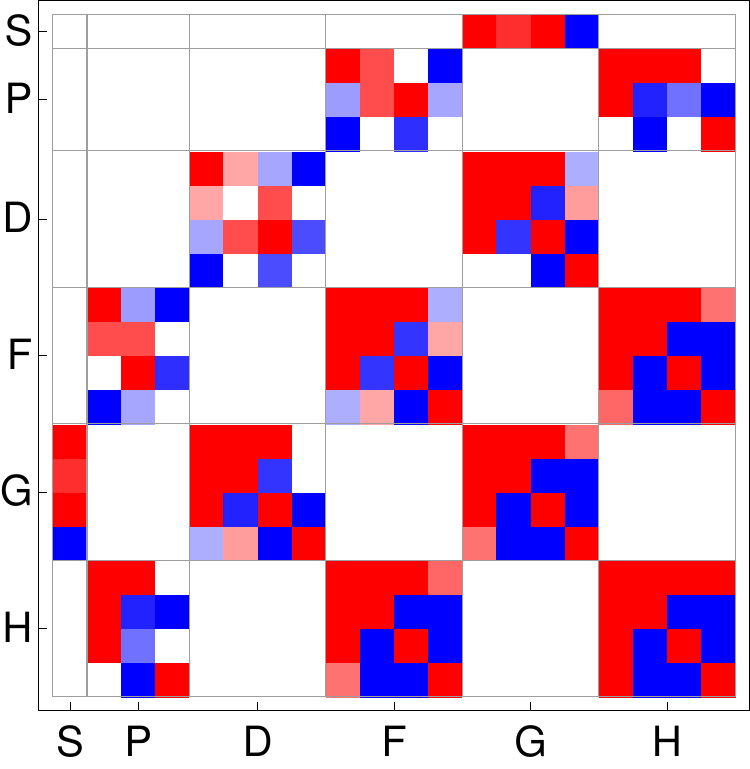} \end{minipage}
\begin{minipage}{.35\linewidth} \vspace*{0.500cm} \hspace*{-0.65cm}\includegraphics[width=1.15\textwidth]{E_l3_coeff_4.pdf}\end{minipage}
\begin{minipage}{.25\linewidth} \begin{align} \left(a_{5}\right)^{\check{E}}_{4} &= \left<D,D\right> + \left<P,F\right> \nonumber \\ & \hspace*{12.5pt} + \left<F,F\right>  + \left<S,G\right>   \nonumber \\ & \hspace*{12.5pt} + \left<D,G\right>  + \left<G,G\right>   \nonumber \\ & \hspace*{12.5pt} + \left<P,H\right>  + \left<F,H\right>   \nonumber  \\ & \hspace*{12.5pt} + \left<H,H\right> \nonumber \end{align} \end{minipage}

\begin{minipage}{.075\linewidth}
\vspace*{-6.5pt}
\hspace*{5pt}
\begin{equation}
\mathcal{C}_{5}^{\check{E}} \equiv \nonumber
\end{equation}
\end{minipage}
\begin{minipage}{.3\linewidth} \vspace*{0.572cm} \includegraphics[width=0.875\textwidth]{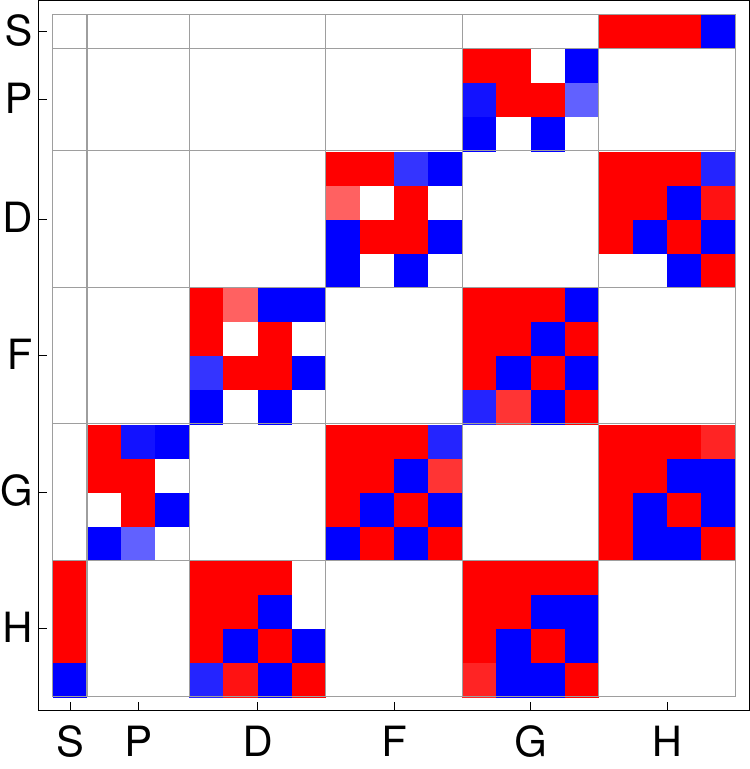} \end{minipage}
\begin{minipage}{.35\linewidth} \vspace*{0.500cm} \hspace*{-0.65cm}\includegraphics[width=1.15\textwidth]{E_l3_coeff_5.pdf}\end{minipage}
\begin{minipage}{.25\linewidth} \begin{align} \left(a_{5}\right)^{\check{E}}_{5} &= \left<D,F\right> + \left<P,G\right> \nonumber \\ & \hspace*{12.5pt} + \left<F,G\right>  + \left<S,H\right>   \nonumber \\ & \hspace*{12.5pt} + \left<D,H\right>  + \left<G,H\right>  \nonumber   \end{align} \end{minipage}

\begin{minipage}{.075\linewidth}
\vspace*{-6.5pt}
\hspace*{5pt}
\begin{equation}
\mathcal{C}_{6}^{\check{E}} \equiv \nonumber
\end{equation}
\end{minipage}
\begin{minipage}{.3\linewidth} \vspace*{0.572cm} \includegraphics[width=0.875\textwidth]{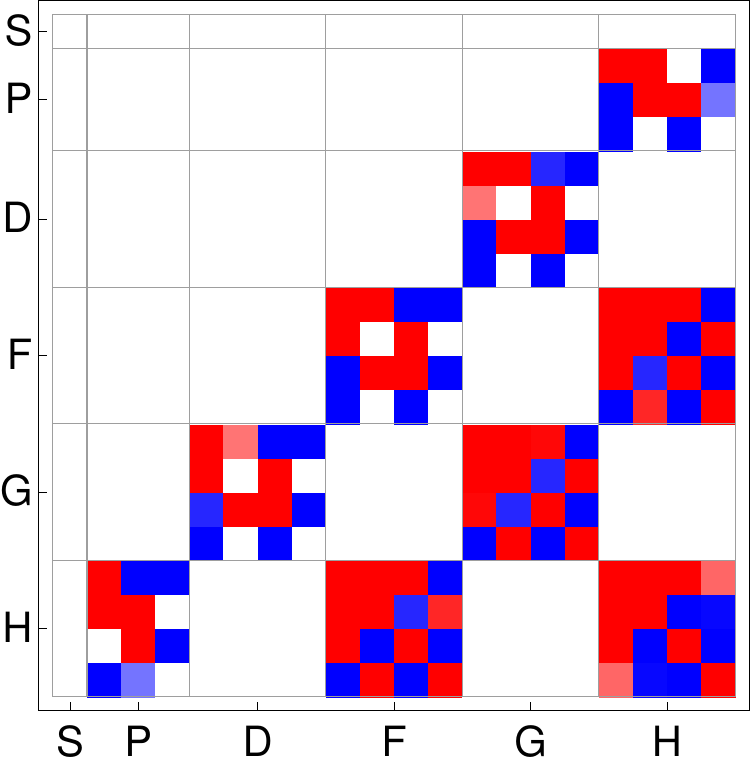} \end{minipage}
\begin{minipage}{.35\linewidth} \vspace*{0.500cm} \hspace*{-0.65cm}\includegraphics[width=1.15\textwidth]{E_l3_coeff_6.pdf}\end{minipage}
\begin{minipage}{.25\linewidth} \begin{align} \left(a_{5}\right)^{\check{E}}_{6} &= \left<F,F\right> + \left<D,G\right> \nonumber \\ & \hspace*{12.5pt} + \left<G,G\right>  + \left<P,H\right>   \nonumber  \\ & \hspace*{12.5pt} + \left<F,H\right>  + \left<H,H\right>   \nonumber  \end{align} \end{minipage}

\begin{minipage}{.075\linewidth}
\vspace*{-6.5pt}
\hspace*{5pt}
\begin{equation}
\mathcal{C}_{7}^{\check{E}} \equiv \nonumber
\end{equation}
\end{minipage}
\begin{minipage}{.3\linewidth} \vspace*{0.572cm} \includegraphics[width=0.875\textwidth]{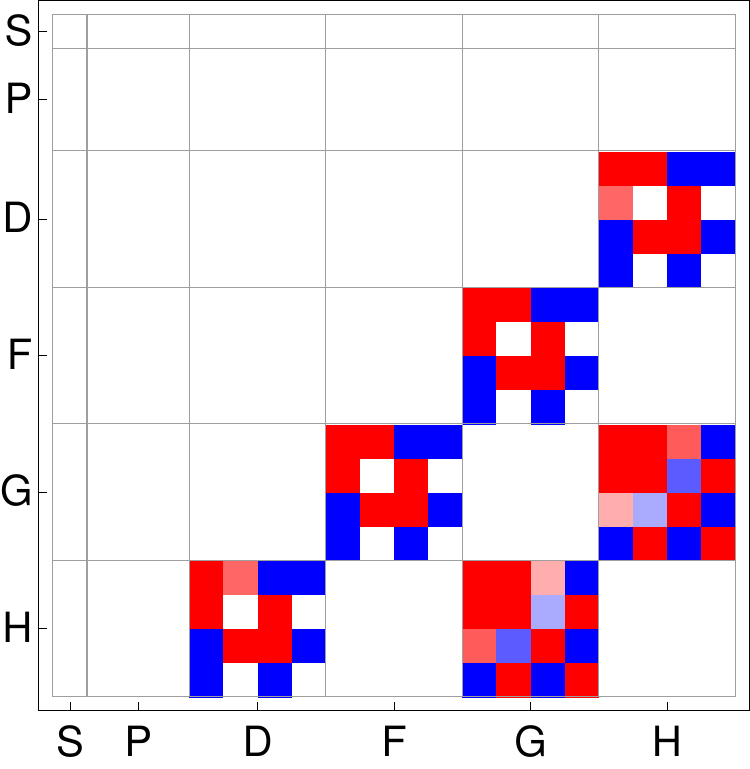} \end{minipage}
\begin{minipage}{.35\linewidth} \vspace*{0.500cm} \hspace*{-0.65cm}\includegraphics[width=1.15\textwidth]{E_l4_coeff_7.pdf}\end{minipage}
\begin{minipage}{.25\linewidth} \begin{align} \left(a_{5}\right)^{\check{E}}_{7} &= \left<F,G\right> + \left<D,H\right> \nonumber \\ & \hspace*{12.5pt} + \left<G,H\right>    \nonumber  \end{align} \end{minipage}
\caption{%
Left: Matrices $\mathcal{C}_{4\cdots 7}^{\check{E}}$, represented here in the color scheme, defines the coefficient $\left(a_{5}\right)_{4\cdots 7}^{\check{E}}$ for an expansion of $\check{E}$ up to $\text{L}_{\text{max}} = 5$. Center: Coefficients $\left(a_{3}\right)_{4\cdots 6}^{\check{E}}$ and $\left(a_{4}\right)_{7}^{\check{E}}$ obtained from a fit to the $\check{E}$-data (black points). For references to the data see Table \ref{tab:DataBasis}. Bonn Gatchina predictions, truncated at different $\text{L}_{\mathrm{max}}$ ($\text{L}_{\mathrm{max}} = 1$ is drawn in green, $\text{L}_{\mathrm{max}} = 2$ in blue, $\text{L}_{\mathrm{max}} = 3$ in red and $\text{L}_{\mathrm{max}} = 4$ in black) are drawn as well. Right: All partial wave interferences for $\text{L}_{\text{max}} = 5$ are indicated.
}
\label{tab:EColorPlots2}
\end{table*}
\begin{table*}[htb]
\RawFloats
\begin{minipage}{.075\linewidth}
\vspace*{-6.5pt}
\hspace*{5pt}
\begin{equation}
\mathcal{C}_{8}^{\check{E}} \equiv \nonumber
\end{equation}
\end{minipage}
\begin{minipage}{.3\linewidth} \vspace*{0.572cm} \includegraphics[width=0.875\textwidth]{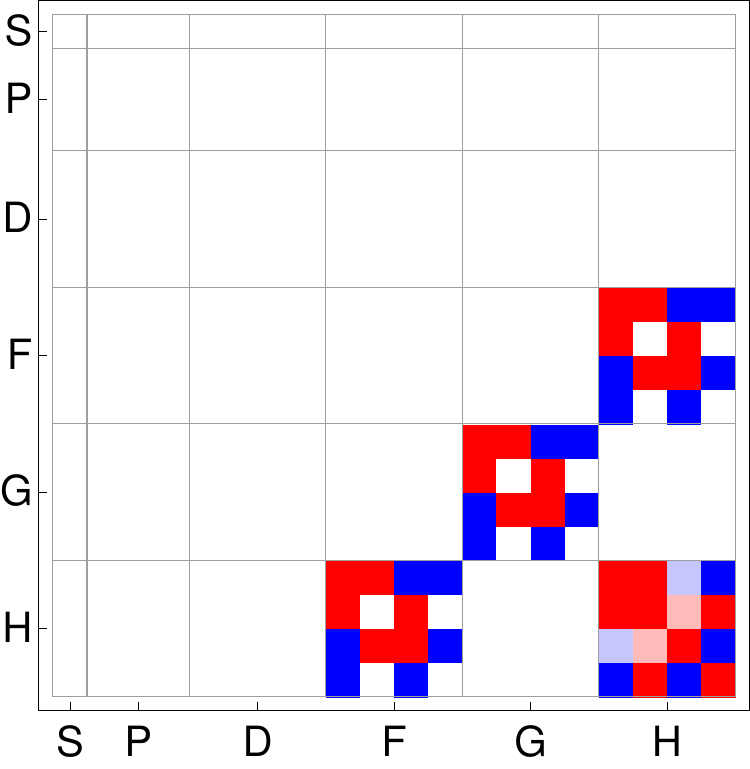} \end{minipage}
\begin{minipage}{.35\linewidth} \vspace*{0.500cm} \hspace*{-0.65cm}\includegraphics[width=1.15\textwidth]{E_l4_coeff_8.pdf}\end{minipage}
\begin{minipage}{.25\linewidth} \begin{align} \left(a_{5}\right)^{\check{E}}_{8} &= \left<G,G\right> + \left<F,H\right> \nonumber \\ & \hspace*{12.5pt} + \left<H,H\right>   \nonumber  \end{align} \end{minipage}

\begin{minipage}{.075\linewidth}
\vspace*{-6.5pt}
\hspace*{5pt}
\begin{equation}
\mathcal{C}_{9}^{\check{E}} \equiv \nonumber
\end{equation}
\end{minipage}
\begin{minipage}{.3\linewidth} \vspace*{0.572cm} \includegraphics[width=0.875\textwidth]{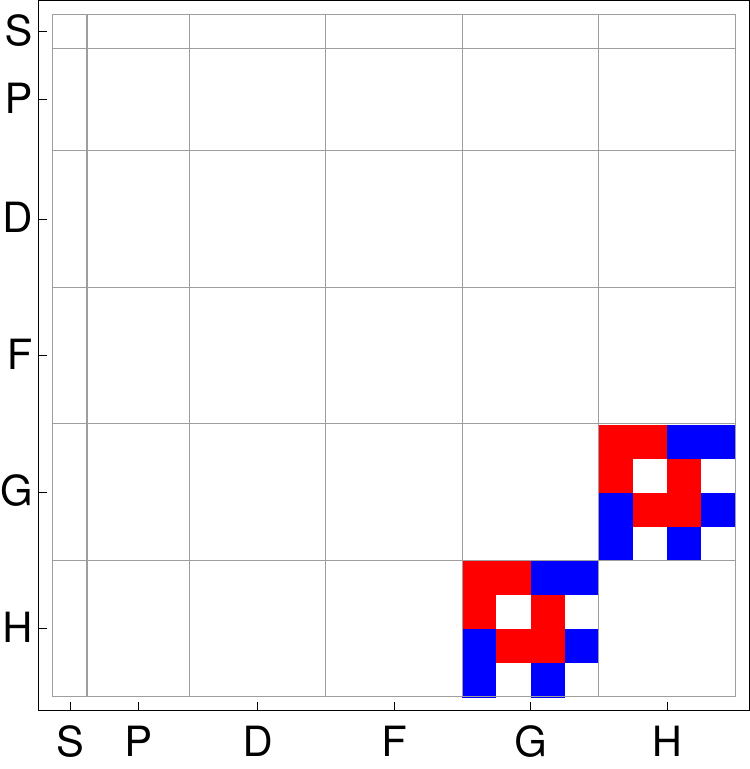} \end{minipage}
\begin{minipage}{.35\linewidth} \vspace*{0.500cm} \hspace*{-0.65cm}\includegraphics[width=1.15\textwidth]{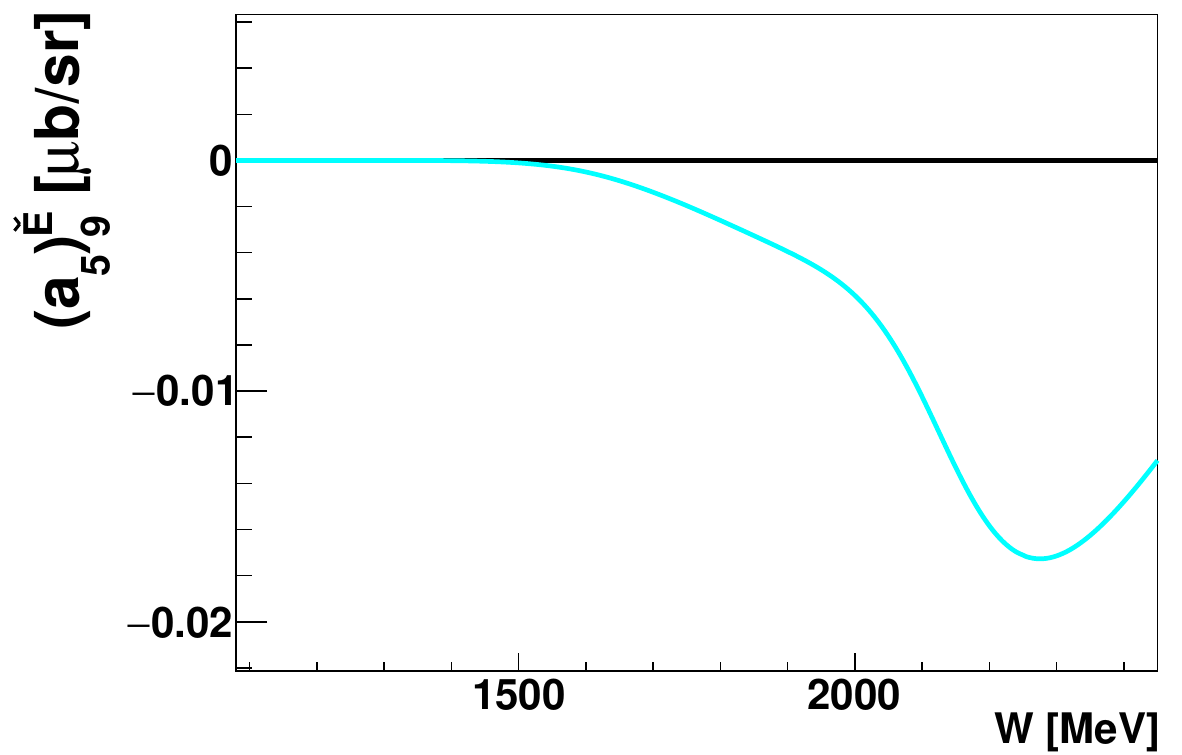}\end{minipage}
\begin{minipage}{.25\linewidth} \begin{align} \left(a_{5}\right)^{\check{E}}_{9} &= \left<G,H\right>   \nonumber   \end{align} \end{minipage}

\begin{minipage}{.075\linewidth}
\vspace*{-6.5pt}
\hspace*{5pt}
\begin{equation}
\mathcal{C}_{10}^{\check{E}} \equiv \nonumber
\end{equation}
\end{minipage}
\begin{minipage}{.3\linewidth} \vspace*{0.572cm} \includegraphics[width=0.875\textwidth]{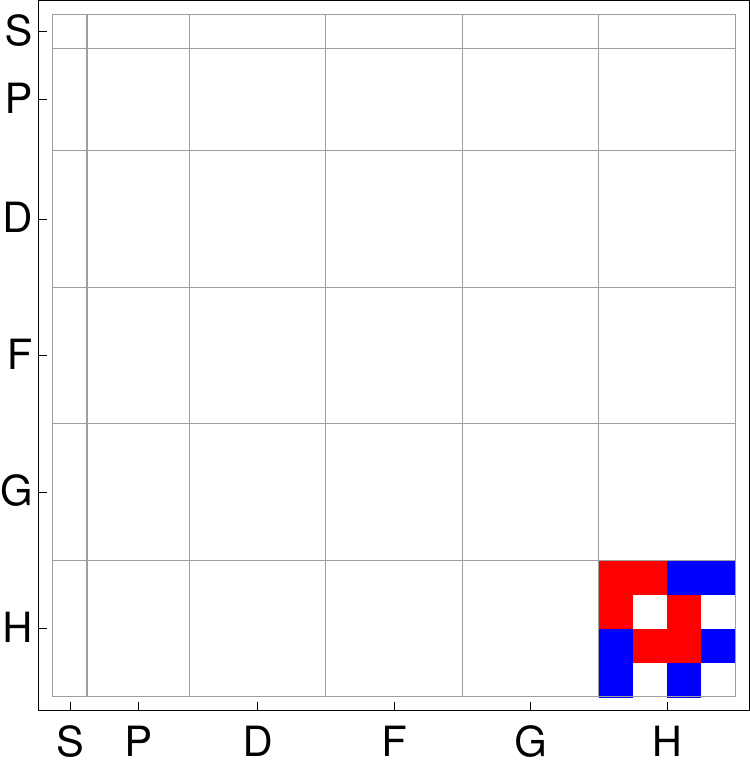} \end{minipage}
\begin{minipage}{.35\linewidth} \vspace*{0.500cm} \hspace*{-0.65cm}\includegraphics[width=1.15\textwidth]{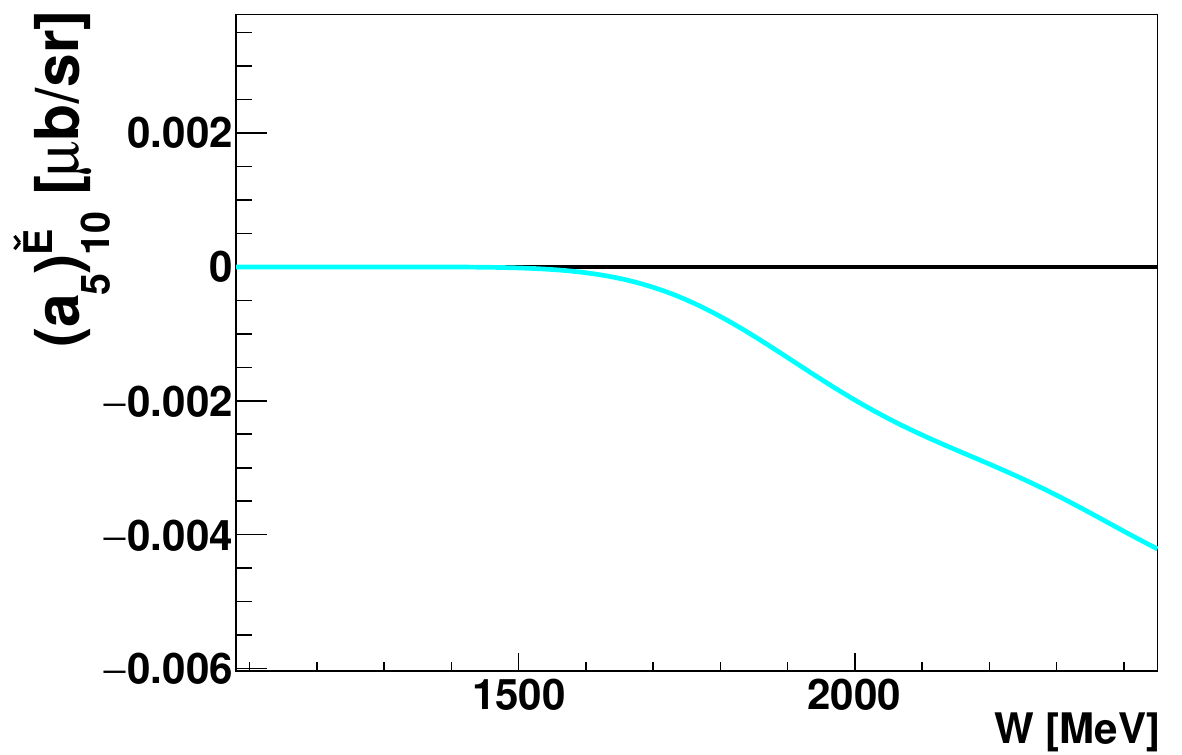}\end{minipage}
\begin{minipage}{.25\linewidth} \begin{align} \left(a_{5}\right)^{\check{E}}_{10} &= \left<H,H\right>  \nonumber   \end{align} \end{minipage}
\caption{%
Left: Matrices $\mathcal{C}_{8\cdots 10}^{\check{E}}$, represented here in the color scheme, defines the coefficient $\left(a_{5}\right)_{8\cdots 10}^{\check{E}}$ for an expansion of $\check{E}$ up to $\text{L}_{\text{max}} = 5$. Center: The coefficient $\left(a_{4}\right)_{8}^{\check{E}}$ obtained from a fit to the $\check{E}$-data (black points). For the highest non-fitted coefficients $\left(a_{5}\right)_{9, 10}^{\check{E}}$, the Bonn Gatchina curves are shown (here, the truncation at $\text{L}_{\mathrm{max}} = 5$ is drawn in cyan). Right: All partial wave interferences for $\text{L}_{\text{max}} = 5$ are indicated.
}
\label{tab:EColorPlots3}
\end{table*}
%


%
\begin{table*}[htb]
\RawFloats
\begin{minipage}{.075\linewidth}
\vspace*{-6.5pt}
\hspace*{5pt}
\begin{equation}
\mathcal{C}_{2}^{\sigma_{3/2}} \equiv \nonumber
\end{equation}
\end{minipage}
\begin{minipage}{.3\linewidth} \vspace*{0.572cm} \includegraphics[width=0.875\textwidth]{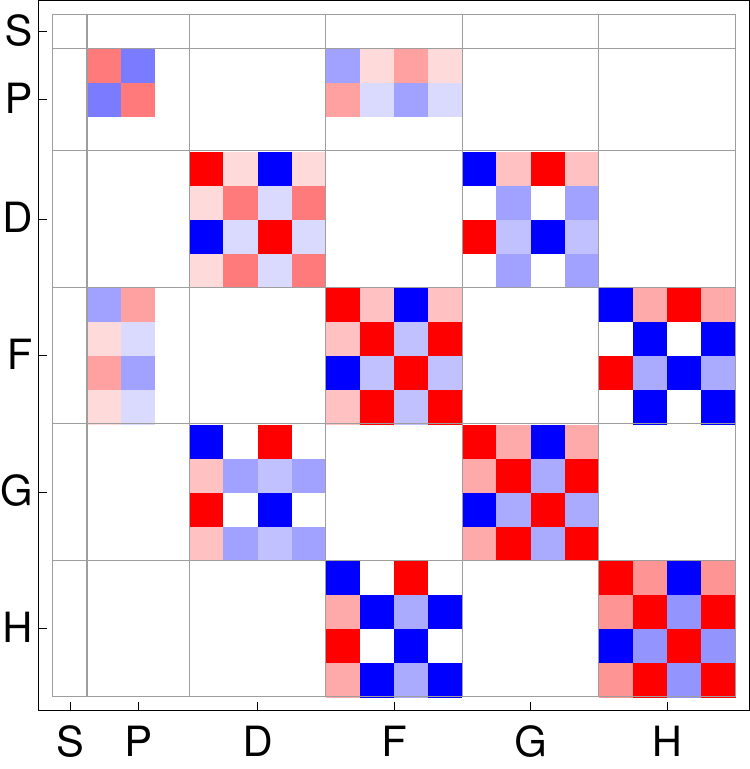} \end{minipage}
\begin{minipage}{.35\linewidth} \vspace*{0.500cm} \hspace*{-0.65cm}\includegraphics[width=1.15\textwidth]{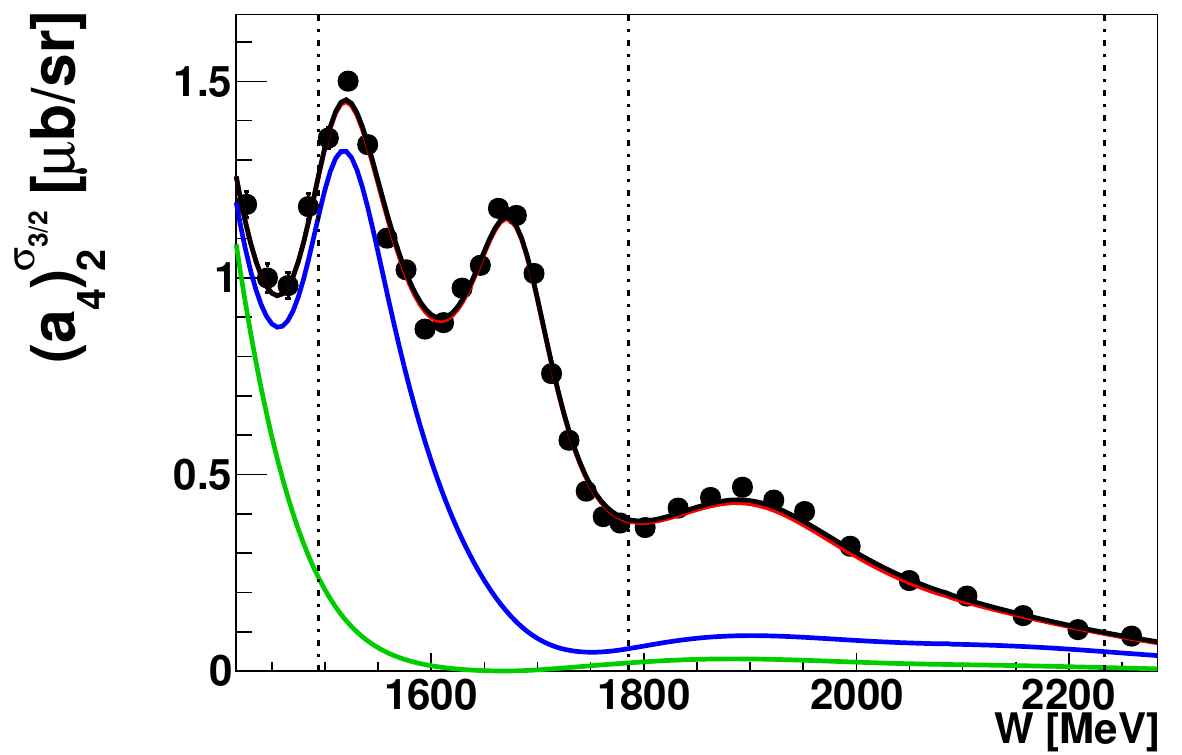}\end{minipage}
\begin{minipage}{.25\linewidth} \begin{align} \left(a_{5}\right)^{\sigma_{3/2}}_{2} &= \left<P,P\right> + \left<D,D\right> \nonumber \\ & \hspace*{12.5pt} + \left<P,F\right>  + \left<F,F\right>   \nonumber \\ & \hspace*{12.5pt} + \left<D,G\right>  + \left<G,G\right>   \nonumber \\ & \hspace*{12.5pt} + \left<F,H\right>  + \left<H,H\right>   \nonumber \end{align} \end{minipage}

\begin{minipage}{.075\linewidth}
\vspace*{-6.5pt}
\hspace*{5pt}
\begin{equation}
\mathcal{C}_{3}^{\sigma_{3/2}} \equiv \nonumber
\end{equation}
\end{minipage}
\begin{minipage}{.3\linewidth} \vspace*{0.572cm} \includegraphics[width=0.875\textwidth]{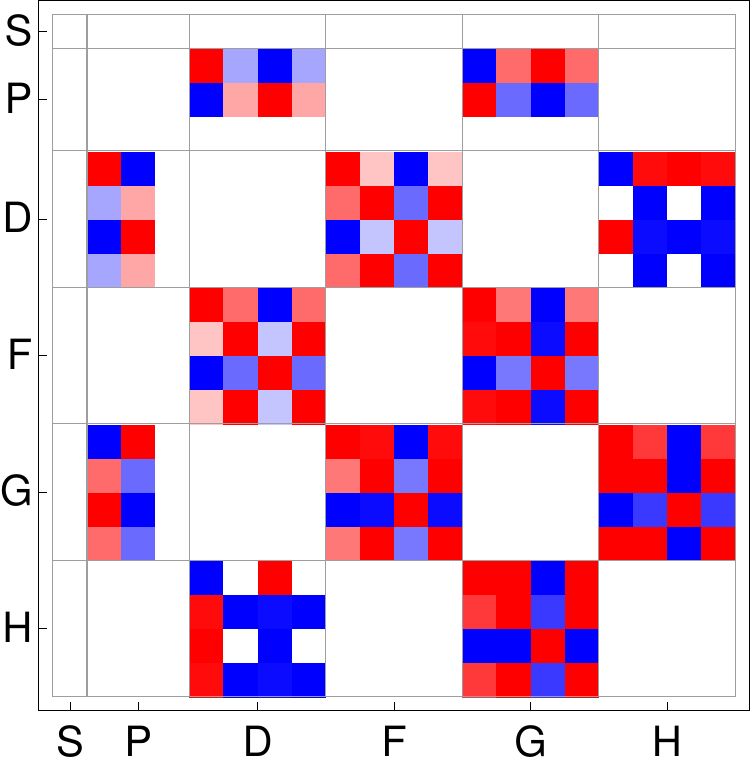} \end{minipage}
\begin{minipage}{.35\linewidth} \vspace*{0.500cm} \hspace*{-0.65cm}\includegraphics[width=1.15\textwidth]{S32_l4_coeff_1.pdf}\end{minipage}
\begin{minipage}{.25\linewidth} \begin{align} \left(a_{5}\right)^{\sigma_{3/2}}_{3} &= \left<P,D\right> + \left<D,F\right> \nonumber \\ & \hspace*{12.5pt} + \left<P,G\right>  + \left<F,G\right>   \nonumber \\ & \hspace*{12.5pt} + \left<D,H\right>  + \left<G,H\right>  \nonumber   \end{align} \end{minipage}

\begin{minipage}{.075\linewidth}
\vspace*{-6.5pt}
\hspace*{5pt}
\begin{equation}
\mathcal{C}_{4}^{\sigma_{3/2}} \equiv \nonumber
\end{equation}
\end{minipage}
\begin{minipage}{.3\linewidth} \vspace*{0.572cm} \includegraphics[width=0.875\textwidth]{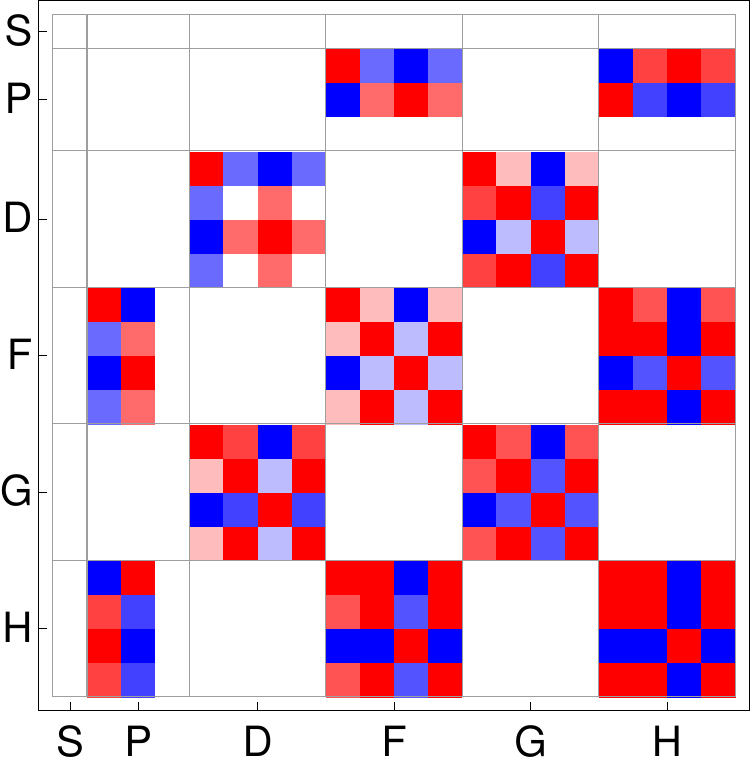} \end{minipage}
\begin{minipage}{.35\linewidth} \vspace*{0.500cm} \hspace*{-0.65cm}\includegraphics[width=1.15\textwidth]{S32_l4_coeff_2.pdf}\end{minipage}
\begin{minipage}{.25\linewidth} \begin{align} \left(a_{5}\right)^{\sigma_{3/2}}_{4} &= \left<D,D\right> + \left<P,F\right> \nonumber \\ & \hspace*{12.5pt} + \left<F,F\right>  + \left<D,G\right>   \nonumber  \\ & \hspace*{12.5pt} + \left<G,G\right>  + \left<P,H\right>   \nonumber \\ & \hspace*{12.5pt} + \left<F,H\right>  + \left<H,H\right>   \nonumber \end{align} \end{minipage}

\begin{minipage}{.075\linewidth}
\vspace*{-6.5pt}
\hspace*{5pt}
\begin{equation}
\mathcal{C}_{5}^{\sigma_{3/2}} \equiv \nonumber
\end{equation}
\end{minipage}
\begin{minipage}{.3\linewidth} \vspace*{0.572cm} \includegraphics[width=0.875\textwidth]{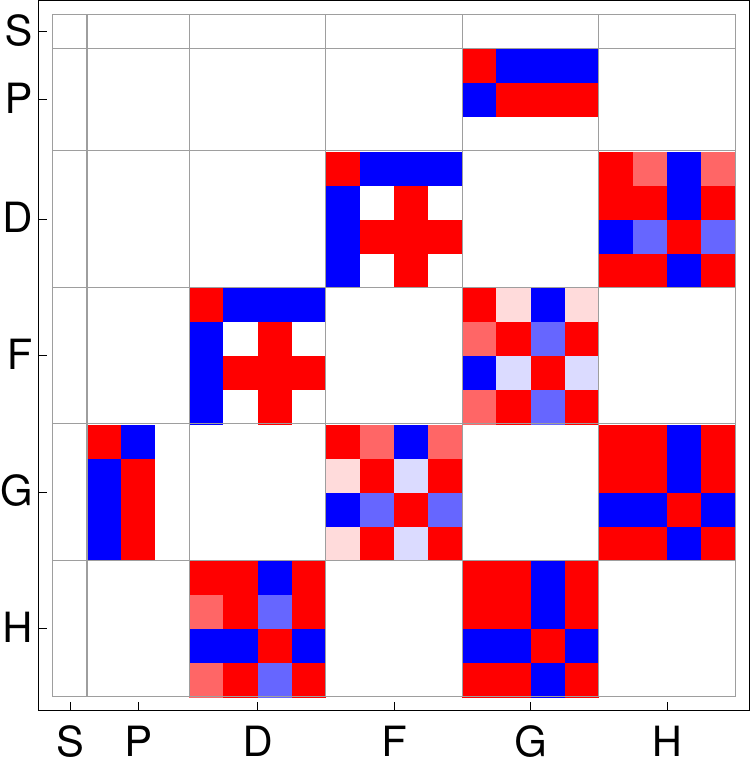} \end{minipage}
\begin{minipage}{.35\linewidth} \vspace*{0.500cm} \hspace*{-0.65cm}\includegraphics[width=1.15\textwidth]{S32_l4_coeff_3.pdf}\end{minipage}
\begin{minipage}{.25\linewidth} \begin{align} \left(a_{5}\right)^{\sigma_{3/2}}_{5} &= \left<D,F\right> + \left<P,G\right> \nonumber \\ & \hspace*{12.5pt} + \left<F,G\right>  + \left<D,H\right> \nonumber \\ & \hspace*{12.5pt}   + \left<G,H\right>       \nonumber  \end{align} \end{minipage}
\caption{%
Left: Matrices $\mathcal{C}_{2\cdots 5}^{\sigma_{3/2}}$, represented here in the color scheme, defines the coefficient $\left(a_{5}\right)_{2\cdots 5}^{\sigma_{3/2}}$ for an expansion of $\sigma_{3/2}$ up to $\text{L}_{\text{max}} = 5$. Center: Coefficients $\left(a_{4}\right)_{2\cdots 5}^{\sigma_{3/2}}$ obtained from a fit to the $\sigma_{3/2}$-data (black points). For references to the data see Table \ref{tab:DataBasis}. Bonn Gatchina predictions, truncated at different $\text{L}_{\mathrm{max}}$ ($\text{L}_{\mathrm{max}} = 1$ is drawn in green, $\text{L}_{\mathrm{max}} = 2$ in blue, $\text{L}_{\mathrm{max}} = 3$ in red and $\text{L}_{\mathrm{max}} = 4$ in black) are drawn as well. Right: All partial wave interferences for $\text{L}_{\text{max}} = 5$ are indicated.
}
\label{tab:DCS32ColorPlots1}
\end{table*}
\begin{table*}[htb]
\RawFloats
\begin{minipage}{.075\linewidth}
\vspace*{-6.5pt}
\hspace*{5pt}
\begin{equation}
\mathcal{C}_{6}^{\sigma_{3/2}} \equiv \nonumber
\end{equation}
\end{minipage}
\begin{minipage}{.3\linewidth} \vspace*{0.572cm} \includegraphics[width=0.875\textwidth]{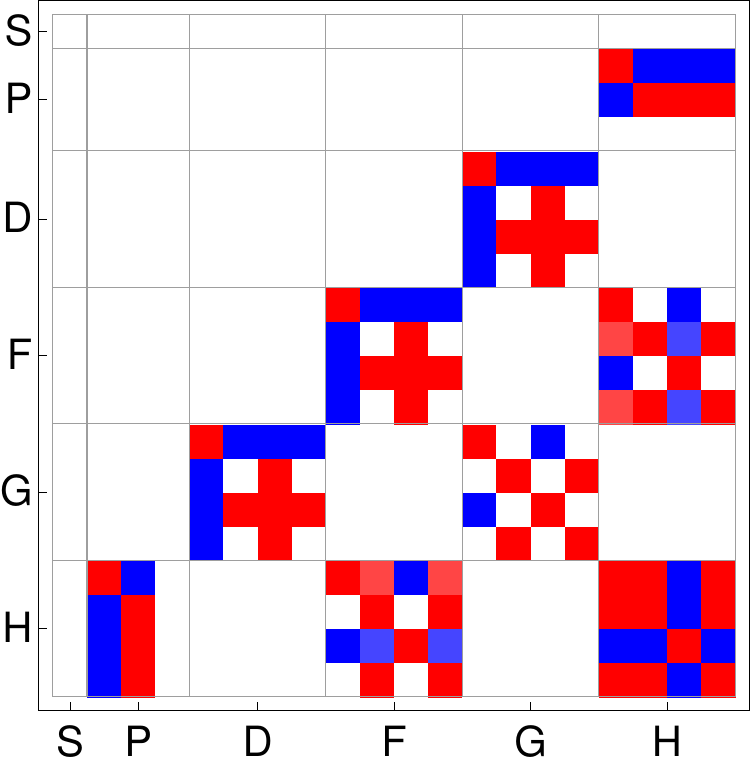} \end{minipage}
\begin{minipage}{.35\linewidth} \vspace*{0.500cm} \hspace*{-0.65cm}\includegraphics[width=1.15\textwidth]{S32_l4_coeff_4.pdf}\end{minipage}
\begin{minipage}{.25\linewidth} \begin{align} \left(a_{5}\right)^{\sigma_{3/2}}_{6} &= \left<F,F\right>  + \left<D,G\right>  \nonumber \\ & \hspace*{12.5pt}   + \left<G,G\right>  + \left<P,H\right>  \nonumber \\ & \hspace*{12.5pt}   + \left<F,H\right>  + \left<H,H\right>  \nonumber \end{align} \end{minipage}

\begin{minipage}{.075\linewidth}
\vspace*{-6.5pt}
\hspace*{5pt}
\begin{equation}
\mathcal{C}_{7}^{\sigma_{3/2}} \equiv \nonumber
\end{equation}
\end{minipage}
\begin{minipage}{.3\linewidth} \vspace*{0.572cm} \includegraphics[width=0.875\textwidth]{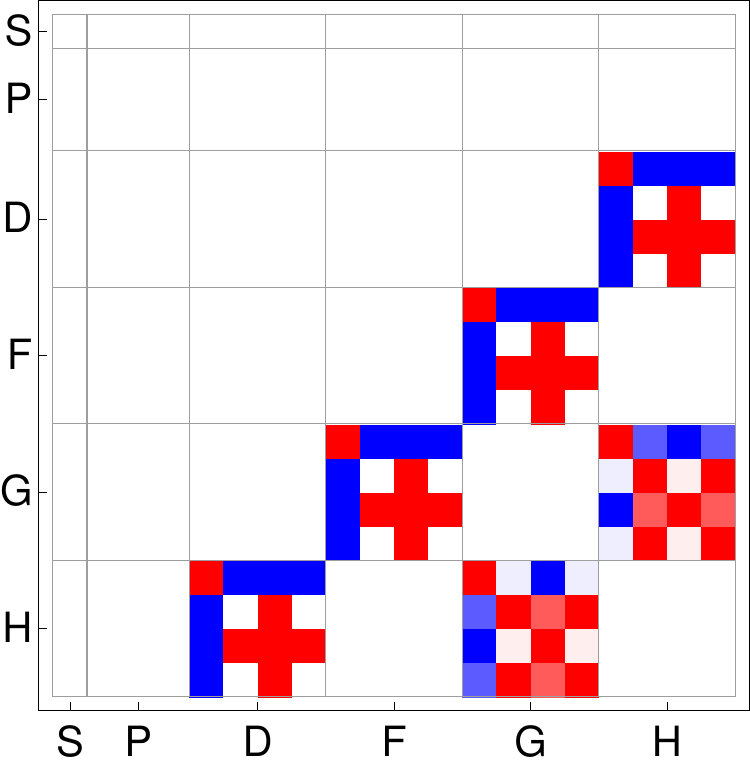} \end{minipage}
\begin{minipage}{.35\linewidth} \vspace*{0.500cm} \hspace*{-0.65cm}\includegraphics[width=1.15\textwidth]{S32_l4_coeff_5.pdf}\end{minipage}
\begin{minipage}{.25\linewidth} \begin{align} \left(a_{5}\right)^{\sigma_{3/2}}_{7} &= \left<F,G\right> + \left<D,H\right> \nonumber \\ & \hspace*{12.5pt} + \left<G,H\right> \nonumber   \end{align} \end{minipage}

\begin{minipage}{.075\linewidth}
\vspace*{-6.5pt}
\hspace*{5pt}
\begin{equation}
\mathcal{C}_{8}^{\sigma_{3/2}} \equiv \nonumber
\end{equation}
\end{minipage}
\begin{minipage}{.3\linewidth} \vspace*{0.572cm} \includegraphics[width=0.875\textwidth]{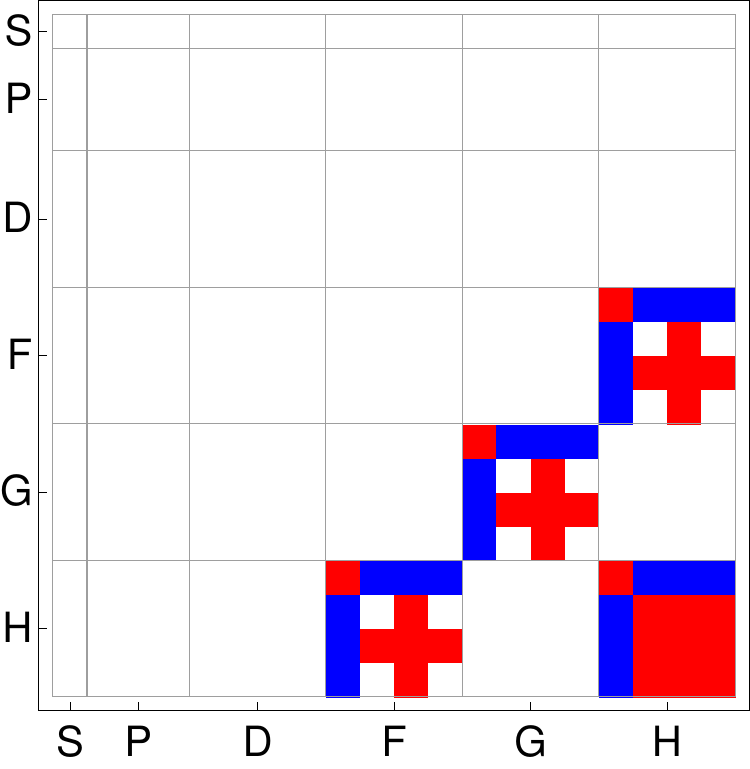} \end{minipage}
\begin{minipage}{.35\linewidth} \vspace*{0.500cm} \hspace*{-0.65cm}\includegraphics[width=1.15\textwidth]{S32_l4_coeff_6.pdf}\end{minipage}
\begin{minipage}{.25\linewidth} \begin{align} \left(a_{5}\right)^{\sigma_{3/2}}_{8} &= \left<G,G\right> + \left<F,H\right> \nonumber \\ & \hspace*{12.5pt} + \left<H,H\right>   \nonumber  \end{align} \end{minipage}

\begin{minipage}{.075\linewidth}
\vspace*{-6.5pt}
\hspace*{5pt}
\begin{equation}
\mathcal{C}_{9}^{\sigma_{3/2}} \equiv \nonumber
\end{equation}
\end{minipage}
\begin{minipage}{.3\linewidth} \vspace*{0.572cm} \includegraphics[width=0.875\textwidth]{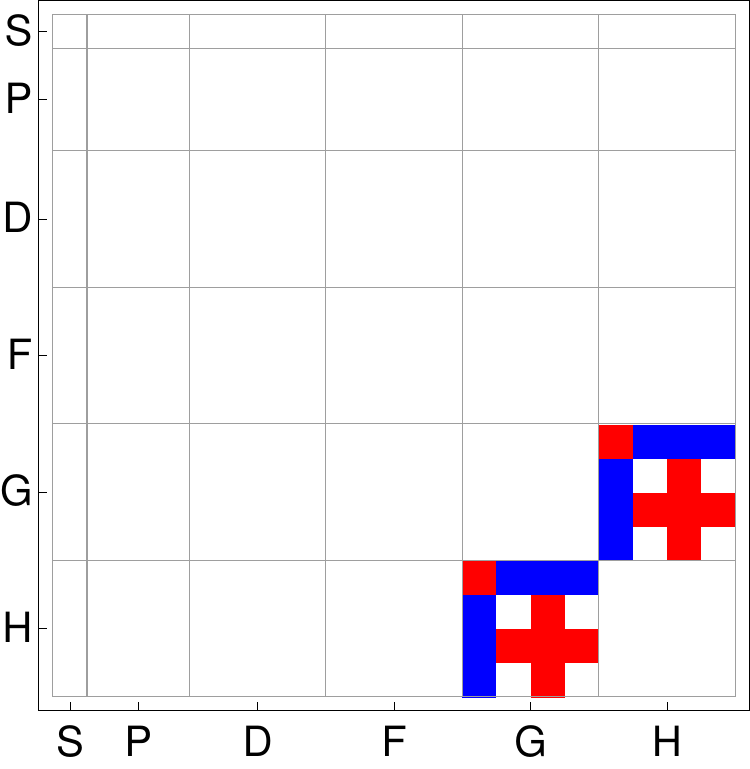} \end{minipage}
\begin{minipage}{.35\linewidth} \vspace*{0.500cm} \hspace*{-0.65cm}\includegraphics[width=1.15\textwidth]{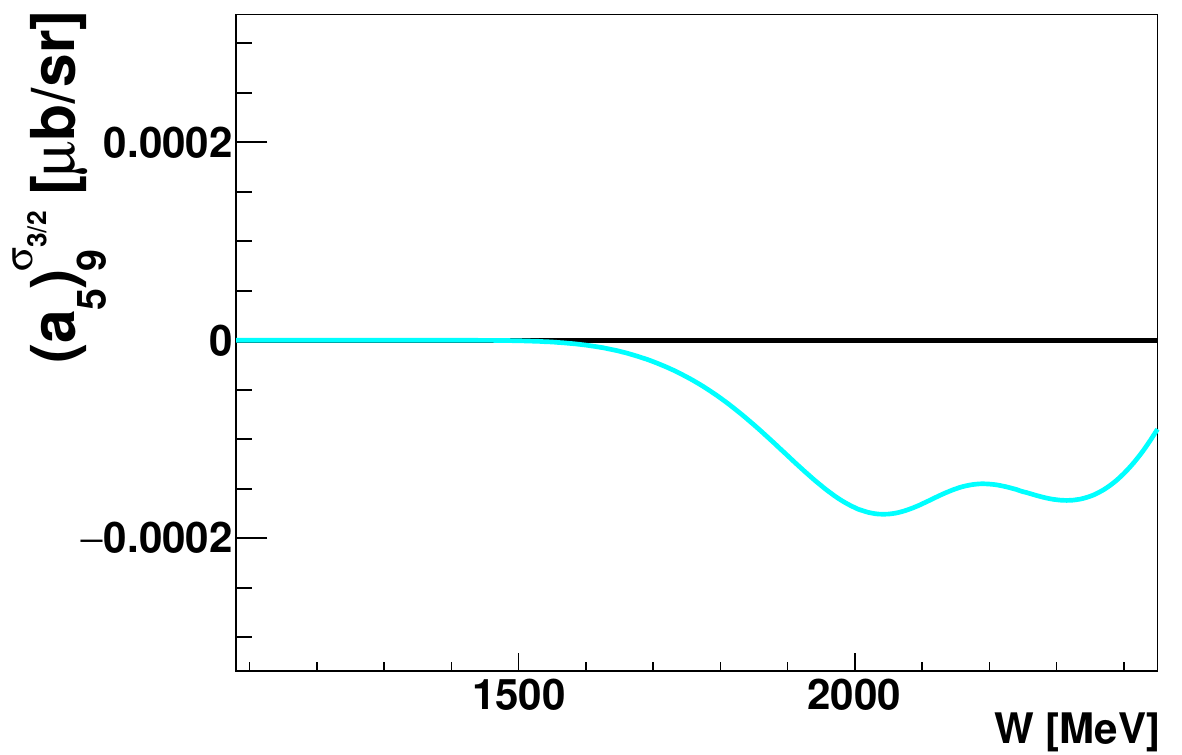}\end{minipage}
\begin{minipage}{.25\linewidth} \begin{align} \left(a_{5}\right)^{\sigma_{3/2}}_{9} &=  \left<G,H\right>    \nonumber  \end{align} \end{minipage}
\caption{%
Left: Matrices $\mathcal{C}_{6\cdots 9}^{\sigma_{3/2}}$, represented here in the color scheme, defines the coefficient $\left(a_{5}\right)_{6\cdots 9}^{\sigma_{3/2}}$ for an expansion of $\sigma_{3/2}$ up to $\text{L}_{\text{max}} = 5$. Center: Coefficients $\left(a_{4}\right)_{6\ldots 8}^{\sigma_{3/2}}$ obtained from a fit to the $\sigma_{3/2}$-data (black points). For references to the data see Table \ref{tab:DataBasis}. Bonn Gatchina predictions, truncated at different $\text{L}_{\mathrm{max}}$ ($\text{L}_{\mathrm{max}} = 1$ is drawn in green, $\text{L}_{\mathrm{max}} = 2$ in blue, $\text{L}_{\mathrm{max}} = 3$ in red and $\text{L}_{\mathrm{max}} = 4$ in black) are drawn as well. For the higher non-fitted coefficient $\left(a_{5}\right)_{9}^{\sigma_{3/2}}$, the Bonn Gatchina curves are shown (here, the truncation at $\text{L}_{\mathrm{max}} = 5$ is drawn in cyan). Right: All partial wave interferences for $\text{L}_{\text{max}} = 5$ are indicated.
}
\label{tab:DCS32ColorPlots2}
\end{table*}
\begin{table*}[htb]
\RawFloats
\begin{minipage}{.075\linewidth}
\vspace*{-6.5pt}
\hspace*{5pt}
\begin{equation}
\mathcal{C}_{10}^{\sigma_{3/2}} \equiv \nonumber
\end{equation}
\end{minipage}
\begin{minipage}{.3\linewidth} \vspace*{0.572cm} \includegraphics[width=0.875\textwidth]{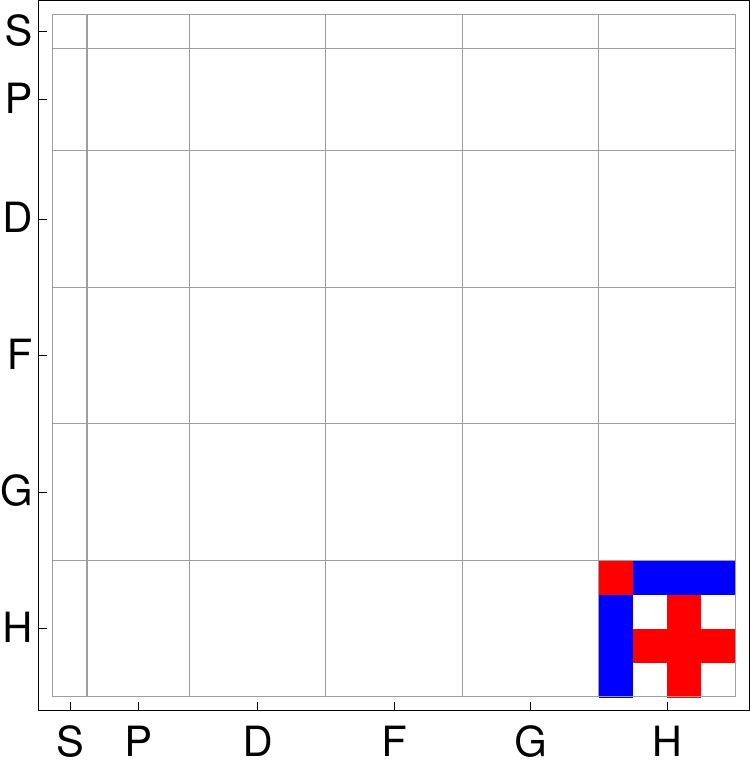} \end{minipage}
\begin{minipage}{.35\linewidth} \vspace*{0.500cm} \hspace*{-0.65cm}\includegraphics[width=1.15\textwidth]{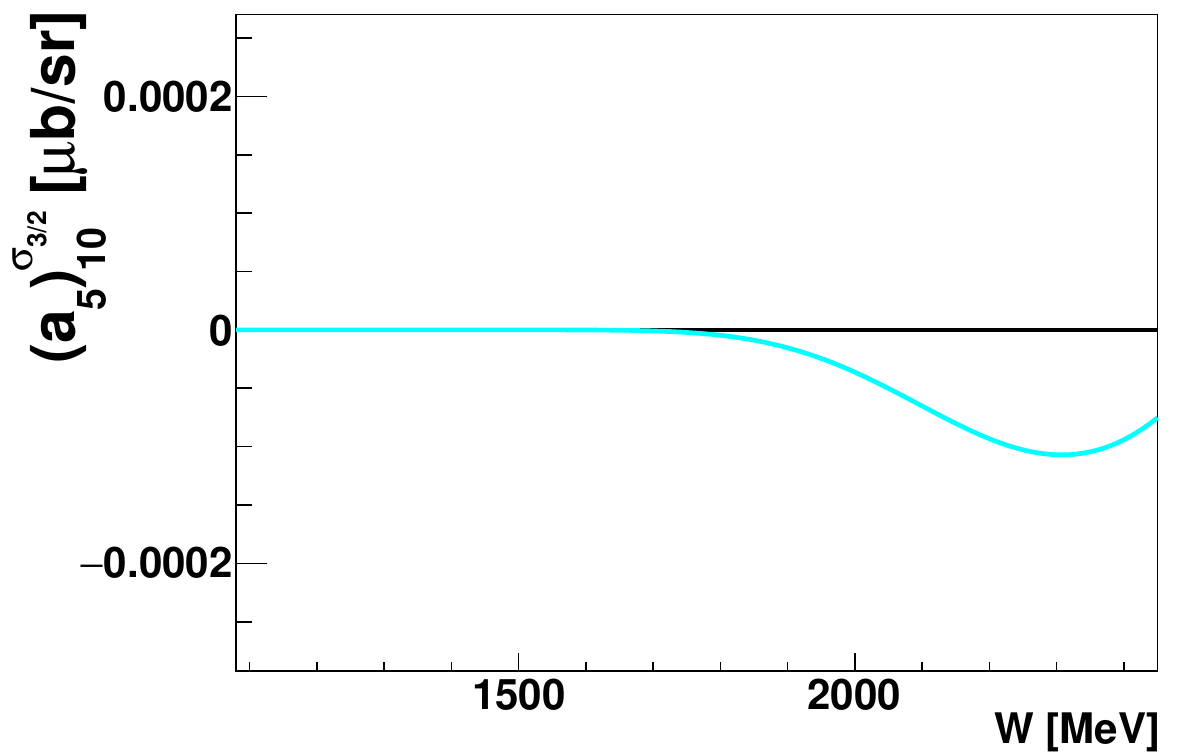}\end{minipage}
\begin{minipage}{.25\linewidth} \begin{align} \left(a_{5}\right)^{\sigma_{3/2}}_{10} &=  \left<H,H\right> \nonumber \end{align} \end{minipage}

\caption{%
Left: Matrix $\mathcal{C}_{10}^{\sigma_{3/2}}$, represented here in the color scheme, defines the coefficient $\left(a_{5}\right)_{10}^{\sigma_{3/2}}$ for an expansion of $\sigma_{3/2}$ up to $\text{L}_{\text{max}} = 5$. Center: For the highest non-fitted coefficient $\left(a_{5}\right)_{10}^{\sigma_{3/2}}$, the Bonn Gatchina curves are shown (here, the truncation at $\text{L}_{\mathrm{max}} = 5$ is drawn in cyan). Right: All partial wave interferences for $\text{L}_{\text{max}} = 5$ are indicated.
}
\label{tab:DCS32ColorPlots3}
\end{table*}
%


%
\begin{table*}[htb]
\RawFloats
\begin{minipage}{.075\linewidth}
\vspace*{-6.5pt}
\hspace*{5pt}
\begin{equation}
\mathcal{C}_{0}^{\sigma_{1/2}} \equiv \nonumber
\end{equation}
\end{minipage}
\begin{minipage}{.3\linewidth} \vspace*{0.572cm} \includegraphics[width=0.875\textwidth]{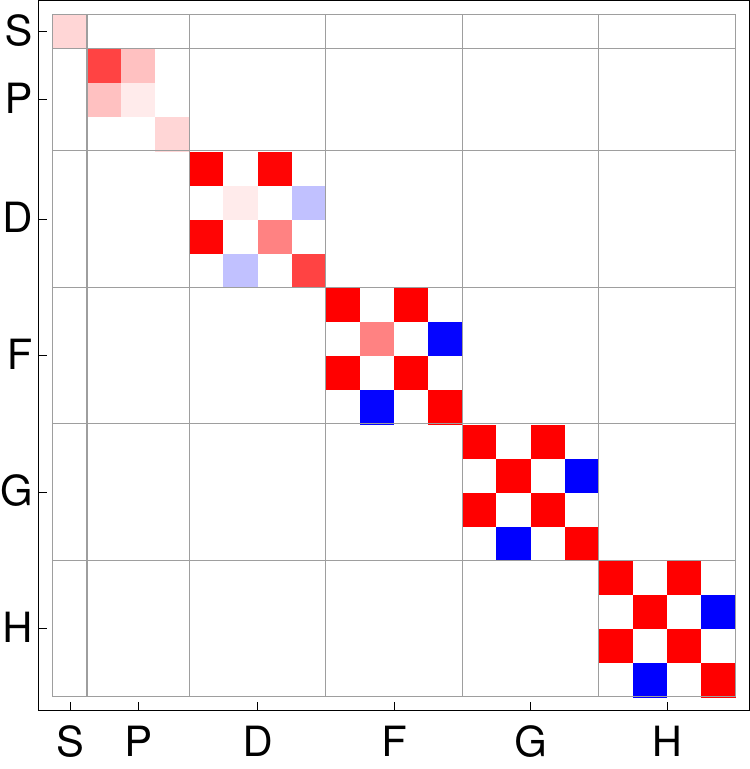} \end{minipage}
\begin{minipage}{.35\linewidth} \vspace*{0.500cm} \hspace*{-0.65cm}\includegraphics[width=1.15\textwidth]{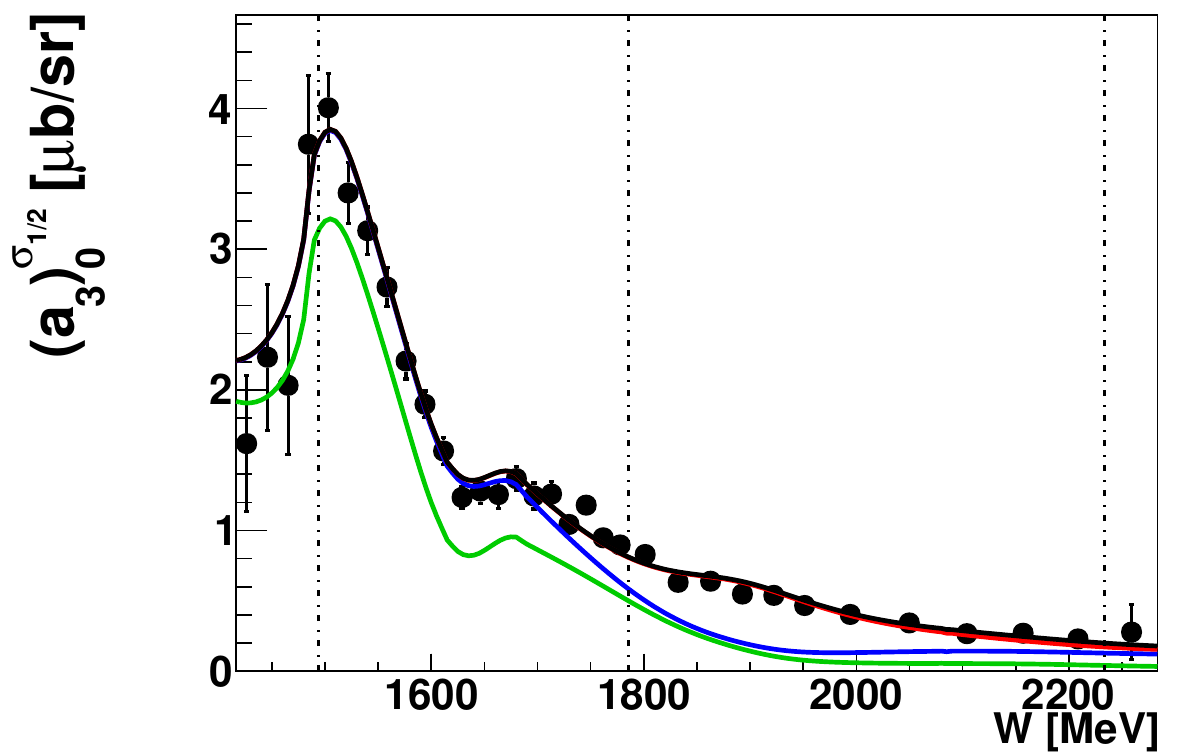}\end{minipage}
\begin{minipage}{.25\linewidth} \begin{align} \left(a_{5}\right)^{\sigma_{1/2}}_{0} &= \left<S,S\right> + \left<P,P\right> \nonumber \\ & \hspace*{12.5pt} + \left<D,D\right>  + \left<F,F\right>   \nonumber \\ & \hspace*{12.5pt} + \left<G,G\right>  + \left<H,H\right>   \nonumber \end{align} \end{minipage}

\begin{minipage}{.075\linewidth}
\vspace*{-6.5pt}
\hspace*{5pt}
\begin{equation}
\mathcal{C}_{1}^{\sigma_{1/2}} \equiv \nonumber
\end{equation}
\end{minipage}
\begin{minipage}{.3\linewidth} \vspace*{0.572cm} \includegraphics[width=0.875\textwidth]{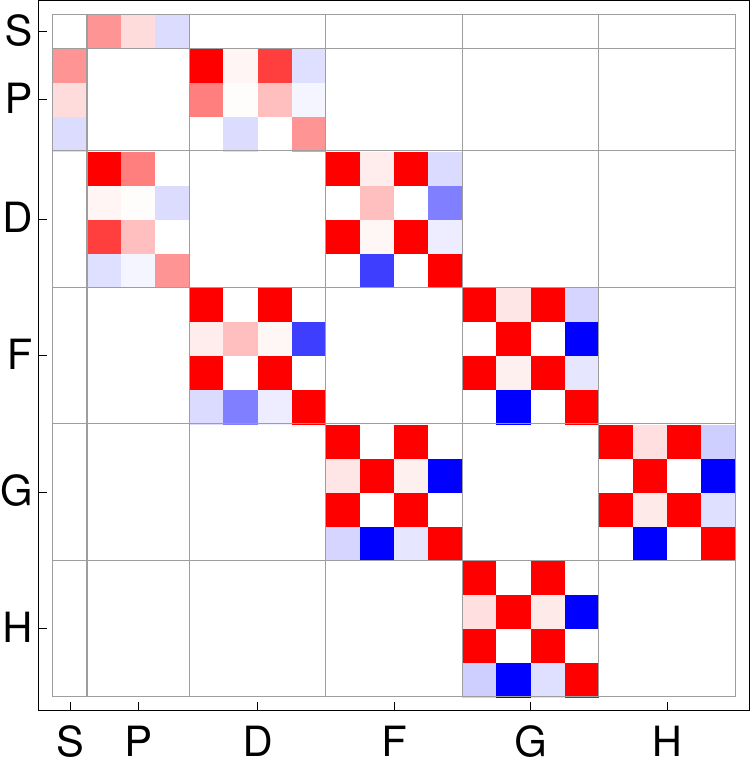} \end{minipage}
\begin{minipage}{.35\linewidth} \vspace*{0.500cm} \hspace*{-0.65cm}\includegraphics[width=1.15\textwidth]{S12_l3_coeff_1.pdf}\end{minipage}
\begin{minipage}{.25\linewidth} \begin{align} \left(a_{5}\right)^{\sigma_{1/2}}_{1} &= \left<S,P\right> + \left<P,D\right> \nonumber \\ & \hspace*{12.5pt} + \left<D,F\right>  + \left<F,G\right>   \nonumber \\ & \hspace*{12.5pt} + \left<G,H\right>    \nonumber   \end{align} \end{minipage}

\begin{minipage}{.075\linewidth}
\vspace*{-6.5pt}
\hspace*{5pt}
\begin{equation}
\mathcal{C}_{2}^{\sigma_{1/2}} \equiv \nonumber
\end{equation}
\end{minipage}
\begin{minipage}{.3\linewidth} \vspace*{0.572cm} \includegraphics[width=0.875\textwidth]{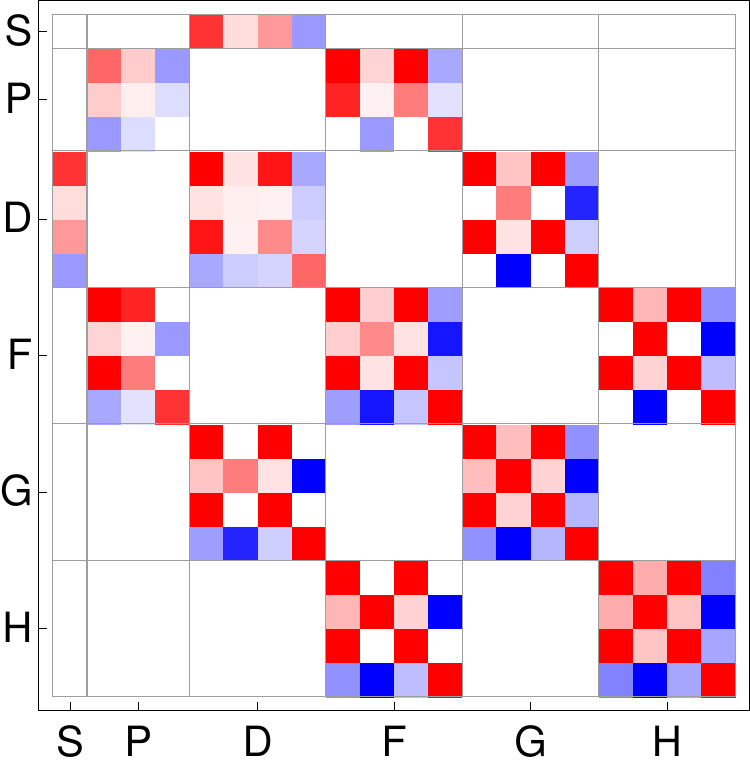} \end{minipage}
\begin{minipage}{.35\linewidth} \vspace*{0.500cm} \hspace*{-0.65cm}\includegraphics[width=1.15\textwidth]{S12_l3_coeff_2.pdf}\end{minipage}
\begin{minipage}{.25\linewidth} \begin{align} \left(a_{5}\right)^{\sigma_{1/2}}_{2} &= \left<P,P\right> + \left<S,D\right> \nonumber \\ & \hspace*{12.5pt} + \left<D,D\right>  + \left<P,F\right>   \nonumber  \\ & \hspace*{12.5pt} + \left<F,F\right>  + \left<D,G\right>   \nonumber \\ & \hspace*{12.5pt} + \left<G,G\right> + \left<F,H\right>  \nonumber \\ & \hspace*{12.5pt} + \left<H,H\right> \nonumber    \nonumber  \end{align} \end{minipage}

\begin{minipage}{.075\linewidth}
\vspace*{-6.5pt}
\hspace*{5pt}
\begin{equation}
\mathcal{C}_{3}^{\sigma_{1/2}} \equiv \nonumber
\end{equation}
\end{minipage}
\begin{minipage}{.3\linewidth} \vspace*{0.572cm} \includegraphics[width=0.875\textwidth]{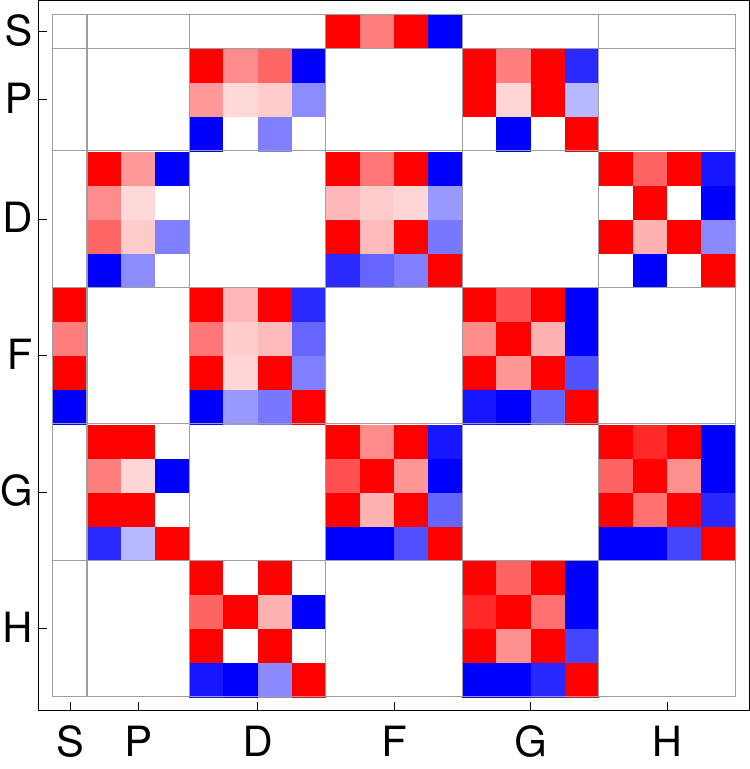} \end{minipage}
\begin{minipage}{.35\linewidth} \vspace*{0.500cm} \hspace*{-0.65cm}\includegraphics[width=1.15\textwidth]{S12_l3_coeff_3.pdf}\end{minipage}
\begin{minipage}{.25\linewidth} \begin{align} \left(a_{5}\right)^{\sigma_{1/2}}_{3} &= \left<P,D\right> + \left<S,F\right> \nonumber \\ & \hspace*{12.5pt} + \left<D,F\right>  + \left<P,G\right>   \nonumber   \\ & \hspace*{12.5pt} + \left<F,G\right>  + \left<D,H\right>   \nonumber \\ & \hspace*{12.5pt} + \left<G,H\right>    \nonumber  \end{align} \end{minipage}
\caption{%
Left: Matrices $\mathcal{C}_{0\cdots 3}^{\sigma_{1/2}}$, represented here in the color scheme, defines the coefficient $\left(a_{5}\right)_{0\cdots 3}^{\sigma_{1/2}}$ for an expansion of $\sigma_{1/2}$ up to $\text{L}_{\text{max}} = 5$. Center: Coefficients $\left(a_{3}\right)_{0\cdots 3}^{\sigma_{1/2}}$ obtained from a fit to the $\sigma_{1/2}$-data (black points). For references to the data see Table \ref{tab:DataBasis}. Bonn Gatchina predictions, truncated at different $\text{L}_{\mathrm{max}}$ ($\text{L}_{\mathrm{max}} = 1$ is drawn in green, $\text{L}_{\mathrm{max}} = 2$ in blue, $\text{L}_{\mathrm{max}} = 3$ in red and $\text{L}_{\mathrm{max}} = 4$ in black) are drawn as well. Right: All partial wave interferences for $\text{L}_{\text{max}} = 5$ are indicated.
}
\label{tab:DCS12ColorPlots1}
\end{table*}
\begin{table*}[htb]
\RawFloats
\begin{minipage}{.075\linewidth}
\vspace*{-6.5pt}
\hspace*{5pt}
\begin{equation}
\mathcal{C}_{4}^{\sigma_{1/2}} \equiv \nonumber
\end{equation}
\end{minipage}
\begin{minipage}{.3\linewidth} \vspace*{0.572cm} \includegraphics[width=0.875\textwidth]{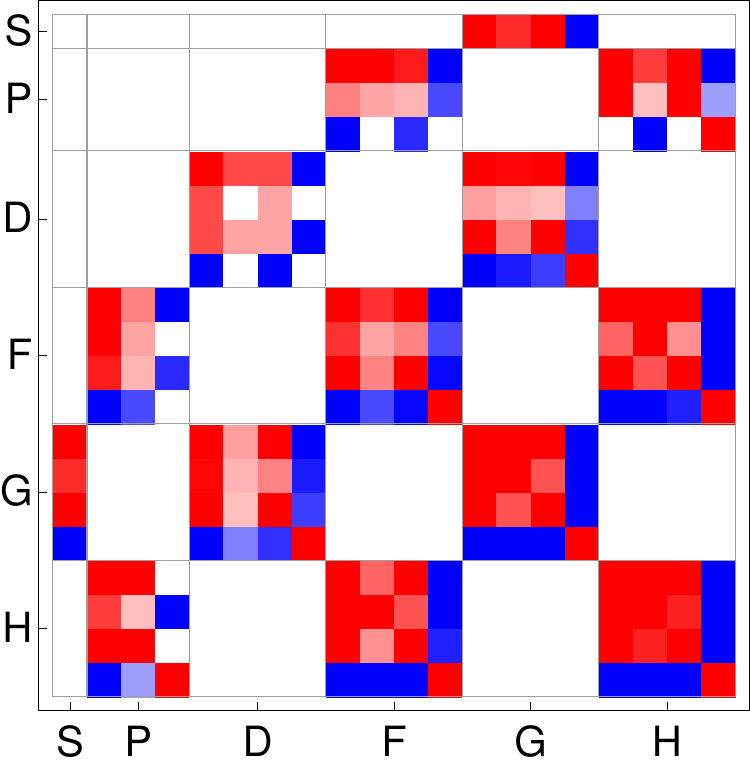} \end{minipage}
\begin{minipage}{.35\linewidth} \vspace*{0.500cm} \hspace*{-0.65cm}\includegraphics[width=1.15\textwidth]{S12_l3_coeff_4.pdf}\end{minipage}
\begin{minipage}{.25\linewidth} \begin{align} \left(a_{5}\right)^{\sigma_{1/2}}_{4} &= \left<D,D\right> + \left<P,F\right> \nonumber \\ & \hspace*{12.5pt} + \left<F,F\right>  + \left<S,G\right>   \nonumber \\ & \hspace*{12.5pt} + \left<D,G\right>  + \left<G,G\right>   \nonumber \\ & \hspace*{12.5pt} + \left<P,H\right>  + \left<F,H\right>   \nonumber  \\ & \hspace*{12.5pt} + \left<H,H\right> \nonumber \end{align} \end{minipage}

\begin{minipage}{.075\linewidth}
\vspace*{-6.5pt}
\hspace*{5pt}
\begin{equation}
\mathcal{C}_{5}^{\sigma_{1/2}} \equiv \nonumber
\end{equation}
\end{minipage}
\begin{minipage}{.3\linewidth} \vspace*{0.572cm} \includegraphics[width=0.875\textwidth]{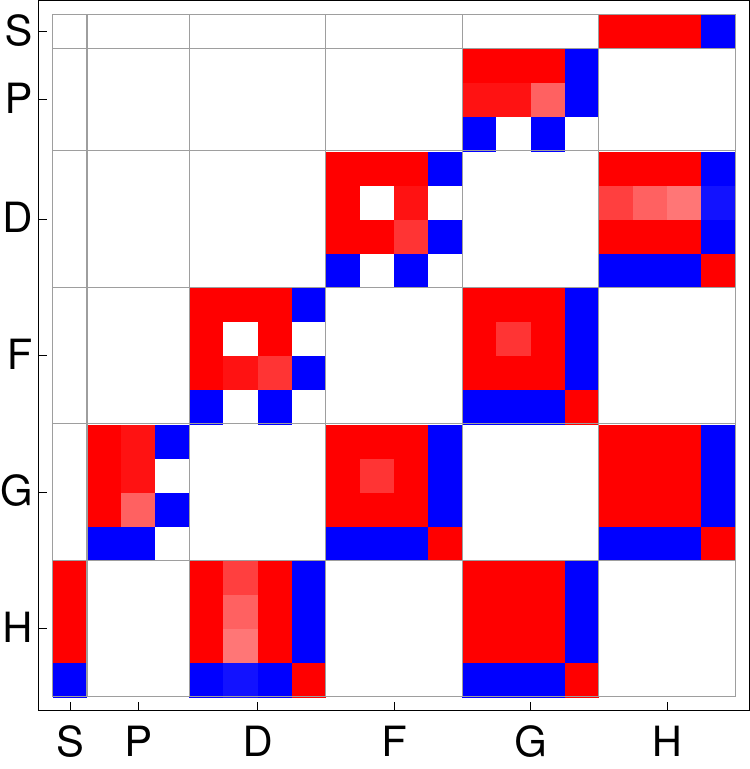} \end{minipage}
\begin{minipage}{.35\linewidth} \vspace*{0.500cm} \hspace*{-0.65cm}\includegraphics[width=1.15\textwidth]{S12_l3_coeff_5.pdf}\end{minipage}
\begin{minipage}{.25\linewidth} \begin{align} \left(a_{5}\right)^{\sigma_{1/2}}_{5} &= \left<D,F\right> + \left<P,G\right> \nonumber \\ & \hspace*{12.5pt} + \left<F,G\right>  + \left<S,H\right>   \nonumber \\ & \hspace*{12.5pt} + \left<D,H\right>  + \left<G,H\right>  \nonumber   \end{align} \end{minipage}

\begin{minipage}{.075\linewidth}
\vspace*{-6.5pt}
\hspace*{5pt}
\begin{equation}
\mathcal{C}_{6}^{\sigma_{1/2}} \equiv \nonumber
\end{equation}
\end{minipage}
\begin{minipage}{.3\linewidth} \vspace*{0.572cm} \includegraphics[width=0.875\textwidth]{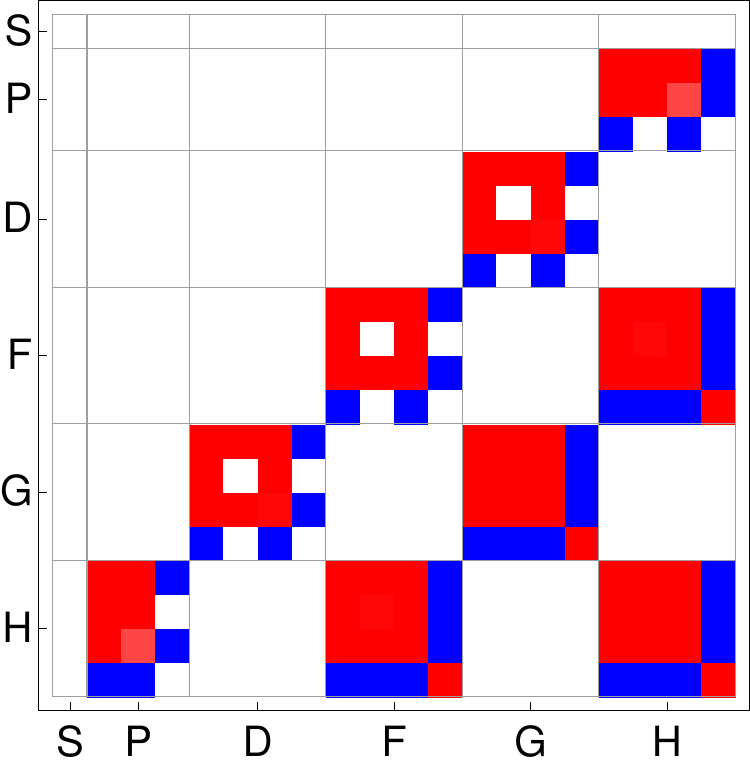} \end{minipage}
\begin{minipage}{.35\linewidth} \vspace*{0.500cm} \hspace*{-0.65cm}\includegraphics[width=1.15\textwidth]{S12_l3_coeff_6.pdf}\end{minipage}
\begin{minipage}{.25\linewidth} \begin{align} \left(a_{5}\right)^{\sigma_{1/2}}_{6} &= \left<F,F\right> + \left<D,G\right> \nonumber \\ & \hspace*{12.5pt} + \left<G,G\right>  + \left<P,H\right>   \nonumber  \\ & \hspace*{12.5pt} + \left<F,H\right>  + \left<H,H\right>   \nonumber  \end{align} \end{minipage}

\begin{minipage}{.075\linewidth}
\vspace*{-6.5pt}
\hspace*{5pt}
\begin{equation}
\mathcal{C}_{7}^{\sigma_{1/2}} \equiv \nonumber
\end{equation}
\end{minipage}
\begin{minipage}{.3\linewidth} \vspace*{0.572cm} \includegraphics[width=0.875\textwidth]{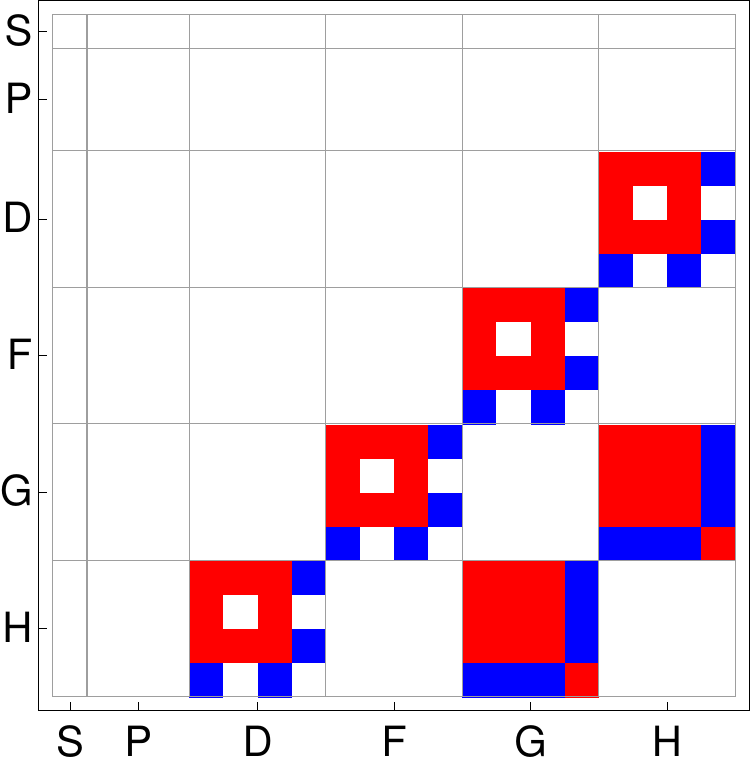} \end{minipage}
\begin{minipage}{.35\linewidth} \vspace*{0.500cm} \hspace*{-0.65cm}\includegraphics[width=1.15\textwidth]{S12_l4_coeff_7.pdf}\end{minipage}
\begin{minipage}{.25\linewidth} \begin{align} \left(a_{5}\right)^{\sigma_{1/2}}_{7} &= \left<F,G\right> + \left<D,H\right> \nonumber \\ & \hspace*{12.5pt} + \left<G,H\right>    \nonumber  \end{align} \end{minipage}
\caption{%
Left: Matrices $\mathcal{C}_{4\cdots 7}^{\sigma_{1/2}}$, represented here in the color scheme, defines the coefficient $\left(a_{5}\right)_{4\cdots 7}^{\sigma_{1/2}}$ for an expansion of $\sigma_{1/2}$ up to $\text{L}_{\text{max}} = 5$. Center: Coefficients $\left(a_{3}\right)_{4\ldots 6}^{\sigma_{1/2}}$ and $\left(a_{4}\right)_{7}^{\sigma_{1/2}}$ obtained from a fit to the $\sigma_{1/2}$-data (black points). For references to the data see Table \ref{tab:DataBasis}. Bonn Gatchina predictions, truncated at different $\text{L}_{\mathrm{max}}$ ($\text{L}_{\mathrm{max}} = 1$ is drawn in green, $\text{L}_{\mathrm{max}} = 2$ in blue, $\text{L}_{\mathrm{max}} = 3$ in red and $\text{L}_{\mathrm{max}} = 4$ in black) are drawn as well. For the higher non-fitted coefficient $\left(a_{5}\right)_{7}^{\sigma_{1/2}}$, the Bonn Gatchina curves are shown (here, the truncation at $\text{L}_{\mathrm{max}} = 5$ is drawn in cyan). Right: All partial wave interferences for $\text{L}_{\text{max}} = 5$ are indicated.
}
\label{tab:DCS12ColorPlots2}
\end{table*}
\begin{table*}[htb]
\RawFloats
\begin{minipage}{.075\linewidth}
\vspace*{-6.5pt}
\hspace*{5pt}
\begin{equation}
\mathcal{C}_{8}^{\sigma_{1/2}} \equiv \nonumber
\end{equation}
\end{minipage}
\begin{minipage}{.3\linewidth} \vspace*{0.572cm} \includegraphics[width=0.875\textwidth]{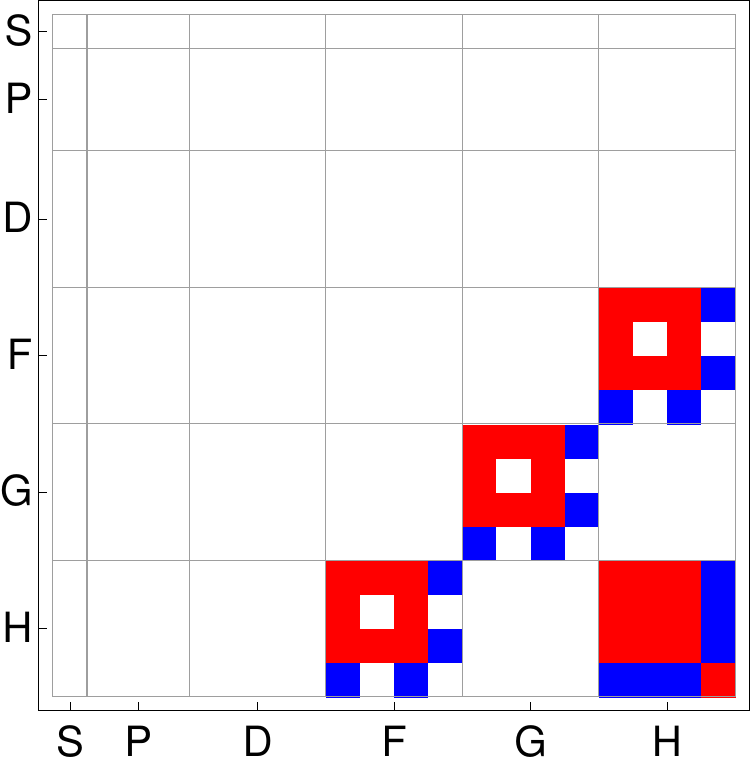} \end{minipage}
\begin{minipage}{.35\linewidth} \vspace*{0.500cm} \hspace*{-0.65cm}\includegraphics[width=1.15\textwidth]{S12_l4_coeff_8.pdf}\end{minipage}
\begin{minipage}{.25\linewidth} \begin{align} \left(a_{5}\right)^{\sigma_{1/2}}_{8} &= \left<G,G\right> + \left<F,H\right> \nonumber \\ & \hspace*{12.5pt} + \left<H,H\right> \nonumber \end{align} \end{minipage}

\begin{minipage}{.075\linewidth}
\vspace*{-6.5pt}
\hspace*{5pt}
\begin{equation}
\mathcal{C}_{9}^{\sigma_{1/2}} \equiv \nonumber
\end{equation}
\end{minipage}
\begin{minipage}{.3\linewidth} \vspace*{0.572cm} \includegraphics[width=0.875\textwidth]{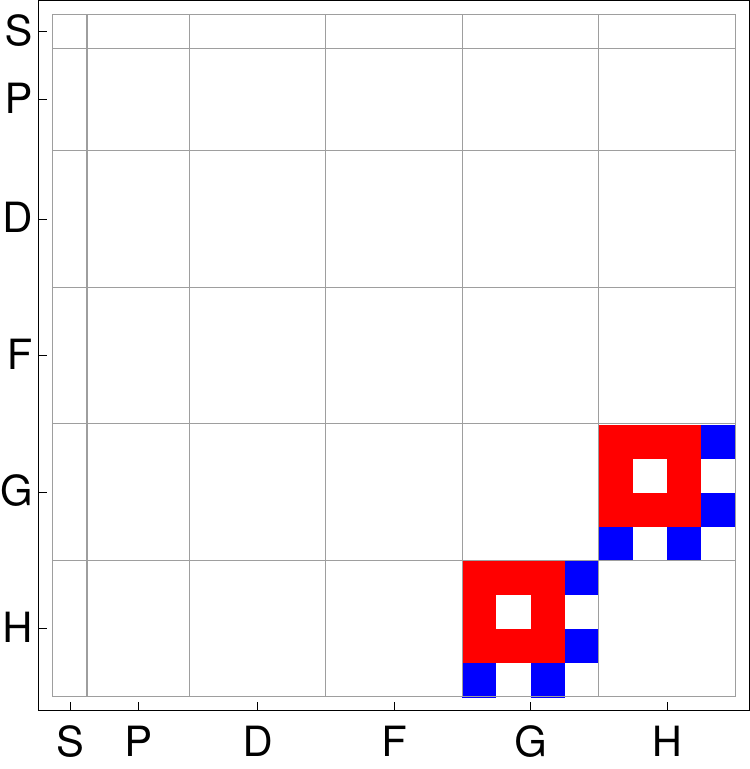} \end{minipage}
\begin{minipage}{.35\linewidth} \vspace*{0.500cm} \hspace*{-0.65cm}\includegraphics[width=1.15\textwidth]{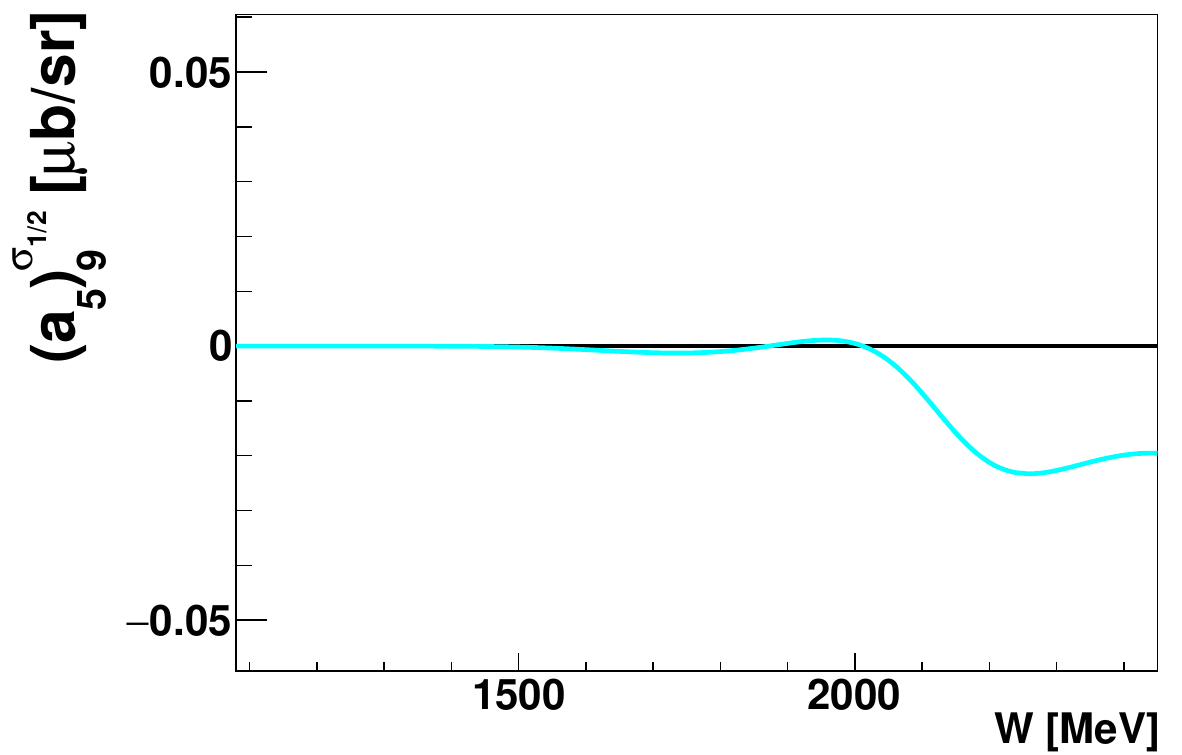}\end{minipage}
\begin{minipage}{.25\linewidth} \begin{align} \left(a_{5}\right)^{\sigma_{1/2}}_{9} &=  \left<G,H\right>  \nonumber   \end{align} \end{minipage}

\begin{minipage}{.075\linewidth}
\vspace*{-6.5pt}
\hspace*{5pt}
\begin{equation}
\mathcal{C}_{10}^{\sigma_{1/2}} \equiv \nonumber
\end{equation}
\end{minipage}
\begin{minipage}{.3\linewidth} \vspace*{0.572cm} \includegraphics[width=0.875\textwidth]{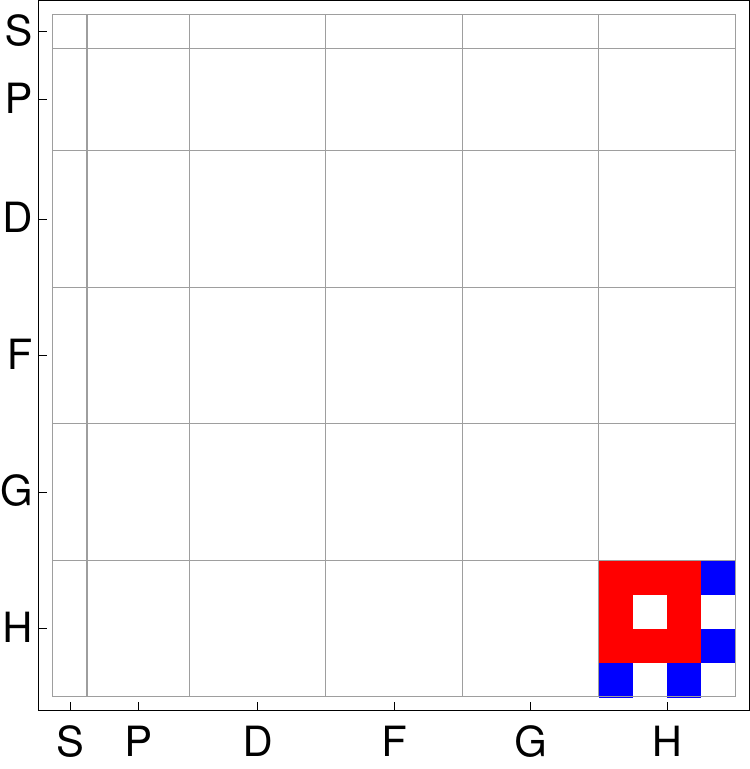} \end{minipage}
\begin{minipage}{.35\linewidth} \vspace*{0.500cm} \hspace*{-0.65cm}\includegraphics[width=1.15\textwidth]{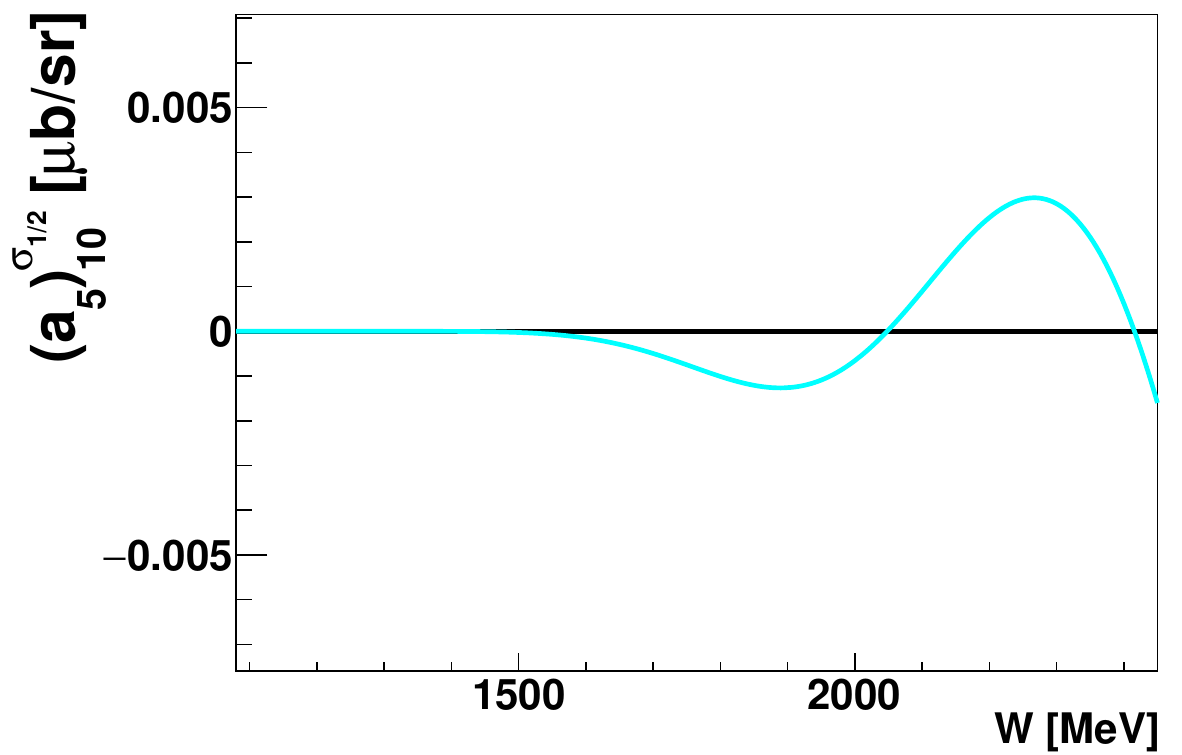}\end{minipage}
\begin{minipage}{.25\linewidth} \begin{align} \left(a_{5}\right)^{\sigma_{1/2}}_{10} &= \left<H,H\right>   \nonumber  \end{align} \end{minipage}

\caption{%
Left: Matrices $\mathcal{C}_{8\cdots 10}^{\sigma_{1/2}}$, represented here in the color scheme, defines the coefficient $\left(a_{5}\right)_{8\cdots 10}^{\sigma_{1/2}}$ for an expansion of $\sigma_{1/2}$ up to $\text{L}_{\text{max}} = 5$. Center: Coefficient $\left(a_{4}\right)_{8}^{\sigma_{1/2}}$ obtained from a fit to the $\sigma_{1/2}$-data (black points). For references to the data see Table \ref{tab:DataBasis}. Bonn Gatchina predictions, truncated at different $\text{L}_{\mathrm{max}}$ ($\text{L}_{\mathrm{max}} = 1$ is drawn in green, $\text{L}_{\mathrm{max}} = 2$ in blue, $\text{L}_{\mathrm{max}} = 3$ in red and $\text{L}_{\mathrm{max}} = 4$ in black) are drawn as well. For the highest non-fitted coefficients $\left(a_{5}\right)_{9, 10}^{\sigma_{1/2}}$, the Bonn Gatchina curves are shown (here, the truncation at $\text{L}_{\mathrm{max}} = 5$ is drawn in cyan). Right: All partial wave interferences for $\text{L}_{\text{max}} = 5$ are indicated.
}
\label{tab:DCS12ColorPlots3}
\end{table*}
%


\begin{table*}[htb]
\RawFloats
\begin{minipage}{.075\linewidth}
\vspace*{-6.5pt}
\hspace*{5pt}
\begin{equation}
\mathcal{C}_{2}^{\check{G}} \equiv \nonumber
\end{equation}
\end{minipage}
\begin{minipage}{.3\linewidth} \vspace*{0.572cm} \includegraphics[width=0.875\textwidth]{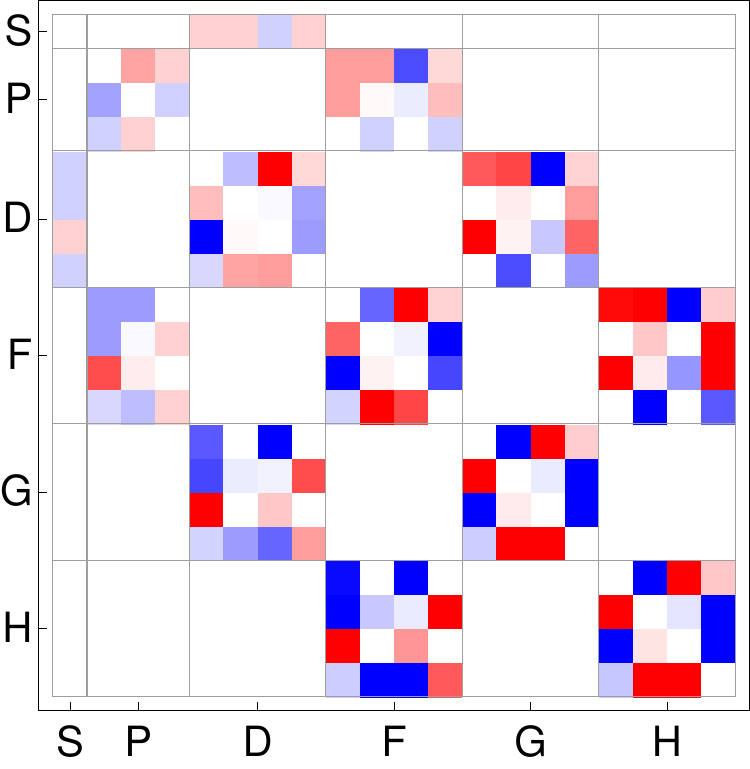} \end{minipage}
\begin{minipage}{.35\linewidth} \vspace*{0.500cm} \hspace*{-0.65cm}\includegraphics[width=1.15\textwidth]{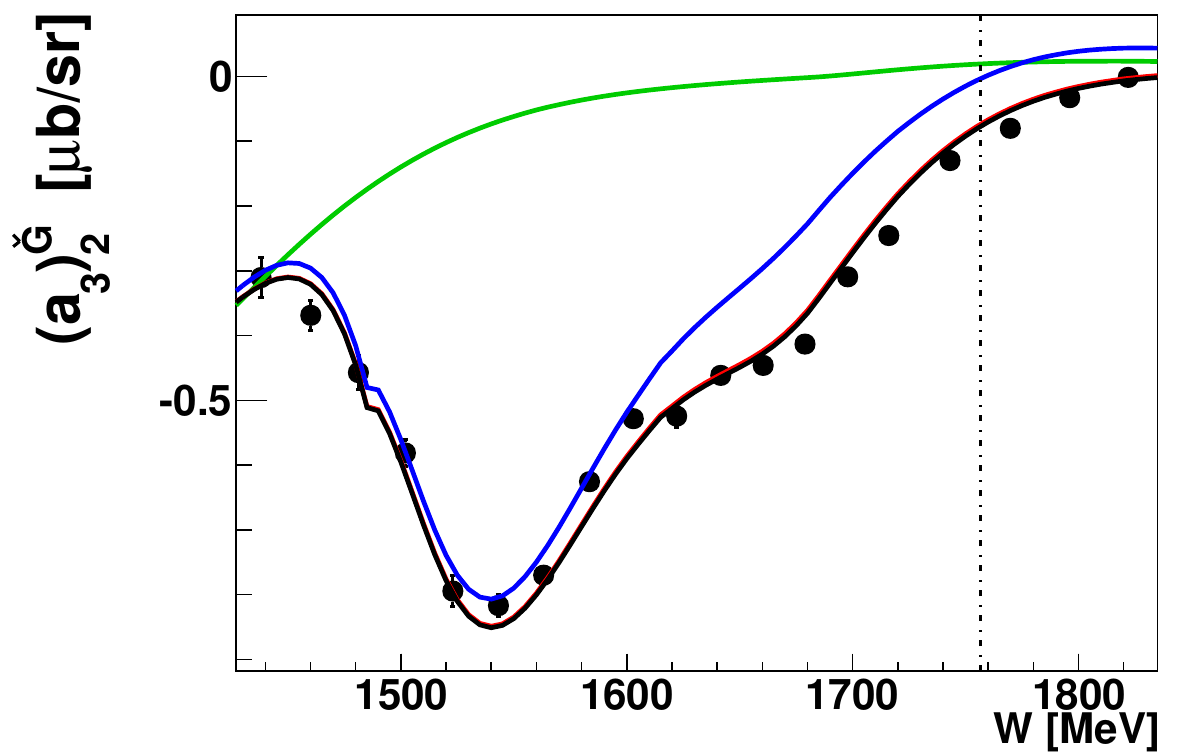}\end{minipage}
\begin{minipage}{.25\linewidth} \begin{align} \left(a_{5}\right)^{\check{G}}_{2} &= \left<S,D\right> + \left<P,P\right> \nonumber \\ & \hspace*{12.5pt} + \left<P,F\right> + \left<D,D\right>  \nonumber \\ & \hspace*{12.5pt}  + \left<D,G\right>   + \left<F,F\right> \nonumber \\ & \hspace*{12.5pt}   + \left<F,H\right>  + \left<G,G\right> \nonumber \\ & \hspace*{12.5pt} + \left<H,H\right>  \nonumber \end{align} \end{minipage}

\begin{minipage}{.075\linewidth}
\vspace*{-6.5pt}
\hspace*{5pt}
\begin{equation}
\mathcal{C}_{3}^{\check{G}} \equiv \nonumber
\end{equation}
\end{minipage}
\begin{minipage}{.3\linewidth} \vspace*{0.572cm} \includegraphics[width=0.875\textwidth]{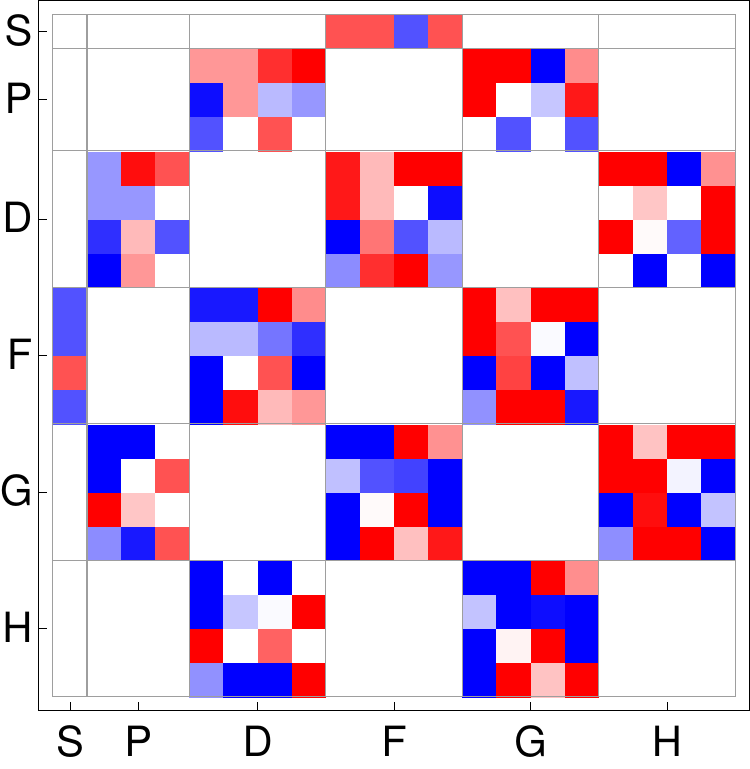} \end{minipage}
\begin{minipage}{.35\linewidth} \vspace*{0.500cm} \hspace*{-0.65cm}\includegraphics[width=1.15\textwidth]{G_l3_coeff_1.pdf}\end{minipage}
\begin{minipage}{.25\linewidth} \begin{align} \left(a_{5}\right)^{\check{G}}_{3} &= \left<S,F\right> + \left<P,D\right>  \nonumber \\ & \hspace*{12.5pt}+ \left<P,G\right> + \left<D,F\right>    \nonumber \\ & \hspace*{12.5pt} + \left<D,H\right> + \left<F,G\right>   \nonumber   \\ & \hspace*{12.5pt}   + \left<G,H\right> \nonumber \end{align} \end{minipage}

\begin{minipage}{.075\linewidth}
\vspace*{-6.5pt}
\hspace*{5pt}
\begin{equation}
\mathcal{C}_{4}^{\check{G}} \equiv \nonumber
\end{equation}
\end{minipage}
\begin{minipage}{.3\linewidth} \vspace*{0.572cm} \includegraphics[width=0.875\textwidth]{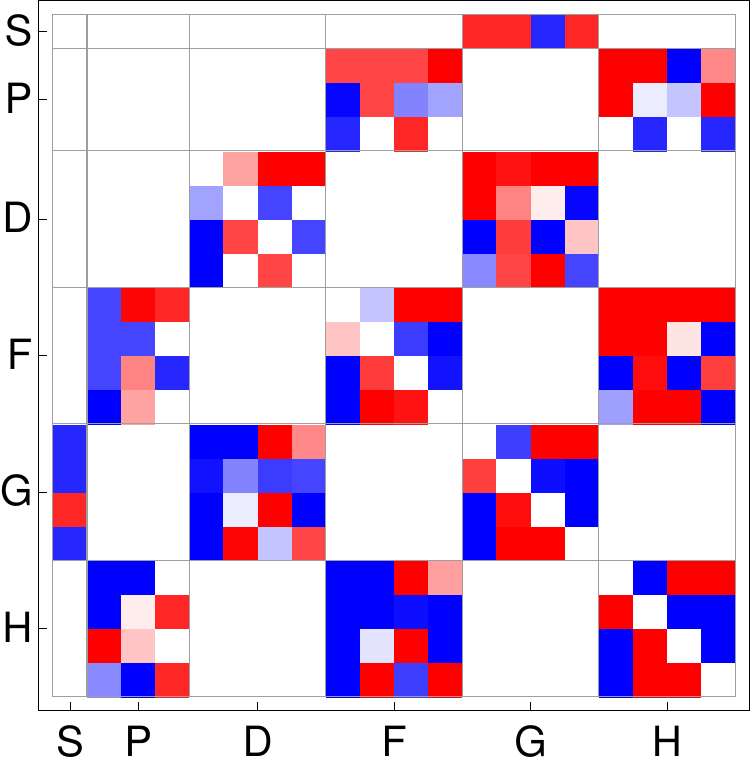} \end{minipage}
\begin{minipage}{.35\linewidth} \vspace*{0.500cm} \hspace*{-0.65cm}\includegraphics[width=1.15\textwidth]{G_l3_coeff_2.pdf}\end{minipage}
\begin{minipage}{.25\linewidth} \begin{align} \left(a_{5}\right)^{\check{G}}_{4} &= \left<S,G\right> + \left<P,F\right> \nonumber \\ & \hspace*{12.5pt} + \left<P,H\right>  + \left<D,D\right>   \nonumber \\ & \hspace*{12.5pt} + \left<D,G\right>  + \left<F,F\right>  \nonumber \\ & \hspace*{12.5pt} + \left<F,H\right>  + \left<G,G\right> \nonumber \\ & \hspace*{12.5pt} + \left<H,H\right>  \nonumber \end{align} \end{minipage}

\begin{minipage}{.075\linewidth}
\vspace*{-6.5pt}
\hspace*{5pt}
\begin{equation}
\mathcal{C}_{5}^{\check{G}} \equiv \nonumber
\end{equation}
\end{minipage}
\begin{minipage}{.3\linewidth} \vspace*{0.572cm} \includegraphics[width=0.875\textwidth]{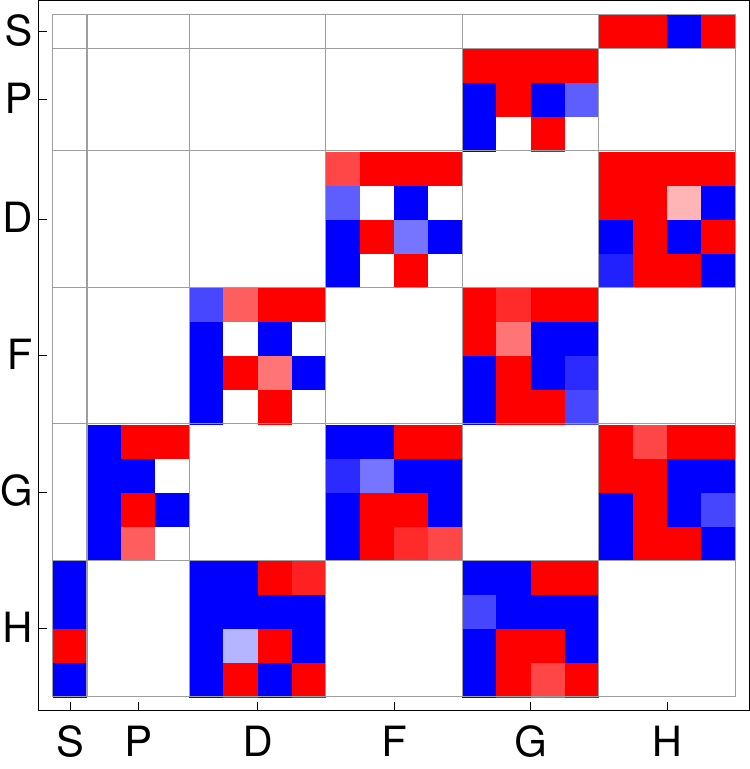} \end{minipage}
\begin{minipage}{.35\linewidth} \vspace*{0.500cm} \hspace*{-0.65cm}\includegraphics[width=1.15\textwidth]{G_l3_coeff_3.pdf}\end{minipage}
\begin{minipage}{.25\linewidth} \begin{align} \left(a_{5}\right)^{\check{G}}_{5} &= \left<S,H\right> + \left<P,G\right> \nonumber \\ & \hspace*{12.5pt} + \left<D,F\right>  + \left<D,H\right>   \nonumber \\ & \hspace*{12.5pt} + \left<F,G\right>  + \left<G,H\right>  \nonumber  \end{align} \end{minipage}
\caption{%
Left: Matrices $\mathcal{C}_{2\cdots 5}^{\check{G}}$, represented here in the color scheme, defines the coefficient $\left(a_{5}\right)_{2\cdots 5}^{\check{G}}$ for an expansion of $\check{G}$ up to $\text{L}_{\text{max}} = 5$. Center: Coefficients $\left(a_{3}\right)_{2\cdots 5}^{\check{G}}$ obtained from a fit to the $\check{G}$-data (black points). For references to the data see Table \ref{tab:DataBasis}. Bonn Gatchina predictions, truncated at different $\text{L}_{\mathrm{max}}$ ($\text{L}_{\mathrm{max}} = 1$ is drawn in green, $\text{L}_{\mathrm{max}} = 2$ in blue, $\text{L}_{\mathrm{max}} = 3$ in red and $\text{L}_{\mathrm{max}} = 4$ in black) are drawn as well. Right: All partial wave interferences for $\text{L}_{\text{max}} = 5$ are indicated.
}
\label{tab:GColorPlots1}
\end{table*}

\begin{table*}[htb]
\RawFloats
\begin{minipage}{.075\linewidth}
\vspace*{-6.5pt}
\hspace*{5pt}
\begin{equation}
\mathcal{C}_{6}^{\check{G}} \equiv \nonumber
\end{equation}
\end{minipage}
\begin{minipage}{.3\linewidth} \vspace*{0.572cm} \includegraphics[width=0.875\textwidth]{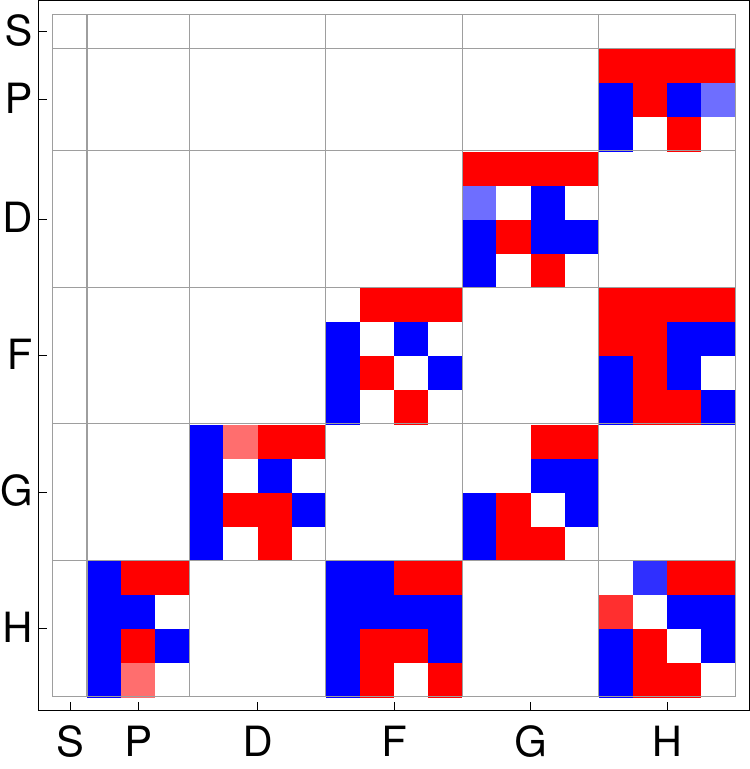} \end{minipage}
\begin{minipage}{.35\linewidth} \vspace*{0.500cm} \hspace*{-0.65cm}\includegraphics[width=1.15\textwidth]{G_l3_coeff_4.pdf}\end{minipage}
\begin{minipage}{.25\linewidth} \begin{align} \left(a_{5}\right)^{\check{G}}_{6} &= \left<P,H\right> + \left<D,G\right> \nonumber \\ & \hspace*{12.5pt} + \left<F,F\right>  + \left<F,H\right>   \nonumber \\ & \hspace*{12.5pt} + \left<G,G\right>  + \left<H,H\right> \nonumber \end{align} \end{minipage}

\begin{minipage}{.075\linewidth}
\vspace*{-6.5pt}
\hspace*{5pt}
\begin{equation}
\mathcal{C}_{7}^{\check{G}} \equiv \nonumber
\end{equation}
\end{minipage}
\begin{minipage}{.3\linewidth} \vspace*{0.572cm} \includegraphics[width=0.875\textwidth]{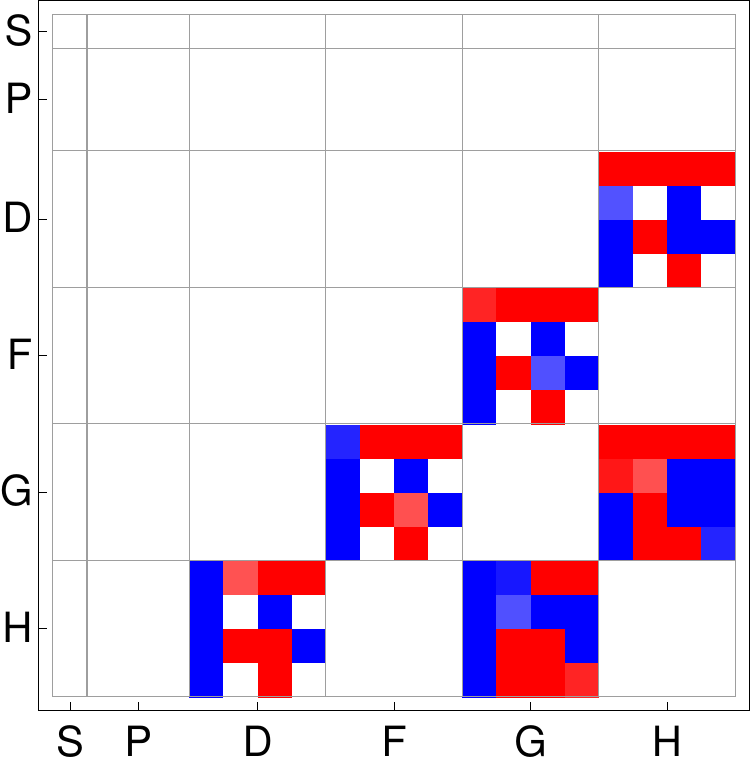} \end{minipage}
\begin{minipage}{.35\linewidth} \vspace*{0.500cm} \hspace*{-0.65cm}\includegraphics[width=1.15\textwidth]{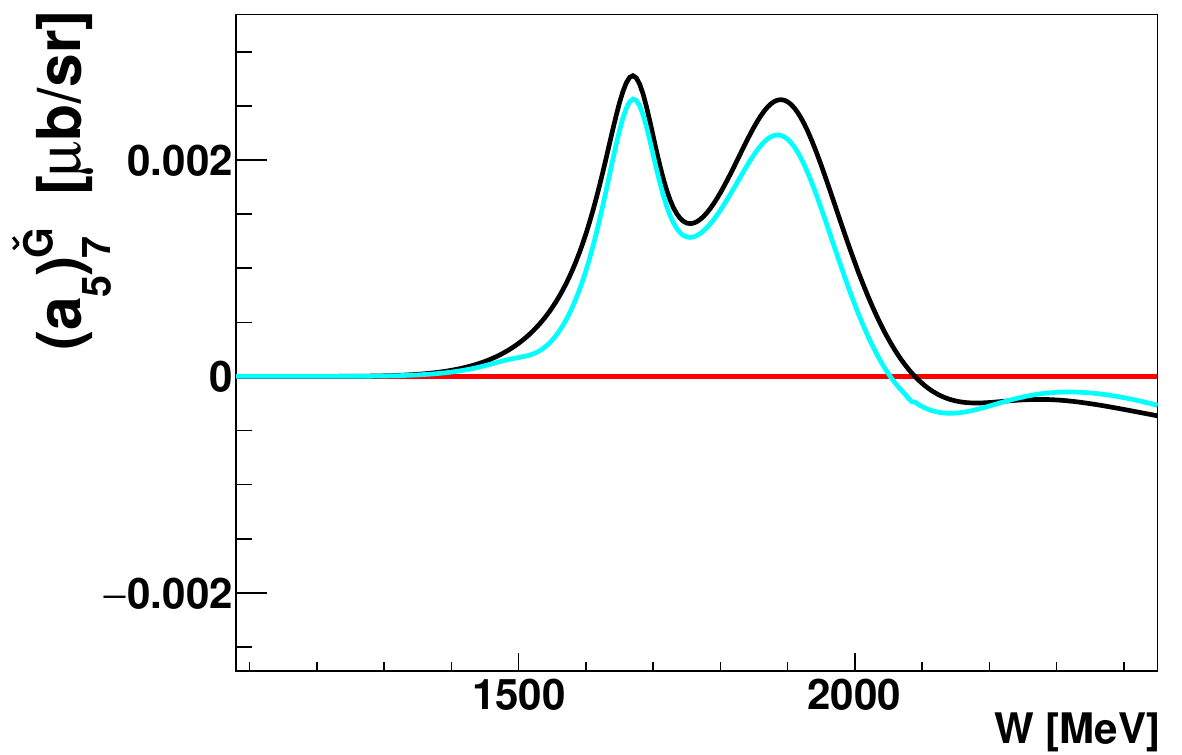}\end{minipage}
\begin{minipage}{.25\linewidth} \begin{align} \left(a_{5}\right)^{\check{G}}_{7} &= \left<D,H\right> + \left<F,G\right> \nonumber \\ & \hspace*{12.5pt} + \left<G,H\right>   \nonumber \end{align} \end{minipage}

\begin{minipage}{.075\linewidth}
\vspace*{-6.5pt}
\hspace*{5pt}
\begin{equation}
\mathcal{C}_{8}^{\check{G}} \equiv \nonumber
\end{equation}
\end{minipage}
\begin{minipage}{.3\linewidth} \vspace*{0.572cm} \includegraphics[width=0.875\textwidth]{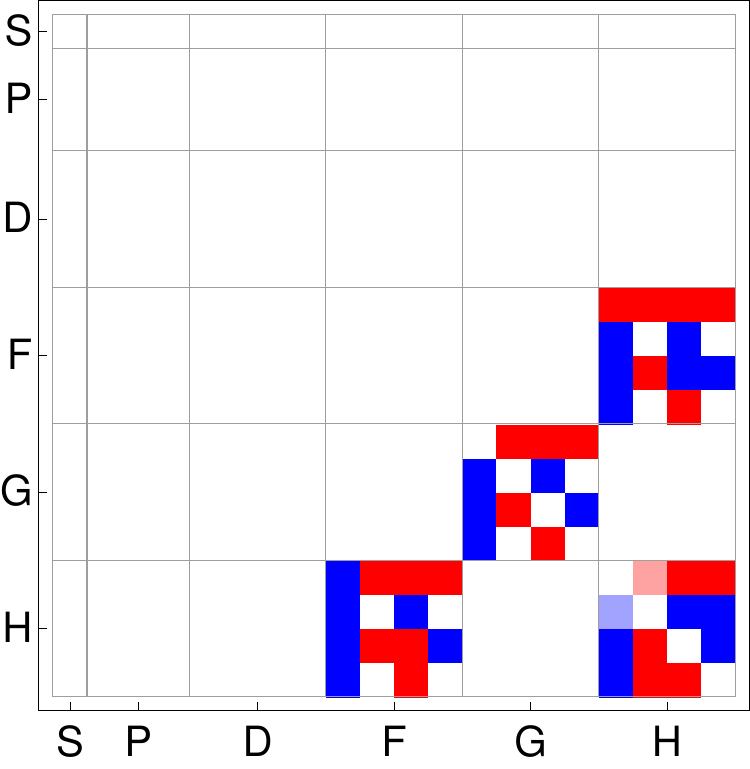} \end{minipage}
\begin{minipage}{.35\linewidth} \vspace*{0.500cm} \hspace*{-0.65cm}\includegraphics[width=1.15\textwidth]{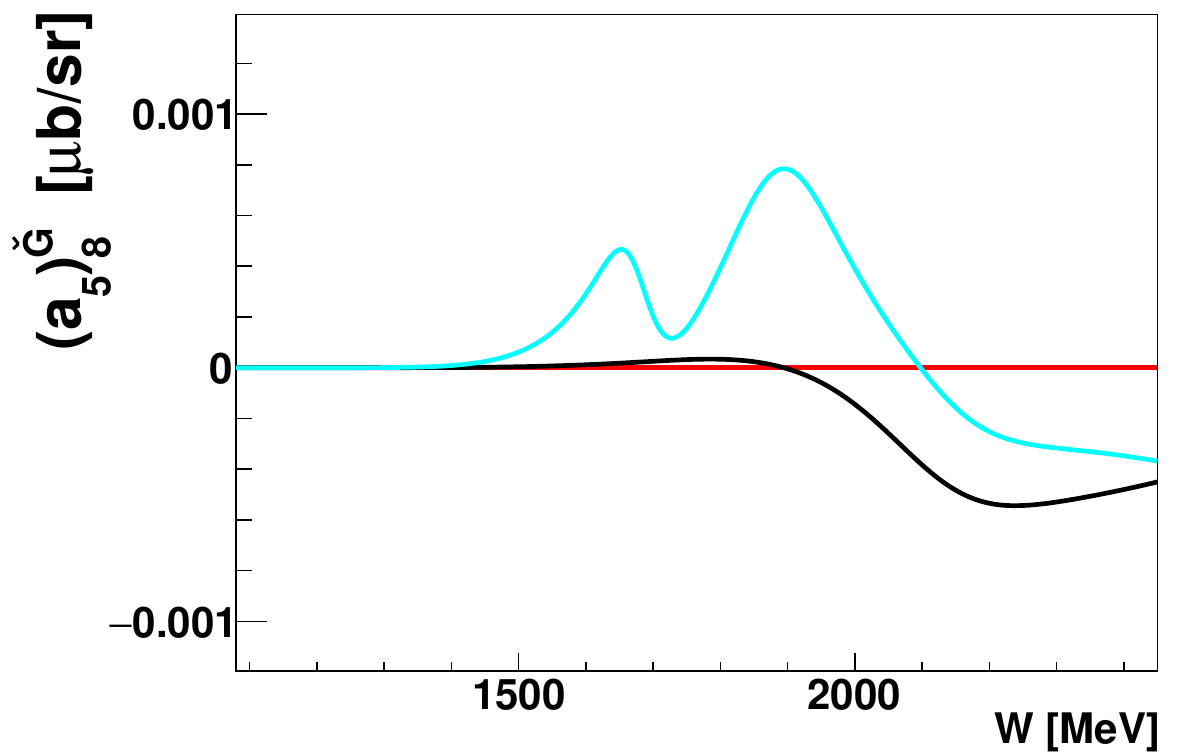}\end{minipage}
\begin{minipage}{.25\linewidth} \begin{align} \left(a_{5}\right)^{\check{G}}_{8} &= \left<F,H\right> + \left<G,G\right> \nonumber \\ & \hspace*{12.5pt} + \left<H,H\right>    \nonumber \end{align} \end{minipage}

\begin{minipage}{.075\linewidth}
\vspace*{-6.5pt}
\hspace*{5pt}
\begin{equation}
\mathcal{C}_{9}^{\check{G}} \equiv \nonumber
\end{equation}
\end{minipage}
\begin{minipage}{.3\linewidth} \vspace*{0.572cm} \includegraphics[width=0.875\textwidth]{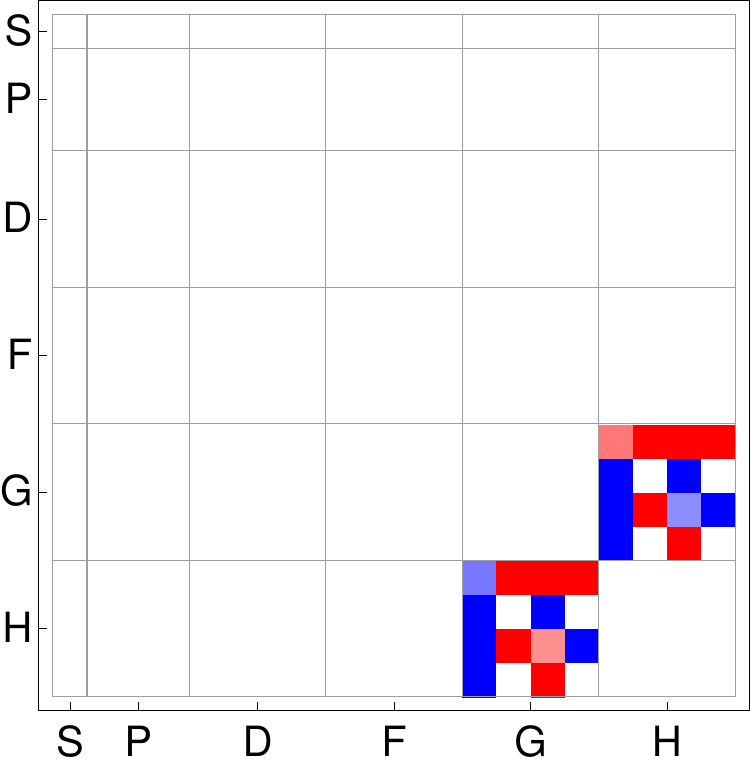} \end{minipage}
\begin{minipage}{.35\linewidth} \vspace*{0.500cm} \hspace*{-0.65cm}\includegraphics[width=1.15\textwidth]{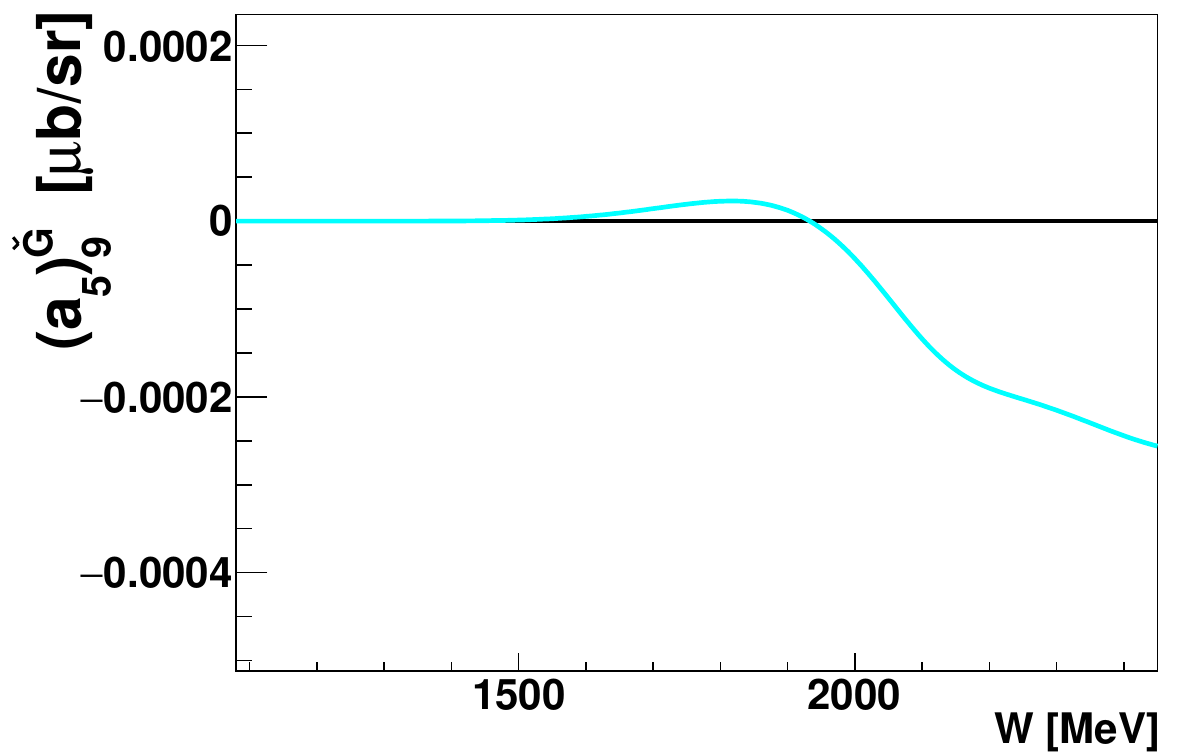}\end{minipage}
\begin{minipage}{.25\linewidth} \begin{align} \left(a_{5}\right)^{\check{G}}_{9} &= \left<G,H\right> \nonumber  \end{align} \end{minipage}
\caption{%
Left: Matrices $\mathcal{C}_{6\cdots 9}^{\check{G}}$, represented here in the color scheme, defines the coefficient $\left(a_{5}\right)_{6\cdots 9}^{\check{G}}$ for an expansion of $\check{G}$ up to $\text{L}_{\text{max}} = 5$. Center: Coefficients $\left(a_{4}\right)_{6\cdots 8}^{\check{G}}$ obtained from a fit to the $\check{G}$-data (black points). For references to the data see Table \ref{tab:DataBasis}. Bonn Gatchina predictions, truncated at different $\text{L}_{\mathrm{max}}$ ($\text{L}_{\mathrm{max}} = 1$ is drawn in green, $\text{L}_{\mathrm{max}} = 2$ in blue, $\text{L}_{\mathrm{max}} = 3$ in red and $\text{L}_{\mathrm{max}} = 4$ in black) are drawn as well. For the higher non-fitted coefficients $\left(a_{5}\right)_{7 \ldots 9}^{\check{G}}$, the Bonn Gatchina curves are shown (here, the truncation at $\text{L}_{\mathrm{max}} = 5$ is drawn in cyan). Right: All partial wave interferences for $\text{L}_{\text{max}} = 5$ are indicated.
}
\label{tab:SigmaColorPlots2Prime}
\end{table*}

 \begin{table*}[htb]
\RawFloats
\begin{minipage}{.075\linewidth}
\vspace*{-6.5pt}
\hspace*{5pt}
\begin{equation}
\mathcal{C}_{10}^{\check{G}} \equiv \nonumber
\end{equation}
\end{minipage}
\begin{minipage}{.3\linewidth} \vspace*{0.572cm} \includegraphics[width=0.875\textwidth]{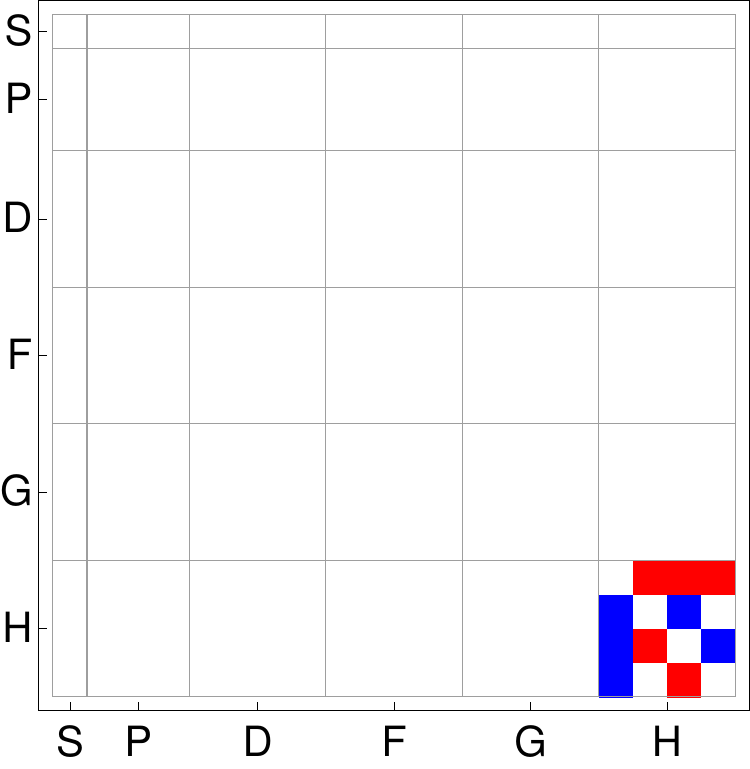} \end{minipage}
\begin{minipage}{.35\linewidth} \vspace*{0.500cm} \hspace*{-0.65cm}\includegraphics[width=1.15\textwidth]{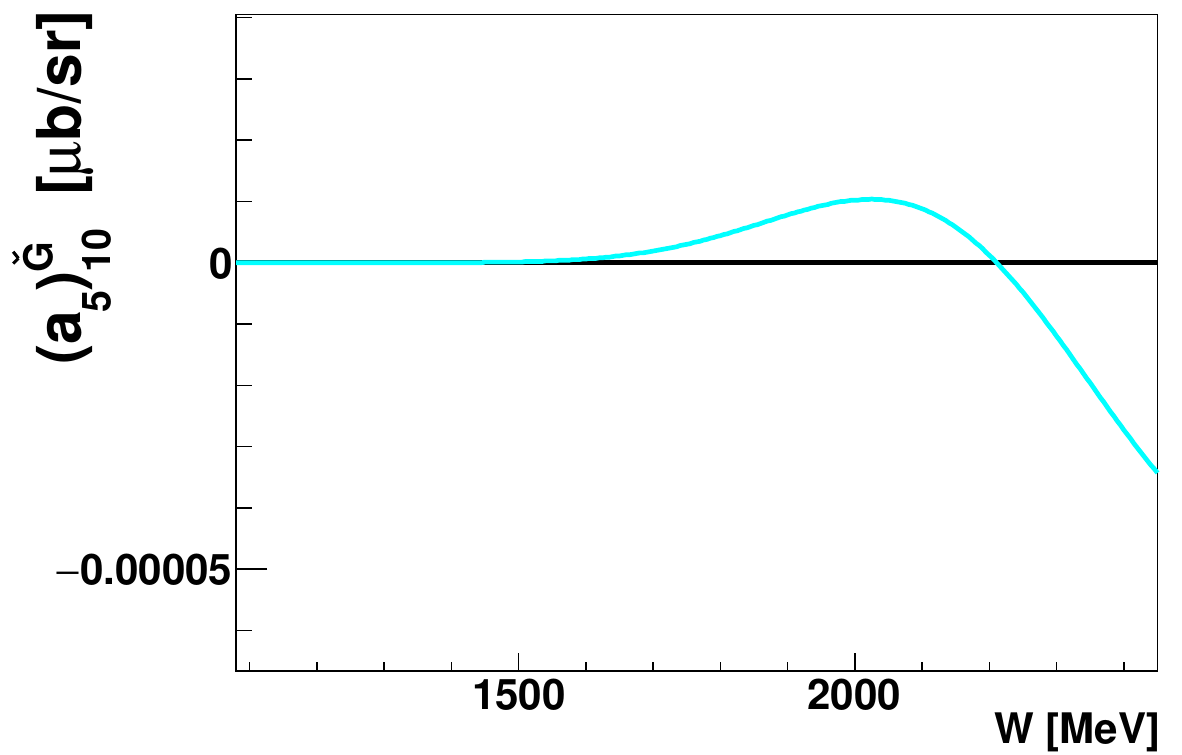}\end{minipage}
\begin{minipage}{.25\linewidth} \begin{align} \left(a_{5}\right)^{\check{G}}_{10} &= \left<H,H\right>  \nonumber \end{align} \end{minipage}
\caption{%
Left: Matrix $\mathcal{C}_{9}^{\check{G}}$, represented here in the color scheme, defines the coefficient $\left(a_{5}\right)_{9}^{\check{G}}$ for an expansion of $\check{G}$ up to $\text{L}_{\text{max}} = 5$. Center: For the highest non-fitted coefficient $\left(a_{5}\right)_{10}^{\check{G}}$, the Bonn Gatchina curves are shown (here, the truncation at $\text{L}_{\mathrm{max}} = 5$ is drawn in cyan). Right: All partial wave interferences for $\text{L}_{\text{max}} = 5$ are indicated.
}
\label{tab:GColorPlots3}
\end{table*}

\begin{table*}[htb]
\RawFloats
\begin{minipage}{.075\linewidth}
\vspace*{-6.5pt}
\hspace*{5pt}
\begin{equation}
\mathcal{C}_{1}^{\check{H}} \equiv \nonumber
\end{equation}
\end{minipage}
\begin{minipage}{.3\linewidth} \vspace*{0.572cm} \includegraphics[width=0.875\textwidth]{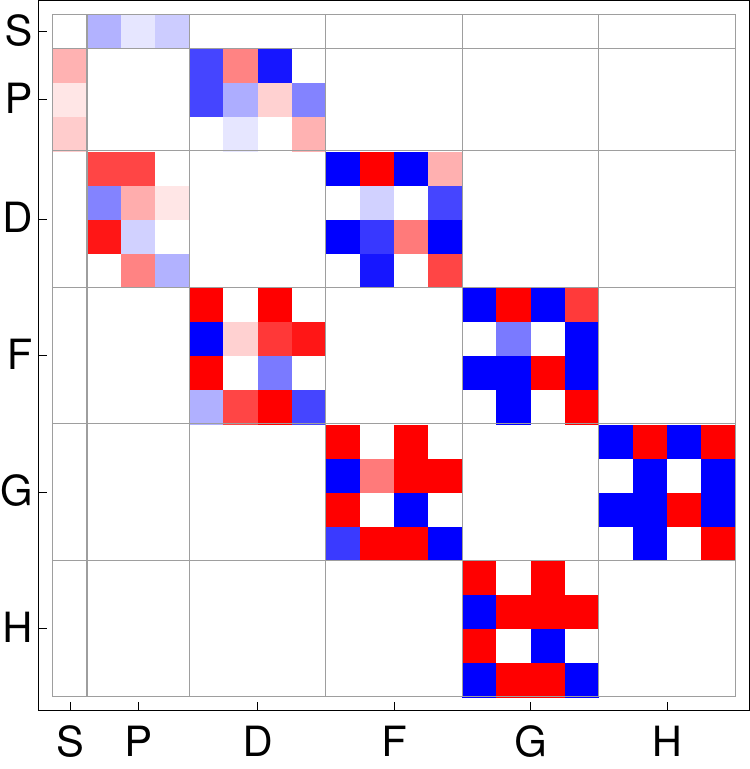} \end{minipage}
\begin{minipage}{.35\linewidth} \vspace*{0.500cm} \hspace*{-0.65cm}\includegraphics[width=1.15\textwidth]{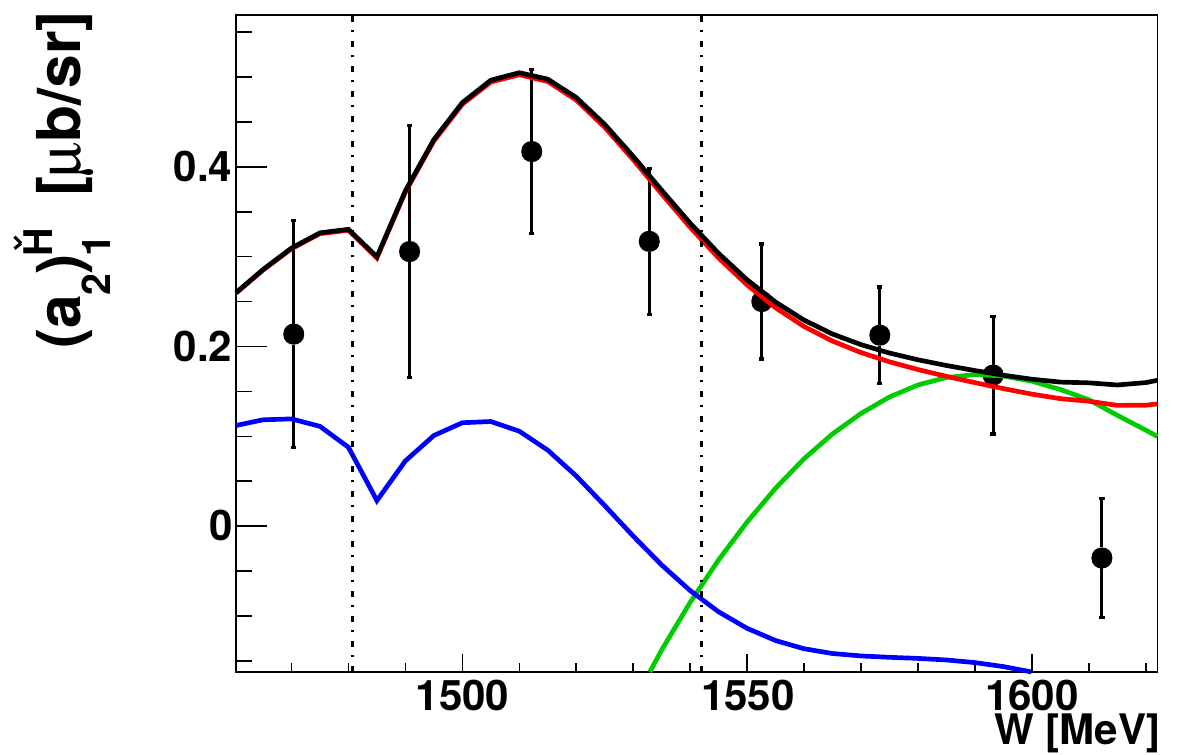}\end{minipage}
\begin{minipage}{.25\linewidth} \begin{align} \left(a_{5}\right)^{\check{H}}_{1} &= \left<S,P\right> + \left<P,D\right> \nonumber \\ & \hspace*{12.5pt} + \left<D,F\right>  + \left<F,G\right> \nonumber \\ & \hspace*{12.5pt}  + \left<G,H\right>   \nonumber  \end{align} \end{minipage}

\begin{minipage}{.075\linewidth}
\vspace*{-6.5pt}
\hspace*{5pt}
\begin{equation}
\mathcal{C}_{2}^{\check{H}} \equiv \nonumber
\end{equation}
\end{minipage}
\begin{minipage}{.3\linewidth} \vspace*{0.572cm} \includegraphics[width=0.875\textwidth]{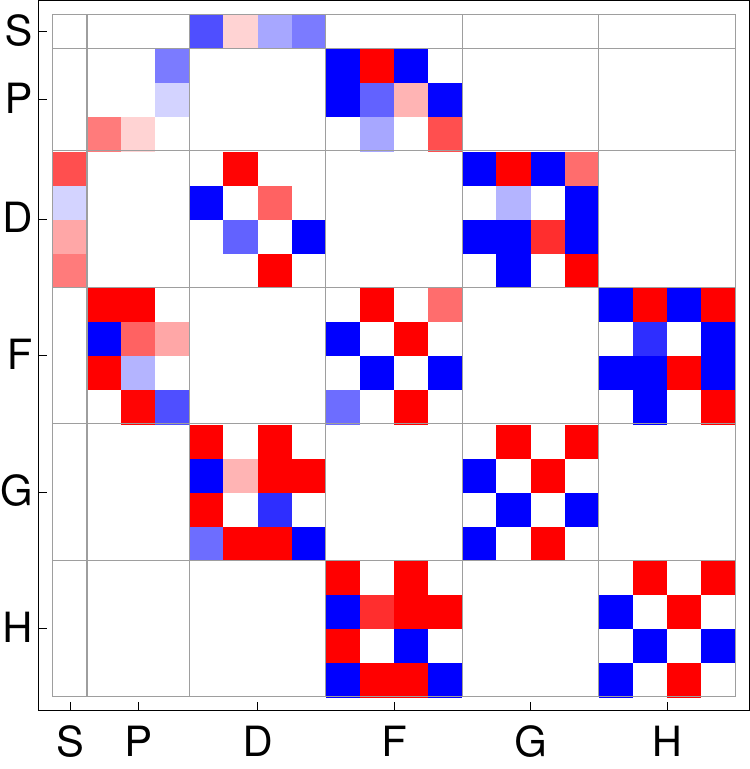} \end{minipage}
\begin{minipage}{.35\linewidth} \vspace*{0.500cm} \hspace*{-0.65cm}\includegraphics[width=1.15\textwidth]{H_l2_coeff_1.pdf}\end{minipage}
\begin{minipage}{.25\linewidth} \begin{align} \left(a_{5}\right)^{\check{H}}_{2} &= \left<S,D\right> + \left<P,P\right> \nonumber \\ & \hspace*{12.5pt} + \left<P,F\right>  + \left<D,D\right>   \nonumber \\ & \hspace*{12.5pt} + \left<D,G\right>  + \left<F,F\right>   \nonumber  \\ & \hspace*{12.5pt} + \left<F,H\right>  + \left<G,G\right>   \nonumber \\ & \hspace*{12.5pt} + \left<H,H\right>    \nonumber  \end{align} \end{minipage}

\begin{minipage}{.075\linewidth}
\vspace*{-6.5pt}
\hspace*{5pt}
\begin{equation}
\mathcal{C}_{3}^{\check{H}} \equiv \nonumber
\end{equation}
\end{minipage}
\begin{minipage}{.3\linewidth} \vspace*{0.572cm} \includegraphics[width=0.875\textwidth]{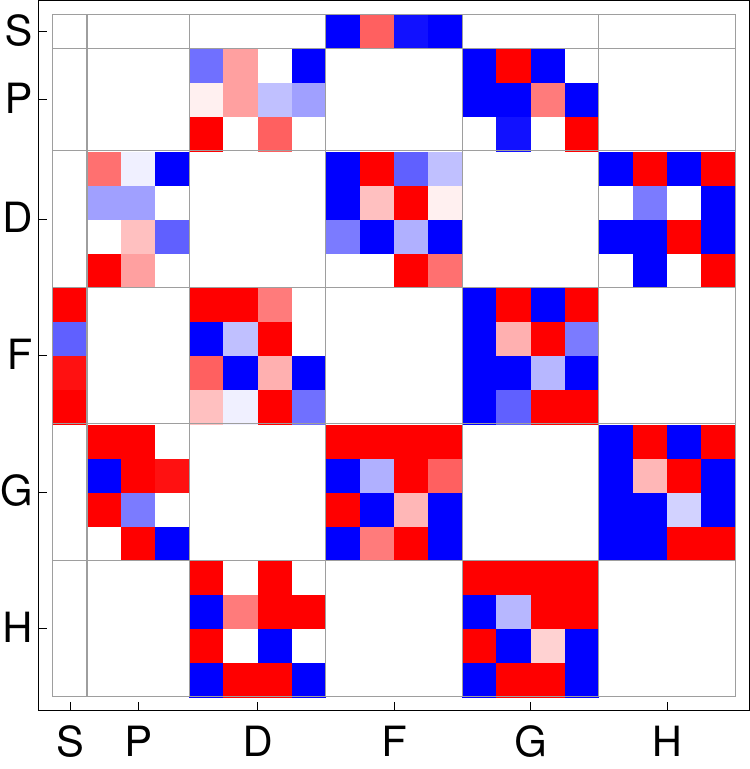} \end{minipage}
\begin{minipage}{.35\linewidth} \vspace*{0.500cm} \hspace*{-0.65cm}\includegraphics[width=1.15\textwidth]{H_l2_coeff_2.pdf}\end{minipage}
\begin{minipage}{.25\linewidth} \begin{align} \left(a_{5}\right)^{\check{H}}_{3} &= \left<S,F\right> + \left<P,D\right> \nonumber \\ & \hspace*{12.5pt} + \left<P,G\right>  + \left<D,F\right>   \nonumber  \\ & \hspace*{12.5pt} + \left<D,H\right>  + \left<F,G\right>   \nonumber \\ & \hspace*{12.5pt} + \left<G,H\right>   \nonumber   \end{align} \end{minipage}

\begin{minipage}{.075\linewidth}
\vspace*{-6.5pt}
\hspace*{5pt}
\begin{equation}
\mathcal{C}_{4}^{\check{H}} \equiv \nonumber
\end{equation}
\end{minipage}
\begin{minipage}{.3\linewidth} \vspace*{0.572cm} \includegraphics[width=0.875\textwidth]{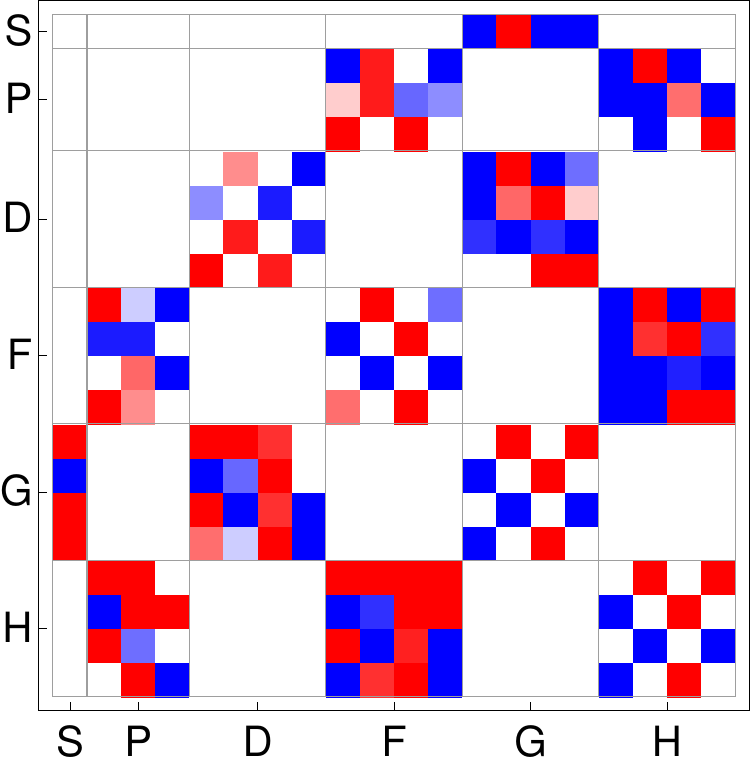} \end{minipage}
\begin{minipage}{.35\linewidth} \vspace*{0.500cm} \hspace*{-0.65cm}\includegraphics[width=1.15\textwidth]{H_l2_coeff_3.pdf}\end{minipage}
\begin{minipage}{.25\linewidth} \begin{align} \left(a_{5}\right)^{\check{H}}_{4} &= \left<S,G\right> + \left<P,F\right> \nonumber \\ & \hspace*{12.5pt} + \left<P,H\right>  + \left<D,D\right>   \nonumber   \\ & \hspace*{12.5pt} + \left<D,G\right>  + \left<F,F\right>   \nonumber \\ & \hspace*{12.5pt} + \left<F,H\right>  + \left<G,G\right>   \nonumber \\ & \hspace*{12.5pt} + \left<H,H\right>    \nonumber  \end{align} \end{minipage}
\caption{%
Left: Matrices $\mathcal{C}_{1\cdots 4}^{\check{H}}$, represented here in the color scheme, defines the coefficient $\left(a_{5}\right)_{1\cdots 4}^{\check{H}}$ for an expansion of $\check{H}$ up to $\text{L}_{\text{max}} = 5$. Center: Coefficients $\left(a_{2}\right)_{1\ldots 4}^{\check{H}}$ obtained from a fit to the $\check{H}$-data (black points). For references to the data see Table \ref{tab:DataBasis}. Bonn Gatchina predictions, truncated at different $\text{L}_{\mathrm{max}}$ ($\text{L}_{\mathrm{max}} = 1$ is drawn in green, $\text{L}_{\mathrm{max}} = 2$ in blue, $\text{L}_{\mathrm{max}} = 3$ in red and $\text{L}_{\mathrm{max}} = 4$ in black) are drawn as well. Right: All partial wave interferences for $\text{L}_{\text{max}} = 5$ are indicated.
}
\label{tab:HColorPlots1}
\end{table*}

\begin{table*}[htb]
\RawFloats
\begin{minipage}{.075\linewidth}
\vspace*{-6.5pt}
\hspace*{5pt}
\begin{equation}
\mathcal{C}_{5}^{\check{H}} \equiv \nonumber
\end{equation}
\end{minipage}
\begin{minipage}{.3\linewidth} \vspace*{0.572cm} \includegraphics[width=0.875\textwidth]{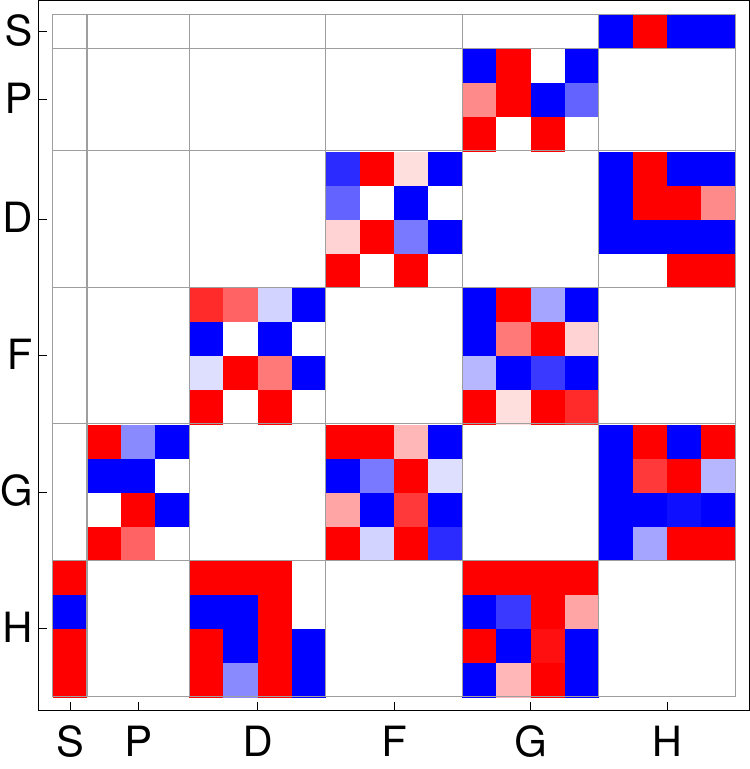} \end{minipage}
\begin{minipage}{.35\linewidth} \vspace*{0.500cm} \hspace*{-0.65cm}\includegraphics[width=1.15\textwidth]{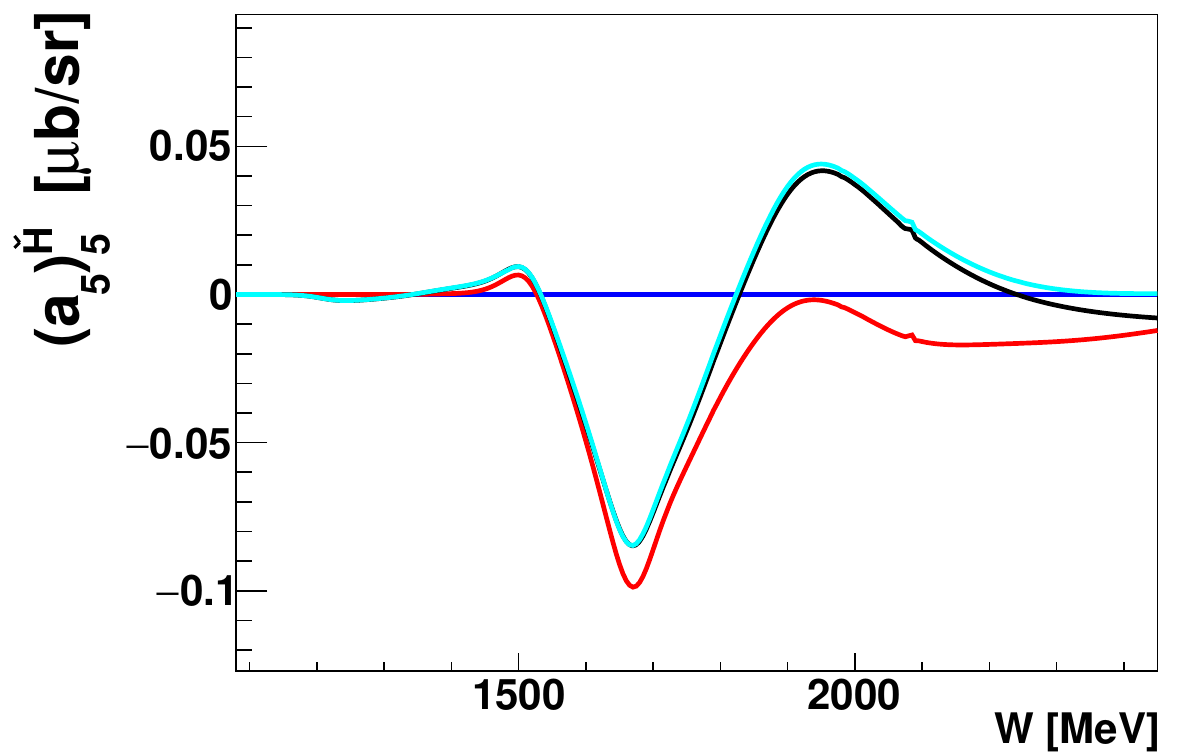}\end{minipage}
\begin{minipage}{.25\linewidth} \begin{align} \left(a_{5}\right)^{\check{H}}_{5} &= \left<S,H\right> + \left<P,G\right> \nonumber \\ & \hspace*{12.5pt} + \left<D,F\right>  + \left<D,H\right>   \nonumber  \\ & \hspace*{12.5pt} + \left<F,G\right>  + \left<G,H\right>   \nonumber\end{align} \end{minipage}

\begin{minipage}{.075\linewidth}
\vspace*{-6.5pt}
\hspace*{5pt}
\begin{equation}
\mathcal{C}_{6}^{\check{H}} \equiv \nonumber
\end{equation}
\end{minipage}
\begin{minipage}{.3\linewidth} \vspace*{0.572cm} \includegraphics[width=0.875\textwidth]{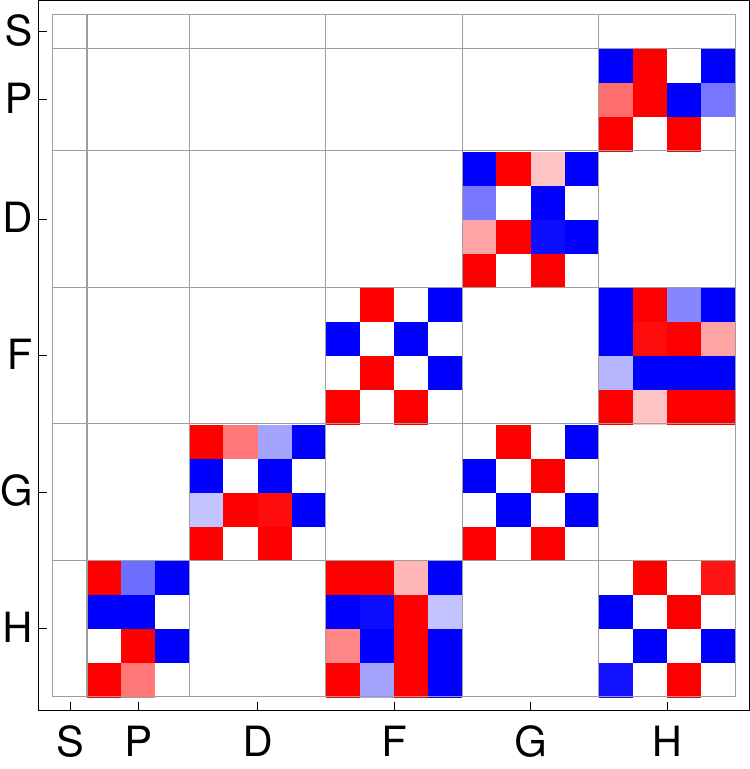} \end{minipage}
\begin{minipage}{.35\linewidth} \vspace*{0.500cm} \hspace*{-0.65cm}\includegraphics[width=1.15\textwidth]{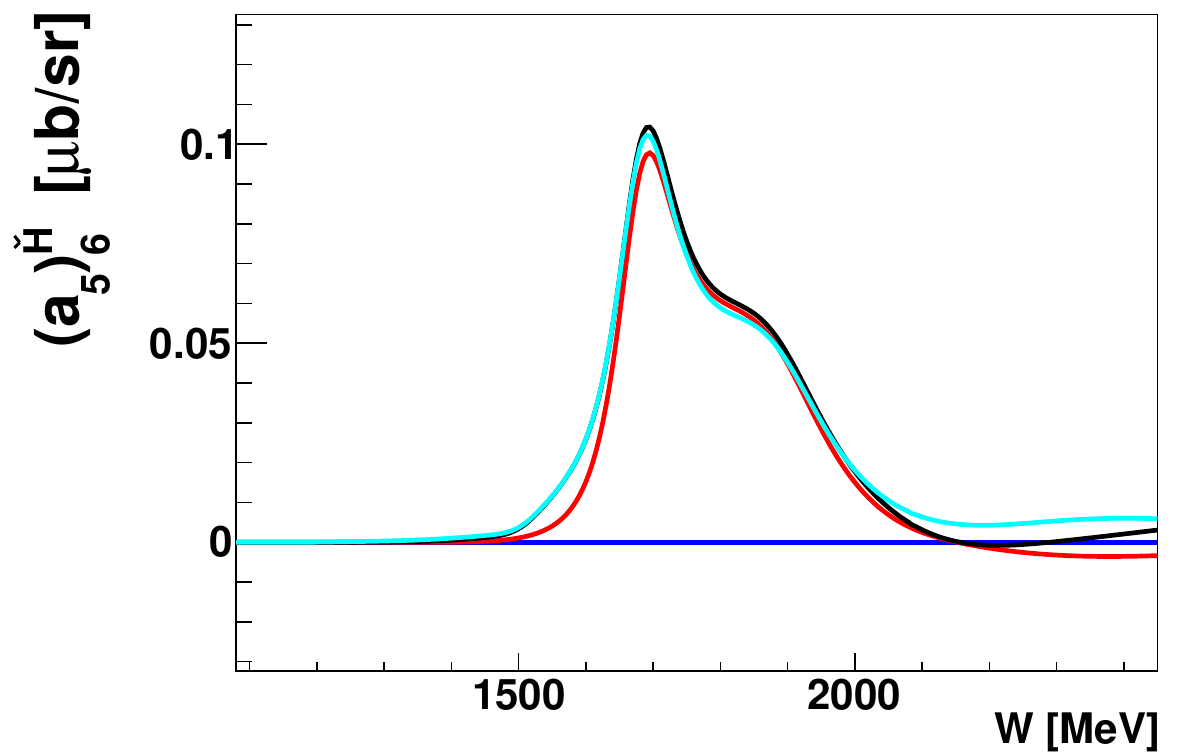}\end{minipage}
\begin{minipage}{.25\linewidth} \begin{align} \left(a_{5}\right)^{\check{H}}_{6} &= \left<P,H\right> + \left<D,G\right> \nonumber \\ & \hspace*{12.5pt} + \left<F,F\right>  + \left<F,H\right>   \nonumber  \\ & \hspace*{12.5pt} + \left<G,G\right>  + \left<H,H\right>   \nonumber   \end{align} \end{minipage}

\begin{minipage}{.075\linewidth}
\vspace*{-6.5pt}
\hspace*{5pt}
\begin{equation}
\mathcal{C}_{7}^{\check{H}} \equiv \nonumber
\end{equation}
\end{minipage}
\begin{minipage}{.3\linewidth} \vspace*{0.572cm} \includegraphics[width=0.875\textwidth]{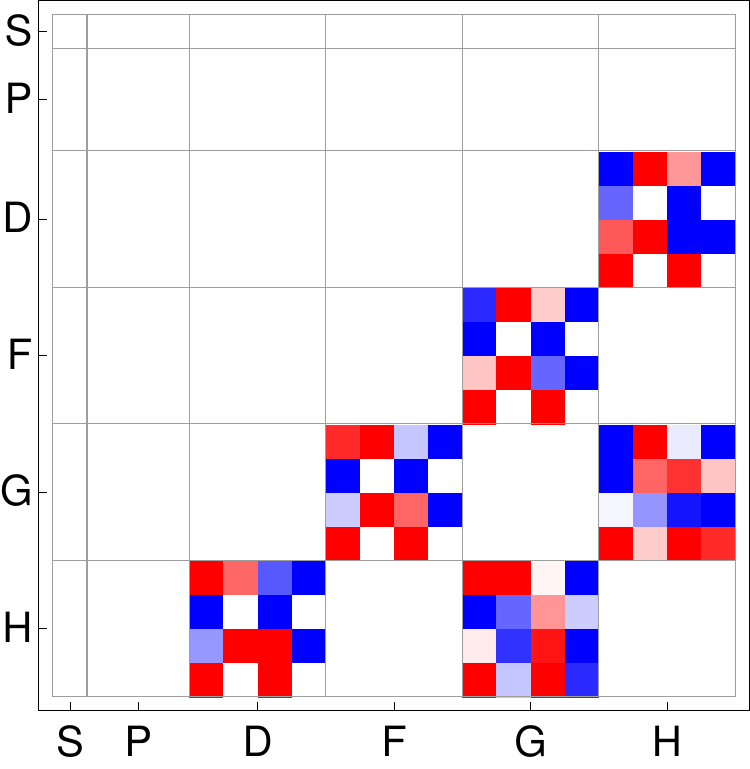} \end{minipage}
\begin{minipage}{.35\linewidth} \vspace*{0.500cm} \hspace*{-0.65cm}\includegraphics[width=1.15\textwidth]{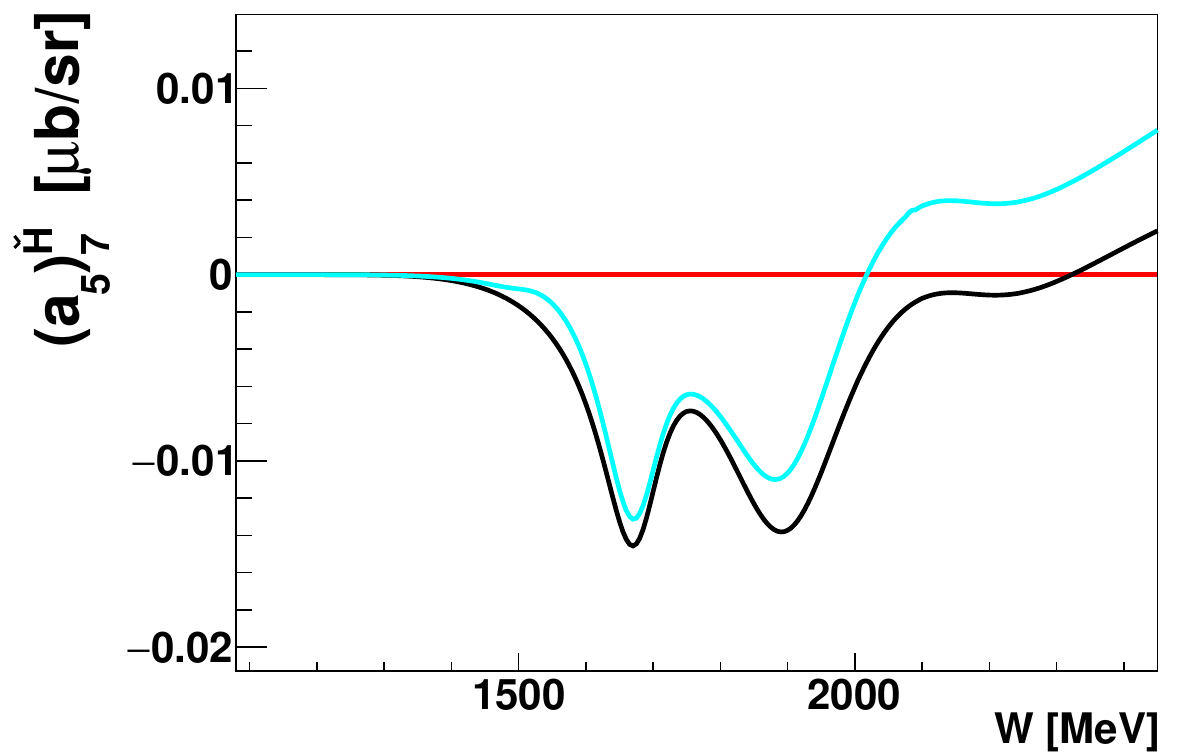}\end{minipage}
\begin{minipage}{.25\linewidth} \begin{align} \left(a_{5}\right)^{\check{H}}_{7} &= \left<D,H\right> + \left<F,G\right> \nonumber \\ & \hspace*{12.5pt} + \left<G,H\right>    \nonumber  \end{align} \end{minipage}

\begin{minipage}{.075\linewidth}
\vspace*{-6.5pt}
\hspace*{5pt}
\begin{equation}
\mathcal{C}_{8}^{\check{H}} \equiv \nonumber
\end{equation}
\end{minipage}
\begin{minipage}{.3\linewidth} \vspace*{0.572cm} \includegraphics[width=0.875\textwidth]{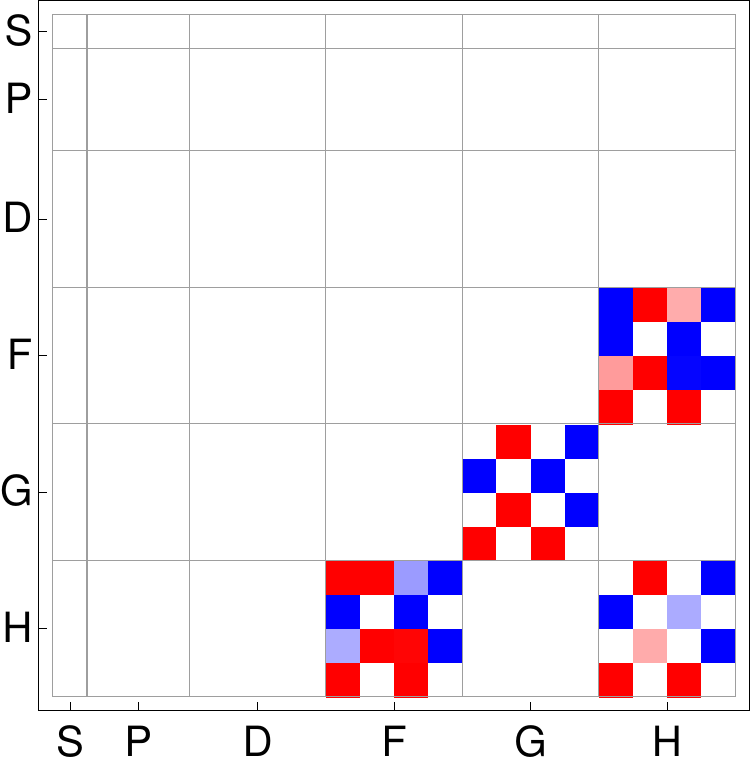} \end{minipage}
\begin{minipage}{.35\linewidth} \vspace*{0.500cm} \hspace*{-0.65cm}\includegraphics[width=1.15\textwidth]{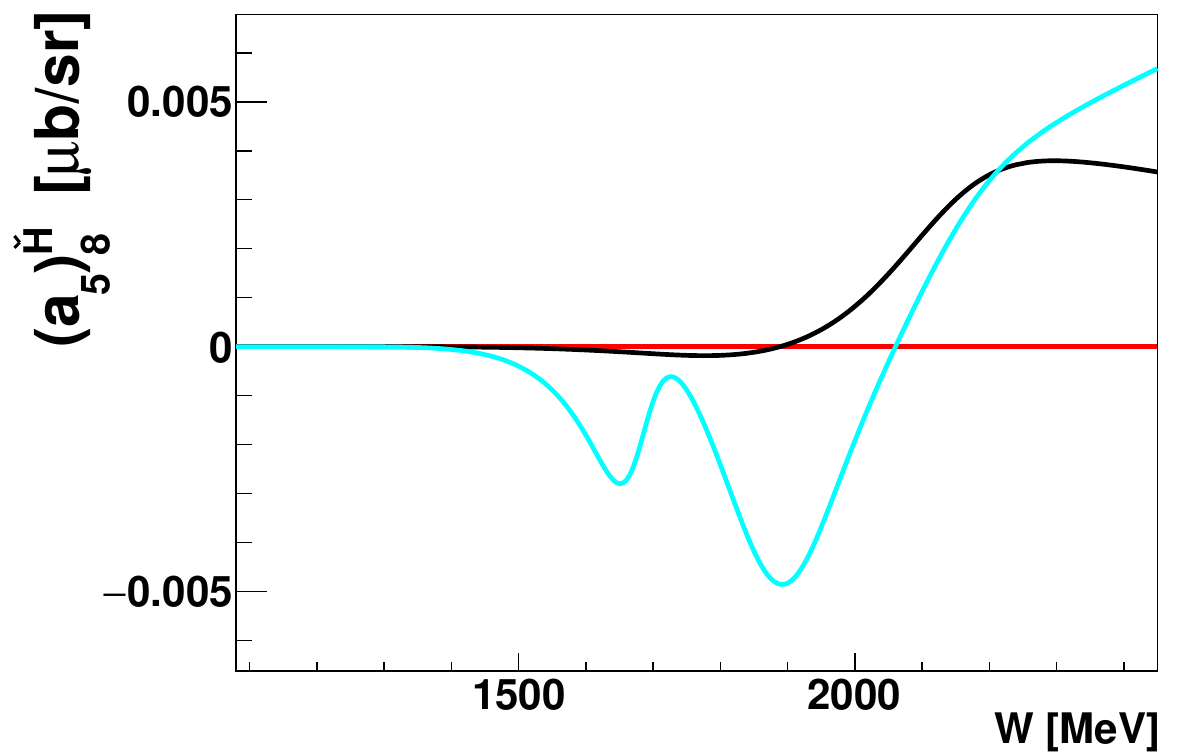}\end{minipage}
\begin{minipage}{.25\linewidth} \begin{align} \left(a_{5}\right)^{\check{H}}_{8} &= \left<F,H\right> + \left<G,G\right> \nonumber \\ & \hspace*{12.5pt} + \left<H,H\right>    \nonumber  \end{align} \end{minipage}
\caption{%
Left: Matrices $\mathcal{C}_{5\cdots 8}^{\check{H}}$, represented here in the color scheme, defines the coefficient $\left(a_{5}\right)_{5\cdots 8}^{\check{H}}$ for an expansion of $\check{H}$ up to $\text{L}_{\text{max}} = 5$. Center: For the higher non-fitted coefficients $\left(a_{5}\right)_{5 \ldots 8}^{\check{H}}$, the Bonn Gatchina curves are shown (here, the truncation at $\text{L}_{\mathrm{max}} = 5$ is drawn in cyan). Right: All partial wave interferences for $\text{L}_{\text{max}} = 5$ are indicated.
}
\label{tab:HColorPlots2}
\end{table*}
 
\begin{table*}[htb]
\RawFloats
\begin{minipage}{.075\linewidth}
\vspace*{-6.5pt}
\hspace*{5pt}
\begin{equation}
\mathcal{C}_{9}^{\check{H}} \equiv \nonumber
\end{equation}
\end{minipage}
\begin{minipage}{.3\linewidth} \vspace*{0.572cm} \includegraphics[width=0.875\textwidth]{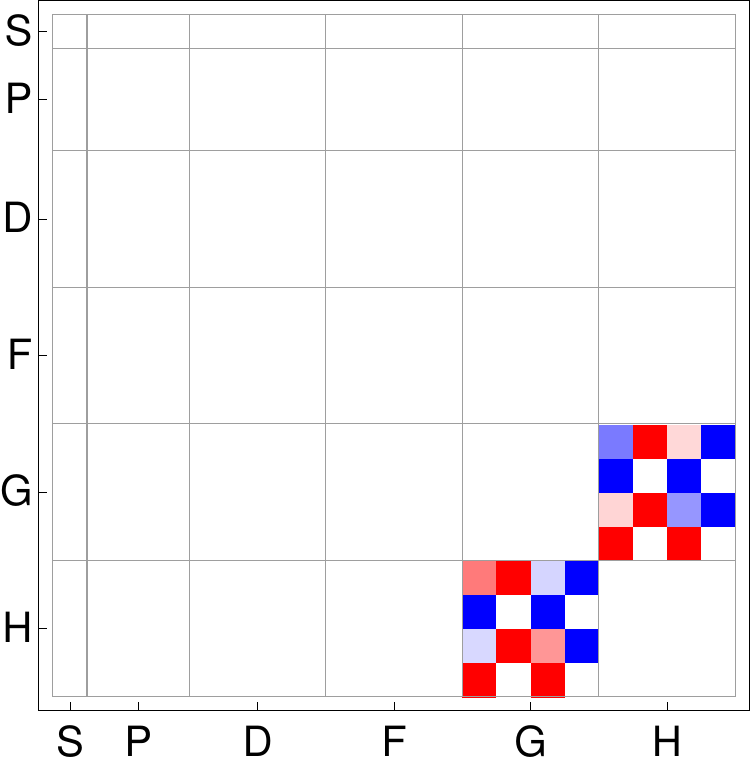} \end{minipage}
\begin{minipage}{.35\linewidth} \vspace*{0.500cm} \hspace*{-0.65cm}\includegraphics[width=1.15\textwidth]{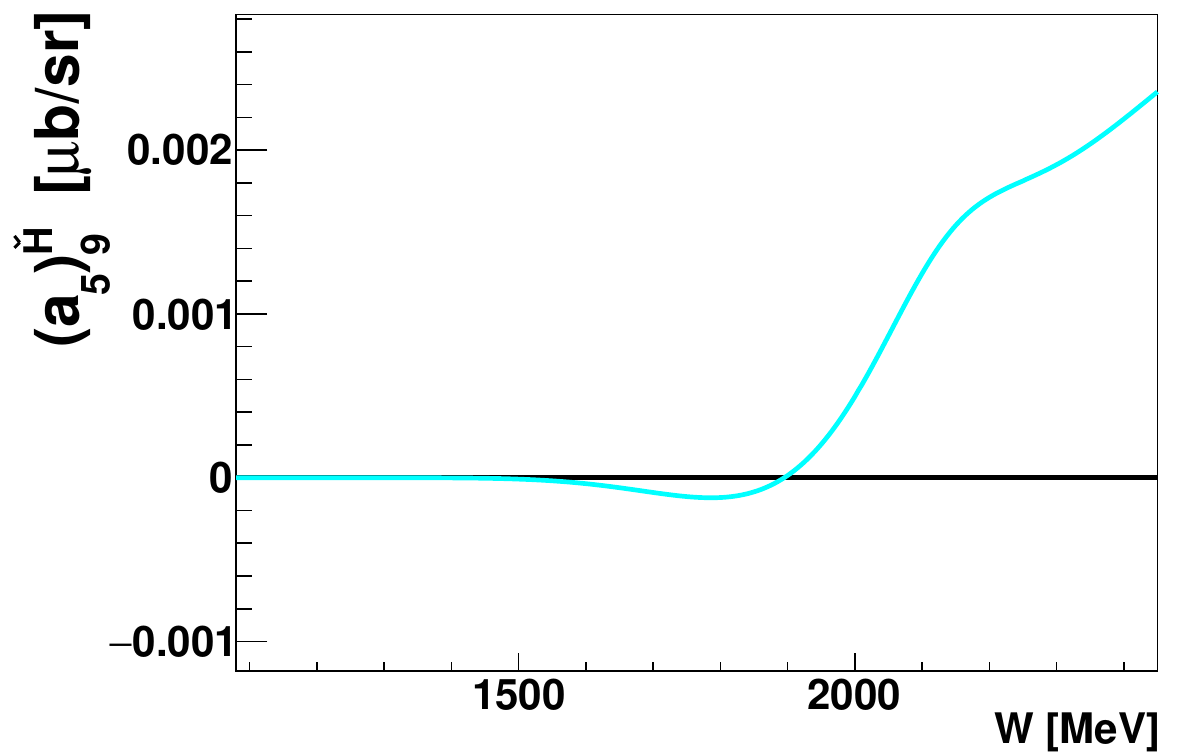}\end{minipage}
\begin{minipage}{.25\linewidth} \begin{align} \left(a_{5}\right)^{\check{H}}_{9} &= \left<G,H\right> \nonumber\end{align} \end{minipage}

\begin{minipage}{.075\linewidth}
\vspace*{-6.5pt}
\hspace*{5pt}
\begin{equation}
\mathcal{C}_{10}^{\check{H}} \equiv \nonumber
\end{equation}
\end{minipage}
\begin{minipage}{.3\linewidth} \vspace*{0.572cm} \includegraphics[width=0.875\textwidth]{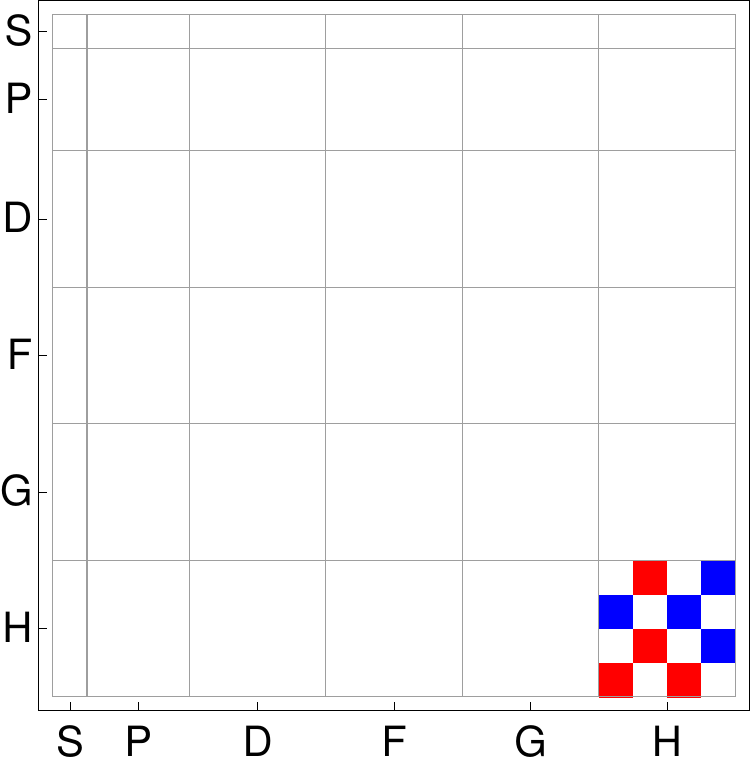} \end{minipage}
\begin{minipage}{.35\linewidth} \vspace*{0.500cm} \hspace*{-0.65cm}\includegraphics[width=1.15\textwidth]{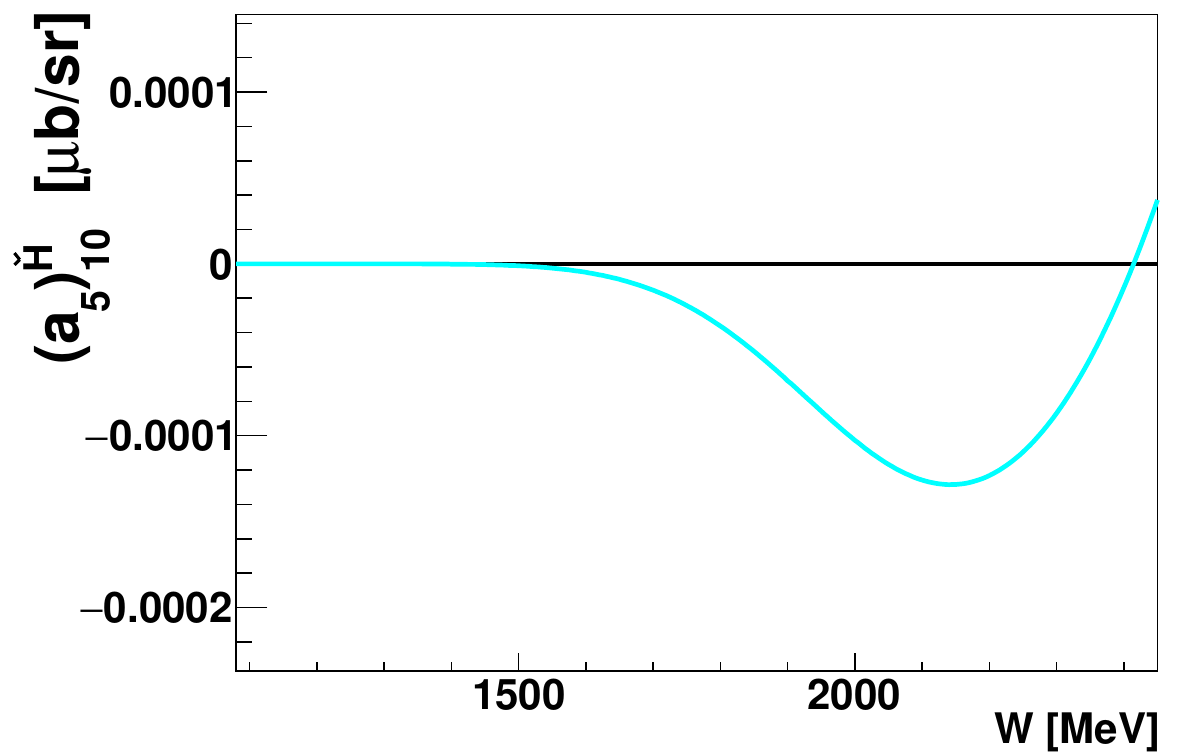}\end{minipage}
\begin{minipage}{.25\linewidth} \begin{align} \left(a_{5}\right)^{\check{H}}_{10} &=  \left<H,H\right>   \nonumber   \end{align} \end{minipage}
\caption{%
Left: Matrices $\mathcal{C}_{9, 10}^{\check{H}}$, represented here in the color scheme, defines the coefficient $\left(a_{5}\right)_{9, 10}^{\check{H}}$ for an expansion of $\check{H}$ up to $\text{L}_{\text{max}} = 5$. Center: For the higher non-fitted coefficients $\left(a_{5}\right)_{9, 10}^{\check{H}}$, the Bonn Gatchina curves are shown (here, the truncation at $\text{L}_{\mathrm{max}} = 5$ is drawn in cyan). Right: All partial wave interferences for $\text{L}_{\text{max}} = 5$ are indicated.
}
\label{tab:HColorPlots3}
\end{table*}
%

\begin{table*}[htb]
\RawFloats
\begin{minipage}{.075\linewidth}
\vspace*{-6.5pt}
\hspace*{5pt}
\begin{equation}
\mathcal{C}_{1}^{\check{F}} \equiv \nonumber
\end{equation}
\end{minipage}
\begin{minipage}{.3\linewidth} \vspace*{0.572cm} \includegraphics[width=0.875\textwidth]{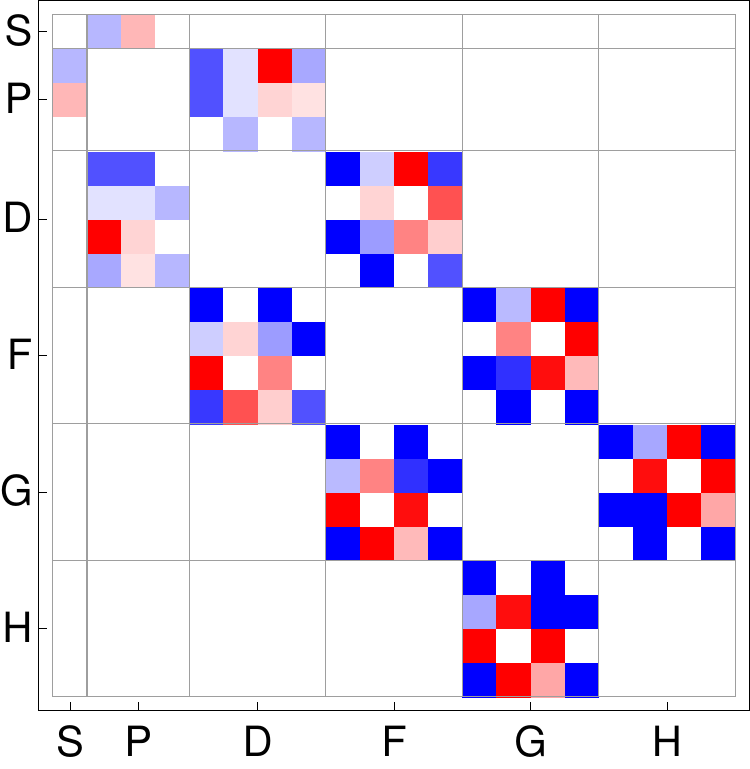} \end{minipage}
\begin{minipage}{.35\linewidth} \vspace*{0.500cm} \hspace*{-0.65cm}\includegraphics[width=1.15\textwidth]{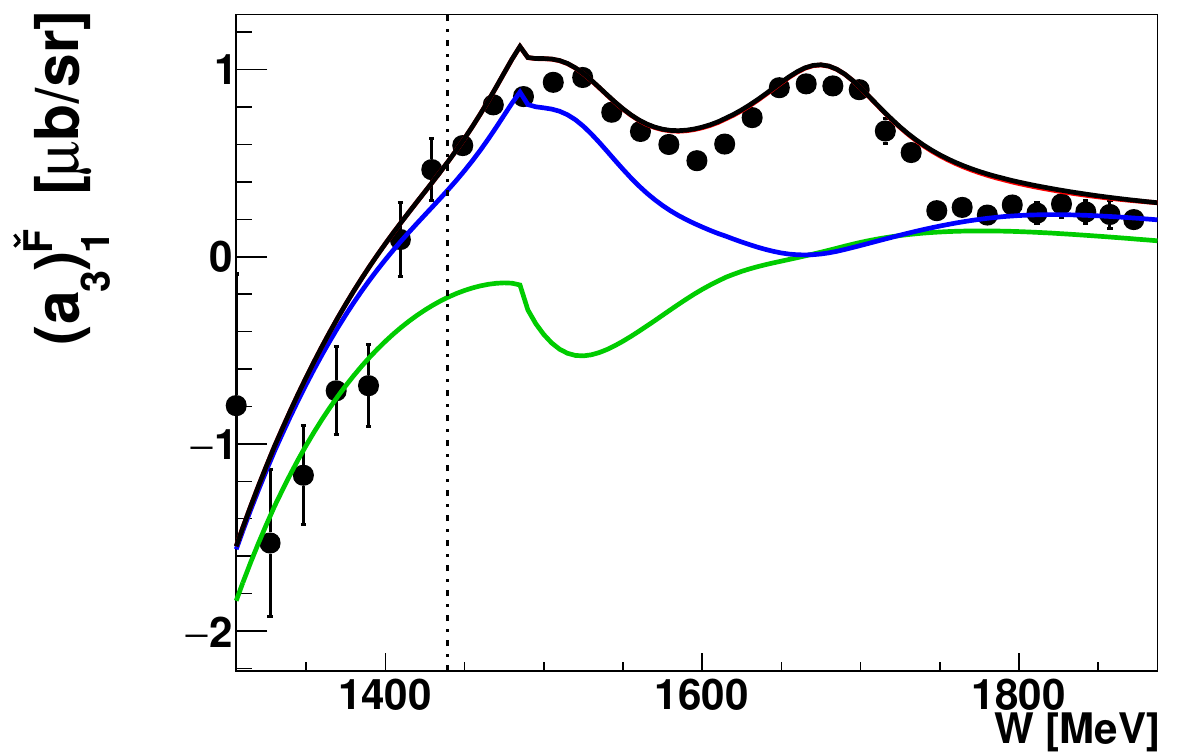}\end{minipage}
\begin{minipage}{.25\linewidth} \begin{align} \left(a_{5}\right)^{\check{F}}_{1} &= \left<S,P\right> + \left<P,D\right> \nonumber \\ & \hspace*{12.5pt} + \left<D,F\right> + \left<F,G\right> \nonumber \\ & \hspace*{12.5pt} + \left<G,H\right>   \nonumber  \end{align} \end{minipage}

\begin{minipage}{.075\linewidth}
\vspace*{-6.5pt}
\hspace*{5pt}
\begin{equation}
\mathcal{C}_{2}^{\check{F}} \equiv \nonumber
\end{equation}
\end{minipage}
\begin{minipage}{.3\linewidth} \vspace*{0.572cm} \includegraphics[width=0.875\textwidth]{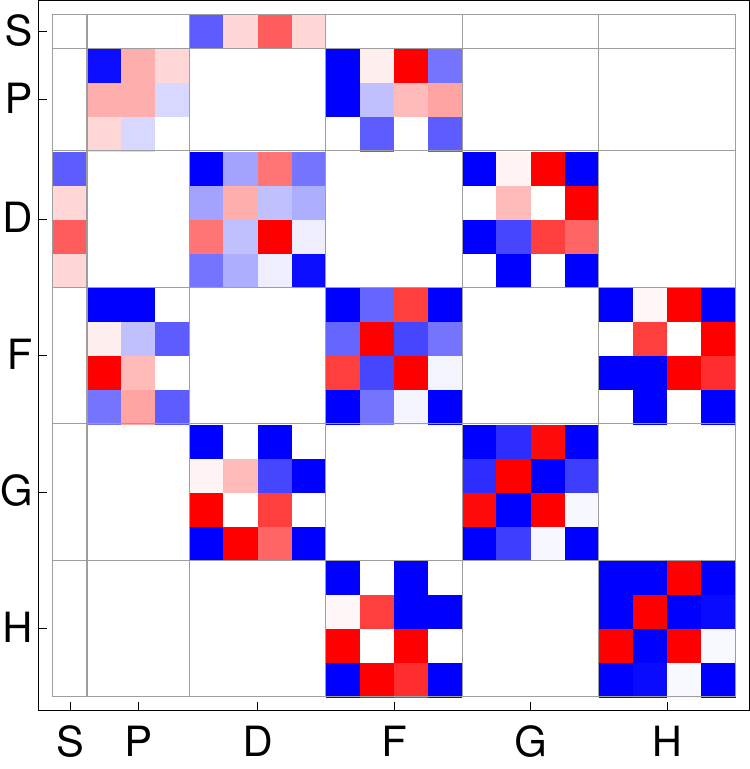} \end{minipage}
\begin{minipage}{.35\linewidth} \vspace*{0.500cm} \hspace*{-0.65cm}\includegraphics[width=1.15\textwidth]{F_l3_coeff_1.pdf}\end{minipage}
\begin{minipage}{.25\linewidth} \begin{align} \left(a_{5}\right)^{\check{F}}_{2} &= \left<S,D\right> + \left<P,P\right> \nonumber \\ & \hspace*{12.5pt} + \left<P,F\right>  + \left<D,D\right>   \nonumber \\ & \hspace*{12.5pt} + \left<D,G\right>  + \left<F,F\right>   \nonumber  \\ & \hspace*{12.5pt} + \left<F,H\right>  + \left<G,G\right>   \nonumber \\ & \hspace*{12.5pt} + \left<H,H\right>    \nonumber  \end{align} \end{minipage}

\begin{minipage}{.075\linewidth}
\vspace*{-6.5pt}
\hspace*{5pt}
\begin{equation}
\mathcal{C}_{3}^{\check{F}} \equiv \nonumber
\end{equation}
\end{minipage}
\begin{minipage}{.3\linewidth} \vspace*{0.572cm} \includegraphics[width=0.875\textwidth]{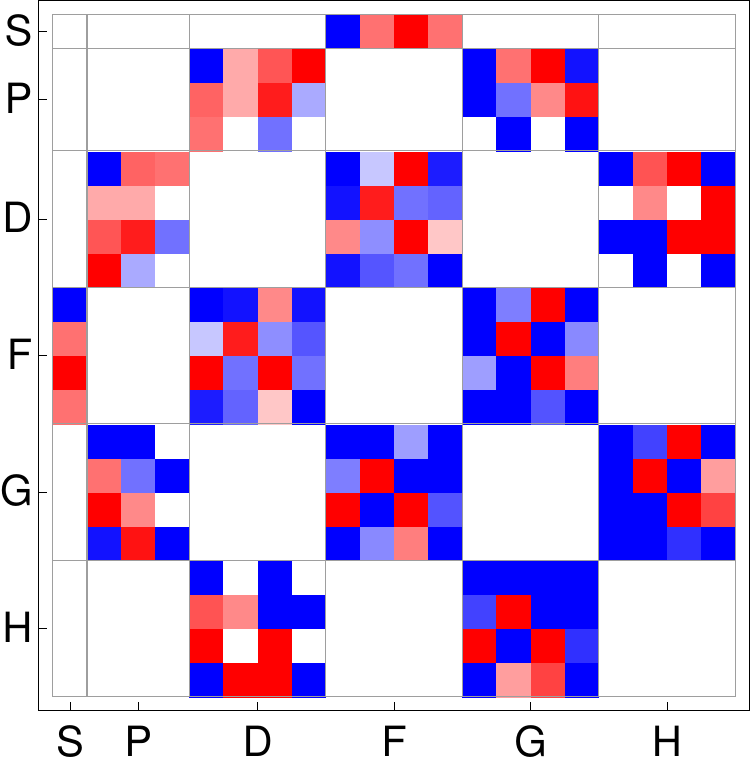} \end{minipage}
\begin{minipage}{.35\linewidth} \vspace*{0.500cm} \hspace*{-0.65cm}\includegraphics[width=1.15\textwidth]{F_l3_coeff_2.pdf}\end{minipage}
\begin{minipage}{.25\linewidth} \begin{align} \left(a_{5}\right)^{\check{F}}_{3} &= \left<S,F\right> + \left<P,D\right> \nonumber \\ & \hspace*{12.5pt} + \left<P,G\right>  + \left<D,F\right>   \nonumber  \\ & \hspace*{12.5pt} + \left<D,H\right>  + \left<F,G\right>   \nonumber \\ & \hspace*{12.5pt} + \left<G,H\right>   \nonumber   \end{align} \end{minipage}

\begin{minipage}{.075\linewidth}
\vspace*{-6.5pt}
\hspace*{5pt}
\begin{equation}
\mathcal{C}_{4}^{\check{F}} \equiv \nonumber
\end{equation}
\end{minipage}
\begin{minipage}{.3\linewidth} \vspace*{0.572cm} \includegraphics[width=0.875\textwidth]{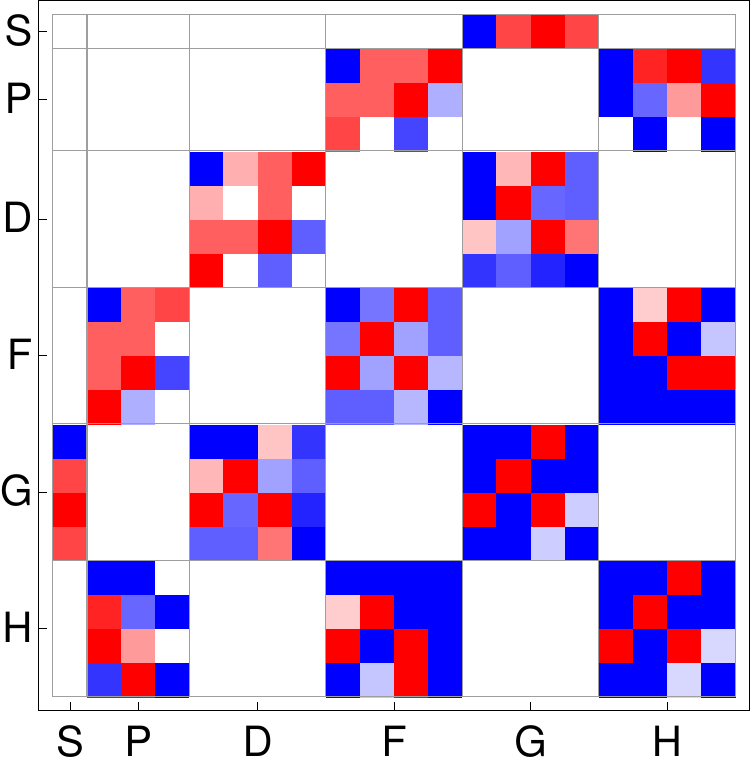} \end{minipage}
\begin{minipage}{.35\linewidth} \vspace*{0.500cm} \hspace*{-0.65cm}\includegraphics[width=1.15\textwidth]{F_l3_coeff_3.pdf}\end{minipage}
\begin{minipage}{.25\linewidth} \begin{align} \left(a_{5}\right)^{\check{F}}_{4} &= \left<S,G\right> + \left<P,F\right> \nonumber \\ & \hspace*{12.5pt} + \left<P,H\right>  + \left<D,D\right>   \nonumber   \\ & \hspace*{12.5pt} + \left<D,G\right>  + \left<F,F\right>   \nonumber \\ & \hspace*{12.5pt} + \left<F,H\right>  + \left<G,G\right>   \nonumber \\ & \hspace*{12.5pt} + \left<H,H\right>    \nonumber  \end{align} \end{minipage}
\caption{%
Left: Matrices $\mathcal{C}_{1\cdots 4}^{\check{F}}$, represented here in the color scheme, defines the coefficient $\left(a_{5}\right)_{1\cdots 4}^{\check{F}}$ for an expansion of $\check{F}$ up to $\text{L}_{\text{max}} = 5$. Center: Coefficients $\left(a_{3}\right)_{1 \cdots 4}^{\check{F}}$ obtained from a fit to the $\check{F}$-data (black points). For references to the data see Table \ref{tab:DataBasis}. Bonn Gatchina predictions, truncated at different $\text{L}_{\mathrm{max}}$ ($\text{L}_{\mathrm{max}} = 1$ is drawn in green, $\text{L}_{\mathrm{max}} = 2$ in blue, $\text{L}_{\mathrm{max}} = 3$ in red and $\text{L}_{\mathrm{max}} = 4$ in black) are drawn as well. Right: All partial wave interferences for $\text{L}_{\text{max}} = 5$ are indicated.
}
\label{tab:FColorPlots1}
\end{table*}

\begin{table*}[htb]
\RawFloats
\begin{minipage}{.075\linewidth}
\vspace*{-6.5pt}
\hspace*{5pt}
\begin{equation}
\mathcal{C}_{5}^{\check{F}} \equiv \nonumber
\end{equation}
\end{minipage}
\begin{minipage}{.3\linewidth} \vspace*{0.572cm} \includegraphics[width=0.875\textwidth]{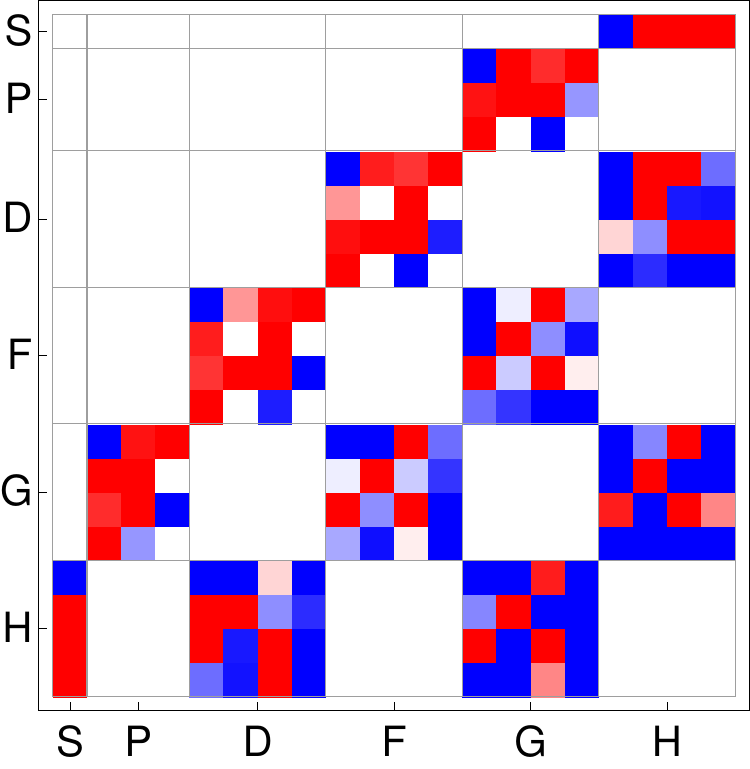} \end{minipage}
\begin{minipage}{.35\linewidth} \vspace*{0.500cm} \hspace*{-0.65cm}\includegraphics[width=1.15\textwidth]{F_l3_coeff_4.pdf}\end{minipage}
\begin{minipage}{.25\linewidth} \begin{align} \left(a_{5}\right)^{\check{F}}_{5} &= \left<S,H\right> + \left<P,G\right> \nonumber \\ & \hspace*{12.5pt} + \left<D,F\right>  + \left<D,H\right>   \nonumber  \\ & \hspace*{12.5pt} + \left<F,G\right>  + \left<G,H\right>   \nonumber\end{align} \end{minipage}

\begin{minipage}{.075\linewidth}
\vspace*{-6.5pt}
\hspace*{5pt}
\begin{equation}
\mathcal{C}_{6}^{\check{F}} \equiv \nonumber
\end{equation}
\end{minipage}
\begin{minipage}{.3\linewidth} \vspace*{0.572cm} \includegraphics[width=0.875\textwidth]{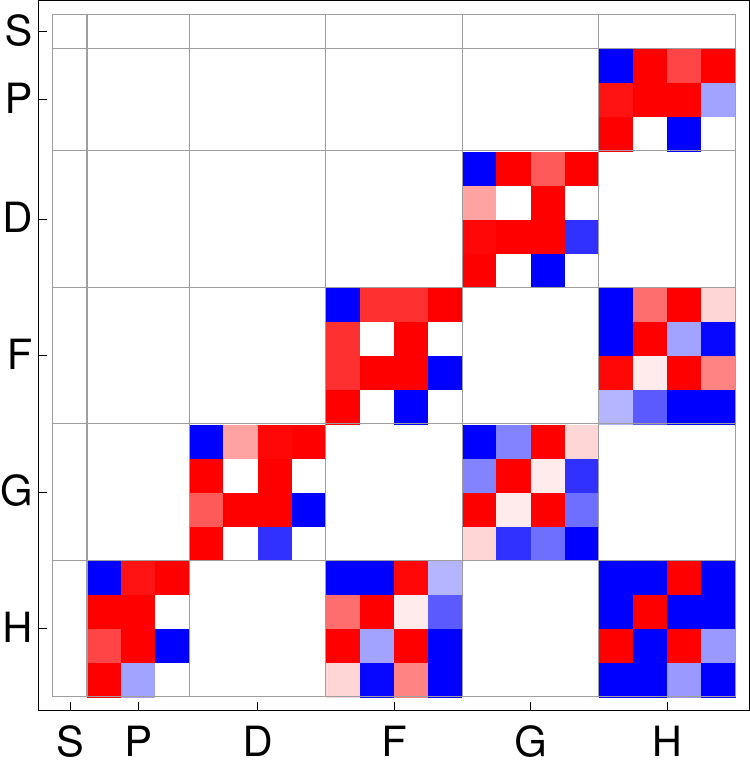} \end{minipage}
\begin{minipage}{.35\linewidth} \vspace*{0.500cm} \hspace*{-0.65cm}\includegraphics[width=1.15\textwidth]{F_l3_coeff_5.pdf}\end{minipage}
\begin{minipage}{.25\linewidth} \begin{align} \left(a_{5}\right)^{\check{F}}_{6} &= \left<P,H\right> + \left<D,G\right> \nonumber \\ & \hspace*{12.5pt} + \left<F,F\right>  + \left<F,H\right>   \nonumber  \\ & \hspace*{12.5pt} + \left<G,G\right>  + \left<H,H\right>   \nonumber   \end{align} \end{minipage}

\begin{minipage}{.075\linewidth}
\vspace*{-6.5pt}
\hspace*{5pt}
\begin{equation}
\mathcal{C}_{7}^{\check{F}} \equiv \nonumber
\end{equation}
\end{minipage}
\begin{minipage}{.3\linewidth} \vspace*{0.572cm} \includegraphics[width=0.875\textwidth]{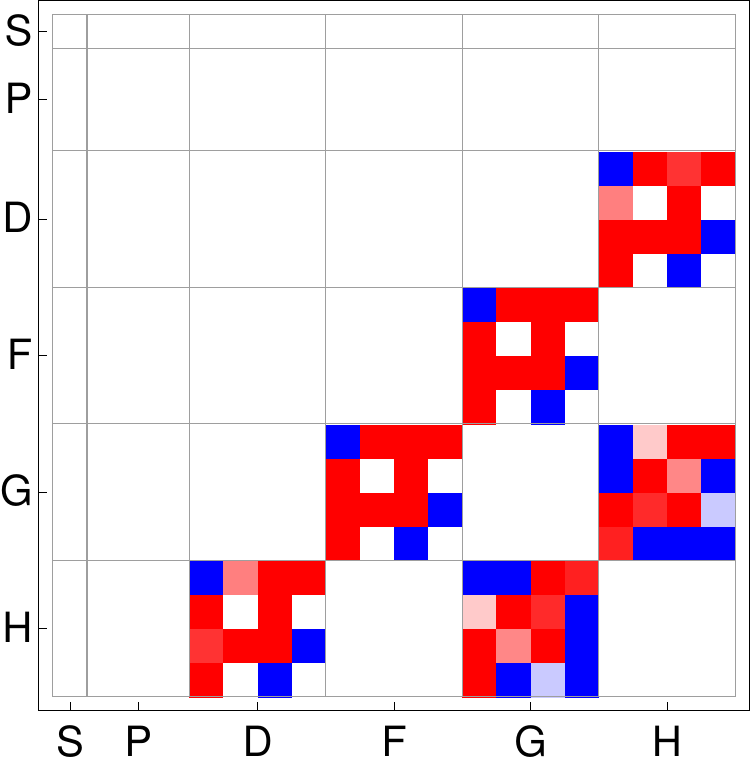} \end{minipage}
\begin{minipage}{.35\linewidth} \vspace*{0.500cm} \hspace*{-0.65cm}\includegraphics[width=1.15\textwidth]{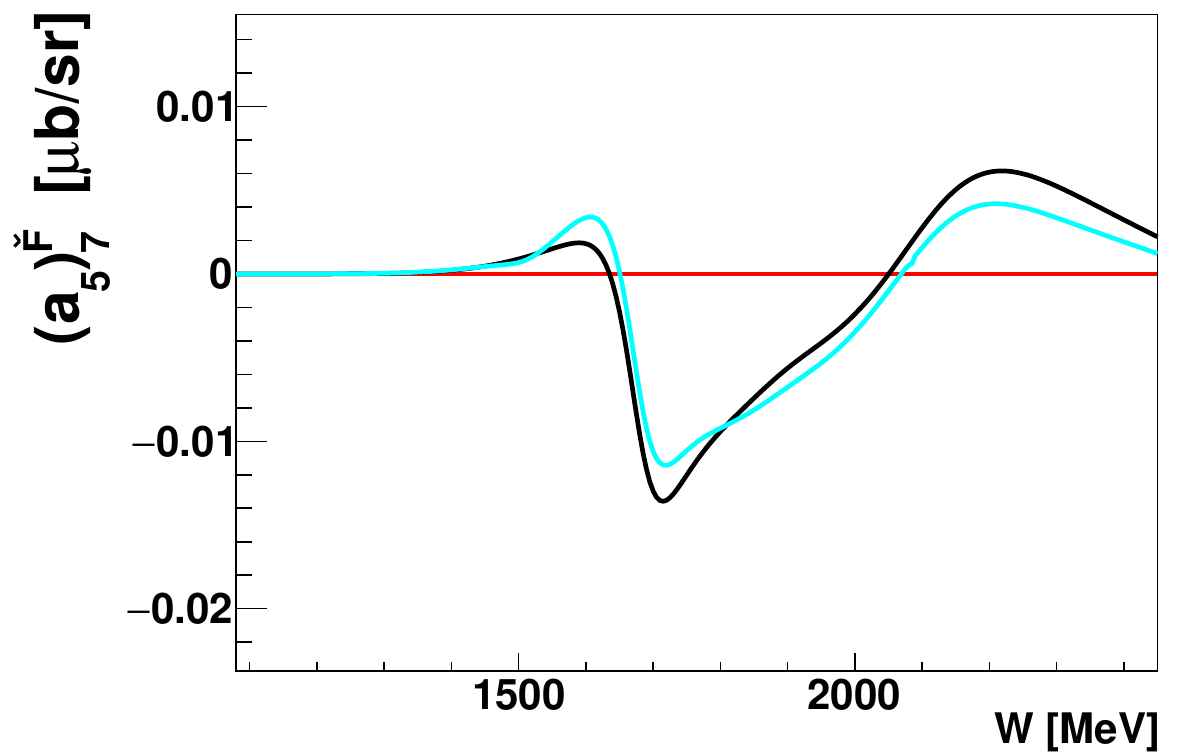}\end{minipage}
\begin{minipage}{.25\linewidth} \begin{align} \left(a_{5}\right)^{\check{F}}_{7} &= \left<D,H\right> + \left<F,G\right> \nonumber \\ & \hspace*{12.5pt} + \left<G,H\right>    \nonumber  \end{align} \end{minipage}

\begin{minipage}{.075\linewidth}
\vspace*{-6.5pt}
\hspace*{5pt}
\begin{equation}
\mathcal{C}_{8}^{\check{F}} \equiv \nonumber
\end{equation}
\end{minipage}
\begin{minipage}{.3\linewidth} \vspace*{0.572cm} \includegraphics[width=0.875\textwidth]{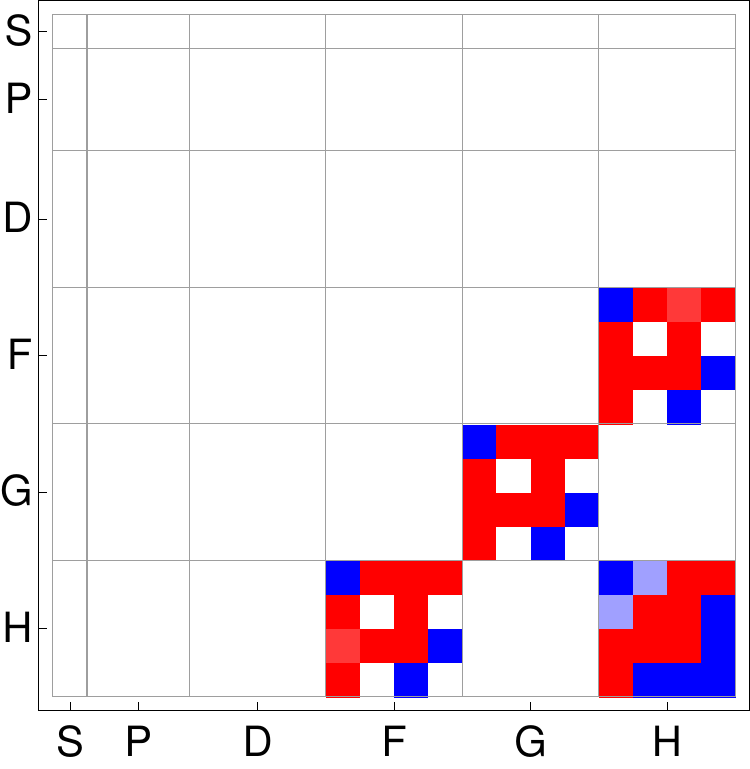} \end{minipage}
\begin{minipage}{.35\linewidth} \vspace*{0.500cm} \hspace*{-0.65cm}\includegraphics[width=1.15\textwidth]{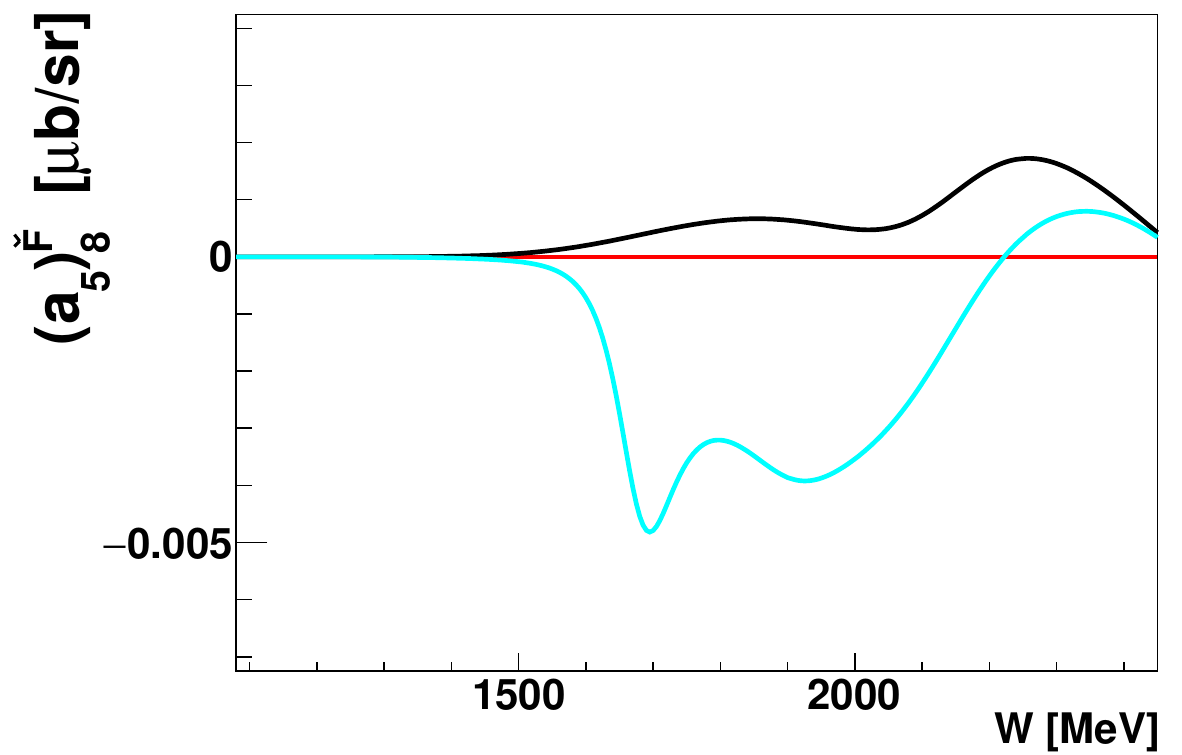}\end{minipage}
\begin{minipage}{.25\linewidth} \begin{align} \left(a_{5}\right)^{\check{F}}_{8} &= \left<F,H\right> + \left<G,G\right> \nonumber \\ & \hspace*{12.5pt} + \left<H,H\right>    \nonumber  \end{align} \end{minipage}
\caption{%
Left: Matrices $\mathcal{C}_{5\cdots 8}^{\check{F}}$, represented here in the color scheme, defines the coefficient $\left(a_{5}\right)_{5\cdots 8}^{\check{F}}$ for an expansion of $\check{F}$ up to $\text{L}_{\text{max}} = 5$. Center: Coefficients $\left(a_{3}\right)_{5, 6}^{\check{F}}$ obtained from a fit to the $\check{F}$-data (black points). For references to the data see Table \ref{tab:DataBasis}. Bonn Gatchina predictions, truncated at different $\text{L}_{\mathrm{max}}$ ($\text{L}_{\mathrm{max}} = 1$ is drawn in green, $\text{L}_{\mathrm{max}} = 2$ in blue, $\text{L}_{\mathrm{max}} = 3$ in red and $\text{L}_{\mathrm{max}} = 4$ in black) are drawn as well. For the higher non-fitted coefficients $\left(a_{5}\right)_{7, 8}^{\check{F}}$, the Bonn Gatchina curves are shown (here, the truncation at $\text{L}_{\mathrm{max}} = 5$ is drawn in cyan). Right: All partial wave interferences for $\text{L}_{\text{max}} = 5$ are indicated.
}
\label{tab:FColorPlots2}
\end{table*}
 
\begin{table*}[htb]
\RawFloats
\begin{minipage}{.075\linewidth}
\vspace*{-6.5pt}
\hspace*{5pt}
\begin{equation}
\mathcal{C}_{9}^{\check{F}} \equiv \nonumber
\end{equation}
\end{minipage}
\begin{minipage}{.3\linewidth} \vspace*{0.572cm} \includegraphics[width=0.875\textwidth]{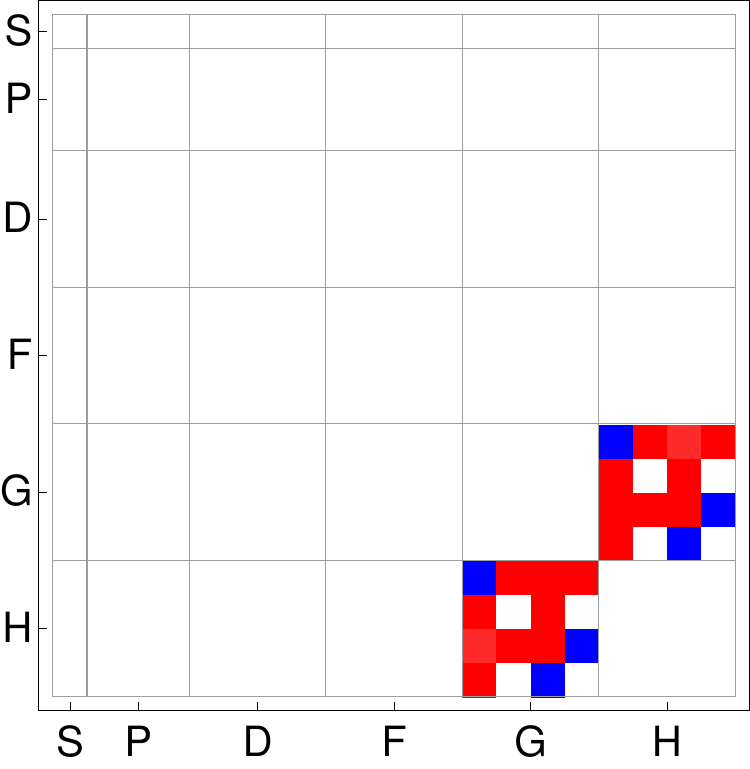} \end{minipage}
\begin{minipage}{.35\linewidth} \vspace*{0.500cm} \hspace*{-0.65cm}\includegraphics[width=1.15\textwidth]{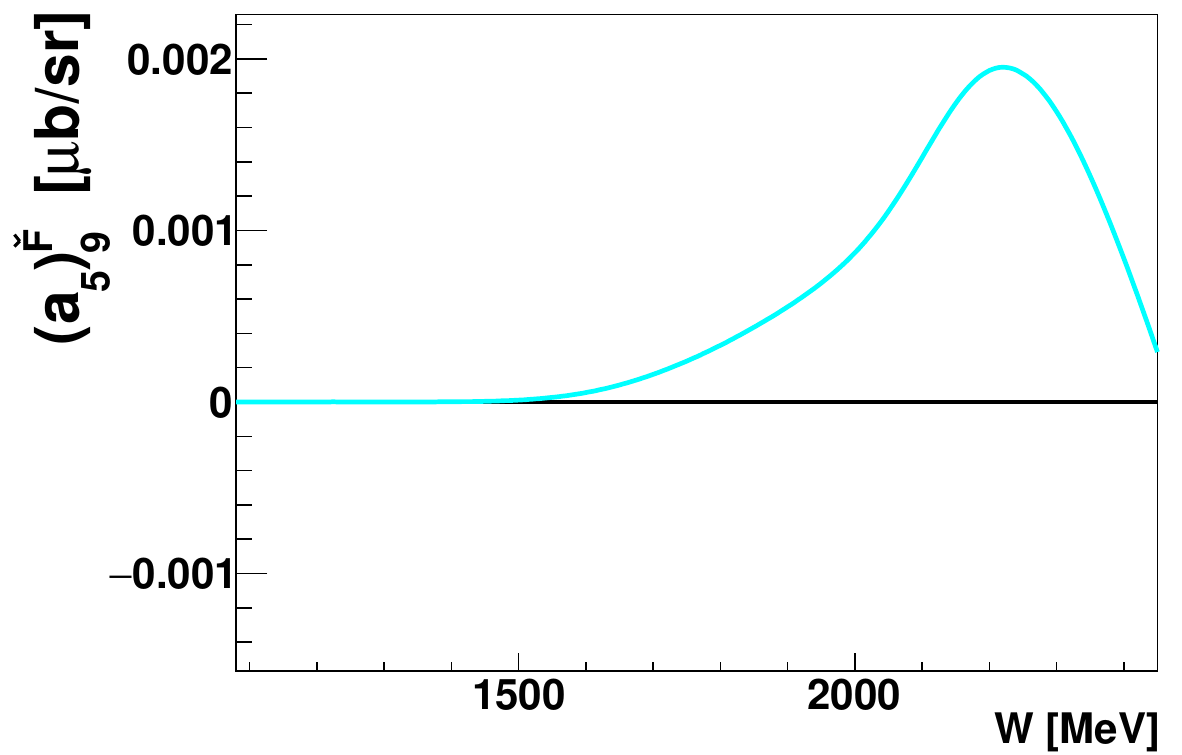}\end{minipage}
\begin{minipage}{.25\linewidth} \begin{align} \left(a_{5}\right)^{\check{F}}_{9} &= \left<G,H\right> \nonumber\end{align} \end{minipage}

\begin{minipage}{.075\linewidth}
\vspace*{-6.5pt}
\hspace*{5pt}
\begin{equation}
\mathcal{C}_{10}^{\check{F}} \equiv \nonumber
\end{equation}
\end{minipage}
\begin{minipage}{.3\linewidth} \vspace*{0.572cm} \includegraphics[width=0.875\textwidth]{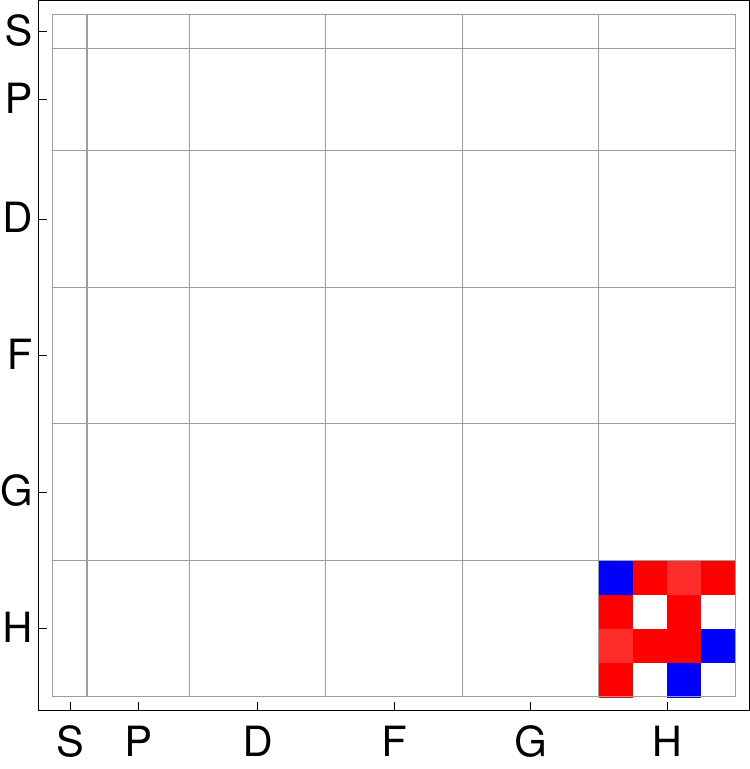} \end{minipage}
\begin{minipage}{.35\linewidth} \vspace*{0.500cm} \hspace*{-0.65cm}\includegraphics[width=1.15\textwidth]{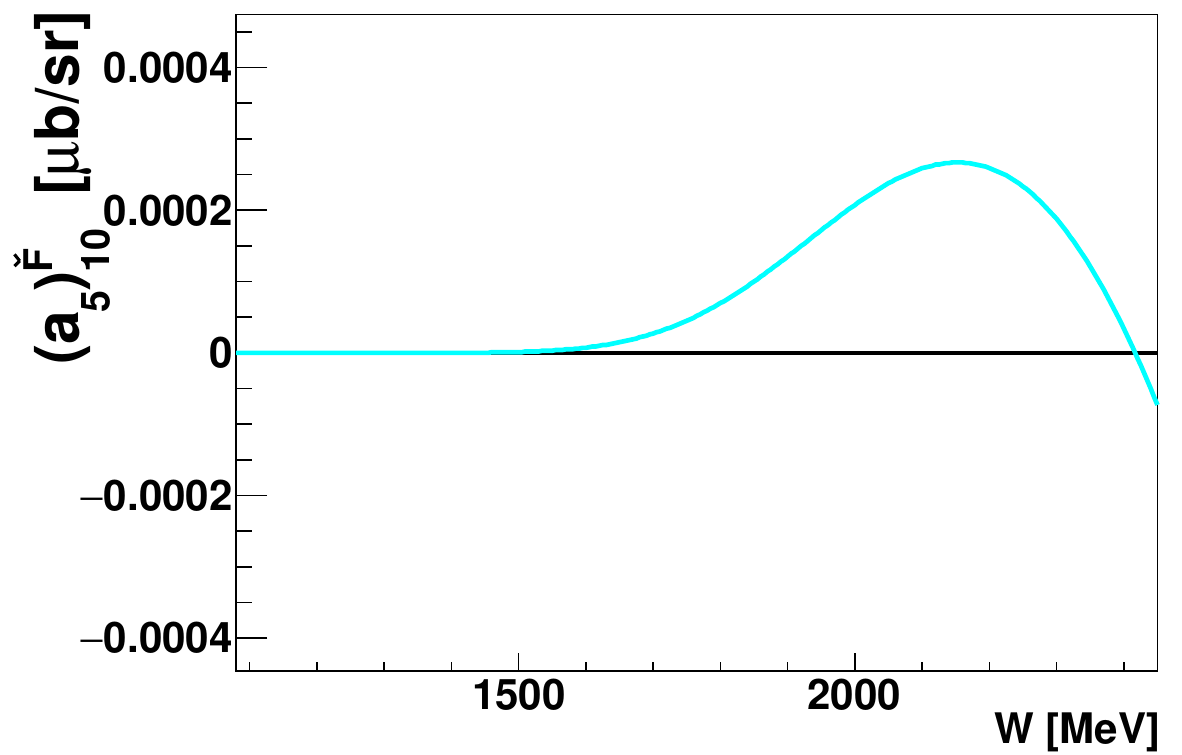}\end{minipage}
\begin{minipage}{.25\linewidth} \begin{align} \left(a_{5}\right)^{\check{F}}_{10} &=  \left<H,H\right>   \nonumber   \end{align} \end{minipage}
\caption{%
Left: Matrices $\mathcal{C}_{9, 10}^{\check{F}}$, represented here in the color scheme, defines the coefficient $\left(a_{5}\right)_{9, 10}^{\check{F}}$ for an expansion of $\check{F}$ up to $\text{L}_{\text{max}} = 5$. Center: For the higher non-fitted coefficients $\left(a_{5}\right)_{9, 10}^{\check{F}}$, the Bonn Gatchina curves are shown (here, the truncation at $\text{L}_{\mathrm{max}} = 5$ is drawn in cyan). Right: All partial wave interferences for $\text{L}_{\text{max}} = 5$ are indicated.
}
\label{tab:FColorPlots3}
\end{table*}
%
\clearpage
%

\begin{table*}
\RawFloats
\centering
\begin{scriptsize}
{ \fontsize{8.5}{11} \selectfont
\begin{tabular}{cc|c|cccccc|c}
\hline
\hline
Resonance & Status & Partial wave  & $ \ell_{\pi} $ & $I$ & $J$ & $ P$ & $ M^{\ast}_{\text{pole}} \left[ \mathrm{MeV} \right] $ & $ \Gamma_{\text{pole}} \left[ \mathrm{MeV} \right] $ & Multipoles \\
\hline
$ N \left( 939 \right) \frac{1}{2}^{+} $ & $ \ast \ast \ast \ast $ & $ P_{11} $ & $ 1 $ & $ 1/2 $ & $ 1/2 $ & $ + $ & $ 939 $ & - & $ M_{1-}^{\left( 1/2 \right)} $ \\
$ \mathbf{\Delta \left( 1232 \right) \frac{3}{2}^{+} }$ & $ \mathbf{\ast \ast \ast \ast }$ & $ \mathbf{P_{33}} $ & $ \mathbf{ 1 }$ & $ \mathbf{ 3/2 } $ & $ \mathbf{ 3/2 } $ & $ \mathbf{ + } $ & $ \mathbf{ 1210 } $ & $ \mathbf{ 100 } $ & $ \mathbf{ E_{1+}^{\left( 3/2 \right)} } $, $ \mathbf{ M_{1+}^{\left( 3/2 \right)} } $ \\
$ \mathbf{N \left( 1440 \right) \frac{1}{2}^{+} } $ & $ \mathbf{ \ast \ast \ast \ast } $  & $ \mathbf{ P_{11} } $ & $ \mathbf{ 1 } $ & $ \mathbf{ 1/2 } $ & $ \mathbf{ 1/2 } $ & $ \mathbf{ + } $ & $ \mathbf{ {\it 1365 } } $ & $ \mathbf{ {\it 183} } $ & $ \mathbf{ M_{1-}^{\left( 1/2 \right)} } $ \\
$\mathbf{ N \left( 1520 \right) \frac{3}{2}^{-} }$ & $ \mathbf{\ast \ast \ast \ast} $  & $\mathbf{ D_{13} }$ & $\mathbf{ 2 }$ & $\mathbf{ 1/2 }$ & $\mathbf{ 3/2 }$ & $\mathbf{ - }$ & $ \mathbf{ 1510 }$ & $ \mathbf{ 110 }$ & $ \mathbf{ E_{2-}^{\left( 1/2 \right)} }$, $ \mathbf{ M_{2-}^{\left( 1/2 \right)} }$ \\
$\mathbf{ N \left( 1535 \right) \frac{1}{2}^{-} }$ & $\mathbf{ \ast \ast \ast \ast }$  & $\mathbf{ S_{11} }$ & $\mathbf{ 0 }$ & $\mathbf{ 1/2 }$ & $\mathbf{ 1/2 }$ & $\mathbf{ - }$ & $ \mathbf{ 1510 }$ & $ \mathbf{ 170 }$ & $\mathbf{ E_{0+}^{\left( 1/2 \right)}} $ \\
$\mathbf{ \Delta \left( 1600 \right) \frac{3}{2}^{+} }$ & $\mathbf{ \ast \ast \ast }$  & $\mathbf{ P_{33} }$ & $\mathbf{ 1 }$ & $\mathbf{ 3/2 }$ & $\mathbf{ 3/2 }$ & $\mathbf{ + }$ & $ \mathbf{ 1510 }$ & $ \mathbf{ 275 }$ & $\mathbf{ E_{1+}^{\left( 3/2 \right)}} $, $\mathbf{ M_{1+}^{\left( 3/2 \right)}} $ \\
$\mathbf{ \Delta \left( 1620 \right) \frac{1}{2}^{-} }$ & $\mathbf{ \ast \ast \ast \ast }$  & $\mathbf{ S_{31} }$ & $\mathbf{ 0 }$ & $\mathbf{ 3/2 }$ & $\mathbf{ 1/2 }$ & $\mathbf{ - }$ & $ \mathbf{ 1600 }$ & $ \mathbf{ 130 }$ & $ \mathbf{E_{0+}^{\left( 3/2 \right)} }$ \\
$\mathbf{ N \left( 1650 \right) \frac{1}{2}^{-} }$ & $\mathbf{ \ast \ast \ast \ast }$  & $\mathbf{ S_{11} }$ & $\mathbf{ 0 }$ & $\mathbf{ 1/2 }$ & $\mathbf{ 1/2 }$ & $\mathbf{ - }$ & $ \mathbf{ 1655 }$ & $ \mathbf{ 135 }$ & $\mathbf{ E_{0+}^{\left( 1/2 \right)}} $ \\
$\mathbf{ N \left( 1675 \right) \frac{5}{2}^{-}} $ & $ \mathbf{\ast \ast \ast \ast }$  & $\mathbf{ D_{15}} $ & $ \mathbf{2} $ & $ \mathbf{1/2} $ & $ \mathbf{5/2} $ & $ \mathbf{-} $ & $ \mathbf{ 1660} $ & $ \mathbf{ 135} $ & $ \mathbf{E_{2+}^{\left( 1/2 \right)}, M_{2+}^{\left( 1/2 \right)}} $ \\
$\mathbf{ N \left( 1680 \right) \frac{5}{2}^{+} }$ & $\mathbf{ \ast \ast \ast \ast }$  & $\mathbf{ F_{15}} $ & $\mathbf{ 3} $ & $\mathbf{ 1/2 }$ & $ \mathbf{5/2} $ & $\mathbf{ +} $ & $ \mathbf{ 1675} $ & $ \mathbf{ 120 }$ & $\mathbf{ E_{3-}^{\left( 1/2 \right)}, M_{3-}^{\left( 1/2 \right)} }$ \\
$\mathbf{ N \left( 1700 \right) \frac{3}{2}^{-} }$ & $\mathbf{ \ast \ast \ast }$  & $\mathbf{ D_{13} }$ & $\mathbf{ 2 }$ & $\mathbf{ 1/2 }$ & $\mathbf{ 3/2 }$ & $\mathbf{ - }$ & $ \mathbf{ 1700 }$ & {\bf $\mathbf{100}$ to $\mathbf{ 300} $ } & $\mathbf{ E_{2-}^{\left( 1/2 \right)}, M_{2-}^{\left( 1/2 \right)} }$ \\
$ \mathbf{ \Delta \left( 1700 \right) \frac{3}{2}^{-} }$ & $\mathbf{ \ast \ast \ast \ast }$  & $\mathbf{ D_{33} }$ & $\mathbf{ 2 }$ & $\mathbf{ 3/2 }$ & $\mathbf{ 3/2 }$ & $\mathbf{ - }$ & $ \mathbf{ 1650 }$ & $ \mathbf{ 230 }$ & $ \mathbf{E_{2-}^{\left( 3/2 \right)}, M_{2-}^{\left( 3/2 \right)}} $ \\
$\mathbf{ N \left( 1710 \right) \frac{1}{2}^{+} }$ & $\mathbf{ \ast \ast \ast \ast }$  & $\mathbf{ P_{11} }$ & $\mathbf{ 1 }$ & $\mathbf{ 1/2 }$ & $\mathbf{ 1/2 }$ & $\mathbf{ + }$ & $ \mathbf{ 1720 }$ & $ \mathbf{ 230 }$ & $\mathbf{ M_{1-}^{\left( 1/2 \right)}} $ \\
$\mathbf{ N \left( 1720 \right) \frac{3}{2}^{+} }$ & $\mathbf{ \ast \ast \ast \ast }$  & $\mathbf{ P_{13} }$ & $\mathbf{ 1 }$ & $\mathbf{ 1/2 }$ & $\mathbf{ 3/2 }$ & $\mathbf{ + }$ & $ \mathbf{ 1675 }$ & $ \mathbf{ 250 }$ & $\mathbf{ E_{1+}^{\left( 1/2 \right)} }$, $\mathbf{ M_{1+}^{\left( 1/2 \right)} }$ \\
$ \Delta \left( 1750 \right) \frac{1}{2}^{+} $ & $ \ast $  & $ P_{31} $ & $ 1 $ & $ 3/2 $ & $ 1/2 $ & $ + $ & $\mathbf{{\it 1748 }}$ & $ \mathbf{{\it 524}} $ & $ M_{1-}^{\left( 3/2 \right)} $ \\
$\mathbf{ N \left( 1860 \right) \frac{5}{2}^{+} }$ & $ \mathbf{ \ast \ast }$  & $\mathbf{ F_{15} }$ & $\mathbf{ 3 }$ & $\mathbf{ 1/2 }$ & $\mathbf{ 5/2 }$ & $\mathbf{ + }$ & $ \mathbf{{\it 1834}} $ & $ \mathrm{{\it 135}} $ & $\mathbf{ E_{3-}^{\left( 1/2 \right)}} $, $\mathbf{ M_{3-}^{\left( 1/2 \right)} }$ \\
$\mathbf{ N \left( 1875 \right) \frac{3}{2}^{-} }$ & $ \mathbf{ \ast \ast \ast }$  & $\mathbf{ D_{13} }$ & $\mathbf{ 2 }$ & $\mathbf{ 1/2 }$ & $\mathbf{ 3/2 }$ & $\mathbf{ - }$ & $ \mathbf{ 1800 }$ {\bf to} $\mathbf{ 1950}$ & $ \mathbf{150} $ {\bf to} $\mathbf{250}$ & $\mathbf{ E_{2-}^{\left( 1/2 \right)}} $, $\mathbf{ M_{2-}^{\left( 1/2 \right)}} $ \\
$\mathbf{ N \left( 1880 \right) \frac{1}{2}^{+} }$ & $ \mathbf{ \ast \ast }$  & $\mathbf{ P_{11} }$ & $\mathbf{ 1 }$ & $\mathbf{ 1/2 }$ & $\mathbf{ 1/2 }$ & $\mathbf{ + }$ & $ \mathbf{{\it 1870}} $ & $ \mathbf{{\it 220}} $ & $\mathbf{ E_{1-}^{\left( 1/2 \right)} }$, $\mathbf{ M_{1-}^{\left( 1/2 \right)} }$ \\
$\mathbf{ N \left( 1895 \right) \frac{1}{2}^{-} }$ & $ \mathbf{ \ast \ast }$  & $\mathbf{ S_{11} }$ & $\mathbf{ 0 }$ & $\mathbf{ 1/2 }$ & $\mathbf{ 1/2 }$ & $\mathbf{ - }$ & $ \mathbf{{\it 1913}} $ & $ \mathbf{{\it 112}} $ & $\mathbf{ E_{0+}^{\left( 1/2 \right)}} $ \\
$\mathbf{ N \left( 1900 \right) \frac{3}{2}^{+}} $ & $\mathbf{ \ast \ast \ast }$  & $\mathbf{ P_{13} }$ & $\mathbf{ 1 }$ & $\mathbf{ 1/2 }$ & $\mathbf{ 3/2 }$ & $\mathbf{ + }$ & $ \mathbf{ 1920 }$ & $ \mathbf{ 130 }$ {\bf to} $ \mathbf{ 300 }$ & $\mathbf{ E_{1+}^{\left( 1/2 \right)} }$, $\mathbf{ M_{1+}^{\left( 1/2 \right)}} $ \\
$\mathbf{ \Delta \left( 1900 \right) \frac{1}{2}^{-} }$ & $\mathbf{ \ast \ast }$  & $\mathbf{ S_{31} }$ & $\mathbf{ 0 }$ & $\mathbf{ 3/2 }$ & $\mathbf{ 1/2 }$ & $\mathbf{ - }$ & $\mathbf{{\it 1853 }}$ & $\mathbf{{\it 240 }}$ & $\mathbf{ E_{0+}^{\left( 3/2 \right)}} $ \\
$\mathbf{ \Delta \left( 1905 \right) \frac{5}{2}^{+} }$ & $\mathbf{ \ast \ast \ast \ast }$  & $\mathbf{ F_{35} }$ & $\mathbf{ 3 }$ & $\mathbf{ 3/2 }$ & $\mathbf{ 5/2 }$ & $\mathbf{ + }$ & $ \mathbf{ 1820 }$ & $ \mathbf{ 280 }$ & $ \mathbf{E_{3-}^{\left( 3/2 \right)}, M_{3-}^{\left( 3/2 \right)}} $ \\
$\mathbf{ \Delta \left( 1910 \right) \frac{1}{2}^{+} }$ & $\mathbf{ \ast \ast \ast \ast }$  & $ \mathbf{ P_{31} }$ & $ \mathbf{ 1 }$ & $ \mathbf{ 3/2 }$ & $ \mathbf{ 1/2 }$ & $ \mathbf{ + }$ & $ \mathbf{ 1855 }$ & $ \mathbf{ 350 }$ & $\mathbf{ M_{1-}^{\left( 3/2 \right)}} $ \\
$\mathbf{ \Delta \left( 1920 \right) \frac{3}{2}^{+} }$ & $\mathbf{ \ast \ast \ast }$  & $\mathbf{ P_{33} }$ & $\mathbf{ 1 }$ & $\mathbf{ 3/2 }$ & $\mathbf{ 3/2 }$ & $\mathbf{ + }$ & $ \mathbf{ 1900 }$ & $ \mathbf{ 300 }$ & $\mathbf{ E_{1+}^{\left( 3/2 \right)}} $, $\mathbf{ M_{1+}^{\left( 3/2 \right)} }$ \\
$ \Delta \left( 1930 \right) \frac{5}{2}^{-} $ & $ \ast \ast \ast $  & $ D_{35} $ & $ 2 $ & $ 3/2 $ & $ 5/2 $ & $ - $ & $  1900 $ & $  270 $ & $ E_{2+}^{\left( 3/2 \right)}, M_{2+}^{\left( 3/2 \right)} $ \\
$\mathbf{ \Delta \left( 1940 \right) \frac{3}{2}^{-} }$ & $\mathbf{ \ast \ast }$  & $\mathbf{ D_{33} }$ & $\mathbf{ 2 }$ & $\mathbf{ 3/2 }$ & $\mathbf{ 3/2 }$ & $\mathbf{ - }$ & $\mathbf{{\it 1886 }}$ & $ \mathbf{{\it 222}} $ & $\mathbf{ E_{2-}^{\left( 3/2 \right)}, M_{2-}^{\left( 3/2 \right)} }$ \\
$ \mathbf{ \Delta \left( 1950 \right) \frac{7}{2}^{+} }$ & $ \mathbf{ \ast \ast \ast \ast} $  & $ \mathbf{ F_{37} }$ & $ \mathbf{ 3 }$ & $ \mathbf{ 3/2 }$ & $ \mathbf{ 7/2 }$ & $ \mathbf{ + }$ & $ \mathbf{ 1880}$ & $ \mathbf{ 240 }$ & $ \mathbf{ E_{3+}^{\left( 3/2 \right)}, M_{3+}^{\left( 3/2 \right)} }$ \\
$ \mathbf{ N \left( 1990 \right) \frac{7}{2}^{+}} $ & $ \mathbf{ \ast \ast } $  & $ \mathbf{ F_{17} } $ & $ \mathbf{ 3 } $ & $ \mathbf{ 1/2 } $ & $ \mathbf{ 7/2 } $ & $ \mathbf{ + } $ & $ \mathbf{{\it 1923}} $ & $ \mathbf{{\it 250}} $ & $ \mathbf{ E_{3+}^{\left( 1/2 \right)}, M_{3+}^{\left( 1/2 \right)} } $ \\
$ \mathbf{ N \left( 2000 \right) \frac{5}{2}^{+}} $ & $ \mathbf{ \ast \ast} $  & $ \mathbf{ F_{15}} $ & $ \mathbf{ 3 }$ & $ \mathbf{ 1/2 } $ & $ \mathbf{ 5/2 } $ & $ \mathbf{ + } $ & $ \mathbf{{\it 2030}} $ & $ \mathbf{{\it 380}} $ & $ \mathbf{ E_{3-}^{\left( 1/2 \right)}, M_{3-}^{\left( 1/2 \right)}} $ \\
$ \Delta \left( 2000 \right) \frac{5}{2}^{+} $ & $ \ast \ast $  & $ F_{35} $ & $ 3 $ & $ 3/2 $ & $ 5/2 $ & $ + $ & $\mathbf{{\it 1998 }}$ & $\mathbf{{\it 403 }}$ & $ E_{3-}^{\left( 3/2 \right)}, M_{3-}^{\left( 3/2 \right)} $ \\
$ N \left( 2040 \right) \frac{3}{2}^{+} $ & $ \ast $  & $ P_{13} $ & $ 1 $ & $ 1/2 $ & $ 3/2 $ & $ + $ & $ \mathbf{{\it 2058}} $ & $ \mathbf{{\it 214}} $ & $ E_{1+}^{\left( 1/2 \right)}, M_{1+}^{\left( 1/2 \right)} $ \\
$\mathbf{ N \left( 2060 \right) \frac{5}{2}^{-}} $ & $\mathbf{ \ast \ast} $  & $\mathbf{ D_{15}} $ & $\mathbf{ 2 }$ & $\mathbf{ 1/2 }$ & $\mathbf{ 5/2 }$ & $\mathbf{ - }$ & $ \mathbf{{\it 2088}} $ & $ \mathbf{{\it 377}} $ & $\mathbf{ E_{2-}^{\left( 1/2 \right)}, M_{2-}^{\left( 1/2 \right)}} $ \\
$ N \left( 2100 \right) \frac{1}{2}^{+} $ & $ \ast $  & $ P_{11} $ & $ 1 $ & $ 1/2 $ & $ 1/2 $ & $ + $ & $ \mathbf{{\it 2057}} $ & $ \mathbf{{\it 331}} $ & $ M_{1-}^{\left( 1/2 \right)} $ \\
$\mathbf{ N \left( 2120 \right) \frac{3}{2}^{-} }$ & $\mathbf{ \ast \ast }$  & $\mathbf{ D_{13} }$ & $\mathbf{ 2 }$ & $\mathbf{ 1/2 }$ & $\mathbf{ 3/2 }$ & $\mathbf{ - }$ & $ \mathbf{{\it 2099}} $ & $ \mathbf{{\it 322}} $ & $\mathbf{ E_{2-}^{\left( 1/2 \right)}} $, $\mathbf{ M_{2-}^{\left( 1/2 \right)} }$ \\
$ \Delta \left( 2150 \right) \frac{1}{2}^{-} $ & $ \ast $  & $ S_{31} $ & $ 0 $ & $ 3/2 $ & $ 1/2 $ & $ - $ & $\mathbf{{\it 2140 }}$ & $\mathbf{{\it 200 }}$ & $ E_{0+}^{\left( 3/2 \right)} $ \\
$\mathbf{ N \left( 2190 \right) \frac{7}{2}^{-}} $ & $\mathbf{ \ast \ast \ast \ast }$  & $\mathbf{ G_{17}} $ & $\mathbf{ 4} $ & $\mathbf{ 1/2} $ & $\mathbf{ 7/2} $ & $\mathbf{ -} $ & $ \mathbf{ 2075} $ & $ \mathbf{ 450} $ & $\mathbf{ E_{4-}^{\left( 1/2 \right)}, M_{4-}^{\left( 1/2 \right)}} $ \\
$ \Delta \left( 2200 \right) \frac{7}{2}^{-} $ & $ \ast $  & $ G_{37} $ & $ 4 $ & $ 3/2 $ & $ 7/2 $ & $ - $ & $\mathbf{{\it 2100 }}$ & $\mathbf{{\it 340 }}$ & $ E_{4-}^{\left( 3/2 \right)}, M_{4-}^{\left( 3/2 \right)} $ \\
$ \mathbf{ N \left( 2220 \right) \frac{9}{2}^{+}} $ & $ \mathbf{ \ast \ast \ast \ast} $  & $ \mathbf{ H_{19}} $ & $ \mathbf{ 5} $ & $ \mathbf{ 1/2 }$ & $ \mathbf{ 9/2 }$ & $ \mathbf{ + }$ & $ \mathbf{ 2170 }$ & $ \mathbf{ 480 }$ & $ \mathbf{ E_{5-}^{\left( 1/2 \right)}, M_{5-}^{\left( 1/2 \right)} }$ \\
$ \mathbf{ N \left( 2250 \right) \frac{9}{2}^{-}} $ & $ \mathbf{ \ast \ast \ast \ast } $  & $ \mathbf{ G_{19} } $ & $ \mathbf{ 4 } $ & $ \mathbf{ 1/2 } $ & $ \mathbf{ 9/2 } $ & $ \mathbf{ - } $ & $ \mathbf{ 2200 } $ & $ \mathbf{ 450 } $ & $ \mathbf{ E_{4+}^{\left( 1/2 \right)}, M_{4+}^{\left( 1/2 \right)} } $ \\
$ N \left( 2300 \right) \frac{1}{2}^{+} $ & $ \ast \ast $  & $  P_{11}  $ & $ 1 $ & $ 1/2  $ & $ 1/2 $ & $ \mathbf{ + } $ & $ {\it 2300 } $ & $ {\it 340 } $ & $ M_{1-}^{\left( 1/2 \right)} $ \\
$ \Delta \left( 2300 \right) \frac{9}{2}^{+} $ & $ \ast \ast $  & $ H_{39} $ & $ 5 $ & $ 3/2 $ & $ 9/2 $ & $ + $ & $\mathbf{{\it 2370 }}$ & $\mathbf{{\it 420 }}$ & $ E_{5-}^{\left( 3/2 \right)}, M_{5-}^{\left( 3/2 \right)} $ \\
$ \Delta \left( 2350 \right) \frac{5}{2}^{-} $ & $ \ast $  & $ D_{35} $ & $ 2 $ & $ 3/2 $ & $ 5/2 $ & $ - $ & $\mathbf{{\it 2400 }}$ & $\mathbf{{\it 400 }}$ & $ E_{2+}^{\left( 3/2 \right)}, M_{2+}^{\left( 3/2 \right)} $ \\
$ \Delta \left( 2390 \right) \frac{7}{2}^{+} $ & $ \ast $  & $ F_{37} $ & $ 3 $ & $ 3/2 $ & $ 7/2 $ & $ + $ & $\mathbf{{\it 2226 }}$ & $\mathbf{{\it 420 }}$ & $ E_{3+}^{\left( 3/2 \right)}, M_{3+}^{\left( 3/2 \right)} $ \\
$ \Delta \left( 2400 \right) \frac{9}{2}^{-} $ & $ \ast \ast $  & $ G_{39} $ & $ 4 $ & $ 3/2 $ & $ 9/2 $ & $ - $ & $\mathbf{{\it 2260 }}$ & $\mathbf{{\it 320 }}$ & $ E_{4+}^{\left( 3/2 \right)}, M_{4+}^{\left( 3/2 \right)} $ \\
$ \Delta \left( 2420 \right) \frac{11}{2}^{+} $ & $ \ast \ast \ast \ast $  & $ H_{3,11} $ & $ 5 $ & $ 3/2 $ & $ 11/2 $ & $ + $ & $  2330 $ & $ 550 $ & $ E_{5+}^{\left( 3/2 \right)}, M_{5+}^{\left( 3/2 \right)} $ \\
$ N \left( 2570 \right) \frac{5}{2}^{-} $ & $ \ast \ast $  & $ D_{15} $ & $ 2 $ & $ 1/2 $ & $ 5/2 $ & $ - $ & $ {\it 2570 }$ & $ {\it 250 } $ & $ E_{2-}^{\left( 1/2 \right)}, M_{2-}^{\left( 1/2 \right)} $ \\
$ N \left( 2600 \right) \frac{11}{2}^{-} $ & $ \ast \ast \ast $  & $ I_{1,11} $ & $ 6 $ & $ 1/2 $ & $ 11/2 $ & $ - $ & $  2600 $ (B.W.) & $  650 $ (B.W.) & $ E_{6-}^{\left( 1/2 \right)}, M_{6-}^{\left( 1/2 \right)} $ \\
$ N \left( 2700 \right) \frac{13}{2}^{+} $ & $ \ast \ast $  & $ K_{1,13} $ & $ 7 $ & $ 1/2 $ & $ 13/2 $ & $ + $ & ${\it 2612 }$ {\it(B.W.)} & ${\it 350 }$ {\it(B.W.)} & $ E_{7-}^{\left( 1/2 \right)}, M_{7-}^{\left( 1/2 \right)} $ \\
$ \Delta \left( 2750 \right) \frac{13}{2}^{-} $ & $ \ast \ast $  & $ I_{3,13} $ & $ 6 $ & $ 3/2 $ & $ 13/2 $ & $ - $ & ${\it 2794 }$ {\it(B.W.)} & ${\it 350 }$ {\it(B.W.)} & $ E_{6+}^{\left( 3/2 \right)}, M_{6+}^{\left( 3/2 \right)} $ \\
$ \Delta \left( 2950 \right) \frac{15}{2}^{+} $ & $ \ast \ast $  & $ K_{3,15} $ & $ 7 $ & $ 3/2 $ & $ 15/2 $ & $ + $ & ${\it 2990 }$ {\it(B.W.)} & ${\it 330 }$ {\it(B.W.)} & $ E_{7+}^{\left( 3/2 \right)}, M_{7+}^{\left( 3/2 \right)} $ \\
\hline
\hline
\end{tabular}
}
\end{scriptsize}
\caption{The $N$ ground state as well as $N$ and $\Delta$ resonances lowest in mass, possible to be examined by pion photoproduction. Name as well as the partial wave notation $ \left(\ell_{\pi}\right)_{2I, 2J} $ are given for every resonance (values $ \ell_{\pi} = 0,1,2,\ldots $ correspond to $ S,P,D,\ldots $). The last column lists multipoles to which the corresponding resonance can give a contribution. All data, especially pole masses $M^{\ast}_{\text{pole}} = \mathrm{Re} \left[ W_{\text{pole}} \right]$ and widths $ \Gamma_{\text{pole}} = - 2 \mathrm{Im} \left[W_{\text{pole}}\right]$ are taken from PDG 2016 \cite{Olive:2016xmw}. For some resonances however, only Breit-Wigner parameters are known (data are marked with "(B.W.)"). \newline Whenever an official average or parameter range is provided by the PDG, we list it. For italicized numbers, official PDG estimates are missing. Here, we evaluated an error weighted mean using all PDG-quoted data above the line. Resonances contained in the Bonn-Gatchina partial wave analysis (according to reference \cite{Anisovich:2011fc}) are written in boldface letters.
}
\label{tab:MultipoleResonanceAssignments}
\end{table*}

\end{document}